\documentclass[12pt,twoside]{article}

\usepackage{amsmath}
\usepackage{graphicx}
\usepackage{epsfig}
\usepackage[figuresright]{rotating}
\usepackage{pstricks}
\usepackage{cite}
\usepackage[dvips]{hyperref}                              
\hypersetup{ pdfborder=0 0 0, urlbordercolor=1 0 0 }      
\usepackage{relsize}                                      
\def\hfurl#1{http://hfag.phys.ntu.edu.tw/b2charm/#1} 
\newcommand{\mysection}[1]{\section{\boldmath #1}}
\newcommand{\mysubsection}[1]{\subsection[#1]{\boldmath #1}}
\newcommand{\mysubsubsection}[1]{\subsubsection[#1]{\boldmath #1}}
\newcommand{\mysubsubsubsection}[1]{\subsubsubsection{\boldmath #1}}

\newcommand{\lesssim}{\ensuremath{\raise-.5ex\hbox{$\buildrel<\over\sim$}\,}} 

\def\dof{{\rm dof}}

\newcommand\VCKM{{V}}
\newcommand\etacpf{{\eta_f}}
\newcommand\etacp{{\eta}}

\renewcommand\Im{{\rm Im}} 
\renewcommand\Re{{\rm Re}}

\newcommand\Abar{\kern 0.18em\overline{\kern -0.18em A}{}}
\newcommand\Af{A_f}
\newcommand\Abarf{\Abar_f}
\newcommand\Afbar{A_{\bar f}}
\newcommand\Abarfbar{\Abar_{\bar f}}
\newcommand\Acp{{\cal A}}
\newcommand\Adirnoncp{\ensuremath{\langle{\cal A}_{f\bar f}\rangle}\xspace}
\newcommand\mc{\multicolumn}

\newcommand {\cbf}{\ensuremath{{\cal B}}}

\newcommand {\vcb}{\ensuremath{|V_{cb}|}}
\newcommand {\vub}{\ensuremath{|V_{ub}|}}

\def\gsl {\ensuremath{\Gamma_{\rm SL}}}

\newcommand {\breco}{\ensuremath{B_{reco}}}

\def\Bp      {\ensuremath{B^{+}}}
\def\Bm      {\ensuremath{B^{-}}}
\def\Bz      {\ensuremath{B^{0}}}
\def\Bs      {\ensuremath{B_{s}}}

\newcommand{\BzbDplnu}    {\ensuremath{\bar{B}^{0}\to D^{+}\ell^{-}\nub}}
\newcommand{\BzbDstarlnu} {\ensuremath{\bar{B}^{0}\to D^{*+}\ell^{-}\nub}}

\newcommand {\rhoz} {\ensuremath{\rho^0}\hbox{ }}

\usepackage{relsize}

\def\beq{\begin{equation}}
\def\eeq#1{\label{#1}\end{equation}}
\def\eeqn{\end{equation}}

\def\beqa{\begin{eqnarray}}
\def\eeqa#1{\label{#1}\end{eqnarray}}
\def\eeqan{\end{eqnarray}}

\let\bar=\overbar

\def\etal{{\it et al.}}
\def\ie{{\it i.e.}}
\def\eg{{\it e.g.}}
\def\etc{{\it etc.}}
\def\cf{{\it cf.}}

\def\D{{\cal D}}

\def\Dslash{\ensuremath{\not{\hbox{\kern-4pt $D$}}}\xspace}
\def\dslash{\not{\hbox{\kern-2pt $\del$}}}

\def\BR{\mbox{\rm BR}}
\def\ee{e^+e^-}

\def\alphas{\alpha_s}
\def\msb{{\bar{\ssstyle M \kern -1pt S}}}

\RequirePackage{xspace}

\usepackage{relsize}
\def\babar{\mbox{\slshape B\kern-0.1em{\smaller A}\kern-0.1em
    B\kern-0.1em{\smaller A\kern-0.2em R}}\xspace}
\def\belle{\mbox{\normalfont Belle}\xspace}
\def\cdf{\mbox{\normalfont CDF}\xspace}
\newcommand{\dzero}{D\O\xspace}

\def\ee         {\ensuremath{e^-e^-}\xspace}

\def\ellp       {\ensuremath{\ell^+}\xspace}

\def\nub        {\ensuremath{\overline{\nu}}\xspace}

\def\nub        {\ensuremath{\overline{\nu}}\xspace}

\def\nul        {\ensuremath{\nu_\ell}\xspace}

\def\Z      {\ensuremath{Z^0}\xspace}

\def\ubar  {\ensuremath{\overline u}\xspace}

\def\dbar  {\ensuremath{\overline d}\xspace}
\def\ddbar {\ensuremath{d\overline d}\xspace}

\def\sbar  {\ensuremath{\overline s}\xspace}

\def\c  {\ensuremath{c}\xspace}

\def\b  {\ensuremath{b}\xspace}
\def\bbar  {\ensuremath{\overline b}\xspace}

\def\piz   {\ensuremath{\pi^0}\xspace}

\def\pip   {\ensuremath{\pi^+}\xspace}
\def\pim   {\ensuremath{\pi^-}\xspace}

\def\etapr {\ensuremath{\eta^{\prime}}\xspace}

\def\Kbar  {\kern 0.2em\overline{\kern -0.2em K}{}\xspace}

\def\Kpm   {\ensuremath{K^\pm}\xspace}
\def\Kmp   {\ensuremath{K^\mp}\xspace}
\def\Kp    {\ensuremath{K^+}\xspace}
\def\Km    {\ensuremath{K^-}\xspace}
\def\KS    {\ensuremath{K^0_{\scriptscriptstyle S}}\xspace} 
\def\KL    {\ensuremath{K^0_{\scriptscriptstyle L}}\xspace}

\def\Kstar   {\ensuremath{K^*}\xspace}

\def\Kstarpm   {\ensuremath{K^{*\pm}}\xspace}
\def\Kstarmp   {\ensuremath{K^{*\mp}}\xspace}
\def\Kz   {\ensuremath{K^0}\xspace}
\def\Kzb   {\ensuremath{\Kbar^0}\xspace}
\def\KzKzb {\ensuremath{K^0 \kern -0.16em \Kzb}\xspace}

\def\Dz    {\ensuremath{D^0}\xspace}
\def\Dbar  {\kern 0.2em\overline{\kern -0.2em D}{}\xspace}

\def\Dzb   {\ensuremath{\Dbar^0}\xspace}
\def\DzDzb {\ensuremath{D^0 {\kern -0.16em \Dzb}}\xspace}
\def\Dp    {\ensuremath{D^+}\xspace}

\def\Dstar   {\ensuremath{D^*}\xspace}
\def\Dstarb  {\ensuremath{\Dbar^*}\xspace}

\def\Dstarzb {\ensuremath{\Dbar^{*0}}\xspace}
\def\Dstarp  {\ensuremath{D^{*+}}}
\def\Dstarm  {\ensuremath{D^{*-}}}
\def\DorDstar   {\ensuremath{D^{(*)}}\xspace}
\def\DorDstarz  {\ensuremath{D^{(*)0}}\xspace}
\def\DorDstarzb {\ensuremath{\Dbar^{(*)0}}\xspace}

\def\Ds    {\ensuremath{D^+_s}\xspace}

\def\Bz    {\ensuremath{B^0}\xspace}
\def\B     {\ensuremath{B}\xspace}
\def\Bbar  {\kern 0.18em\overline{\kern -0.18em B}{}\xspace}
\def\Bb    {\ensuremath{\Bbar}\xspace}
\def\Bzb   {\ensuremath{\Bbar^0}\xspace}
\def\Bu    {\ensuremath{B^+}\xspace}

\def\Bpm   {\ensuremath{B^\pm}\xspace}
\def\Bmp   {\ensuremath{B^\mp}\xspace}
\def\Bs    {\ensuremath{B_s}\xspace}
\def\Bsb   {\ensuremath{\Bbar_s}\xspace}
\def\BB    {\ensuremath{B\Bbar}\xspace} 
\def\BzBzb {\ensuremath{B^0 {\kern -0.16em \Bzb}}\xspace}

\def\jpsi  {\ensuremath{{J\mskip -3mu/\mskip -2mu\psi\mskip 2mu}}\xspace}

\mathchardef\Upsilon="7107
\def\Y#1S{\ensuremath{\Upsilon{(#1S)}}\xspace}

\def\FourS {\Y4S}

\mathchardef\Deltares="7101
\mathchardef\Xi="7104
\mathchardef\Lambda="7103
\mathchardef\Sigma="7106
\mathchardef\Omega="710A
\def\Deltabar   {\kern 0.25em\overline{\kern -0.25em \Deltares}{}\xspace}
\def\Lbar {\kern 0.2em\overline{\kern -0.2em\Lambda\kern 0.05em}\kern-0.05em{}\xspace}
\def\Sigbar{\kern 0.2em\overline{\kern -0.2em \Sigma}{}\xspace}
\def\Xibar{\kern 0.2em\overline{\kern -0.2em \Xi}{}\xspace}
\def\Obar{\kern 0.2em\overline{\kern -0.2em \Omega}{}\xspace}
\def\Nbar{\kern 0.2em\overline{\kern -0.2em N}{}\xspace}
\def\Xb{\kern 0.2em\overline{\kern -0.2em X}{}}

\def\BR{{\ensuremath{\cal B}}}

\newcommand{\tev}{\ensuremath{\mathrm{Te\kern -0.1em V}}\xspace}
\newcommand{\gev}{\ensuremath{\mathrm{Ge\kern -0.1em V}}\xspace}
\newcommand{\mev}{\ensuremath{\mathrm{Me\kern -0.1em V}}\xspace}
\newcommand{\kev}{\ensuremath{\mathrm{ke\kern -0.1em V}}\xspace}
\newcommand{\ev}{\ensuremath{\mathrm{e\kern -0.1em V}}\xspace}
\newcommand{\gevc}{\ensuremath{{\mathrm{Ge\kern -0.1em V\!/}c}}\xspace}
\newcommand{\mevc}{\ensuremath{{\mathrm{Me\kern -0.1em V\!/}c}}\xspace}
\newcommand{\gevcc}{\ensuremath{{\mathrm{Ge\kern -0.1em V\!/}c^2}}\xspace}
\newcommand{\mevcc}{\ensuremath{{\mathrm{Me\kern -0.1em V\!/}c^2}}\xspace}

\def\invfb   {\ensuremath{\mbox{\,fb}^{-1}}\xspace}
\def\mus  {\ensuremath{\rm \,\mus}\xspace}

\def\ps   {\ensuremath{\rm \,ps}\xspace}

\def\mus        {\ensuremath{\,\mu{\rm s}}\xspace}    
\def\ps         {\ensuremath{{\rm \,ps}}\xspace}  
\def\gsim{{~\raise.15em\hbox{$>$}\kern-.85em
          \lower.35em\hbox{$\sim$}~}\xspace}
\def\lsim{{~\raise.15em\hbox{$<$}\kern-.85em
          \lower.35em\hbox{$\sim$}~}\xspace}

\def\CP                 {\ensuremath{C\!P}\xspace}
\def\CPT                {\ensuremath{C\!PT}\xspace}
\def\ra                 {\ensuremath{\to}\xspace}

\def\pep2{PEP-II}

\def\rhobar {\ensuremath{\overline{\rho}}\xspace}
\def\etabar {\ensuremath{\overline{\eta}}\xspace}

\def\Vub  {\ensuremath{|V_{ub}|}\xspace}
\def\Vcb  {\ensuremath{|V_{cb}|}\xspace}

\def\stwob{\ensuremath{\sin\! 2 \beta   }\xspace}

\def\deltamd{\ensuremath{{\rm \Delta}m_d}\xspace}

\xspace

\newcommand{\plb}       [1]  {{Phys.\ Lett.\ B~{\bf #1}}}   

\def\jetset74   {\mbox{\tt Jetset \hspace{-0.5em}7.\hspace{-0.2em}4}}

\newcommand{\aerr}[4]   {\mbox{${{#1}^{+ #2}_{- #3}\pm #4}$}}
\newcommand{\berr}[4]   {\mbox{${{#1}\pm #2^{+ #3}_{- #4}}$}}
\newcommand{\cerr}[3]   {\mbox{${{#1}^{+ #2}_{- #3}}$}}
\newcommand{\aerrsy}[5] {\mbox{${{#1}^{+ #2 + #4}_{- #3 - #5}}$}}

\newcommand{\err}[3]   {\mbox{${{#1}\pm{#2}\pm{#3}}$}}

\newcommand{\nodata}{$$}
\newcommand{\vs}{\mbox{$vs.$}}
\def\etapr{{\eta^{\prime}}}

\def\sgline{\noalign{\vskip 0.10truecm\hrule\vskip 0.10truecm}}
\def\sglinespt{\noalign{\vskip 0.05truecm\hrule}}
\def\sglinespb{\noalign{\hrule\vskip 0.05truecm}}
\newcommand{\prl}      [1]  {{Phys.\ Rev.\ Lett.\ {\bf #1}}}
\newcommand{\prd}      [1]  {{Phys.\ Rev.\ D~{\bf #1}}}

\newcommand{\kz}    {\mbox{$K^0$}}

\newcommand{\RPP}{}

\renewcommand{\mysection}[1]{\section[#1]{#1}} 

\begin{document}

\setcounter{page}{1}

\title{
Averages of $b$-hadron and $c$-hadron Properties \\
at the End of 2007
}
\author{Heavy Flavor Averaging Group (HFAG)\footnote{%
The HFAG members involved in producing the averages for the end of 2007
update are:
   E.~Barberio,       
   R.~Bernhard,       
   S.~Blyth,          
   O.~Buchmueller,    
   G.~Cavoto,         
   P.~Chang,          
   F.~Di~Lodovico,    
   H.~Flaecher,       
   T.~Gershon,        
   L.~Gibbons,        
   R.~Godang,         
   B.~Golob,          
   G.~Gomez-Ceballos, 
   R.~Harr,           
   R.~Kowalewski,     
   H.~Lacker,         
   C.-J.~Lin,         
   D.~Lopes-Pegna,    
   V.~L\"{u}th,       
   D.~Pedrini,        
   B.~Petersen,       
   M.~Purohit,        
   O.~Schneider,      
   C.~Schwanda,       
   A.~J.~Schwartz,    
   J.~Smith,          
   A.~Snyder,         
   D.~Tonelli,        
   S.~Tosi,           
   K.~Trabelsi,       
   P.~Urquijo,        
   R.~Van~Kooten,     
   C.~Voena,          
   and C.~Weiser.     
  } 
 }

\date{8 August 2008}
\maketitle
\thispagestyle{empty}
\begin{abstract}
This article reports world averages for measurements of $b$-hadron and $c$-hadron 
properties obtained by the Heavy Flavor Averaging Group (HFAG) using the results
available at the end of 2007. For the averaging, common input parameters used in the 
various analyses are adjusted (rescaled) to common values, and known correlations 
are taken into account. The averages include branching fractions, lifetimes, 
neutral meson mixing parameters, \CP~violation parameters, and parameters 
of semileptonic decays.
\end{abstract}

\newpage
\tableofcontents
\newpage


\mysection{Introduction}
\label{sec:intro}

Flavor dynamics is an important element in understanding the nature of
particle physics.  The accurate knowledge of properties of heavy flavor
hadrons, especially $b$ hadrons, plays an essential role for
determination of the Cabibbo-Kobayashi-Maskawa (CKM)
matrix~\cite{ref_ckm}. Since asymmetric-energy $e^+e^-$ $B$ factories
started their operation, the size of available $B$ meson samples has
dramatically increased and the accuracies of measurements have been
improved. Tevatron experiments have also provided important results on
$B$ and $D$ decays with increased Run~II data samples, including
confirmation of $D^0$-$\overline{D}{}^{\,0}$ mixing in $D^0\ra K^+\pi^-$
decays.
 
The Heavy Flavor Averaging Group (HFAG) was formed in 2002 to continue
the activities of the LEP Heavy Flavor Steering group~\cite{LEPHFS}. This
group was responsible for averages of measurements of $b$-flavor related quantities. 
HFAG currently consists of six subgroups:
\begin{itemize}
\item the ``Lifetime and Mixing'' subgroup provides averages for $b$-hadron
  lifetimes, $b$-hadron fractions in $\Upsilon(4S)$ decay and high
  energy collisions, and various parameters in $B^0$ and $B_s^0$
  oscillations (mixing);

\item the ``$\CP(t)$ and Unitarity Triangle Angles'' subgroup provides
  averages for time-dependent $\CP$ asymmetry parameters and angles of
  the $B$ unitarity triangle;

\item the ``Semileptonic $B$ Decays'' subgroup provides averages
  for inclusive and exclusive $B$-decay branching fractions, and
  estimates of the CKM matrix elements $|V_{cb}|$ and $|V_{ub}|$;

\item the ``$B$ to Charm Decays'' subgroup provides averages of branching
  fractions for $B$ decays to final states involving open charm mesons
  or charmonium.

\item the ``Rare Decays'' subgroup provides averages of branching fractions
  and their asymmetries between $B$ and $\bar B$ for charmless mesonic, 
  radiative, leptonic, and baryonic $B$ decays;

\item the ``Charm Physics'' subgroup provides averages of branching fractions
for charm decays and averages of $D^0$-$\overline{D}{}^{\,0}$ mixing and 
direct and indirect $\CP$ violation parameters. 
\end{itemize}
The first and third subgroups continue the activities of the LEP working groups
with some reorganization (merging four groups into two groups). The second and
latter three groups were newly formed to provide averages for
results that became available from the $B$ factory experiments
(and now from the Fermilab Tevatron experiments).
All subgroups consist of representatives and contact persons from
\babar, \belle, CDF, CLEO, \dzero, and LEP. 

This article is an update of the ``End of 2006'' HFAG preprint~\cite{hfag_hepex_endof2006}, 
and here we report world averages using the results available at the end of 2007. 
In general we use all publicly available results, including preliminary results
presented at conferences; however, we do not use preliminary results that
have remained unpublished for an extended period time or for which no publication 
is planned. Close contacts have been established between representatives from
the experiments and members of different subgroups in charge of the
averages to ensure that the data are prepared in a form suitable
for combinations.  

In the case of obtaining a world average for which $\chi^2/\dof > 1$,
where $\dof$ is the number of degrees of freedom in the average
calculation, we do not scale the resulting error (as is presently 
done by the Particle Data Group~\cite{PDG_2007}). Rather, 
we examine the systematics of each measurement to better understand them. 
Unless we find possible systematic discrepancies between the measurements, 
we do not apply any additional correction to the calculated error. 
We provide the confidence
level of the fit as an indicator for the consistency of the measurements
included in the average. We include a warning message in case some
special treatment was necessary to calculate an average, or if an
approximation used in the average calculation may not be good enough 
(\eg, assuming Gaussian errors when the likelihood function indicates 
non-Gaussian behavior).

Section~\ref{sec:method} describes the methodology for calculating
averages for various quantities used by HFAG. In the averaging, common
input parameters used in the various analyses are adjusted (rescaled) to
common values, and, where possible, known correlations are taken into account. 
Sections~\ref{sec:life_mix}--\ref{sec:charm_physics} present world average 
values from each of the subgroups mentioned above.

The complete listing of averages and plots described in this article
are also available on the HFAG web page:
 
 {\tt http://www.slac.stanford.edu/xorg/hfag } and 

 {\tt http://belle.kek.jp/mirror/hfag } (KEK mirror site).

\section{Methodology } \label{sec:method} 

The general averaging problem that HFAG faces is to combine 
information provided by different measurements of the same parameter
to obtain our best estimate of the parameter's value and
uncertainty. The methodology described here focuses on the problems of
combining measurements performed with different systematic assumptions
and with potentially-correlated systematic uncertainties. Our methodology
relies on the close involvement of the people performing the
measurements in the averaging process.

Consider two hypothetical measurements of a parameter $x$, which might
be summarized as
\begin{align*}
x &= x_1 \pm \delta x_1 \pm \Delta x_{1,1} \pm \Delta x_{2,1} \ldots \\
x &= x_2 \pm \delta x_2 \pm \Delta x_{1,2} \pm \Delta x_{2,2} \ldots
\; ,
\end{align*}
where the $\delta x_k$ are statistical uncertainties, and
the $\Delta x_{i,k}$ are contributions to the systematic
uncertainty. One popular approach is to combine statistical and
systematic uncertainties in quadrature
\begin{align*}
x &= x_1 \pm \left(\delta x_1 \oplus \Delta x_{1,1} \oplus \Delta
x_{2,1} \oplus \ldots\right) \\
x &= x_2 \pm \left(\delta x_2 \oplus \Delta x_{1,2} \oplus \Delta
x_{2,2} \oplus \ldots\right)
\end{align*}
and then perform a weighted average of $x_1$ and $x_2$, using their
combined uncertainties, as if they were independent. This approach
suffers from two potential problems that we attempt to address. First,
the values of the $x_k$ may have been obtained using different
systematic assumptions. For example, different values of the \Bz
lifetime may have been assumed in separate measurements of the
oscillation frequency $\deltamd$. The second potential problem is that
some contributions of the systematic uncertainty may be correlated
between experiments. For example, separate measurements of $\deltamd$
may both depend on an assumed Monte-Carlo branching fraction used to
model a common background.

The problems mentioned above are related since, ideally, any quantity $y_i$
that $x_k$ depends on has a corresponding contribution $\Delta x_{i,k}$ to the
systematic error which reflects the uncertainty $\Delta y_i$ on $y_i$
itself. We assume that this is the case and use the values of $y_i$ and
$\Delta y_i$ assumed by each measurement explicitly in our
averaging (we refer to these values as $y_{i,k}$ and $\Delta y_{i,k}$
below). Furthermore, since we do not lump all the systematics
together,
we require that each measurement used in an average have a consistent
definition of the various contributions to the systematic uncertainty.
Different analyses often use different decompositions of their systematic
uncertainties, so achieving consistent definitions for any potentially
correlated contributions requires close coordination between HFAG and
the experiments. In some cases, a group of
systematic uncertainties must be lumped to obtain a coarser
description that is consistent between measurements. Systematic uncertainties
that are uncorrelated with any other sources of uncertainty appearing
in an average are lumped with the statistical error, so that the only
systematic uncertainties treated explicitly are those that are
correlated with at least one other measurement via a consistently-defined
external parameter $y_i$. When asymmetric statistical or systematic
uncertainties are quoted, we symmetrize them since our combination
method implicitly assumes parabolic likelihoods for each measurement.

\begin{figure}
\begin{center}
\includegraphics[width=6.0in]{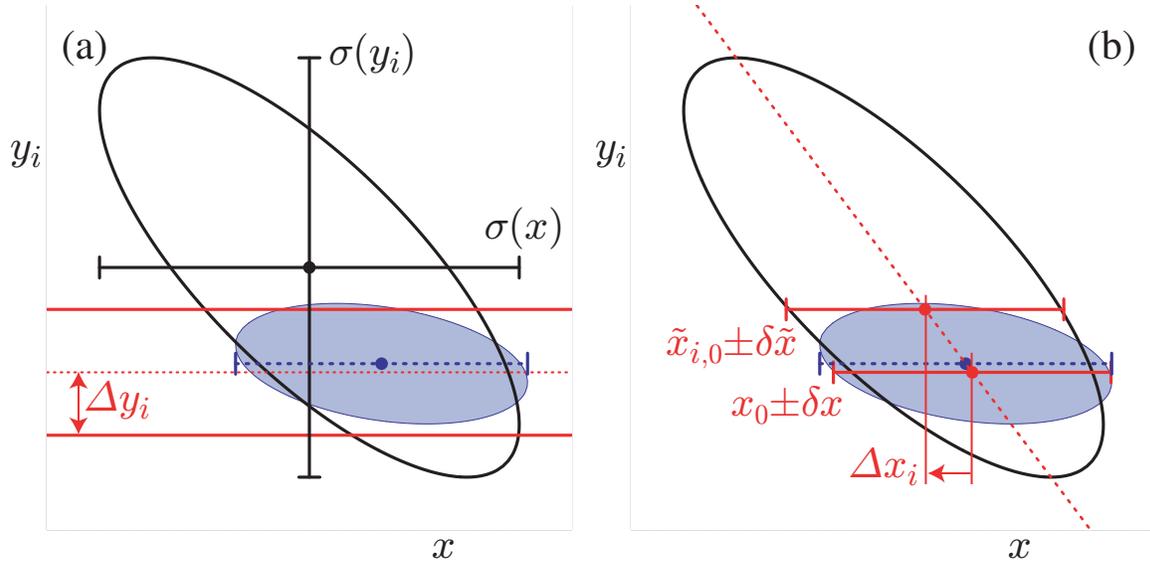}
\end{center}
\caption{The left-hand plot (a) compares the 68\% confidence-level
  contours of a
  hypothetical measurement's unconstrained (large ellipse) and
  constrained (filled ellipse) likelihoods, using the Gaussian
  constraint on $y_i$ represented by the horizontal band. The solid
  error bars represent the statistical uncertainties $\sigma(x)$ and
  $\sigma(y_i)$ of the unconstrained likelihood. The dashed
  error bar shows the statistical error on $x$ from a
  constrained simultaneous fit to $x$ and $y_i$. The right-hand plot
  (b) illustrates the method described in the text of performing fits
  to $x$ with $y_i$ fixed at different values. The dashed
  diagonal line between these fit results has the slope
  $\rho(x,y_i)\sigma(y_i)/\sigma(x)$ in the limit of a parabolic
  unconstrained likelihood. The result of the constrained simultaneous
  fit from (a) is shown as a dashed error bar on $x$.}
\label{fig:singlefit}
\end{figure}

The fact that a measurement of $x$ is sensitive to the value of $y_i$
indicates that, in principle, the data used to measure $x$ could
equally-well be used for a simultaneous measurement of $x$ and $y_i$, as
illustrated by the large contour in Fig.~\ref{fig:singlefit}(a) for a hypothetical
measurement. However, we often have an external constraint $\Delta
y_i$ on the value of $y_i$ (represented by the horizontal band in
Fig.~\ref{fig:singlefit}(a)) that is more precise than the constraint
$\sigma(y_i)$ from
our data alone. Ideally, in such cases we would perform a simultaneous
fit to $x$ and $y_i$, including the external constraint, obtaining the
filled $(x,y)$ contour and corresponding dashed one-dimensional estimate of
$x$ shown in Fig.~\ref{fig:singlefit}(a). Throughout, we assume that
the external constraint $\Delta y_i$ on $y_i$ is Gaussian.

In practice, the added technical complexity of a constrained fit with
extra free parameters is not justified by the small increase in
sensitivity, as long as the external constraints $\Delta y_i$ are
sufficiently precise when compared with the sensitivities $\sigma(y_i)$
to each $y_i$ of the data alone. Instead, the usual procedure adopted
by the experiments is to perform a baseline fit with all $y_i$ fixed
to nominal values $y_{i,0}$, obtaining $x = x_0 \pm \delta
x$. This baseline fit neglects the uncertainty due to $\Delta y_i$, but
this error can be mostly recovered by repeating the fit separately for
each external parameter $y_i$ with its value fixed at $y_i = y_{i,0} +
\Delta y_i$ to obtain $x = \tilde{x}_{i,0} \pm \delta\tilde{x}$, as
illustrated in Fig.~\ref{fig:singlefit}(b). The absolute shift,
$|\tilde{x}_{i,0} - x_0|$, in the central value of $x$ is what the
experiments usually quote as their systematic uncertainty $\Delta x_i$
on $x$ due to the unknown value of $y_i$. Our procedure requires that
we know not only the magnitude of this shift but also its sign. In the
limit that the unconstrained data is represented by a parabolic
likelihood, the signed shift is given by
\begin{equation}
\Delta x_i = \rho(x,y_i)\frac{\sigma(x)}{\sigma(y_i)}\,\Delta y_i \;,
\end{equation}
where $\sigma(x)$ and $\rho(x,y_i)$ are the statistical uncertainty on
$x$ and the correlation between $x$ and
$y_i$ in the unconstrained data.
While our procedure is not
equivalent to the constrained fit with extra parameters, it yields (in
the limit of a parabolic unconstrained likelihood) a central value
$x_0$ that agrees 
to ${\cal O}(\Delta y_i/\sigma(y_i))^2$ and an uncertainty $\delta x
\oplus \Delta x_i$ that agrees to ${\cal O}(\Delta y_i/\sigma(y_i))^4$.

In order to combine two or more measurements that share systematics
due to the same external parameters $y_i$, we would ideally perform a
constrained simultaneous fit of all data samples to obtain values of
$x$ and each $y_i$, being careful to only apply the constraint on each
$y_i$ once. This is not practical since we generally do not have
sufficient information to reconstruct the unconstrained likelihoods
corresponding to each measurement. Instead, we perform the two-step
approximate procedure described below.

\begin{figure}
\begin{center}
\includegraphics[width=6.0in]{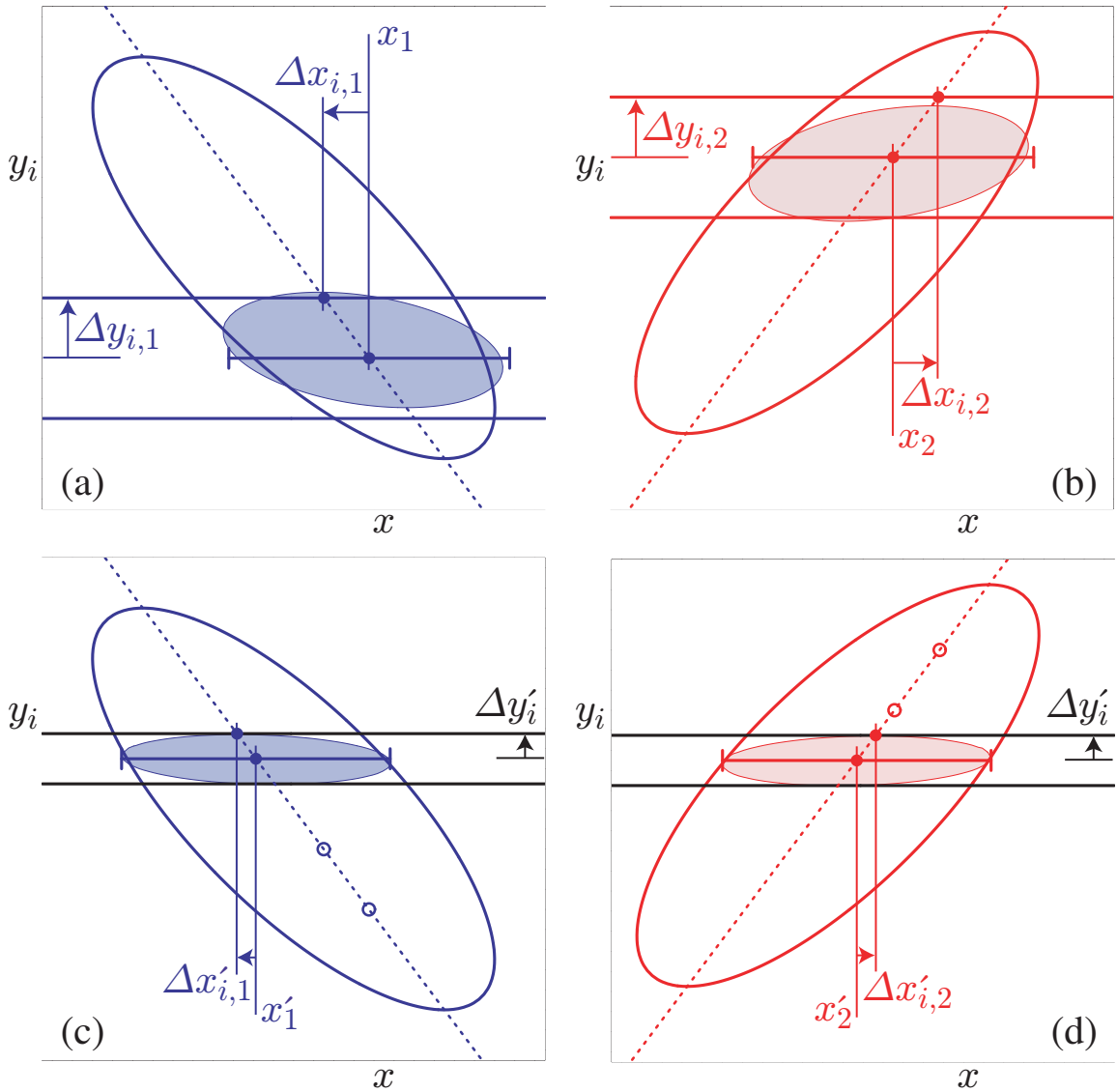}
\end{center}
\caption{The upper plots (a) and (b) show examples of two individual
  measurements to be combined. The large ellipses represent their
  unconstrained likelihoods, and the filled ellipses represent their
  constrained likelihoods. Horizontal bands indicate the different
  assumptions about the value and uncertainty of $y_i$ used by each
  measurement. The error bars show the results of the approximate
  method described in the text for obtaining $x$ by performing fits
  with $y_i$ fixed to different values. The lower plots (c) and (d)
  illustrate the adjustments to accommodate updated and consistent
  knowledge of $y_i$ as described in the text. Open circles mark the
  central values of the unadjusted fits to $x$ with $y$ fixed; these
  determine the dashed line used to obtain the adjusted values. }
\label{fig:multifit}
\end{figure}

Figs.~\ref{fig:multifit}(a,b) illustrate two
statistically-independent measurements, $x_1 \pm (\delta x_1 \oplus
\Delta x_{i,1})$ and $x_2\pm(\delta x_i\oplus \Delta x_{i,2})$, of the same
hypothetical quantity $x$ (for simplicity, we only show the
contribution of a single correlated systematic due to an external
parameter $y_i$). As our knowledge of the external parameters $y_i$
evolves, it is natural that the different measurements of $x$ will
assume different nominal values and ranges for each $y_i$. The first
step of our procedure is to adjust the values of each measurement to
reflect the current best knowledge of the values $y_i'$ and ranges
$\Delta y_i'$ of the external parameters $y_i$, as illustrated in
Figs.~\ref{fig:multifit}(c,b). We adjust the
central values $x_k$ and correlated systematic uncertainties $\Delta
x_{i,k}$ linearly for each measurement (indexed by $k$) and each
external parameter (indexed by $i$):
\begin{align}
x_k' &= x_k + \sum_i\,\frac{\Delta x_{i,k}}{\Delta y_{i,k}}
\left(y_i'-y_{i,k}\right)\\
\Delta x_{i,k}'&= \Delta x_{i,k}\cdot \frac{\Delta y_i'}{\Delta
  y_{i,k}} \; .
\end{align}
This procedure is exact in the limit that the unconstrained
likelihoods of each measurement is parabolic.

\begin{figure}
\begin{center}
\includegraphics[width=3.5in]{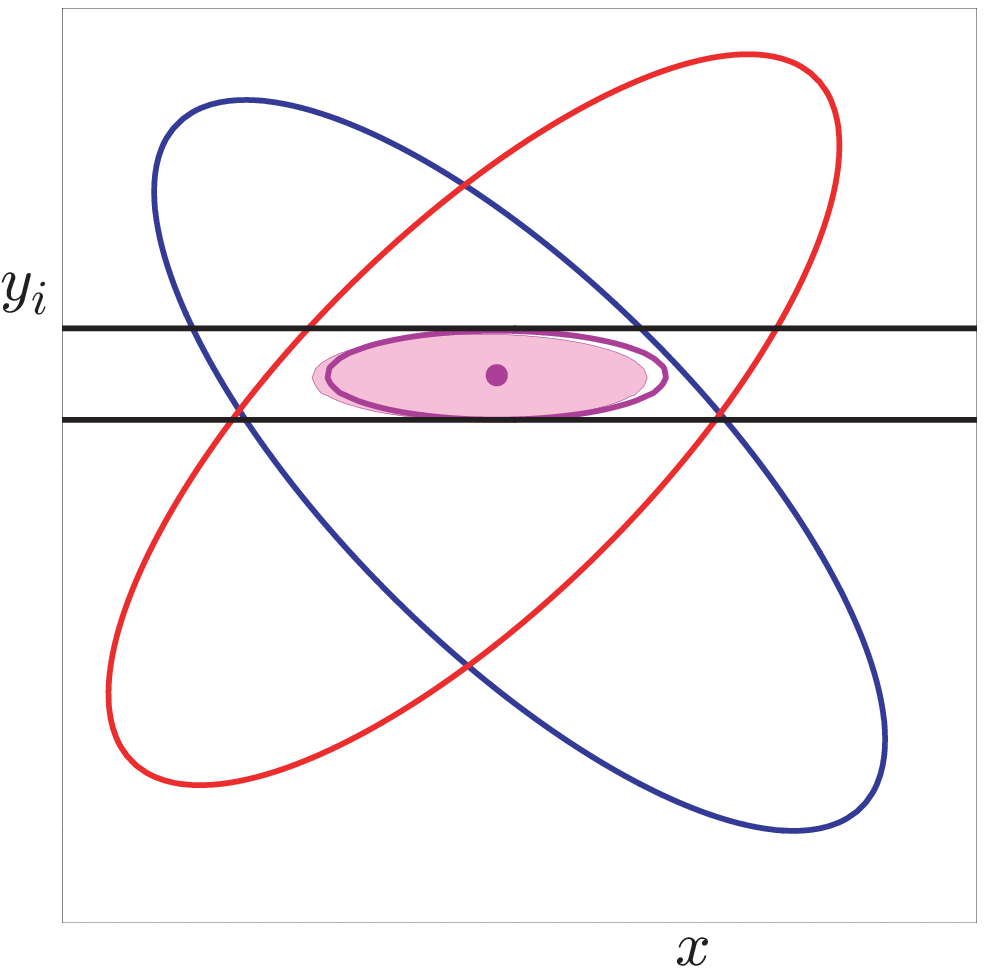}
\end{center}
\caption{An illustration of the combination of two hypothetical
  measurements of $x$ using the method described in the text. The
  ellipses represent the unconstrained likelihoods of each measurement,
  and the horizontal band represents the latest knowledge about $y_i$ 
  that is used to adjust the individual measurements. The filled small
  ellipse shows the result of the exact method using 
  ${\cal L}_{\text{comb}}$, and the hollow small ellipse and dot show 
  the result of the approximate method using $\chi^2_{\text{comb}}$.}
\label{fig:fit12}
\end{figure}

The second step of our procedure is to combine the adjusted
measurements, $x_k'\pm (\delta x_k\oplus \Delta x_{k,1}'\oplus \Delta
x_{k,2}'\oplus\ldots)$ using the chi-square 
\begin{equation}
\chi^2_{\text{comb}}(x,y_1,y_2,\ldots) \equiv \sum_k\,
\frac{1}{\delta x_k^2}\left[
x_k' - \left(x + \sum_i\,(y_i-y_i')\frac{\Delta x_{i,k}'}{\Delta y_i'}\right)
\right]^2 + \sum_i\,
\left(\frac{y_i - y_i'}{\Delta y_i'}\right)^2 \; ,
\end{equation}
and then minimize this $\chi^2$ to obtain the best values of $x$ and
$y_i$ and their uncertainties, as illustrated in
Fig.~\ref{fig:fit12}. Although this method determines new values for
the $y_i$, we do not report them since the $\Delta x_{i,k}$ reported
by each experiment are generally not intended for this purpose (for
example, they may represent a conservative upper limit rather than a
true reflection of a 68\% confidence level).

For comparison, the exact method we would
perform if we had the unconstrained likelihoods ${\cal L}_k(x,y_1,y_2,\ldots)$
available for each
measurement is to minimize the simultaneous constrained likelihood
\begin{equation}
{\cal L}_{\text{comb}}(x,y_1,y_2,\ldots) \equiv \prod_k\,{\cal
  L}_k(x,y_1,y_2,\ldots)\,\prod_{i}\,{\cal 
  L}_i(y_i) \; ,
\end{equation}
with an independent Gaussian external constraint on each $y_i$
\begin{equation}
{\cal L}_i(y_i) \equiv \exp\left[-\frac{1}{2}\,\left(\frac{y_i-y_i'}{\Delta
 y_i'}\right)^2\right] \; .
\end{equation}
The results of this exact method are illustrated by the filled ellipses
in Figs.~\ref{fig:fit12}(a,b) and agree with our method in the limit that
each ${\cal L}_k$ is parabolic and that each $\Delta
y_i' \ll \sigma(y_i)$. In the case of a non-parabolic unconstrained
likelihood, experiments would have to provide a description of ${\cal
  L}_k$ itself to allow an improved combination. In the case of
$\sigma(y_i)\simeq \Delta y_i'$, experiments are advised to perform a
simultaneous measurement of both $x$ and $y$ so that their data will
improve the world knowledge about $y$. 

 The algorithm described above is used as a default in the averages
reported in the following sections.  For some cases, somewhat simplified
or more complex algorithms are used and noted in the corresponding 
sections. Some examples for extensions of the standard method for extracting
averages are given here. These include the case where measurement errors
depend on the measured value, i.e. are relative errors, unknown
correlation coefficients and the breakdown of error sources.

For measurements with Gaussian errors, the usual estimator for the
average of a set of measurements is obtained by minimizing the following
$\chi^2$:
\begin{equation}
\chi^2(t) = \sum_i^N \frac{\left(y_i-t\right)^2}{\sigma^2_i} ,
\label{eq:chi2t}
\end{equation}
where $y_i$ is the measured value for input $i$ and $\sigma_i^2$ is the
variance of the distribution from which $y_i$ was drawn.  The value $\hat{t}$
of $t$ at minimum $\chi^2$ is our estimator for the average.  (This
discussion is given for independent measurements for the sake of
simplicity; the generalization to correlated measurements is
straightforward, and has been used when averaging results.) 
The true $\sigma_i$ are unknown but typically the error as assigned by the
experiment $\sigma_i^{\mathrm{raw}}$ is used as an estimator for it.
Caution is advised,
however, in the case where $\sigma_i^{\mathrm{raw}}$
depends on the value measured for $y_i$. Examples of this include
an uncertainty in any multiplicative factor (like
an acceptance) that enters the determination of $y_i$, i.e. the $\sqrt{N}$
dependence of Poisson statistics, where $y_i \propto N$
and $\sigma_i \propto \sqrt{N}$.
Failing to account for this type of
dependence when averaging leads to a biased average.
Biases in the average can be avoided (or at least reduced)
by minimizing the following
$\chi^2$:
\begin{equation}
\chi^2(t) = \sum_i^N \frac{\left(y_i-t\right)^2}{\sigma^2_i(\hat{t})} .
\label{eq:chi2that}
\end{equation}
In the above $\sigma_i(\hat{t})$ is the uncertainty
assigned to input $i$ that includes the assumed dependence of the
stated error on the value measured.  As an example, consider 
a pure acceptance error, for which
$\sigma_i(\hat{t}) = (\hat{t} / y_i)\times\sigma_i^{\mathrm{raw}}$ .
It is easily verified that solving Eq.~\ref{eq:chi2that} 
leads to the correct behavior, namely
$$ 
\hat{t} = \frac{\sum_i^N y_i^3/(\sigma_i^{\mathrm{raw}})^2}{\sum_i^N y_i^2/(\sigma_i^{\mathrm{raw}})^2},
$$
i.e. weighting by the inverse square of the 
fractional uncertainty, $\sigma_i^{\mathrm{raw}}/y_i$.
It is sometimes difficult to assess the dependence of $\sigma_i^{\mathrm{raw}}$ on
$\hat{t}$ from the errors quoted by experiments.  


Another issue that needs careful treatment is the question of correlation
among different measurements, e.g. due to using the same theory for
calculating acceptances.  A common practice is to set the correlation
coefficient to unity to indicate full correlation.  However, this is
not a ``conservative'' thing to do, and can in fact lead to a significantly
underestimated uncertainty on the average.  In the absence of
better information, the most conservative choice of correlation coefficient
between two measurements $i$ and $j$
is the one that maximizes the uncertainty on $\hat{t}$
due to that pair of measurements:
\begin{equation}
\sigma_{\hat{t}(i,j)}^2 = \frac{\sigma_i^2\,\sigma_j^2\,(1-\rho_{ij}^2)}
   {\sigma_i^2 + \sigma_j^2 - 2\,\rho_{ij}\,\sigma_i\,\sigma_j} ,
\label{eq:correlij}
\end{equation}
namely
\begin{equation}
\rho_{ij} = \mathrm{min}\left(\frac{\sigma_i}{\sigma_j},\frac{\sigma_j}{\sigma_i}\right) ,
\label{eq:correlrho}
\end{equation}
which corresponds to setting $\sigma_{\hat{t}(i,j)}^2=\mathrm{min}(\sigma_i^2,\sigma_j^2)$.
Setting $\rho_{ij}=1$ when $\sigma_i\ne\sigma_j$ can lead to a significant
underestimate of the uncertainty on $\hat{t}$, as can be seen
from Eq.~\ref{eq:correlij}.

Finally, we carefully consider the various sources of error
contributing to the overall uncertainty of an average.
The overall covariance matrix is constructed from a number of
individual sources, e.g.
$\mathbf{V} = \mathbf{V_{stat}+V_{sys}+V_{th}}$.
The variance on the average $\hat{t}$ can be written
\begin{eqnarray}
\sigma^2_{\hat{t}} 
 &=& 
\frac{ \sum_{i,j}\left(\mathbf{V^{-1}}\, 
\mathbf{[V_{stat}+V_{sys}+V_{th}]}\, \mathbf{V^{-1}}\right)_{ij}}
{\left(\sum_{i,j} V^{-1}_{ij}\right)^2}
= \sigma^2_{stat} + \sigma^2_{sys} + \sigma^2_{th} .
\end{eqnarray}
Written in this form, one can readily determine the 
contribution of each source of uncertainty to the overall uncertainty
on the average.  This breakdown of the uncertainties is used 
in the following sections.

Following the prescription described above, the central values and
errors are rescaled to a common set of input parameters in the averaging
procedures according to the dependency on any of these input parameters.
We try to use the most up-to-date values for these common inputs and 
the same values among the HFAG subgroups. For the parameters whose
averages are produced by HFAG, we use the values in the current 
update cycle.  For other external parameters, we use the most
recent PDG values\cite{PDG_2007}. The parameters and values 
used are listed in each subgroup section.

\clearpage
%
%
%
%

%

%
%
%
%

%

\renewcommand{\floatpagefraction}{0.8}
\renewcommand{\topfraction}{0.9}

\newcommand{\comment}[1]{}

\newcommand{\auth}[1]{#1,}
\newcommand{\coll}[1]{#1 Collaboration,}
\newcommand{\authcoll}[2]{#1 \etal\ (#2 Collaboration),}
\newcommand{\authgrp}[2]{#1 \etal\ (#2),}
\newcommand{\titl}[1]{``#1'',} 
\newcommand{\J}[4]{{#1} {\bf #2}, #3 (#4)}
\newcommand{\subJ}[1]{submitted to #1}
\newcommand{\PRL}[3]{\J{Phys.\ Rev.\ Lett.}{#1}{#2}{#3}}
\newcommand{\subPRL}{\subJ{Phys.\ Rev.\ Lett.}}
\newcommand{\PRD}[3]{\J{Phys.\ Rev.\ D}{#1}{#2}{#3}}
\newcommand{\subPRD}{\subJ{Phys.\ Rev.\ D}}
\newcommand{\PREP}[3]{\J{Phys.\ Reports}{#1}{#2}{#3}}
\newcommand{\ZPC}[3]{\J{Z.\ Phys.\ C}{#1}{#2}{#3}}
\newcommand{\PLB}[3]{\J{Phys.\ Lett.\ B}{#1}{#2}{#3}}
\newcommand{\subPLB}{\subJ{Phys.\ Lett.\ B}}
\newcommand{\EPJC}[3]{\J{Eur.\ Phys.\ J.\ C}{#1}{#2}{#3}}
\newcommand{\NPB}[3]{\J{Nucl.\ Phys.\ B}{#1}{#2}{#3}}
\newcommand{\subNPB}{\subJ{Nucl.\ Phys.\ B}}
\newcommand{\NIMA}[3]{\J{Nucl.\ Instrum.\ Methods A}{#1}{#2}{#3}}
\newcommand{\subNIMA}{\subJ{Nucl.\ Instrum.\ Methods A}}
\newcommand{\JHEP}[3]{\J{J.\ of High Energy Physics }{#1}{#2}{#3}}
\newcommand{\JPG}[3]{\J{J.\ of Physics G}{#1}{#2}{#3}}
\newcommand{\ARNS}[3]{\J{Ann.\ Rev.\ Nucl.\ Sci.}{#1}{#2}{#3}}
\newcommand{\newref}{\\}

\newcommand{\particle}[1]{\ensuremath{#1}\xspace}
\renewcommand{\ee}{\particle{e^+e^-}}
\newcommand{\Ups}{\particle{\Upsilon(4S)}}
\newcommand{\Upsfive}{\particle{\Upsilon(5S)}}
\renewcommand{\b}{\particle{b}}
\renewcommand{\B}{\particle{B}}
\newcommand{\Bd}{\particle{B^0}}
\renewcommand{\Bs}{\particle{B^0_s}}
\renewcommand{\Bu}{\particle{B^+}}
\newcommand{\Bc}{\particle{B^+_c}}
\newcommand{\Bdbar}{\particle{\bar{B}^0}}
\newcommand{\Bsbar}{\particle{\bar{B}^0_s}}
\newcommand{\Lb}{\particle{\Lambda_b^0}}
\newcommand{\Xib}{\particle{\Xi_b}}
\newcommand{\Lc}{\particle{\Lambda_c^+}}

\newcommand{\fBs}{\ensuremath{f_{\particle{s}}}\xspace}
\newcommand{\fBd}{\ensuremath{f_{\particle{d}}}\xspace}
\newcommand{\fBu}{\ensuremath{f_{\particle{u}}}\xspace}
\newcommand{\fbb}{\ensuremath{f_{\rm baryon}}\xspace}

\newcommand{\dmd}{\ensuremath{\Delta m_{\particle{d}}}\xspace}
\newcommand{\dms}{\ensuremath{\Delta m_{\particle{s}}}\xspace}
\newcommand{\xd}{\ensuremath{x_{\particle{d}}}\xspace}
\newcommand{\xs}{\ensuremath{x_{\particle{s}}}\xspace}
\newcommand{\yd}{\ensuremath{y_{\particle{d}}}\xspace}
\newcommand{\ys}{\ensuremath{y_{\particle{s}}}\xspace}
\newcommand{\chibar}{\ensuremath{\overline{\chi}}\xspace}
\newcommand{\chid}{\ensuremath{\chi_{\particle{d}}}\xspace}
\newcommand{\chis}{\ensuremath{\chi_{\particle{s}}}\xspace}
\newcommand{\Gd}{\ensuremath{\Gamma_{\particle{d}}}\xspace}
\newcommand{\DGd}{\ensuremath{\Delta\Gd}\xspace}
\newcommand{\DGGd}{\ensuremath{\DGd/\Gd}\xspace}
\newcommand{\Gs}{\ensuremath{\Gamma_{\particle{s}}}\xspace}
\newcommand{\DGs}{\ensuremath{\Delta\Gs}\xspace}
\newcommand{\DGGs}{\ensuremath{\Delta\Gs/\Gs}\xspace}
\newcommand{\ASLd}{\ensuremath{{\cal A}_{\rm SL}^\particle{d}}\xspace}
\newcommand{\ASLs}{\ensuremath{{\cal A}_{\rm SL}^\particle{s}}\xspace}

\renewcommand{\BR}[1]{\particle{{\cal B}(#1)}}
\newcommand{\CL}[1]{#1\%~\mbox{CL}}
\newcommand{\Qjet}{\ensuremath{Q_{\rm jet}}\xspace}

\newcommand{\labe}[1]{\label{equ:#1}}
\newcommand{\labs}[1]{\label{sec:#1}}
\newcommand{\labf}[1]{\label{fig:#1}}
\newcommand{\labt}[1]{\label{tab:#1}}
\newcommand{\refe}[1]{\ref{equ:#1}}
\newcommand{\refs}[1]{\ref{sec:#1}}
\newcommand{\reff}[1]{\ref{fig:#1}}
\newcommand{\reft}[1]{\ref{tab:#1}}
\newcommand{\Ref}[1]{Ref.~\cite{#1}}
\newcommand{\Refs}[1]{Refs.~\cite{#1}}
\newcommand{\Refss}[2]{Refs.~\cite{#1} and \cite{#2}}
\newcommand{\Refsss}[3]{Refs.~\cite{#1}, \cite{#2} and \cite{#3}}
\newcommand{\eq}[1]{(\refe{#1})}
\newcommand{\Eq}[1]{Eq.~(\refe{#1})}
\newcommand{\Eqs}[1]{Eqs.~(\refe{#1})}
\newcommand{\Eqss}[2]{Eqs.~(\refe{#1}) and (\refe{#2})}
\newcommand{\Eqssor}[2]{Eqs.~(\refe{#1}) or (\refe{#2})}
\newcommand{\Eqsss}[3]{Eqs.~(\refe{#1}), (\refe{#2}), and (\refe{#3})}
\newcommand{\Figure}[1]{Figure~\reff{#1}}
\newcommand{\Figuress}[2]{Figures~\reff{#1} and \reff{#2}}
\newcommand{\Fig}[1]{Fig.~\reff{#1}}
\newcommand{\Figs}[1]{Figs.~\reff{#1}}
\newcommand{\Figss}[2]{Figs.~\reff{#1} and \reff{#2}}
\newcommand{\Figsss}[3]{Figs.~\reff{#1}, \reff{#2}, and \reff{#3}}
\newcommand{\Section}[1]{Section~\refs{#1}}
\newcommand{\Sec}[1]{Sec.~\refs{#1}}
\newcommand{\Secs}[1]{Secs.~\refs{#1}}
\newcommand{\Secss}[2]{Secs.~\refs{#1} and \refs{#2}}
\newcommand{\Secsss}[3]{Secs.~\refs{#1}, \refs{#2}, and \refs{#3}}
\newcommand{\Table}[1]{Table~\reft{#1}}
\newcommand{\Tables}[1]{Tables~\reft{#1}}
\newcommand{\Tabless}[2]{Tables~\reft{#1} and \reft{#2}}
\newcommand{\Tablesss}[3]{Tables~\reft{#1}, \reft{#2}, and \reft{#3}}

\newcommand{\subsubsubsection}[1]{\vspace{2ex}\par\noindent {\bf\boldmath\em #1} \vspace{2ex}\par}


\newcommand{\definemath}[2]{\newcommand{#1}{\ensuremath{#2}\xspace}}

\definemath{\hfagCHIBARLEPval}{0.1259}
\definemath{\hfagCHIBARLEPerr}{\pm0.0042}
\definemath{\hfagTAUBDval}{1.530}
\definemath{\hfagTAUBDerr}{\pm0.008}
\definemath{\hfagTAUBUval}{1.639}
\definemath{\hfagTAUBUerr}{\pm0.009}
\definemath{\hfagRTAUBUval}{1.073}
\definemath{\hfagRTAUBUerr}{\pm0.008}
\definemath{\hfagTAUBSval}{1.459}
\definemath{\hfagTAUBSerr}{\pm0.030}
\definemath{\hfagRTAUBSval}{0.953}
\definemath{\hfagRTAUBSerr}{\pm0.020}
\definemath{\hfagTAULBval}{1.379}
\definemath{\hfagTAULBerr}{\pm0.051}
\definemath{\hfagTAUXBval}{1.42}
\definemath{\hfagTAUXBerp}{^{+0.28}}
\definemath{\hfagTAUXBern}{_{-0.24}}
\definemath{\hfagTAUBBval}{1.311}
\definemath{\hfagTAUBBerr}{\pm0.040}
\definemath{\hfagRTAUBBval}{0.857}
\definemath{\hfagRTAUBBerr}{\pm0.026}
\definemath{\hfagTAUBval}{1.568}
\definemath{\hfagTAUBerr}{\pm0.009}
\definemath{\hfagTAUBCval}{0.461}
\definemath{\hfagTAUBCerr}{\pm0.036}
\definemath{\hfagTAUBSSLval}{1.456}
\definemath{\hfagTAUBSSLerr}{\pm0.030}
\definemath{\hfagTAUBSSLXval}{1.458}
\definemath{\hfagTAUBSSLXerr}{\pm0.031}
\definemath{\hfagTAUBSMEANCONval}{1.478}
\definemath{\hfagTAUBSMEANCONerp}{^{+0.020}}
\definemath{\hfagTAUBSMEANCONern}{_{-0.022}}
\definemath{\hfagTAUBSJFval}{1.477}
\definemath{\hfagTAUBSJFerr}{\pm0.046}
\definemath{\hfagRTAUBSSLval}{0.951}
\definemath{\hfagRTAUBSSLerr}{\pm0.020}
\definemath{\hfagRTAUBSMEANCONval}{0.966}
\definemath{\hfagRTAUBSMEANCONerr}{\pm0.015}
\definemath{\hfagRTAUBSMEANCONsig}{2.4}
\definemath{\hfagONEMINUSRTAUBSMEANCONpercent}{(3.4\pm1.5)\%}
\definemath{\hfagRTAULBval}{0.901}
\definemath{\hfagRTAULBerr}{\pm0.034}
\definemath{\hfagTAUBVTXval}{1.572}
\definemath{\hfagTAUBVTXerr}{\pm0.009}
\definemath{\hfagTAUBLEPval}{1.537}
\definemath{\hfagTAUBLEPerr}{\pm0.020}
\definemath{\hfagTAUBJPval}{1.533}
\definemath{\hfagTAUBJPerp}{^{+0.038}}
\definemath{\hfagTAUBJPern}{_{-0.034}}
\definemath{\hfagNSIGMATAULBCDFTWO}{3.4}
\definemath{\hfagSDGDGDval}{0.009}
\definemath{\hfagSDGDGDerr}{\pm0.037}
\definemath{\hfagTAUBSMEANval}{1.515}
\definemath{\hfagTAUBSMEANerp}{^{+0.034}}
\definemath{\hfagTAUBSMEANern}{_{-0.034}}
\definemath{\hfagDGSGSval}{+0.154}
\definemath{\hfagDGSGSerp}{^{+0.067}}
\definemath{\hfagDGSGSern}{_{-0.065}}
\definemath{\hfagDGSGSlow}{+0.024}
\definemath{\hfagDGSGSupp}{+0.290}
\definemath{\hfagRHODGSGSTAUBSMEAN}{+9.999999}
\definemath{\hfagDGSval}{+0.102}
\definemath{\hfagDGSerp}{^{+0.043}}
\definemath{\hfagDGSern}{_{-0.043}}
\definemath{\hfagDGSlow}{+0.016}
\definemath{\hfagDGSupp}{+0.192}
\definemath{\hfagRHODGSTAUBSMEAN}{+9.999999}
\definemath{\hfagTAUBSLval}{1.407}
\definemath{\hfagTAUBSLerp}{^{+0.035}}
\definemath{\hfagTAUBSLern}{_{-0.034}}
\definemath{\hfagTAUBSHval}{1.642}
\definemath{\hfagTAUBSHerp}{^{+0.091}}
\definemath{\hfagTAUBSHern}{_{+0.083}}
\definemath{\hfagDGSGSCONBDval}{N/A}
\definemath{\hfagDGSGSCONBDerp}{^{N/A}}
\definemath{\hfagDGSGSCONBDern}{_{N/A}}
\definemath{\hfagTAUBSMEANCONXval}{1.478}
\definemath{\hfagTAUBSMEANCONXerp}{^{+0.020}}
\definemath{\hfagTAUBSMEANCONXern}{_{-0.022}}
\definemath{\hfagDGSGSCONval}{+0.099}
\definemath{\hfagDGSGSCONerp}{^{+0.048}}
\definemath{\hfagDGSGSCONern}{_{-0.053}}
\definemath{\hfagDGSGSCONlow}{-0.003}
\definemath{\hfagDGSGSCONupp}{+0.191}
\definemath{\hfagRHODGSGSTAUBSMEANCON}{+9.999999}
\definemath{\hfagDGSCONval}{+0.067}
\definemath{\hfagDGSCONerp}{^{+0.031}}
\definemath{\hfagDGSCONern}{_{-0.035}}
\definemath{\hfagDGSCONlow}{-0.002}
\definemath{\hfagDGSCONupp}{+0.130}
\definemath{\hfagRHODGSTAUBSMEANCON}{+9.999999}
\definemath{\hfagTAUBSLCONval}{1.408}
\definemath{\hfagTAUBSLCONerp}{^{+0.035}}
\definemath{\hfagTAUBSLCONern}{_{-0.033}}
\definemath{\hfagTAUBSHCONval}{1.554}
\definemath{\hfagTAUBSHCONerp}{^{+0.051}}
\definemath{\hfagTAUBSHCONern}{_{+0.053}}
\definemath{\hfagDGSGSCONBDCONval}{N/A}
\definemath{\hfagDGSGSCONBDCONerp}{^{N/A}}
\definemath{\hfagDGSGSCONBDCONern}{_{N/A}}
\definemath{\hfagBRDSDSval}{0.046}
\definemath{\hfagBRDSDSerr}{\pm0.022}
\definemath{\hfagDGSGSBRDSDSval}{+0.096}
\definemath{\hfagDGSGSBRDSDSerr}{\pm0.048}
\definemath{\hfagFCWval}{0.515}
\definemath{\hfagFCWerr}{\pm0.006}
\definemath{\hfagFNWval}{0.485}
\definemath{\hfagFNWerr}{\pm0.006}
\definemath{\hfagFFWval}{1.062}
\definemath{\hfagFFWerr}{\pm0.025}
\definemath{\hfagNSIGMAFFW}{2.4}
\definemath{\hfagFCNval}{0.513}
\definemath{\hfagFCNerr}{\pm0.013}
\definemath{\hfagFNNval}{0.487}
\definemath{\hfagFNNerr}{\pm0.013}
\definemath{\hfagFFNval}{1.053}
\definemath{\hfagFFNerr}{\pm0.054}
\definemath{\hfagFCval}{0.515}
\definemath{\hfagFCerr}{\pm0.007}
\definemath{\hfagFNval}{0.485}
\definemath{\hfagFNerr}{\pm0.007}
\definemath{\hfagFFval}{1.064}
\definemath{\hfagFFerr}{\pm0.029}
\definemath{\hfagNSIGMAFF}{2.2}
\definemath{\hfagFPRODval}{0.518}
\definemath{\hfagFPRODerr}{\pm0.019}
\definemath{\hfagFSUMval}{1.005}
\definemath{\hfagFSUMerr}{\pm0.030}
\definemath{\hfagFSFIVEval}{0.194}
\definemath{\hfagFSFIVEsta}{\pm0.011}
\definemath{\hfagFSFIVEsys}{\pm0.027}
\definemath{\hfagFSFIVEerr}{\pm0.029}
\definemath{\hfagLFSFACTOR}{}
\definemath{\hfagLFBSNOMIXval}{0.087}
\definemath{\hfagLFBSNOMIXerr}{\pm0.014}
\definemath{\hfagLFBBNOMIXval}{0.100}
\definemath{\hfagLFBBNOMIXerr}{\pm0.017}
\definemath{\hfagLFBDNOMIXval}{0.406}
\definemath{\hfagLFBDNOMIXerr}{\pm0.009}
\definemath{\hfagWFSFACTOR}{1.2}
\definemath{\hfagWFBSNOMIXval}{0.097}
\definemath{\hfagWFBSNOMIXerr}{\pm0.016}
\definemath{\hfagWFBBNOMIXval}{0.099}
\definemath{\hfagWFBBNOMIXerr}{\pm0.020}
\definemath{\hfagWFBDNOMIXval}{0.402}
\definemath{\hfagWFBDNOMIXerr}{\pm0.011}
\definemath{\hfagCHIBARTEVval}{0.147}
\definemath{\hfagCHIBARTEVerr}{\pm0.011}
\definemath{\hfagCHIBARSFACTOR}{1.8}
\definemath{\hfagCHIBARval}{0.1284}
\definemath{\hfagCHIBARerr}{\pm0.0069}
\definemath{\hfagWFBSMIXval}{0.119}
\definemath{\hfagWFBSMIXerr}{\pm0.019}
\definemath{\hfagLFBSMIXval}{0.112}
\definemath{\hfagLFBSMIXerr}{\pm0.012}
\definemath{\hfagCHIDUval}{0.182}
\definemath{\hfagCHIDUerr}{\pm0.015}
\definemath{\hfagCHIDWUval}{0.1881}
\definemath{\hfagCHIDWUerr}{\pm0.0023}
\definemath{\hfagXDWval}{0.777}
\definemath{\hfagXDWerr}{\pm0.008}
\definemath{\hfagXDWUval}{0.777}
\definemath{\hfagXDWUerr}{\pm0.008}
\definemath{\hfagDMDWval}{0.508}
\definemath{\hfagDMDWsta}{\pm0.003}
\definemath{\hfagDMDWsys}{\pm0.003}
\definemath{\hfagDMDWerr}{\pm0.004}
\definemath{\hfagDMDWUval}{0.507}
\definemath{\hfagDMDWUerr}{\pm0.004}
\definemath{\hfagLFBSval}{0.101}
\definemath{\hfagLFBSerr}{\pm0.009}
\definemath{\hfagLFBBval}{0.092}
\definemath{\hfagLFBBerr}{\pm0.015}
\definemath{\hfagLFBDval}{0.403}
\definemath{\hfagLFBDerr}{\pm0.009}
\definemath{\hfagLRHOFBBFBS}{+0.032}
\definemath{\hfagLRHOFBDFBS}{-0.515}
\definemath{\hfagLRHOFBDFBB}{-0.873}
\definemath{\hfagWFBSval}{0.106}
\definemath{\hfagWFBSerr}{\pm0.012}
\definemath{\hfagWFBBval}{0.093}
\definemath{\hfagWFBBerr}{\pm0.019}
\definemath{\hfagWFBDval}{0.401}
\definemath{\hfagWFBDerr}{\pm0.011}
\definemath{\hfagWRHOFBBFBS}{-0.087}
\definemath{\hfagWRHOFBDFBS}{-0.475}
\definemath{\hfagWRHOFBDFBB}{-0.836}
\definemath{\hfagDMDHval}{0.495}
\definemath{\hfagDMDHsta}{\pm0.010}
\definemath{\hfagDMDHsys}{\pm0.009}
\definemath{\hfagDMDHerr}{\pm0.013}
\definemath{\hfagDMDBval}{0.508}
\definemath{\hfagDMDBsta}{\pm0.003}
\definemath{\hfagDMDBsys}{\pm0.003}
\definemath{\hfagDMDBerr}{\pm0.005}
\definemath{\hfagDMDTWODval}{0.509}
\definemath{\hfagDMDTWODsta}{\pm0.005}
\definemath{\hfagDMDTWODsys}{\pm0.003}
\definemath{\hfagDMDTWODerr}{\pm0.006}
\definemath{\hfagTAUBDTWODval}{1.527}
\definemath{\hfagTAUBDTWODsta}{\pm0.007}
\definemath{\hfagTAUBDTWODsys}{\pm0.007}
\definemath{\hfagTAUBDTWODerr}{\pm0.010}
\definemath{\hfagRHOstaDMDTAUBD}{-0.19}
\definemath{\hfagRHOsysDMDTAUBD}{-0.29}
\definemath{\hfagRHODMDTAUBD}{-0.23}
\definemath{\hfagQPDBval}{1.0024}
\definemath{\hfagQPDBerr}{\pm0.0023}
\definemath{\hfagQPDAval}{1.0030}
\definemath{\hfagQPDAerr}{\pm0.0017}
\definemath{\hfagASLDBval}{-0.0047}
\definemath{\hfagASLDBerr}{\pm0.0046}
\definemath{\hfagASLDAval}{-0.0058}
\definemath{\hfagASLDAerr}{\pm0.0034}
\definemath{\hfagREBDBval}{-0.0012}
\definemath{\hfagREBDBerr}{\pm0.0011}
\definemath{\hfagREBDAval}{-0.0015}
\definemath{\hfagREBDAerr}{\pm0.0008}
\definemath{\hfagASLSval}{+0.0016}
\definemath{\hfagASLSsta}{\pm0.0054}
\definemath{\hfagASLSsys}{\pm0.0066}
\definemath{\hfagASLSerr}{\pm0.0085}
\definemath{\hfagQPSval}{0.9992}
\definemath{\hfagQPSsta}{\pm0.0027}
\definemath{\hfagQPSsys}{\pm0.0033}
\definemath{\hfagQPSerr}{\pm0.0042}
\definemath{\hfagDMSWLIMval}{17.2}
\definemath{\hfagDMSWSENSval}{31.3}
\definemath{\hfagDMSWUPPval}{18.4}
\definemath{\hfagXSWLIMval}{25.1}
\definemath{\hfagCHISWLIMval}{0.49921}
\definemath{\hfagDMSval}{17.78}
\definemath{\hfagDMSerr}{\pm0.12}
\definemath{\hfagXSval}{26.3}
\definemath{\hfagXSerr}{\pm0.4}
\definemath{\hfagCHISval}{0.49928}
\definemath{\hfagCHISerr}{\pm0.00002}
\definemath{\hfagRATIODMDDMSval}{0.0285}
\definemath{\hfagRATIODMDDMSerr}{\pm0.0003}
\definemath{\hfagVTDVTSval}{0.2061}
\definemath{\hfagVTDVTSexx}{\pm0.0011}
\definemath{\hfagVTDVTSthp}{^{+0.0080}}
\definemath{\hfagVTDVTSthn}{_{-0.0060}}
\definemath{\hfagVTDVTSerp}{^{+0.0081}}
\definemath{\hfagVTDVTSern}{_{-0.0061}}
\definemath{\hfagXIval}{1.210}
\definemath{\hfagXIerp}{^{+0.047}}
\definemath{\hfagXIern}{_{-0.035}}
\definemath{\hfagBETASCOMBAval}{+0.39}
\definemath{\hfagBETASCOMBAerp}{^{+0.18}}
\definemath{\hfagBETASCOMBAern}{_{-0.14}}
\definemath{\hfagBETASCOMBAlow}{+0.14}
\definemath{\hfagBETASCOMBAupp}{+0.73}
\definemath{\hfagBETASCOMBBval}{+1.18}
\definemath{\hfagBETASCOMBBerp}{^{+0.14}}
\definemath{\hfagBETASCOMBBern}{_{-0.18}}
\definemath{\hfagBETASCOMBBlow}{+0.82}
\definemath{\hfagBETASCOMBBupp}{+1.43}
\definemath{\hfagPHISCOMBAval}{-0.77}
\definemath{\hfagPHISCOMBAerp}{^{+0.29}}
\definemath{\hfagPHISCOMBAern}{_{-0.37}}
\definemath{\hfagPHISCOMBAlow}{-1.47}
\definemath{\hfagPHISCOMBAupp}{-0.29}
\definemath{\hfagPHISCOMBBval}{-2.36}
\definemath{\hfagPHISCOMBBerp}{^{+0.37}}
\definemath{\hfagPHISCOMBBern}{_{-0.29}}
\definemath{\hfagPHISCOMBBlow}{-2.85}
\definemath{\hfagPHISCOMBBupp}{-1.65}
\definemath{\hfagDGSCOMBAval}{+0.154}
\definemath{\hfagDGSCOMBAerp}{^{+0.054}}
\definemath{\hfagDGSCOMBAern}{_{-0.070}}
\definemath{\hfagDGSCOMBAlow}{+0.036}
\definemath{\hfagDGSCOMBAupp}{+0.264}
\definemath{\hfagDGSCOMBBval}{-0.154}
\definemath{\hfagDGSCOMBBerp}{^{+0.070}}
\definemath{\hfagDGSCOMBBern}{_{-0.054}}
\definemath{\hfagDGSCOMBBlow}{-0.264}
\definemath{\hfagDGSCOMBBupp}{-0.036}
\definemath{\hfagNSIGMASM}{2.2}
\definemath{\hfagBETASCOMBACONval}{+0.38}
\definemath{\hfagBETASCOMBACONerp}{^{+0.17}}
\definemath{\hfagBETASCOMBACONern}{_{-0.18}}
\definemath{\hfagBETASCOMBACONlow}{+0.07}
\definemath{\hfagBETASCOMBACONupp}{+0.63}
\definemath{\hfagBETASCOMBBCONval}{+1.19}
\definemath{\hfagBETASCOMBBCONerp}{^{+0.18}}
\definemath{\hfagBETASCOMBBCONern}{_{-0.17}}
\definemath{\hfagBETASCOMBBCONlow}{+0.94}
\definemath{\hfagBETASCOMBBCONupp}{+1.50}
\definemath{\hfagPHISCOMBACONval}{-0.76}
\definemath{\hfagPHISCOMBACONerp}{^{+0.37}}
\definemath{\hfagPHISCOMBACONern}{_{-0.33}}
\definemath{\hfagPHISCOMBACONlow}{-1.26}
\definemath{\hfagPHISCOMBACONupp}{-0.13}
\definemath{\hfagPHISCOMBBCONval}{-2.37}
\definemath{\hfagPHISCOMBBCONerp}{^{+0.33}}
\definemath{\hfagPHISCOMBBCONern}{_{-0.37}}
\definemath{\hfagPHISCOMBBCONlow}{-3.00}
\definemath{\hfagPHISCOMBBCONupp}{-1.88}
\definemath{\hfagDGSCOMBACONval}{+0.054}
\definemath{\hfagDGSCOMBACONerp}{^{+0.060}}
\definemath{\hfagDGSCOMBACONern}{_{-0.033}}
\definemath{\hfagDGSCOMBABCONlow}{-0.148}
\definemath{\hfagDGSCOMBABCONupp}{+0.148}
\definemath{\hfagDGSCOMBBCONval}{-0.054}
\definemath{\hfagDGSCOMBBCONerp}{^{+0.033}}
\definemath{\hfagDGSCOMBBCONern}{_{-0.060}}
\definemath{\hfagNSIGMASMCON}{2.4}

\newcommand{\unit}[1]{~\ensuremath{\rm #1}\xspace}
\renewcommand{\ps}{\unit{ps}}
\newcommand{\invps}{\unit{ps^{-1}}}
\newcommand{\TeV}{\unit{TeV}}
\newcommand{\MeVcc}{\unit{MeV/\mbox{$c$}^2}}
\newcommand{\MeV}{\unit{MeV}}

\definemath{\hfagCHIBARLEP}{\hfagCHIBARLEPval\hfagCHIBARLEPerr}
\definemath{\hfagTAUBD}{\hfagTAUBDval\hfagTAUBDerr\ps}
\definemath{\hfagTAUBDnounit}{\hfagTAUBDval\hfagTAUBDerr}
\definemath{\hfagTAUBU}{\hfagTAUBUval\hfagTAUBUerr\ps}
\definemath{\hfagTAUBUnounit}{\hfagTAUBUval\hfagTAUBUerr}
\definemath{\hfagRTAUBU}{\hfagRTAUBUval\hfagRTAUBUerr}
\definemath{\hfagTAUBS}{\hfagTAUBSval\hfagTAUBSerr\ps}
\definemath{\hfagTAUBSnounit}{\hfagTAUBSval\hfagTAUBSerr}
\definemath{\hfagRTAUBS}{\hfagRTAUBSval\hfagRTAUBSerr}
\definemath{\hfagTAULB}{\hfagTAULBval\hfagTAULBerr\ps}
\definemath{\hfagTAULBnounit}{\hfagTAULBval\hfagTAULBerr}
\definemath{\hfagTAUXBerr}{\hfagTAUXBerp\hfagTAUXBern}
\definemath{\hfagTAUXB}{\hfagTAUXBval\hfagTAUXBerr\ps}
\definemath{\hfagTAUXBnounit}{\hfagTAUXBval\hfagTAUXBerr}
\definemath{\hfagTAUBB}{\hfagTAUBBval\hfagTAUBBerr\ps}
\definemath{\hfagTAUBBnounit}{\hfagTAUBBval\hfagTAUBBerr}
\definemath{\hfagRTAUBB}{\hfagRTAUBBval\hfagRTAUBBerr}
\definemath{\hfagTAUB}{\hfagTAUBval\hfagTAUBerr\ps}
\definemath{\hfagTAUBnounit}{\hfagTAUBval\hfagTAUBerr}
\definemath{\hfagTAUBC}{\hfagTAUBCval\hfagTAUBCerr\ps}
\definemath{\hfagTAUBCnounit}{\hfagTAUBCval\hfagTAUBCerr}
\definemath{\hfagTAUBSSL}{\hfagTAUBSSLval\hfagTAUBSSLerr\ps}
\definemath{\hfagTAUBSSLnounit}{\hfagTAUBSSLval\hfagTAUBSSLerr}
\definemath{\hfagTAUBSSLX}{\hfagTAUBSSLXval\hfagTAUBSSLXerr\ps}
\definemath{\hfagTAUBSSLXnounit}{\hfagTAUBSSLXval\hfagTAUBSSLXerr}
\definemath{\hfagTAUBSMEANCONerr}{\hfagTAUBSMEANCONerp\hfagTAUBSMEANCONern}
\definemath{\hfagTAUBSMEANCON}{\hfagTAUBSMEANCONval\hfagTAUBSMEANCONerr\ps}
\definemath{\hfagTAUBSMEANCONnounit}{\hfagTAUBSMEANCONval\hfagTAUBSMEANCONerr}
\definemath{\hfagTAUBSJF}{\hfagTAUBSJFval\hfagTAUBSJFerr\ps}
\definemath{\hfagTAUBSJFnounit}{\hfagTAUBSJFval\hfagTAUBSJFerr}
\definemath{\hfagRTAUBSSL}{\hfagRTAUBSSLval\hfagRTAUBSSLerr}
\definemath{\hfagRTAUBSMEANCON}{\hfagRTAUBSMEANCONval\hfagRTAUBSMEANCONerr}
\definemath{\hfagRTAULB}{\hfagRTAULBval\hfagRTAULBerr}
\definemath{\hfagTAUBVTX}{\hfagTAUBVTXval\hfagTAUBVTXerr\ps}
\definemath{\hfagTAUBVTXnounit}{\hfagTAUBVTXval\hfagTAUBVTXerr}
\definemath{\hfagTAUBLEP}{\hfagTAUBLEPval\hfagTAUBLEPerr\ps}
\definemath{\hfagTAUBLEPnounit}{\hfagTAUBLEPval\hfagTAUBLEPerr}
\definemath{\hfagTAUBJPerr}{\hfagTAUBJPerp\hfagTAUBJPern}
\definemath{\hfagTAUBJP}{\hfagTAUBJPval\hfagTAUBJPerr\ps}
\definemath{\hfagTAUBJPnounit}{\hfagTAUBJPval\hfagTAUBJPerr}
\definemath{\hfagSDGDGD}{\hfagSDGDGDval\hfagSDGDGDerr}
\definemath{\hfagTAUBSMEANerr}{\hfagTAUBSMEANerp\hfagTAUBSMEANern}
\definemath{\hfagTAUBSMEAN}{\hfagTAUBSMEANval\hfagTAUBSMEANerr\ps}
\definemath{\hfagTAUBSMEANnounit}{\hfagTAUBSMEANval\hfagTAUBSMEANerr}
\definemath{\hfagDGSGSerr}{\hfagDGSGSerp\hfagDGSGSern}
\definemath{\hfagDGSGS}{\hfagDGSGSval\hfagDGSGSerr}
\definemath{\hfagDGSerr}{\hfagDGSerp\hfagDGSern}
\definemath{\hfagDGS}{\hfagDGSval\hfagDGSerr\invps}
\definemath{\hfagDGSnounit}{\hfagDGSval\hfagDGSerr}
\definemath{\hfagTAUBSLerr}{\hfagTAUBSLerp\hfagTAUBSLern}
\definemath{\hfagTAUBSL}{\hfagTAUBSLval\hfagTAUBSLerr\ps}
\definemath{\hfagTAUBSLnounit}{\hfagTAUBSLval\hfagTAUBSLerr}
\definemath{\hfagTAUBSHerr}{\hfagTAUBSHerp\hfagTAUBSHern}
\definemath{\hfagTAUBSH}{\hfagTAUBSHval\hfagTAUBSHerr\ps}
\definemath{\hfagTAUBSHnounit}{\hfagTAUBSHval\hfagTAUBSHerr}
\definemath{\hfagDGSGSCONBDerr}{\hfagDGSGSCONBDerp\hfagDGSGSCONBDern}
\definemath{\hfagDGSGSCONBD}{\hfagDGSGSCONBDval\hfagDGSGSCONBDerr}
\definemath{\hfagTAUBSMEANCONXerr}{\hfagTAUBSMEANCONXerp\hfagTAUBSMEANCONXern}
\definemath{\hfagTAUBSMEANCONX}{\hfagTAUBSMEANCONXval\hfagTAUBSMEANCONXerr\ps}
\definemath{\hfagTAUBSMEANCONXnounit}{\hfagTAUBSMEANCONXval\hfagTAUBSMEANCONXerr}
\definemath{\hfagDGSGSCONerr}{\hfagDGSGSCONerp\hfagDGSGSCONern}
\definemath{\hfagDGSGSCON}{\hfagDGSGSCONval\hfagDGSGSCONerr}
\definemath{\hfagDGSCONerr}{\hfagDGSCONerp\hfagDGSCONern}
\definemath{\hfagDGSCON}{\hfagDGSCONval\hfagDGSCONerr\invps}
\definemath{\hfagDGSCONnounit}{\hfagDGSCONval\hfagDGSCONerr}
\definemath{\hfagTAUBSLCONerr}{\hfagTAUBSLCONerp\hfagTAUBSLCONern}
\definemath{\hfagTAUBSLCON}{\hfagTAUBSLCONval\hfagTAUBSLCONerr\ps}
\definemath{\hfagTAUBSLCONnounit}{\hfagTAUBSLCONval\hfagTAUBSLCONerr}
\definemath{\hfagTAUBSHCONerr}{\hfagTAUBSHCONerp\hfagTAUBSHCONern}
\definemath{\hfagTAUBSHCON}{\hfagTAUBSHCONval\hfagTAUBSHCONerr\ps}
\definemath{\hfagTAUBSHCONnounit}{\hfagTAUBSHCONval\hfagTAUBSHCONerr}
\definemath{\hfagDGSGSCONBDCONerr}{\hfagDGSGSCONBDCONerp\hfagDGSGSCONBDCONern}
\definemath{\hfagDGSGSCONBDCON}{\hfagDGSGSCONBDCONval\hfagDGSGSCONBDCONerr}
\definemath{\hfagBRDSDS}{\hfagBRDSDSval\hfagBRDSDSerr}
\definemath{\hfagDGSGSBRDSDS}{\hfagDGSGSBRDSDSval\hfagDGSGSBRDSDSerr}
\definemath{\hfagFCW}{\hfagFCWval\hfagFCWerr}
\definemath{\hfagFNW}{\hfagFNWval\hfagFNWerr}
\definemath{\hfagFFW}{\hfagFFWval\hfagFFWerr}
\definemath{\hfagFCN}{\hfagFCNval\hfagFCNerr}
\definemath{\hfagFNN}{\hfagFNNval\hfagFNNerr}
\definemath{\hfagFFN}{\hfagFFNval\hfagFFNerr}
\definemath{\hfagFC}{\hfagFCval\hfagFCerr}
\definemath{\hfagFN}{\hfagFNval\hfagFNerr}
\definemath{\hfagFF}{\hfagFFval\hfagFFerr}
\definemath{\hfagFPROD}{\hfagFPRODval\hfagFPRODerr}
\definemath{\hfagFSUM}{\hfagFSUMval\hfagFSUMerr}
\definemath{\hfagFSFIVE}{\hfagFSFIVEval\hfagFSFIVEerr}
\definemath{\hfagFSFIVEfull}{\hfagFSFIVEval\hfagFSFIVEsta\hfagFSFIVEsys}
\definemath{\hfagLFBSNOMIX}{\hfagLFBSNOMIXval\hfagLFBSNOMIXerr}
\definemath{\hfagLFBBNOMIX}{\hfagLFBBNOMIXval\hfagLFBBNOMIXerr}
\definemath{\hfagLFBDNOMIX}{\hfagLFBDNOMIXval\hfagLFBDNOMIXerr}
\definemath{\hfagWFBSNOMIX}{\hfagWFBSNOMIXval\hfagWFBSNOMIXerr}
\definemath{\hfagWFBBNOMIX}{\hfagWFBBNOMIXval\hfagWFBBNOMIXerr}
\definemath{\hfagWFBDNOMIX}{\hfagWFBDNOMIXval\hfagWFBDNOMIXerr}
\definemath{\hfagCHIBARTEV}{\hfagCHIBARTEVval\hfagCHIBARTEVerr}
\definemath{\hfagCHIBAR}{\hfagCHIBARval\hfagCHIBARerr}
\definemath{\hfagWFBSMIX}{\hfagWFBSMIXval\hfagWFBSMIXerr}
\definemath{\hfagLFBSMIX}{\hfagLFBSMIXval\hfagLFBSMIXerr}
\definemath{\hfagCHIDU}{\hfagCHIDUval\hfagCHIDUerr}
\definemath{\hfagCHIDWU}{\hfagCHIDWUval\hfagCHIDWUerr}
\definemath{\hfagXDW}{\hfagXDWval\hfagXDWerr}
\definemath{\hfagXDWU}{\hfagXDWUval\hfagXDWUerr}
\definemath{\hfagDMDW}{\hfagDMDWval\hfagDMDWerr\invps}
\definemath{\hfagDMDWnounit}{\hfagDMDWval\hfagDMDWerr}
\definemath{\hfagDMDWfull}{\hfagDMDWval\hfagDMDWsta\hfagDMDWsys\invps}
\definemath{\hfagDMDWnounitfull}{\hfagDMDWval\hfagDMDWsta\hfagDMDWsys}
\definemath{\hfagDMDWU}{\hfagDMDWUval\hfagDMDWUerr\invps}
\definemath{\hfagDMDWUnounit}{\hfagDMDWUval\hfagDMDWUerr}
\definemath{\hfagLFBS}{\hfagLFBSval\hfagLFBSerr}
\definemath{\hfagLFBB}{\hfagLFBBval\hfagLFBBerr}
\definemath{\hfagLFBD}{\hfagLFBDval\hfagLFBDerr}
\definemath{\hfagWFBS}{\hfagWFBSval\hfagWFBSerr}
\definemath{\hfagWFBB}{\hfagWFBBval\hfagWFBBerr}
\definemath{\hfagWFBD}{\hfagWFBDval\hfagWFBDerr}
\definemath{\hfagDMDH}{\hfagDMDHval\hfagDMDHerr\invps}
\definemath{\hfagDMDHnounit}{\hfagDMDHval\hfagDMDHerr}
\definemath{\hfagDMDHfull}{\hfagDMDHval\hfagDMDHsta\hfagDMDHsys\invps}
\definemath{\hfagDMDHnounitfull}{\hfagDMDHval\hfagDMDHsta\hfagDMDHsys}
\definemath{\hfagDMDB}{\hfagDMDBval\hfagDMDBerr\invps}
\definemath{\hfagDMDBnounit}{\hfagDMDBval\hfagDMDBerr}
\definemath{\hfagDMDBfull}{\hfagDMDBval\hfagDMDBsta\hfagDMDBsys\invps}
\definemath{\hfagDMDBnounitfull}{\hfagDMDBval\hfagDMDBsta\hfagDMDBsys}
\definemath{\hfagDMDTWOD}{\hfagDMDTWODval\hfagDMDTWODerr\invps}
\definemath{\hfagDMDTWODnounit}{\hfagDMDTWODval\hfagDMDTWODerr}
\definemath{\hfagDMDTWODfull}{\hfagDMDTWODval\hfagDMDTWODsta\hfagDMDTWODsys\invps}
\definemath{\hfagDMDTWODnounitfull}{\hfagDMDTWODval\hfagDMDTWODsta\hfagDMDTWODsys}
\definemath{\hfagTAUBDTWOD}{\hfagTAUBDTWODval\hfagTAUBDTWODerr\ps}
\definemath{\hfagTAUBDTWODnounit}{\hfagTAUBDTWODval\hfagTAUBDTWODerr}
\definemath{\hfagTAUBDTWODfull}{\hfagTAUBDTWODval\hfagTAUBDTWODsta\hfagTAUBDTWODsys\ps}
\definemath{\hfagTAUBDTWODnounitfull}{\hfagTAUBDTWODval\hfagTAUBDTWODsta\hfagTAUBDTWODsys}
\definemath{\hfagQPDB}{\hfagQPDBval\hfagQPDBerr}
\definemath{\hfagQPDA}{\hfagQPDAval\hfagQPDAerr}
\definemath{\hfagASLDB}{\hfagASLDBval\hfagASLDBerr}
\definemath{\hfagASLDA}{\hfagASLDAval\hfagASLDAerr}
\definemath{\hfagREBDB}{\hfagREBDBval\hfagREBDBerr}
\definemath{\hfagREBDA}{\hfagREBDAval\hfagREBDAerr}
\definemath{\hfagASLS}{\hfagASLSval\hfagASLSerr}
\definemath{\hfagASLSfull}{\hfagASLSval\hfagASLSsta\hfagASLSsys}
\definemath{\hfagQPS}{\hfagQPSval\hfagQPSerr}
\definemath{\hfagQPSfull}{\hfagQPSval\hfagQPSsta\hfagQPSsys}
\definemath{\hfagDMSWLIM}{\hfagDMSWLIMval\invps}
\definemath{\hfagDMSWSENS}{\hfagDMSWSENSval\invps}
\definemath{\hfagDMSWUPP}{\hfagDMSWUPPval\invps}
\definemath{\hfagXSWLIM}{\hfagXSWLIMval}
\definemath{\hfagCHISWLIM}{\hfagCHISWLIMval}
\definemath{\hfagDMS}{\hfagDMSval\hfagDMSerr\invps}
\definemath{\hfagDMSnounit}{\hfagDMSval\hfagDMSerr}
\definemath{\hfagXS}{\hfagXSval\hfagXSerr}
\definemath{\hfagCHIS}{\hfagCHISval\hfagCHISerr}
\definemath{\hfagRATIODMDDMS}{\hfagRATIODMDDMSval\hfagRATIODMDDMSerr}
\definemath{\hfagVTDVTSerr}{\hfagVTDVTSerp\hfagVTDVTSern}
\definemath{\hfagVTDVTSthe}{\hfagVTDVTSthp\hfagVTDVTSthn}
\definemath{\hfagVTDVTS}{\hfagVTDVTSval\hfagVTDVTSerr}
\definemath{\hfagVTDVTSfull}{\hfagVTDVTSval\hfagVTDVTSexx\hfagVTDVTSthe}
\definemath{\hfagXIerr}{\hfagXIerp\hfagXIern}
\definemath{\hfagXI}{\hfagXIval\hfagXIerr}
\definemath{\hfagBETASCOMBAerr}{\hfagBETASCOMBAerp\hfagBETASCOMBAern}
\definemath{\hfagBETASCOMBA}{\hfagBETASCOMBAval\hfagBETASCOMBAerr}
\definemath{\hfagBETASCOMBBerr}{\hfagBETASCOMBBerp\hfagBETASCOMBBern}
\definemath{\hfagBETASCOMBB}{\hfagBETASCOMBBval\hfagBETASCOMBBerr}
\definemath{\hfagPHISCOMBAerr}{\hfagPHISCOMBAerp\hfagPHISCOMBAern}
\definemath{\hfagPHISCOMBA}{\hfagPHISCOMBAval\hfagPHISCOMBAerr}
\definemath{\hfagPHISCOMBBerr}{\hfagPHISCOMBBerp\hfagPHISCOMBBern}
\definemath{\hfagPHISCOMBB}{\hfagPHISCOMBBval\hfagPHISCOMBBerr}
\definemath{\hfagDGSCOMBAerr}{\hfagDGSCOMBAerp\hfagDGSCOMBAern}
\definemath{\hfagDGSCOMBA}{\hfagDGSCOMBAval\hfagDGSCOMBAerr\invps}
\definemath{\hfagDGSCOMBAnounit}{\hfagDGSCOMBAval\hfagDGSCOMBAerr}
\definemath{\hfagDGSCOMBBerr}{\hfagDGSCOMBBerp\hfagDGSCOMBBern}
\definemath{\hfagDGSCOMBB}{\hfagDGSCOMBBval\hfagDGSCOMBBerr\invps}
\definemath{\hfagDGSCOMBBnounit}{\hfagDGSCOMBBval\hfagDGSCOMBBerr}
\definemath{\hfagBETASCOMBACONerr}{\hfagBETASCOMBACONerp\hfagBETASCOMBACONern}
\definemath{\hfagBETASCOMBACON}{\hfagBETASCOMBACONval\hfagBETASCOMBACONerr}
\definemath{\hfagBETASCOMBBCONerr}{\hfagBETASCOMBBCONerp\hfagBETASCOMBBCONern}
\definemath{\hfagBETASCOMBBCON}{\hfagBETASCOMBBCONval\hfagBETASCOMBBCONerr}
\definemath{\hfagPHISCOMBACONerr}{\hfagPHISCOMBACONerp\hfagPHISCOMBACONern}
\definemath{\hfagPHISCOMBACON}{\hfagPHISCOMBACONval\hfagPHISCOMBACONerr}
\definemath{\hfagPHISCOMBBCONerr}{\hfagPHISCOMBBCONerp\hfagPHISCOMBBCONern}
\definemath{\hfagPHISCOMBBCON}{\hfagPHISCOMBBCONval\hfagPHISCOMBBCONerr}
\definemath{\hfagDGSCOMBACONerr}{\hfagDGSCOMBACONerp\hfagDGSCOMBACONern}
\definemath{\hfagDGSCOMBACON}{\hfagDGSCOMBACONval\hfagDGSCOMBACONerr\invps}
\definemath{\hfagDGSCOMBACONnounit}{\hfagDGSCOMBACONval\hfagDGSCOMBACONerr}
\definemath{\hfagDGSCOMBBCONerr}{\hfagDGSCOMBBCONerp\hfagDGSCOMBBCONern}
\definemath{\hfagDGSCOMBBCON}{\hfagDGSCOMBBCONval\hfagDGSCOMBBCONerr\invps}
\definemath{\hfagDGSCOMBBCONnounit}{\hfagDGSCOMBBCONval\hfagDGSCOMBBCONerr}


\mysection{\b-hadron production fractions, lifetimes and mixing parameters}
\labs{life_mix}


Quantities such as \b-hadron production fractions, \b-hadron lifetimes, 
and neutral \B-meson oscillation frequencies have been studied
in the nineties at LEP and SLC 
(\ee colliders at $\sqrt{s}=m_{\particle{Z}}$) 
as well as at the 
first version of the Tevatron
(\particle{p\bar{p}} collider at $\sqrt{s}=1.8\TeV$). 
Since then 
precise measurements of the \Bd and \Bu lifetimes, as well as of the 
\Bd oscillation frequency, have also been performed at the 
asymmetric \B factories, KEKB and PEPII
(\ee colliders at $\sqrt{s}=m_{\Ups}$) while measurements related 
to the other \b-hadrons, in particular \Bs, \Bc and \Lb, 
are being performed at the upgraded Tevatron ($\sqrt{s}=1.96\TeV$).
In most cases, these basic quantities, although interesting by themselves,
became 
necessary ingredients for the more complicated and 
refined analyses 
at the asymmetric \B factories
and at the Tevatron, 
in particular the time-dependent \CP asymmetry measurements.
It is therefore important that the best experimental
values of these quantities continue to be kept up-to-date and improved. 

In several cases, the averages presented in this chapter are 
needed and used as input for the results given in the subsequent chapters. 
Within this chapter, some averages need the knowledge of other 
averages in a circular way. This coupling, which appears through the 
\b-hadron fractions whenever inclusive or semi-exclusive measurements 
have to be considered, has been reduced significantly in the last several years 
with increasingly precise exclusive measurements becoming available. 

In addition to \b-hadron fractions, lifetimes and 
mixing parameters, this chapter also deals with the 
CP-violating phase $\beta_s$, which is the phase 
difference between the \Bs mixing amplitude and the 
$b\to c\bar{c}s$ decay amplitude. The angle $\beta$, 
which is the equivalent of $\beta_s$ for the \Bd 
system, is discussed in Chapter~\ref{sec:cp_uta}. 

\mysubsection{\b-hadron production fractions}
\labs{fractions}
 
We consider here the relative fractions of the different \b-hadron 
species found in an unbiased sample of weakly-decaying \b hadrons 
produced under some specific conditions. The knowledge of these fractions
is useful to characterize the signal composition in inclusive \b-hadron 
analyses, or to predict the background composition in exclusive analyses.
Many analyses in \B physics need these fractions as input. We distinguish 
here the following three conditions: \Ups decays, \Upsfive decays, and 
high-energy collisions (including \Z decays). 

\mysubsubsection{\b-hadron production fractions in \Ups decays}
\labs{fraction_Ups4S}

Only pairs of the two lightest (charged and neutral) \B mesons 
can be produced in \Ups decays, 
and it is enough to determine the following branching 
fractions:
\begin{eqnarray}
f^{+-} & = & \Gamma(\Ups \to \particle{B^+B^-})/
             \Gamma_{\rm tot}(\Ups)  \,, \\
f^{00} & = & \Gamma(\Ups \to \particle{B^0\bar{B}^0})/
             \Gamma_{\rm tot}(\Ups) \,.
\end{eqnarray}
In practice, most analyses measure their ratio
\begin{equation}
R^{+-/00} = f^{+-}/f^{00} = \Gamma(\Ups \to \particle{B^+B^-})/
             \Gamma(\Ups \to \particle{B^0\bar{B}^0}) \,,
\end{equation}
which is easier to access experimentally.
Since an inclusive (but separate) reconstruction of 
\Bu and \Bd is difficult, specific exclusive decay modes, 
${\Bu} \to x^+$ and ${\Bd} \to x^0$, are usually considered to perform 
a measurement of $R^{+-/00}$, whenever they can be related by 
isospin symmetry (for example \particle{\Bu \to J/\psi K^+} and 
\particle{\Bd \to J/\psi K^0}).
Under the assumption that $\Gamma(\Bu \to x^+) = \Gamma(\Bd \to x^0)$, 
\ie\ that isospin invariance holds in these \B decays,
the ratio of the number of reconstructed
$\Bu \to x^+$ and $\Bd \to x^0$ mesons is proportional to
\begin{equation}
\frac{f^{+-}\, \BR{\Bu\to x^+}}{f^{00}\, \BR{\Bd\to x^0}}
= \frac{f^{+-}\, \Gamma({\Bu}\to x^+)\, \tau(\Bu)}%
{f^{00}\, \Gamma({\Bd}\to x^0)\,\tau(\Bd)}
= \frac{f^{+-}}{f^{00}} \, \frac{\tau(\Bu)}{\tau(\Bd)}  \,, 
\end{equation} 
where $\tau(\Bu)$ and $\tau(\Bd)$ are the \Bu and \Bd 
lifetimes respectively.
Hence the primary quantity measured in these analyses 
is $R^{+-/00} \, \tau(\Bu)/\tau(\Bd)$, 
and the extraction of $R^{+-/00}$ with this method therefore 
requires the knowledge of the $\tau(\Bu)/\tau(\Bd)$ lifetime ratio. 

\begin{table}
\caption{Published measurements of the $\Bu/\Bd$ production ratio
in \Ups decays, together with their average (see text).
Systematic uncertainties due to the imperfect knowledge of 
$\tau(\Bu)/\tau(\Bd)$ are included. The latest \babar result\cite{BABAR_R2005}
supersedes  the earlier \babar measurements \cite{BABAR_R2002,BABAR_R2004}.}
\labt{R_data}
\begin{center}
\begin{tabular}{lccll}
\hline
Experiment & Ref. & Decay modes & Published value of & Assumed value \\
and year & & or method & $R^{+-/00}=f^{+-}/f^{00}$ & of $\tau(\Bu)/\tau(\Bd)$ \\
\hline
CLEO,   2001 & \cite{CLEO_R2001}  & \particle{J/\psi K^{(*)}} 
             & $1.04 \pm0.07 \pm0.04$ & $1.066 \pm0.024$ \\
\babar, 2002 & \cite{BABAR_R2002} & \particle{(c\bar{c})K^{(*)}}
             & $1.10 \pm0.06 \pm0.05$ & $1.062 \pm0.029$\\ 
CLEO,   2002 & \cite{CLEO_R2002}  & \particle{D^*\ell\nu}
             & $1.058 \pm0.084 \pm0.136$ & $1.074 \pm0.028$\\
\belle, 2003 & \cite{BELLE_dmd_dilepton} & dilepton events 
             & $1.01 \pm0.03 \pm0.09$ & $1.083 \pm0.017$\\
\babar, 2004 & \cite{BABAR_R2004} & \particle{J/\psi K}
             & $1.006 \pm0.036 \pm0.031$ & $1.083 \pm0.017$ \\
\babar, 2005 & \cite{BABAR_R2005} & \particle{(c\bar{c})K^{(*)}}
             & $1.06 \pm0.02 \pm0.03$ & $1.086 \pm0.017$\\ 
\hline
Average      & & & \hfagFF~(tot) & \hfagRTAUBU \\
\hline
\end{tabular}
\end{center}
\end{table}

The published measurements of $R^{+-/00}$ are listed 
in \Table{R_data} together with the corresponding assumed values of 
$\tau(\Bu)/\tau(\Bd)$.
All measurements are based on the above-mentioned method, 
except the one from \belle, which is a by-product of the 
\Bd mixing frequency analysis using dilepton events
(but note that it also assumes isospin invariance, 
namely $\Gamma(\Bu \to \ell^+{\rm X}) = \Gamma(\Bd \to \ell^+{\rm X})$).
The latter is therefore treated in a slightly different 
manner in the following procedure used to combine 
these measurements:
\begin{itemize} 
\item each published value of $R^{+-/00}$ from CLEO and \babar
      is first converted back to the original measurement of 
      $R^{+-/00} \, \tau(\Bu)/\tau(\Bd)$, using the value of the 
      lifetime ratio assumed in the corresponding analysis;
\item a simple weighted average of these original
      measurements of $R^{+-/00} \, \tau(\Bu)/\tau(\Bd)$ from 
      CLEO and \babar (which do not depend on the assumed value 
      of the lifetime ratio) is then computed, assuming no 
      statistical or systematic correlations between them;


\item the weighted average of $R^{+-/00} \, \tau(\Bu)/\tau(\Bd)$ 
      is converted into a value of $R^{+-/00}$, using the latest 
      average of the lifetime ratios, $\tau(\Bu)/\tau(\Bd)=\hfagRTAUBU$ 
      (see \Sec{lifetime_ratio});
\item the \belle measurement of $R^{+-/00}$ is adjusted to the 
      current values of $\tau(\Bd)=\hfagTAUBD$ and 
      $\tau(\Bu)/\tau(\Bd)=\hfagRTAUBU$ (see \Sec{lifetime_ratio}),
      using the quoted systematic uncertainties due to these parameters;
\item the combined value of $R^{+-/00}$ from CLEO and \babar is averaged 
      with the adjusted value of $R^{+-/00}$ from \belle, assuming a 100\% 
      correlation of the systematic uncertainty due to the limited 
      knowledge on $\tau(\Bu)/\tau(\Bd)$; no other correlation is considered. 
\end{itemize} 
The resulting global average, 
\begin{equation}
R^{+-/00} = \frac{f^{+-}}{f^{00}} =  \hfagFF \,,
\labe{Rplusminus}
\end{equation}
is consistent with an equal production of charged and neutral \B mesons, 
although only at the $\hfagNSIGMAFF \sigma$ level.

On the other hand, the \babar collaboration has 
performed a direct measurement of the $f^{00}$ fraction 
using a novel method, which does not rely on isospin symmetry nor requires 
the knowledge of $\tau(\Bu)/\tau(\Bd)$. Its analysis, 
based on a comparison between the number of events where a single 
$B^0 \to D^{*-} \ell^+ \nu$ decay could be reconstructed and the number 
of events where two such decays could be reconstructed, yields~\cite{BABAR_f00}
\begin{equation}
f^{00}= 0.487 \pm 0.010\,\mbox{(stat)} \pm 0.008\,\mbox{(syst)} \,.
\labe{fzerozero}
\end{equation}

The two results of \Eqss{Rplusminus}{fzerozero} are of very different natures 
and completely independent of each other. 
Their product is equal to $f^{+-} = \hfagFPROD$, 
while another combination of them gives $f^{+-} + f^{00}= \hfagFSUM$, 
compatible with unity.
Assuming\footnote{A few non-$\B\bar{B}$
decay modes of the $\Upsilon(4S)$ 
($\Upsilon(1S)\pi^+\pi^-$,
$\Upsilon(2S)\pi^+\pi^-$, $\Upsilon(1S)\eta$) 
have been observed with branching fractions
of the order of $10^{-4}$~\cite{BABAR_BELLE_Ups},
corresponding to a partial
width several times larger than that in the \ee channel.
However, this can still be
neglected and the assumption $f^{+-}+f^{00}=1$ remains valid
in the present context of the determination of $f^{+-}$ and $f^{00}$.}
 $f^{+-}+f^{00}= 1$, also consistent with 
CLEO's observation that the fraction of \Ups decays 
to \BB pairs is larger than 0.96 at \CL{95}~\cite{CLEO_frac_limit},
the results of \Eqss{Rplusminus}{fzerozero}
can be averaged (first converting \Eq{Rplusminus} 
into a value of $f^{00}=1/(R^{+-/00}+1)$) 
to yield the following more precise estimates:
\begin{equation}
f^{00} = \hfagFNW  \,,~~~ f^{+-} = 1 -f^{00} =  \hfagFCW \,,~~~
\frac{f^{+-}}{f^{00}} =  \hfagFFW \,.
\end{equation}
The latter ratio differs from one by $\hfagNSIGMAFFW \sigma$.

\mysubsubsection{\b-hadron production fractions in \Upsfive decays}
\labs{fraction_Ups5S}

Hadronic events produced in $e^+e^-$ collisions at the \Upsfive energy
can be classified into three categories: 
light-quark continuum events, $b\bar{b}$ continuum events,
and \Upsfive events. The latter two cannot be distinguished and 
are expected to always produce one of the following final states 
with a pair of $b$-flavored mesons:
$B\bar{B}$,
$B\bar{B}^*$, $B^*\bar{B}$, $B^*\bar{B}^*$, 
$B\bar{B}\pi$, $B\bar{B}^*\pi$, $B^*\bar{B}\pi$, 
$B^*\bar{B}^*\pi$, $B\bar{B}\pi\pi$,
$B_s^0\bar{B}_s^0$, $B_s^0\bar{B}_s^{0*}$, $B_s^{0*}\bar{B}_s^0$ or 
$B_s^{0*}\bar{B}_s^{0*}$, where 
$B$ denotes a $B^0$ or $B^+$ meson and 
$\bar{B}$ denotes a $\bar{B}^0$ or $B^-$ meson.
The excited states decay via $B^* \to B \gamma$ and
$B_s^{0*} \to B_s^0 \gamma$.
We define here $f_s(\Upsfive)$ as the fraction of 
$B_s^{0(*)}\bar{B}_s^{0(*)}$ events over all 
events with a pair of $b$-flavored mesons 
at the \Upsfive energy:
\begin{equation}
f_s(\Upsfive) =
\frac{\sigma(e^+e^- \to B_s^{0(*)}\bar{B}_s^{0(*)})}%
{\sigma(e^+e^- \to \Upsfive ~ \mbox{or} ~ b\bar{b}X)}
~~ \mbox{at} ~ \sqrt{s}=m(\Upsfive) \,.
\end{equation}

\begin{table}
\caption{Published values of $f_s(\Upsfive)$ and their average.}
\labt{fsFiveS}
\begin{center}
\begin{tabular}{lccl}
\hline
Experiment,   & Ref.                      & Decay modes              & Published value of          \\
year and dataset &                           & or method                & $f_s(\Upsfive)$         \\ \hline
CLEO, 2006, 0.42\invfb   & \cite{CLEO_huang2007}     & $\Upsfive\to D_{s}X$     & $0.168 \pm 0.026^{+0.067}_{-0.034}$  \\
             & \cite{CLEO_huang2007}     & $\Upsfive \to \phi X$    & $0.246 \pm 0.029^{+0.110}_{-0.053}$ \\
             & \cite{CLEO_huang2007}     & $\Upsfive \to B\bar{B}X$ & $0.411 \pm 0.100 \pm 0.092$ \\  
             & \cite{CLEO_huang2007}     & CLEO average of above 3  & $0.21^{+0.06}_{-0.03}$      \\  \hline
Belle, 2006, 1.86\invfb  & \cite{Belle_drutskoy2007} & $\Upsfive \to D_s X$     & $0.179 \pm 0.014 \pm 0.041$ \\
             & \cite{Belle_drutskoy2007} & $\Upsfive \to D^0 X$     & $0.181 \pm 0.036 \pm 0.075$ \\  
             & \cite{Belle_drutskoy2007} & Belle average of above 2 & $0.180 \pm 0.013 \pm 0.032$ \\  \hline 
\multicolumn{3}{l}{Average of all above
after adjustments to inputs of \Table{fsFiveS_external}} & 
\hfagFSFIVEfull             \\  \hline 
\end{tabular}
\end{center}
\end{table}

\begin{table}
\caption{External inputs on which the $f_s(\Upsfive)$ average is based.}
\labt{fsFiveS_external}
\begin{center}
\begin{tabular}{lcc}
\hline
Branching fraction   & Value     & Explanation and reference \\
\hline
${\cal B}(B\to D_s X)\times {\cal B}(D_s \to \phi\pi)$ & 
$0.00374\pm 0.00014$ & derived from~\cite{PDG_2007} \\
${\cal B}(B^0_s \to D_s X)$ & 
$0.92\pm0.11$ & model-dependent estimate~\cite{CLEO_artuso2005} \\
${\cal B}(D_s \to \phi\pi)$ & 
$0.045\pm0.004$ & \cite{PDG_2007} \\
${\cal B}(B\to D^0 X)\times {\cal B}(D^0 \to K\pi)$ & 
$0.0243\pm0.0011$ & derived from~\cite{PDG_2007} \\
${\cal B}(B^0_s \to D^0 X)$ & 
$0.08\pm0.07$ & model-dependent estimate~\cite{Belle_drutskoy2007,CLEO_artuso2005} \\
${\cal B}(D^0 \to K\pi)$ & 
$0.0382\pm0.0007$ & \cite{PDG_2007} \\
${\cal B}(B \to \phi X)$ & 
$0.0344\pm0.0012$ & world average~\cite{PDG_2007,CLEO_huang2007} \\
${\cal B}(B^0_s \to \phi X)$ &
$0.161\pm0.024$ & model-dependent estimate~\cite{CLEO_huang2007} \\
\hline
\end{tabular}
\end{center}
\end{table}

The CLEO and Belle collaborations have recently published 
measurements of several inclusive \Upsfive branching fractions, 
${\cal B}(\Upsfive\to D_s X)$, 
${\cal B}(\Upsfive\to \phi X)$, 
${\cal B}(\Upsfive\to D^0 X)$, and 
${\cal B}(\Upsfive\to B\bar{B} X)$, 
from which they extract the
model-dependent estimates of $f_s(\Upsfive)$
reported in \Table{fsFiveS}.
This extraction requires the knowledge of several other 
branching fractions, which are listed in \Table{fsFiveS_external}
together with their most recent values. 
Before being averaged, the CLEO and Belle results are 
adjusted to these new external inputs. The world average of $f_s(\Upsfive)$
taking into account all systematic correlations introduced by the 
use of common external inputs, as well as the experiment-specific 
correlations due to the estimated number of $b\bar{b}$ events, is
\begin{equation}
f_s(\Upsfive) = \hfagFSFIVE \,.
\end{equation}
This production of $B^0_s$ mesons at the \Upsfive
is observed to be dominated by the $B_s^{0*}\bar{B}_s^{0*}$
channel~\cite{CLEO_bonvicini2006,Belle_drutskoy2007e}, with 
$\sigma(e^+e^- \to B_s^{0*}\bar{B}_s^{0*})/%
\sigma(e^+e^- \to B_s^{0(*)}\bar{B}_s^{0(*)})
= (93^{+7}_{-9}\pm 1)\%$~\cite{Belle_drutskoy2007e}.

\mysubsubsection{\b-hadron production fractions at high energy}
\labs{fractions_high_energy}
\labs{chibar}

At high energy, all species of weakly-decaying \b hadrons 
can be produced, either directly or in strong and electromagnetic 
decays of excited \b hadrons.
We assume here that the fractions of these different species 
are the same in unbiased samples of high-$p_{\rm T}$ \b jets 
originating from \particle{Z^0} decays or from \particle{p\bar{p}} 
collisions at the Tevatron.
This hypothesis is plausible considering that, in both cases, 
the last step of the jet hadronization is a non-perturbative
QCD process occurring at the scale of $\Lambda_{\rm QCD}$.
On the other hand, there is no strong argument to claim that these 
fractions should be strictly equal, so this assumption 
should be checked experimentally.
Although the available data is not sufficient at 
this time to perform a significant check, 
it is expected that more data from 
Tevatron Run~II may improve this situation and 
allow one to confirm or disprove this assumption with reasonable 
confidence. Meanwhile, the attitude adopted here is that these 
fractions are assumed to be equal at all high-energy colliders
until demonstrated otherwise by experiment.\footnote{It is not unlikely
that the \b-hadron fractions in low-$p_{\rm T}$ jets 
at a hadronic machine be different; in particular, beam-remnant effects may
enhance the \b-baryon production.}
However, as explained below, the measurements performed at LEP and at 
the Tevatron show slight discrepancies. Therefore we present two sets 
of averages: one set including only measurements performed at LEP, 
and a second set including measurements performed at both LEP and Tevatron. 

Contrary to what happens in the charm sector where the fractions of \particle{D^+} 
and \particle{D^0} are different, the relative amount of \Bu and \Bd is not affected by the 
electromagnetic decays of excited ${\Bu}^*$ and ${\Bd}^*$ states and strong decays of excited
${\Bu}^{**}$ and ${\Bd}^{**}$ states. Decays of the type \particle{{\Bs}^{**} \to B^{(*)}K}
also contribute to the \Bu and \Bd rates, but with the same magnitude if mass effects
can be neglected.
We therefore assume equal production of \Bu and \Bd. We also  
neglect the production of weakly-decaying states
made of several heavy quarks (like \Bc and other heavy baryons) 
which is known to be very small. Hence, for the purpose of determining 
the \b-hadron fractions, we use the constraints
\begin{equation}
\fBu = \fBd ~~~~\mbox{and}~~~ \fBu + \fBd + \fBs + \fbb = 1 \,,
\labe{constraints}
\end{equation}
where \fBu, \fBd, \fBs and \fbb
are the unbiased fractions of \Bu, \Bd, \Bs and \b baryons, respectively.

The LEP experiments have measured
$\fBs \times \BR{\Bs\to\particle{D_s^-} \ell^+ \nu_\ell \mbox{$X$}}$~\cite{LEP_fs}, 
$\BR{\b\to\Lb} \times \BR{\Lb\to\Lc\ell^-\bar{\nu}_\ell \mbox{$X$}}$~\cite{DELPHI_fla,ALEPH_fla}
and $\BR{\b\to\Xib^-} \times \BR{\Xi_b^- \to \Xi^-\ell^-\overline\nu_\ell 
\mbox{$X$}}$~\cite{ALEPH_fxi,DELPHI_fxi2}\footnote{The DELPHI result 
of \Ref{DELPHI_fxi2} is considered to supersede an older one~\cite{DELPHI_fxi}.}
from partially reconstructed final states 
including a lepton, \fbb
from protons identified in \b events~\cite{ALEPH-fbar}, and the 
production rate of charged \b hadrons~\cite{DELPHI-fch}. 
The various \b-hadron fractions 
have also been measured at CDF using electron-charm final states~\cite{CDF_f_ec}
and double semileptonic decays with
\particle{K^*\mu\mu} and \particle{\phi\mu\mu}
final states~\cite{CDF_f_phil_Kstl}.
All these 
results\footnote{Recent
measurements~\cite{CDF2_fractions} performed by CDF with Run~II data 
have not been included here.}
have been combined
following the procedure and assumptions described in~\cite{LEPHFS},
to yield $\fBu=\fBd=\hfagWFBDNOMIX$, 
$\fBs=\hfagWFBSNOMIX$ and $\fbb=\hfagWFBBNOMIX$
under the constraints of \Eq{constraints}. 
Following the PDG prescription, we have scaled the
combined uncertainties on these fractions 
by \hfagWFSFACTOR to account for slight 
discrepancies in the input data.
Repeating the combination using LEP data only, 
we obtain $\fBu=\fBd=\hfagLFBDNOMIX$,
$\fBs=\hfagLFBSNOMIX$ and $\fbb=\hfagLFBBNOMIX$
and find that no scaling factor is necessary. 
For these combinations other external inputs are used, \eg\ the branching 
ratios of \B mesons to final states with a \particle{D}, \particle{D^*} or 
\particle{D^{**}} in semileptonic decays, which are needed to evaluate the 
fraction of semileptonic \Bs decays with a \particle{D_s^-} in the final state.


Time-integrated mixing analyses performed with lepton pairs 
from \particle{b\bar{b}} 
events produced at high-energy colliders measure the quantity 
\begin{equation}
\chibar = f'_{\particle{d}} \,\chid + f'_{\particle{s}} \,\chis \,,
\labe{chibar}
\end{equation}
where $f'_{\particle{d}}$ and $f'_{\particle{s}}$ are 
the fractions of \Bd and \Bs hadrons 
in a sample of semileptonic \b-hadron decays, and where \chid and \chis 
are the \Bd and \Bs time-integrated mixing probabilities.
Assuming that all \b hadrons have the same semileptonic decay width implies 
$f'_i = f_i R_i$, where $R_i = \tau_i/\tau_{\b}$ is the ratio of the lifetime 
$\tau_i$ of species $i$ to the average \b-hadron lifetime 
$\tau_{\b} = \sum_i f_i \tau_i$.
Hence measurements of the mixing probabilities
\chibar, \chid and \chis can be used to improve our 
knowledge of \fBu, \fBd, \fBs and \fbb.
In practice, the above relations yield another determination of 
\fBs obtained from \fbb and mixing information, 
\begin{equation}
\fBs = \frac{1}{R_{\particle{s}}}
\frac{(1+r)\overline{\chi}-(1-\fbb R_{\rm baryon}) \chid}{(1+r)\chis - \chid} \,,
\labe{fBs-mixing}
\end{equation}
where $r=R_{\particle{u}}/R_{\particle{d}} = \tau(\Bu)/\tau(\Bd)$.

The published measurements of \chibar performed by the LEP
experiments have been combined by the LEP Electroweak Working Group to yield 
$\chibar = \hfagCHIBARLEP$~\cite{LEPEWWG}.
This can be compared with the Tevatron average, 
$\chibar = \hfagCHIBARTEV$,
obtained from a CDF measurement with Run~I data~\cite{CDF_chibar}
and from a recent \dzero measurement with Run~II data~\cite{D0_qp_chibar}.
The two averages deviate
from each other by $\hfagCHIBARSFACTOR\,\sigma$; 
this could be an indication that the production fractions of \b hadrons 
at the \particle{Z} peak or at the Tevatron are not the same. 
Although this discrepancy 
is not very significant it should be carefully monitored in the future. 
We choose to combine these two results in a simple weighted average,
assuming no correlations, and, following the PDG prescription, we 
multiply the combined uncertainty by \hfagCHIBARSFACTOR to account 
for the discrepancy. Our world average is then
$\chibar = \hfagCHIBAR$.

\begin{table}
\caption{Time-integrated mixing probability \chibar (defined in \Eq{chibar}), 
and fractions of the different \b-hadron species in an unbiased sample of 
weakly-decaying \b hadrons, obtained from both direct
and mixing measurements.
Measurements performed in $Z$ decays are included 
in both sets of averages.}
\labt{fractions}
\begin{center}
\begin{tabular}{llcc}
\hline
Quantity            &                      & in $Z$ decays   & at high energy   \\
\hline
Mixing probability  & $\overline{\chi}$    & \hfagCHIBARLEP  & \hfagCHIBAR \\
\Bu or \Bd fraction & $f_u = f_d$          & \hfagLFBD       & \hfagWFBD   \\
\Bs fraction        & $f_s$                & \hfagLFBS       & \hfagWFBS   \\
\b baryon fraction  & $f_{\rm baryon}$     & \hfagLFBB       & \hfagWFBB   \\
\multicolumn{2}{l}{Correlation between $f_s$ and $f_u = f_d$}            & \hfagLRHOFBDFBS & \hfagWRHOFBDFBS \\
\multicolumn{2}{l}{Correlation between $f_{\rm baryon}$ and $f_u = f_d$} & \hfagLRHOFBDFBB & \hfagWRHOFBDFBB \\
\multicolumn{2}{l}{Correlation between $f_{\rm baryon}$ and $f_s$}       & \hfagLRHOFBBFBS & \hfagWRHOFBBFBS \\
\hline
\end{tabular}
\end{center}
\end{table}

Introducing the \chibar average in \Eq{fBs-mixing}, together with our world average 
$\chid = \hfagCHIDWU$ (see \Eq{chid} of \Sec{dmd}), the assumption $\chis= 1/2$ 
(justified by \Eq{chis} in \Sec{dms}), the 
best knowledge of the lifetimes (see \Sec{lifetimes}) and the estimate of \fbb given above, 
yields $\fBs = \hfagWFBSMIX$ 
(or $\fBs = \hfagLFBSMIX$ using only LEP data),
an estimate dominated by the mixing information. 
Taking into account all known correlations (including the one introduced by \fbb), 
this result is then combined with the set of fractions obtained from direct measurements 
(given above), to yield the 
improved estimates of \Table{fractions}, 
still under the constraints of \Eq{constraints}. 
As can be seen, our knowledge on the mixing parameters 
substantially reduces the uncertainty on \fBs, and this 
even in the case of the world averages where a rather strong 
deweighting was introduced in the computation of \chibar.
It should be noted that the results 
are correlated, as indicated in \Table{fractions}.

New CDF measurements performed with Run~II data, and not included in the above averages,
have recently been published~\cite{CDF2_fractions},
\begin{eqnarray}
\frac{\fBu}{\fBd} &=& 1.054 \pm 0.018\mbox{(stat)}\, ^{+0.025}_{-0.041}\mbox{(syst)} \pm 0.058({\cal B})  \,, \\
\frac{\fBs}{\fBd+\fBu} &=& 0.160 \pm 0.005\mbox{(stat)}\, ^{+0.011}_{-0.010}\mbox{(syst)}\, ^{+0.057}_{-0.034}({\cal B})  \,, \\
\frac{f_{\Lambda_{\b}}}{\fBd+\fBu} &=& 0.281 \pm 0.012\mbox{(stat)}\, ^{+0.058}_{-0.056}\mbox{(syst)}\, ^{+0.128}_{-0.087}({\cal B})  \,,
\end{eqnarray}
where $f_{\Lambda_{\b}}$ is the fraction of $\Lambda_{\b}$
in an unbiased sample of weakly-decaying \b-hadrons, and
where the third quoted error (${\cal B}$) is due to uncertainties on measured branching fractions. 
According to the authors of Ref.~\cite{CDF2_fractions}, these results on $\fBu/\fBd$ and
$\fBs/(\fBd+\fBu)$ are in agreement with the 
averages computed from LEP data, but the central value of 
$f_{\Lambda_{\b}}/(\fBd+\fBu)$ is twice as large as that of 
$\fbb/(\fBd+\fBu)$ measured at LEP, showing a $\sim 2.3\sigma$
discrepancy attributed to a possible momentum dependence 
of the \b-baryon fragmentation. 


%
%

\mysubsection{\b-hadron lifetimes}
\labs{lifetimes}

In the spectator model the decay of \b-flavored hadrons $H_b$ is
governed entirely by the flavor changing \particle{b\to Wq} transition
($\particle{q}=\particle{c,u}$).  For this very reason, lifetimes of all
\b-flavored hadrons are the same in the spectator approximation
regardless of the (spectator) quark content of the $H_b$.  In the early
1990's experiments became sophisticated enough to start seeing the
differences of the lifetimes among various $H_b$ species.  The first
theoretical calculations of the spectator quark effects on $H_b$
lifetime emerged only few years earlier.

Currently, most of such calculations are performed in the framework of
the Heavy Quark Expansion, HQE.  In the HQE, under certain assumptions
(most important of which is that of quark-hadron duality), the decay
rate of an $H_b$ to an inclusive final state $f$ is expressed as the sum
of a series of expectation values of operators of increasing dimension,
multiplied by the correspondingly higher powers of $\Lambda_{\rm
QCD}/m_b$:
\begin{equation}
\Gamma_{H_b\to f} = |CKM|^2\sum_n c_n^{(f)}
\Bigl(\frac{\Lambda_{\rm QCD}}{m_b}\Bigr)^n\langle H_b|O_n|H_b\rangle,
\labe{hqe}
\end{equation}
where $|CKM|^2$ is the relevant combination of the CKM matrix elements.
Coefficients $c_n^{(f)}$ of this expansion, known as Operator Product
Expansion~\cite{OPE}, can be calculated perturbatively.  Hence, the HQE
predicts $\Gamma_{H_b\to f}$ in the form of an expansion in both
$\Lambda_{\rm QCD}/m_{\b}$ and $\alpha_s(m_{\b})$.  The precision of
current experiments makes it mandatory to go to the next-to-leading
order in QCD, {\em i.e.}\ to include correction of the order of
$\alpha_s(m_{\b})$ to the $c_n^{(f)}$'s.  All non-perturbative physics
is shifted into the expectation values $\langle H_b|O_n|H_b\rangle$ of
operators $O_n$.  These can be calculated using lattice QCD or QCD sum
rules, or can be related to other observables via the
HQE~\cite{Bigi_1995}.  One may reasonably expect that powers of
$\Lambda_{\rm QCD}/m_{\b}$ provide enough suppression that only the
first few terms of the sum in \Eq{hqe} matter.

Theoretical predictions are usually made for the ratios of the lifetimes
(with $\tau(\Bd)$ chosen as the common denominator) rather than for the
individual lifetimes, for this allows several uncertainties to cancel.
The precision of the current HQE calculations (see
\Refs{nlo_lifetimes,tarantino,Gabbiani_et_al} for the latest updates)
is in some instances already surpassed by the measurements,
\eg\ in the case of $\tau(\Bu)/\tau(\Bd)$.  Also, HQE calculations are
not assumption-free.  More accurate predictions are a matter of progress
in the evaluation of the non-perturbative hadronic matrix elements and
verifying the assumptions that the calculations are based upon.
However, the HQE, even in its present shape, draws a number of important
conclusions, which are in agreement with experimental observations:
\begin{itemize}
\item The heavier the mass of the heavy quark the smaller is the
  variation in the lifetimes among different hadrons containing this
  quark, which is to say that as $m_{\b}\to\infty$ we retrieve the
  spectator picture in which the lifetimes of all $H_b$'s are the same.
   This is well illustrated by the fact that lifetimes are rather
   similar in the \b sector, while they differ by large factors
   in the \particle{c} sector ($m_{\particle{c}}<m_{\b}$).
\item The non-perturbative corrections arise only at the order of
  $\Lambda_{\rm QCD}^2/m_{\b}^2$, which translates into 
  differences among $H_b$ lifetimes of only a few percent.
\item It is only the difference between meson and baryon lifetimes that
  appears at the $\Lambda_{\rm QCD}^2/m_{\b}^2$ level.  The splitting of the
  meson lifetimes occurs at the $\Lambda_{\rm QCD}^3/m_{\b}^3$ level, yet it is
  enhanced by a phase space factor $16\pi^2$ with respect to the leading
  free \b decay.
\end{itemize}

To ensure that certain sources of systematic uncertainty cancel, 
lifetime analyses are sometimes designed to measure a 
ratio of lifetimes.  However, because of the differences in decay
topologies, abundance (or lack thereof) of decays of a certain kind,
{\em etc.}, measurements of the individual lifetimes are more 
common.  In the following section we review the most common
types of the lifetime measurements.  This discussion is followed by the
presentation of the averaging of the various lifetime measurements, each
with a brief description of its particularities.



\mysubsubsection{Lifetime measurements, uncertainties and correlations}

In most cases lifetime of an $H_b$ is estimated from a flight distance
and a $\beta\gamma$ factor which is used to convert the geometrical
distance into the proper decay time.  Methods of accessing lifetime
information can roughly be divided in the following five categories:
\begin{enumerate}
\item {\bf\em Inclusive (flavor-blind) measurements}.  These
  measurements are aimed at extracting the lifetime from a mixture of
  \b-hadron decays, without distinguishing the decaying species.  Often
  the knowledge of the mixture composition is limited, which makes these
  measurements experiment-specific.  Also, these
  measurements have to rely on Monte Carlo for estimating the
  $\beta\gamma$ factor, because the decaying hadrons are not fully
  reconstructed.  On the bright side, these usually are the largest
  statistics \b-hadron lifetime measurements that are accessible to a
  given experiment, and can, therefore, serve as an important
  performance benchmark.
\item {\bf\em Measurements in semileptonic decays of a specific
  {\boldmath $H_b$\unboldmath}}.  \particle{W}from \particle{\b\to Wc}
  produces $\ell\nu_l$ pair (\particle{\ell=e,\mu}) in about 21\% of the
  cases.  Electron or muon from such decays is usually a well-detected
  signature, which provides for clean and efficient trigger.
  \particle{c} quark from \particle{b\to Wc} transition and the other
  quark(s) making up the decaying $H_b$ combine into a charm hadron,
  which is reconstructed in one or more exclusive decay channels.
  Knowing what this charmed hadron is allows one to separate, at least
  statistically, different $H_b$ species.  The advantage of these
  measurements is in statistics, which usually is superior to that of the
  exclusively reconstructed $H_b$ decays.  Some of the main
  disadvantages are related to the difficulty of estimating lepton+charm
  sample composition and Monte Carlo reliance for the $\beta\gamma$
  factor estimate.
\item {\bf\em Measurements in exclusively reconstructed hadronic decays}.
  These
  have the advantage of complete reconstruction of decaying $H_b$, which
  allows one to infer the decaying species as well as to perform precise
  measurement of the $\beta\gamma$ factor.  Both lead to generally
  smaller systematic uncertainties than in the above two categories.
  The downsides are smaller branching ratios, larger combinatoric
  backgrounds, especially in $H_b\rightarrow H_c\pi(\pi\pi)$ and
  multi-body $H_c$ decays, or in a hadron collider environment with
  non-trivial underlying event.  $H_b\to J/\psi H_s$ are relatively
  clean and easy to trigger on $J/\psi\to \ell^+\ell^-$, but their
  branching fraction is only about 1\%.
\item {\bf\em Measurements at asymmetric B factories}. 

In the $\Ups\rightarrow B \bar{B}$ decay, the \B mesons (\Bu or \Bd) are
essentially at rest in the \Ups frame.  This makes direct lifetime
measurements impossible in experiments at symmetric colliders producing 
\Ups at rest. 
At asymmetric \B factories the \Ups meson is boosted
resulting in \B and \particle{\bar{B}} moving nearly parallel to each 
other with the same boost. The lifetime is inferred from the distance $\Delta z$        
separating the \B and \particle{\bar{B}} decay vertices along the beam axis 
and from the \Ups boost known from the beam energies. This boost is equal to 
$\beta \gamma \approx 0.55$ (0.43) in the \babar (\belle) experiment,
resulting in an average \B decay length of approximately 250~(190)~$\mu$m. 
In order to determine the charge of the \B mesons in each event, one of the them is
fully reconstructed in a semileptonic or hadronic decay mode.
The other \B is typically not fully reconstructed, only the position
of its decay vertex is determined from the remaining tracks in the event.
These measurements benefit from large statistics, but suffer from poor proper time 
resolution, comparable to the \B lifetime itself. This resolution is dominated by the 
uncertainty on the decay vertices, which is typically 50~(100)~$\mu$m for a
fully (partially) reconstructed \B meson. 
With very large future statistics,
the resolution and purity could be improved (and hence the systematics reduced)
by fully reconstructing both \B mesons in the event. 
 
\item {\bf\em Direct measurement of lifetime ratios}.  This method has
  so far been only applied in the measurement of $\tau(\Bu)/\tau(\Bd)$.
  The ratio of the lifetimes is extracted from the dependence of the
  observed relative number of \Bu and \Bd candidates (both reconstructed
  in semileptonic decays) on the proper decay time.
\end{enumerate}

In some of the latest analyses, measurements of two (\eg\ $\tau(\Bu)$ and
$\tau(\Bu)/\tau(\Bd)$) or three (\eg\ $\tau(\Bu)$,
$\tau(\Bu)/\tau(\Bd)$, and \dmd) quantities are combined.  This
introduces correlations among measurements.  Another source of
correlations among the measurements are the systematic effects, which
could be common to an experiment or to an analysis technique across the
experiments.  When calculating the averages, such correlations are taken
into account per general procedure, described in
\Ref{lifetime_details}.

\mysubsubsection{Inclusive \b-hadron lifetimes}

The inclusive \b hadron lifetime is defined as $\tau_{\b} = \sum_i f_i
\tau_i$ where $\tau_i$ are the individual species lifetimes and $f_i$ are
the fractions of the various species present in an unbiased sample of
weakly-decaying \b hadrons produced at a high-energy
collider.\footnote{In principle such a quantity could be slightly
different in \particle{Z} decays and a the Tevatron, in case the
fractions of \b-hadron species are not exactly the same; see the
discussion in \Sec{fractions_high_energy}.}  This quantity is certainly
less fundamental than the lifetimes of the individual species, the
latter being much more useful in comparisons of the measurements with
the theoretical predictions.  Nonetheless, we perform the averaging of
the inclusive lifetime measurements for completeness as well as for the
reason that they might be of interest as ``technical numbers.''

\begin{table}[tp]
\caption{Measurements of average \b-hadron lifetimes.}
\labt{lifeincl}
\begin{center}
\begin{tabular}{lcccl} \hline
Experiment &Method           &Data set & $\tau_{\b}$ (ps)       &Ref.\\
\hline
ALEPH  &Dipole               &91     &$1.511\pm 0.022\pm 0.078$ &\cite{ALEIN2}\\
DELPHI &All track i.p.\ (2D) &91--92 &$1.542\pm 0.021\pm 0.045$ &\cite{DELIN0}$^a$\\
DELPHI &Sec.\ vtx            &91--93 &$1.582\pm 0.011\pm 0.027$ &\cite{DELIN}$^a$\\
DELPHI &Sec.\ vtx            &94--95 &$1.570\pm 0.005\pm 0.008$ &\cite{DELB04}\\
L3     &Sec.\ vtx + i.p.     &91--94 &$1.556\pm 0.010\pm 0.017$ &\cite{L3IN1}$^b$\\
OPAL   &Sec.\ vtx            &91--94 &$1.611\pm 0.010\pm 0.027$ &\cite{OPAIN2}\\
SLD    &Sec.\ vtx            &93     &$1.564\pm 0.030\pm 0.036$ &\cite{SLDIN}\\ 
\hline
\multicolumn{2}{l}{Average set 1 (\b vertex)} && \hfagTAUBVTXnounit &\\
\hline\hline
ALEPH  &Lepton i.p.\ (3D)    &91--93 &$1.533\pm 0.013\pm 0.022$ &\cite{ALEIN1}\\
L3     &Lepton i.p.\ (2D)    &91--94 &$1.544\pm 0.016\pm 0.021$ &\cite{L3IN1}$^b$\\
OPAL   &Lepton i.p.\ (2D)    &90--91 &$1.523\pm 0.034\pm 0.038$ &\cite{OPAIN1}\\ 
\hline
\multicolumn{2}{l}{Average set 2 ($\b\to\ell$)} && \hfagTAUBLEPnounit &\\
\hline\hline
CDF1   &\particle{J/\psi} vtx&92--95 &$1.533\pm 0.015^{+0.035}_{-0.031}$ &\cite{CDFIN_BS1} \\ 
\hline\hline
\multicolumn{2}{l}{Average of all above} && \hfagTAUBnounit & \\
\hline
\multicolumn{5}{l}{$^a$ \footnotesize The combined DELPHI result quoted in
\cite{DELIN} is 1.575 $\pm$ 0.010 $\pm$ 0.026 ps.} \\[-0.5ex]
\multicolumn{5}{l}{$^b$ \footnotesize The combined L3 result quoted in \cite{L3IN1} 
is 1.549 $\pm$ 0.009 $\pm$ 0.015 ps.}
\end{tabular}
\end{center}
\end{table}

In practice, an unbiased measurement of the inclusive lifetime is
difficult to achieve, because it would imply an efficiency which is
guaranteed to be the same across species.  So most of the measurements
are biased.  In an attempt to group analyses which are expected to
select the same mixture of \b hadrons, the available results (given in
\Table{lifeincl}) are divided into the following three sets:
\begin{enumerate}
\item measurements at LEP and SLD that accept any \b-hadron decay, based 
      on topological reconstruction (secondary vertex or track impact
      parameters);
\item measurements at LEP based on the identification
      of a lepton from a \b decay; and
\item measurements at the Tevatron based on inclusive 
      \particle{H_b\to J/\psi X} reconstruction, where the
      \particle{J/\psi} is fully reconstructed.
\end{enumerate}

The measurements of the first set are generally considered as estimates
of $\tau_{\b}$, although the efficiency to reconstruct a secondary
vertex most probably depends, in an analysis-specific way, on the number
of tracks coming from the vertex, thereby depending on the type of the
$H_b$.  Even though these efficiency variations can in principle be
accounted for using Monte Carlo simulations (which inevitably contain
assumptions on branching fractions), the $H_b$ mixture in that case can
remain somewhat ill-defined and could be slightly different among
analyses in this set.

On the contrary, the mixtures corresponding to the other two sets of
measurements are better defined in the limit where the reconstruction
and selection efficiency of a lepton or a \particle{J/\psi} from an
$H_b$ does not depend on the decaying hadron type.  These mixtures are
given by the production fractions and the inclusive branching fractions
for each $H_b$ species to give a lepton or a \particle{J/\psi}.  In
particular, under the assumption that all \b hadrons have the same
semileptonic decay width, the analyses of the second set should measure
$\tau(\b\to\ell) = (\sum_i f_i \tau_i^2) /(\sum_i f_i \tau_i)$ which is
necessarily larger than $\tau_{\b}$ if lifetime differences exist.
Given the present knowledge on $\tau_i$ and $f_i$,
$\tau(\b\to\ell)-\tau_{\b}$ is expected to be of the order of 0.01\ps.

Measurements by SLC and LEP experiments are subject to a number of
common systematic uncertainties, such as those due to (lack of knowledge
of) \b and \particle{c} fragmentation, \b and \particle{c} decay models,
\BR{B\to\ell}, \BR{B\to c\to\ell}, \BR{c\to\ell}, $\tau_{\particle{c}}$,
and $H_b$ decay multiplicity.  In the averaging, these systematic
uncertainties are assumed to be 100\% correlated.  The averages for the
sets defined above (also given in \Table{lifeincl}) are
\begin{eqnarray}
\tau(\b~\mbox{vertex}) &=& \hfagTAUBVTX \,,\\
\tau(\b\to\ell) &=& \hfagTAUBLEP  \,, \\
\tau(\b\to\particle{J/\psi}) &=& \hfagTAUBJP\,,
\end{eqnarray}
whereas an average of all measurements, ignoring mixture differences, 
yields \hfagTAUB.

\mysubsubsection{\Bd and \Bu lifetimes and their ratio}
\labs{taubd}
\labs{taubu}
\labs{lifetime_ratio}

\begin{table}[tp]
\caption{Measurements of the \Bd lifetime.}
\labt{lifebd}
\begin{center}
\begin{tabular}{lcccl} \hline
Experiment &Method                    &Data set &$\tau(\Bd)$ (ps)                  &Ref.\\
\hline
ALEPH  &\particle{D^{(*)} \ell}       &91--95 &$1.518\pm 0.053\pm 0.034$          &\cite{ALEB01}\\
ALEPH  &Exclusive                     &91--94 &$1.25^{+0.15}_{-0.13}\pm 0.05$     &\cite{ALEB0}\\
ALEPH  &Partial rec.\ $\pi^+\pi^-$    &91--94 &$1.49^{+0.17+0.08}_{-0.15-0.06}$   &\cite{ALEB0}\\
DELPHI &\particle{D^{(*)} \ell}       &91--93 &$1.61^{+0.14}_{-0.13}\pm 0.08$     &\cite{DELB01}\\
DELPHI &Charge sec.\ vtx              &91--93 &$1.63 \pm 0.14 \pm 0.13$           &\cite{DELB02}\\
DELPHI &Inclusive \particle{D^* \ell} &91--93 &$1.532\pm 0.041\pm 0.040$          &\cite{DELB03}\\
DELPHI &Charge sec.\ vtx              &94--95 &$1.531 \pm 0.021\pm0.031$          &\cite{DELB04}\\
\end{tabular}
\end{center}
\end{table}

After a number of years of dominating these averages the LEP experiments
yielded the scene to the asymmetric \B~factories and
the Tevatron experiments.  The \B~factories have been very successful in
utilizing their potential -- in only a few years of running, \babar and,
to a greater extent, \belle, have struck a balance between the
statistical and the systematic uncertainties, with both being close to
(or even better than) the impressive 1\%.  In the meanwhile, CDF and
\dzero have emerged as significant contributors to the field as the
Tevatron Run~II data flowed in.  Both appear to enjoy relatively small
systematic effects, and while current statistical uncertainties of their
measurements are factors of 2 to 4 larger than those of their \B-factory
counterparts, both Tevatron experiments stand to increase their samples
by almost an order of magnitude.

\begin{table}[tbp]
\caption{Measurements of the \Bu lifetime.}
\labt{lifebu}
\begin{center}
\begin{tabular}{lcccl} \hline
Experiment &Method                 &Data set &$\tau(\Bu)$ (ps)                 &Ref.\\
\hline
ALEPH  &\particle{D^{(*)} \ell}    &91--95 &$1.648\pm 0.049\pm 0.035$          &\cite{ALEB01}\\
ALEPH  &Exclusive                  &91--94 &$1.58^{+0.21+0.04}_{-0.18-0.03}$   &\cite{ALEB0}\\
DELPHI &\particle{D^{(*)} \ell}    &91--93 &$1.61\pm 0.16\pm 0.12$             &\cite{DELB01}$^a$\\
DELPHI &Charge sec.\ vtx           &91--93 &$1.72\pm 0.08\pm 0.06$             &\cite{DELB02}$^a$\\
DELPHI &Charge sec.\ vtx           &94--95 &$1.624\pm 0.014\pm 0.018$          &\cite{DELB04}\\
L3     &Charge sec.\ vtx           &94--95 &$1.66\pm  0.06\pm 0.03$            &\cite{L3B01}\\
OPAL   &\particle{D^{(*)} \ell}    &91--93 &$1.52 \pm 0.14\pm 0.09$            &\cite{OPAB0}\\
\end{tabular}
\end{center}
\end{table}

At present time we are in an interesting position of having three sets
of measurements (from LEP/SLC, \B factories and the Tevatron) that
originate from different environments, obtained using substantially
different techniques and are precise enough for incisive comparison.


\begin{table}[tb]
\caption{Measurements of the ratio $\tau(\Bu)/\tau(\Bd)$.}
\labt{liferatioBuBd}
\begin{center}
\begin{tabular}{lcccl} 
\hline
Experiment &Method                 &Data set &Ratio $\tau(\Bu)/\tau(\Bd)$      &Ref.\\
\hline
ALEPH  &\particle{D^{(*)} \ell}    &91--95 &$1.085\pm 0.059\pm 0.018$          &\cite{ALEB01}\\
ALEPH  &Exclusive                  &91--94 &$1.27^{+0.23+0.03}_{-0.19-0.02}$   &\cite{ALEB0}\\
DELPHI &\particle{D^{(*)} \ell}    &91--93 &$1.00^{+0.17}_{-0.15}\pm 0.10$     &\cite{DELB01}\\
DELPHI &Charge sec.\ vtx           &91--93 &$1.06^{+0.13}_{-0.11}\pm 0.10$     &\cite{DELB02}\\
DELPHI &Charge sec.\ vtx           &94--95 &$1.060\pm 0.021 \pm 0.024$         &\cite{DELB04}\\
L3     &Charge sec.\ vtx           &94--95 &$1.09\pm 0.07  \pm 0.03$           &\cite{L3B01}\\
OPAL   &\particle{D^{(*)} \ell}    &91--93 &$0.99\pm 0.14^{+0.05}_{-0.04}$     &\cite{OPAB0}\\
\end{tabular}
\end{center}
\end{table}

The averaging of $\tau(\Bu)$, $\tau(\Bd)$ and $\tau(\Bu)/\tau(\Bd)$
measurements is summarized in \Tablesss{lifebd}{lifebu}{liferatioBuBd}.
For $\tau(\Bu)/\tau(\Bd)$ we averaged only the measurements of this
quantity provided by experiments rather than using all available
knowledge, which would have included, for example, $\tau(\Bu)$ and
$\tau(\Bd)$ measurements which did not contribute to any of the ratio
measurements.

The following sources of correlated (within experiment/machine)
systematic uncertainties have been considered:
\begin{itemize}
\item for SLC/LEP measurements -- \particle{D^{**}} branching ratio uncertainties~\cite{LEPHFS},
momentum estimation of \b mesons from \particle{Z^0} decays
(\b-quark fragmentation parameter $\langle X_E \rangle = 0.702 \pm 0.008$~\cite{LEPHFS}),
\Bs and \b baryon lifetimes (see \Secss{taubs}{taulb}),
and \b-hadron fractions at high energy (see \Table{fractions}); 
\item for \babar measurements -- alignment, $z$ scale, PEP-II boost,
sample composition (where applicable);
\item for \dzero and CDF Run~II measurements -- alignment (separately
within each experiment).
\end{itemize}
The resultant averages are:
\begin{eqnarray}
\tau(\Bd) & = & \hfagTAUBD \,, \\
\tau(\Bu) & = & \hfagTAUBU \,, \\
\tau(\Bu)/\tau(\Bd) & = & \hfagRTAUBU \,.
\end{eqnarray}
%
%
%

\mysubsubsection{\Bs lifetime}
\labs{taubs}

Similar to the kaon system, neutral \B mesons contain
short- and long-lived components, since the
light (L) and heavy  (H)
eigenstates, $\B_{\rm L}$ and $\B_{\rm H}$, differ not only
in their masses, but also in their widths 
with $\Delta\Gamma = \Gamma_{\rm L} - \Gamma_{\rm H}$. 
In the case of the \Bs system, $\DGs$ can
be particularly large. The current theoretical
prediction in the Standard Model for
the fractional width difference is
$\DGs = 0.096 \pm 0.039$~\cite{delta_gams},
where $\Gs = (\Gamma_{\rm L} + \Gamma_{\rm H})/2$.
Specific measurements of \DGs and \Gs are explained
in \Sec{DGs}, but the result for
\Gs is quoted here.

Neglecting \CP violation in $\Bs-\Bsbar$ mixing, 
which is expected to be small~\cite{delta_gams}, the
\Bs mass eigenstates are also \CP eigenstates. In
the Standard Model assuming no \CP violation in
the \Bs system,
$\Gamma_{\rm L}$ is the width of
the \CP-even state and
$\Gamma_{\rm H}$ the width of
the \CP-odd state.
Final states can be decomposed into
\CP-even and \CP-odd components, each with a different
lifetime.

In view of a possibly substantial width difference,
and the fact that various
decay channels will have different proportions of 
the $\B_{\rm L}$ and $\B_{\rm H}$ eigenstates,
the straight average of all available 
\Bs lifetime measurements
is rather ill-defined.  Therefore,
the \Bs lifetime measurements are broken down into
four categories and averaged separately.

\begin{itemize}
\item 
{\bf\em Flavor-specific decays}, such as semileptonic
$\particle{B_s} \to \particle{D_s \ell \nu}$
or $\particle{B_s} \to \particle {D_s \pi}$, will
have equal 
fractions of $\B_{\rm L}$ and $\B_{\rm H}$ at time
zero, where
$\tau_{\rm L} = 1/\Gamma_{\rm L}$ 
is expected to be the shorter-lived component and
$\tau_{\rm H} = 1/\Gamma_{\rm H}$ 
expected
to be the longer-lived component.  A superposition
of two exponentials thus results with decay
widths $\Gs \pm \DGs /2$.
Fitting to a single exponential one obtains a
measure of the flavor-specific 
lifetime~\cite{Hartkorn_Moser}:
\begin{equation}
\tau(\Bs)_{\rm fs} = \frac{1}{\Gs}
\frac{{1+\left(\frac{\DGs}{2\Gs}\right)^2}}{{1-\left(\frac{\DGs}{2\Gs}\right)^2}
}.
\end{equation}
As given in \Table{lifebs}, the flavor-specific 
\Bs lifetime world average is:
\begin{equation}
\tau(\Bs)_{\rm fs} = \hfagTAUBSSL \,.
\labe{fslife_const2}
\end{equation}
This world average will be used later in \Sec{DGs} in combination
with other measurements to find
$\bar{\tau}(\Bs) = 1/\Gs$ and $\DGs$.

The following correlated systematic errors were considered:
average \B lifetime used in backgrounds,
\Bs decay multiplicity, and branching ratios used to determine 
backgrounds (\eg\ \BR{B\to D_s D}).
A knowledge of the multiplicity of \Bs decays is important for
measurements that partially reconstruct the final state such as 
\particle{\B\to D_s \mbox{$X$}} (where $X$ is not a lepton). 
The boost deduced from Monte Carlo simulation depends on the multiplicity used.
Since this is not well known, the multiplicity in the simulation is
varied and this range of values observed is taken to be a systematic.
Similarly not all the branching ratios for the potential background
processes are measured. Where they are available, the PDG values are
used for the error estimate. Where no measurements are available
estimates can usually be made by using measured branching ratios of
related processes and using some reasonable extrapolation.
\end{itemize}



\begin{table}[tb]
\caption{Measurements of the \Bs lifetime.}
\labt{lifebs}
\begin{center}
\begin{tabular}{lcccl} \hline
Experiment & Method           & Data set & $\tau(\Bs)$ (ps)               & Ref. \\
\hline
ALEPH  & \particle{D_s \ell}  & 91--95 & $1.54^{+0.14}_{-0.13}\pm 0.04$   & \cite{ALEBS1}          \\
CDF1   & \particle{D_s \ell}  & 92--96 & $1.36\pm 0.09 ^{+0.06}_{-0.05}$  & \cite{CDFBS}           \\
DELPHI & \particle{D_s \ell}  & 91--95 & $1.42^{+0.14}_{-0.13}\pm 0.03$   & \cite{DELBS0_dms_dgs}          \\
OPAL   & \particle{D_s \ell}  & 90--95 & $1.50^{+0.16}_{-0.15}\pm 0.04$   & \cite{OPABS1_OPALAM2}  \\
\dzero & \particle{D_s \mu}  & 02--04 & $1.398 \pm 0.044 ^{+0.028}_{-0.025}   $   & \cite{D0BS2}       \\ 
CDF2   & \particle{D_s \pi (X)} 
                              & 02--06 & $1.518 \pm 0.041 \pm 0.025     $   & \cite{CDFBS2}$^p$      \\
CDF2   & \particle{D_s \ell}  & 02--04 & $1.381 \pm 0.055 ^{+0.052}_{-0.046} $ &
\cite{CDFBS3}$^p$ \\
\end{tabular}
\end{center}
\end{table}

\begin{itemize}
\item
{\bf\em \boldmath $\Bs\to\Ds X$ decays}.
Included in \Table{lifebs} are measurements
of lifetimes using samples of \particle{\Bs} decays to
\particle{D_s} plus
hadrons, and hence into a less known mixture
of \CP-states.  A lifetime
weighted this way can still be a useful input
for analyses examining such an inclusive sample.
These are separated in \Table{lifebs} and combined
with the semileptonic lifetime to obtain:
\begin{equation}
\tau(\Bs)_{\particle{D_s {\rm X}}} = \hfagTAUBS \,.
\end{equation}

\item
{\bf\em Fully exclusive 
{\boldmath \Bs $\to J/\psi \phi$ \unboldmath}decays}
are expected to be
dominated by the \CP-even state and its lifetime.
First measurements of the \CP mix for this decay mode
are outlined in \Sec{DGs}.
CDF and \dzero measurements from this particular mode
\particle{\Bs\to J/\psi\phi} are combined into an
average
given in \Table{lifebs}.  There are no correlations
between the measurements for this fully exclusive
channel, and the world average for this 
specific decay is:
\begin{equation}
\tau(\Bs)_{\particle{J/\psi \phi}} = \hfagTAUBSJF \,.
\end{equation}
A caveat is that different experimental acceptances
will likely lead to different admixtures of the 
\CP-even and \CP-odd states, and fits to a single
exponential may result in inherently different 
measurements of these quantities.

\item
{\bf\em Fully exclusive 
{\boldmath \Bs $\to K^+ K^-$ \unboldmath}decays}
are expected to be
\CP even to within 5\%, and hence measures the lifetime
of the ``light" mass eigenstate
$\tau_L = 1/\Gamma_L$.  The measurement of this 
lifetime from CDF in Run~II~\cite{CDFBS4}
is:
\begin{equation}
\tau(\Bs)_{\particle{K^+K^-}} = 1.53 \pm 0.18 \pm 0.02 
\thinspace {\mathrm{ps}}, 
\end{equation}
and will be used as an input in \Sec{DGs} for the
average described below.
\end{itemize}

Finally, as will be shown in \Sec{DGs}, measurements
of $\DGs$, including separation into
\CP-even and \CP-odd components, give
\begin{equation}
\bar{\tau}(\Bs) = 1/\Gs = \hfagTAUBSMEAN \,,
\end{equation}
and when combined with the flavor-specific lifetime
measurements:
\begin{equation}
\bar{\tau}(\Bs) = 1/\Gs = \hfagTAUBSMEANCON \,.
\end{equation}

\mysubsubsection{\Bc lifetime}
\labs{taubc}

There are currently three measurements of the lifetime of the \Bc meson
from CDF~\cite{CDFBC1,CDFBC2} and \dzero~\cite{D0BC1} using the semileptonic decay
mode \particle{\Bc \to J/\psi \ell} and fitting
simultaneously to the mass and lifetime using the vertex formed
with the leptons from the decay of the \particle{J/\psi} and
the third lepton. Correction factors
to estimate the boost due to the missing neutrino are used.
In the analysis of the CDF Run~I data~\cite{CDFBC1},
a mass value of 
$6.40 \pm 0.39 \pm 0.13$~GeV/$c^2$ 
is found by fitting
to the tri-lepton invariant mass spectrum. 
In the CDF and \dzero Run~II results~\cite{CDFBC2,D0BC1}, 
the \Bc mass is assumed to be 
$6285.7 \pm 5.3 \pm 1.2$~MeV/$c^2$, taken from a 
CDF result~\cite{BCmass}. 
These mass measurements
are consistent within uncertainties, and also consistent with the
most recent precision determination from CDF of 
$6275.6 \pm 2.9 \pm 2.5$~MeV/$c^2$~\cite{BCmass2007}.
Correlated systematic errors include the impact
of the uncertainty of the \Bc $p_T$ spectrum on the correction
factors, the level of feed-down from $\psi(2S)$, 
MC modeling of the decay model varying from phase space
to the ISGW model, and mass variations.
Values of the \particle{\Bc} lifetime are given
in \Table{lifebc} and the world average is
determined to be:
\begin{equation}
\tau(\Bc) = \hfagTAUBC \,.
\end{equation}

\begin{table}[tb]
\caption{Measurements of the \Bc lifetime.}
\labt{lifebc}
\begin{center}
\begin{tabular}{lcccl} \hline
Experiment & Method                    & Data set  & $\tau(\Bc)$ (ps)
      & Ref.\\   \hline
CDF1       & \particle{J/\psi \ell} & 92--95  & $0.46^{+0.18}_{-0.16} \pm
 0.03$   & \cite{CDFBC1}  \\ 
CDF2       & \particle{J/\psi \ell} & 02--06  & $0.475^{+0.053}_{-0.049} \pm 0.018$   & \cite{CDFBC2}$^p$ \\
 \dzero & \particle{J/\psi \mu} & 02--06  & $0.448^{+0.038}_{-0.036} \pm 0.032$
   & \cite{D0BC1}  \\ \hline
  \multicolumn{2}{l}{Average} &   &  \hfagTAUBCnounit
                 &    \\   \hline
\multicolumn{5}{l}{$^p$ \footnotesize Preliminary.}
\end{tabular}
\end{center}
\end{table}

\mysubsubsection{\Lb and \b-baryon lifetimes}
\labs{taulb}

The most precise measurements of the \b-baryon lifetime
originate from two classes of partially reconstructed decays.
In the first class, decays with an exclusively 
reconstructed \Lc baryon
and a lepton of opposite charge are used. These products are
more likely to occur in the decay of \Lb baryons.
In the second class, more inclusive final states with a baryon
(\particle{p}, \particle{\bar{p}}, $\Lambda$, or $\bar{\Lambda}$) 
and a lepton have been used, and these final states can generally
arise from any \b baryon.

The following sources of correlated systematic uncertainties have 
been considered:
experimental time resolution within a given experiment, \b-quark
fragmentation distribution into weakly decaying \b baryons,
\Lb polarization, decay model,
and evaluation of the \b-baryon purity in the selected event samples.
In computing the averages
the central values of the masses are scaled to 
$M(\Lb) = 5620 \pm 2\MeVcc$~\cite{PDGmass} and
$M(\mbox{\b-baryon}) = 5670 \pm 100\MeVcc$.

For the semi-inclusive lifetime measurements, 
the meaning of decay model
systematic uncertainties
and the correlation of these uncertainties between measurements
are not always clear.
Uncertainties related to the decay model are dominated by
assumptions on the fraction of $n$-body semileptonic decays.
To be conservative it is assumed
that these are 100\%  correlated whenever given as an error.
DELPHI varies the fraction of 4-body decays from 0.0 to 0.3. 
In computing the average, the DELPHI
result is corrected to a value of  $0.2 \pm 0.2$ for this fraction.

Furthermore, in computing the average,
the semileptonic decay results from LEP are corrected for a polarization of 
$-0.45^{+0.19}_{-0.17}$~\cite{LEPHFS} and  a 
\Lb fragmentation parameter
$\langle X_E \rangle =0.70\pm 0.03$~\cite{LBFRAG}.




\begin{table}[t]
\caption{Measurements of the \b-baryon lifetimes.
}
\labt{lifelb}
\begin{center}
\begin{tabular}{lcccl} 
\hline
Experiment&Method                &Data set& Lifetime (ps) & Ref. \\\hline
ALEPH  &$\Lc\ell$             & 91--95 &$1.18^{+0.13}_{-0.12} \pm 0.03$ & \cite{ALEPH_fla}$^a$\\
ALEPH  &$\Lambda\ell^-\ell^+$ & 91--95 &$1.30^{+0.26}_{-0.21} \pm 0.04$ & \cite{ALEPH_fla}$^a$\\
CDF1   &$\Lc\ell$             & 91--95 &$1.32 \pm 0.15        \pm 0.07$ & \cite{CDFLAM}\\
CDF2   &$J/\psi \Lambda$      & 02--06 &$1.580 \pm 0.077 \pm 0.012$ & \cite{CDFBJPSIXprel2007}$^p$ \\
\dzero &$J/\psi \Lambda$      & 02--06 &$1.218^{+0.130}_{-0.115} \pm 0.042$ & \cite{D0LAMB}$^b$ \\
\dzero &$\Lc\mu$              & 02--06 &$1.290^{+0.119+0.087}_{-0.110-0.091}$ & \cite{D0LAMB2}$^b$ \\
DELPHI &$\Lc\ell$             & 91--94 &$1.11^{+0.19}_{-0.18} \pm 0.05$ & \cite{DELLAM0}$^c$\\
\end{tabular}
\end{center}
\end{table}

Inputs to the averages are given in \Table{lifelb}.
Note that the CDF $\Lambda_b \to J/\psi \Lambda$
lifetime result~\cite{CDFBJPSIXprel2007} is
$4.0\sigma$ larger than the world average computed excluding this result. 
%
%
It is nonetheless combined with the rest 
without adjustment of input errors.
The world average lifetime of \b baryons is then:
\begin{equation}
\langle\tau(\mbox{\b-baryon})\rangle = \hfagTAUBB \,.
\end{equation}
Keeping only \particle{\Lambda^{\pm}_c \ell^{\mp}}, 
$\Lambda \ell^- \ell^+$, and fully exclusive
final states, as representative of
the \Lb baryon, the following lifetime is obtained:
\begin{equation}
\tau(\Lb) = \hfagTAULB \,. 
\end{equation}

Averaging the measurements based on the $\Xi^{\mp} \ell^{\mp}$
final states~\cite{ALEPH_fxi,DELPHI_fxi,DELPHI_fxi2} gives
a lifetime value for a sample of events
containing $\Xib^0$ and $\Xib^-$ baryons:
\begin{equation}
\langle\tau(\Xib)\rangle = \hfagTAUXB \,.
\end{equation}

\mysubsubsection{Summary and comparison with theoretical predictions}
\labs{lifesummary}

Averages of lifetimes of specific \b-hadron species are collected
in \Table{sumlife}.
\begin{table}[t]
\caption{Summary of lifetimes of different \b-hadron species.}
\labt{sumlife}
\begin{center}
\begin{tabular}{lc} \hline
\b-hadron species & Measured lifetime \\ \hline
\Bu                         & \hfagTAUBU   \\
\Bd                         & \hfagTAUBD   \\
\Bs ($\to$ flavor specific) & \hfagTAUBSSL \\
\Bs ($\to J/\psi\phi$)      & \hfagTAUBSJF \\
\Bs ($1/\Gs$)               & \hfagTAUBSMEANCON \\
\Bc                         & \hfagTAUBC   \\ 
\Lb                         & \hfagTAULB   \\
\Xib mixture                & \hfagTAUXB   \\
\b-baryon mixture           & \hfagTAUBB   \\
\b-hadron mixture           & \hfagTAUB    \\
\hline
\end{tabular}
\end{center}
\caption{Measured ratios of \b-hadron lifetimes relative to
the \Bd lifetime and ranges predicted
by theory~\cite{tarantino,Gabbiani_et_al}.}
\labt{liferatio}
\begin{center}
\begin{tabular}{lcc} \hline
Lifetime ratio & Measured value & Predicted range \\ \hline
$\tau(\Bu)/\tau(\Bd)$ & \hfagRTAUBU & 1.04 -- 1.08 \\
$\bar{\tau}(\Bs)/\tau(\Bd)^a$ & \hfagRTAUBSMEANCON & 0.99 -- 1.01 \\
$\tau(\Lb)/\tau(\Bd)$ & \hfagRTAULB & 0.86 -- 0.95    \\
$\tau(\mbox{\b-baryon})/\tau(\Bd)$  & \hfagRTAUBB & 0.86 -- 0.95 \\
\hline
\multicolumn{3}{l}{$^a$ \footnotesize 
Using $\bar{\tau}(\Bs) = 1/\Gs = 2/(\Gamma_{\rm L} + \Gamma_{\rm H})$.
}
\end{tabular}
\end{center}
\end{table}
As described in \Sec{lifetimes},
Heavy Quark Effective Theory
can be employed to explain the hierarchy of
$\tau(\Bc) \ll \tau(\Lb) < \bar{\tau}(\Bs) \approx \tau(\Bd) < \tau(\Bu)$,
and used to predict the ratios between lifetimes.
Typical predictions are compared to the measured 
lifetime ratios in \Table{liferatio}.
A recent prediction of the ratio between the \Bu and \Bd lifetimes,
is $1.06 \pm 0.02$~\cite{tarantino}, in good agreement with experiment. 


The total widths of the \Bs and \Bd mesons
are expected to be very close and differ by at most 
1\%~\cite{equal_lifetimes,Gabbiani_et_al}.
However, the experimental ratio $\bar{\tau}(\Bs)/\tau(\Bd)$,
where $\bar{\tau}(\Bs)=1/\Gs$ is obtained from \DGs and 
flavour-specific lifetime measurements, appears to be 
smaller than 1 by 
\hfagONEMINUSRTAUBSMEANCONpercent, 
at deviation with respect to the prediction. 

The ratio $\tau(\Lb)/\tau(\Bd)$ has particularly
been the source of theoretical
scrutiny since earlier calculations~\cite{OPE,lblife_early}
predicted a value greater than 0.90, almost two sigma higher
than the world average at the time. 
Many predictions cluster around a most likely central value
of 0.94~\cite{lblife_mid}.
More recent calculations
of this ratio that include higher-order effects predict a
lower ratio between the
\Lb and \Bd lifetimes~\cite{tarantino,Gabbiani_et_al}
and reduce this difference.
References~\cite{tarantino,Gabbiani_et_al} present probability density functions
of their predictions with variation of theoretical inputs, and the
indicated ranges in \Table{liferatio}
are the RMS of the distributions from the most probable values.
Again, the CDF measurement of the $\Lambda_b$ lifetime
in the exclusive decay mode $J/\psi \Lambda$~\cite{CDFBJPSIXprel2007} is significantly 
higher than the world average before inclusion, with a ratio
to the $\tau(\Bd)$ world average of 
$\tau(\Lb)/\tau(\Bd) = 1.042 \pm 0.057$, 
%
resulting in continued interest in lifetimes of $b$ baryons.



\mysubsection{Neutral \B-meson mixing}
\labs{mixing}

The $\Bd-\Bdbar$ and $\Bs-\Bsbar$ systems
both exhibit the phenomenon of particle-antiparticle mixing. For each of them, 
there are two mass eigenstates which are linear combinations of the two flavour states,
\B and $\bar{\B}$. 
The heaviest (lightest) of the these mass states is denoted
$\B_{\rm H}$ ($\B_{\rm L}$),
with mass $m_{\rm H}$ ($m_{\rm L}$)
and total decay width $\Gamma_{\rm H}$ ($\Gamma_{\rm L}$). We define
\begin{eqnarray}
\Delta m = m_{\rm H} - m_{\rm L} \,, &~~~~&  x = \Delta m/\Gamma \,, \labe{dm} \\
\Delta \Gamma \, = \Gamma_{\rm L} - \Gamma_{\rm H} \,, ~ &~~~~&  y= \Delta\Gamma/(2\Gamma) \,, \labe{dg}
\end{eqnarray}
where 
$\Gamma = (\Gamma_{\rm H} + \Gamma_{\rm L})/2 =1/\bar{\tau}(\B)$ 
is the average decay width.
$\Delta m$ is positive by definition, and 
$\Delta \Gamma$ is expected to be positive within
the Standard Model.\footnote{For reason of symmetry in 
\Eqss{dm}{dg}, $\Delta \Gamma$ is sometimes defined with 
the opposite sign. The definition adopted here, \ie\
\Eq{dg}, is the one used by most experimentalists and many
phenomenologists in \B physics.}

There are four different time-dependent probabilities describing the 
case of a neutral \B meson produced 
as a flavour state and decaying to a flavour-specific final state.
If \CPT is conserved (which  
will be assumed throughout), they can be written as 
\begin{equation}
\left\{
\begin{array}{rcl}
{\cal P}(\B\to\B) & = &  \frac{e^{-\Gamma t}}{2} 
\left[ \cosh\!\left(\frac{\Delta\Gamma}{2}t\right) + \cos\!\left(\Delta m t\right)\right]  \\
{\cal P}(\B\to\bar{\B}) & = &  \frac{e^{-\Gamma t}}{2} 
\left[ \cosh\!\left(\frac{\Delta\Gamma}{2}t\right) - \cos\!\left(\Delta m t\right)\right] 
\left|\frac{q}{p}\right|^2 \\
{\cal P}(\bar{\B}\to\B) & = &  \frac{e^{-\Gamma t}}{2} 
\left[ \cosh\!\left(\frac{\Delta\Gamma}{2}t\right) - \cos\!\left(\Delta m t\right)\right] 
\left|\frac{p}{q}\right|^2 \\
{\cal P}(\bar{\B}\to\bar{\B}) & = &  \frac{e^{-\Gamma t}}{2} 
\left[ \cosh\!\left(\frac{\Delta\Gamma}{2}t\right) + \cos\!\left(\Delta m t\right)\right] 
\end{array} \right. \,,
\labe{oscillations}
\end{equation}
where $t$ is the proper time of the system (\ie\ the time interval between the production 
and the decay in the rest frame of the \B meson). 
At the \B factories, only the proper-time difference $\Delta t$ between the decays
of the two neutral \B mesons from the \Ups can be determined, but, 
because the two \B mesons evolve coherently (keeping opposite flavours as long as
none of them has decayed), the 
above formulae remain valid 
if $t$ is replaced with $\Delta t$ and the production flavour is replaced by the flavour 
at the time of the decay of the accompanying \B meson in a flavour-specific state.
As can be seen in the above expressions,
the mixing probabilities 
depend on three mixing observables:
$\Delta m$, $\Delta\Gamma$,
and $|q/p|^2$ which signals \CP violation in the mixing if $|q/p|^2 \ne 1$.

In the next sections we review in turn the experimental knowledge
on these three parameters, separately 
for the \Bd meson (\dmd, \DGd, $|q/p|_{\particle{d}}$) 
and the \Bs meson (\dms, \DGs, $|q/p|_{\particle{s}}$). 

\mysubsubsection{\Bd mixing parameters}
\labs{qpd}
\labs{dmd}
\labs{DGd}

\subsubsubsection{\boldmath \CP violation parameter $|q/p|_{\particle{d}}$}

Evidence for \CP violation in \Bd mixing
has been searched for,
both with flavor-specific and inclusive \Bd decays, 
in samples where the initial 
flavor state is tagged. In the case of semileptonic 
(or other flavor-specific) decays, 
where the final state tag is 
also available, the following asymmetry
\begin{equation} 
\ASLd = \frac{
N(\hbox{\Bdbar}(t) \to \ell^+      \nu_{\ell} X) -
N(\hbox{\Bd}(t)    \to \ell^- \bar{\nu}_{\ell} X) }{
N(\hbox{\Bdbar}(t) \to \ell^+      \nu_{\ell} X) +
N(\hbox{\Bd}(t)    \to \ell^- \bar{\nu}_{\ell} X) } 
= \frac{|p/q|_{\particle{d}}^2 - |q/p|_{\particle{d}}^2}%
{|p/q|_{\particle{d}}^2 + |q/p|_{\particle{d}}^2}
\labe{ASL}
\end{equation} 
has been measured, either in time-integrated analyses at 
CLEO~\cite{CLEO_chid_CP,CLEO_chid_CP_y,CLEO_CP_semi},
CDF~\cite{CDF_CP_semi,CDF2_ASL_prel} and \dzero~\cite{D0_qp_chibar},
or in time-dependent analyses at 
OPAL~\cite{OPAL_CP_semi}, ALEPH~\cite{ALEPH_CP}, 
\babar~\cite{BABAR_DGd_qp,BABAR_CP_semi,BABAR_qp_dileptons,BABAR_qp_dstarlnu}
and \belle~\cite{BELLE_CP_semi}.
In the inclusive case, also investigated and published
at ALEPH~\cite{ALEPH_CP} and OPAL~\cite{OPAL_CP_incl},
no final state tag is used, and the asymmetry~\cite{incl_asym}
\begin{equation} 
\frac{
N(\hbox{\Bd}(t) \to {\rm all}) -
N(\hbox{\Bdbar}(t) \to {\rm all}) }{
N(\hbox{\Bd}(t) \to {\rm all}) +
N(\hbox{\Bdbar}(t) \to {\rm all}) } 
\simeq
\ASLd \left[ \frac{\dmd}{2\Gd} \sin(\dmd \,t) - 
\sin^2\left(\frac{\dmd \,t}{2}\right)\right] 
\labe{ASLincl}
\end{equation} 
must be measured as a function of the proper time to extract information 
on \CP violation.
In all cases asymmetries compatible with zero have been found,  
with a precision limited by the available statistics. 

A simple average of all measurements performed at 
\B factories~\cite{CLEO_chid_CP_y,CLEO_CP_semi,BABAR_DGd_qp,BABAR_qp_dileptons,BABAR_qp_dstarlnu,BELLE_CP_semi}
yields 
\begin{equation}
\ASLd = \hfagASLDB 
\labe{ASLDB}
\end{equation}
or, equivalently through \Eq{ASL},
\begin{equation}
|q/p|_{\particle{d}} = \hfagQPDB \,.
\end{equation}
Analyses performed at higher energy, either at LEP or at the Tevatron,
can't separate the contributions from the 
\Bd and \Bs mesons. Under the assumption of no \CP violation in \Bs mixing, a number of 
these analyses~\cite{D0_qp_chibar,OPAL_CP_semi,ALEPH_CP,OPAL_CP_incl}
quote a measurement of $\ASLd$ or $|q/p|_{\particle{d}}$ for the \Bd meson. Combining 
these results, as well as that of a recent preliminary CDF analysis\cite{CDF2_ASL_prel}%
\footnote{A low-statistics analysis published by CDF using the Run I data\cite{CDF_CP_semi} 
has not been included.},
with the above \B factory averages leads to 
\begin{equation}
\left.
\begin{array}{l}
\ASLd = \hfagASLDA \\
|q/p|_{\particle{d}} = \hfagQPDA 
\end{array} \right\} ~~ \mbox{if $\ASLs =0$, $|q/p|_{\particle{s}}=1$.}
\end{equation}
These results\footnote{Early analyses and (perhaps hence) the PDG use the complex
parameter $\epsilon_{\B} = (p-q)/(p+q)$; if \CP violation in the mixing in small, 
$\ASLd \cong 4 {\rm Re}(\epsilon_{\B})/(1+|\epsilon_{\B}|^2)$ and our 
current averages  
are ${\rm Re}(\epsilon_{\B})/(1+|\epsilon_{\B}|^2)=\hfagREBDB$ (\B factory
measurements only) and $\hfagREBDA$ (all measurements).}, 
summarized in \Table{qoverp},
are compatible 
with no \CP violation in the \Bd mixing, an assumption we make for the rest 
of this section.

\begin{table}
\caption{Measurements of \CP violation in \Bd mixing and their average
in terms of both \ASLd and $|q/p|_{\particle{d}}$.
The individual results are listed as quoted in the original publications, 
or converted\addtocounter{footnote}{-1}\protect\footnotemark\
to an \ASLd value.
When two errors are quoted, the first one is statistical and the 
second one systematic. The second group of measurements, performed 
at high-energy colliders, assume no \CP violation in \Bs mixing, 
\ie\ $|q/p|_{\particle{s}}=1$.}
\labt{qoverp}
\begin{center}
\begin{tabular}{@{}rcl@{$\,\pm$}l@{$\pm$}ll@{$\,\pm$}l@{$\pm$}l@{}}
\hline
Exp.\ \& Ref. & Method & \multicolumn{3}{c}{Measured \ASLd} 
                       & \multicolumn{3}{c}{Measured $|q/p|_{\particle{d}}$} \\
\hline
CLEO   \cite{CLEO_chid_CP_y} & partial hadronic rec. 
                             & $+0.017$ & 0.070 & 0.014 
                             & \multicolumn{3}{c}{} \\
CLEO   \cite{CLEO_CP_semi}   & dileptons 
                             & $+0.013$ & 0.050 & 0.005 
                             & \multicolumn{3}{c}{} \\
CLEO   \cite{CLEO_CP_semi}   & average of above two 
                             & $+0.014$ & 0.041 & 0.006 
                             & \multicolumn{3}{c}{} \\
\babar \cite{BABAR_DGd_qp}   & full hadronic rec. 
                             & \multicolumn{3}{c}{}  
                             & 1.029 & 0.013 & 0.011  \\
\babar \cite{BABAR_qp_dileptons}  & dileptons
                             & \multicolumn{3}{c}{}
                             & 0.9992 & 0.0027 & 0.0019 \\ 
\babar \cite{BABAR_qp_dstarlnu}$^p$  & part.\ rec.\ $D^{*}\ell\nu$ 
                             & $-0.0130$ & 0.0068 & 0.0040 
                             & 1.0065 & 0.0034 & 0.0020 \\ 
\belle \cite{BELLE_CP_semi}  & dileptons 
                             & $-0.0011$ & 0.0079 & 0.0085 
                             & 1.0005 & 0.0040 & 0.0043 \\
\end{tabular}
\end{center}
\end{table}

\subsubsubsection{\boldmath Mass and decay width differences \dmd and \DGd}

\begin{table}
\caption{Time-dependent measurements included in the \dmd average.
The results obtained from multi-dimensional fits involving also 
the \Bd (and \Bu) lifetimes
as free parameter(s)~\cite{BABAR3,BABAR5,BELLE2} 
have been converted into one-dimensional measurements of \dmd.
All the measurements have then been adjusted to a common set of physics
parameters before being combined. 
The CDF results from Run~II are preliminary.}
\labt{dmd}
\begin{center}
\begin{tabular}{@{}rc@{}cc@{}c@{}cc@{}c@{}c@{}}
\hline
Experiment & \multicolumn{2}{c}{Method} & \multicolumn{3}{l}{\dmd in\invps}   
                                        & \multicolumn{3}{l}{\dmd in\invps}     \\
and Ref.   &  rec. & tag                & \multicolumn{3}{l}{before adjustment} 
                                        & \multicolumn{3}{l}{after adjustment} \\
\hline
 ALEPH~\cite{ALEPH_dmd}  & \particle{ \ell  } & \particle{ \Qjet  } & $  0.404 $ & $ \pm  0.045 $ & $ \pm  0.027 $ & & & \\
 ALEPH~\cite{ALEPH_dmd}  & \particle{ \ell  } & \particle{ \ell  } & $  0.452 $ & $ \pm  0.039 $ & $ \pm  0.044 $ & & & \\
 ALEPH~\cite{ALEPH_dmd}  & \multicolumn{2}{c}{above two combined} & $  0.422 $ & $ \pm  0.032 $ & $ \pm  0.026 $ & $  0.439 $ & $ \pm  0.032 $ & $ ^{+  0.020 }_{-  0.019 } $ \\
 ALEPH~\cite{ALEPH_dmd}  & \particle{ D^*  } & \particle{ \ell,\Qjet  } & $  0.482 $ & $ \pm  0.044 $ & $ \pm  0.024 $ & $  0.482 $ & $ \pm  0.044 $ & $ \pm  0.024 $ \\
 DELPHI~\cite{DELPHI_dmd}  & \particle{ \ell  } & \particle{ \Qjet  } & $  0.493 $ & $ \pm  0.042 $ & $ \pm  0.027 $ & $  0.502 $ & $ \pm  0.042 $ & $ \pm  0.024 $ \\
 DELPHI~\cite{DELPHI_dmd}  & \particle{ \pi^*\ell  } & \particle{ \Qjet  } & $  0.499 $ & $ \pm  0.053 $ & $ \pm  0.015 $ & $  0.501 $ & $ \pm  0.053 $ & $ \pm  0.015 $ \\
\hline
\end{tabular}
\end{center}
\end{table}

Many time-dependent \Bd--\Bdbar oscillation analyses have been performed by the 
ALEPH, \babar, \belle, CDF, \dzero, DELPHI, L3 and OPAL collaborations. 
The corresponding measurements of \dmd are summarized in 
\Table{dmd},
where only the most recent results
are listed (\ie\ measurements superseded by more recent ones have been omitted). 
Although a variety of different techniques have been used, the 
individual \dmd
results obtained at high-energy colliders have remarkably similar precision.
Their average is compatible with the recent and more precise measurements 
from the asymmetric \B factories.
The systematic uncertainties are not negligible; 
they are often dominated by sample composition, mistag probability,
or \b-hadron lifetime contributions.
Before being combined, the measurements are adjusted on the basis of a 
common set of input values, including the averages of the 
\b-hadron fractions and lifetimes given in this report 
(see \Secss{fractions}{lifetimes}).
Some measurements are statistically correlated. 
Systematic correlations arise both from common physics sources 
(fractions, lifetimes, branching ratios of \b hadrons), and from purely 
experimental or algorithmic effects (efficiency, resolution, flavour tagging, 
background description). Combining all published measurements
listed in \Table{dmd}
and accounting for all identified correlations
as described in~\cite{LEPHFS} yields $\dmd = \hfagDMDWfull$.

On the other hand, ARGUS and CLEO have published 
measurements of the time-integrated mixing probability 
\chid~\cite{ARGUS_chid,CLEO_chid_CP,CLEO_chid_CP_y}, 
which average to $\chid =\hfagCHIDU$.
Following \Ref{CLEO_chid_CP_y}, 
the width difference \DGd could 
in principle be extracted from the
measured value of $\Gd=1/\tau(\Bd)$ and the above averages for 
\dmd and \chid 
(provided that \DGd has a negligible impact on 
the \dmd $\tau(\Bd)$ analyses that have assumed $\DGd=0$), 
using the relation
\begin{equation}
\chid = \frac{\xd^2+\yd^2}{2(\xd^2+1)} ~~~ \mbox{with} ~~ \xd=\frac{\dmd}{\Gd} 
~~~ \mbox{and} ~~ \yd=\frac{\DGd}{2\Gd} \,.
\labe{chid_definition}
\end{equation}
However, direct time-dependent studies provide much stronger constraints: 
$|\DGd|/\Gd < 18\%$ at \CL{95} from DELPHI~\cite{DELPHI_dmd_dms_vtx},
and $-6.8\% < {\rm sign}({\rm Re} \lambda_{\CP}) \DGGd < 8.4\%$
at \CL{90} from \babar~\cite{BABAR_DGd_qp},
where $\lambda_{\CP} = (q/p)_{\particle{d}} (\bar{A}_{\CP}/A_{\CP})$
is defined for a \CP-even final state 
(the sensitivity to the overall sign of 
${\rm sign}({\rm Re} \lambda_{\CP}) \DGGd$ comes
from the use of \Bd decays to \CP final states).
Combining these two results after adjustment to 
$1/\Gd=\tau(\Bd)=\hfagTAUBD$ yields
\begin{equation}
{\rm sign}({\rm Re} \lambda_{\CP}) \DGGd  = \hfagSDGDGD \,.
\end{equation}
The sign of ${\rm Re} \lambda_{\CP}$ is not measured,
but expected to be positive from the global fits
of the Unitarity Triangle within the Standard Model.

Assuming $\DGd=0$ 
and using $1/\Gd=\tau(\Bd)=\hfagTAUBD$,
the \dmd and \chid results are combined through \Eq{chid_definition} 
to yield the 
world average
\begin{equation} 
\dmd = \hfagDMDWU \,,
\labe{dmd}
\end{equation} 
or, equivalently,
\begin{equation} 
\xd= \hfagXDWU ~~~ \mbox{and} ~~~ \chid=\hfagCHIDWU \,.  
\labe{chid}
\end{equation}
\Figure{dmd} compares the \dmd values obtained by the different experiments.

\begin{figure}
\begin{center}
\epsfig{figure=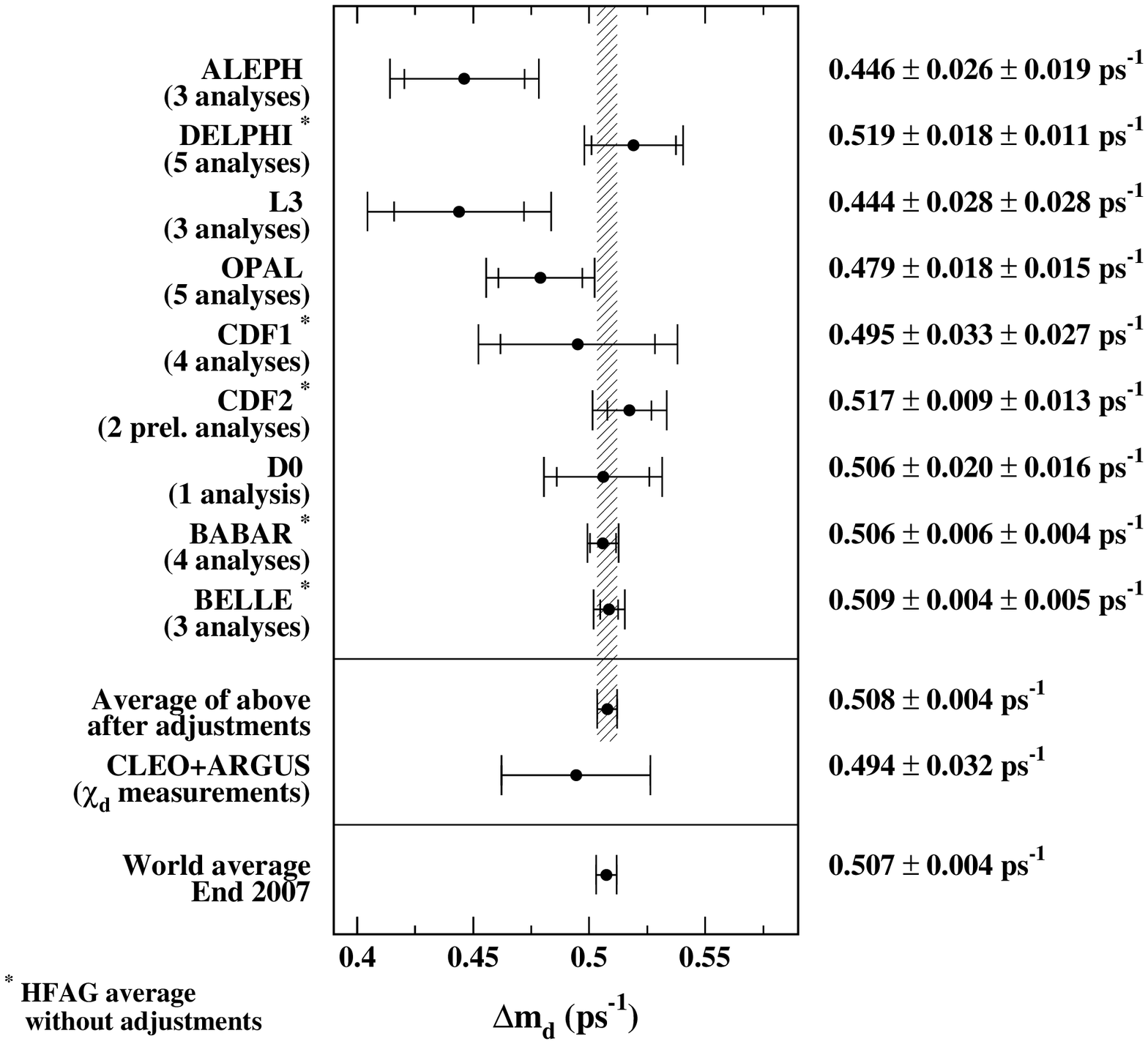,width=\textwidth}
\caption{The \Bd--\Bdbar oscillation frequency \dmd as measured by the different experiments. 
The averages quoted for ALEPH, L3 and OPAL are taken from the original publications, while the 
ones for DELPHI, CDF, \babar, and \belle have been computed from the individual results 
listed in \Table{dmd} without performing any adjustments. The time-integrated measurements 
of \chid from the symmetric \B factory experiments ARGUS and CLEO have been converted 
to a \dmd value using $\tau(\Bd)=\hfagTAUBD$. The two global averages have been obtained 
after adjustments of all the individual \dmd results of \Table{dmd} (see text).}
\labf{dmd}
\end{center}
\end{figure}

The \Bd mixing averages given in \Eqss{dmd}{chid}
and the \b-hadron fractions of \Table{fractions} have been obtained in a fully 
consistent way, taking into account the fact that the fractions are computed using 
the \chid value of \Eq{chid} and that many individual measurements of \dmd
at high energy depend on the assumed values for the \b-hadron fractions.
Furthermore, this set of averages is consistent with the lifetime averages 
of \Sec{lifetimes}.

\begin{table}
\caption{Simultaneous measurements of \dmd and $\tau(\Bd)$, and their average.
The \belle analysis also 
measures $\tau(\Bu)$ at the same time, but it is converted here into a two-dimensional measurement 
of \dmd and $\tau(\Bd)$, for an assumed value of $\tau(\Bu)$. 
The first quoted error on the measurements is statistical
and the second one systematic; in the case of adjusted measurements, the 
latter includes a contribution obtained from the variation of $\tau(\Bu)$ or 
$\tau(\Bu)/\tau(\Bd)$ in the indicated range. Units are\invps\ for \dmd
and\unit{ps} for lifetimes. 
The three different values of $\rho(\dmd,\tau(\Bd))$ correspond 
to the statistical, systematic and total correlation coefficients
between the adjusted measurements of \dmd and $\tau(\Bd)$.}
\labt{dmd2D}
\begin{center}
\begin{tabular}{@{}r@{~}c@{}c@{}c@{~}c@{}c@{}c@{~}c@{}c@{}c@{\hspace{0ex}}c@{}}
\hline
Exp.\ \& Ref.
& \multicolumn{3}{c}{Measured \dmd}   
& \multicolumn{3}{c}{Measured $\tau(\Bd)$}   
& \multicolumn{3}{c}{Measured $\tau(\Bu)$}   
&  Assumed $\tau(\Bu)$ \\
\hline
\babar \cite{BABAR3}  
      & $0.492$ & $\pm 0.018$ & $\pm 0.013$ 
      & $1.523$ & $\pm 0.024$ & $\pm 0.022$ 
      & \multicolumn{3}{c}{---}
      & $(1.083\pm 0.017)\tau(\Bd)$ \\  
\babar \cite{BABAR5}  
      & $0.511$ & $\pm 0.007$ & $^{+0.007}_{-0.006}$ 
      & $1.504$ & $\pm 0.013$ & $^{+0.018}_{-0.013}$
      & \multicolumn{3}{c}{---}
      & $1.671\pm 0.018$ \\  
\belle \cite{BELLE2}  
      & $0.511$ & $\pm 0.005$ & $\pm 0.006$
      & $1.534$ & $\pm 0.008$ & $\pm 0.010$
      & $1.635$ & $\pm 0.011$ & $\pm 0.011$
      & --- \\  
\cline{2-10}
& \multicolumn{3}{c}{Adjusted \dmd}   
& \multicolumn{3}{c}{Adjusted $\tau(\Bd)$}   
& \multicolumn{3}{c}{$\rho(\dmd,\Bd)$} 
&  Assumed $\tau(\Bu)$ \\
\cline{2-10}
\babar \cite{BABAR3}  
      & $0.492$ & $\pm 0.018$ & $\pm 0.013$  
      & $1.524$ & $\pm 0.025$ & $\pm 0.022$  
      & $-0.22$ & $+0.74$ & $+0.16$ 
      & $(\hfagRTAUBUval$$\hfagRTAUBUerr)\tau(\Bd)$ \\  
\babar \cite{BABAR5} 
      & $0.512$ & $\pm 0.007$ & $\pm 0.007$  
      & $1.506$ & $\pm 0.013$ & $\pm 0.018$ 
      & $+0.01$ & $-0.85$ & $-0.48$ 
      & $\hfagTAUBUval$$\hfagTAUBUerr$ \\
\hline 
\end{tabular}
\end{center}
\end{table}
\begin{figure}
\begin{center}
\vspace{-0.5cm}
\epsfig{figure=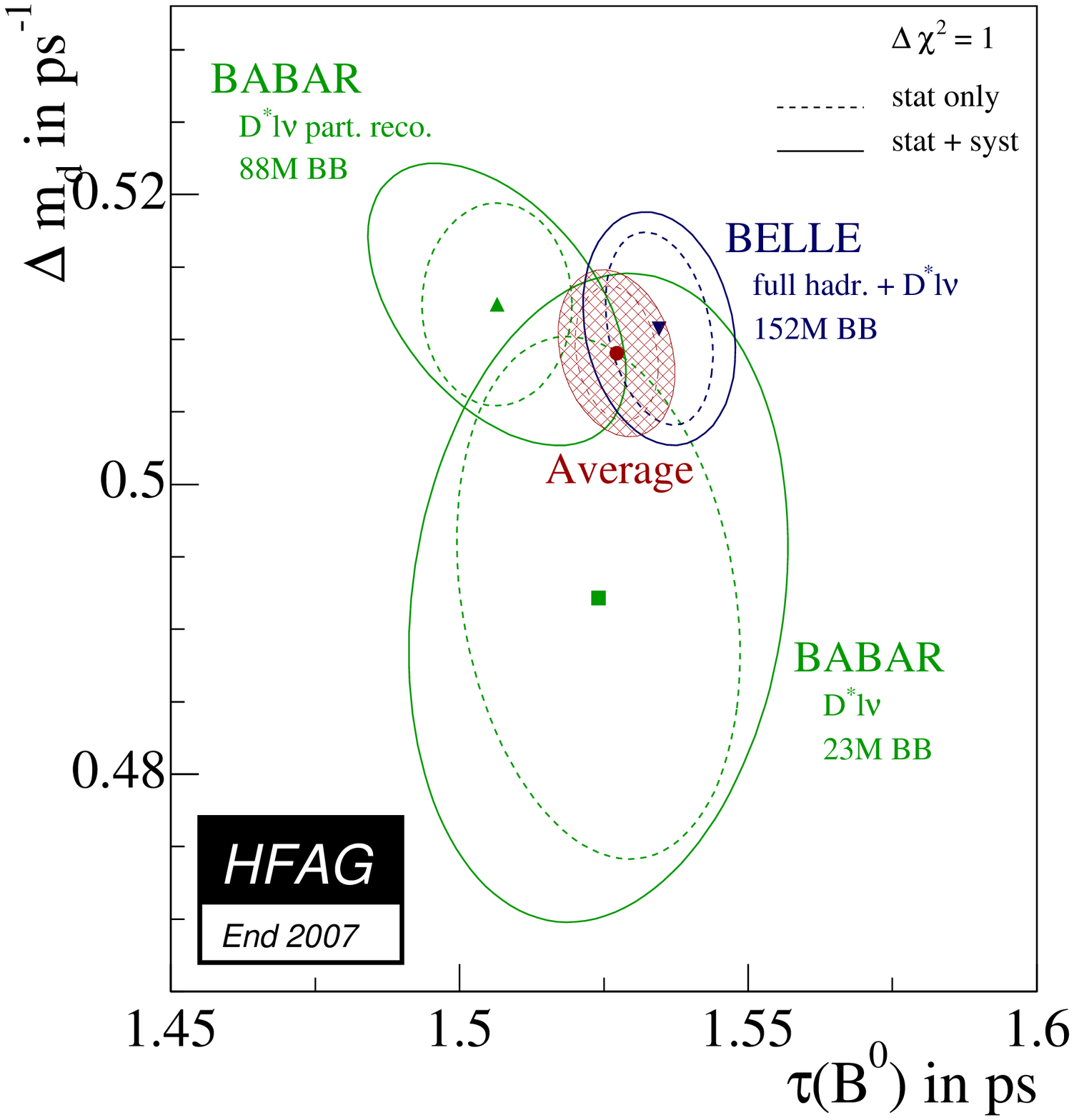,width=0.6\textwidth}
\vspace{-0.5cm}
\caption{Simultaneous measurements of
\dmd and $\tau(\Bd)$~\cite{BABAR3,BABAR5,BELLE2}, 
after adjustment to a common set of parameters (see text). 
Statistical and total uncertainties are represented as dashed and
solid contours respectively.
The average of the three measurements
is indicated by a hatched ellipse.}
\labf{dmd2D}
\end{center}
\end{figure}

It should be noted that the most recent (and precise) analyses at the 
asymmetric \B factories measure \dmd
as a result of a multi-dimensional fit. 
Two \babar analyses~\cite{BABAR3,BABAR5},  
based on fully and partially reconstructed $\Bd \to D^*\ell\nu$ decays
respectively, 
extract simultaneously \dmd and $\tau(\Bd)$
while the latest \belle analysis~\cite{BELLE2},  
based on fully reconstructed hadronic \Bd decays and $\Bd \to D^*\ell\nu$ decays, 
extracts simultaneously \dmd, $\tau(\Bd)$ and $\tau(\Bu)$.
The measurements of \dmd and $\tau(\Bd)$ of these three analyses 
are displayed in \Table{dmd2D} and in \Fig{dmd2D}. Their two-dimensional average, 
taking into account all statistical and systematic correlations, and expressed
at $\tau(\Bu)=\hfagTAUBU$, is
\begin{equation}
\left.
\begin{array}{r@{}l}
\dmd = \hfagDMDTWODnounit & \invps \\
\tau(\Bd) = \hfagTAUBDTWODnounit & \ps
\end{array}
\right\}
~\mbox{with a total correlation of \hfagRHODMDTAUBD.}
\end{equation}

\mysubsubsection{\Bs mixing parameters}
\labs{qps}
\labs{DGs}
\labs{dms}

\subsubsubsection{\boldmath \CP violation parameter $|q/p|_{\particle{s}}$}

Constraints on a combination of $|q/p|_{\particle{d}}$ and 
$|q/p|_{\particle{s}}$ (or equivalently \ASLd and \ASLs) 
have been explicitly quoted by the Tevatron 
experiments, using inclusive semileptonic decays of \b hadrons:
\begin{eqnarray}
\frac{1}{4}\left(f'_{\particle{d}} \,\chid \ASLd +
                 f'_{\particle{s}} \,\chis \ASLs \right) = 
+0.0015 \pm 0.0038 \mbox{(stat)} \pm 0.0020 \mbox{(syst)}
&~~& \mbox{CDF1~\cite{CDF_CP_semi}} \,, 
\labe{CDF_ASLDS} \\
\frac{f'_{\particle{d}}Z_{\particle{d}} \ASLd + f'_{\particle{s}}Z_{\particle{s}} \ASLs}%
{f'_{\particle{d}}Z_{\particle{d}} + f'_{\particle{s}}Z_{\particle{s}}} =
+0.0080 \pm 0.0090 \mbox{(stat)} \pm 0.0068 \mbox{(syst)}
&~~& \mbox{CDF2~\cite{CDF2_ASL_prel}} \,,
\labe{CDF2_ASLDS} \\
\frac{1}{4}\left(\ASLd +
\ASLs \frac{f'_\particle{s}\chis}{f'_\particle{d}\chid} \right) =
-0.0023 \pm 0.0011 \mbox{(stat)} \pm 0.0008 \mbox{(syst)}
&~~& \mbox{\dzero~\cite{D0_qp_chibar}} \,,
\labe{CDF_Dzero_ASLDS}
\end{eqnarray}
where\footnote{In Ref.~\cite{D0_combined_Bs_CPV}, the \dzero result~\cite{D0_qp_chibar}
was reinterpreted by replacing $\chi_{\particle{s}}/\chi_{\particle{d}}$
with $Z_{\particle{s}}/Z_{\particle{d}}$.
For simplicity, and since this has anyway a negligible numerical effect on our
combined result of \Eq{ASLs}, we 
follow the same interpretation and set $\chi_{\particle{q}}=Z_{\particle{q}}/2$
in \Eqss{CDF_ASLDS}{CDF_Dzero_ASLDS}. We also set $f'_{\particle{q}}=f_{\particle{q}}$.}
$Z_{\particle{q}} = 1/(1-y_{\particle{q}}^2)-1/(1+x_{\particle{q}}^2)
= 2 \chi_{\particle{q}}/(1-y_{\particle{q}}^2)$, $q=d,s$.
In addition a first direct determination of \ASLs
and hence $|q/p|_{\particle{s}}$ has
been obtained by \dzero by measuring the charge asymmetry of
$\Bs \rightarrow D_s \mu \nu$ decays:
\begin{eqnarray}
\mbox{\hspace{3cm}}
\ASLs = +0.0245 \pm 0.0193 \mbox{(stat)} \pm 0.0035 \mbox{(syst)}
&~~& \mbox{\dzero~\cite{D0_semi_asym}}\,.
\end{eqnarray}

Given the average $\ASLd = \hfagASLDB$ of \Eq{ASLDB},
obtained from results at \B factories,
as well as other averages presented in this chapter
for the quantities appearing in 
\Eqsss{CDF_ASLDS}{CDF2_ASLDS}{CDF_Dzero_ASLDS}, 
these four results are turned into measurements of \ASLs 
(displayed in \Fig{ASLs}) and combined to yield
\begin{figure}
\begin{center}
\epsfig{figure=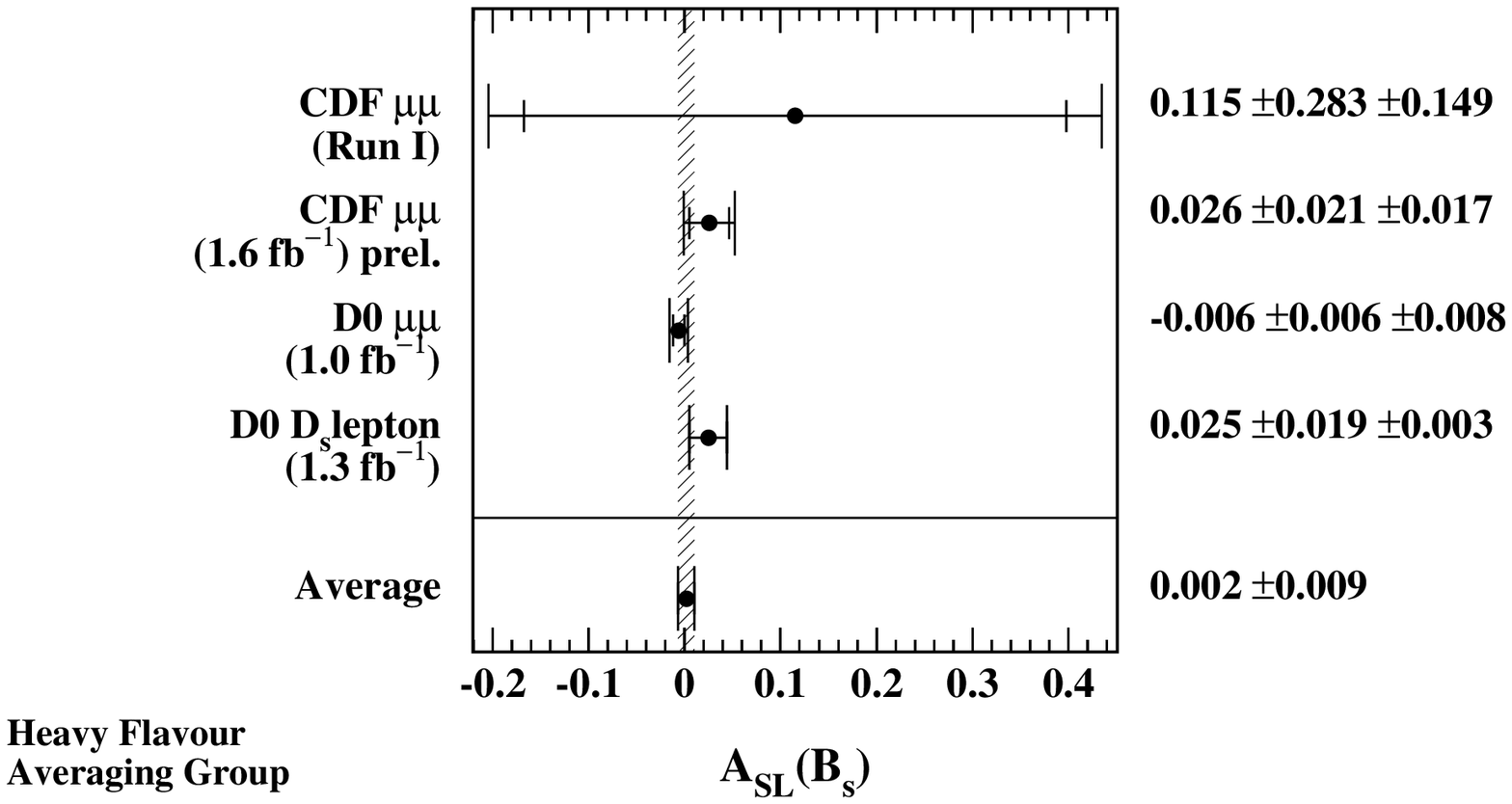,width=0.8\textwidth}
\caption{Measurements of \ASLs, derived from CDF~\cite{CDF_CP_semi,CDF2_ASL_prel}
and \dzero~\cite{D0_qp_chibar,D0_semi_asym} analyses and ajusted to the latest 
averages of \ASLd, \b-hadron fractions and mixing parameters. 
The combined value of \ASLs is also shown.}
\labf{ASLs}
\end{center}
\end{figure}
\begin{equation}
\ASLs = \hfagASLSval\hfagASLSsta\mbox{(stat)}\hfagASLSsys\mbox{(syst)} = \hfagASLS 
\labe{ASLs}
\end{equation}
or, equivalently through \Eq{ASL},
\begin{equation}
|q/p|_{\particle{s}} = \hfagQPSval\hfagQPSsta\mbox{(stat)}\hfagQPSsys\mbox{(syst)} = \hfagQPS \,.
\end{equation}
The quoted systematic errors include experimental systematics as well as the correlated dependence on external 
parameters. These results are compatible with no \CP violation in \Bs
mixing, an assumption made in almost all of the results described below.


%
%
%

\subsubsubsection{Decay width difference \DGs}



Definitions and an introduction to \DGs can also 
be found in \Sec{taubs}.
Neglecting \CP violation, the mass eigenstates are
also \CP eigenstates, with the short-lived state being
\CP-even and the long-lived one being \CP-odd.
Information on \DGs can be obtained by studying the proper time 
distribution of untagged data samples enriched in 
\Bs mesons~\cite{Hartkorn_Moser}.
In the case of an inclusive \Bs selection~\cite{L3B01} or a semileptonic 
\Bs decay selection~\cite{DELBS0_dms_dgs,CDFBS,D0BS2}, 
both the short- and long-lived
components are present, and the proper time distribution is a superposition 
of two exponentials with decay constants 
$\Gs\pm\DGs/2$.
In principle, this provides sensitivity to both \Gs and 
$(\DGGs)^2$. Ignoring \DGs and fitting for 
a single exponential leads to an estimate of \Gs with a 
relative bias proportional to $(\DGGs)^2$. 
An alternative approach, which is directly sensitive to first order in \DGGs, 
is to determine the lifetime of \Bs candidates decaying to \CP
eigenstates; measurements exist for 
\particle{\Bs\to J/\psi\phi}~\cite{CDFIN_BS1,CDFBJPSIXprel2007,D0BS1} and
\particle{\Bs\to D_s^{(*)+} D_s^{(*)-}}, discussed later, which are 
mostly \CP-even states~\cite{Aleksan}. 
However, more recent
time-dependent angular analyses of \particle{\Bs\to J/\psi\phi} 
allow the simultaneous extraction of \DGs and the \CP-even and \CP-odd 
amplitudes~\cite{CDF2_DGs,CDF2_DGs_FT,D01_DGs}.
Flavor tagging the \Bs (or $\bar{B}^0_s$)
that subsequently decays to \particle{J/\psi\phi}
allows for a more effective
extraction of the weak mixing phase as discussed later.
The CDF analysis~\cite{CDF2_DGs} that does not employ flavor tagging
under the assumption of no \CP violation provides a better measurement
of \DGs and is used here, while the CDF analysis~\cite{CDF2_DGs_FT}
that does use flavor tagging is used as an input for determining
an average weak mixing phase in the next subsection.
The \dzero flavor-tagged 
\particle{\Bs\to J/\psi\phi} analysis~\cite{D01_DGs} gives separate results
both assuming the very small SM value of mixing-induced 
CP violation in the \Bs system
(effectively zero compared to current experimental resolution), and 
also allowing for large CP violation.

\begin{table}
\caption{Experimental constraints on \DGGs from lifetime analyses,
assuming no (or very small SM) \CP violation.
The upper limits,
which have been obtained by the working group, are quoted at the \CL{95}.}
\labt{dgammat}
\begin{center}
\begin{tabular}{l|c|c|c}
\hline
Experiment & Method            & $\Delta \Gs/\Gs$ & Ref.  \\
\hline
L3         & lifetime of inclusive \b-sample              
           & $<0.67$   & \cite{L3B01}      \\
DELPHI     & $\Bsb\to D_s^+\ell^- \overline{\nu_{\ell}} X$, lifetime
	   & $<0.46$   & \cite{DELBS0_dms_dgs} \\
DELPHI     & $\Bsb \to D_s^+$ hadron, lifetime
           & $<0.69$ & \cite{DELBS1_dms_excl}   \\
CDF1       & $\Bs \to J/\psi\phi$, lifetime
	   & $0.33^{+0.45}_{-0.42}$ & \cite{CDFIN_BS1} \\ \hline
	   &                   & $ \DGs$  \\ \hline
CDF2       & $\Bs \to J/\psi\phi$, time-dependent angular analysis
           & $0.076^{+0.059}_{-0.063}{\pm 0.006\invps}$ & \cite{CDF2_DGs} \\ 
\dzero     & $\Bs \to J/\psi\phi$, time-dependent angular analysis
           & $0.14{\pm 0.07\invps}$ & \cite{D01_DGs} \\
	 \hline
	 \end{tabular}
	 \end{center}
	 \end{table}

Measurements quoting \DGs results from lifetime analyses 
are listed in \Table{dgammat} under the hypothesis of no
(or very small SM) \CP violation.
There is significant correlation
between \DGs and $1/\Gamma_s$. In order to combine these measurements,
the two-dimensional log-likelihood for each measurement
in the $(1/\Gs,\,\DGs)$ plane is summed and the total
normalized with respect to its minimum.  The one-sigma contour (corresponding
to 0.5 units of log-likelihood greater than the minimum) and
95\% contour are found. 
Inputs as indicated in \Table{dgammat} were used in the combination, 
with the exception of the L3~\cite{L3B01} result since the likelihood
in this case was not available.

Results of the combination are shown as the one-sigma contour
labeled ``Direct" in both plots of \Fig{DGs}.  Transformation
of variables from $(1/\Gs,\,\DGs)$ space to other pairs
of variables such as $(1/\Gs,\,\DGGs)$ and 
$(\tau_{\rm L} = 1/\Gamma_{\rm L},\,\tau_{\rm H} = 1/\Gamma_{\rm H})$
are also made.
The resulting one-sigma contour for the latter is shown in
\Fig{DGs}(b). 

\begin{figure}
\begin{center}
\epsfig{figure=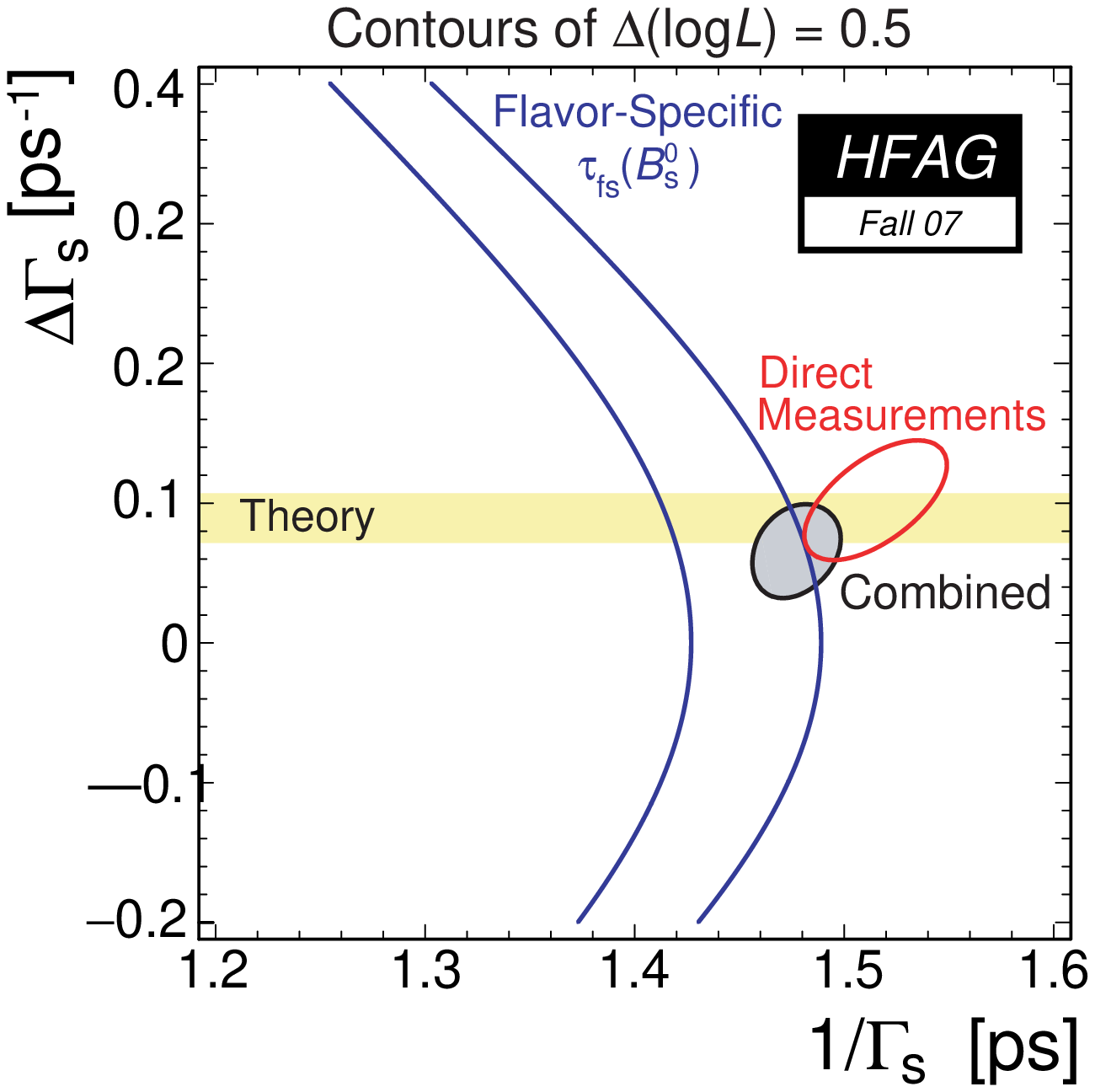,width=0.45\textwidth}
\hfill
\epsfig{figure=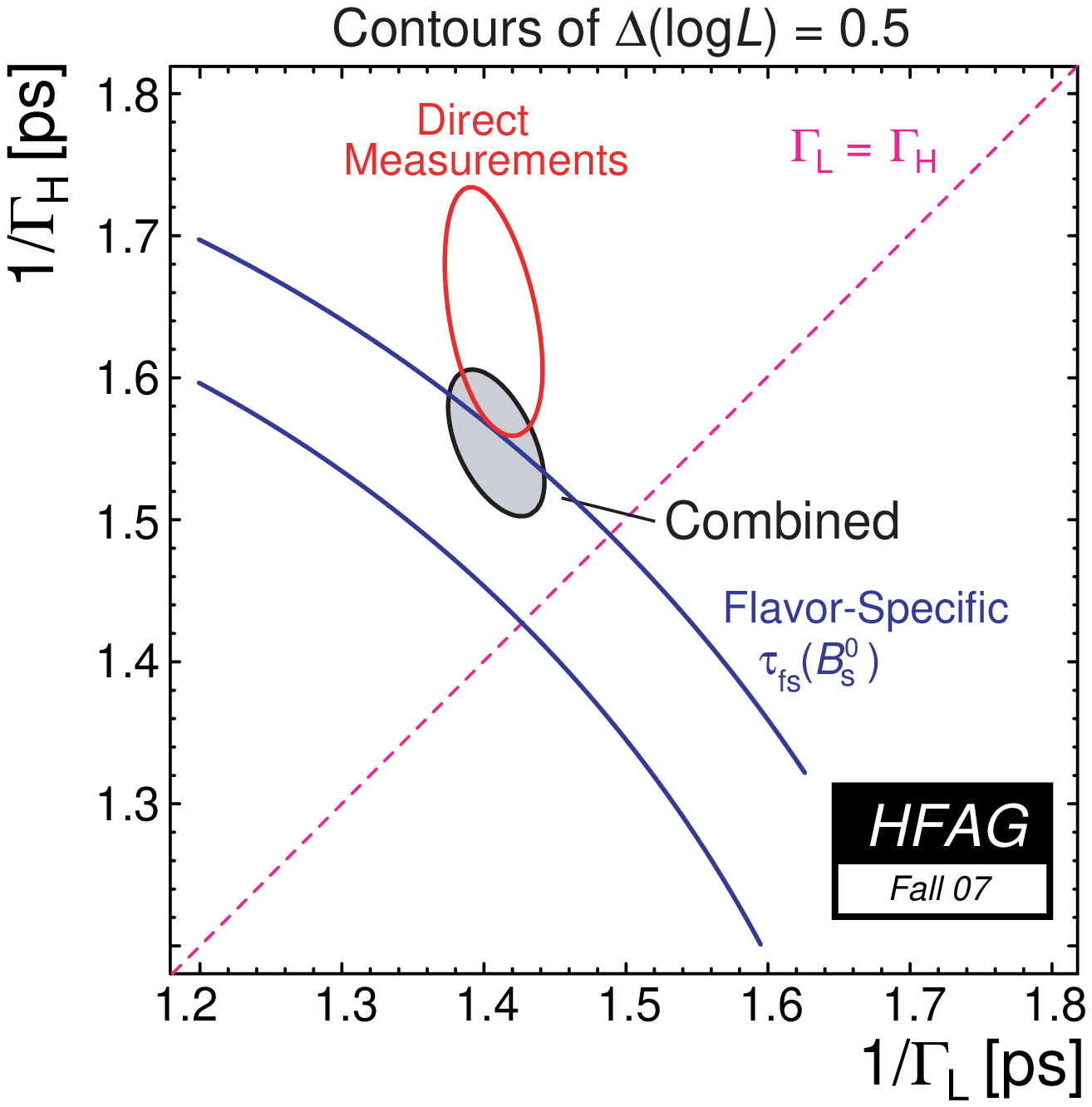,width=0.45\textwidth}
\caption{\DGs combination results with one-sigma contours
($\Delta\log\mathcal{L} = 0.5$) shown for (a) \DGs versus
$\bar{\tau}(\Bs) = 1/\Gs$  and (b)
$\tau_{\rm H} = 1/\Gamma_{\rm H}$ versus $\tau_{\rm L} = 1/\Gamma_{\rm L}$.
The red contours labeled ``Direct" are the result of the combination of
most measurements of \Table{dgammat}, the blue bands are the one-sigma
contours due to the world average of flavor-specific 
\Bs lifetime measurements,
and the shaded region the combination constraints described in the
text.
In (b), the diagonal dashed line indicates 
$\Gamma_{\rm L} = \Gamma_{\rm H}$, \ie, where $\DGs = 0$.}
\labf{DGs}
\end{center}
\end{figure}




\comment{

{\em This text below taken by Donatella from previous \DGs notes \ldots}

\newcommand{\Bh}{B^{\rm heavy}_{d,s}}
\newcommand{\Bl}{B^{\rm light}_{d,s}}
\newcommand{\Mh}{m^{\rm heavy}_{d,s}}
\newcommand{\Ml}{m^{\rm light}_{d,s}}
\newcommand{\Gh}{\Gamma^{\rm heavy}_{d,s}}
\newcommand{\Gl}{\Gamma^{\rm light}_{d,s}}
\newcommand{\Gsho}{\Gamma^{\rm short}_{d,s}}
\newcommand{\Glon}{\Gamma^{\rm long}_{d,s}}
\newcommand{\G}{\Gamma_{d,s}}
\newcommand{\tb}{\tau(B^0_{d,s})}
\newcommand{\dg}{\Delta\Gamma_{d,s}}
\newcommand{\tbssemi}{\tau(B_s)_{\rm semi}}
\newcommand{\Bssh}{B^{\rm short}_s}
\newcommand{\tbsshort}{\tau(\Bssh)}

The Standard Model predicts that the \Bs and \Bd can mix before decay. 
The phenomenology of this interaction can be 
described in terms of a $2\times 2$ effective Hamiltonian matrix, 
$M - i\Gamma /2$.
This results in new states called heavy and light, 
$\Bh$  and $\Bl$, for \Bs and \Bd with masses $\Mh$ ,
$\Ml$. Also the  widths  $\Gh$ and $\Gl$ could be different.

Neglecting \CP violation, the mass eigenstates are also \CP eigenstates, the ``long''  state being 
\CP even and the short one being \CP odd.  For convenience of notation, in the following
we therefore substitute 
$\Gl \equiv \Gsho$ and $\Gh \equiv \Glon$, and 
define $\G=1/\tb=(\Glon+\Gsho)/2$ and 
$\Delta \G = \Gsho-\Glon$ which is positive.

$\dg$ is related to the off-diagonal matrix elements, which have been recently 
calculated at the NLO including NLO QCD correction~\cite{Ciuchini2}. 
The theoretical values are:
\begin{equation}
\DGGs = (7.4\pm 2.4 )\times 10^{-2} \,, \hspace{1truecm} \DGGd = (2.42 \pm 0.59 ) \times 10^{-3} \,.
\end{equation}

In the same work the ratio $\DGd/\DGs$ is evaluated since the uncertainties 
coming from higher orders of QCD and $\Lambda_{\rm QCD}/m_{\b}$ corrections cancel out:
\begin{equation}
\DGd/\DGs = (3.2\pm 0.8)\times10^{-2}
\end{equation}

Experimentally \DGs can be measured fitting the lifetime of the 
light and heavy component of the \Bs.
An alternative method is based on the measurement of the  
branching fraction \particle{\Bs\to D_s^{(*)+}D_s^{(*)-}}.
Methods based on lifetime measurements have two different approaches.
Double exponential lifetime fits to samples containing a mixture of \CP eigenstates like
inclusive or semileptonic \Bs decays or $\Bs\to D_s$-hadron have a quadratic sensitivity 
to \DGs.
Whereas the isolation of a single \CP eigenstate as $\Bs\to\phi\phi$ or 
\particle{\Bs\to J/\psi\phi} to extract the lifetime of the \CP-even or odd state have 
a linear dependence on \DGs and it is more sensitive to \DGs but tend 
to suffer from reduced statistics.
The branching fraction method, exploited by ALEPH~\cite{ALEPH-phiphi}, 
is based on several theoretical assumptions~\cite{theoBR}, and allows to have 
information on \DGs only through the branching fraction measurement:
\begin{equation}
\BR{B_s\to D_s^{(*)+}
D_s^{(*)-}} = \frac{\DGs}{\Gs\left(1+\frac{\DGs}{2\Gs}\right)} \,.
\label{eq:dg_ratio}
\end{equation}

The available results are summarized in \Table{dgammat}. 
The values of the limit on \DGGs quoted in the last column of this 
table have been obtained by the working group.

Details on how these measurements are included in the average can be found  
in the previous summaries~\cite{Pcomb}.

\begin{table}
\caption{Experimental constraints on \DGGs. The upper limits,
which have been obtained by the working group, are quoted at the \CL{95}.}
\labt{dgammat}
\begin{center}
\begin{tabular}{|l|c|c|c|} 
\hline
Experiment & Selection        & Measurement            & $\Delta \Gs/\Gs$ \\ 
\hline
L3~\cite{L3B01}         & inclusive \b-sample              &                               & $<0.67$         \\
DELPHI~\cite{DELBS0_dms_dgs}     & $\Bsb\to D_s^+\ell^- \overline{\nu_{\ell}} X$ & $\tbssemi=(1.42^{+0.14}_{-0.13}\pm0.03)$~ps  & $<0.46$ \\
others~\cite{ref:others}& $\Bsb\to D_s^+\ell^-  \overline{\nu_{\ell}} X$  & $\tbssemi=(1.46\pm{0.07})$~ps & $<0.30$ \\
ALEPH~\cite{ALEPH-phiphi}      & $\Bs\to\phi\phi X$      & 
$\BR{\Bssh \to D_s^{(*)+} D_s^{(*)-}} =(23\pm10^{+19}_{-~9})\%$       & $0.26^{+0.30}_{-0.15}$ \\
ALEPH~\cite{ALEPH-phiphi}      & $\Bs\to\phi\phi X$      & $\tbsshort=(1.27\pm0.33\pm0.07)$~ps           & 
$0.45^{+0.80}_{-0.49}$ \\ 
DELPHI~\cite{DELBS0_dms_dgs}$^a$    & $\Bsb \to D_s^+$ hadron    
&  $\tau_{\rm B^{D_s-had.}_s}=(1.53^{+0.16}_{-0.15}\pm0.07)$~ps                          & $<0.69$         \\
CDF~\cite{CDFB01} & $\Bs \to {\rm J}/\psi\phi$        
& $\tau_{\rm B^{{\rm J}/\psi \phi}_s}=(1.34^{+0.23}_{-0.19}\pm0.05)$~ps & $0.33^{+0.45}_{-0.42}$ \\ 
\hline
\multicolumn{4}{l}{$^a$ \footnotesize 
The value quoted for the measured lifetime differs
slightly from the one quoted in \Table{bs} because it} \\[-1ex]
\multicolumn{4}{l}{~~ \footnotesize 
corresponds to the present status of the analysis in which the information
on \DGs has been obtained.}
\end{tabular}
\end{center}
\end{table}

Here only a short description will be given.

L3 and DELPHI use inclusively reconstructed
\Bs and $\Bs\to \particle{D_s} \ell\nu X$ events respectively.
If those sample are fitted assuming a single exponential lifetime then,
assuming \DGGs is small, the measured lifetime is given by:
\begin{equation}
\tau(\Bs)_{\rm incl.} = \frac{1}{\Gs} \frac{1}{1-\left(\frac{\DGs}{2\Gs}\right)^2}
\quad \quad ; \quad \quad
\tau(\Bs)_{\rm semi.} = \frac{1}{\Gs} 
\frac{{1+\left(\frac{\DGs}{2\Gs}\right)^2}}{{1-\left(\frac{\DGs}{2\Gs}\right)^2}}.     
\end{equation}

The single lifetime fit is thus more sensitive to the effects of 
\DGs in the semileptonic case than in the fully inclusive case.

The same method is used for the \Bs world average lifetime
(recomputed without the DELPHI measurement) obtained
by using only the semileptonic decays and referenced in \Table{dgammat} as {\it others}.

The technique of reconstructing only decays at defined \CP
has been exploited by ALEPH, DELPHI and CDF.

ALEPH reconstructs the decay
$\Bs\to \particle{D_s^{(*)+}D_s^{(*)-}} \to \phi\phi X$
which is predominantly \CP even.
The proper time dependence of the \Bs component is a simple
exponential and the lifetime is related to \DGs via
\begin{equation}
\frac{\Delta \Gs}{\Gs}=2(\frac{1}{\Gs~\tbsshort}-1).  
\end{equation} 
 The same data have been used by ALEPH to exploit the branching fraction method.

DELPHI uses a sample of $\Bs\to D_s$-hadron,
which is expected to have an increased \CP-even component as the contribution
due to \particle{D_s^{(*)+}_s^{(*)-}} events is enhanced by
selection criteria.

CDF reconstructs \particle{\Bs\to J/\psi\phi} with
\particle{J/\psi\to\mu^+\mu^-} and \particle{\phi\to K^+K^-}
where the \CP-even component is equal to $0.84\pm 0.16$ obtained by
combining CLEO~\cite{cleo} measurement of \CP-even fraction in
\particle{\Bd\to J/\psi K^{*0}} and possible SU(3) symmetry
correction.

In order to combine all the measurements~\footnote{L3 is not 
included since the likelihood for the results
was not available} the two-dimensional log-likelihood in the ($1/\Gs$, \DGGs) 
plane is summed and normalized with respect to its minimum.
The 68\%, 95\% and \CL{99} contours of the combined negative 
log-likelihood are shown in \Fig{dgplot} (left)
The corresponding limit on \DGGs is:
\begin{eqnarray}
\DGGs & = & 0.16^{+0.15}_{-0.16}  \,, \\
\DGGs & < & 0.54~\mbox{at \CL{95}} \,. 
\end{eqnarray}

\begin{figure}
\begin{center}
\epsfig{figure=figures/osc/dg_w_notaubd_bw_2d.eps,width=\textwidth}
\epsfig{figure=figures/osc/dg_w_taubd_bw_2d.eps,width=\textwidth}
\end{center}
\caption{Top: 68\%, 95\% and \CL{99} contours of the negative log-likelihood 
distribution in the plane ($1/\Gs$, \DGGs).
Bottom: Same, but with the constraint $1/\Gs \equiv\tau_{\Bd}$} 
\labf{dgplot}
\end{figure}

\begin{figure}
\begin{center}
\epsfig{figure=figures/osc/dg_w_taubd_col_1d,width=\textwidth}
\end{center}
\caption{Probability density distribution for \DGGs after applying the constraint; 
the three shaded regions show the limits at the 68\%, 95\% and \CL{99} respectively.} 
\labf{dgprobplot}
\end{figure}

An improved limit on \DGGs can be obtained by applying the $\tau_{\Bd}=\HFAGtauBd$ constraint.
The world average \Bs lifetime is not used, as its meaning 
is not clear if $\Delta \Gs$ is non-zero.
This is well motivated theoretically, as 
the total widths of the \Bs and \Bd mesons
are expected to be 
equal within less than one percent~\cite{bigilife}, \cite{Beneke}
and \DGd is expected to be small. 
 
The two-dimensional log-likelihood obtained, after including the constraint is shown in 
\Fig{dgplot} (right). The resulting probability density distribution for \DGGs is 
shown in \Fig{dgprobplot}. The corresponding limit on \DGGs is:
\begin{eqnarray}
\DGGs & = & 0.07^{+0.09}_{-0.07} \,, \\
\DGGs & < & 0.29~~\mbox{at \CL{95}} \,.
\end{eqnarray}

} 




Numerical results of the combination of the described inputs
of \Table{dgammat} are:
\begin{eqnarray}
\DGGs &\in& [\hfagDGSGSlow,\hfagDGSGSupp] ~ \mbox{at \CL{95}} \,, \\
\DGGs &=& \hfagDGSGS \,, \\
\DGs &\in& [\hfagDGSlow,\hfagDGSupp]\invps ~ \mbox{at \CL{95}} \,, \\
\DGs &=& \hfagDGS \,, \\
\bar{\tau}(\Bs) = 1/\Gs &=& \hfagTAUBSMEAN \,, \\
1/\Gamma_{\rm L} = \tau_{\rm short} &=& \hfagTAUBSL \,, \\
1/\Gamma_{\rm H} = \tau_{\rm long}  &=& \hfagTAUBSH \,. 
\end{eqnarray}

Flavor-specific lifetime measurements are of an equal mix
of \CP-even and \CP-odd states at time zero, and  
if a single exponential function is used in the likelihood
lifetime fit of such a sample~\cite{Hartkorn_Moser}, 
\begin{equation}
\tau(\Bs)_{\rm fs} = \frac{1}{\Gs}
\frac{{1+\left(\frac{\DGs}{2\Gs}\right)^2}}{{1-\left(\frac{\DGs}{2\Gs}\right)^2}
} \,.
\labe{fslife_const}
\end{equation}
Using the world average flavor-specific 
lifetime\footnote{The world average of all \Bs lifetime 
measurements using flavour-specific final states is \hfagTAUBSSL; however,
for the purpose of the \DGs extraction, we remove from this average one
DELPHI analysis~\cite{DELBS0_dms_dgs} 
that is already included in the set of ``direct 
measurements'' and obtain \hfagTAUBSSLX, 
shown as the blue bands on the two plots of \Fig{DGs}.} of \Sec{taubs}
the one-sigma blue bands shown in \Fig{DGs} are obtained. 
Higher-order corrections were checked to be negligible in the
combination.

As described earlier, \Bs $\to K^+ K^-$ decays
can be used to
measure the lifetime
of the ``light" mass eigenstate
$\tau_L = 1/\Gamma_L = 
\tau(\Bs)_{\particle{K^+K^-}} = 1.53 \pm 0.18 \pm 0.02 
\thinspace {\mathrm{ps}}$~\cite{CDFBS4}, and this 
additional constraint is included in the combination shown in 
\Fig{DGs}.

When the flavor-specific lifetime measurements and $\tau_L$ 
measurements
are combined with the measurements of \Table{dgammat}, the shaded
regions of \Fig{DGs} are obtained, with numerical results:
\begin{eqnarray}
\DGGs &\in& [\hfagDGSGSCONlow,\hfagDGSGSCONupp] ~ \mbox{at \CL{95}} \,, \\
\DGGs &=& \hfagDGSGSCON \,, \labe{DGGs_ave} \\
\DGs &\in& [\hfagDGSCONlow,\hfagDGSCONupp]\invps ~ \mbox{at \CL{95}} \,, \\
\DGs &=& \hfagDGSCON \,, \\
\bar{\tau}(\Bs) = 1/\Gs &=& \hfagTAUBSMEANCON \,, \labe{oneoverGs} \\
1/\Gamma_{\rm L} = \tau_{\rm short} &=& \hfagTAUBSLCON \,, \\
1/\Gamma_{\rm H} = \tau_{\rm long}  &=& \hfagTAUBSHCON \,. 
\end{eqnarray}
These results can
be compared with the theoretical prediction of 
$\DGs = 0.096 \pm 0.039\invps$
(or $\DGs = 0.088 \pm 0.017\invps$ if there is no new physics in
\dms)~\cite{delta_gams}.

Measurements of $\BR{B^0_s \rightarrow D_s^{(*)+} D_s^{(*)-}}$ can 
also be sensitive to \DGs.
The decay $\Bs \rightarrow D_s^{+} D_s^{-}$ is into
a final state that is purely \CP even. 
Under various theoretical assumptions~\cite{Aleksan}, the
inclusive decay into this plus the excited states
$\Bs \rightarrow D_s^{(*)+} D_s^{(*)-}$ is also \CP even
to within 5\%, and 
$\Bs \rightarrow D_s^{(*)+} D_s^{(*)-}$ saturates
$\Gamma_s^{\CP \thinspace {\mathrm{even}}}$.
Under these assumptions, for no \CP violation, we have: 
\begin{equation}
\DGGs \approx
\frac{2 \BR{\Bs \rightarrow D_s^{(*)+} D_s^{(*)-}}}
{1 - \BR{\Bs \rightarrow D_s^{(*)+} D_s^{(*)-}}} \,.
\labe{dGsBr}
\end{equation}
However, there are concerns~\cite{nierste} that the assumptions needed
for the above are overly restrictive and that the inclusive branching
ratio may be \CP even to only 30\%.
Due to this uncertainty, extracted values of \DGGs 
from this branching ratio are not included in the overall combination 
but are only extracted here to compare with the world average result.

\begin{table}
\caption{Measurements of $\BR{\Bs \rightarrow D_s^{(*)+} D_s^{(*)-}}$.}
\labt{dGsBr}
\begin{center}
\begin{tabular}{l|c|c|c}
\hline
Experiment & Method & Value & Ref.  \\
\hline
\belle         & \Bs-pair production at $\Upsilon(5S)$
              & $< 0.257$ at 90\% CL & \cite{Belle_drutskoy2007e}$^a$  \\
ALEPH         & $\phi$-$\phi$ correlations              
           & $0.077 \pm 0.034^{+0.038}_{-0.026}$  & \cite{ALEPH_DGs}$^b$     \\
\dzero        & $D_s \rightarrow \phi \pi$, $D_s \rightarrow \phi \mu \nu$            
           & $0.039^{+0.019 + 0.016}_{-0.017 - 0.015}$  & \cite{D0_DsDs}   \\
	 \hline
Average       &      &   \hfagBRDSDS  &   \\
      \hline
\multicolumn{4}{l}{
$^a$ \footnotesize This limit is for $\Bs \rightarrow D_s^{*+} D_s^{*-}$.} \\[-0.5ex]
\multicolumn{4}{l}{
$^b$ \footnotesize Recalculated using the PDG 2006 
value of $\BR{D_s \rightarrow \phi \pi}$.} 
\end{tabular}
\end{center}
\end{table}

Measurements for the branching fraction for this
decay channel are shown in \Table{dGsBr}.
Using their average value of \hfagBRDSDS with \Eq{dGsBr} yields
\begin{equation}
\DGGs = \hfagDGSGSBRDSDS \,,
\end{equation}
consistent with the value given in \Eq{DGGs_ave}, but with the above
caveat.
CDF has also measured the exclusive branching fraction 
$\BR{B^0_s \rightarrow D^+_s D^-_s} = 
(9.4^{+4.4}_{-4.2}) \times 10^{-3}$~\cite{CDF_DsDs}, and
they use this to set a lower bound of
$\Delta\Gamma_s^{CP}/\Gamma_s \geq 0.012$ at \CL{95} (since
on its own it does not saturate the CP-even states).

\subsubsubsection{Weak phase in \Bs mixing}
\label{phasebs}
In general there will be a \CP-violating weak phase difference:
\begin{equation}
\phi_s = \arg \left[ -{M_{12}}/{\Gamma_{12}} \right], 
\end{equation}
where $M_{12}$ and $\Gamma_{12}$ are the off-diagonal
elements of the mass and decay matrices of the 
\Bs-\Bsbar system.
This is related to the observed \DGs through the relation:
\begin{equation}
\DGs = 2|\Gamma_{12}|\cos\phi_s.
\labe{new_phys_phase}
\end{equation}
The SM prediction for this phase is tiny,
$\phi_s^{\mathrm{SM}} = 0.004$~\cite{betaSM}; however,
new physics in \Bs mixing could change this observed phase to
\begin{equation}
\phi_s = \phi_s^{\mathrm{SM}} + \phi_s^{\mathrm{NP}}.
\end{equation}
The relative phase between the \Bs mixing amplitude and that of
specific $b \rightarrow c\bar{c}s$ quark transitions such as 
for \Bs or \Bsbar $\rightarrow J/\psi \phi$ in the SM 
is~\cite{betaSM,CKMUTfit}: 
\begin{equation}
2\beta_s^{SM} = 2\arg\left[-\left(V_{ts}V^*_{tb}\right)/\left(V_{cs}V^*_{cb}\right)\right] 
= 0.037 \pm 0.002 \approx 0.04.
\end{equation}
This angle is analogous to the $\beta$ angle in the usual CKM
unitarity triangle aside from the negative sign (resulting in a
positive angle in the SM).
The same additional contribution due to new physics would show up in this
observed phase~\cite{betaSM}, i.e.:
\begin{equation}
2\beta_s = 2\beta_s^{\mathrm{SM}} - \phi_s^{\mathrm{NP}}.
\end{equation}
The current experimental precision does not allow these small
\CP-violating phases $\phi_s^{\mathrm{SM}}$ and
$\beta_s^{\mathrm{SM}}$ to be resolved, 
and for large new physics effect, we can approximate
$\phi_s \approx -2\beta_s \approx \phi_s^{\mathrm{NP}}$, i.e., a significantly 
large observed phase would indicate new physics.

For non-zero $|\Gamma_{12}|$, analysis of the time-dependent
decay \particle{\Bs \rightarrow J/\psi\phi} can measure
the weak phase.  Including information on the \Bs flavor at production
time via flavor tagging improves precision and also resolves the 
sign ambiguity on the weak phase angle for a given \DGs.
Both CDF~\cite{CDF2_DGs_FT} and \dzero~\cite{D01_DGs} have performed 
such analyses and measure the same observed phase that we denote
$\phi_s^{J/\psi \phi} = -2\beta_s^{J/\psi \phi}$ to reflect
the different conventions of the experiments.

Under the assumption of non-zero $\phi_s^{J/\psi\phi}$, 
in addition to the result listed
in \Table{dgammat}, 
the \dzero collaboration~\cite{D01_DGs}  has also made simultaneous
fits allowing $\phi_s^{J/\psi\phi}$ to float while weakly 
constraining the strong phases, $\delta_i$ to find: 
\begin{eqnarray}
\DGs &=& +0.19 \pm 0.07 ^{+0.02}_{-0.01}~{\mathrm{ps}}^{-1}\,,  \\ 
\bar{\tau}(\Bs) &= &1/\Gs = 1.52 \pm 0.06~{\mathrm{ps}}\,,  \\
\phi_s^{J/\psi\phi} &=& -0.57 ^{+0.24+0.07}_{-0.30-0.02} \,. 
\end{eqnarray}
If the SM value of $\phi_s^{J/\psi\phi} = -0.04$ is assumed, a probability of 
6.6\% to obtain a value of $\phi_s^{J/\psi\phi}$ lower than $-0.57$ is
found.

The CDF analysis~\cite{CDF2_DGs_FT} reports confidence regions
in the two-dimensional space of $2\beta_s^{J/\psi\phi}$ and \DGs.
They present a Feldman-Cousins confidence interval of $2\beta_s^{J/\psi\phi}$
where \DGs is treated as a nuisance parameter:
\begin{equation}
2\beta_s^{J/\psi\phi} = -\phi_s^{J/\psi\phi} \in [0.32,2.82]~{\mathrm{at~95\%~CL}}.
\end{equation}
Note that only a confidence range is quoted and a  point 
estimate is not given since biases were observed in the analysis.
Assuming the SM predictions for $2\beta_s$ and \DGs, they find
that the probability of a deviation as large as the level of the 
observed data is 15\%.

Given the consistency of these two measurements of the weak phase,
as well as their
deviations from the SM, there is interest in combining the results and
using in global fits, e.g., see Ref.~\cite{UTfit_DGs}.
To allow a combination on equal footing, the \dzero collaboration
has redone their fits~\cite{D01_DGs_float} 
allowing  strong phase values, $\delta_i$, to float
as in the CDF analysis.
Ensemble studies to test confidence level coverage were performed 
by both collaborations and used to adjust likelihood
values to correspond to the usual Gaussian confidence levels. 
Two-dimensional likelihoods were combined with the result shown in 
\Fig{DGs_phase}(a).  
After the combination, consistency  
of the best fit values for $\phi_s^{J/\psi\phi} = -2\beta_s^{J/\psi\phi}$ with
SM predictions is at the level of $\hfagNSIGMASM\sigma$, with numerical results
for the two solutions given below.
Despite possible biases in the CDF input, point estimates are still
presented and the confidence level regions are straight projections
onto the \DGs or phase angle axes.
\begin{eqnarray}
\DGs &=& \hfagDGSCOMBA 
~ \mbox{or} ~ \hfagDGSCOMBB \,, \\
     &\in& [\hfagDGSCOMBAlow,\hfagDGSCOMBAupp]
     \cup  [\hfagDGSCOMBBlow,\hfagDGSCOMBBupp]\invps ~ \mbox{at \CL{90}} \,, \\
\phi_s^{J/\psi\phi} = -2\beta_s^{J/\psi\phi} &=& \hfagPHISCOMBA 
~ \mbox{or} ~ \hfagPHISCOMBB \,, \\
     &\in& [\hfagPHISCOMBAlow,\hfagPHISCOMBAupp]
     \cup  [\hfagPHISCOMBBlow,\hfagPHISCOMBBupp]~ \mbox{at \CL{90}} \,.
\end{eqnarray}

Further constraints were applied to the above sum of the
CDF and \dzero likelihoods, i.e., to the flavor-specific
\Bs lifetime world average of \Eq{fslife_const2} 
through \Eq{fslife_const} and to
the world average \Bs semileptonic asymmetry of 
\Eq{ASLs} through~\cite{ASLs_const}:
\begin{equation}
\ASLs = \frac{\DGs}{\dms}\tan\phi_s.
\end{equation}
Confidence level regions  following these 
constraints are shown in \Fig{DGs_phase}(b)
including the region delineated by new physics traced by 
the relation of \Eq{new_phys_phase}. Numerical results for the 
two solutions are:
\begin{eqnarray}
\DGs &=& \hfagDGSCOMBACON ~ \mbox{or} ~
         \hfagDGSCOMBBCON \,, \\
    &\in& [\hfagDGSCOMBABCONlow,\hfagDGSCOMBABCONupp]\invps 
    ~ \mbox{at \CL{90}} \,, \\
\phi_s^{J/\psi\phi} = -2\beta_s^{J/\psi\phi} &=& \hfagPHISCOMBACON 
    ~ \mbox{or} \hfagPHISCOMBBCON \,, \\
    &\in& [\hfagPHISCOMBACONlow,\hfagPHISCOMBACONupp]
    \cup  [\hfagPHISCOMBBCONlow,\hfagPHISCOMBBCONupp]~ \mbox{at \CL{90}} \,.
\end{eqnarray}
with a consistency
of the best fit values with
SM predictions of $2\beta_s$ at the level of $\hfagNSIGMASMCON\sigma$.

\begin{figure}
\begin{center}
\epsfig{figure=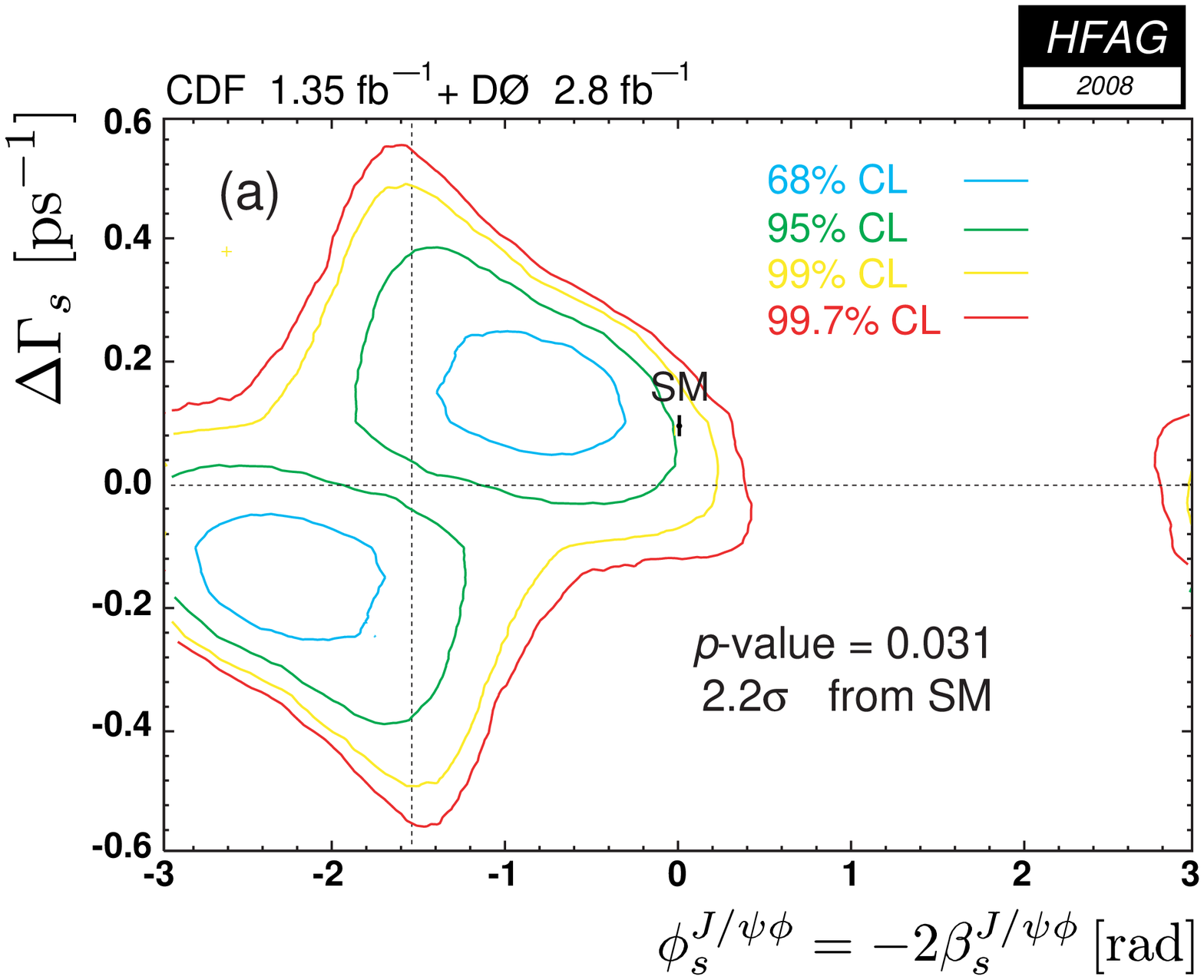,width=0.45\textwidth}
\hfill    
\epsfig{figure=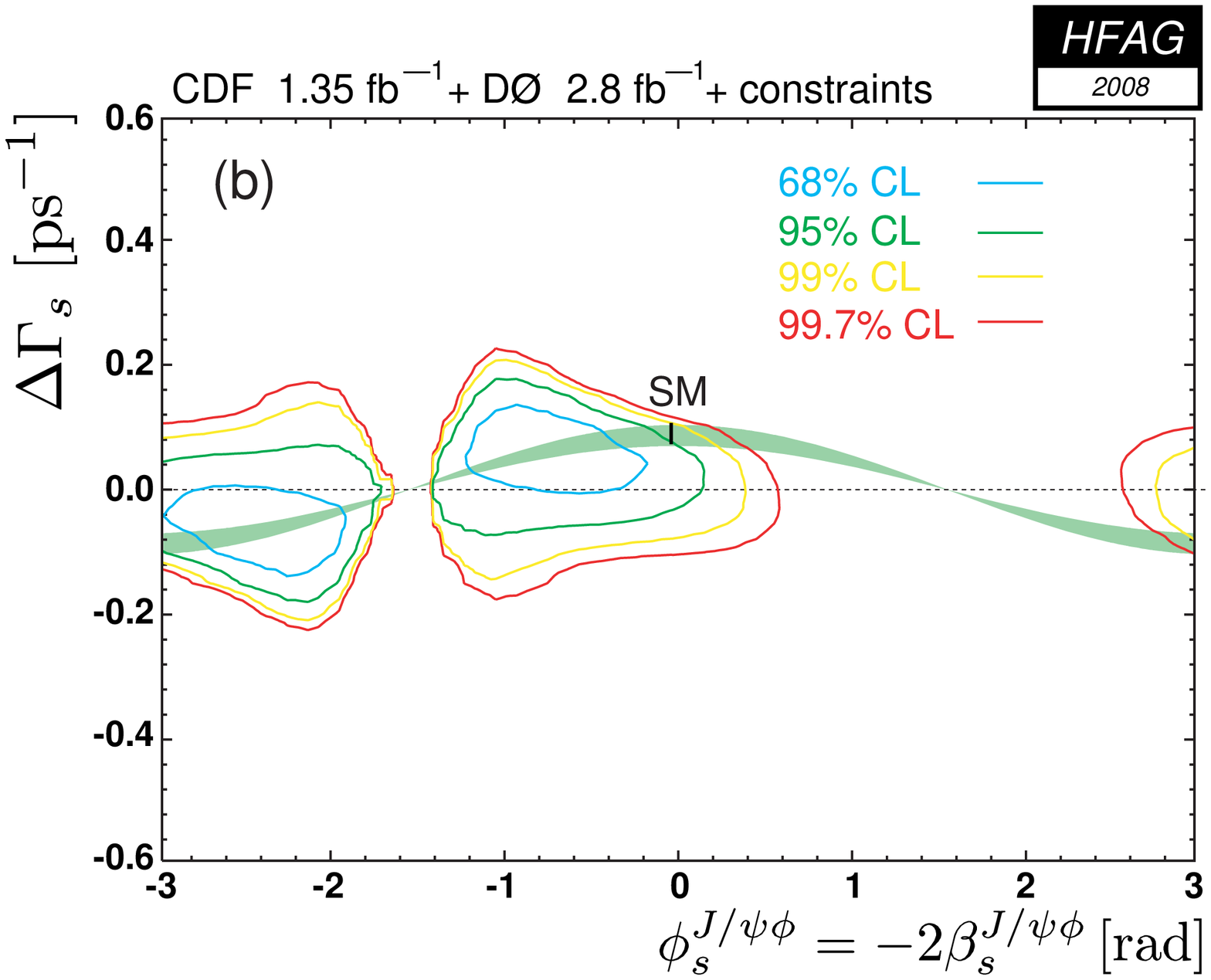,width=0.45\textwidth}
\caption{(a) Confidence regions
in \Bs width difference \DGs and
weak phase angle $\phi_s^{J/\psi\phi} = -2\beta_s^{J/\psi\phi}$
from combined CDF and \dzero
likelihoods determined in  flavor-tagged
\particle{\Bs \rightarrow J/\psi \phi} time-dependent
angular analyses~\cite{CDF2_DGs_FT,D01_DGs_float} compared
to the SM value of $-2\beta_s^{SM}$; (b) after adding
constraints from the world average flavor-specific \Bs lifetime and
\Bs semileptonic asymmetry, \ASLs.
The region allowed in new physics models given by
$\DGs = 2|\Gamma_{12}|\cos\phi_s$ is also shown (light green band).}
\labf{DGs_phase}
\end{center}
\end{figure}

\subsubsubsection{\boldmath Mass difference \dms}

\Bs oscillations have been observed for the first time in 2006
by the CDF collaboration~\cite{CDF2_dms_observation},
based on samples of flavour-tagged hadronic and semileptonic \Bs decays
(in flavour-specific final states), partially or fully reconstructed in 
$1\invfb$ of data collected during Tevatron's Run~II. 
From the proper-time dependence of these \Bs candidates, CDF
observe \Bs oscillations with a significance of at least $5\sigma$ 
and measure $\dms = 17.77 \pm 0.10 \pm 0.07\invps$~\cite{CDF2_dms_observation}.
More recently, the \dzero collaboration has obtained with $2.4\invfb$ an independent  
$\sim 3\sigma$ preliminary evidence for \Bs oscillations;
combining all their results~\cite{D0_dms_prel_evidence}
they obtain $\dms = 18.53 \pm 0.93 \pm 0.30\invps$~\cite{HEvans_dms}.
To a good approximation, both the CDF and \dzero results have Gaussian 
errors, and the world average value of \dms can be obtained as a simple weighted 
average:
\begin{equation}
\dms = \hfagDMS \,.  \labe{dms}
\end{equation}

Multiplying this result with the 
mean \Bs lifetime of \Eq{oneoverGs}, $1/\Gs=\hfagTAUBSMEANCON$,
yields
\begin{equation}
\xs = \frac{\dms}{\Gs} = \hfagXS \,. \labe{xs}
\end{equation}
With $2\ys = \DGGs=\hfagDGSGSCON$ 
(see \Eq{DGGs_ave})
and under the assumption of no \CP violation in \Bs mixing,
this corresponds to
\begin{equation}
\chis = \frac{\xs^2+\ys^2}{2(\xs^2+1)} = \hfagCHIS \,. \labe{chis}
\end{equation}
The ratio of the \Bd and \Bs oscillation frequencies, 
obtained from \Eqss{dmd}{dms}, 
\begin{equation}
\frac{\dmd}{\dms} = \hfagRATIODMDDMS \,, \labe{dmd_over_dms}
\end{equation}
can be used to extract the following ratio of CKM matrix elements, 
\begin{equation}
\left|\frac{V_{td}}{V_{ts}}\right| =
\xi \sqrt{\frac{\dmd}{\dms}\frac{m(\Bs)}{m(\Bd)}} = 
\hfagVTDVTSfull \,, \labe{Vtd_over_Vts}
\end{equation}
where the first quoted error is from experimental uncertainties 
(with the masses $m(\Bs)$ and $m(\Bd)$ taken from~\cite{PDG_2007}),
and where the second quoted error is from theoretical uncertainties 
in the estimation of the SU(3) flavor-symmetry breaking factor
$\xi 
= \hfagXI$
obtained from lattice QCD calculations~\cite{xi_lattice_QCD}.

\Bs mesons were known to mix since many years. Indeed 
the time-integrated measurements of \chibar (see \Sec{chibar}),
when compared to our knowledge
of \chid and the \b-hadron fractions, indicated that \Bs mixing was large,
with a value of \chis close to its maximal possible value of $1/2$.
However, the time dependence of this mixing 
could not be observed until recently, 
mainly because of lack of proper-time resolution to resolve 
the small period of the \Bs\ oscillations.

The statistical significance ${\cal S}$ of a \Bs oscillation signal can be
approximated as~\cite{amplitude}
\begin{equation}
{\cal S} \approx \sqrt{\frac{N}{2}} \,f_{\rm sig}\, (1-2w)\,
\exp{\left(-\left(\dms\sigma_t\right)^2/2\right)}\,,
\labe{significance}
\end{equation}
where $N$ is 
the number of selected and tagged \Bs candidates, 
$f_{\rm sig}$ is the fraction of \Bs signal
in the selected and tagged sample, $w$ is the total mistag probability, 
and $\sigma_t$ is the resolution on proper time.
As can be seen, the quantity ${\cal S}$ decreases very quickly as 
\dms increases: this dependence is controlled by $\sigma_t$, 
which is therefore the most critical parameter for \dms analyses. 
The method widely used for \Bs oscillation searches
consists of measuring a \Bs oscillation amplitude ${\cal A}$
at several different test values of \dms, 
using a maximum likelihood fit based on the functions 
of \Eq{oscillations} where the cosine terms have been multiplied 
by ${\cal A}$.
One expects ${\cal A}=1$ at the true 
value of \dms and ${\cal A}=0$ at a test value of \dms 
(far) below the true value.
To a good approximation, the statistical uncertainty on ${\cal A}$
is Gaussian and equal to $1/{\cal S}$~\cite{amplitude}.
In any analysis, a particular value of \dms
can be excluded at \CL{95} if ${\cal A}+ 1.645\,\sigma_{\cal A} < 1$, 
where $\sigma_{\cal A}$ is the total uncertainty on ${\cal A}$.
Because of the proper time resolution, the quantity $\sigma_{\cal A}(\dms)$
is an increasing function of \dms (see \Eq{significance} which merely models  
$1/\sigma_{\cal A}(\dms)$ in an analysis limited 
by the available statistics). Therefore, 
if the true value of \dms were infinitely large, one 
expects to be able to exclude all values of \dms up to $\dms^{\rm sens}$, 
where $\dms^{\rm sens}$, called here the
sensitivity of the analysis, is defined by
$1.645\,\sigma_{\cal A}(\dms^{\rm sens}) = 1$. 

\begin{figure}
\begin{center}
\epsfig{figure=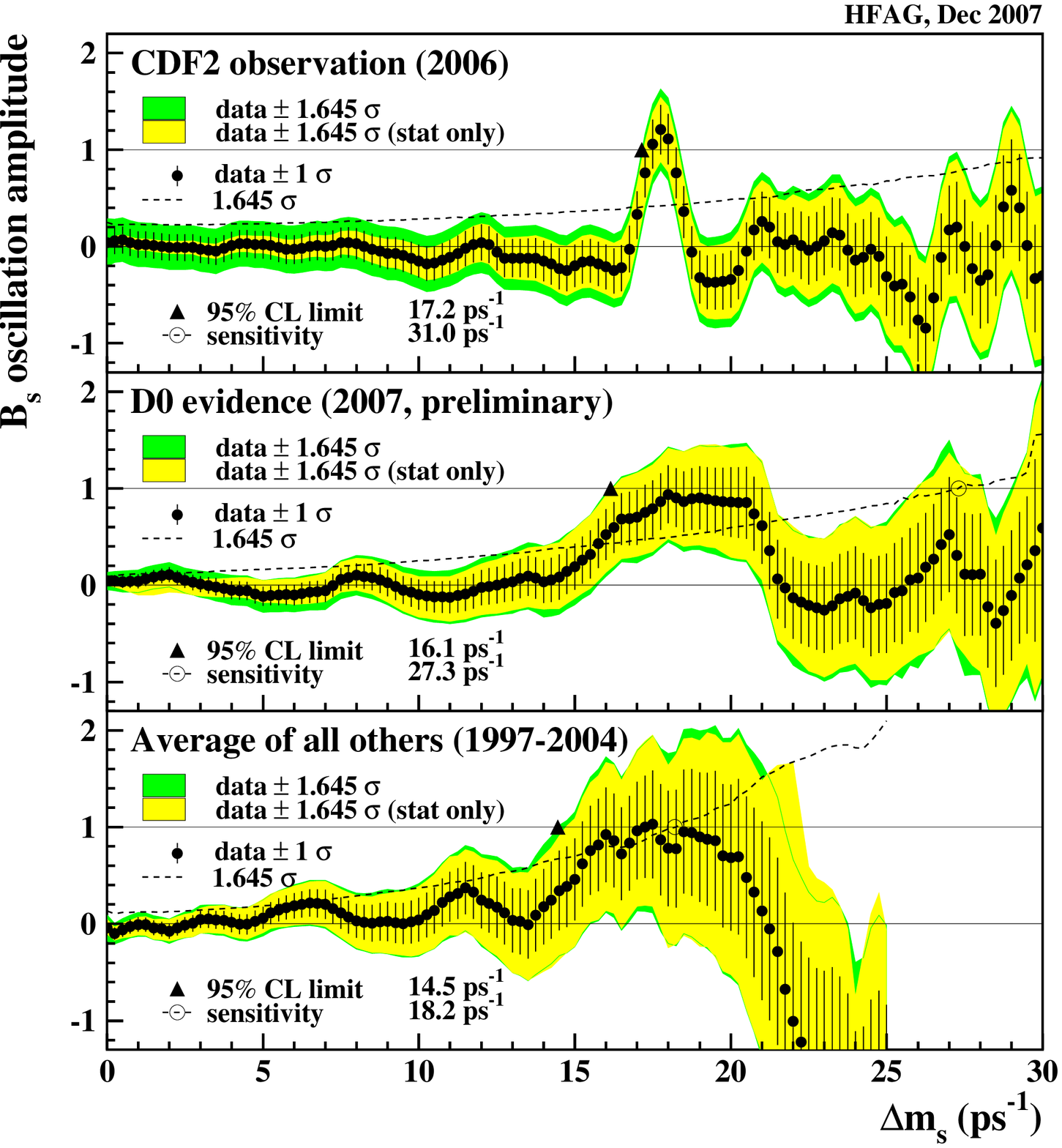,width=\textwidth}
\caption{\Bs oscillation amplitude as a function of \dms.
\underline{Top:} CDF result based on Run II data, published in 2006~\cite{CDF2_dms_observation}.
\underline{Middle:} Average of all preliminary \dzero results available at the end
of 2007~\cite{D0_dms_prel_evidence}.
\underline{Bottom:} Average of all
ALEPH~\cite{ALEPH_dms},
DELPHI~\cite{DELBS0_dms_dgs,DELBS1_dms_excl,DELPHI_dmd_dms_vtx,DELPHI_dms_last},
OPAL~\cite{OPAL_dms_l,OPAL_dms_dsl},
SLD~\cite{SLD_dms_dipole,SLD_dms_ds},
and CDF Run~I~\cite{CDF1_dms} results 
published between 1997 and 2004.
Statistical uncertainties dominate. 
Neighboring points are statistically correlated.}
\labf{amplitude}
\end{center}
\end{figure}

\Figure{amplitude} shows the measured \Bs amplitude as a function of \dms, 
as obtained by CDF (top) and \dzero (middle) using Run II data. 
The recent \dzero evidence of a \Bs oscillation signal is consistent
with the 2006 observation by CDF. 
A large number of \Bs oscillation searches,
already based on the amplitude method, 
had been performed previously by ALEPH~\cite{ALEPH_dms},
CDF (Run~I)~\cite{CDF1_dms},
DELPHI~\cite{DELBS0_dms_dgs,DELBS1_dms_excl,DELPHI_dmd_dms_vtx,DELPHI_dms_last},
OPAL~\cite{OPAL_dms_l,OPAL_dms_dsl} and 
SLD~\cite{SLD_dms_dipole,SLD_dms_ds,SLD_dms_leptDvtx_unpublished}
(we omit references to searches that have been superseded
by more recent ones). All the results published by these experiments 
(between 1997 and 2004) have been combined 
by averaging the measured amplitudes ${\cal A}$ at each test value 
of \dms.
The individual results have been adjusted to common physics inputs, 
and all known correlations have been accounted for; 
in the case of the inclusive (lepton) analyses, performed at LEP and SLC, 
the sensitivities (\ie\ 
the statistical uncertainties on ${\cal A}$), which depend directly 
through \Eq{significance} on the assumed fraction $f_{\rm sig}\sim\fBs$
of \Bs mesons in an unbiased sample of weakly-decaying \b hadrons, 
have also been rescaled to the LEP average $\fBs = \hfagLFBS$.
The resulting average amplitude spectrum, completely dominated 
by the $e^+e^-\to\particle{Z}$ experiments, is displayed 
as the bottom plot of \Fig{amplitude}. Although no significant signal 
is seen, it is interesting to note the hint in the region 15--20\invps, 
consistent with the recent results from the Tevatron. 


\comment{ 
\begin{figure}
\begin{center}
\epsfig{figure=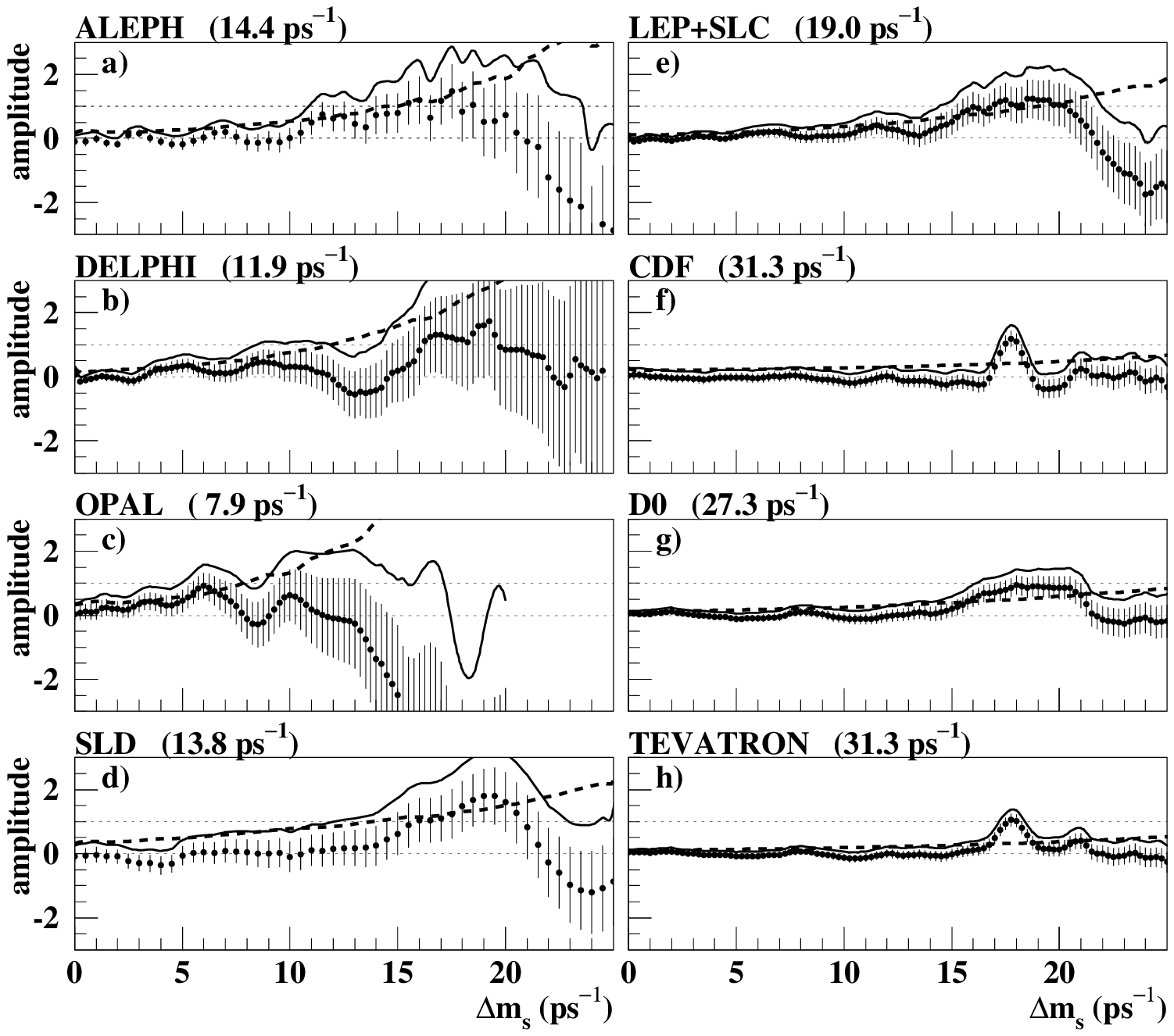,width=\textwidth,%
bbllx=55,bblly=55,bburx=490,bbury=490}
\caption{Combined \Bs-oscillation amplitude spectra, displayed separately
for each experiment. 
The points and error bars represent the measured
amplitude ${\cal A}$ and its total
uncertainty $\sigma_{\cal A}$, adjusted to a set of physics parameters
common to all analyses (including $\fBs=\hfagLFBS$ for LEP and SLC analyses).
Values of \dms for which the solid curve
(${\cal A}+1.645\,\sigma_{\cal A}$) is below 1 are excluded at \CL{95}.
The dashed curve shows $1.645\,\sigma_{\cal A}$; the number in parenthesis
indicates where this curve is equal to 1, and is a measure of the sensitivity 
for \CL{95} exclusion.
\newline
a) ALEPH~\cite{ALEPH_dms}, 
b) DELPHI~\cite{DELBS0_dms_dgs,DELBS1_dms_excl,DELPHI_dmd_dms_vtx,DELPHI_dms_last},
c) OPAL~\cite{OPAL_dms_l,OPAL_dms_dsl}, 
d) SLD~\cite{SLD_dms_dipole,SLD_dms_ds,SLD_dms_leptDvtx_unpublished},
\newline
e) ALEPH, DELPHI, OPAL and SLD combined,
f) CDF~\cite{CDF2_dms_observation,CDF1_dms}, 
g) \dzero~\cite{D0_dms_pub2006,D0_dms_prel2006}, 
h) CDF and \dzero combined.
}
\labf{individual_amplitudes_3}
\end{center}
\end{figure}

\begin{figure}
\begin{center}
\epsfig{figure=figures/osc/dms_history.eps,width=\textwidth}
\caption{World averages of all measurements
of the \Bs oscillation amplitude as a function of \dms.
\underline{Top:} situation in Summer 2000~\cite{LEPHFS}.
\underline{Middle:} situation at the end of 2005~\cite{hfag_hepex_endof2005}.
\underline{Bottom:} situation at the end of 2006, combining 
all published results~\cite{%
ALEPH_dms,DELBS0_dms_dgs,DELBS1_dms_excl,DELPHI_dmd_dms_vtx,DELPHI_dms_last,%
OPAL_dms_l,OPAL_dms_dsl,SLD_dms_dipole,SLD_dms_ds,%
CDF2_dms_observation,CDF1_dms,D0_dms_pub2006} 
as well as all recent preliminary results 
from Tevatron Run~II~\cite{D0_dms_prel2006}. 
The new results from CDF~\cite{CDF2_dms_observation} and
\dzero~\cite{D0_dms_pub2006,D0_dms_prel2006}, 
which became available in 2006,  
supersede all previous analyses
of Run~II data~\cite{CDF2_dms_fall2005_prel,D0_dms_dsmux_prel}. 
Statistical uncertainties dominate. 
Neighboring points are statistically correlated.}
\labf{amplitude}
\end{center}
\end{figure}

A large number of \Bs oscillation searches,
all based on the amplitude method, 
have been performed over the years by ALEPH~\cite{ALEPH_dms},
CDF~\cite{CDF2_dms_observation,CDF1_dms},
\dzero~\cite{D0_dms_prel_evidence},
DELPHI~\cite{DELBS0_dms_dgs,DELBS1_dms_excl,DELPHI_dmd_dms_vtx,DELPHI_dms_last},
OPAL~\cite{OPAL_dms_l,OPAL_dms_dsl} and 
SLD~\cite{SLD_dms_dipole,SLD_dms_ds,SLD_dms_leptDvtx_unpublished}
(we omit references to searches that have been superseded
by more recent ones). They have been combined 
by averaging the measured amplitudes ${\cal A}$ at each test value 
of \dms.
The individual results have been adjusted to common physics inputs, 
and all known correlations have been accounted for; 
in the case of the inclusive (lepton) analyses, performed at LEP and SLC, 
the sensitivities (\ie\ 
the statistical uncertainties on ${\cal A}$), which depend directly 
through \Eq{significance} on the assumed fraction $f_{\rm sig}\sim\fBs$
of \Bs mesons in an unbiased sample of weakly-decaying \b hadrons, 
have also been rescaled to the LEP average $\fBs = \hfagLFBS$.

The combined amplitude spectra for the individual experiments are 
displayed in \Fig{individual_amplitudes_3}, and the world average spectrum is 
displayed in \Fig{amplitude}.
The appearance of the \Bs oscillation signal in 2006, which can clearly 
be seen by comparing the last two plots on \Fig{amplitude}, has been made 
possible by the latest analysis of the CDF data~\cite{CDF2_dms_observation},
which is, by far, the most sensitive one. 
It is interesting to note that a hint of a signal 
in the region 15--20\invps has been around since many years
at $e^+e^-\to\particle{Z}$ experiments,
and more recently at the \dzero experiment as well. 
} 

\clearpage
\mysection{Measurements related to Unitarity Triangle angles
}
\label{sec:cp_uta}

The charge of the ``$\CP(t)$ and Unitarity Triangle angles'' group
is to provide averages of measurements 
from time-dependent asymmetry analyses,
and other quantities that are related 
to the angles of the Unitarity Triangle (UT).
In cases where considerable theoretical input is required to 
extract the fundamental quantities, no attempt is made to do so at 
this stage. However, straightforward interpretations of the averages 
are given, where possible.

In Sec.~\ref{sec:cp_uta:introduction} 
a brief introduction to the relevant phenomenology is given.
In Sec.~\ref{sec:cp_uta:notations}
an attempt is made to clarify the various different notations in use.
In Sec.~\ref{sec:cp_uta:common_inputs}
the common inputs to which experimental results are rescaled in the
averaging procedure are listed. 
We also briefly introduce the treatment of experimental errors. 
In the remainder of this section,
the experimental results and their averages are given,
divided into subsections based on the underlying quark-level decays.

\mysubsection{Introduction
}
\label{sec:cp_uta:introduction}

The Standard Model Cabibbo-Kobayashi-Maskawa (CKM) quark mixing matrix $\VCKM$ 
must be unitary. A $3 \times 3$ unitary matrix has four free parameters,\footnote{
  In the general case there are nine free parameters,
  but five of these are absorbed into unobservable quark phases.}
and these are conventionally written by the product
of three (complex) rotation matrices~\cite{Chau:1984fp}, 
where the rotations are characterized by the Euler angles 
$\theta_{12}$, $\theta_{13}$ and $\theta_{23}$, which are the mixing angles
between the generations, and one overall phase $\delta$,
\begin{equation}
\label{eq:ckmPdg}
\VCKM =
        \left(
          \begin{array}{ccc}
            V_{ud} & V_{us} & V_{ub} \\
            V_{cd} & V_{cs} & V_{cb} \\
            V_{td} & V_{ts} & V_{tb} \\
          \end{array}
        \right)
        =
        \left(
        \begin{array}{ccc}
        c_{12}c_{13}    
                &    s_{12}c_{13}   
                        &   s_{13}e^{-i\delta}  \\
        -s_{12}c_{23}-c_{12}s_{23}s_{13}e^{i\delta} 
                &  c_{12}c_{23}-s_{12}s_{23}s_{13}e^{i\delta} 
                        & s_{23}c_{13} \\
        s_{12}s_{23}-c_{12}c_{23}s_{13}e^{i\delta}  
                &  -c_{12}s_{23}-s_{12}c_{23}s_{13}e^{i\delta} 
                        & c_{23}c_{13} 
        \end{array}
        \right)
\end{equation}
where $c_{ij}=\cos\theta_{ij}$, $s_{ij}=\sin\theta_{ij}$ for 
$i<j=1,2,3$. 

Following the observation of a hierarchy between the different
matrix elements, the Wolfenstein parameterization~\cite{Wolfenstein:1983yz}
is an expansion of $\VCKM$ in terms of the four real parameters $\lambda$
(the expansion parameter), $A$, $\rho$ and $\eta$. Defining to 
all orders in $\lambda$~\cite{Buras:1994ec}
\begin{eqnarray}
  \label{eq:burasdef}
  s_{12}             &\equiv& \lambda,\nonumber \\ 
  s_{23}             &\equiv& A\lambda^2, \\
  s_{13}e^{-i\delta} &\equiv& A\lambda^3(\rho -i\eta),\nonumber
\end{eqnarray}
and inserting these into the representation of Eq.~(\ref{eq:ckmPdg}), 
unitarity of the CKM matrix is achieved to all orders.
A Taylor expansion of $\VCKM$ leads to the familiar approximation
\begin{equation}
  \label{eq:cp_uta:ckm}
  \VCKM
  = 
  \left(
    \begin{array}{ccc}
      1 - \lambda^2/2 & \lambda & A \lambda^3 ( \rho - i \eta ) \\
      - \lambda & 1 - \lambda^2/2 & A \lambda^2 \\
      A \lambda^3 ( 1 - \rho - i \eta ) & - A \lambda^2 & 1 \\
    \end{array}
  \right) + {\cal O}\left( \lambda^4 \right) \, .
\end{equation}
At order $\lambda^{5}$, the obtained CKM matrix in this extended
Wolfenstein parametrization is:
{\small
  \begin{equation}
    \label{eq:cp_uta:ckm_lambda5}
    \VCKM
    =
    \left(
      \begin{array}{ccc}
        1 - \frac{1}{2}\lambda^{2} - \frac{1}{8}\lambda^4 &
        \lambda &
        A \lambda^{3} (\rho - i \eta) \\
        - \lambda + \frac{1}{2} A^2 \lambda^5 \left[ 1 - 2 (\rho + i \eta) \right] &
        1 - \frac{1}{2}\lambda^{2} - \frac{1}{8}\lambda^4 (1+4A^2) &
        A \lambda^{2} \\
        A \lambda^{3} \left[ 1 - (1-\frac{1}{2}\lambda^2)(\rho + i \eta) \right] &
        -A \lambda^{2} + \frac{1}{2}A\lambda^4 \left[ 1 - 2(\rho + i \eta) \right] &
        1 - \frac{1}{2}A^2 \lambda^4
      \end{array} 
    \right) + {\cal O}\left( \lambda^{6} \right).
  \end{equation}
}
The non-zero imaginary part of the CKM matrix,
which is the origin of $\CP$ violation in the Standard Model,
is encapsulated in a non-zero value of $\eta$.



The unitarity relation $\VCKM^\dagger\VCKM = {\mathit 1}$
results in a total of nine expressions,
that can be written as
$\sum_{i=u,c,t} V^*_{ij}V_{ik} = \delta_{jk}$,
where $\delta_{jk}$ is the Kronecker symbol.
Of the off-diagonal expressions ($j \neq k$),
three can be transformed into the other three 
leaving six relations, in which three complex numbers sum to zero,
which therefore can be expressed as triangles in the complex plane.
More details about unitarity triangles can be found in~\cite{Jarlskog:1985ht,Jarlskog:2005uq,Bjorken:2005rm}.

\begin{figure}[t]
  \begin{center}
    \resizebox{0.55\textwidth}{!}{\includegraphics{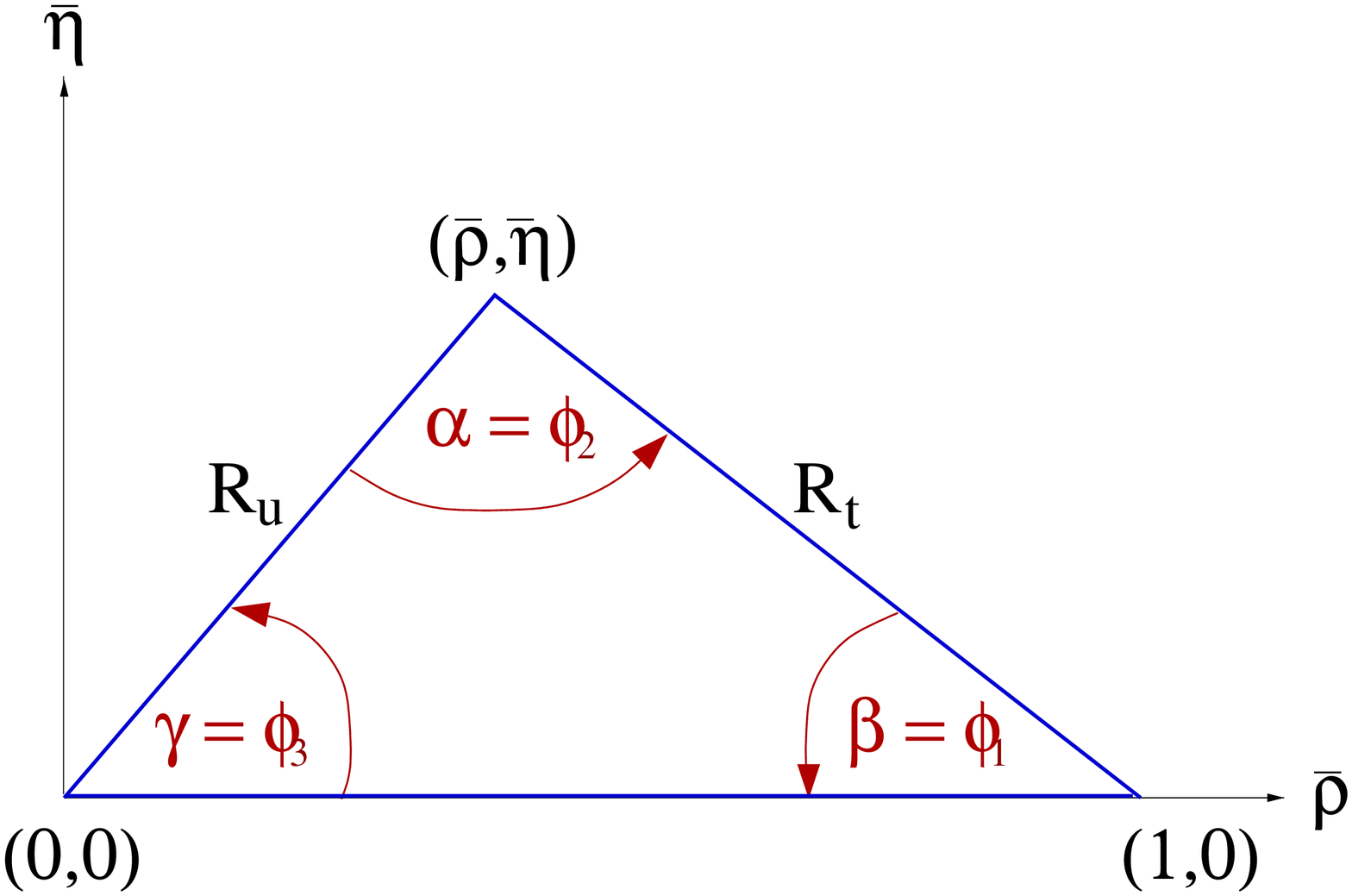}}
    \caption{The Unitarity Triangle.}
    \label{fig:cp_uta:ut}
  \end{center}
\end{figure}

One of these relations,
\begin{equation}
  \label{eq:cp_uta:ut}
  V_{ud}V^*_{ub} + V_{cd}V^*_{cb} + V_{td}V^*_{tb} = 0,
\end{equation}
is of particular importance to the $\B$ system, 
being specifically related to flavour changing 
neutral current $b \to d$ transitions.
The three terms in Eq.~(\ref{eq:cp_uta:ut}) are of the same order 
(${\cal O}\left( \lambda^3 \right)$),
and this relation is commonly known as the Unitarity Triangle.
For presentational purposes,
it is convenient to rescale the triangle by $(V_{cd}V^*_{cb})^{-1}$,
as shown in Fig.~\ref{fig:cp_uta:ut}.

Two popular naming conventions for the UT angles exist in the literature:
\begin{equation}
  \label{eq:cp_uta:abc}
  \alpha  \equiv  \phi_2  = 
  \arg\left[ - \frac{V_{td}V_{tb}^*}{V_{ud}V_{ub}^*} \right],
  \hspace{0.5cm}
  \beta   \equiv   \phi_1 =  
  \arg\left[ - \frac{V_{cd}V_{cb}^*}{V_{td}V_{tb}^*} \right],
  \hspace{0.5cm}
  \gamma  \equiv   \phi_3  =  
  \arg\left[ - \frac{V_{ud}V_{ub}^*}{V_{cd}V_{cb}^*} \right].
\end{equation}
In this document the $\left( \alpha, \beta, \gamma \right)$ set is used.\footnote{
  The relevant unitarity triangle for the $\Bs$ system is obtained 
  by replacing $d \leftrightarrow s$ in Eq.~\ref{eq:cp_uta:ut}.
  Definitions of the set of angles $( \alpha_s, \beta_s, \gamma_s )$ 
  can be obtained using equivalant relations to those of Eq.~\ref{eq:cp_uta:abc},
  for example $\beta_s = \arg\left[ - (V_{cs}V_{cb}^*) / (V_{ts}V_{tb}^*) \right]$.
  This definition gives a value of $\beta_s$ that is negative in the Standard Model,
  so that the sign is often flipped in the literature.
}
The sides $R_u$ and $R_t$ of the Unitarity Triangle 
(the third side being normalized to unity) 
are given by
\begin{equation}
  \label{eq:ru_rt}
  R_u =
  \left|\frac{V_{ud}V_{ub}^*}{V_{cd}V_{cb}^*} \right|
  = \sqrt{\rhobar^2+\etabar^2},
  \hspace{0.5cm}
  R_t = 
  \left|\frac{V_{td}V_{tb}^*}{V_{cd}V_{cb}^*}\right| 
  = \sqrt{(1-\rhobar)^2+\etabar^2}.
\end{equation} 
where $\rhobar$ and $\etabar$ define 
the apex of the Unitarity Triangle~\cite{Buras:1994ec} 
\begin{equation}
  \label{eq:rhoetabar}
  \rhobar + i\etabar
  \equiv -\frac{V_{ud}V_{ub}^*}{V_{cd}V_{cb}^*}
  \equiv 1 + \frac{V_{td}V_{tb}^*}{V_{cd}V_{cb}^*}
  = \frac{\sqrt{1-\lambda^2}\,(\rho + i \eta)}{\sqrt{1-A^2\lambda^4}+\sqrt{1-\lambda^2}A^2\lambda^4(\rho+i\eta)}
\end{equation}
The exact relation between $\left( \rho, \eta \right)$ and 
$\left( \rhobar, \etabar \right)$ is
\begin{equation}
  \label{eq:rhoetabarinv}
  \rho + i\eta \;=\; 
  \frac{ 
    \sqrt{ 1-A^2\lambda^4 }(\rhobar+i\etabar) 
  }{
    \sqrt{ 1-\lambda^2 } \left[ 1-A^2\lambda^4(\rhobar+i\etabar) \right]
  } \, .
\end{equation}

By expanding in powers of $\lambda$, several useful approximate expressions
can be obtained, including
\begin{equation}
  \label{eq:rhoeta_approx}
  \rhobar = \rho (1 - \frac{1}{2}\lambda^{2}) + {\cal O}(\lambda^4) \, ,
  \hspace{0.5cm}
  \etabar = \eta (1 - \frac{1}{2}\lambda^{2}) + {\cal O}(\lambda^4) \, ,
  \hspace{0.5cm}
  V_{td} = A \lambda^{3} (1-\rhobar -i\etabar) + {\cal O}(\lambda^6) \, .
\end{equation}

\mysubsection{Notations
}
\label{sec:cp_uta:notations}

Several different notations for $\CP$ violation parameters
are commonly used.
This section reviews those found in the experimental literature,
in the hope of reducing the potential for confusion, 
and to define the frame that is used for the averages.

In some cases, when $\B$ mesons decay into 
multibody final states via broad resonances ($\rho$, $\Kstar$, \etc),
the experimental analyses ignore the effects of interference 
between the overlapping structures.
This is referred to as the quasi-two-body (Q2B) approximation
in the following.

\mysubsubsection{$\CP$ asymmetries
}
\label{sec:cp_uta:notations:pra}

The $\CP$ asymmetry is defined as the difference between the rate 
involving a $b$ quark and that involving a $\bar b$ quark, divided 
by the sum. For example, the partial rate (or charge) asymmetry for 
a charged $\B$ decay would be given as 
\begin{equation}
  \label{eq:cp_uta:pra}
  \Acp_{f} \;\equiv\; 
  \frac{\Gamma(\Bm \to f)-\Gamma(\Bp \to \bar{f})}{\Gamma(\Bm \to f)+\Gamma(\Bp \to \bar{f})}.
\end{equation}

\mysubsubsection{Time-dependent \CP asymmetries in decays to $\CP$ eigenstates
}
\label{sec:cp_uta:notations:cp_eigenstate}

If the amplitudes for $\Bz$ and $\Bzb$ to decay to a final state $f$, 
which is a $\CP$ eigenstate with eigenvalue $\etacpf$,
are given by $\Af$ and $\Abarf$, respectively, 
then the decay distributions for neutral $\B$ mesons, 
with known flavour at time $\Delta t =0$,
are given by
\begin{eqnarray}
  \label{eq:cp_uta:td_cp_asp1}
  \Gamma_{\Bzb \to f} (\Delta t) & = &
  \frac{e^{-| \Delta t | / \tau(\Bz)}}{4\tau(\Bz)}
  \left[ 
    1 +
    \frac{2\, \Im(\lambda_f)}{1 + |\lambda_f|^2} \sin(\Delta m \Delta t) -
    \frac{1 - |\lambda_f|^2}{1 + |\lambda_f|^2} \cos(\Delta m \Delta t)
  \right], \\
  \label{eq:cp_uta:td_cp_asp2}
  \Gamma_{\Bz \to f} (\Delta t) & = &
  \frac{e^{-| \Delta t | / \tau(\Bz)}}{4\tau(\Bz)}
  \left[ 
    1 -
    \frac{2\, \Im(\lambda_f)}{1 + |\lambda_f|^2} \sin(\Delta m \Delta t) +
    \frac{1 - |\lambda_f|^2}{1 + |\lambda_f|^2} \cos(\Delta m \Delta t)
  \right].
\end{eqnarray}
Here $\lambda_f = \frac{q}{p} \frac{\Abarf}{\Af}$ 
contains terms related to $\Bz$\textendash$\Bzb$ mixing and to the decay amplitude
(the eigenstates of the effective Hamiltonian in the $\BzBzb$ system 
are $\left| B_\pm \right> = p \left| \Bz \right> \pm q \left| \Bzb \right>$).
This formulation assumes $\CPT$ invariance, 
and neglects possible lifetime differences 
(between the eigenstates of the effective Hamiltonian;
see Section~\ref{sec:mixing} where the mass difference $\Delta m$ is also defined)
in the neutral $\B$ meson system.
The case where non-zero lifetime differences are taken into account is 
discussed in Section~\ref{sec:cp_uta:notations:Bs}.
The time-dependent $\CP$ asymmetry,
again defined as the difference between the rate 
involving a $b$ quark and that involving a $\bar b$ quark,
is then given by
\begin{equation}
  \label{eq:cp_uta:td_cp_asp}
  \Acp_{f} \left(\Delta t\right) \; \equiv \;
  \frac{
    \Gamma_{\Bzb \to f} (\Delta t) - \Gamma_{\Bz \to f} (\Delta t)
  }{
    \Gamma_{\Bzb \to f} (\Delta t) + \Gamma_{\Bz \to f} (\Delta t)
  } \; = \;
  \frac{2\, \Im(\lambda_f)}{1 + |\lambda_f|^2} \sin(\Delta m \Delta t) -
  \frac{1 - |\lambda_f|^2}{1 + |\lambda_f|^2} \cos(\Delta m \Delta t).
\end{equation}

While the coefficient of the $\sin(\Delta m \Delta t)$ term in 
Eq.~(\ref{eq:cp_uta:td_cp_asp}) is everywhere\footnote
{
  Occasionally one also finds Eq.~(\ref{eq:cp_uta:td_cp_asp}) written as
  $\Acp_{f} \left(\Delta t\right) = 
  {\cal A}^{\rm mix}_f \sin(\Delta m \Delta t) + {\cal A}^{\rm dir}_f \cos(\Delta m \Delta t)$,
  or similar.
} denoted $S_f$:
\begin{equation}
  \label{eq:cp_uta:s_def}
  S_f \;\equiv\; \frac{2\, \Im(\lambda_f)}{1 + \left|\lambda_f\right|^2},
\end{equation}
different notations are in use for the
coefficient of the $\cos(\Delta m \Delta t)$ term:
\begin{equation}
  \label{eq:cp_uta:c_def}
  C_f \;\equiv\; - A_f \;\equiv\; \frac{1 - \left|\lambda_f\right|^2}{1 + \left|\lambda_f\right|^2}.
\end{equation}
The $C$ notation is used by the \babar\  collaboration 
(see \eg~\cite{Aubert:2001sp}), 
and also in this document.
The $A$ notation is used by the \belle\ collaboration
(see \eg~\cite{Abe:2001xe}).

Neglecting effects due to $\CP$ violation in mixing 
(by taking $|q/p| = 1$),
if the decay amplitude contains terms with 
a single weak (\ie, $\CP$ violating) phase
then $\left|\lambda_f\right| = 1$ and one finds
$S_f = -\etacpf \sin(\phi_{\rm mix} + \phi_{\rm dec})$, $C_f = 0$,
where $\phi_{\rm mix}=\arg(q/p)$ and $\phi_{\rm dec}=\arg(\Abarf/\Af)$.
Note that the $\Bz$--$\Bzb$ mixing phase $\phi_{\rm mix}\approx2\beta$
in the Standard Model (in the usual phase convention)~\cite{Carter:1980tk,Bigi:1981qs}. 

If amplitudes with different weak phases contribute to the decay, 
no clean interpretation of $S_f$ is possible. If the decay amplitudes
have in addition different $\CP$ conserving strong phases,
then $\left| \lambda_f \right| \neq 1$ and no clean interpretation is possible.
The coefficient of the cosine term becomes non-zero,
indicating direct $\CP$ violation.
The sign of $A_f$ as defined above is consistent with that of $\Acp_{f}$ in 
Eq.~(\ref{eq:cp_uta:pra}).

Frequently, we are interested in combining measurements 
governed by similar or identical short-distance physics,
but with different final states
(\eg, $\Bz \to \jpsi \KS$ and $\Bz \to \jpsi \KL$).
In this case, we remove the dependence on the $\CP$ eigenvalue 
of the final state by quoting $-\etacp S_f$.
In cases where the final state is not a $\CP$ eigenstate but has
an effective $\CP$ content (see below),
the reported $-\etacp S$ is corrected by the effective $\CP$.

\mysubsubsection{Time-dependent \CP asymmetries in decays to vector-vector final states
}
\label{sec:cp_uta:notations:vv}

Consider \B decays to states consisting of two spin-1 particles,
such as $\jpsi K^{*0}(\to\KS\piz)$, $D^{*+}D^{*-}$ and $\rho^+\rho^-$,
which are eigenstates of charge conjugation but not of parity.\footnote{
  \noindent
  This is not true of all vector-vector final states,
  \eg, $D^{*\pm}\rho^{\mp}$ is clearly not an eigenstate of 
  charge conjugation.
}
In fact, for such a system, there are three possible final states;
in the helicity basis these can be written $h_{-1}, h_0, h_{+1}$.
The $h_0$ state is an eigenstate of parity, and hence of $\CP$;
however, $\CP$ transforms $h_{+1} \leftrightarrow h_{-1}$ (up to 
an unobservable phase). In the transversity basis, these states 
are transformed into  $h_\parallel =  (h_{+1} + h_{-1})/2$ and 
$h_\perp = (h_{+1} - h_{-1})/2$.
In this basis all three states are $\CP$ eigenstates, 
and $h_\perp$ has the opposite $\CP$ to the others.

The amplitudes to these states are usually given by $A_{0,\perp,\parallel}$
(here we use a normalization such that 
$| A_0 |^2 + | A_\perp |^2 + | A_\parallel |^2 = 1$).
Then the effective $\CP$ of the vector-vector state is known if 
$| A_\perp |^2$ is measured.
An alternative strategy is to measure just the longitudinally polarized 
component,  $| A_0 |^2$
(sometimes denoted by $f_{\rm long}$), 
which allows a limit to be set on the effective $\CP$ since
$| A_\perp |^2 \leq | A_\perp |^2 + | A_\parallel |^2 = 1 - | A_0 |^2$.
The most complete treatment for 
neutral $\B$ decays to vector-vector final states
is time-dependent angular analysis 
(also known as time-dependent transversity analysis).
In such an analysis, 
the interference between the $\CP$-even and $\CP$-odd states 
provides additional sensitivity to the weak and strong phases involved.

In most analyses of time-dependent \CP asymmetries in decays to 
vector-vector final states carried out to date,
an assumption has been made that each helicity (or transversity) amplitude
has the same weak phase.
This is a good approximation for decays that are dominated by 
amplitudes with a single weak phase, such $\Bz \to \jpsi K^{*0}$,
and is a reasonable approximation in any mode for which only 
very limited statistics are available.
However, for modes that have contributions from amplitudes with different 
weak phases, the relative size of these contributions can be different 
for each helicity (or transversity) amplitude,
and therefore the time-dependent \CP asymmetry parameters can also differ.
The most generic analysis, suitable for modes with sufficient statistics,
would allow for this effect;
an intermediate analysis can allow different parameters for the 
$\CP$-even and $\CP$-odd components.
Such an analysis has been carried out by \babar\ for the decay
$\Bz \to D^{*+}D^{*-}$~\cite{Aubert:2007rr}.

\mysubsubsection{Time-dependent asymmetries: self-conjugate multiparticle final states
}
\label{sec:cp_uta:notations:dalitz}

Amplitudes for neutral \B decays into 
self-conjugate multiparticle final states
such as $\pi^+\pi^-\pi^0$, $K^+K^-\KS$,
$\jpsi \pi^+\pi^-$ or $D\pi^0$ with $D \to \KS\pi^+\pi^-$
may be written in terms of \CP-even and \CP-odd amplitudes.
As above, the interference between these terms 
provides additional sensitivity to the weak and strong phases
involved in the decay,
and the time-dependence depends on both the sine and cosine
of the weak phase difference.
In order to perform unbinned maximum likelihood fits,
and thereby extract as much information as possible from the distributions,
it is necessary to select a model for the multiparticle decay,
and therefore the results acquire some model dependence
(binned, model independent methods are also possible,
though are not as statistically powerful).
The number of observables depends on the final state (and on the model used);
the key feature is that as long as there are regions where both
\CP-even and \CP-odd amplitudes contribute,
the interference terms will be sensitive to the cosine 
of the weak phase difference.
Therefore, these measurements allow distinction between multiple solutions
for, \eg, the four values of $\beta$ from the measurement of $\sin(2\beta)$.

We now consider the various notations which have been used 
in experimental studies of
time-dependent asymmetries in decays to self-conjugate multiparticle final states.

\mysubsubsubsection{$\Bz \to D^{(*)}h^0$ with $D \to \KS\pi^+\pi^-$
}
\label{sec:cp_uta:notations:dalitz:dh0}

The states $D\pi^0$, $D^*\pi^0$, $D\eta$, $D^*\eta$, $D\omega$
are collectively denoted $D^{(*)}h^0$.
When the $D$ decay model is fixed,
fits to the time-dependent decay distributions can be performed
to extract the weak phase difference.
However, it is experimentally advantageous to use the sine and cosine of 
this phase as fit parameters, since these behave as essentially 
independent parameters, with low correlations and (potentially)
rather different uncertainties.
A parameter representing direct $\CP$ violation in the $B$ decay 
can also be floated.  
For consistency with other analyses, this could be chosen to be $C_f$,
but could equally well be $\left| \lambda_f \right|$, or other possibilities.

\belle\ performed an analysis of these channels
with $\sin(2\phi_1)$ and $\cos(2\phi_1)$ as free parameters~\cite{Krokovny:2006sv}.
\babar\ have performed an analysis floating also $\left| \lambda_f \right|$~\cite{Aubert:2007rp}
(and, of course, replacing $\phi_1 \Leftrightarrow \beta$).

\mysubsubsubsection{$\Bz \to D^{*+}D^{*-}\KS$
}
\label{sec:cp_uta:notations:dalitz:dstardstarks}

The hadronic structure of the $\Bz \to D^{*+}D^{*-}\KS$ decay
is not sufficiently well understood to perform a full 
time-dependent Dalitz plot analysis.
Instead, following Browder {\it et al.}~\cite{Browder:1999ng},
\babar~\cite{Aubert:2006fh} divide the Dalitz plane in two:
$m(D^{*+}\KS)^2 > m(D^{*-}\KS)^2$ $(\eta_y = +1)$ and 
$m(D^{*+}\KS)^2 < m(D^{*-}\KS)^2$ $(\eta_y = -1)$;
and then fit to a decay time distribution with asymmetry given by
\begin{equation}
  \Acp_{f} \left(\Delta t\right) =
  \eta_y \frac{J_c}{J_0} \cos(\Delta m \Delta t) -  
  \left[ 
    \frac{2J_{s1}}{J_0} \sin(2\beta) + \eta_y \frac{2J_{s2}}{J_0} \cos(2\beta) 
  \right] \sin(\Delta m \Delta t) \, .
\end{equation}
A similar analysis has also been carried out by \belle~\cite{Dalseno:2007hx}.
The measured values are $\frac{J_c}{J_0}$, $\frac{2J_{s1}}{J_0} \sin(2\beta)$
and $\frac{2J_{s2}}{J_0} \cos(2\beta)$, 
where the parameters $J_0$, $J_c$, $J_{s1}$ and $J_{s2}$ are the integrals 
over the half Dalitz plane $m(D^{*+}\KS)^2 < m(D^{*-}\KS)^2$ 
of the functions $|a|^2 + |\bar{a}|^2$, $|a|^2 - |\bar{a}|^2$, 
$\Re(\bar{a}a^*)$ and $\Im(\bar{a}a^*)$ respectively, 
where $a$ and $\bar{a}$ are the decay amplitudes of 
$\Bz \to D^{*+}D^{*-}\KS$ and $\Bzb \to D^{*+}D^{*-}\KS$ respectively. 
The parameter $J_{s2}$ (and hence $J_{s2}/J_0$) is predicted to be positive;
with this assumption is it possible to determine the sign of $\cos(2\beta)$.

\mysubsubsubsection{$\Bz \to K^+K^-\Kz$
}
\label{sec:cp_uta:notations:dalitz:kkk0}

Studies of $\Bz \to K^+K^-\Kz$~\cite{Aubert:2007sd} 
and of the related decay 
$\Bp \to K^+K^-K^+$~\cite{Garmash:2004wa,Aubert:2006nu},
show that the decay is dominated by components from the 
intermediate $K^+K^-$ resonances $\phi(1020)$, $f_0(980)$,
a poorly understood scalar structure that peaks near 
$m(K^+K^-) \sim 1550 \ {\rm MeV}/c^2$ and is denoted $X_0(1550)$,
as well as a large nonresonant contribution.
There is also a contribution from $\chi_{c0}$.

The full time-dependent Dalitz plot analysis allows 
the complex amplitudes of each contributing term to be determined from data,
including $\CP$ violation effects
(\ie\ allowing the complex amplitude for the $\Bz$ decay to be independent
from that for $\Bzb$ decay), although one amplitude must be fixed 
to give a reference point.
There are several choices for parametrization of the complex amplitudes 
(\eg\ real and imaginary part, or magnitude and phase).
Similarly, there are various approaches to include $\CP$ violation effects.
Note that positive definite parameters such as magnitudes are
disfavoured in certain circumstances 
(they inevitably lead to biases for small values).
In order to compare results between analyses,
it is useful for each experiment to present results in terms of the 
parameters that can be measured in a Q2B analysis
(such as $\Acp_{f}$, $S_f$, $C_f$, 
$\sin(2\beta^{\rm eff})$, $\cos(2\beta^{\rm eff})$, \etc)

In the \babar\ analysis of $\Bz \to K^+K^-\Kz$~\cite{Aubert:2007sd},
the complex amplitude for each resonant contribution is written as
\begin{equation}
  A_f = c_f ( 1 + b_f ) e^{i ( \phi_f + \delta_f )} 
  \ , \ \ \ \ 
  \bar{A}_f = c_f ( 1 - b_f ) e^{i ( \phi_f - \delta_f )} \, ,
\end{equation}
where $b_f$ and $\delta_f$ introduce $\CP$ violation in the magnitude 
and phase respectively.
[The weak phase in $B^0$--$\bar{B}^0$ mixing ($2\beta$) also appears 
in the full formula for the time-dependent decay distribution.]
The Q2B direct $\CP$ violation parameter is directly related to $b_f$
\begin{equation}
  \Acp_{f} = \frac{-2b_f}{1+b_f^2} \, .
\end{equation}

\babar\ present results for $c_f$, $\phi_f$, $\Acp_{f}$ and $\beta^{\rm eff}$
for each resonant contribution,
as well as averaged values of $\Acp_{f}$ and $\beta^{\rm eff}$
for the entire $K^+K^-\Kz$ Dalitz plot.

\mysubsubsubsection{$\Bz \to \pi^+\pi^-\KS$
}
\label{sec:cp_uta:notations:dalitz:pipik0}

Studies of $\Bz \to \pi^+\pi^-\KS$~\cite{Aubert:2007vi,Garmash:2006fh}
and of the related decay
$\Bp \to \pi^+\pi^-K^+$~\cite{Garmash:2004wa,Aubert:2008bj}
show that the decay is dominated by components from intermediate resonances 
in the $K\pi$ ($K^*(892)$, $K^*_0(1430)$) 
and $\pi\pi$ ($\rho(770)$, $f_0(980)$, $f_2(1270)$) spectra,
together with a poorly understood scalar structure that peaks near 
$m(\pi\pi) \sim 1300 \ {\rm MeV}/c^2$ and is denoted $f_X(1300)$
(that could be identified as either the $f_0(1370)$ or $f_0(1500)$),
and a large nonresonant component.
There is also a contribution from $\chi_{c0}$.

The full time-dependent Dalitz plot analysis allows 
the complex amplitudes of each contributing term to be determined from data,
including $\CP$ violation effects.
In the \babar\ analysis~\cite{Aubert:2007vi}, 
the magnitude and phase of each component (for both $\Bz$ and $\Bzb$ decays) 
are measured relative to $\Bz \to f_0(980)\KS$.
These results are translated into quasi-two-body parameters 
such as $2\beta^{\rm eff}_f$, $S_f$, $C_f$ for each \CP\ eigenstate $f$,
and direct \CP\ asymmetries for each flavour-specific state.
Relative phase differences between resonant terms are also extracted.

\mysubsubsubsection{$\Bz \to \pi^+\pi^-\pi^0$
}
\label{sec:cp_uta:notations:dalitz:pipipi0}

The $\Bz \to \pi^+\pi^-\pi^0$ decay is dominated by 
intermediate $\rho$ resonances.
Though it is possible, as above, 
to determine directly the complex amplitudes for each component,
an alternative approach, 
suggested by Quinn and Silva~\cite{Quinn:2000by},
has been used by both \babar~\cite{Aubert:2007jn}
and \belle~\cite{Kusaka:2007dv,:2007mj}.
The amplitudes for $\Bz$ and $\Bzb$ to $\pi^+\pi^-\pi^0$ are written
\begin{equation}
  A_{3\pi} = f_+ A_+ + f_- A_- + f_0 A_0
  \ , \ \ \ 
  \bar{A}_{3\pi} = f_+ \bar{A}_+ + f_- \bar{A}_- + f_0 \bar{A}_0
\end{equation}
respectively.
$A_+$, $A_-$ and $A_0$
represent the complex decay amplitudes for 
$\Bz \to \rho^+\pi^-$, $\Bz \to \rho^-\pi^+$ and $\Bz \to \rho^0\pi^0$
while 
$\bar{A}_+$, $\bar{A}_-$ and $\bar{A}_0$
represent those for 
$\Bzb \to \rho^+\pi^-$, $\Bzb \to \rho^-\pi^+$ and $\Bzb \to \rho^0\pi^0$
respectively.
$f_+$, $f_-$ and $f_0$ incorporate kinematical and dynamical factors
and depend on the Dalitz plot coordinates.
The full time-dependent decay distribution can then be written 
in terms of 27 free parameters,
one for each coefficient of the form factor bilinears,
as listed in Table~\ref{tab:cp_uta:pipipi0:uandi}.
These parameters are often referred to as ``the $U$s and $I$s'',
and can be expressed in terms of 
$A_+$, $A_-$, $A_0$, $\bar{A}_+$, $\bar{A}_-$ and $\bar{A}_0$.
If the full set of parameters is determined,
together with their correlations,
other parameters, such as weak and strong phases,
direct $\CP$ violation parameters, \etc, 
can be subsequently extracted.
Note that one of the parameters (typically $U_+^+$)
is often fixed to unity to provide a reference point;
this does not affect the analysis.


\begin{table}[htb]
  \begin{center}
    \setlength{\tabcolsep}{0.3pc}
\begin{tabular}{l@{\extracolsep{5mm}}l}
      \hline
      Parameter   & Description \\
      \hline
      $U_+^+$          & Coefficient of $|f_+|^2$ \\
      $U_0^+$          & Coefficient of $|f_0|^2$ \\
      $U_-^+$          & Coefficient of $|f_-|^2$ \\
      [0.15cm]
      $U_0^-$          & Coefficient of $|f_0|^2\cos(\Delta m\Delta t)$ \\
      $U_-^-$          & Coefficient of $|f_-|^2\cos(\Delta m\Delta t)$ \\
      $U_+^-$          & Coefficient of $|f_+|^2\cos(\Delta m\Delta t)$ \\
      [0.15cm]
      $I_0$            & Coefficient of $|f_0|^2\sin(\Delta m\Delta t)$ \\
\hline
    \end{tabular}
    \caption{
      Definitions of the $U$ and $I$ coefficients.
      Modified from~\cite{Aubert:2007jn}.
    }
    \label{tab:cp_uta:pipipi0:uandi}
  \end{center}
\end{table}

\mysubsubsection{Time-dependent \CP asymmetries in decays to non-$\CP$ eigenstates
}
\label{sec:cp_uta:notations:non_cp}

Consider a non-$\CP$ eigenstate $f$, and its conjugate $\bar{f}$. 
For neutral $\B$ decays to these final states,
there are four amplitudes to consider:
those for $\Bz$ to decay to $f$ and $\bar{f}$
($\Af$ and $\Afbar$, respectively),
and the equivalents for $\Bzb$
($\Abarf$ and $\Abarfbar$).
If $\CP$ is conserved in the decay, then
$\Af = \Abarfbar$ and $\Afbar = \Abarf$.


The time-dependent decay distributions can be written in many different ways.
Here, we follow Sec.~\ref{sec:cp_uta:notations:cp_eigenstate}
and define $\lambda_f = \frac{q}{p}\frac{\Abarf}{\Af}$ and
$\lambda_{\bar f} = \frac{q}{p}\frac{\Abarfbar}{\Afbar}$.
The time-dependent \CP asymmetries then follow Eq.~(\ref{eq:cp_uta:td_cp_asp}):
\begin{eqnarray}
\label{eq:cp_uta:non-cp-obs}
  {\cal A}_f (\Delta t) \; \equiv \;
  \frac{
    \Gamma_{\Bzb \to f} (\Delta t) - \Gamma_{\Bz \to f} (\Delta t)
  }{
    \Gamma_{\Bzb \to f} (\Delta t) + \Gamma_{\Bz \to f} (\Delta t)
  } & = & S_f \sin(\Delta m \Delta t) - C_f \cos(\Delta m \Delta t), \\
  {\cal A}_{\bar{f}} (\Delta t) \; \equiv \;
  \frac{
    \Gamma_{\Bzb \to \bar{f}} (\Delta t) - \Gamma_{\Bz \to \bar{f}} (\Delta t)
  }{
    \Gamma_{\Bzb \to \bar{f}} (\Delta t) + \Gamma_{\Bz \to \bar{f}} (\Delta t)
  } & = & S_{\bar{f}} \sin(\Delta m \Delta t) - C_{\bar{f}} \cos(\Delta m \Delta t),
\end{eqnarray}
with the definitions of the parameters 
$C_f$, $S_f$, $C_{\bar{f}}$ and $S_{\bar{f}}$,
following Eqs.~(\ref{eq:cp_uta:s_def}) and~(\ref{eq:cp_uta:c_def}).

The time-dependent decay rates are given by
\begin{eqnarray}
  \Gamma_{\Bzb \to f} (\Delta t) & = &
  \frac{e^{-\left| \Delta t \right| / \tau(\Bz)}}{8\tau(\Bz)} 
  ( 1 + \Adirnoncp ) 
  \left\{ 
    1 + S_f \sin(\Delta m \Delta t) - C_f \cos(\Delta m \Delta t) 
  \right\},
  \\
  \Gamma_{\Bz \to f} (\Delta t) & = &
  \frac{e^{-\left| \Delta t \right| / \tau(\Bz)}}{8\tau(\Bz)} 
  ( 1 + \Adirnoncp ) 
  \left\{ 
    1 - S_f \sin(\Delta m \Delta t) + C_f \cos(\Delta m \Delta t) 
  \right\},
  \\
  \Gamma_{\Bzb \to \bar{f}} (\Delta t) & = &
  \frac{e^{-\left| \Delta t \right| / \tau(\Bz)}}{8\tau(\Bz)} 
  ( 1 - \Adirnoncp ) 
  \left\{ 
    1 + S_{\bar{f}} \sin(\Delta m \Delta t) - C_{\bar{f}} \cos(\Delta m \Delta t) 
  \right\},
  \\
  \Gamma_{\Bz \to \bar{f}} (\Delta t) & = &
    \frac{e^{-\left| \Delta t \right| / \tau(\Bz)}}{8\tau(\Bz)} 
  ( 1 - \Adirnoncp ) 
  \left\{ 
    1 - S_{\bar{f}} \sin(\Delta m \Delta t) + C_{\bar{f}} \cos(\Delta m \Delta t) 
  \right\},
\end{eqnarray}
where the time-independent parameter \Adirnoncp
represents an overall asymmetry in the production of the 
$f$ and $\bar{f}$ final states,\footnote{
  This parameter is often denoted ${\cal A}_f$ (or ${\cal A}_{\CP}$),
  but here we avoid this notation to prevent confusion with the
  time-dependent $\CP$ asymmetry.
}
\begin{equation}
  \Adirnoncp = 
  \frac{
    \left( 
      \left| \Af \right|^2 + \left| \Abarf \right|^2
    \right) - 
    \left( 
      \left| \Afbar \right|^2 + \left| \Abarfbar \right|^2
    \right)
  }{
    \left( 
      \left| \Af \right|^2 + \left| \Abarf \right|^2
    \right) +
    \left( 
      \left| \Afbar \right|^2 + \left| \Abarfbar \right|^2
    \right)
  }.
\end{equation}
Assuming $|q/p| = 1$,
the parameters $C_f$ and $C_{\bar{f}}$
can also be written in terms of the decay amplitudes as follows:
\begin{equation}
  C_f = 
  \frac{
    \left| \Af \right|^2 - \left| \Abarf \right|^2 
  }{
    \left| \Af \right|^2 + \left| \Abarf \right|^2
  }
  \hspace{5mm}
  {\rm and}
  \hspace{5mm}
  C_{\bar{f}} = 
  \frac{
    \left| \Afbar \right|^2 - \left| \Abarfbar \right|^2
  }{
    \left| \Afbar \right|^2 + \left| \Abarfbar \right|^2
  },
\end{equation}
giving asymmetries in the decay amplitudes of $\Bz$ and $\Bzb$
to the final states $f$ and $\bar{f}$ respectively.
In this notation, the direct $\CP$ invariance conditions are
$\Adirnoncp = 0$ and $C_f = - C_{\bar{f}}$.
Note that $C_f$ and $C_{\bar{f}}$ are typically non-zero;
\eg, for a flavour-specific final state, 
$\Abarf = \Afbar = 0$ ($\Af = \Abarfbar = 0$), they take the values
$C_f = - C_{\bar{f}} = 1$ ($C_f = - C_{\bar{f}} = -1$).

The coefficients of the sine terms
contain information about the weak phase. 
In the case that each decay amplitude contains only a single weak phase
(\ie, no direct $\CP$ violation),
these terms can be written
\begin{equation}
  S_f = 
  \frac{ 
    - 2 \left| \Af \right| \left| \Abarf \right| 
    \sin( \phi_{\rm mix} + \phi_{\rm dec} - \delta_f )
  }{
    \left| \Af \right|^2 + \left| \Abarf \right|^2
  } 
  \hspace{5mm}
  {\rm and}
  \hspace{5mm}
  S_{\bar{f}} = 
  \frac{
    - 2 \left| \Afbar \right| \left| \Abarfbar \right| 
    \sin( \phi_{\rm mix} + \phi_{\rm dec} + \delta_f )
  }{
    \left| \Afbar \right|^2 + \left| \Abarfbar \right|^2
  },
\end{equation}
where $\delta_f$ is the strong phase difference between the decay amplitudes.
If there is no $\CP$ violation, the condition $S_f = - S_{\bar{f}}$ holds.
If decay amplitudes with different weak and strong phases contribute,
no clean interpretation of $S_f$ and $S_{\bar{f}}$ is possible.

Since two of the $\CP$ invariance conditions are 
$C_f = - C_{\bar{f}}$ and $S_f = - S_{\bar{f}}$,
there is motivation for a rotation of the parameters:
\begin{equation}
\label{eq:cp_uta:non-cp-s_and_deltas}
  S_{f\bar{f}} = \frac{S_{f} + S_{\bar{f}}}{2},
  \hspace{4mm}
  \Delta S_{f\bar{f}} = \frac{S_{f} - S_{\bar{f}}}{2},
  \hspace{4mm}
  C_{f\bar{f}} = \frac{C_{f} + C_{\bar{f}}}{2},
  \hspace{4mm}
  \Delta C_{f\bar{f}} = \frac{C_{f} - C_{\bar{f}}}{2}.
\end{equation}
With these parameters, the $\CP$ invariance conditions become
$S_{f\bar{f}} = 0$ and $C_{f\bar{f}} = 0$. 
The parameter $\Delta C_{f\bar{f}}$ gives a measure of the ``flavour-specificity''
of the decay:
$\Delta C_{f\bar{f}}=\pm1$ corresponds to a completely flavour-specific decay,
in which no interference between decays with and without mixing can occur,
while $\Delta C_{f\bar{f}} = 0$ results in 
maximum sensitivity to mixing-induced $\CP$ violation.
The parameter $\Delta S_{f\bar{f}}$ is related to the strong phase difference 
between the decay amplitudes of $\Bz$ to $f$ and to $\bar f$. 
We note that the observables of Eq.~(\ref{eq:cp_uta:non-cp-s_and_deltas})
exhibit experimental correlations 
(typically of $\sim 20\%$, depending on the tagging purity, and other effects)
between $S_{f\bar{f}}$ and  $\Delta S_{f\bar{f}}$, 
and between $C_{f\bar{f}}$ and $\Delta C_{f\bar{f}}$. 
On the other hand, 
the final state specific observables of Eq.~(\ref{eq:cp_uta:non-cp-obs})
tend to have low correlations.

Alternatively, if we recall that the $\CP$ invariance
conditions at the decay amplitude level are
$\Af = \Abarfbar$ and $\Afbar = \Abarf$,
we are led to consider the parameters~\cite{Charles:2004jd}
\begin{equation}
  \label{eq:cp_uta:non-cp-directcp}
  {\cal A}_{f\bar{f}} = 
  \frac{
    \left| \Abarfbar \right|^2 - \left| \Af \right|^2 
  }{
    \left| \Abarfbar \right|^2 + \left| \Af \right|^2
  }
  \hspace{5mm}
  {\rm and}
  \hspace{5mm}
  {\cal A}_{\bar{f}f} = 
  \frac{
    \left| \Abarf \right|^2 - \left| \Afbar \right|^2
  }{
    \left| \Abarf \right|^2 + \left| \Afbar \right|^2
  }.
\end{equation}
These are sometimes considered more physically intuitive parameters
since they characterize direct $\CP$ violation 
in decays with particular topologies.
For example, in the case of $\Bz \to \rho^\pm\pi^\mp$
(choosing $f =  \rho^+\pi^-$ and $\bar{f} = \rho^-\pi^+$),
${\cal A}_{f\bar{f}}$ (also denoted ${\cal A}^{+-}_{\rho\pi}$)
parameterizes direct $\CP$ violation
in decays in which the produced $\rho$ meson does not contain the 
spectator quark,
while ${\cal A}_{\bar{f}f}$ (also denoted ${\cal A}^{-+}_{\rho\pi}$)
parameterizes direct $\CP$ violation 
in decays in which it does.
Note that we have again followed the sign convention that the asymmetry 
is the difference between the rate involving a $b$ quark and that
involving a $\bar{b}$ quark, \cf\ Eq.~(\ref{eq:cp_uta:pra}). 
Of course, these parameters are not independent of the 
other sets of parameters given above, and can be written
\begin{equation}
  {\cal A}_{f\bar{f}} =
  - \frac{
    \Adirnoncp + C_{f\bar{f}} + \Adirnoncp \Delta C_{f\bar{f}} 
  }{
    1 + \Delta C_{f\bar{f}} + \Adirnoncp C_{f\bar{f}} 
  }
  \hspace{5mm}
  {\rm and}
  \hspace{5mm}
  {\cal A}_{\bar{f}f} =
  \frac{
    - \Adirnoncp + C_{f\bar{f}} + \Adirnoncp \Delta C_{f\bar{f}} 
  }{
    - 1 + \Delta C_{f\bar{f}} + \Adirnoncp C_{f\bar{f}}  
  }.
\end{equation}
They usually exhibit strong correlations.

We now consider the various notations which have been used 
in experimental studies of
time-dependent $\CP$ asymmetries in decays to non-$\CP$ eigenstates.

\mysubsubsubsection{$\Bz \to D^{*\pm}D^\mp$
}
\label{sec:cp_uta:notations:non_cp:dstard}

The above set of parameters 
($\Adirnoncp$, $C_f$, $S_f$, $C_{\bar{f}}$, $S_{\bar{f}}$),
has been used by both \babar~\cite{Aubert:2007pa} 
and \belle~\cite{Aushev:2004uc} in the $D^{*\pm}D^{\mp}$ system
($f = D^{*+}D^-$, $\bar{f} = D^{*-}D^+$).
However, slightly different names for the parameters are used:
\babar\ uses 
(${\cal A}$, $C_{+-}$, $S_{+-}$, $C_{-+}$, $S_{-+}$);
\belle\ uses
(${\cal A}$, $C_{+}$,  $S_{+}$,  $C_{-}$,  $S_{-}$).
In this document, we follow the notation used by \babar.

\mysubsubsubsection{$\Bz \to \rho^{\pm}\pi^\mp$
}
\label{sec:cp_uta:notations:non_cp:rhopi}

In the $\rho^\pm\pi^\mp$ system, the 
($\Adirnoncp$, $C_{f\bar{f}}$, $S_{f\bar{f}}$, $\Delta C_{f\bar{f}}$, 
$\Delta S_{f\bar{f}}$)
set of parameters has been used 
originally by \babar~\cite{Aubert:2003wr} and \belle~\cite{Wang:2004va}, 
in the Q2B approximation; 
the exact names\footnote{
  \babar\ has used the notations
  $A_{\CP}^{\rho\pi}$~\cite{Aubert:2003wr} and 
  ${\cal A}_{\rho\pi}$~\cite{Aubert:2007jn}
  in place of ${\cal A}_{\CP}^{\rho\pi}$.
}
used in this case are
$\left( 
  {\cal A}_{\CP}^{\rho\pi}, C_{\rho\pi}, S_{\rho\pi}, \Delta C_{\rho\pi}, \Delta S_{\rho\pi}
\right)$,
and these names are also used in this document.

Since $\rho^\pm\pi^\mp$ is reconstructed in the final state $\pi^+\pi^-\pi^0$,
the interference between the $\rho$ resonances
can provide additional information about the phases 
(see Sec.~\ref{sec:cp_uta:notations:dalitz}).
Both \babar~\cite{Aubert:2007jn} 
and \belle~\cite{Kusaka:2007dv,:2007mj}
have performed time-dependent Dalitz plot analyses, 
from which the weak phase $\alpha$ is directly extracted.
In such an analysis, the measured Q2B parameters are 
also naturally corrected for interference effects.
See Sec.~\ref{sec:cp_uta:notations:dalitz:pipipi0}.

\mysubsubsubsection{$\Bz \to D^{\pm}\pi^{\mp}, D^{*\pm}\pi^{\mp}, D^{\pm}\rho^{\mp}$
}
\label{sec:cp_uta:notations:non_cp:dstarpi}

Time-dependent $\CP$ analyses have also been performed for the
final states $D^{\pm}\pi^{\mp}$, $D^{*\pm}\pi^{\mp}$ and $D^{\pm}\rho^{\mp}$.
In these theoretically clean cases, no penguin contributions are possible,
so there is no direct $\CP$ violation.
Furthermore, due to the smallness of the ratio of the magnitudes of the 
suppressed ($b \to u$) and favoured ($b \to c$) amplitudes (denoted $R_f$),
to a very good approximation, $C_f = - C_{\bar{f}} = 1$
(using $f = D^{(*)-}h^+$, $\bar{f} = D^{(*)+}h^-$ $h = \pi,\rho$),
and the coefficients of the sine terms are given by
\begin{equation}
  S_f = - 2 R_f \sin( \phi_{\rm mix} + \phi_{\rm dec} - \delta_f )
  \hspace{5mm}
  {\rm and}
  \hspace{5mm}
  S_{\bar{f}} = - 2 R_f \sin( \phi_{\rm mix} + \phi_{\rm dec} + \delta_f ).
\end{equation}
Thus weak phase information can be cleanly obtained from measurements
of $S_f$ and $S_{\bar{f}}$, 
although external information on at least one of $R_f$ or $\delta_f$ is necessary.
(Note that $\phi_{\rm mix} + \phi_{\rm dec} = 2\beta + \gamma$ for all the decay modes 
in question, while $R_f$ and $\delta_f$ depend on the decay mode.)

Again, different notations have been used in the literature.
\babar~\cite{Aubert:2006tw,Aubert:2005yf}
defines the time-dependent probability function by
\begin{equation}
  f^\pm (\eta, \Delta t) = \frac{e^{-|\Delta t|/\tau}}{4\tau} 
  \left[  
    1 \mp S_\zeta \sin (\Delta m \Delta t) \mp \eta C_\zeta \cos(\Delta m \Delta t) 
  \right],
\end{equation} 
where the upper (lower) sign corresponds to 
the tagging meson being a $\Bz$ ($\Bzb$). 
[Note here that a tagging $\Bz$ ($\Bzb$) corresponds to $-S_\zeta$ ($+S_\zeta$).]
The parameters $\eta$ and $\zeta$ take the values $+1$ and $+$ ($-1$ and $-$) 
when the final state is, \eg, $D^-\pi^+$ ($D^+\pi^-$). 
However, in the fit, the substitutions $C_\zeta = 1$ and 
$S_\zeta = a \mp \eta b_i - \eta c_i$ are made.\footnote{
  The subscript $i$ denotes tagging category.
}
[Note that, neglecting $b$ terms, $S_+ = a - c$ and $S_- = a + c$, 
so that $a = (S_+ + S_-)/2$, $c = (S_- - S_+)/2$, in analogy to 
the parameters of Eq.~(\ref{eq:cp_uta:non-cp-s_and_deltas}).] 
The subscript $i$ denotes the tagging category. 
These are motivated by the possibility of 
$\CP$ violation on the tag side~\cite{Long:2003wq}, 
which is absent for semileptonic $\B$ decays (mostly lepton tags). 
The parameter $a$ is not affected by tag side $\CP$ violation. 
The parameter $b$ only depends on tag side $\CP$ violation parameters 
and is not directly useful for determining UT angles.
A clean interpretation of the $c$ parameter is only possible for 
lepton-tagged events,
so the \babar\ measurements report $c$ measured with those events only.

The parameters used by \belle\ in the analysis using 
partially reconstructed $\B$ decays~\cite{Abe:2004mu}, 
are similar to the $S_\zeta$ parameters defined above. 
However, in the \belle\ convention, 
a tagging $\Bz$ corresponds to a $+$ sign in front of the sine coefficient; 
furthermore the correspondence between the super/subscript 
and the final state is opposite, so that $S_\pm$ (\babar) = $- S^\mp$ (\belle). 
In this analysis, only lepton tags are used, 
so there is no effect from tag side $\CP$ violation. 
In the \belle\ analysis using 
fully reconstructed $\B$ decays~\cite{Abe:2003gn}, 
this effect is measured and taken into account using $\Dstar l \nu$ decays; 
in neither \belle\ analysis are the $a$, $b$ and $c$ parameters used. 
In the latter case, the measured parameters are 
$2 R_{D^{(*)}\pi} \sin( 2\phi_1 + \phi_3 \pm \delta_{D^{(*)}\pi} )$; 
the definition is such that 
$S^\pm$ (\belle) = $- 2 R_{\Dstar \pi} \sin( 2\phi_1 + \phi_3 \pm \delta_{\Dstar \pi} )$. 
However, the definition includes an 
angular momentum factor $(-1)^L$~\cite{Fleischer:2003yb}, 
and so for the results in the $D\pi$ system, 
there is an additional factor of $-1$ in the conversion.

Explicitly, the conversion then reads as given in 
Table~\ref{tab:cp_uta:notations:non_cp:dstarpi}, 
where we have neglected the $b_i$ terms used by \babar
(which are zero in the absence of tag side $\CP$ violation).
For the averages in this document,
we use the $a$ and $c$ parameters,
and give the explicit translations used in 
Table~\ref{tab:cp_uta:notations:non_cp:dstarpi2}.
It is to be fervently hoped that the experiments will
converge on a common notation in future.

\begin{table}
  \begin{center} 
    \caption{
      Conversion between the various notations used for 
      $\CP$ violation parameters in the 
      $D^{\pm}\pi^{\mp}$, $D^{*\pm}\pi^{\mp}$ and $D^{\pm}\rho^{\mp}$ systems.
      The $b_i$ terms used by \babar\ have been neglected.
      Recall that $\left( \alpha, \beta, \gamma \right) = \left( \phi_2, \phi_1, \phi_3 \right)$.
    }
    \vspace{0.2cm}
    \setlength{\tabcolsep}{0.0pc}
\begin{tabular*}{\textwidth}{@{\extracolsep{\fill}}cccc} \hline 
      & \babar\ & \belle\ partial rec. & \belle\ full rec. \\
      \hline
      $S_{D^+\pi^-}$    & $- S_- = - (a + c_i)$ &  N/A  &
      $2 R_{D\pi} \sin( 2\phi_1 + \phi_3 + \delta_{D\pi} )$ \\
      $S_{D^-\pi^+}$    & $- S_+ = - (a - c_i)$ &  N/A  &
      $2 R_{D\pi} \sin( 2\phi_1 + \phi_3 - \delta_{D\pi} )$ \\
      $S_{D^{*+}\pi^-}$ & $- S_- = - (a + c_i)$ & $S^+$ &   
      $- 2 R_{\Dstar \pi} \sin( 2\phi_1 + \phi_3 + \delta_{\Dstar \pi} )$ \\
      $S_{D^{*-}\pi^+}$ & $- S_+ = - (a - c_i)$ & $S^-$ &
      $- 2 R_{\Dstar \pi} \sin( 2\phi_1 + \phi_3 - \delta_{\Dstar \pi} )$ \\
      $S_{D^+\rho^-}$    & $- S_- = - (a + c_i)$ &  N/A  &  N/A  \\
      $S_{D^-\rho^+}$    & $- S_+ = - (a - c_i)$ &  N/A  &  N/A  \\
      \hline 
    \end{tabular*}
    \label{tab:cp_uta:notations:non_cp:dstarpi}
  \end{center}
\end{table}
   
\begin{table}
  \begin{center} 
    \caption{
      Translations used to convert the parameters measured by \belle
      to the parameters used for averaging in this document.
      The angular momentum factor $L$ is $-1$ for $\Dstar\pi$ and $+1$ for $D\pi$.
      Recall that $\left( \alpha, \beta, \gamma \right) = \left( \phi_2, \phi_1, \phi_3 \right)$.
    }
    \vspace{0.2cm}
    \setlength{\tabcolsep}{0.0pc}
    \begin{tabular*}{\textwidth}{@{\extracolsep{\fill}}ccc} \hline 
        & $\Dstar\pi$ partial rec. & $D^{(*)}\pi$ full rec. \\
        \hline
        $a$ & $- (S^+ + S^-)$ &
        $\frac{1}{2} (-1)^{L+1}
        \left(
          2 R_{D^{(*)}\pi} \sin( 2\phi_1 + \phi_3 + \delta_{D^{(*)}\pi} ) + 
          2 R_{D^{(*)}\pi} \sin( 2\phi_1 + \phi_3 - \delta_{D^{(*)}\pi} )
        \right)$ \\
        $c$ & $- (S^+ - S^-)$ & 
        $\frac{1}{2} (-1)^{L+1}
        \left(
          2 R_{D^{(*)}\pi} \sin( 2\phi_1 + \phi_3 + \delta_{D^{(*)}\pi} ) -
          2 R_{D^{(*)}\pi} \sin( 2\phi_1 + \phi_3 - \delta_{D^{(*)}\pi} )
        \right)$ \\
        \hline 
      \end{tabular*}
    \label{tab:cp_uta:notations:non_cp:dstarpi2}
  \end{center}
\end{table}

\mysubsubsubsection{Time-dependent asymmetries in radiative $\B$ decays
}
\label{sec:cp_uta:notations:non_cp:radiative}

As a special case of decays to non-$\CP$ eigenstates,
let us consider radiative $\B$ decays.
Here, the emitted photon has a distinct helicity,
which is in principle observable, but in practice is not usually measured.
Thus the measured time-dependent decay rates 
are given by~\cite{Atwood:1997zr,Atwood:2004jj}
\begin{eqnarray}
  \Gamma_{\Bzb \to X \gamma} (\Delta t) & = &
  \Gamma_{\Bzb \to X \gamma_L} (\Delta t) + \Gamma_{\Bzb \to X \gamma_R} (\Delta t) \\ \nonumber
  & = &
  \frac{e^{-\left| \Delta t \right| / \tau(\Bz)}}{4\tau(\Bz)} 
  \left\{ 
    1 + 
    \left( S_L + S_R \right) \sin(\Delta m \Delta t) - 
    \left( C_L + C_R \right) \cos(\Delta m \Delta t) 
  \right\},
  \\
  \Gamma_{\Bz \to X \gamma} (\Delta t) & = & 
  \Gamma_{\Bz \to X \gamma_L} (\Delta t) + \Gamma_{\Bz \to X \gamma_R} (\Delta t) \\ \nonumber 
  & = &
  \frac{e^{-\left| \Delta t \right| / \tau(\Bz)}}{4\tau(\Bz)} 
  \left\{ 
    1 - 
    \left( S_L + S_R \right) \sin(\Delta m \Delta t) + 
    \left( C_L + C_R \right) \cos(\Delta m \Delta t) 
  \right\},
\end{eqnarray}
where in place of the subscripts $f$ and $\bar{f}$ we have used $L$ and $R$
to indicate the photon helicity.
In order for interference between decays with and without $\Bz$-$\Bzb$ mixing
to occur, the $X$ system must not be flavour-specific,
\eg, in case of $\Bz \to K^{*0}\gamma$, the final state must be $\KS \pi^0 \gamma$.
The sign of the sine term depends on the $C$ eigenvalue of the $X$ system.
At leading order, the photons from 
$b \to q \gamma$ ($\bar{b} \to \bar{q} \gamma$) are predominantly
left (right) polarized, with corrections of order of $m_q/m_b$,
thus interference effects are suppressed.
Higher order effects can lead to corrections of order 
$\Lambda_{\rm QCD}/m_b$~\cite{Grinstein:2004uu,Grinstein:2005nu},
though explicit calculations indicate such corrections are small
for exclusive final states~\cite{Matsumori:2005ax,Ball:2006cva}.
The predicted smallness of the $S$ terms in the Standard Model
results in sensitivity to new physics contributions.

\mysubsubsection{Time-dependent \CP asymmetries in the $B_s$ System}
\label{sec:cp_uta:notations:Bs}

A complete analysis of the time-dependent decay rates of 
neutral $B$ mesons must also take into account the lifetime difference
between the eigenstates of the effective Hamiltonian, 
denoted by $\Delta \Gamma$.
This is particularly important in the $B_s$ system,
since non-negligible values of  $\Delta \Gamma_s$ are expected 
(see Section~\ref{sec:mixing} for the latest experimental constraints).
Neglecting $\CP$ violation in mixing,
the relevant replacements for 
Eqs.~\ref{eq:cp_uta:td_cp_asp1}~\&~\ref{eq:cp_uta:td_cp_asp2} 
are~\cite{Dunietz:2000cr}
\begin{equation}
  \label{eq:cp_uta:td_cp_bs_asp1}
  \begin{array}{lcr}
    \mc{2}{l}{
      \Gamma_{\Bsb \to f} (\Delta t) = 
      {\cal N} 
      \frac{e^{-| \Delta t | / \tau(\Bs)}}{4\tau(\Bs)}
      \Big[ 
      \cosh(\frac{\Delta \Gamma \Delta t}{2}) +
    } & \hspace{40mm} \\
    \hspace{40mm} &
    \mc{2}{r}{
      \frac{2\, \Im(\lambda_f)}{1 + |\lambda_f|^2} \sin(\Delta m \Delta t) -
      \frac{1 - |\lambda_f|^2}{1 + |\lambda_f|^2} \cos(\Delta m \Delta t) -
      \frac{2\, \Re(\lambda_f)}{1 + |\lambda_f|^2} \sinh(\frac{\Delta \Gamma \Delta t}{2})
      \Big],
    } \\
  \end{array}
\end{equation}
and
\begin{equation}
  \label{eq:cp_uta:td_cp_bs_asp2}
  \begin{array}{lcr}
    \mc{2}{l}{
      \Gamma_{\Bs \to f} (\Delta t) =
      {\cal N} 
      \frac{e^{-| \Delta t | / \tau(\Bs)}}{4\tau(\Bs)}
      \Big[ 
      \cosh(\frac{\Delta \Gamma \Delta t}{2}) -
    } & \hspace{40mm} \\
    \hspace{40mm} & 
    \mc{2}{r}{
      \frac{2\, \Im(\lambda_f)}{1 + |\lambda_f|^2} \sin(\Delta m \Delta t) +
      \frac{1 - |\lambda_f|^2}{1 + |\lambda_f|^2} \cos(\Delta m \Delta t) -
      \frac{2\, \Re(\lambda_f)}{1 + |\lambda_f|^2} \sinh(\frac{\Delta \Gamma \Delta t}{2})
      \Big]. 
    } \\
  \end{array}
\end{equation}

To be consistent with our earlier notation,\footnote{
  As ever, alternative and conflicting notations appear in the literature.
  One popular alternative notation for this parameter is 
  ${\cal A}_{\Delta \Gamma}$.
  Particular care must be taken over the signs.
}
we write here the coefficient of the $\sinh$ term as
\begin{equation}
  A^{\Delta \Gamma}_f = - \frac{2\, \Re(\lambda_f)}{1 + |\lambda_f|^2} \, .
\end{equation}
A complete, tagged, time-dependent analysis of \CP asymmetries in 
$B_s$ decays to a \CP eigenstate $f$ can thus obtain the parameters 
$S_f$, $C_f$ and $A^{\Delta \Gamma}_f$.
Note that 
\begin{equation}
  \left( S_f \right)^2 + \left( C_f \right)^2 + \left( A^{\Delta \Gamma}_f \right)^2 = 1 \, .
\end{equation}
Since these parameters have sensitivity to both
$\Im(\lambda_f)$ and $\Re(\lambda_f)$,
alternative choices of parametrization, 
including those directly involving \CP violating phases (such as $\beta_s$), 
are possible.
These can also be adopted for vector-vector final states.

The {\it untagged} time-dependent decay rate is given by
\begin{equation}
  \Gamma_{\Bsb \to f} (\Delta t) + \Gamma_{\Bs \to f} (\Delta t)
  = 
  {\cal N} 
  \frac{e^{-| \Delta t | / \tau(\Bs)}}{2\tau(\Bs)}
  \Big[ 
  \cosh(\frac{\Delta \Gamma \Delta t}{2}) -
  \frac{2\, \Re(\lambda_f)}{1 + |\lambda_f|^2} \sinh(\frac{\Delta \Gamma \Delta t}{2})
  \Big] \, .
\end{equation}
With the requirement
$\int_{-\infty}^{+\infty} \Gamma_{\Bsb \to f} (\Delta t) + \Gamma_{\Bs \to f} (\Delta t) d(\Delta t) = 1$,
the normalization factor ${\cal N}$ 
is fixed to $1 - (\frac{\Delta \Gamma}{2\Gamma})^2$.
Note that an untagged time-dependent analysis can probe
$\lambda_f$, through $\Re(\lambda_f)$, when $\Delta \Gamma \neq 0$.
The tagged analysis is, of course, more sensitive.

Other expressions can be similarly modified to take into account 
non-zero lifetime differences.
Note that when the final state contains 
a mixture of $\CP$-even and $\CP$-odd states
(as, for example, for vector-vector or multibody self-conjugate states),
that $\Re(\lambda_f)$ contains terms proportional to 
both the sine and cosine of the weak phase difference, 
albeit with rather different sensitivities.

\mysubsubsection{Asymmetries in $\B \to \DorDstar K^{(*)}$ decays
}
\label{sec:cp_uta:notations:cus}

$\CP$ asymmetries in $\B \to \DorDstar K^{(*)}$ decays are sensitive to $\gamma$.
The neutral $D^{(*)}$ meson produced 
is an admixture of $\DorDstarz$ (produced by a $b \to c$ transition) and 
$\DorDstarzb$ (produced by a colour-suppressed $b \to u$ transition) states.
If the final state is chosen so that both $\DorDstarz$ and $\DorDstarzb$ 
can contribute, the two amplitudes interfere,
and the resulting observables are sensitive to $\gamma$, 
the relative weak phase between 
the two $\B$ decay amplitudes~\cite{Bigi:1988ym}.
Various methods have been proposed to exploit this interference,
including those where the neutral $D$ meson is reconstructed 
as a $\CP$ eigenstate (GLW)~\cite{Gronau:1990ra,Gronau:1991dp},
in a suppressed final state (ADS)~\cite{Atwood:1996ci,Atwood:2000ck},
or in a self-conjugate three-body final state, 
such as $\KS \pi^+\pi^-$ (Dalitz)~\cite{Giri:2003ty,Poluektov:2004mf}.
It should be emphasised that while each method 
differs in the choice of $D$ decay,
they are all sensitive to the same parameters of the $B$ decay,
and can be considered as variations of the same technique.

Consider the case of $\Bmp \to D \Kmp$,
with $D$ decaying to a final state $f$,
which is accessible to both $\Dz$ and $\Dzb$.
We can write the decay rates for $\Bm$ and $\Bp$ ($\Gamma_\mp$), 
the charge averaged rate ($\Gamma = (\Gamma_- + \Gamma_+)/2$)
and the charge asymmetry 
(${\cal A} = (\Gamma_- - \Gamma_+)/(\Gamma_- + \Gamma_+)$, see Eq.~(\ref{eq:cp_uta:pra})) as 
\begin{eqnarray}
  \label{eq:cp_uta:dk:rate_def}
  \Gamma_\mp  & \propto & 
  r_B^2 + r_D^2 + 2 r_B r_D \cos \left( \delta_B + \delta_D \mp \gamma \right), \\
  \label{eq:cp_uta:dk:av_rate_def}
  \Gamma & \propto &  
  r_B^2 + r_D^2 + 2 r_B r_D \cos \left( \delta_B + \delta_D \right) \cos \left( \gamma \right), \\
  \label{eq:cp_uta:dk:acp_def}
  {\cal A} & = & 
  \frac{
    2 r_B r_D \sin \left( \delta_B + \delta_D \right) \sin \left( \gamma \right)
  }{
    r_B^2 + r_D^2 + 2 r_B r_D \cos \left( \delta_B + \delta_D \right) \cos \left( \gamma \right),  
  }
\end{eqnarray}
where the ratio of $\B$ decay amplitudes\footnote{
  Note that here we use the notation $r_B$ to denote the ratio
  of $\B$ decay amplitudes, 
  whereas in Sec.~\ref{sec:cp_uta:notations:non_cp:dstarpi} 
  we used, \eg, $R_{D\pi}$, for a rather similar quantity.
  The reason is that here we need to be concerned also with 
  $D$ decay amplitudes,
  and so it is convenient to use the subscript to denote the decaying particle.
  Hopefully, using $r$ in place of $R$ will help reduce potential confusion.
} 
is usually defined to be less than one,
\begin{equation}
  \label{eq:cp_uta:dk:rb_def}
  r_B = 
  \frac{
    \left| A\left( \Bm \to \Dzb K^- \right) \right|
  }{
    \left| A\left( \Bm \to \Dz  K^- \right) \right|
  },
\end{equation}
and the ratio of $D$ decay amplitudes is correspondingly defined by
\begin{equation}
  \label{eq:cp_uta:dk:rd_def}
  r_D = 
  \frac{
    \left| A\left( \Dz  \to f \right) \right|
  }{
    \left| A\left( \Dzb \to f \right) \right|
  }.
\end{equation}
The strong phase differences between the $\B$ and $D$ decay amplitudes 
are given by $\delta_B$ and $\delta_D$, respectively.
The values of $r_D$ and $\delta_D$ depend on the final state $f$:
for the GLW analysis, $r_D = 1$ and $\delta_D$ is trivial (either zero or $\pi$),
in the Dalitz plot analysis $r_D$ and $\delta_D$ vary across the Dalitz plot,
and depend on the $D$ decay model used,
for the ADS analysis, the values of $r_D$ and $\delta_D$ are not trivial.

Note that, for given values of $r_B$ and $r_D$, 
the maximum size of ${\cal A}$ (at $\sin \left( \delta_B + \delta_D \right) = 1$)
is $2 r_B r_D \sin \left( \gamma \right) / \left( r_B^2 + r_D^2 \right)$.
Thus even for $D$ decay modes with small $r_D$, 
large asymmetries, and hence sensitivity to $\gamma$, 
may occur for $B$ decay modes with similar values of $r_B$.
For this reason, the ADS analysis of the decay $B^\mp \to D \pi^\mp$ 
is also of interest.

In the GLW analysis, the measured quantities are the 
partial rate asymmetry, and the charge averaged rate,
which are measured both for $\CP$-even and $\CP$-odd $D$ decays.
For the latter, it is experimentally convenient to measure a double ratio,
\begin{equation}
  \label{eq:cp_uta:dk:double_ratio}
  R_{\CP} = 
  \frac{
    \Gamma\left( \Bm \to D_{\CP} \Km  \right) \, / \, \Gamma\left( \Bm \to \Dz \Km \right)
  }{
    \Gamma\left( \Bm \to D_{\CP} \pim \right) \, / \, \Gamma\left( \Bm \to \Dz \pim \right)
  }
\end{equation}
that is normalized both to the rate for the favoured $\Dz \to \Km\pip$ decay, 
and to the equivalent quantities for $\Bm \to D\pim$ decays
(charge conjugate modes are implicitly 
included in Eq.~(\ref{eq:cp_uta:dk:double_ratio})).
In this way the constant of proportionality drops out of 
Eq.~(\ref{eq:cp_uta:dk:av_rate_def}).

For the ADS analysis, using a suppressed $D \to f$ decay,
the measured quantities are again the partial rate asymmetry, 
and the charge averaged rate.
In this case it is sufficient to measure the rate in a single ratio
(normalized to the favoured $D \to \bar{f}$ decay)
since detection systematics cancel naturally;
the observed quantity is then
\begin{equation}
  \label{eq:cp_uta:dk:r_ads}
  R_{\rm ADS} = 
  \frac{
    \Gamma \left( \Bm \to \left[ f \right]_D \Km \right)
  }{
    \Gamma \left( \Bm \to \left[ \bar{f} \right]_D \Km \right)
  }.
\end{equation}
In the ADS analysis, there are an additional two unknowns ($r_D$ and $\delta_D$)
compared to the GLW case.  
However, the value of $r_D$ can be measured using 
decays of $D$ mesons of known flavour.

In the Dalitz plot analysis,
once a model is assumed for the $D$ decay, 
which gives the values of $r_D$ and $\delta_D$ across the Dalitz plot,
it is possible to perform a simultaneous fit to the $B^+$ and $B^-$ samples 
and directly extract $\gamma$, $r_B$ and $\delta_B$.
However, the uncertainties on the phases depend approximately inversely on $r_B$.
Furthermore, $r_B$ is positive definite (and small), 
and therefore tends to be overestimated,
which can lead to an underestimation of the uncertainty.
Some statistical treatment is necessary to correct for this bias.
An alternative approach is to extract from the data the ``Cartesian''
variables
\begin{equation}
  \left( x_\pm, y_\pm \right) = 
  \left( \Re(r_B e^{i(\delta_B\pm\gamma)}), \Im(r_B e^{i(\delta_B\pm\gamma)}) \right) = 
  \left( r_B \cos(\delta_B\pm\gamma), r_B \sin(\delta_B\pm\gamma) \right).
\end{equation}
These are (a) approximately statistically uncorrelated 
and (b) almost Gaussian.
The pairs of variables $\left( x_\pm, y_\pm \right)$ can be extracted
from independent fits of the $B^\pm$ data samples.
Use of these variables makes the combination of results much simpler.

However, if the Dalitz plot is effectively dominated by one $CP$ state,
there will be additional sensitivity to $\gamma$ in the numbers of events
in the $B^\pm$ data samples.
This can be taken into account in various ways.
One possibility is to extract GLW-like variables 
in addition to the $\left( x_\pm, y_\pm \right)$ parameters.
An alternative proceeds by defining $z_\pm = x_\pm + i y_\pm$
and $x_0 = - \int \Re \left[ f(s_1,s_2)f^*(s_2,s_1) \right] ds_1ds_2$,
where $s_1, s_2$ are the coordinates of invariant mass squared that
define the Dalitz plot and $f$ is the complex amplitude for $D$ decay
as a function of the Dalitz plot coordinates.\footnote{
  The $x_0$ parameter is closely related to the $c_i$ parameters of 
  the model dependent Dalitz plot analysis~\cite{Giri:2003ty,Bondar:2005ki,Bondar:2008hh},
  and the coherence factor of inclusive ADS-type analyses~\cite{Atwood:2003mj},
  integrated over the entire Dalitz plot.
}
The fitted parameters ($\rho^\pm, \theta^\pm$) are then defined by
\begin{equation}
  \rho^\pm e^{i \theta^\pm} = z_\pm - x_0 \, .
\end{equation}
Note that the yields of $B^\pm$ decays are proportional 
to $1 + (\rho^\pm)^2 - (x_0)^2$. 
This choice of variables has been used by \babar\ in the analysis of
$\Bmp \to D\Kmp$ with $D \to \pi^+\pi^-\pi^0$~\cite{Aubert:2007ii};
for this $D$ decay, $x_0 = 0.850$.

The relations between the measured quantities and the
underlying parameters are summarized in Table~\ref{tab:cp_uta:notations:dk}.
Note carefully that the hadronic factors $r_B$ and $\delta_B$ 
are different, in general, for each $\B$ decay mode.

\begin{table}[htb]
  \begin{center} 
    \caption{
      Summary of relations between measured and physical parameters 
      in GLW, ADS and Dalitz analyses of $\B \to \DorDstar K^{(*)}$.
    }
    \vspace{0.2cm}
    \setlength{\tabcolsep}{1.0pc}
    \begin{tabular}{cc} \hline 
      \mc{2}{c}{GLW analysis} \\
      $R_{\CP\pm}$ & $1 + r_B^2 \pm 2 r_B \cos \left( \delta_B \right) \cos \left( \gamma \right)$ \\
      $A_{\CP\pm}$ & $\pm 2 r_B \sin \left( \delta_B \right) \sin \left( \gamma \right) / R_{\CP\pm}$ \\
      \hline
      \mc{2}{c}{ADS analysis} \\
      $R_{\rm ADS}$ & $r_B^2 + r_D^2 + 2 r_B r_D \cos \left( \delta_B + \delta_D \right) \cos \left( \gamma \right)$ \\
      $A_{\rm ADS}$ & $2 r_B r_D \sin \left( \delta_B + \delta_D \right) \sin \left( \gamma \right) / R_{\rm ADS}$ \\
      \hline
      \mc{2}{c}{Dalitz analysis ($D \to \KS \pi^+\pi^-$)} \\
      $x_\pm$ & $r_B \cos(\delta_B\pm\gamma)$ \\
      $y_\pm$ & $r_B \sin(\delta_B\pm\gamma)$ \\
      \hline
      \mc{2}{c}{Dalitz analysis ($D \to \pi^+\pi^-\pi^0$)} \\
      $\rho^\pm$ & $|z_\pm - x_0|$ \\
      $\theta^\pm$ & $\tan^{-1}(\Im(z_\pm)/(\Re(z_\pm) - x_0))$ \\
      \hline
    \end{tabular}
    \label{tab:cp_uta:notations:dk}
  \end{center}
\end{table}

\mysubsection{Common inputs and error treatment
}
\label{sec:cp_uta:common_inputs}

The common inputs used for rescaling are listed in 
Table~\ref{tab:cp_uta:common_inputs}.
The $\Bz$ lifetime ($\tau(\Bz)$) and mixing parameter ($\Delta m_d$)
averages are provided by the HFAG Lifetimes and Oscillations 
subgroup (Sec.~\ref{sec:life_mix}).
The fraction of the perpendicularly polarized component 
($\left| A_{\perp} \right|^2$) in $\B \to \jpsi \Kstar(892)$ decays,
which determines the $\CP$ composition, 
is averaged from results by 
\babar~\cite{Aubert:2007hz} and \belle~\cite{Itoh:2005ks}.
See also HFAG $B$ to Charm Decay Parameters subgroup 
(Sec.~\ref{sec:BtoCharm}).

At present, we only rescale to a common set of input parameters
for modes with reasonably small statistical errors
($b \to c\bar{c}s$ transitions).
Correlated systematic errors are taken into account
in these modes as well.
For all other modes, the effect of such a procedure is 
currently negligible.

\begin{table}[htb]
  \begin{center}
    \caption{
      Common inputs used in calculating the averages.
    }
    \vspace{0.2cm}
    \setlength{\tabcolsep}{1.0pc}
    \begin{tabular}{cc} \hline 
      $\tau(\Bz)$ $({\rm ps})$  & $1.527 \pm 0.008$  \\
      $\Delta m_d$ $({\rm ps}^{-1})$ & $0.508 \pm 0.004$ \\
      $\left| A_{\perp} \right|^2 (\jpsi \Kstar)$ & $0.219 \pm 0.009$ \\
      \hline
    \end{tabular}
    \label{tab:cp_uta:common_inputs}
  \end{center}
\end{table}

As explained in Sec.~\ref{sec:intro},
we do not apply a rescaling factor on the error of an average
that has $\chi^2/\dof > 1$ 
(unlike the procedure currently used by the PDG~\cite{PDG_2007}).
We provide a confidence level of the fit so that
one can know the consistency of the measurements included in the average,
and attach comments in case some care needs to be taken in the interpretation.
Note that, in general, results obtained from data samples with low statistics
will exhibit some non-Gaussian behaviour.
We average measurements with asymmetric errors 
using the PDG~\cite{PDG_2007} prescription.
In cases where several measurements are correlated
(\eg\ $S_f$ and $C_f$ in measurements of time-dependent $\CP$ violation
in $B$ decays to a particular $\CP$ eigenstate)
we take these into account in the averaging procedure
if the uncertainties are sufficiently Gaussian.
For measurements where one error is given, 
it represents the total error, 
where statistical and systematic uncertainties have been added in quadrature.
If two errors are given, the first is statistical and the second systematic.
If more than two errors are given,
the origin of the additional uncertainty will be explained in the text.


\clearpage
\mysubsection{Time-dependent asymmetries in $b \to c\bar{c}s$ transitions
}
\label{sec:cp_uta:ccs}

\mysubsubsection{Time-dependent $\CP$ asymmetries in $b \to c\bar{c}s$ decays to $\CP$ eigenstates
}
\label{sec:cp_uta:ccs:cp_eigen}

\begin{figure}[htb]
  \begin{center}
    \resizebox{0.55\textwidth}{!}{
      \includegraphics{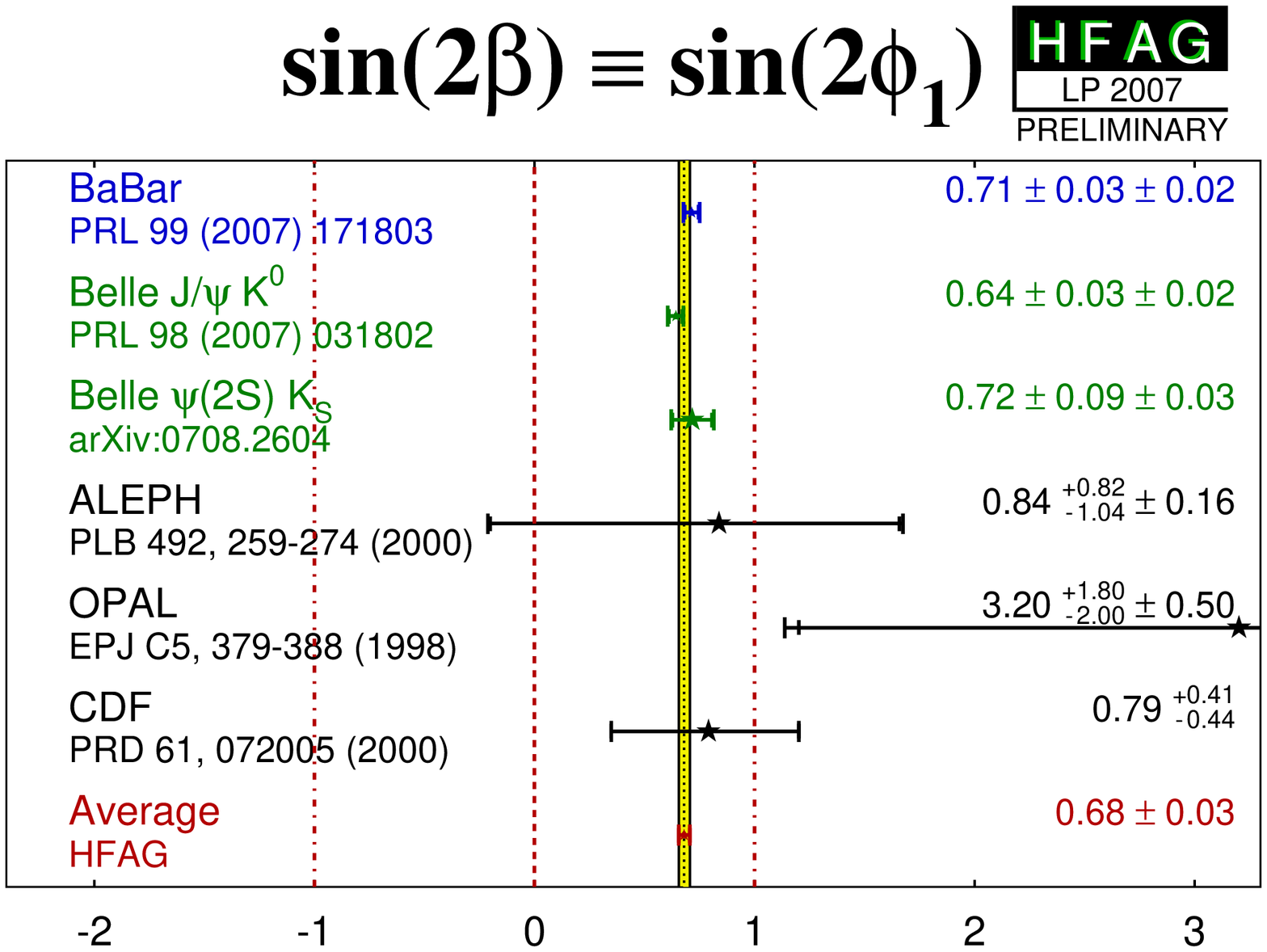}
    }
    \hfill
    \resizebox{0.44\textwidth}{!}{
      \includegraphics{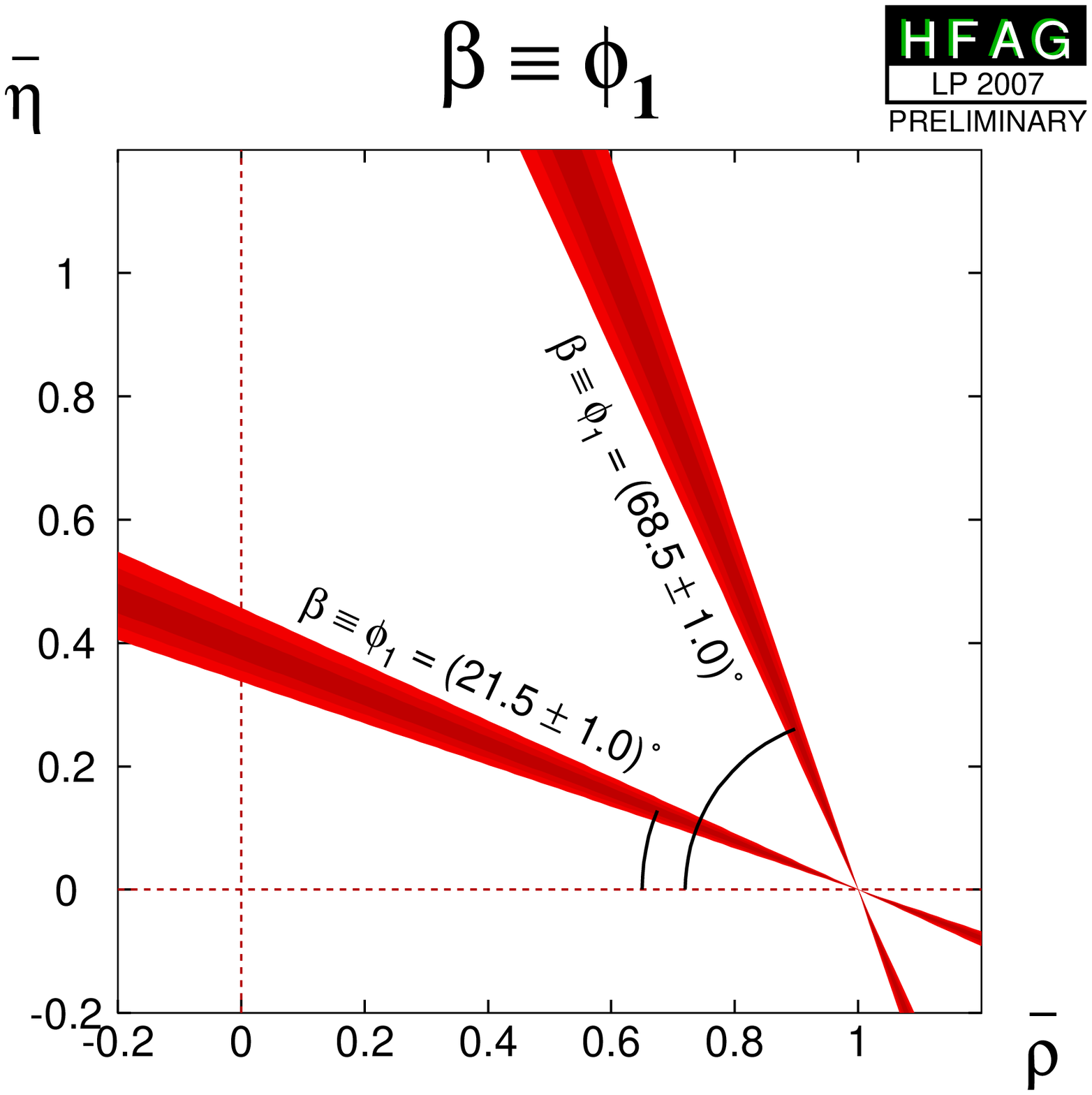}
    }
  \end{center}
  \vspace{-0.5cm}
  \caption{
    (Left) Average of measurements of $S_{b \to c\bar c s}$.
    (Right) Constraints on the $(\rhobar,\etabar)$ plane,
    obtained from the average of $-\etacp S_{b \to c\bar c s}$ 
    and Eq.~\ref{eq:cp_uta:sin2beta}.
  }
  \label{fig:cp_uta:ccs}
\end{figure}

In the Standard Model, the time-dependent parameters for
$b \to c\bar c s$ transitions are predicted to be: 
$S_{b \to c\bar c s} = - \etacp \sin(2\beta)$,
$C_{b \to c\bar c s} = 0$ to very good accuracy.
The averages for $-\etacp S_{b \to c\bar c s}$ and $C_{b \to c\bar c s}$
are provided in Table~\ref{tab:cp_uta:ccs}.
The averages for $-\etacp S_{b \to c\bar c s}$ 
are shown in Fig.~\ref{fig:cp_uta:ccs}.

Both \babar\  and \belle\ have used the $\etacp = -1$ modes
$\jpsi \KS$, $\psi(2S) \KS$, $\chi_{c1} \KS$ and $\eta_c \KS$, 
as well as $\jpsi \KL$, which has $\etacp = +1$
and $\jpsi K^{*0}(892)$, which is found to have $\etacp$ close to $+1$
based on the measurement of $\left| A_\perp \right|$ 
(see Sec.~\ref{sec:cp_uta:common_inputs}).
ALEPH, OPAL and CDF used only the $\jpsi \KS$ final state.
In the latest result from \belle~\cite{Chen:2006nk}, 
only $\jpsi \KS$ and $\jpsi \KL$ are used,
while results from $\psi(2{\rm S}) \KS$ have been presented separately~\cite{Sahoo:2008zz}.
A breakdown of results in each charmonium-kaon final state is given in 
Table~\ref{tab:cp_uta:ccs-BF}.

\begin{table}[htb]
	\begin{center}
		\caption{
                        $S_{b \to c\bar c s}$ and $C_{b \to c\bar c s}$.
                }
		\vspace{0.2cm}
		\setlength{\tabcolsep}{0.0pc}
\begin{tabular*}{\textwidth}{@{\extracolsep{\fill}}lrccc} \hline
      \mc{2}{l}{Experiment} & $N(B\bar{B})$ & $- \etacp S_{b \to c\bar c s}$ & $C_{b \to c\bar c s}$ \\
      \hline
	\babar & \cite{Aubert:2007hm} & 384M & $0.714 \pm 0.032 \pm 0.018$ & $0.049 \pm 0.022 \pm 0.017$ \\
	\belle $J/\psi K^0$ & \cite{Chen:2006nk} & 535M & $0.642 \pm 0.031 \pm 0.017$ & $-0.018 \pm 0.021 \pm 0.014$ \\
	\belle $\psi(2{\rm S}) \KS$ & \cite{Sahoo:2008zz} & 657M & $0.718 \pm 0.090 \pm 0.033$ & $-0.039 \pm 0.069 \pm 0.049$ \\
	\mc{3}{l}{\bf \boldmath $\B$ factory average} & $0.680 \pm 0.025$ & $0.012 \pm 0.020$ \\
	\mc{3}{l}{\small Confidence level} & {\small $0.33$} & {\small $0.10$} \\
        \hline
        ALEPH & \cite{Barate:2000tf} & \textendash{} & $0.84 \, ^{+0.82}_{-1.04} \pm 0.16$ &  \textendash{} \\
        OPAL  & \cite{Ackerstaff:1998xz} & \textendash{} & $3.2 \, ^{+1.8}_{-2.0} \pm 0.5$ &  \textendash{} \\
        CDF   & \cite{Affolder:1999gg} & \textendash{} & $0.79 \, ^{+0.41}_{-0.44}$ &  \textendash{} \\
        \mc{3}{l}{\bf Average} & $0.681 \pm 0.025$ & $0.012 \pm 0.020 $ \\
		\hline
		\end{tabular*}
                \label{tab:cp_uta:ccs}
        \end{center}
\end{table}

\begin{table}[htb]
	\begin{center}
		\caption{
                        Breakdown of $B$ factory results on $S_{b \to c\bar c s}$ and $C_{b \to c\bar c s}$.
                }
		\vspace{0.2cm}
		\setlength{\tabcolsep}{0.0pc}
		\begin{tabular*}{\textwidth}{@{\extracolsep{\fill}}lrccc} \hline
        \mc{2}{l}{Mode} & $N(B\bar{B})$ & $- \etacp S_{b \to c\bar c s}$ & $C_{b \to c\bar c s}$ \\
        \hline
        \mc{5}{c}{\babar} \\
        $J/\psi \KS$ & \cite{Aubert:2007hm} & 384M & $0.686 \pm 0.039 \pm 0.015$ & $0.051 \pm 0.027 \pm 0.015$ \\
        $J/\psi \KL$ & \cite{Aubert:2007hm} & 384M & $0.735 \pm 0.074 \pm 0.067$ & $-0.063 \pm 0.062 \pm 0.030$ \\
        {\bf \boldmath $J/\psi K^0$} & \cite{Aubert:2007hm} & 384M & $0.697 \pm 0.035 \pm 0.016$ & $0.035 \pm 0.025 \pm 0.018$ \\
        $\psi(2S) \KS$ & \cite{Aubert:2007hm} & 384M & $0.947 \pm 0.112 \pm 0.062$ & $0.142 \pm 0.079 \pm 0.047$ \\
        $\chi_{c1} \KS$ & \cite{Aubert:2007hm} & 384M & $0.759 \pm 0.170 \pm 0.037$ & $0.339 \pm 0.102 \pm 0.104$ \\
        $\eta_c \KS$ & \cite{Aubert:2007hm} & 384M & $0.778 \pm 0.195 \pm 0.093$ & $0.053 \pm 0.141 \pm 0.037$ \\
        $\jpsi K^{*0}(892)$ & \cite{Aubert:2007hm} & 384M & $0.477 \pm 0.271 \pm 0.155$ & $0.047 \pm 0.083 \pm 0.026$ \\
        {\bf All} & \cite{Aubert:2007hm} & 384M & $0.714 \pm 0.032 \pm 0.018$ & $0.049 \pm 0.022 \pm 0.017$ \\
	\hline
	\mc{5}{c}{\bf \belle} \\
        $J/\psi \KS$ & \cite{Chen:2006nk} & 535M & $0.643 \pm 0.038_{\rm stat}$ & $0.001 \pm 0.028_{\rm stat}$ \\
        $J/\psi \KL$ & \cite{Chen:2006nk} & 535M & $0.641 \pm 0.057_{\rm stat}$ &$-0.045 \pm 0.033_{\rm stat}$ \\
        {\bf \boldmath $J/\psi K^0$} & \cite{Chen:2006nk} & 535M & $0.642 \pm 0.031 \pm 0.017$ & $-0.018 \pm 0.021 \pm 0.014$ \\
	$\psi(2{\rm S}) \KS$ & \cite{Sahoo:2008zz} & 657M & $0.718 \pm 0.090 \pm 0.033$ & $-0.039 \pm 0.069 \pm 0.049$ \\
        {\bf All} & \cite{Sahoo:2008zz} & \textendash{} & $0.650 \pm 0.029 \pm 0.015$ & \textendash{} \\
	\hline
        \mc{3}{l}{\bf \boldmath $J/\psi K^0$ average} & $0.668 \pm 0.026$ & $0.002 \pm 0.021$ \\ 
        \mc{3}{l}{\bf \boldmath $\psi(2{\rm S}) K^0$ average} & $0.802 \pm 0.077$ & $0.032 \pm 0.068$ \\ 
		\hline
		\end{tabular*}
                \label{tab:cp_uta:ccs-BF}
        \end{center}
\end{table}

It should be noted that, while the uncertainty in the average for 
$-\etacp S_{b \to c\bar c s}$ is still limited by statistics,
that for $C_{b \to c\bar c s}$ is close to being dominated by systematics.
This occurs due to the possible effect of tag side interference on the
$C_{b \to c\bar c s}$ measurement, an effect which is correlated between
the different experiments.
Understanding of this effect may continue to improve in future,
allowing the uncertainty to reduce.

From the average for $-\etacp S_{b \to c\bar c s}$ above, 
we obtain the following solutions for $\beta$
(in $\left[ 0, \pi \right]$):
\begin{equation}
  \beta = \left( 21.5 \pm 1.0 \right)^\circ
  \hspace{5mm}
  {\rm or}
  \hspace{5mm}
  \beta = \left( 68.5 \pm 1.0 \right)^\circ
  \label{eq:cp_uta:sin2beta}
\end{equation}
In radians, these values are 
$\beta = \left( 0.38 \pm 0.02 \right)$, $\beta = \left( 1.20 \pm 0.02 \right)$.

This result gives a precise constraint on the $(\rhobar,\etabar)$ plane,
as shown in Fig.~\ref{fig:cp_uta:ccs}.
The measurement is in remarkable agreement with other constraints from 
$\CP$ conserving quantities, 
and with $\CP$ violation in the kaon system, in the form of the parameter $\epsilon_K$.
Such comparisons have been performed by various phenomenological groups,
such as CKMfitter~\cite{Charles:2004jd} 
and UTFit~\cite{Bona:2005vz}.

\mysubsubsection{Time-dependent transversity analysis of $\Bz \to J/\psi K^{*0}$
}
\label{sec:cp_uta:ccs:vv}

$\B$ meson decays to the vector-vector final state $J/\psi K^{*0}$
are also mediated by the $b \to c \bar c s$ transition.
When a final state which is not flavour-specific ($K^{*0} \to \KS \pi^0$) is used,
a time-dependent transversity analysis can be performed 
allowing sensitivity to both 
$\sin(2\beta)$ and $\cos(2\beta)$~\cite{Dunietz:1990cj}.
Such analyses have been performed by both $\B$ factory experiments.
In principle, the strong phases between the transversity amplitudes
are not uniquely determined by such an analysis, 
leading to a discrete ambiguity in the sign of $\cos(2\beta)$.
The \babar\ collaboration resolves 
this ambiguity using the known variation~\cite{Aston:1987ir}
of the P-wave phase (fast) relative to the S-wave phase (slow) 
with the invariant mass of the $K\pi$ system 
in the vicinity of the $K^*(892)$ resonance. 
The result is in agreement with the prediction from 
$s$ quark helicity conservation,
and corresponds to Solution II defined by Suzuki~\cite{Suzuki:2001za}.
We use this phase convention for the averages given in 
Table~\ref{tab:cp_uta:ccs:psi_kstar}.

\begin{table}[htb]
	\begin{center}
		\caption{
			Averages from $\Bz \to J/\psi K^{*0}$ transversity analyses.
		}
		\vspace{0.2cm}
		\setlength{\tabcolsep}{0.0pc}
		\begin{tabular*}{\textwidth}{@{\extracolsep{\fill}}lrcccc} \hline
		\mc{2}{l}{Experiment} & $N(B\bar{B})$ & $\sin 2\beta$ & $\cos 2\beta$ & Correlation \\
		\hline
	\babar & \cite{Aubert:2004cp} & 88M & $-0.10 \pm 0.57 \pm 0.14$ & $3.32 ^{+0.76}_{-0.96} \pm 0.27$ & $-0.37$ \\
	\belle & \cite{Itoh:2005ks} & 275M & $0.24 \pm 0.31 \pm 0.05$ & $0.56 \pm 0.79 \pm 0.11$ & $0.22$ \\
	\mc{3}{l}{\bf Average} & $0.16 \pm 0.28$ & $1.64 \pm 0.62$ &  \hspace{-8mm} {\small uncorrelated averages}  \\
        \mc{3}{l}{\small Confidence level} & {\small $0.61~(0.5\sigma)$} & {\small $0.03~(2.2\sigma)$} & \\
		\hline
		\end{tabular*}
		\label{tab:cp_uta:ccs:psi_kstar}
	\end{center}
\end{table}

At present the results are dominated by 
large and non-Gaussian statistical errors,
and exhibit significant correlations.
We perform uncorrelated averages, 
the interpretation of which has to be done with the greatest care. 
Nonetheless, it is clear that $\cos(2\beta)>0$ is preferred 
by the experimental data in $J/\psi \Kstar$.
[\babar~\cite{Aubert:2004cp} 
find a confidence level for $\cos(2\beta)>0$ of $89\%$.]

\mysubsubsection{Time-dependent $\CP$ asymmetries in $\Bz \to \Dstarp \Dstarm \KS$ decays
}
\label{sec:cp_uta:ccs:DstarDstarKs}

\begin{figure}[htb]
  \begin{center}
    \begin{tabular}{c@{\hspace{-1mm}}c@{\hspace{-1mm}}c}
      \resizebox{0.32\textwidth}{!}{
        \includegraphics{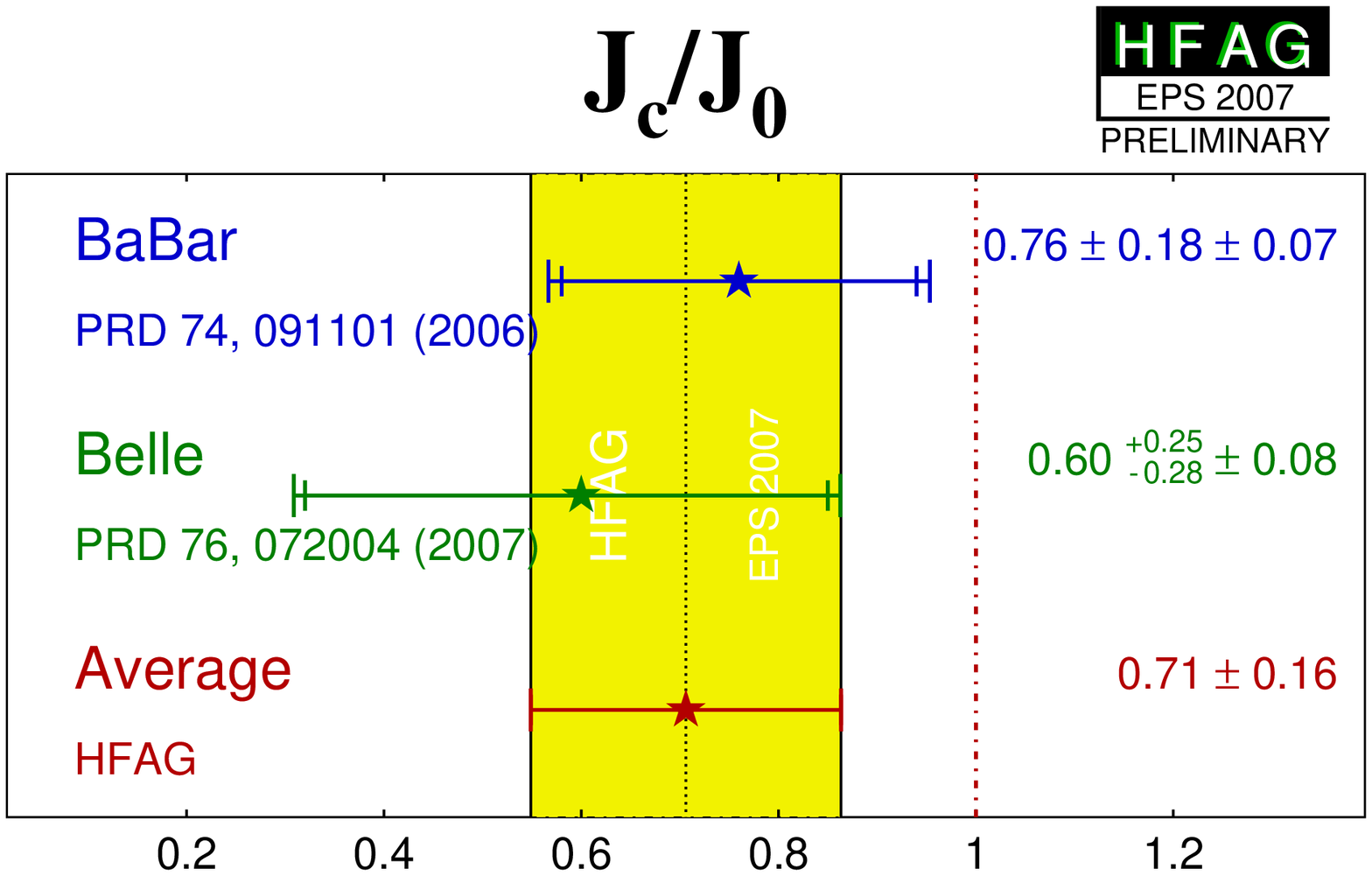}
      }
      &
      \resizebox{0.32\textwidth}{!}{
        \includegraphics{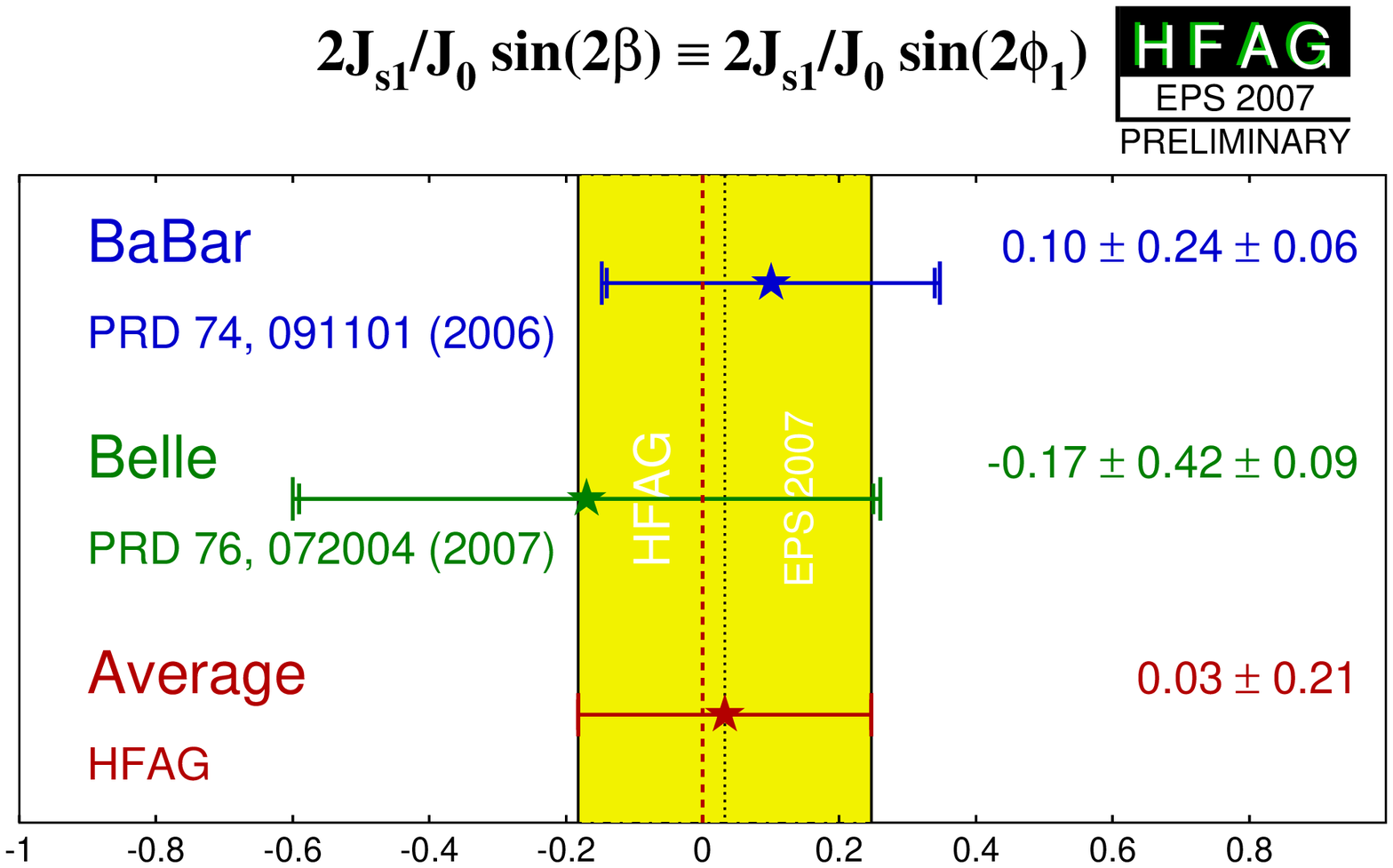}
      }
      &
      \resizebox{0.32\textwidth}{!}{
        \includegraphics{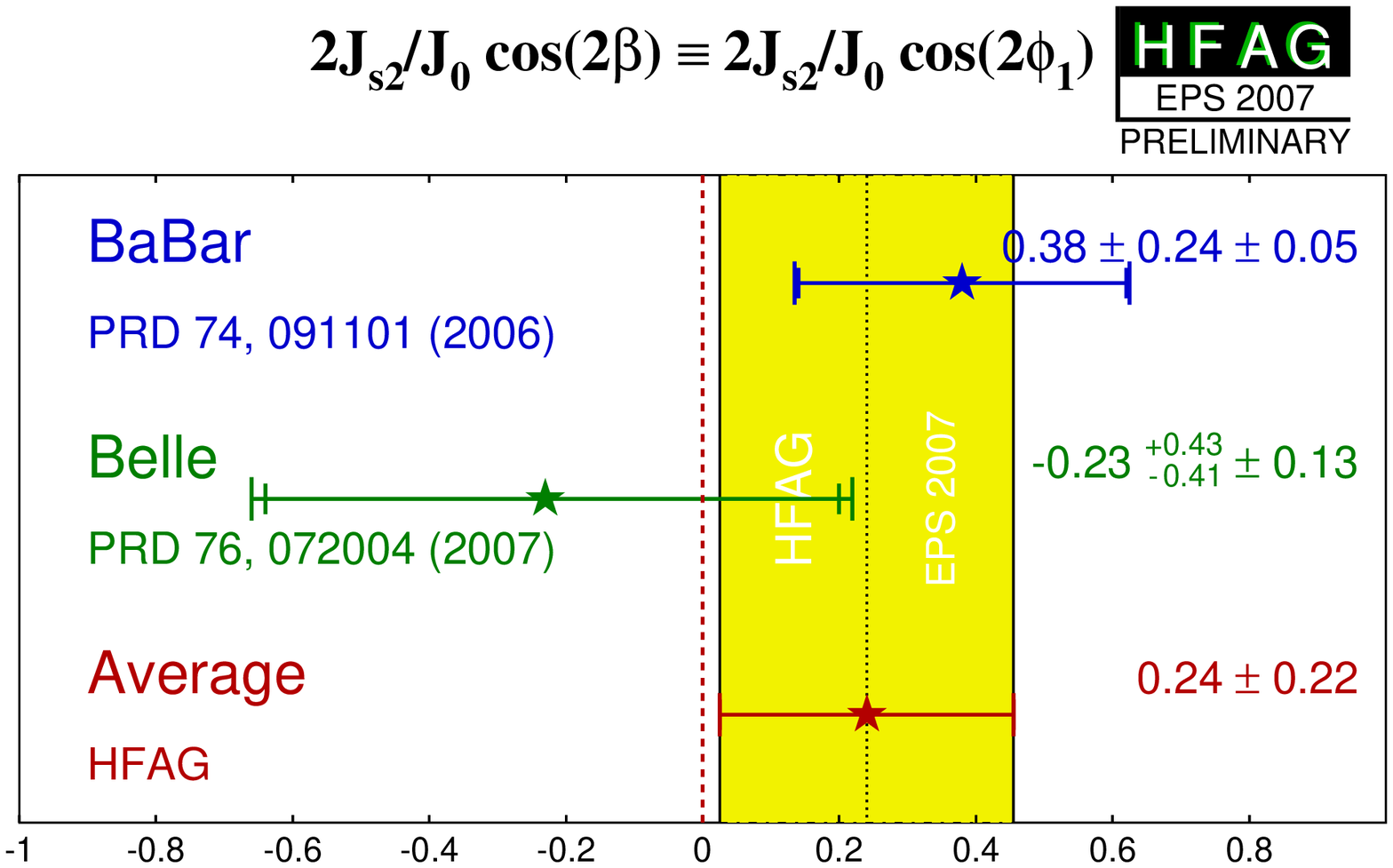}
      }
    \end{tabular}
  \end{center}
  \vspace{-0.8cm}
  \caption{
    Averages of 
    (left) $(J_c/J_0)$, (middle) $(2J_{s1}/J_0) \sin(2\beta)$ and 
    (right) $(2J_{s2}/J_0) \cos(2\beta)$
    from time-dependent analyses of $\Bz \to \Dstarp \Dstarm \KS$ decays.
  }
  \label{fig:cp_uta:ccs:dstardstarks}
\end{figure}

Both \babar~\cite{Aubert:2006fh} and \belle~\cite{Dalseno:2007hx} have performed
time-dependent analyses of the $\Bz \to \Dstarp \Dstarm \KS$ decay,
to obtain information on the sign of $\cos(2\beta)$.
More information can be found in 
Sec.~\ref{sec:cp_uta:notations:dalitz:dstardstarks}.
The results are shown in Table~\ref{tab:cp_uta:ccs:dstardstarks}, 
and Fig.~\ref{fig:cp_uta:ccs:dstardstarks}.

\begin{table}[htb]
	\begin{center}
		\caption{
                        Results from time-dependent analysis of $\Bz \to \Dstarp \Dstarm \KS$.
		}
		\vspace{0.2cm}
		\setlength{\tabcolsep}{0.0pc}
		\begin{tabular*}{\textwidth}{@{\extracolsep{\fill}}lrcccc} \hline
                \mc{2}{l}{Experiment} & $N(B\bar{B})$ & $\frac{J_c}{J_0}$ & $\frac{2J_{s1}}{J_0} \sin(2\beta)$ &  $\frac{2J_{s2}}{J_0} \cos(2\beta)$ \\
		\hline
	\babar & \cite{Aubert:2006fh} & 230M & $0.76 \pm 0.18 \pm 0.07$ & $0.10 \pm 0.24 \pm 0.06$ & $0.38 \pm 0.24 \pm 0.05$ \\
	\belle & \cite{Dalseno:2007hx} & 449M & $0.60 \,^{+0.25}_{-0.28} \pm 0.08$ & $-0.17 \pm 0.42 \pm 0.09$ & $-0.23 \,^{+0.43}_{-0.41} \pm 0.13$ \\
	\mc{3}{l}{\bf Average} & $0.71 \pm 0.16$ & $0.03 \pm 0.21$ & $0.24 \pm 0.22$ \\
	\mc{3}{l}{\small Confidence level} & {\small $0.63~(0.5\sigma)$} & {\small $0.59~(0.5\sigma)$} & {\small $0.23~(1.2\sigma)$} \\
		\hline
		\end{tabular*}
		\label{tab:cp_uta:ccs:dstardstarks}
	\end{center}
\end{table}

From the above result and the assumption that $J_{s2}>0$, 
\babar\ infer that $\cos(2\beta)>0$ at the $94\%$ confidence level.

\mysubsubsection{Time-dependent analysis of $\Bs \to J/\psi \phi$
}
\label{sec:cp_uta:ccs:jpsiphi}

As described in Sec.~\ref{sec:cp_uta:notations:Bs},
time-dependent analysis of $\Bs \to J/\psi \phi$ probes the 
$\CP$ violating phase of $\Bs$--$\Bsb$ oscillations, $\phi_s$.
Within the Standard Model, this parameter is predicted to be small.\footnote{
   We make the approximation $\phi_s = œôø²2 \beta_s$, 
   where $\phi_s \equiv \arg\left[ -M_{12}/\Gamma_{12} \right]$ 
   and $2\beta_s \equiv 2 \arg\left[ -(V_{ts}V_{tb}^*)/(V_{cs}V_{cb}^*) \right]$
   (see Section~\ref{sec:cp_uta:introduction}). 
   This is a reasonable approximation since, 
   although the equality does not hold in the Standard Model~\cite{Lenz:2006hd}, 
   both are much smaller than the current experimental resolution, 
   whereas new physics contributions add a phase $\phi_{\rm NP}$ to $\phi_s$
   and subtract the same phase from $2\beta_s$, 
   so that the approximation remains valid.
}

Both \cdf~\cite{Aaltonen:2007he} and \dzero~\cite{:2008fj} 
have performed full tagged, time-dependent angular analyses of 
$\Bs \to J/\psi \phi$ decays,
superceding their previous untagged results~\cite{:2007gf,Abazov:2007tx}.

Both experiments perform analyses that take into account the correlations
between the average $\Bs$ lifetime $\tau(\Bs)$, 
$\Delta \Gamma_s$, $\phi_s$, 
the magnitude of the perpendicularly polarized component $A_\perp$, 
the difference in the fractions of the two $CP$-even components $|A_0|^2 - |A|_\parallel^2$, 
and the strong phases associated with the two $CP$-even components 
$\delta_0$ and $\delta_\parallel$.
The \cdf\ analysis~\cite{Aaltonen:2007he} uses $1.35 \ {\rm fb}^{œôø²1}$ of data.
The likelihood function is found to have a highly non-Gaussian shape,
so that central values and uncertainties are not presented.
The \dzero\ analysis~\cite{:2008fj} uses $2.8 \ {\rm fb}^{œôø²1}$ of data,
and constrains the strong phase differences to take equivalent values 
to those measured in $\Bz \to J/\psi \Kstar$~\cite{Aubert:2007hz},
up to an uncertainty of $\pi/5$ that allows for SU(3) breaking effects.

The results are given in Table~\ref{tab:cp_uta:ccs:psiphi} below.
See also HFAG Lifetimes and Oscillations, Sec.~\ref{sec:life_mix}.

\begin{table}[htb]
	\begin{center}
		\caption{
			 Results from time-dependent analysis of $\Bs \to J/\psi \phi$.
		}
		\vspace{0.2cm}
		\setlength{\tabcolsep}{0.0pc}
    \resizebox{\textwidth}{!}{
      \begin{tabular}{@{\extracolsep{2mm}}lrccccc} 
        \hline
        \mc{2}{l}{Experiment} & $\tau(\Bs)$ & $\Delta\Gamma$ & $\phi_s$ & $A_\perp$ & $|A_0|^2 - |A|_\parallel^2$ \\
	\dzero & \cite{:2008fj} & $1.52 \pm 0.06 \pm 0.01$ & $0.19 \pm 0.07 \,^{+0.02}_{-0.01}$ & $-0.57 \,^{+0.24}_{-0.30} \,^{+0.07}_{-0.02}$ & $0.41 \pm 0.04 \,^{+0.01}_{-0.02}$ & $0.34 \pm 0.05 \pm 0.03$ \\
        \hline
      \end{tabular}
    }

		\label{tab:cp_uta:ccs:psiphi}
	\end{center}
\end{table}

Note the implicit convention above is that 
$|A_\perp|^2 + |A_0|^2 + |A_\parallel|^2 = 1$, 
and the strong phases are measured relative to that of the $A_\perp$ component 
(which is set to zero). 
The polarization components are defined at time $t=0$, 
\ie\ at the production (primary) vertex of the $\Bs$. 
Note also that there is an ambiguity in the result for $\phi_s$.

\vspace{3ex}

\noindent
\underline{\large Constraints on $\phi_s$}

\begin{itemize}\setlength{\itemsep}{0.5ex}
\item 
  \cdf~\cite{Aaltonen:2007he} present a confidence region in the 
  $\phi_s$--$\Delta \Gamma_s$ plane, from which they obtain 
  $\phi_s \in \left[-2.82, -0.32 \right]$ at the 68\% confidence level. 
  The consistency with the Standard Model expectation for 
  $(\phi_s, \Delta \Gamma_s)$ is 15\%.

\item 
  \dzero~\cite{:2008fj} obtain a 90\% CL allowed interval 
  $\phi_s \in \left[-1.20, +0.06\right]$. 

\item 
  The UTFit group have performed a preliminary average of the above two results.

\end{itemize}

For more details, 
see the HFAG Lifetimes and Oscillations group, Sec.~\ref{sec:life_mix}.

\clearpage
\mysubsection{Time-dependent $\CP$ asymmetries in colour-suppressed $b \to c\bar{u}d$ transitions
}
\label{sec:cp_uta:cud_beta}

Decays of $\B$ mesons to final states such as $D\pi^0$ are 
governed by $b \to c\bar{u}d$ transitions. 
If the final state is a $\CP$ eigenstate, \eg\ $D_{\CP}\pi^0$, 
the usual time-dependence formulae are recovered, 
with the sine coefficient sensitive to $\sin(2\beta)$. 
Since there is no penguin contribution to these decays, 
there is even less associated theoretical uncertainty 
than for $b \to c\bar{c}s$ decays like $\B \to \jpsi \KS$.
Such measurements therefore allow to test the Standard Model prediction
that the $\CP$ violation parameters in $b \to c\bar{u}d$ transitions
are the same as those in $b \to c\bar{c}s$~\cite{Grossman:1996ke}.

Note that there is an additional contribution from CKM suppressed
$b \to u \bar{c} d$ decays.
The effect of this contribution is small, and can be taken into 
account in the analysis~\cite{Fleischer:2003ai,Fleischer:2003aj}.

Results of such an analysis are available from \babar~\cite{Aubert:2007mn}.
The decays $\Bz \to D\pi^0$, $\Bz \to D\eta$, $\Bz \to D\omega$,
$\Bz \to D^*\pi^0$ and $\Bz \to D^*\eta$ are used.
The daughter decay $D^* \to D\pi^0$ is used.
The $\CP$-even $D$ decay to $K^+K^-$ is used for all decay modes,
with the $\CP$-odd $D$ decay to $\KS\omega$ also used in $\Bz \to D^{(*)}\pi^0$
and the additional $\CP$-odd $D$ decay to $\KS\pi^0$ 
also used in $\Bz \to D\omega$.
Results are presented separately for $\CP$-even and $\CP$-odd 
$D^{(*)}$ decays (denoted $D^{(*)}_+ h^0$ and $D^{(*)}_- h^0$ respectively),
and for both combined, with the different $\CP$ factors accounted for
(denoted $D^{(*)}_{CP} h^0$).
The results are summarized in Table~\ref{tab:cp_uta:cud_cp_beta}.

\begin{table}[htb]
	\begin{center}
		\caption{
			Results from analyses of $\Bz \to D^{(*)}h^0$, $D \to CP$ eigenstates decays.
		}
		\vspace{0.2cm}
		\setlength{\tabcolsep}{0.0pc}
		\begin{tabular*}{\textwidth}{@{\extracolsep{\fill}}lrcccc} \hline
	\mc{2}{l}{Experiment} & $N(B\bar{B})$ & $S_{CP}$ & $C_{CP}$ & Correlation \\
	\hline
        \mc{6}{c}{$D^{(*)}_+ h^0$}  \\
	\babar & \cite{Aubert:2007mn} & 383M & $-0.65 \pm 0.26 \pm 0.06$ & $-0.33 \pm 0.19 \pm 0.04$ & $0.04$ \\
	\hline

        \mc{6}{c}{$D^{(*)}_- h^0$} \\
	\babar & \cite{Aubert:2007mn} & 383M & $-0.46 \pm 0.46 \pm 0.13$ & $-0.03 \pm 0.28 \pm 0.07$ & $-0.14$ \\
	\hline

        \mc{6}{c}{$D^{(*)}_{CP} h^0$} \\
	\babar & \cite{Aubert:2007mn} & 383M & $-0.56 \pm 0.23 \pm 0.05$ & $-0.23 \pm 0.16 \pm 0.04$ & $-0.02$ \\
	\hline
		\end{tabular*}
		\label{tab:cp_uta:cud_cp_beta}
	\end{center}
\end{table}

When multibody $D$ decays, such as $D \to \KS\pi^+\pi^-$ are used, 
a time-dependent analysis of the Dalitz plot of the neutral $D$ decay 
allows a direct determination of the weak phase: $2\beta$. 
(Equivalently, both $\sin(2\beta)$ and $\cos(2\beta)$ can be measured.)
This information allows to resolve the ambiguity in the 
measurement of $2\beta$ from $\sin(2\beta)$~\cite{Bondar:2005gk}.

\begin{figure}[htb]
  \begin{center}
    \begin{tabular}{cc}
      \resizebox{0.46\textwidth}{!}{
        \includegraphics{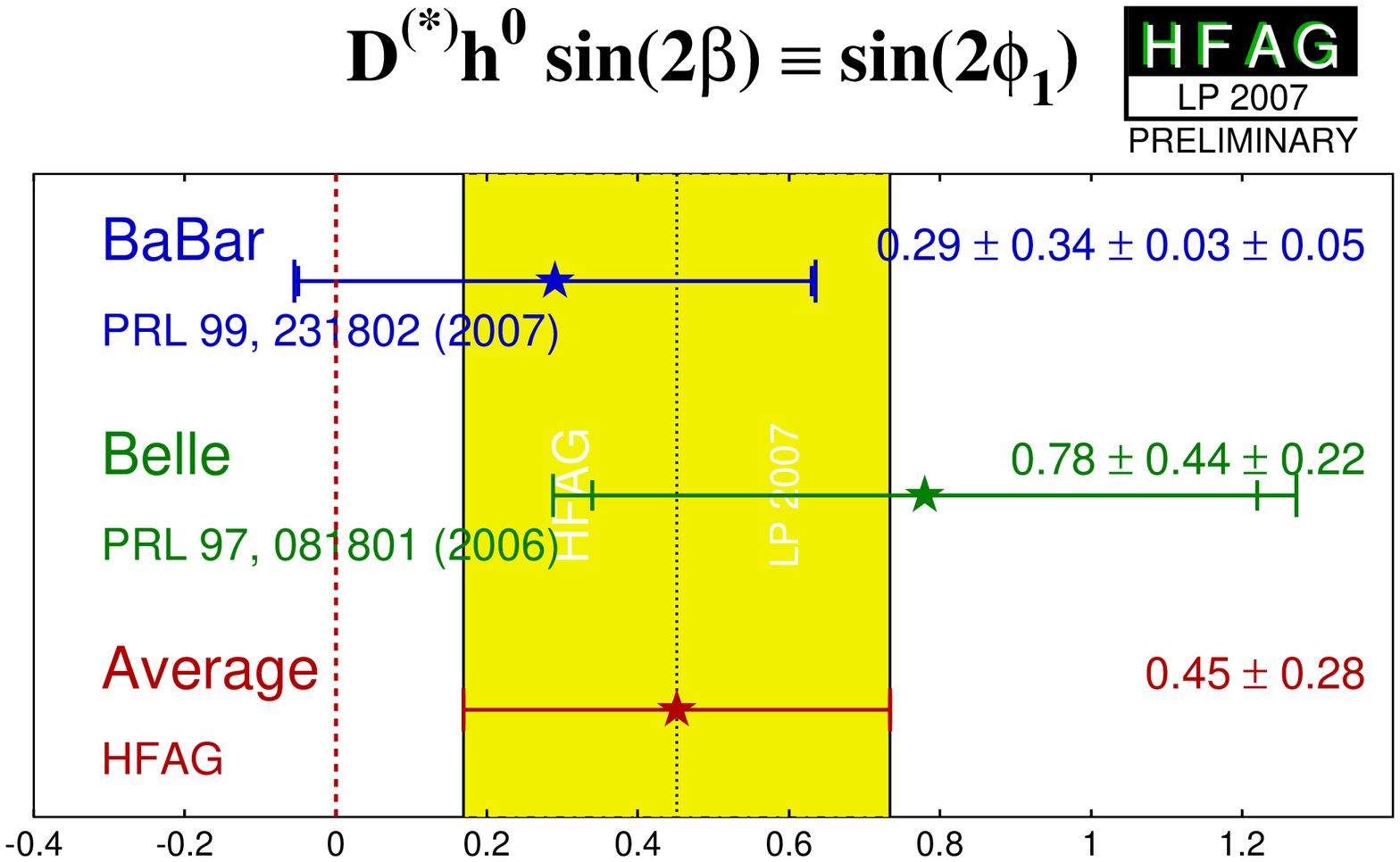}
      }
      &
      \resizebox{0.46\textwidth}{!}{
        \includegraphics{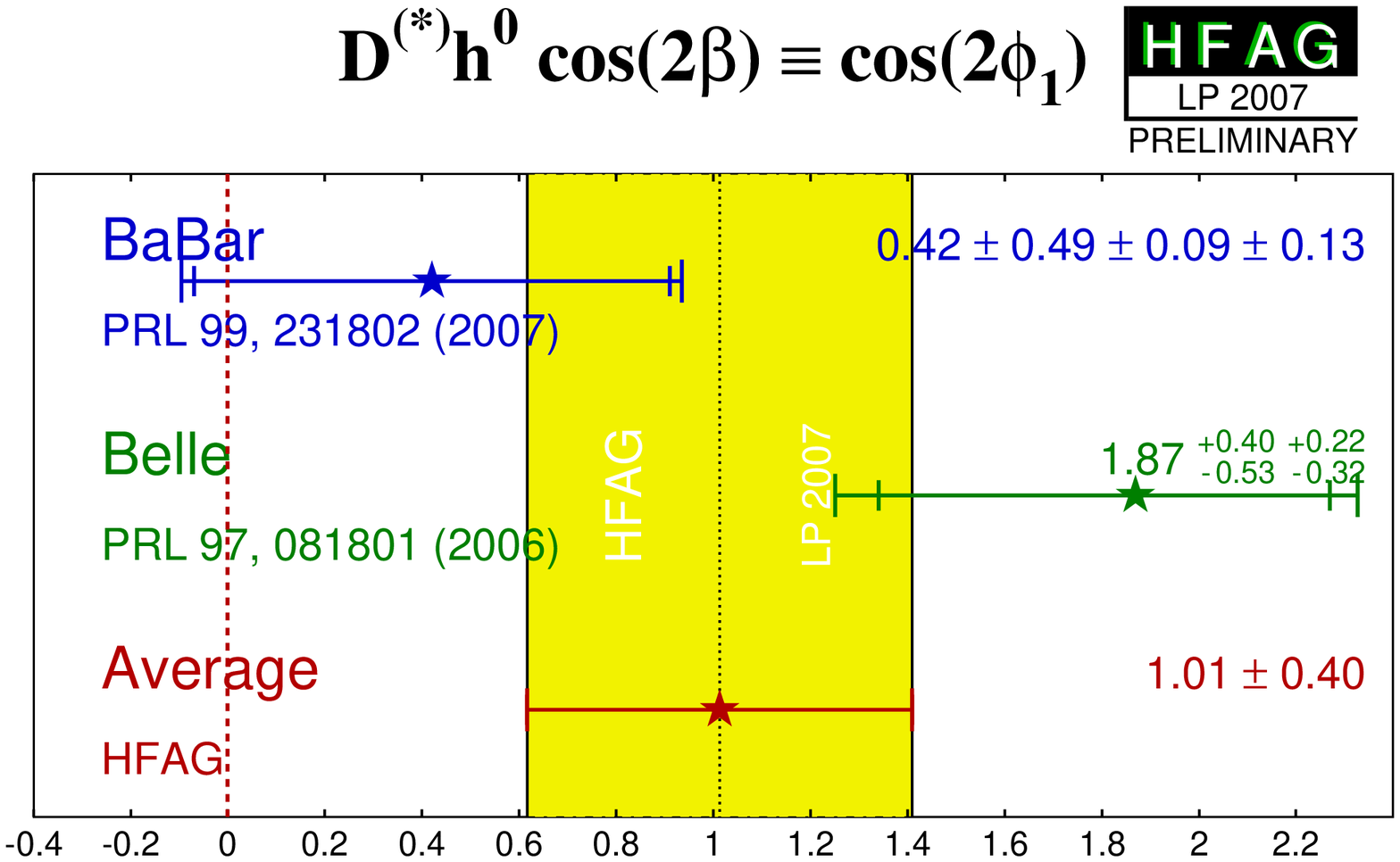}
      }
    \end{tabular}
  \end{center}
  \vspace{-0.8cm}
  \caption{
    Averages of 
    (left) $\sin(2\beta)$ and (right) $\cos(2\beta)$
    measured in colour-suppressed $b \to c\bar{u}d$ transitions.
  }
  \label{fig:cp_uta:cud_beta}
\end{figure}

Results of such analyses are available from both 
\belle~\cite{Krokovny:2006sv} and \babar~\cite{Aubert:2007rp}.
The decays $\B \to D\pi^0$, $\B \to D\eta$, $\B \to D\omega$, 
$\B \to D^*\pi^0$ and $\B \to D^*\eta$ are used. 
[This collection of states is denoted by $D^{(*)}h^0$.]
The daughter decays are $D^* \to D\pi^0$ and $D \to \KS\pi^+\pi^-$.
The results are shown in Table~\ref{tab:cp_uta:cud_beta},
and Fig.~\ref{fig:cp_uta:cud_beta}.
Note that \babar\ quote uncertainties due to the $D$ decay model 
separately from other systematic errors, while \belle\ do not.

\begin{table}[htb]
	\begin{center}
		\caption{
			Averages from $\Bz \to D^{(*)}h^0$, $D \to K_S\pi^+\pi^-$ analyses.
		}
		\vspace{0.2cm}
		\setlength{\tabcolsep}{0.0pc}
    \resizebox{\textwidth}{!}{
      		\begin{tabular*}{\textwidth}{@{\extracolsep{\fill}}lrcccc} \hline
	\mc{2}{l}{Experiment} & $N(B\bar{B})$ & $\sin 2\beta$ & $\cos 2\beta$ & $|\lambda|$ \\
		\hline
	\babar & \cite{Aubert:2007rp} & 383M & $0.29 \pm 0.34 \pm 0.03 \pm 0.05$ & $0.42 \pm 0.49 \pm 0.09 \pm 0.13$ & $1.01 \pm 0.08 \pm 0.02$ \\
	\belle & \cite{Krokovny:2006sv} & 386M & $0.78 \pm 0.44 \pm 0.22$ & $1.87 \,^{+0.40}_{-0.53} \,^{+0.22}_{-0.32}$ & \textendash{} \\
	\mc{3}{l}{\bf Average} & $0.45 \pm 0.28$ & $1.01 \pm 0.40$ & $1.01 \pm 0.08$ \\
	\mc{3}{l}{\small Confidence level} & {\small $0.59~(0.5\sigma)$} & {\small $0.12~(1.6\sigma)$} & \textendash{} \\
		\hline
		\end{tabular*}
    }
		\label{tab:cp_uta:cud_beta}
	\end{center}
\end{table}

Again, it is clear that the data prefer $\cos(2\beta)>0$.
Indeed, \belle~\cite{Krokovny:2006sv} 
determine the sign of $\cos(2\phi_1)$ to be positive at $98.3\%$ confidence level,
while \babar~\cite{Aubert:2007rp} 
favour the solution of $\beta$ with $\cos(2\beta)>0$ at $87\%$ confidence level.
Note, however, that the Belle measurement has strongly non-Gaussian behaviour. 
Therefore, we perform uncorrelated averages, 
from which any interpretation has to be done with the greatest care. 

\clearpage
\mysubsection{Time-dependent $\CP$ asymmetries in charmless $b \to q\bar{q}s$ transitions
}
\label{sec:cp_uta:qqs}

The flavour changing neutral current $b \to s$ penguin
can be mediated by any up-type quark in the loop, 
and hence the amplitude can be written as
\begin{equation}
  \label{eq:cp_uta:b_to_s}
  \begin{array}{ccccc}
    A_{b \to s} & = & 
    \mc{3}{l}{F_u V_{ub}V^*_{us} + F_c V_{cb}V^*_{cs} + F_t V_{tb}V^*_{ts}} \\
    & = & (F_u - F_c) V_{ub}V^*_{us} & + & (F_t - F_c) V_{tb}V^*_{ts} \\
    & = & {\cal O}(\lambda^4) & + & {\cal O}(\lambda^2) \\
  \end{array}
\end{equation}
using the unitarity of the CKM matrix.
Therefore, in the Standard Model, 
this amplitude is dominated by $V_{tb}V^*_{ts}$, 
and to within a few degrees 
($\delta\beta \lesssim 2^\circ$ for $\beta \approx 20^\circ$) 
the time-dependent parameters can be written\footnote
{
  The presence of a small (${\cal O}(\lambda^2)$) weak phase in 
  the dominant amplitude of the $s$ penguin decays introduces 
  a phase shift given by
  $S_{b \to q\bar q s} = -\eta\sin(2\beta)\cdot(1 + \Delta)$. 
  Using the CKMfitter results for the Wolfenstein 
  parameters~\cite{Charles:2004jd}, one finds: 
  $\Delta \simeq 0.033$, which corresponds to a shift of 
  $2\beta$ of $+2.1$ degrees. 
  Nonperturbative contributions can alter this result.
}
$S_{b \to q\bar q s} \approx - \etacp \sin(2\beta)$,
$C_{b \to q\bar q s} \approx 0$,
assuming $b \to s$ penguin contributions only ($q = u,d,s$).

Due to the large virtual mass scales occurring in the penguin loops,
additional diagrams from physics beyond the Standard Model,
with heavy particles in the loops, may contribute.
In general, these contributions will affect the values of 
$S_{b \to q\bar q s}$ and $C_{b \to q\bar q s}$.
A discrepancy between the values of 
$S_{b \to c\bar c s}$ and $S_{b \to q\bar q s}$
can therefore provide a clean indication of new physics~\cite{Grossman:1996ke,Fleischer:1996bv,London:1997zk,Ciuchini:1997zp}.

However, there is an additional consideration to take into account.
The above argument assumes only the $b \to s$ penguin contributes
to the $b \to q\bar q s$ transition.
For $q = s$ this is a good assumption, which neglects only rescattering effects.
However, for $q = u$ there is a colour-suppressed $b \to u$ tree diagram
(of order ${\cal O}(\lambda^4)$), 
which has a different weak (and possibly strong) phase.
In the case $q = d$, any light neutral meson that is formed from $d \bar{d}$ 
also has a $u \bar{u}$ component, and so again there is ``tree pollution''. 
The \Bz decays to $\piz\KS$, $\rhoz\KS$ and $\omega\KS$ belong to this category.
The mesons $f_0$ and $\etapr$ are expected to have predominant $s\bar s$ parts,
which reduces the relative size of the possible tree pollution. 
If the inclusive decay $\Bz\to\Kp\Km\Kz$ (excluding $\phi\Kz$) is dominated by
a nonresonant three-body transition, 
an OZI-rule suppressed tree-level diagram can occur 
through insertion of an $s\sbar$ pair. 
The corresponding penguin-type transition 
proceeds via insertion of a $u\ubar$ pair, which is expected
to be favored over the $s\sbar$ insertion by fragmentation models.
Neglecting rescattering, the final state $\Kz\Kzb\Kz$ 
(reconstructed as $\KS\KS\KS$) has no tree pollution~\cite{Gershon:2004tk}.
Various estimates, using different theoretical approaches,
of the values of $\Delta S = S_{b \to q\bar q s} - S_{b \to c\bar c s}$
exist in the literature~\cite{Grossman:2003qp,Gronau:2003ep,Gronau:2003kx,Gronau:2004hp,Cheng:2005bg,Gronau:2005gz,Buchalla:2005us,Beneke:2005pu,Engelhard:2005hu,Cheng:2005ug,Engelhard:2005ky,Gronau:2006qh,Silvestrini:2007yf,Dutta:2008xw}.
In general, there is agreement that the modes
$\phi\Kz$, $\etapr\Kz$ and $\Kz\Kzb\Kz$ are the cleanest,
with values of $\left| \Delta S \right|$ at or below the few percent level 
($\Delta S$ is usually positive).

\mysubsubsection{Time-dependent $\CP$ asymmetries: $b \to q\bar{q}s$ decays to $\CP$ eigenstates
}
\label{sec:cp_uta:qqs:cp_eigen}

\begin{figure}[htb]
  \begin{center}
    \resizebox{0.45\textwidth}{!}{
      \includegraphics{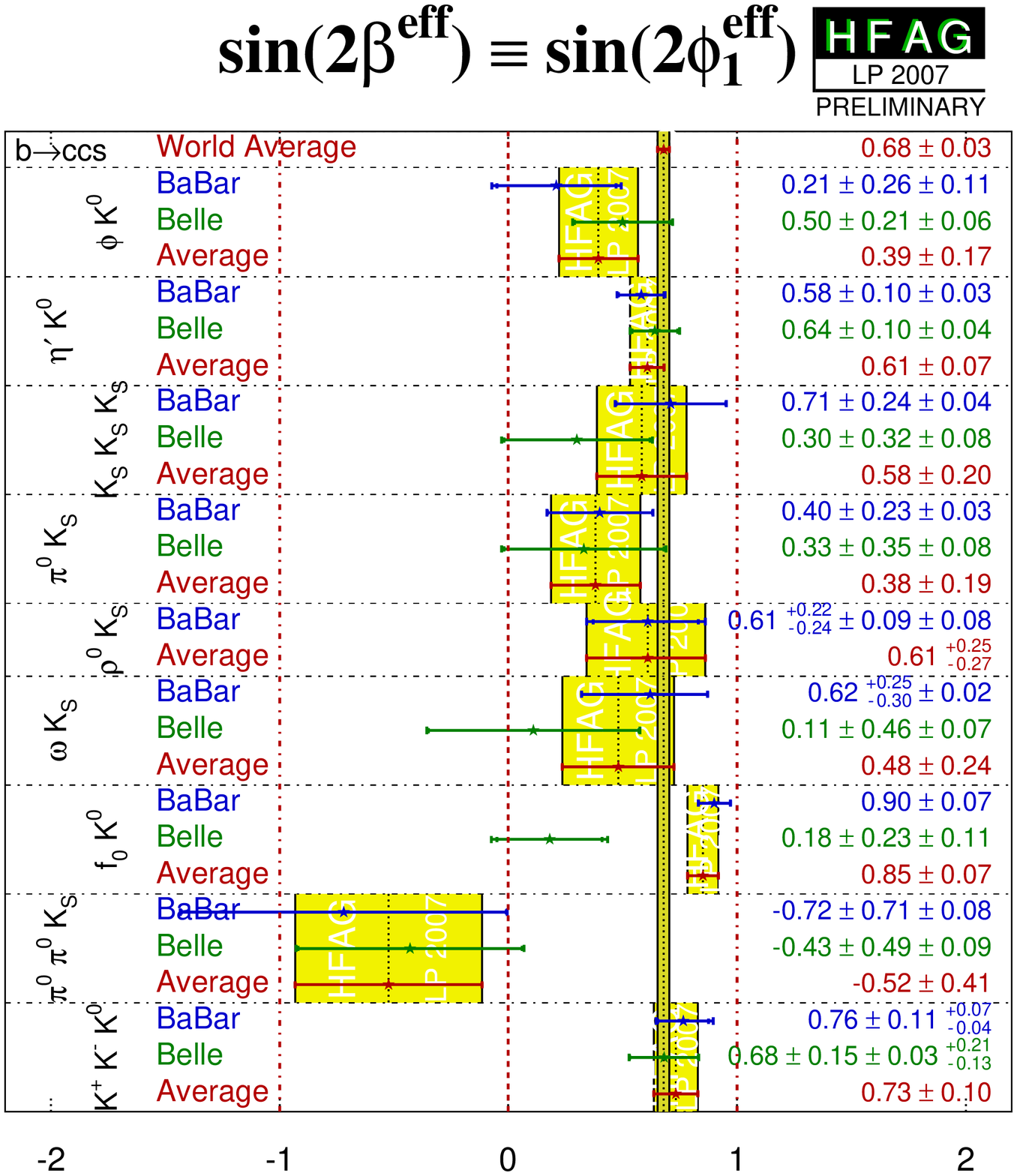}
    }
    \hfill
    \resizebox{0.45\textwidth}{!}{
      \includegraphics{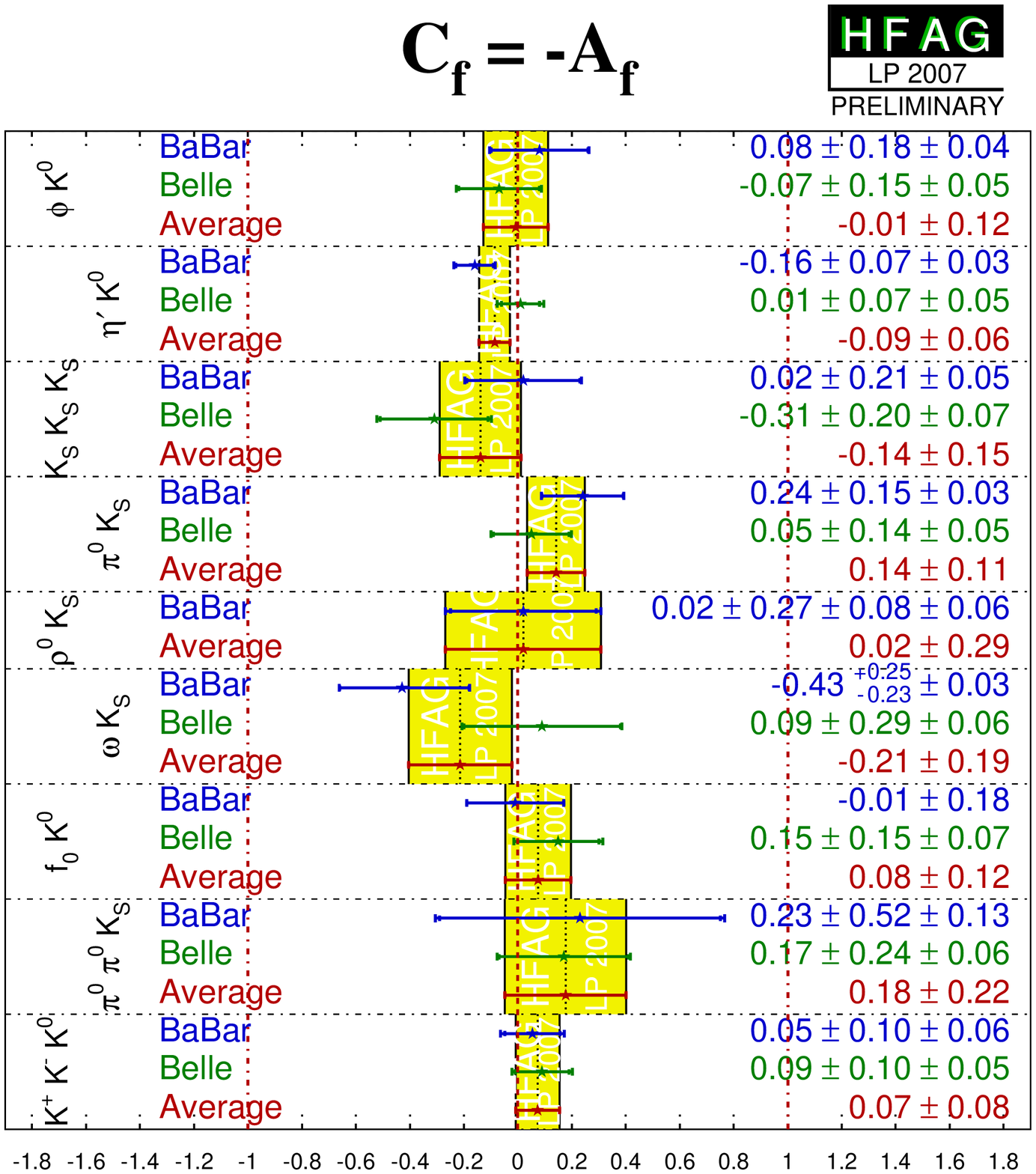}
    }
    \\
    \resizebox{0.45\textwidth}{!}{
      \includegraphics{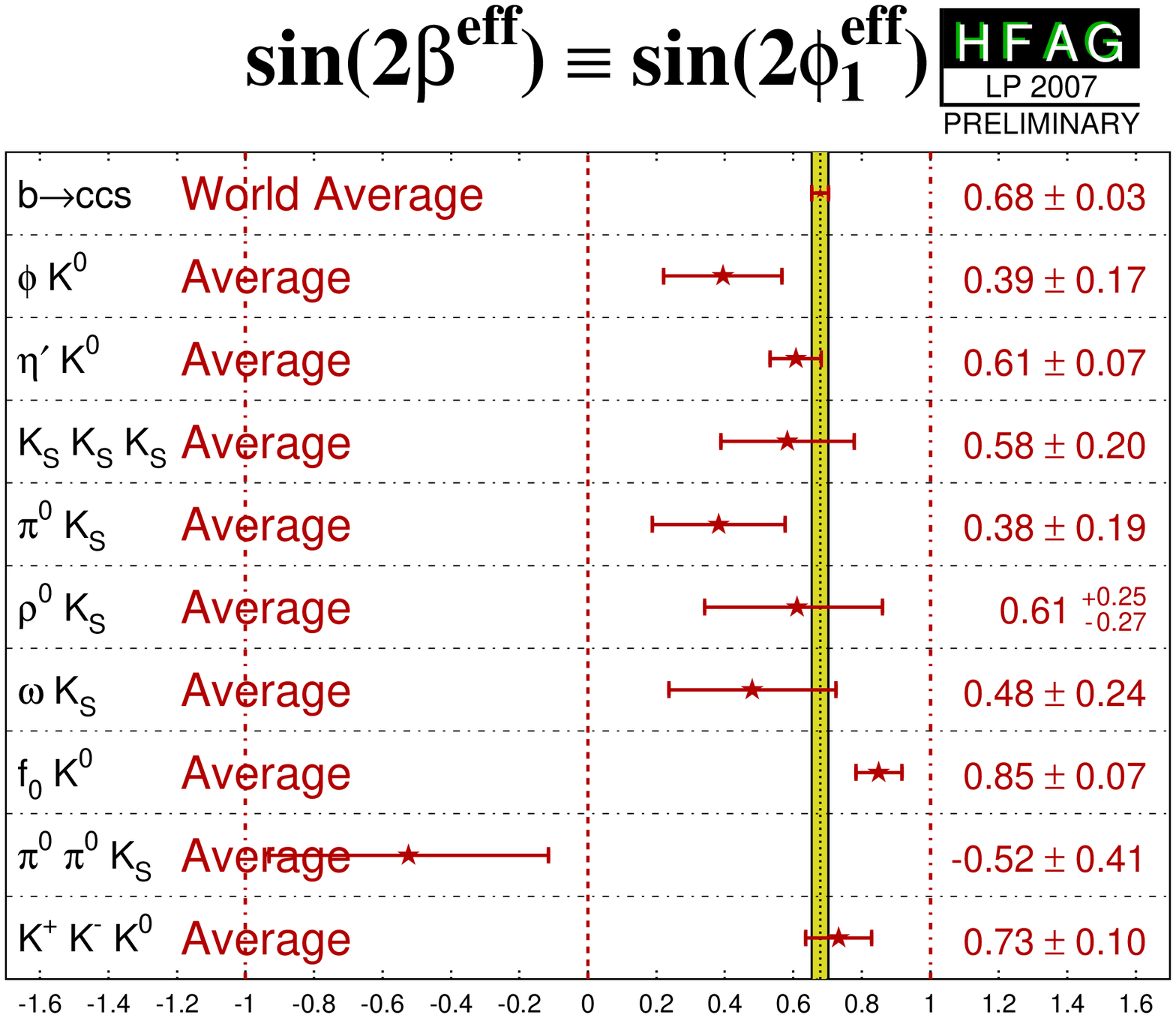}
    }
    \hfill
    \resizebox{0.45\textwidth}{!}{
      \includegraphics{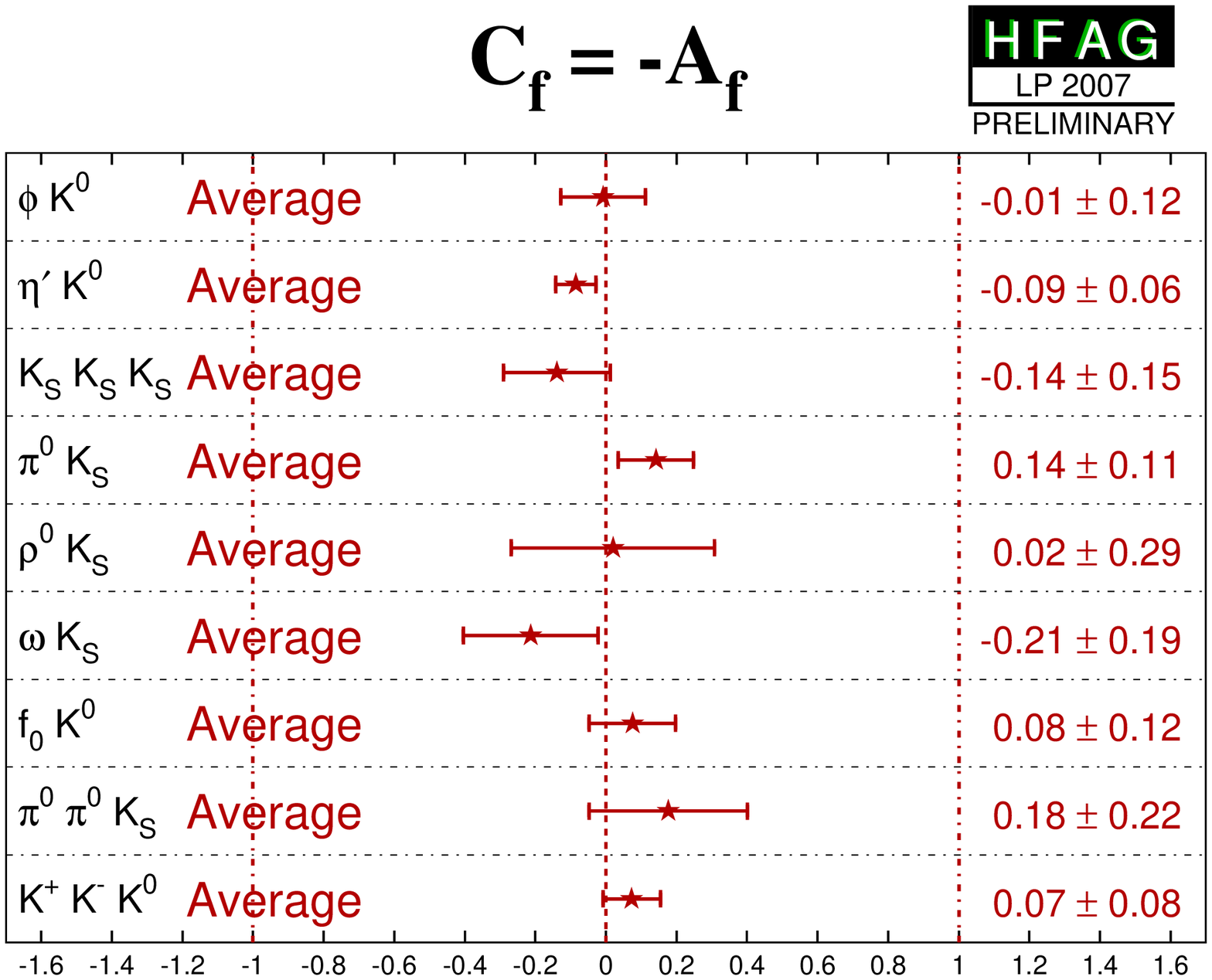}
    }
  \end{center}
  \vspace{-0.8cm}
  \caption{
    (Top)
    Averages of 
    (left) $-\etacp S_{b \to q\bar q s}$ and (right) $C_{b \to q\bar q s}$.
    The $-\etacp S_{b \to q\bar q s}$ figure compares the results to 
    the world average 
    for $-\etacp S_{b \to c\bar c s}$ (see Section~\ref{sec:cp_uta:ccs:cp_eigen}).
    (Bottom) Same, but only averages for each mode are shown.
    More figures are available from the HFAG web pages.
  }
  \label{fig:cp_uta:qqs}
\end{figure}

\begin{figure}[htb]
  \begin{center}
    \resizebox{0.33\textwidth}{!}{
      \includegraphics{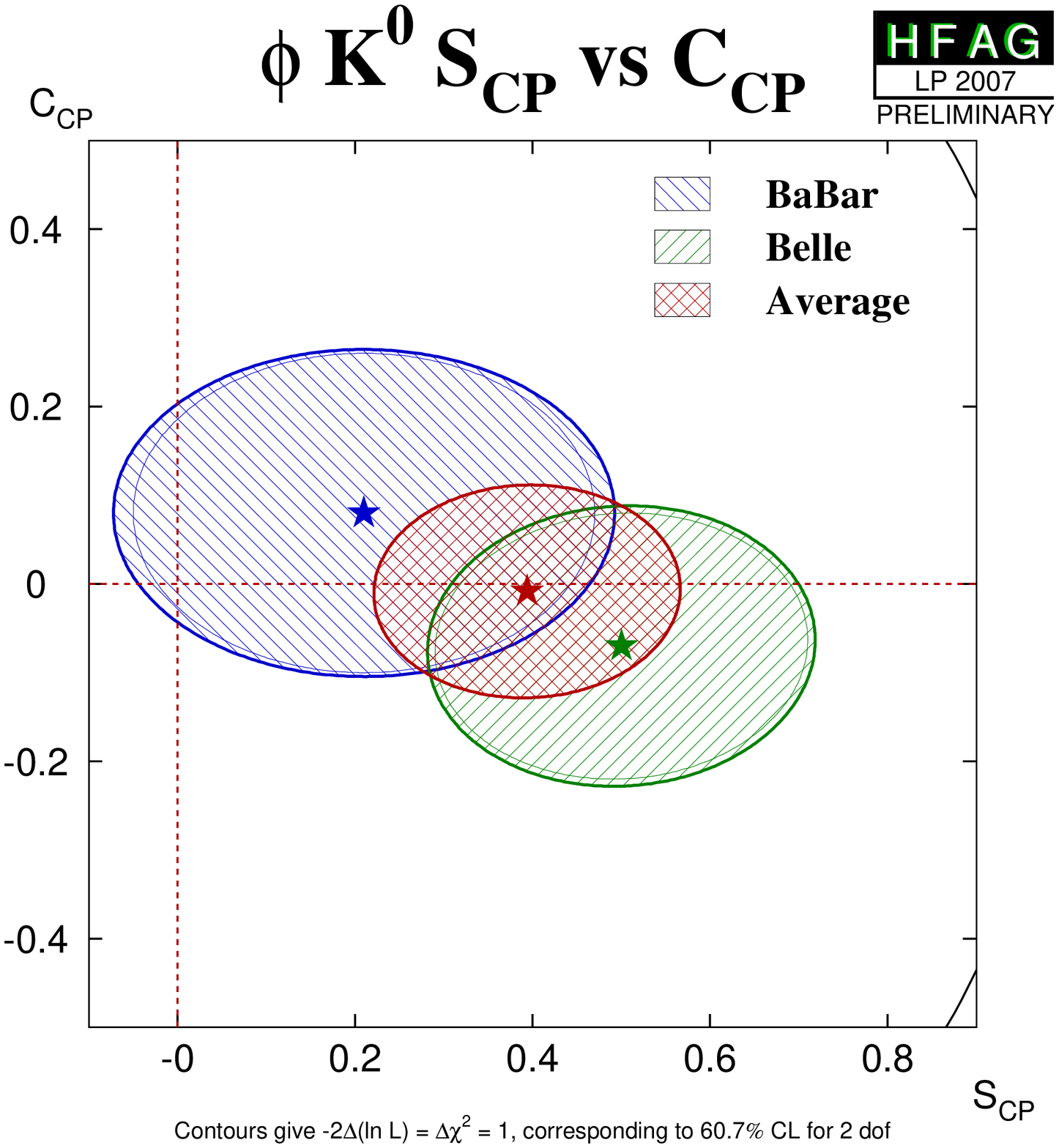}
    }
    \hspace{0.08\textwidth}
    \resizebox{0.33\textwidth}{!}{
      \includegraphics{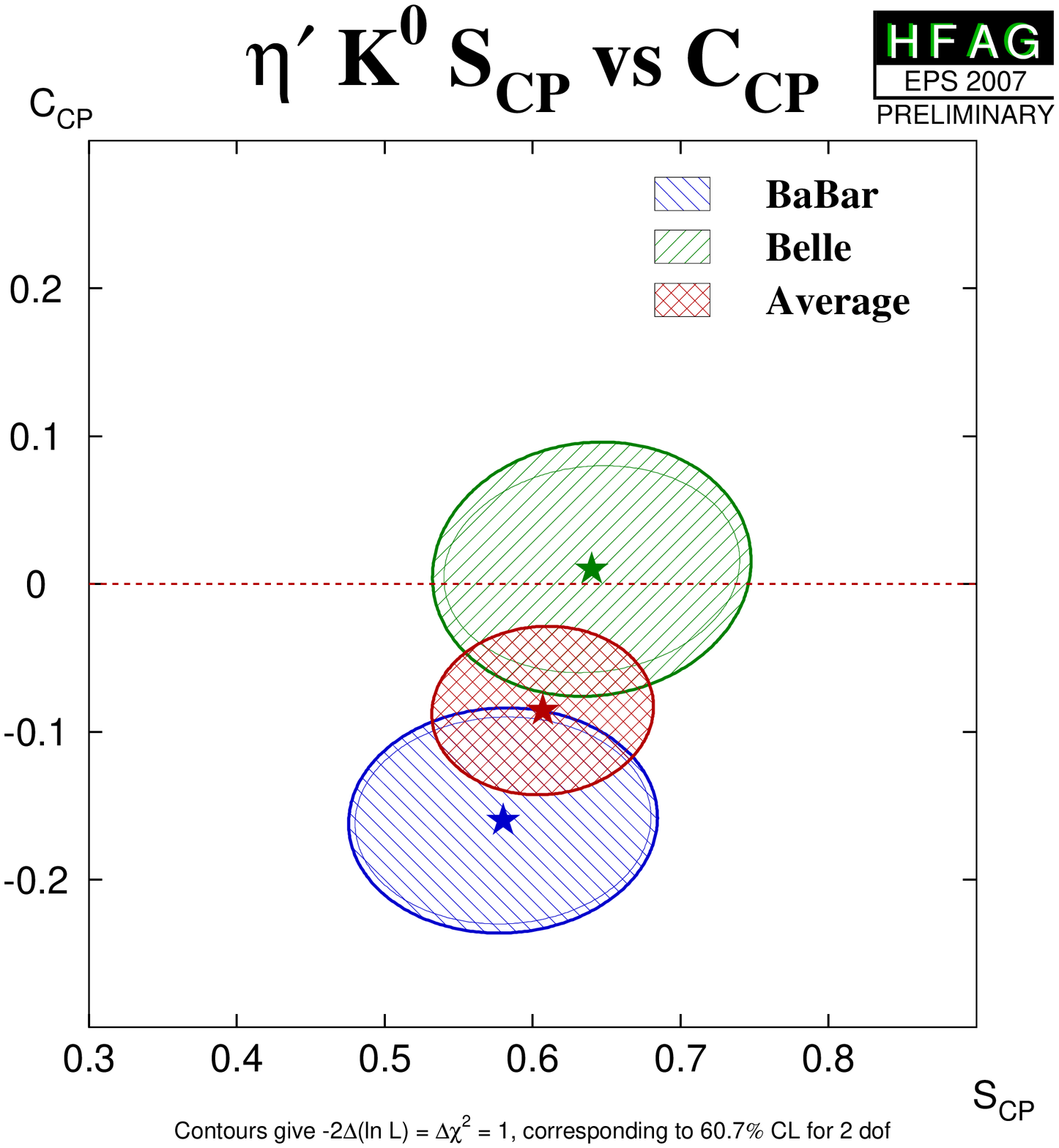}
    }
    \\
    \resizebox{0.33\textwidth}{!}{
      \includegraphics{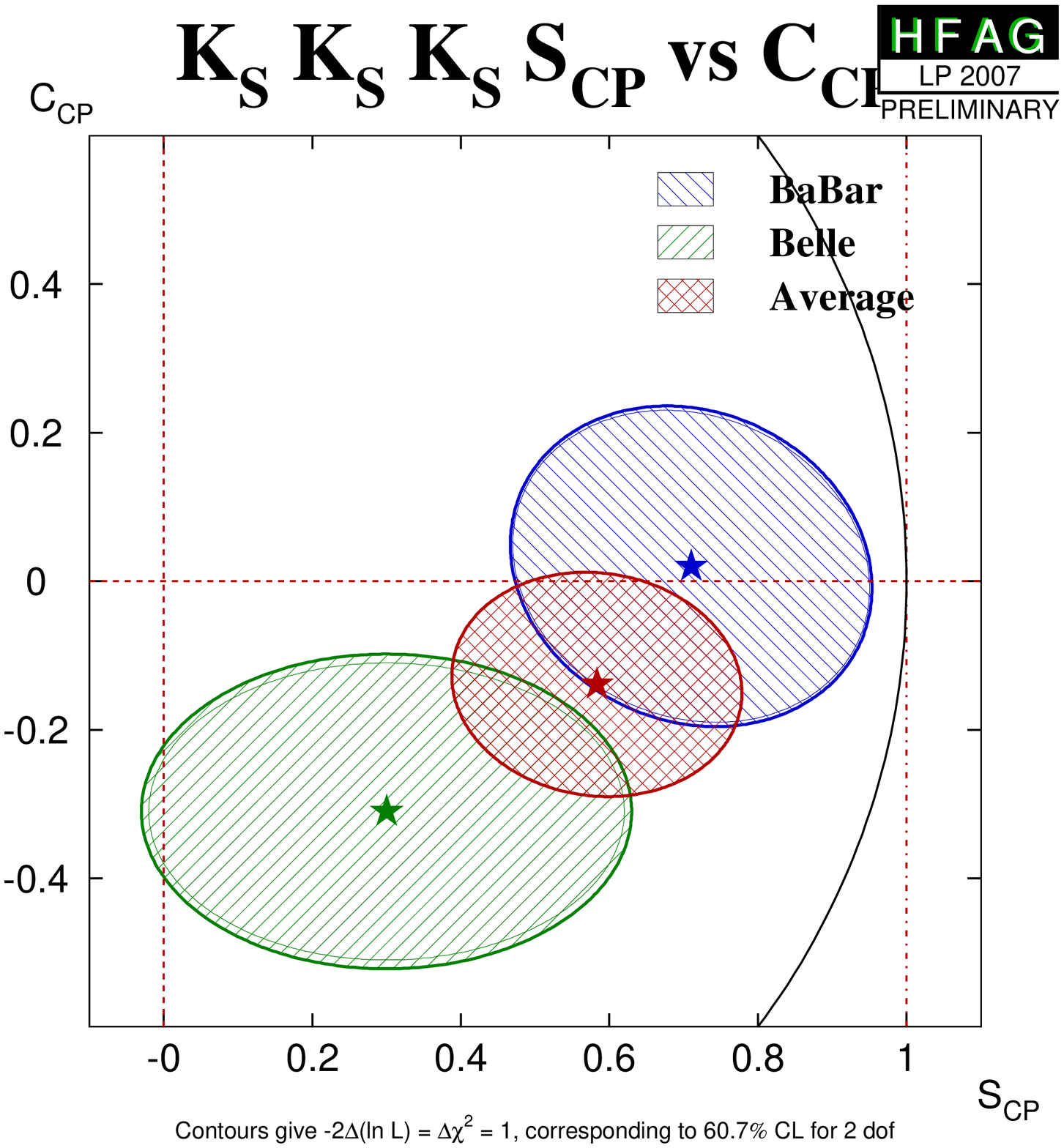}
    }
    \hspace{0.08\textwidth}
    \resizebox{0.33\textwidth}{!}{
      \includegraphics{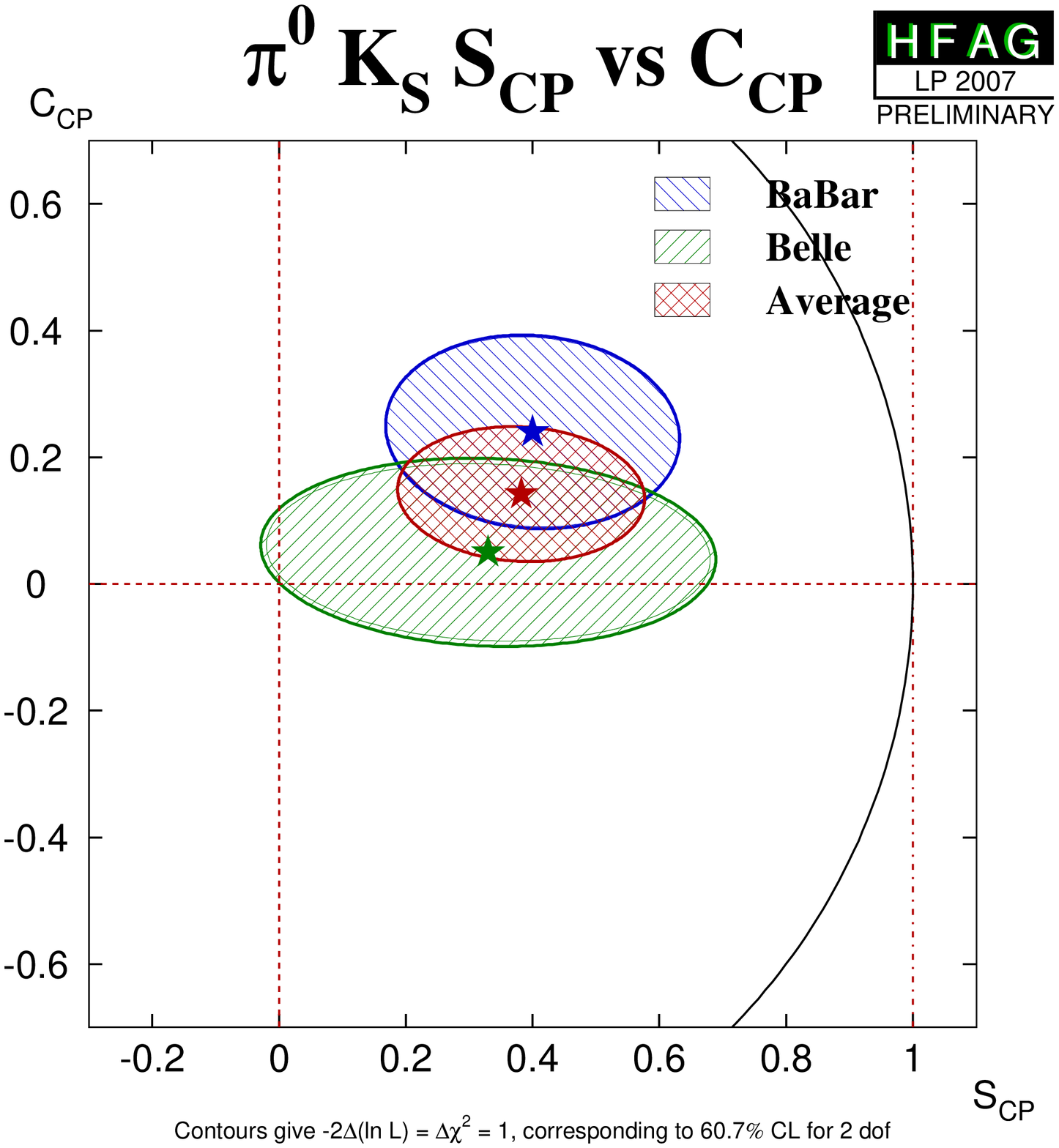}
    }
  \end{center}
  \vspace{-0.8cm}
  \caption{
    Averages of four $b \to q\bar q s$ dominated channels,
    for which correlated averages are performed,
    in the $S_{\CP}$ \vs\ $C_{\CP}$ plane,
    where $S_{\CP}$ has been corrected by the $\CP$ eigenvalue to give
    $\sin(2\beta^{\rm eff})$.
    (Top left) $\Bz \to \phi\Kz$,
    (top right) $\Bz \to \eta^\prime\Kz$,
    (bottom left) $\Bz \to \KS\KS\KS$,
    (bottom right) $\Bz \to \pi^0\KS$.
    More figures are available from the HFAG web pages.
  }
  \label{fig:cp_uta:qqs_SvsC}
\end{figure}

\begin{figure}[htb]
  \begin{center}
    \resizebox{0.66\textwidth}{!}{
      \includegraphics{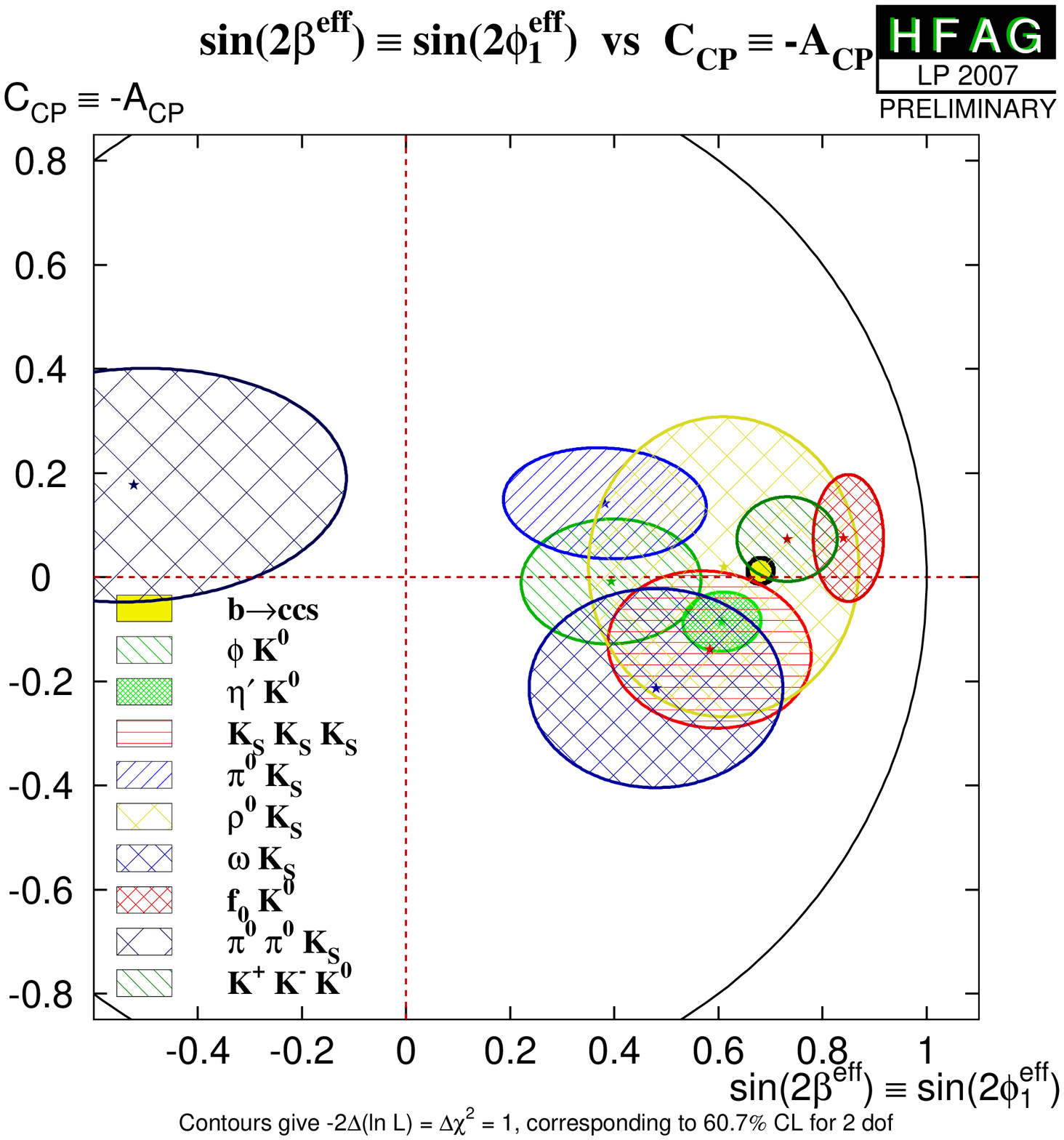}
    }
  \end{center}
  \vspace{-0.8cm}
  \caption{
    Compilation of constraints in the 
    $-\etacp S_{b \to q\bar q s}$ \vs\ $C_{b \to q\bar q s}$ plane.
  }
  \label{fig:cp_uta:qqs_SvsC-all}
\end{figure}

The averages for $-\etacp S_{b \to q\bar q s}$ and $C_{b \to q\bar q s}$
can be found in Table~\ref{tab:cp_uta:qqs},
and are shown in Figs.~\ref{fig:cp_uta:qqs},~\ref{fig:cp_uta:qqs_SvsC} 
and~\ref{fig:cp_uta:qqs_SvsC-all}.
Results from both \babar\  and \belle\ are averaged for the modes
$\phi\Kz$, $\etapr\Kz$, $f_0\Kz$ and $K^+K^-\Kz$
($\Kz$ indicates that both $\KS$ and $\KL$ are used, 
although \belle\ use neither $f_0\KL$ nor $K^+K^-\KL$), 
$\KS\KS\KS$, $\pi^0 \KS$, $\omega\KS$ and $\pi^0\pi^0\KS$. 
\babar\ also has results for the mode $\rho^0\KS$.
Of these,
$\phi\KS$, $\etapr\KS$, $\pi^0 \KS$, $\omega\KS$ and $\rho^0\KS$
have $\CP$ eigenvalue $\etacp = -1$, 
while $\phi\KL$, $\etapr\KL$, $f_0 \KS$, $\pi^0\pi^0\KS$ and $\KS\KS\KS$ 
have $\etacp = +1$.

\begin{table}[!htb]
	\begin{center}
		\caption{
      Averages of $-\etacp S_{b \to q\bar q s}$ and $C_{b \to q\bar q s}$.
		}
		\vspace{0.2cm}
		\setlength{\tabcolsep}{0.0pc}
\begin{tabular*}{\textwidth}{@{\extracolsep{\fill}}lrcccc} \hline
        \mc{2}{l}{Experiment} & $N(B\bar{B})$ & $- \etacp S_{b \to q\bar q s}$ & $C_{b \to q\bar q s}$ & Correlation \\
	\hline
      \mc{6}{c}{$\phi \Kz$} \\
	\babar & \cite{Aubert:2007sd} & 383M & $0.21 \pm 0.26 \pm 0.11$ & $0.08 \pm 0.18 \pm 0.04$ & \textendash{} \\
	\belle & \cite{Chen:2006nk} & 535M & $0.50 \pm 0.21 \pm 0.06$ & $-0.07 \pm 0.15 \pm 0.05$ & $0.05$ \\
	\mc{3}{l}{\bf Average} & $0.39 \pm 0.17$ & $-0.01 \pm 0.12$ & $0.03$ \\
	\mc{3}{l}{\small Confidence level} & \mc{2}{c}{\small $0.59~(0.5\sigma)$} & \\
		\hline

      \mc{6}{c}{$\etapr \Kz$} \\
	\babar & \cite{Aubert:2006wv} & 384M & $0.58 \pm 0.10 \pm 0.03$ & $-0.16 \pm 0.07 \pm 0.03$ & $0.03$ \\
	\belle & \cite{Chen:2006nk} & 535M & $0.64 \pm 0.10 \pm 0.04$ & $0.01 \pm 0.07 \pm 0.05$ & $0.09$ \\
	\mc{3}{l}{\bf Average} & $0.61 \pm 0.07$ & $-0.09 \pm 0.06$ & $0.04$ \\
	\mc{3}{l}{\small Confidence level} & \mc{2}{c}{\small $0.32~(1.0\sigma)$} & \\
		\hline

      \mc{6}{c}{$\KS\KS\KS$} \\
	\babar & \cite{Aubert:2007me} & 384M & $0.71 \pm 0.24 \pm 0.04$ & $0.02 \pm 0.21 \pm 0.05$ & $-0.14$ \\
	\belle & \cite{Chen:2006nk} & 535M & $0.30 \pm 0.32 \pm 0.08$ & $-0.31 \pm 0.20 \pm 0.07$ & \textendash{} \\
	\mc{3}{l}{\bf Average} & $0.58 \pm 0.20$ & $-0.14 \pm 0.15$ & $-0.08$ \\
	\mc{3}{l}{\small Confidence level} & \mc{2}{c}{\small $0.31~(1.0\sigma)$} & \\
		\hline

      \mc{6}{c}{$\pi^0 \KS$} \\
	\babar & \cite{Aubert:2007mgb} & 383M & $0.40 \pm 0.23 \pm 0.03$ & $0.24 \pm 0.15 \pm 0.03$ & $-0.07$ \\
	\belle & \cite{Abe:2006gy} & 535M & $0.33 \pm 0.35 \pm 0.08$ & $0.05 \pm 0.14 \pm 0.05$ & $-0.08$ \\
	\mc{3}{l}{\bf Average} & $0.38 \pm 0.19$ & $0.14 \pm 0.11$ & $-0.07$ \\
	\mc{3}{l}{\small Confidence level} & \mc{2}{c}{\small $0.66~(0.5\sigma)$} & \\
		\hline

		\hline
      \mc{6}{c}{$\rho^0 \KS$} \\
	\babar & \cite{Aubert:2007vi} & 383M & $0.61 \,^{+0.22}_{-0.24} \pm 0.09 \pm 0.08$ & $0.02 \pm 0.27 \pm 0.08 \pm 0.06$ & \textendash{} \\
	\mc{3}{l}{\bf Average} & $0.61 \,^{+0.25}_{-0.27}$ & $0.02 \pm 0.29$ & \textendash{} \\
		\hline

      \mc{6}{c}{$\omega \KS$} \\
	\babar & \cite{Aubert:2006ai} & 347M & $0.62 \,^{+0.25}_{-0.30} \pm 0.02$ & $-0.43 \,^{+0.25}_{-0.23} \pm 0.03$ & \textendash{} \\
	\belle & \cite{Abe:2006gy} & 535M & $0.11 \pm 0.46 \pm 0.07$ & $0.09 \pm 0.29 \pm 0.06$ & \textendash{} \\
	\mc{3}{l}{\bf Average} & $0.48 \pm 0.24$ & $-0.21 \pm 0.19$ & \textendash{} \\
	\mc{3}{l}{\small Confidence level} & {\small $0.35~(0.9\sigma)$} & {\small $0.18~(1.3\sigma)$} & \\
		\hline

      \mc{6}{c}{$f_0 \Kz$} \\
	\babar & \cite{Aubert:2007sd,Aubert:2007vi} & \textendash{} & $0.90 \pm 0.07$ & $-0.01 \pm 0.18$ & \textendash{} \\
	\belle & \cite{Abe:2006gy} & 535M & $0.18 \pm 0.23 \pm 0.11$ & $0.15 \pm 0.15 \pm 0.07$ & $-0.01$ \\
	\mc{3}{l}{\bf Average} & $0.85 \pm 0.07$ & $0.08 \pm 0.12$ & $-0.00$ \\
	\mc{3}{l}{\small Confidence level} & \mc{2}{c}{\small $0.02~(2.3\sigma)$} & \\
		\hline

      \mc{6}{c}{$\pi^0 \pi^0 \KS$} \\
	\babar & \cite{Aubert:2007ub} & 227M & $-0.72 \pm 0.71 \pm 0.08$ & $0.23 \pm 0.52 \pm 0.13$ & $-0.02$ \\
	\belle & \cite{:2007xd} & 657M & $-0.43 \pm 0.49 \pm 0.09$ & $0.17 \pm 0.24 \pm 0.06$ & $0.09$ \\
	\mc{3}{l}{\bf Average} & $-0.52 \pm 0.41$ & $0.18 \pm 0.22$ & $0.06$ \\
	\mc{3}{l}{\small Confidence level} & \mc{2}{c}{\small $0.94~(0.1\sigma)$} & \\
		\hline

      \mc{6}{c}{$K^+K^- \Kz$} \\
	\babar & \cite{Aubert:2007sd} & 383M & $0.76 \pm 0.11 \,^{+0.07}_{-0.04}$ & $0.05 \pm 0.10 \pm 0.06$ & \textendash{} \\
	\belle & \cite{Abe:2006gy} & 535M & $0.68 \pm 0.15 \pm 0.03 \,^{+0.21}_{-0.13}$ & $0.09 \pm 0.10 \pm 0.05$ & \textendash{} \\
	\mc{3}{l}{\bf Average} & $0.73 \pm 0.10$ & $0.07 \pm 0.08$ & \textendash{} \\
	\mc{3}{l}{\small Confidence level} & {\small $0.67~(0.4\sigma)$} & {\small $0.82~(0.2\sigma)$} & \\
		\hline



		\hline
		\end{tabular*}
		\label{tab:cp_uta:qqs}
	\end{center}
\end{table}

The final state $K^+K^-\Kz$ 
(contributions from $\phi\Kz$ are implicitly excluded) 
is not a $\CP$ eigenstate.
However, it can be treated as a quasi-two-body decay, 
with the $\CP$ composition determined using either an 
isospin argument (used by \belle\ to determine a $\CP$-even fraction of 
$0.93 \pm 0.09 \pm 0.05$~\cite{Abe:2006gy})
or a moments analysis 
(previously used by \babar\ to find a $\CP$-even fractions of 
$0.89 \pm 0.08 \pm 0.06$ in $K^+K^-\KS$~\cite{Aubert:2005ja}).
Note that uncertainty in the $\CP$ composition of the final state leads to 
a third source of uncertainty on the \belle\ results for $-\etacp S_{K^+K^-\Kz}$.

\babar\ have performed time-dependent Dalitz plot analyses of
$\Bz \to K^+K^-\Kz$ and $\Bz \to \pi^+\pi^-\KS$
(see subsection~\ref{sec:cp_uta:qqs:dp}).
Their results for $\phi\Kz$ and $K^+K^-\Kz$ are determined from the 
$\Bz \to K^+K^-\Kz$ analysis,
the latter taken from the inclusive ``high-mass'' 
($m_{K^+K^-} > 1.1 \ {\rm GeV}/c^2$) region 
(this approach automatically corrects for the 
$\CP$ composition of the final state).
\babar\ results for $\rho^0\KS$ are taken from the $\Bz \to \pi^+\pi^-\KS$ analysis,
and their results for $f_0\Kz$ are a combination of results from the 
both time-dependent Dalitz plot analyses 
($-\etacp S_{b \to q\bar q s} = 0.25 \pm 0.26 \pm 0.10$, 
$C_{b \to q\bar q s} = -0.41 \pm 0.23 \pm 0.07$
from $\Bz \to f_0\Kz$ with $f_0 \to K^+K^-$~\cite{Aubert:2007sd};
$-\etacp S_{b \to q\bar q s} = 0.94\, ^{+0.02}_{-0.07} \, ^{+0.03}_{-0.05} \pm 0.02$, 
$C_{b \to q\bar q s} = 0.35 \pm 0.27 \pm 0.07 \pm 0.04$
from $\Bz \to f_0\Kz$ with $f_0 \to \pi^+\pi^-$~\cite{Aubert:2007vi}). 

It must be noted that Q2B parameters extracted from Dalitz plot analyses 
are constrained to lie within the physical boundary ($S_{\CP}^2 + C_{\CP}^2 < 1$)
and consequenty the obtained errors are highly non-Gaussian when
the central value is close to the boundary.  
This is particularly evident in the \babar\ results for 
$\Bz \to f_0\Kz$ with $f_0 \to \pi^+\pi^-$~\cite{Aubert:2007vi}.
These results must be treated with extreme caution.

As explained above,
each of the modes listed in Table~\ref{tab:cp_uta:qqs} has
different uncertainties within the Standard Model,
and so each may have a different value of $-\etacp S_{b \to q\bar q s}$.
Therefore, there is no strong motivation to make a combined average
over the different modes.
We refer to such an average as a ``na\"\i ve $s$-penguin average.''
It is na\"\i ve not only because of the neglect of the theoretical uncertainty,
but also since possible correlations of systematic effects 
between different modes are neglected.
In spite of these caveats, there remains substantial interest 
in the value of this quantity,
and therefore it is given here:
$\langle -\etacp S_{b \to q\bar q s} \rangle = 0.64 \pm 0.04$,
with confidence level $0.008~(2.7\sigma)$.
This value is in very good agreement with the average 
$-\etacp S_{b \to c\bar c s}$ given in Sec.~\ref{sec:cp_uta:ccs:cp_eigen}.
However, this result is strongly effected by the highly non-Gaussian errors
of the \babar\ result for 
$\Bz \to f_0\Kz$ with $f_0 \to \pi^+\pi^-$~\cite{Aubert:2007vi},
which is only marginally consistent with \babar\ result for 
$\Bz \to f_0\Kz$ with $f_0 \to K^+K^-$~\cite{Aubert:2007sd}
and the \belle\ result for $\Bz \to f_0\Kz$ with $f_0 \to \pi^+\pi^-$ 
(treated as a Q2B decay)~\cite{Abe:2006gy}.
If the na\"\i ve $s$-penguin average is recalculated without this result,
we find $\langle -\etacp S_{b \to q\bar q s} \rangle = 0.56 \pm 0.05$,
with confidence level $0.25~(1.1\sigma)$.
Again treating the uncertainties as Gaussian and neglecting correlations,
this value is found to be $2.2\sigma$ below the average 
$-\etacp S_{b \to c\bar c s}$ given in Sec.~\ref{sec:cp_uta:ccs:cp_eigen}.

(The average for $C_{b \to q\bar q s}$ is 
$\langle C_{b \to q\bar q s} \rangle = -0.01 \pm 0.04$
with confidence level $0.54~(0.6\sigma)$.
This result is not strongly effected by the inclusion of the \babar\ result for 
$\Bz \to f_0\Kz$ with $f_0 \to \pi^+\pi^-$.)

We emphasise again that we do not advocate the use of these averages,
and that the values should be treated with {\it extreme caution}, if at all.

From Table~\ref{tab:cp_uta:qqs} it may be noted 
that the average for $-\etacp S_{b \to q\bar q s}$ in $\etapr \Kz$ 
($0.61 \pm 0.07$),
is now more than $5\sigma$ away from zero, 
so that $\CP$ violation in this mode is well established.
Amongst other modes,
$\CP$ violation effects in both $f_0 \Kz$ and $K^+K^- \Kz$ appear 
to be established --
\babar\ have claimed $5.1\sigma$ observation of $\CP$ violation in 
$\Bz \to K^+K^- \Kz$~\cite{Aubert:2007sd} and 
$4.3\sigma$ evidence of $\CP$ violation in 
$\Bz \to f_0\KS$ with $f_0 \to \pi^+\pi^-$~\cite{Aubert:2007vi}.
Due to possible non-Gaussian errors in these results
it may be prudent to defer any strong conclusions 
on the numerical significance of the averages.
There is no evidence (above $2\sigma$) for direct $\CP$ violation 
in any $b \to q \bar q s$ mode.

\mysubsubsection{Time-dependent Dalitz plot analyses: $\Bz \to K^+K^-\Kz$ and $\Bz \to \pi^+\pi^-\KS$}
\label{sec:cp_uta:qqs:dp}

As mentioned in Sec.~\ref{sec:cp_uta:notations:dalitz:kkk0} and above,
\babar\ have performed time-dependent Dalitz plot analysis of
$\Bz \to K^+K^-\Kz$~\cite{Aubert:2007sd} and 
$\Bz \to \pi^+\pi^-\KS$~\cite{Aubert:2007mgb} decays.
The results are summarized in Tabs.~\ref{tab:cp_uta:kkk0_tddp} 
and~\ref{tab:cp_uta:pipik0_tddp}.
For $\Bz \to K^+K^-\Kz$, results are presented 
in terms of the effective weak phase (from mixing and decay)
difference $\beta^{\rm eff}$ and the direct $\CP$ violation parameter $\Acp$
($\Acp = -C$) for each of the resonant contributions 
$\phi\Kz$ and $f_0\Kz$, together with averaged values of those parameters 
(taking $\CP$ properties into account) over the entire 
high-mass ($m_{K^+K^-} > 1.1 \ {\rm GeV}/c^2$) region of the Dalitz plot.
For $\Bz \to \pi^+\pi^-\KS$, results for $2\beta^{\rm eff}$ are presented 
in Tab.~\ref{tab:cp_uta:pipik0_tddp}.
Results on flavour specific amplitudes that contribute to the Dalitz plot
(such as $K^{*+}\pi^-$) are averaged by the HFAG Rare Decays subgroup 
(Sec.~\ref{sec:rare}).

 \begin{table}[htb]
  \begin{center}
    \caption{
      Results from time-dependent Dalitz plot analysis of 
      the $\Bz \to K^+K^-\Kz$ decay.
    }
    \vspace{0.2cm}
    \setlength{\tabcolsep}{0.0pc}
    \resizebox{\textwidth}{!}{
      \begin{tabular}{l@{\hspace{2mm}}r@{\hspace{2mm}}c@{\hspace{2mm}}|@{\hspace{2mm}}c@{\hspace{2mm}}c@{\hspace{2mm}}|@{\hspace{2mm}}c@{\hspace{2mm}}c} 
        \hline 
        \mc{2}{l}{Experiment} & $N(B\bar{B})$ &
        \mc{2}{c}{$\phi\Kz$} & \mc{2}{c}{$f_0\Kz$} \\
        & & & $\beta^{\rm eff}$ & $\Acp$ & $\beta^{\rm eff}$ & $\Acp$ \\
        \babar & \cite{Aubert:2007sd} & 383M &
        $0.11 \pm 0.14 \pm 0.06$ & $-0.08 \pm 0.18 \pm 0.14$ &
        $0.14 \pm 0.15 \pm 0.05$ & $0.41 \pm 0.23 \pm 0.07$ \\
        \hline
      \end{tabular}
    }

    \vspace{2ex}

      \begin{tabular}{l@{\hspace{2mm}}r@{\hspace{2mm}}c@{\hspace{2mm}}|@{\hspace{2mm}}c@{\hspace{2mm}}c} 
        \hline 
        \mc{2}{l}{Experiment} & $N(B\bar{B})$ &
        \mc{2}{c}{$K^+K^-\Kz$} \\
        & & & $\beta^{\rm eff}$ & $\Acp$ \\
        \babar & \cite{Aubert:2007sd} & 383M &
        $0.436 \pm 0.087 \, ^{+0.055}_{-0.031}$ & $-0.054 \pm 0.102 \pm 0.060$ \\
        \hline
      \end{tabular}

      \label{tab:cp_uta:kkk0_tddp}
  \end{center}
\end{table}

\begin{table}[htb]
  \begin{center}
    \caption{
      Results from time-dependent Dalitz plot analysis of 
      the $\Bz \to \pi^+\pi^-\Kz$ decay.
    }
    \vspace{0.2cm}
    \setlength{\tabcolsep}{0.0pc}
    \begin{tabular}{l@{\hspace{2mm}}r@{\hspace{2mm}}c@{\hspace{2mm}}|@{\hspace{2mm}}c@{\hspace{2mm}}c} 
      \hline 
      \mc{2}{l}{Experiment} & $N(B\bar{B})$ & $2\beta^{\rm eff}\ (f_0\KS)$ & $2\beta^{\rm eff}\ (\rho^0 \KS)$ \\
      \babar & \cite{Aubert:2007mgb} & 383M & $(89\, ^{+22}_{-20} \pm 5 \pm 8)^\circ$ & $(37\, ^{+19}_{-17} \pm 5 \pm 6)^\circ$ \\
      \hline
    \end{tabular}

      \label{tab:cp_uta:pipik0_tddp}
  \end{center}
\end{table}

From the results in Tab.~\ref{tab:cp_uta:pipik0_tddp},
\babar\ infer that the trigonometric reflection 
at $\pi/2 - \beta^{\rm eff}$ in $\Bz \to K^+K^-\Kz$,
which is inconsistent with the Standard Model expectation,
is disfavoured at $4.6\sigma$.

\clearpage
\mysubsection{Time-dependent $\CP$ asymmetries in $b \to c\bar{c}d$ transitions
}
\label{sec:cp_uta:ccd}

The transition $b \to c\bar c d$ can occur via either a $b \to c$ tree
or a $b \to d$ penguin amplitude.  
Similarly to Eq.~(\ref{eq:cp_uta:b_to_s}), the amplitude for 
the $b \to d$ penguin can be written
\begin{equation}
  \label{eq:cp_uta:b_to_d}
  \begin{array}{ccccc}
    A_{b \to d} & = & 
    \mc{3}{l}{F_u V_{ub}V^*_{ud} + F_c V_{cb}V^*_{cd} + F_t V_{tb}V^*_{td}} \\
    & = & (F_u - F_c) V_{ub}V^*_{ud} & + & (F_t - F_c) V_{tb}V^*_{td} \\
    & = & {\cal O}(\lambda^3) & + & {\cal O}(\lambda^3). \\
  \end{array}
\end{equation}
From this it can be seen that the $b \to d$ penguin amplitude 
contains terms with different weak phases at the same order of
CKM suppression.

In the above, we have followed Eq.~(\ref{eq:cp_uta:b_to_s}) 
by eliminating the $F_c$ term using unitarity.
However, we could equally well write
\begin{equation}
  \label{eq:cp_uta:b_to_d_alt}
  \begin{array}{ccccc}
    A_{b \to d} 
    & = & (F_u - F_t) V_{ub}V^*_{ud} & + & (F_c - F_t) V_{cb}V^*_{cd}, \\
    & = & (F_c - F_u) V_{cb}V^*_{cd} & + & (F_t - F_u) V_{tb}V^*_{td}. \\
  \end{array}
\end{equation}
Since the $b \to c\bar{c}d$ tree amplitude 
has the weak phase of $V_{cb}V^*_{cd}$,
either of the above expressions allow the penguin to be decomposed into 
parts with weak phases the same and different to the tree amplitude
(the relative weak phase can be chosen to be either $\beta$ or $\gamma$).
However, if the tree amplitude dominates,
there is little sensitivity to any phase 
other than that from $\Bz$\textendash$\Bzb$ mixing.

\begin{figure}[htb]
  \begin{center}
\begin{tabular}{cc}
      \resizebox{0.46\textwidth}{!}{
        \includegraphics{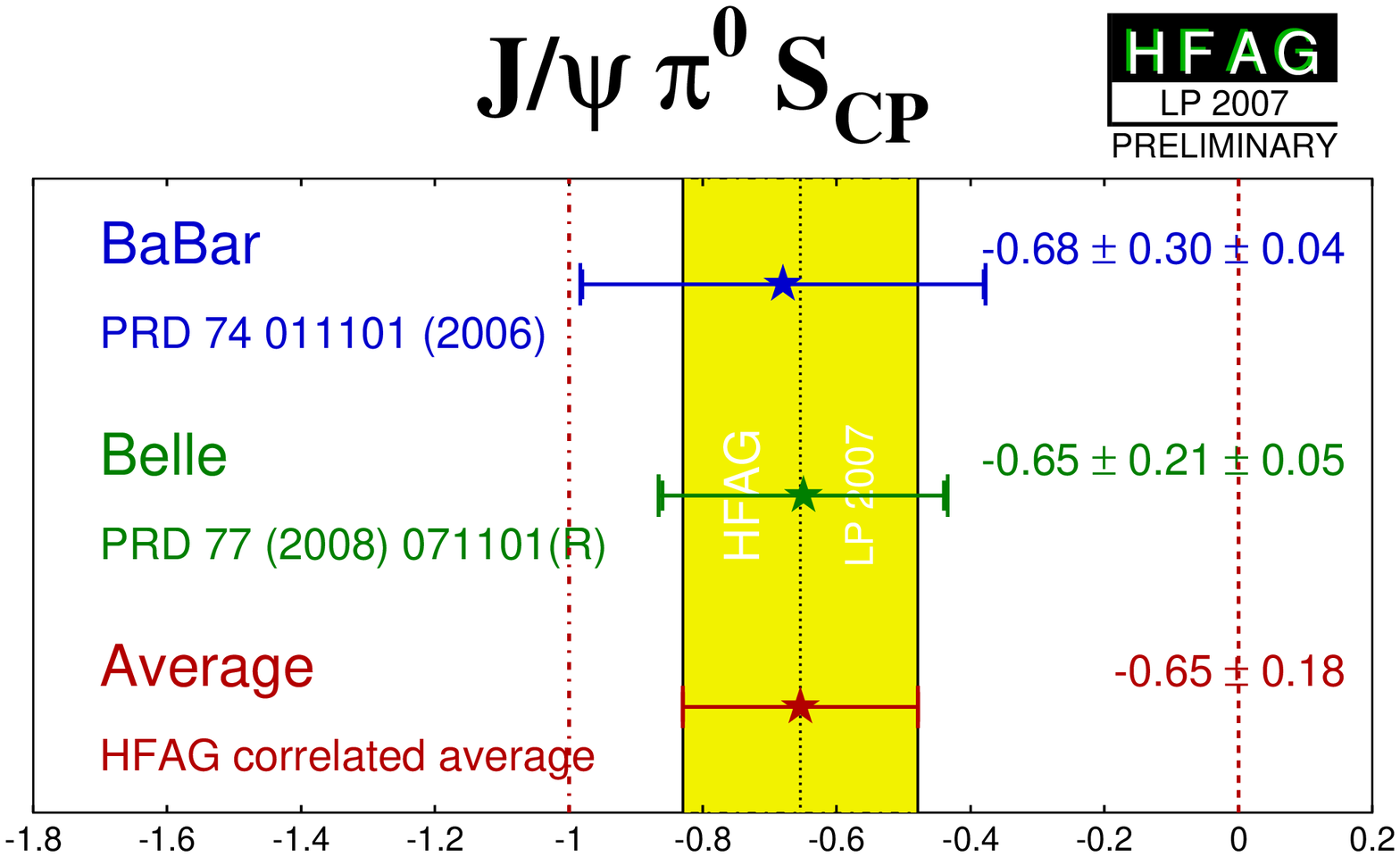}
      }
      &
      \resizebox{0.46\textwidth}{!}{
        \includegraphics{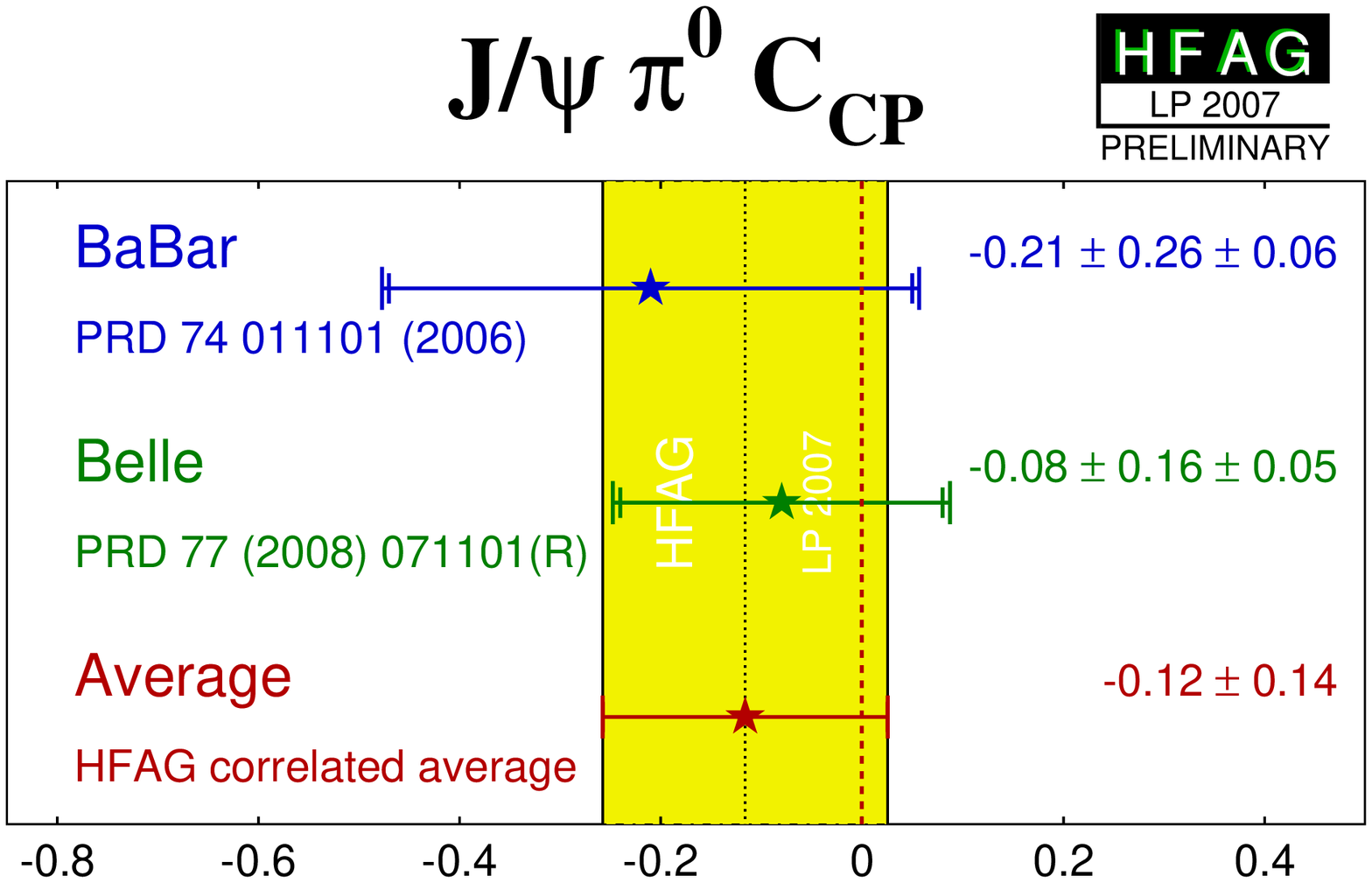}
      }
    \end{tabular}
  \end{center}
  \vspace{-0.8cm}
  \caption{
    Averages of 
    (left) $S_{b \to c\bar c d}$ and (right) $C_{b \to c\bar c d}$ 
    for the mode $\Bz \to J/ \psi \pi^0$.
  }
  \label{fig:cp_uta:ccd:psipi0}
\end{figure}

\begin{figure}[htb]
  \begin{center}
    \begin{tabular}{cc}
      \resizebox{0.46\textwidth}{!}{
        \includegraphics{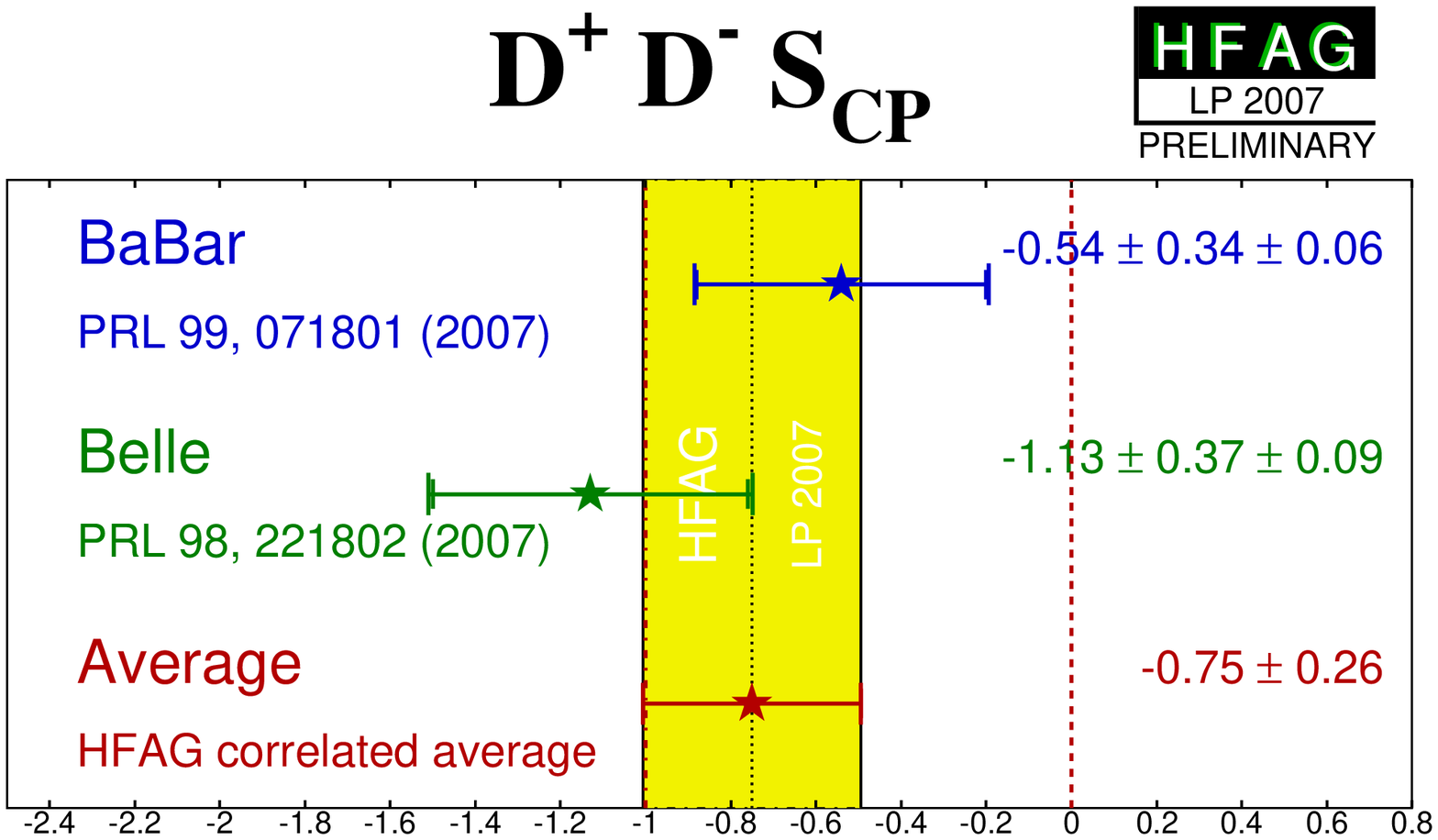}
      }
      &
      \resizebox{0.46\textwidth}{!}{
        \includegraphics{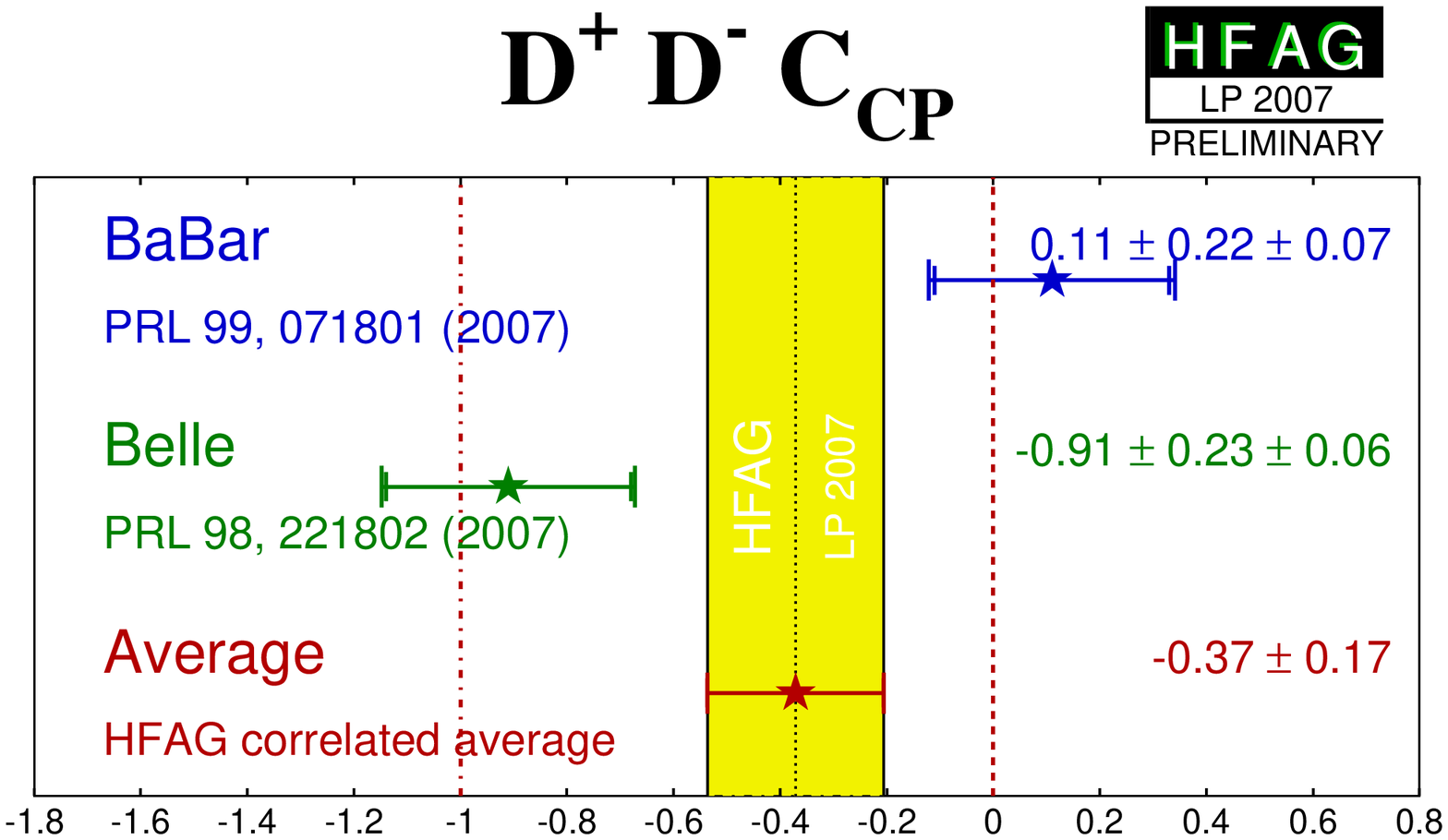}
      }
    \end{tabular}
  \end{center}
  \vspace{-0.8cm}
  \caption{
    Averages of 
    (left) $S_{b \to c\bar c d}$ and (right) $C_{b \to c\bar c d}$ 
    for the mode $\Bz \to D^+D^-$.
  }
  \label{fig:cp_uta:ccd:dd}
\end{figure}

\begin{figure}[htb]
  \begin{center}
    \begin{tabular}{cc}
      \resizebox{0.46\textwidth}{!}{
        \includegraphics{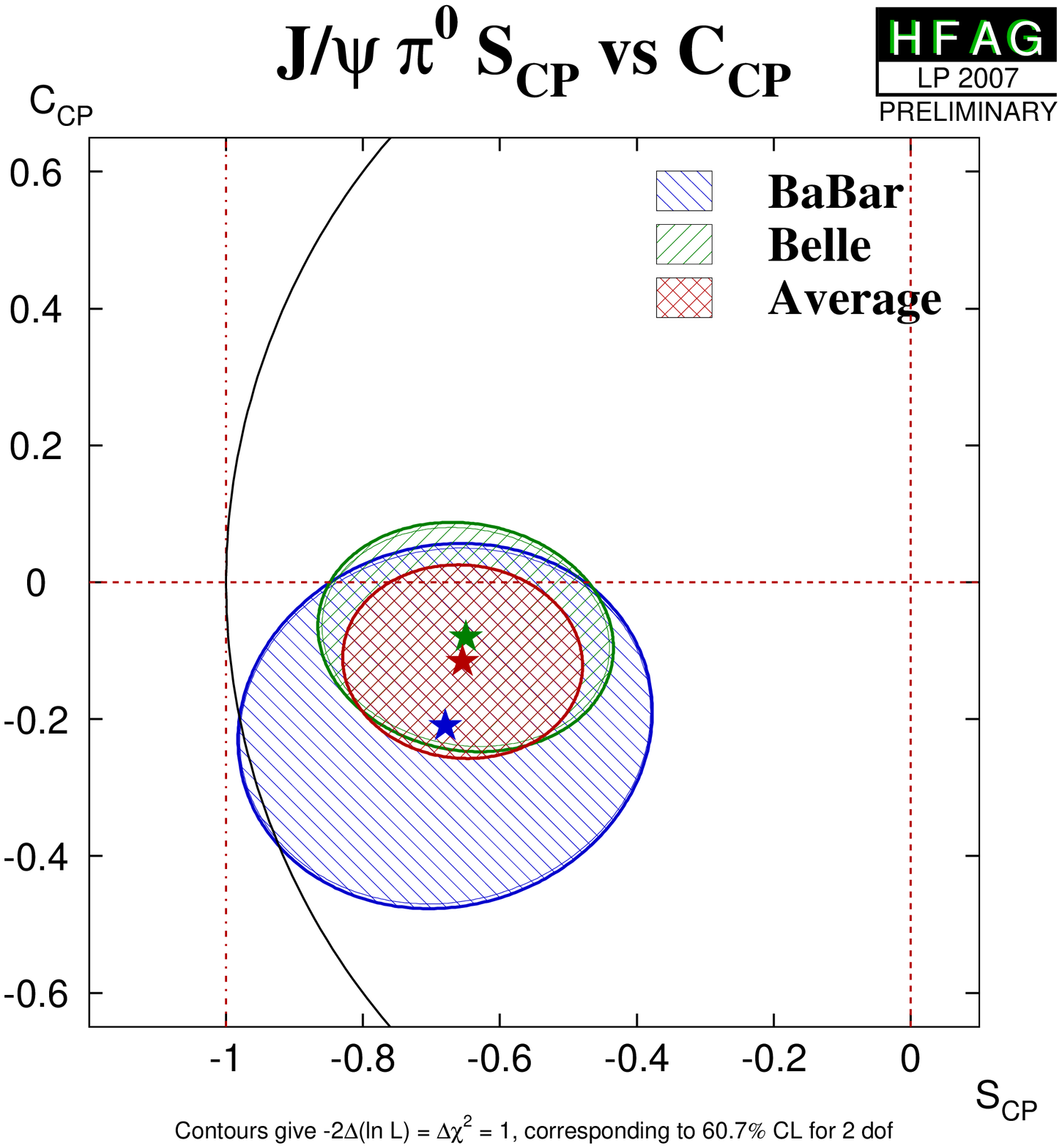}
      }
      &
      \resizebox{0.46\textwidth}{!}{
        \includegraphics{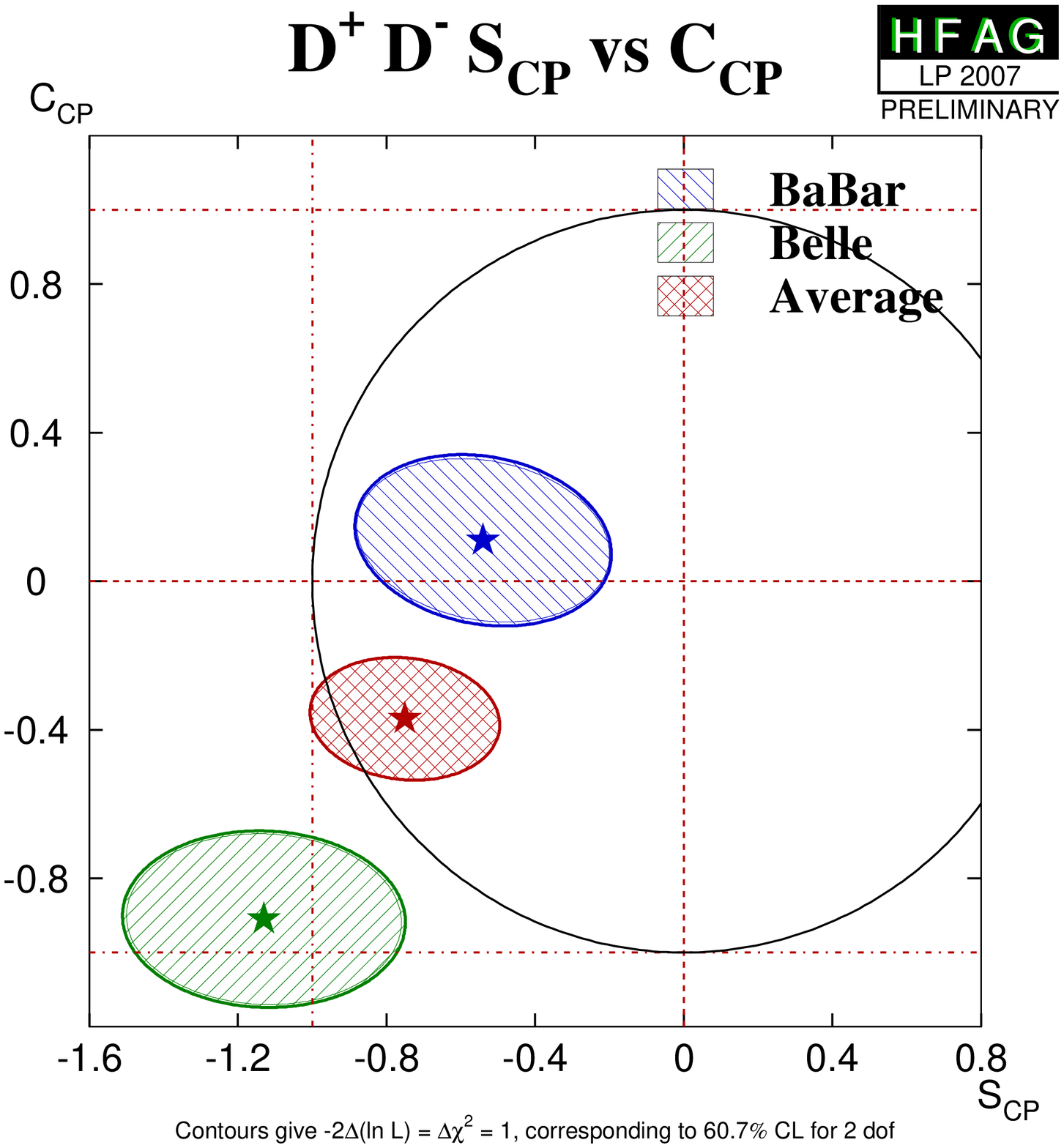}
      }
    \end{tabular}
  \end{center}
  \vspace{-0.8cm}
  \caption{
    Averages of two $b \to c\bar c d$ dominated channels,
    for which correlated averages are performed,
    in the $S_{\CP}$ \vs\ $C_{\CP}$ plane.
    (Left) $\Bz \to J/ \psi \pi^0$ and (right) $\Bz \to D^+D^-$.
  }
  \label{fig:cp_uta:ccd_SvsC}
\end{figure}

The $b \to c\bar{c}d$ transitions can be investigated with studies 
of various different final states. 
Results are available from both \babar\  and \belle\ 
using the final states $\jpsi \pi^0$, $D^+D^-$, 
$D^{*+}D^{*-}$ and $D^{*\pm}D^{\mp}$,
the averages of these results are given in Table~\ref{tab:cp_uta:ccd}.
The results using the $\CP$ eigenstate ($\etacp = +1$) modes
$\jpsi \pi^0$ and $D^+D^-$
are shown in Fig.~\ref{fig:cp_uta:ccd:psipi0} and 
Fig.~\ref{fig:cp_uta:ccd:dd} respectively,
with two-dimensional constraints shown in Fig.~\ref{fig:cp_uta:ccd_SvsC}.

\begin{figure}[htb]
  \begin{center}
    \begin{tabular}{cc}
      \resizebox{0.46\textwidth}{!}{
        \includegraphics{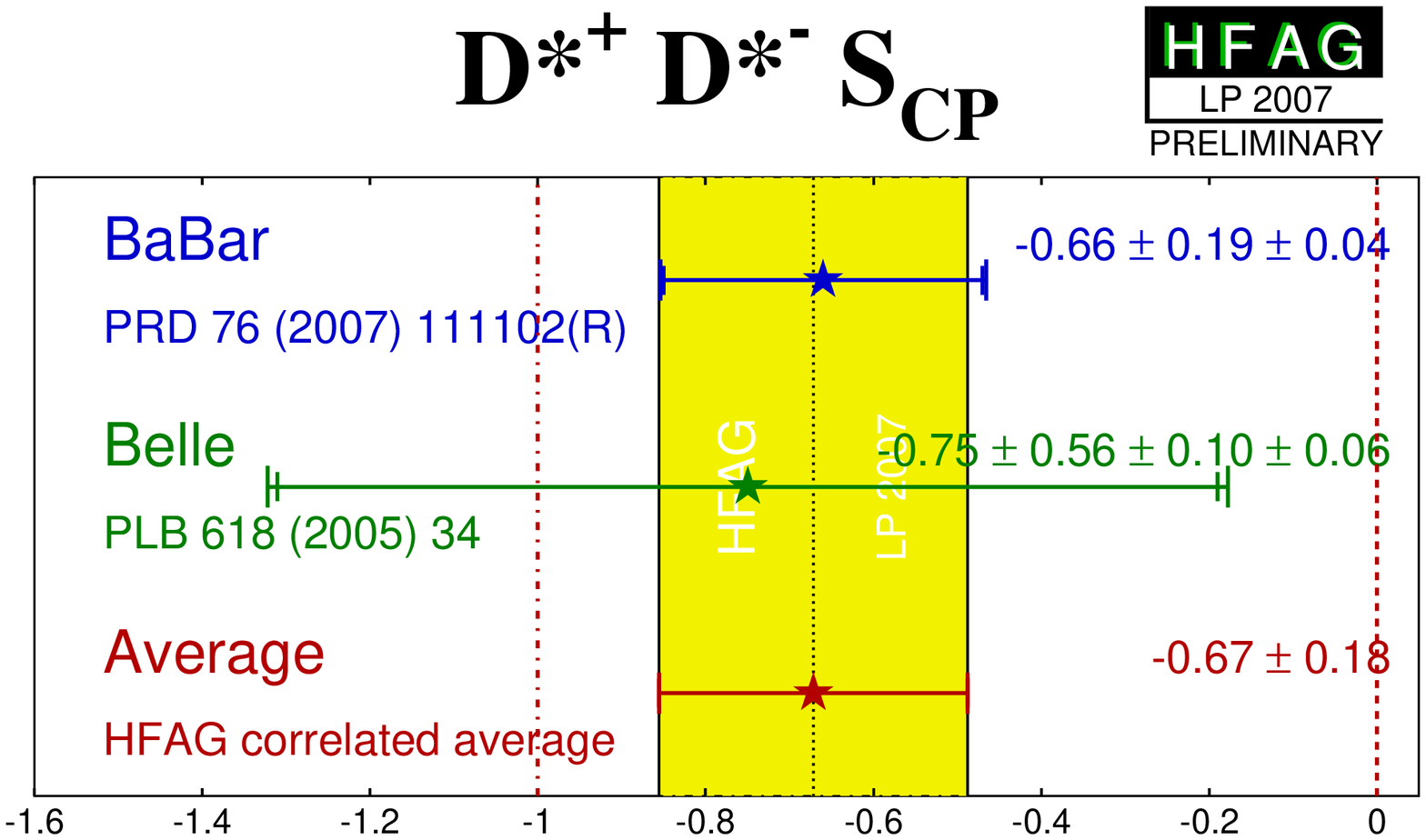}
      }
      &
      \resizebox{0.46\textwidth}{!}{
        \includegraphics{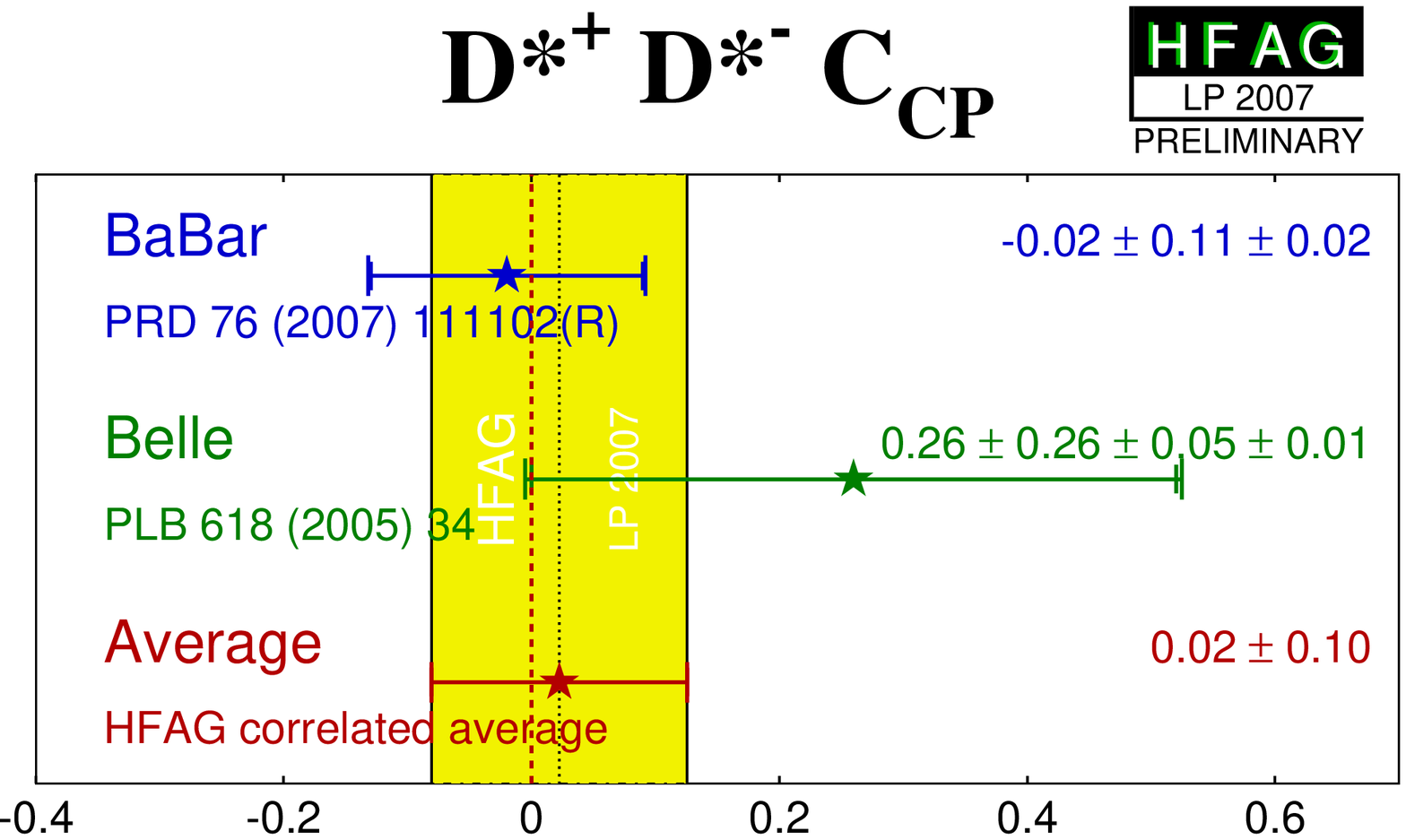}
      }
    \end{tabular}
  \end{center}
  \vspace{-0.8cm}
  \caption{
    Averages of 
    (left) $S_{b \to c\bar c d}$ and (right) $C_{b \to c\bar c d}$ 
    for the mode $\Bz \to D^{*+}D^{*-}$.
  }
  \label{fig:cp_uta:ccd:dstardstar}
\end{figure}

The vector-vector mode $D^{*+}D^{*-}$ 
is found to be dominated by the $\CP$-even longitudinally polarized component;
\babar\ measures a $\CP$-odd fraction of 
$0.143 \pm 0.034 \pm 0.008$~\cite{Aubert:2007rr} while
\belle\ measures a $\CP$-odd fraction of 
$0.19  \pm 0.08  \pm 0.01 $~\cite{Miyake:2005qb}
(here we do not average these fractions and rescale the inputs,
however the average is almost independent of the treatment).
\babar\ allows the $\CP$-odd fraction 
(actually, the transversely polarized fraction, $R_\perp$, which is equivalent)
to float in the fit, so that its uncertainty is automatically incorporated 
into the statistical error.
\belle\ quotes a third uncertainty due to the polarization.
\babar\ have also performed an additional fit in which the 
$\CP$-even and $\CP$-odd components are allowed to have different 
$\CP$ violation parameters $S$ and $C$.  
These results are included in Table~\ref{tab:cp_uta:ccd}.
Results using $D^{*+}D^{*-}$ are shown in Fig.~\ref{fig:cp_uta:ccd:dstardstar}.

For the non-$\CP$ eigenstate mode $D^{*\pm}D^{\mp}$
\babar\ uses fully reconstructed events while 
\belle\ combines both fully and partially reconstructed samples.
At present we perform uncorrelated averages of the parameters in the 
$D^{*\pm}D^{\mp}$ system.

\begin{table}[htb]
	\begin{center}
		\caption{
     Averages for the $b \to c\bar{c}d$ modes,
     $\Bz \to J/\psi \pi^{0}$, $D^+D^-$, $D^{*+}D^{*-}$ and $D^{*\pm}D^\mp$.
		}
		\vspace{0.2cm}
		\setlength{\tabcolsep}{0.0pc}
\begin{tabular*}{\textwidth}{@{\extracolsep{\fill}}lrcccc} \hline
	\mc{2}{l}{Experiment} & $N(B\bar{B})$ & $S_{CP}$ & $C_{CP}$ & Correlation \\
	\hline
        \mc{6}{c}{$J/\psi \pi^{0}$} \\
	\babar & \cite{Aubert:2006qca} & 232M & $-0.68 \pm 0.30 \pm 0.04$ & $-0.21 \pm 0.26 \pm 0.06$ & $0.08$ \\
	\belle & \cite{:2007wd} & 535M & $-0.65 \pm 0.21 \pm 0.05$ & $-0.08 \pm 0.16 \pm 0.05$ & $-0.10$ \\
	\mc{3}{l}{\bf Average} & $-0.65 \pm 0.18$ & $-0.12 \pm 0.14$ & $-0.04$ \\
	\mc{3}{l}{\small Confidence level} & \mc{2}{c}{\small $0.92~(0.1\sigma)$} & \\
		\hline

        \mc{6}{c}{$D^{+} D^{-}$} \\
	\babar & \cite{Aubert:2007pa} & 364M & $-0.54 \pm 0.34 \pm 0.06$ & $0.11 \pm 0.22 \pm 0.07$ & $-0.17$ \\
	\belle & \cite{Fratina:2007zk} & 535M & $-1.13 \pm 0.37 \pm 0.09$ & $-0.91 \pm 0.23 \pm 0.06$ & $-0.04$ \\
	\mc{3}{l}{\bf Average} & $-0.75 \pm 0.26$ & $-0.37 \pm 0.17$ & $-0.10$ \\
	\mc{3}{l}{\small Confidence level} & \mc{2}{c}{\small $0.003~(3.0\sigma)$} & \\
		\hline

        \mc{6}{c}{$D^{*+} D^{*-}$} \\
	\babar & \cite{Aubert:2007rr} & 383M & $-0.66 \pm 0.19 \pm 0.04$ & $-0.02 \pm 0.11 \pm 0.02$ & $-0.04$ \\
	\belle & \cite{Miyake:2005qb} & 152M & $-0.75 \pm 0.56 \pm 0.10 \pm 0.06$ & $0.26 \pm 0.26 \pm 0.05 \pm 0.01$ & \textendash{} \\
	\mc{3}{l}{\bf Average} & $-0.67 \pm 0.18$ & $0.02 \pm 0.10$ & $-0.03$ \\
	\mc{3}{l}{\small Confidence level} & \mc{2}{c}{\small $0.62~(0.5\sigma)$} & \\
		\hline
		\end{tabular*}

                \vspace{2ex}

    \resizebox{\textwidth}{!}{
		\begin{tabular}{@{\extracolsep{2mm}}lrccccc} \hline
	\mc{2}{l}{Experiment} & $N(B\bar{B})$ & $S_{CP+}$ & $C_{CP+}$ & $S_{CP-}$ & $C_{CP-}$ \\
	\hline
        \mc{7}{c}{$D^{*+} D^{*-}$} \\
	\babar & \cite{Aubert:2007rr} & 383M & $-0.72 \pm 0.19 \pm 0.05$ & $-0.05 \pm 0.14 \pm 0.02$ & $-1.83 \pm 1.04 \pm 0.23$ & $0.23 \pm 0.67 \pm 0.10$ \\
		\hline
		\end{tabular}
    }

                \vspace{2ex}

    \resizebox{\textwidth}{!}{
		\begin{tabular}{@{\extracolsep{2mm}}lrcccccc} \hline
	\mc{2}{l}{Experiment} & $N(B\bar{B})$ & $S_{+-}$ & $C_{+-}$ & $S_{-+}$ & $C_{-+}$ & ${\cal A}$ \\
        \hline
        \mc{8}{c}{$D^{*\pm} D^{\mp}$} \\
	\babar & \cite{Aubert:2007pa} & 364M & $-0.79 \pm 0.21 \pm 0.06$ & $0.18 \pm 0.15 \pm 0.04$ & $-0.44 \pm 0.22 \pm 0.06$ & $0.23 \pm 0.15 \pm 0.04$ & $0.12 \pm 0.06 \pm 0.02$ \\
	\belle & \cite{Aushev:2004uc} & 152M & $-0.55 \pm 0.39 \pm 0.12$ & $-0.37 \pm 0.22 \pm 0.06$ & $-0.96 \pm 0.43 \pm 0.12$ & $0.23 \pm 0.25 \pm 0.06$ & $0.07 \pm 0.08 \pm 0.04$ \\
	\mc{3}{l}{\bf Average} & $-0.74 \pm 0.19$ & $0.01 \pm 0.13$ & $-0.55 \pm 0.20$ & $0.23 \pm 0.13$ & $0.10 \pm 0.05$ \\
	\mc{3}{l}{\small Confidence level} & {\small $0.60~(0.5\sigma)$} & {\small $0.05~(2.0\sigma)$} & {\small $0.30~(1.0\sigma)$} & {\small $1.00~(0.0\sigma)$} & {\small $0.65~(0.5\sigma)$} \\
        \hline
                \end{tabular}
    }
		\label{tab:cp_uta:ccd}
	\end{center}
\end{table}

In the absence of the penguin contribution (tree dominance),
the time-dependent parameters would be given by
$S_{b \to c\bar c d} = - \etacp \sin(2\beta)$,
$C_{b \to c\bar c d} = 0$,
$S_{+-} = \sin(2\beta + \delta)$,
$S_{-+} = \sin(2\beta - \delta)$,
$C_{+-} = - C_{-+}$ and 
${\cal A} = 0$,
where $\delta$ is the strong phase difference between the 
$D^{*+}D^-$ and $D^{*-}D^+$ decay amplitudes.
In the presence of the penguin contribution,
there is no clean interpretation in terms of CKM parameters,
however
direct $\CP$ violation may be observed as any of
$C_{b \to c\bar c d} \neq 0$, $C_{+-} \neq - C_{-+}$ or $A_{+-} \neq 0$.

\begin{figure}[htb]
  \begin{center}
    \begin{tabular}{cc}
      \resizebox{0.46\textwidth}{!}{
        \includegraphics{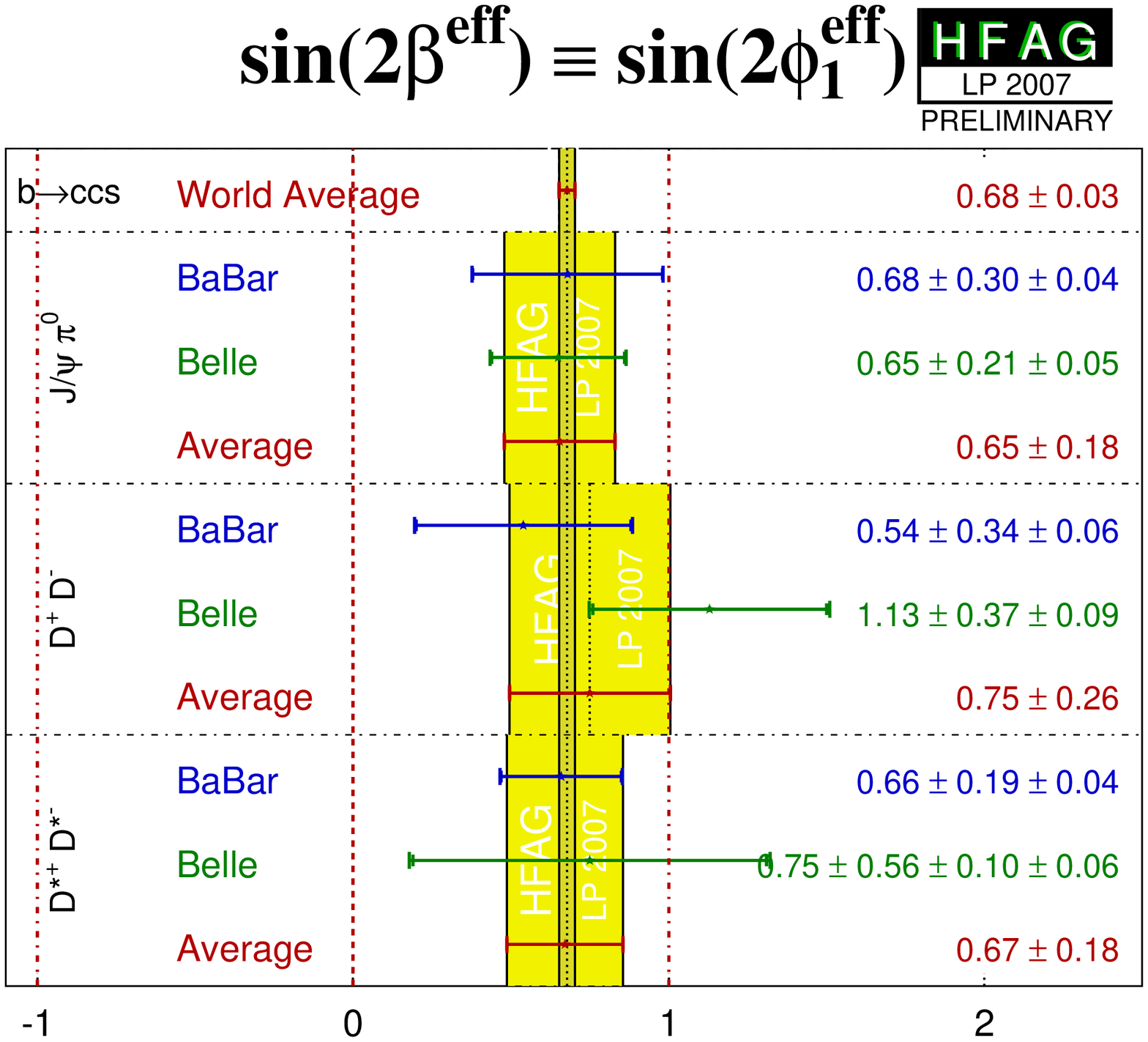}
      }
      &
      \resizebox{0.46\textwidth}{!}{
        \includegraphics{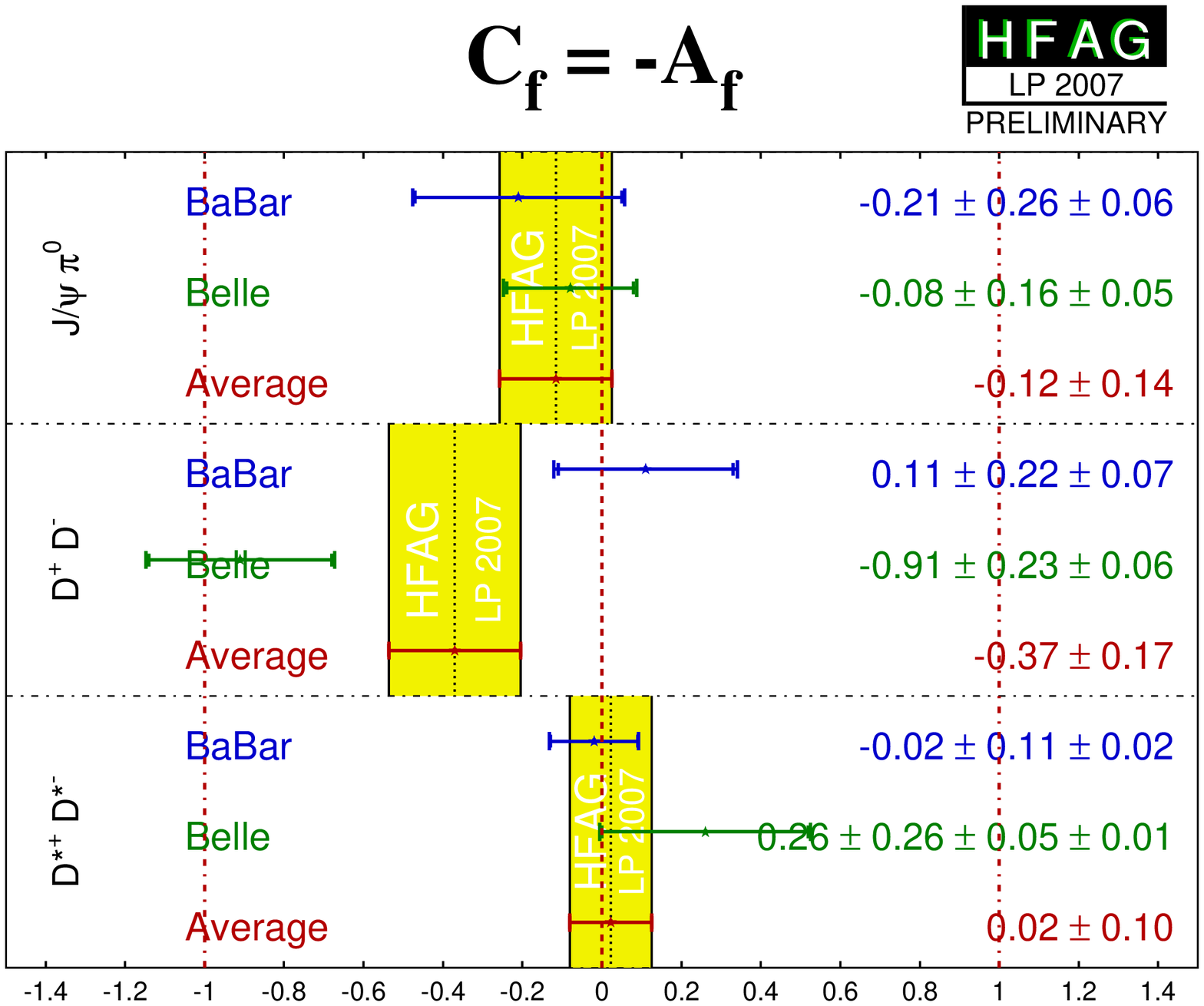}
      }
    \end{tabular}
  \end{center}
  \vspace{-0.8cm}
  \caption{
    Averages of 
    (left) $-\etacp S_{b \to c\bar c d}$ and (right) $C_{b \to c\bar c d}$.
    The $-\etacp S_{b \to q\bar q s}$ figure compares the results to 
    the world average 
    for $-\etacp S_{b \to c\bar c s}$ (see Section~\ref{sec:cp_uta:ccs:cp_eigen}).
  }
  \label{fig:cp_uta:ccd}
\end{figure}

\begin{figure}[htb]
  \begin{center}
    \resizebox{0.66\textwidth}{!}{
      \includegraphics{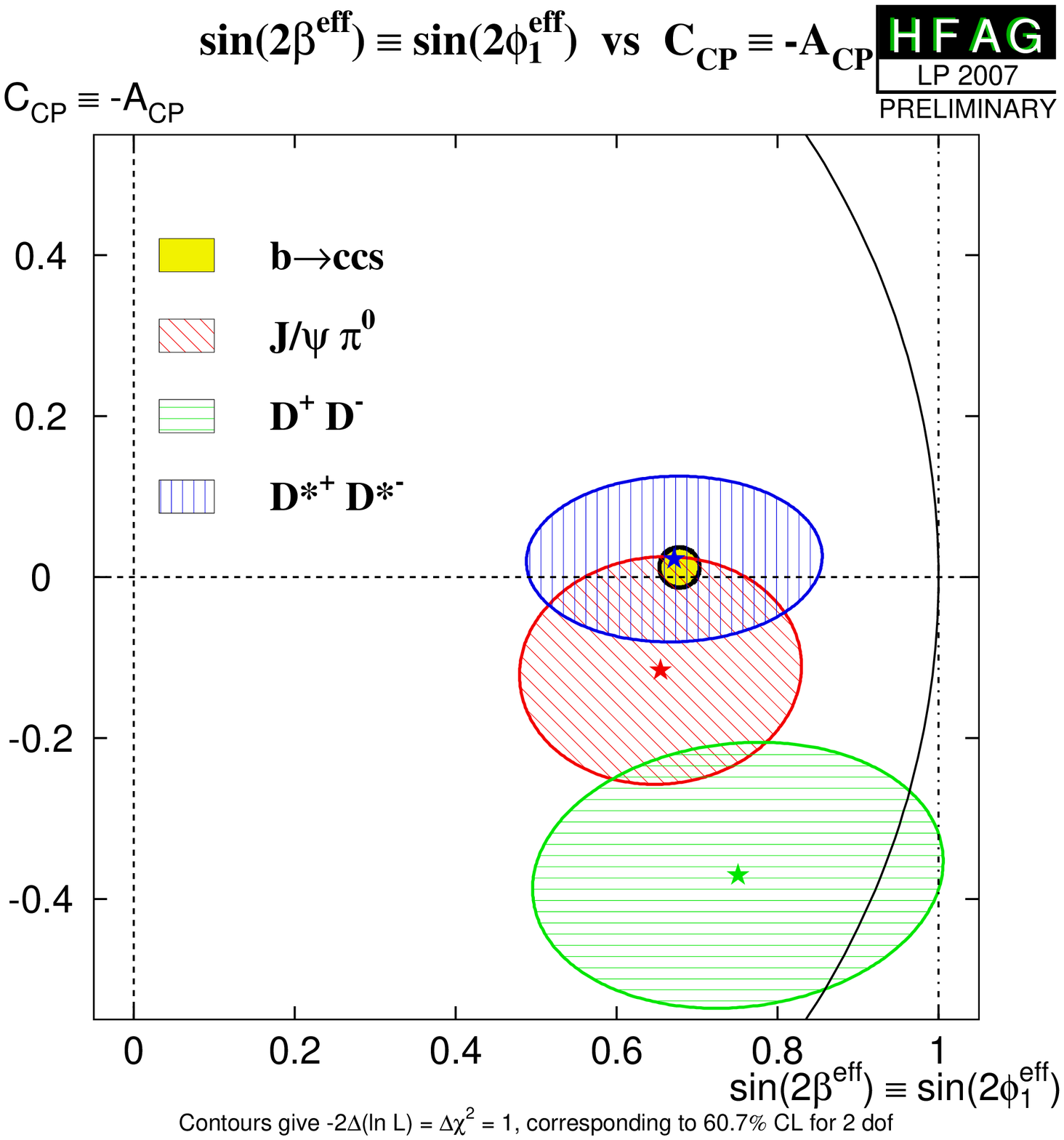}
    }
  \end{center}
  \vspace{-0.8cm}
  \caption{
    Compilation of constraints in the 
    $-\etacp S_{b \to c\bar c d}$ \vs\ $C_{b \to c\bar c d}$ plane.
  }
  \label{fig:cp_uta:ccd_SvsC-all}
\end{figure}

The averages for the $b \to c\bar c d$ modes 
are shown in Figs.~\ref{fig:cp_uta:ccd} and~\ref{fig:cp_uta:ccd_SvsC-all}.
Results are consistent with tree dominance,
and with the Standard Model,
though the \belle\ results in $\Bz \to D^+D^-$~\cite{Fratina:2007zk}
show an indication of direct $\CP$ violation,
and hence a non-zero penguin contribution.
The average of $S_{b \to c\bar c d}$ in both $J/\psi \pi^{0}$ and 
$D^{*+}D^{*-}$ final states are about $3\sigma$ from zero;
however, due to the large uncertainty and possible non-Gaussian effects,
any strong conclusion should be deferred.


\clearpage
\mysubsection{Time-dependent $\CP$ asymmetries in $b \to q\bar{q}d$ transitions
}
\label{sec:cp_uta:qqd}

Decays such as $\Bz\to\KS\KS$ are pure $b \to q\bar{q}d$ penguin transitions.
As shown in Eq.~\ref{eq:cp_uta:b_to_d},
this diagram has different contributing weak phases,
and therefore the observables are sensitive to the difference 
(which can be chosen to be either $\beta$ or $\gamma$).
Note that if the contribution with the top quark in the loop dominates,
the weak phase from the decay amplitudes should cancel that from mixing,
so that no $\CP$ violation (neither mixing-induced nor direct) occurs.
Non-zero contributions from loops with intermediate up and charm quarks
can result in both types of effect 
(as usual, a strong phase difference is required for direct $\CP$ violation
to occur).

\begin{figure}[htb]
  \begin{center}
    \begin{tabular}{cc}
      \resizebox{0.46\textwidth}{!}{
        \includegraphics{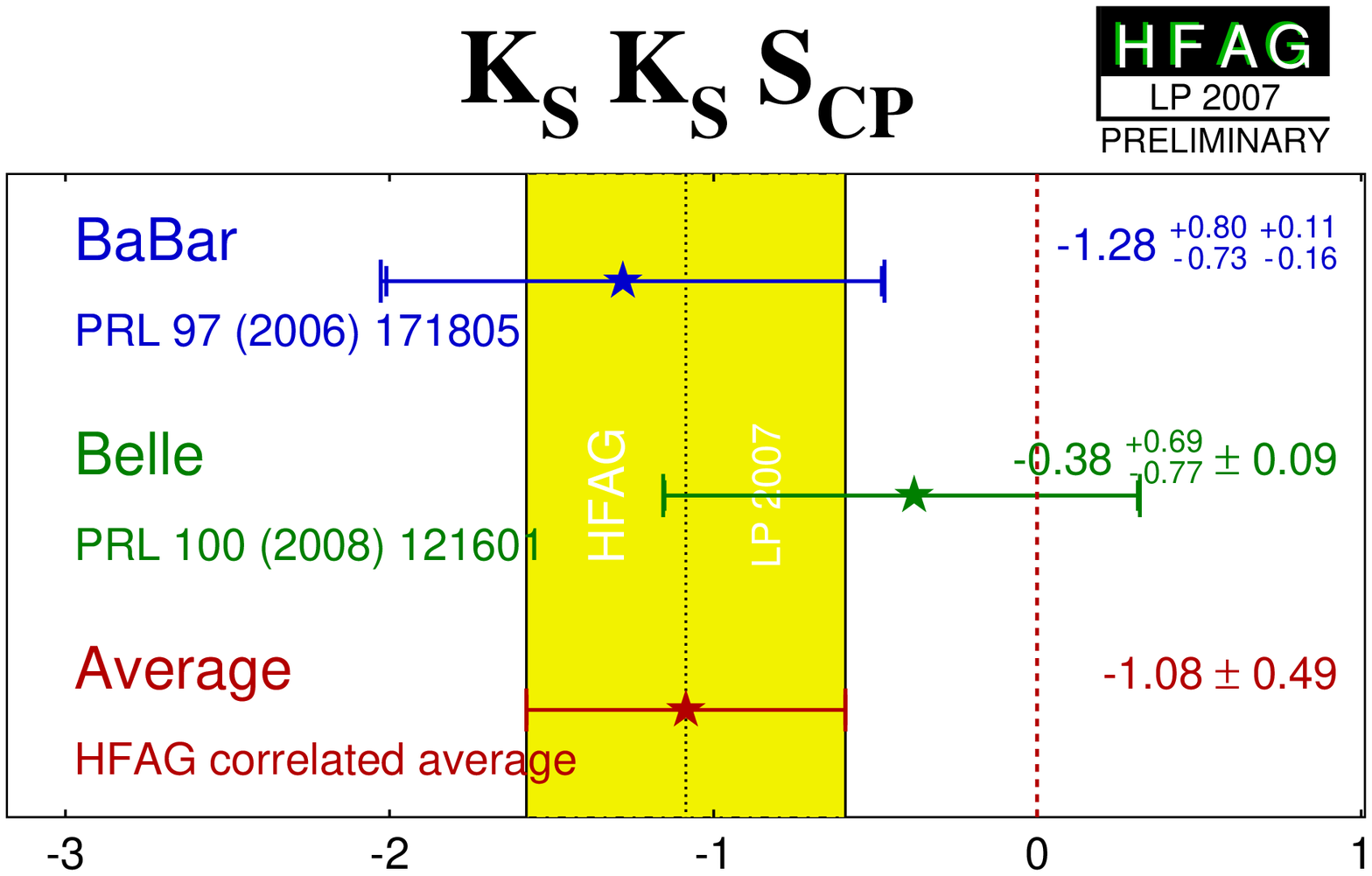}
      }
      &
      \resizebox{0.46\textwidth}{!}{
        \includegraphics{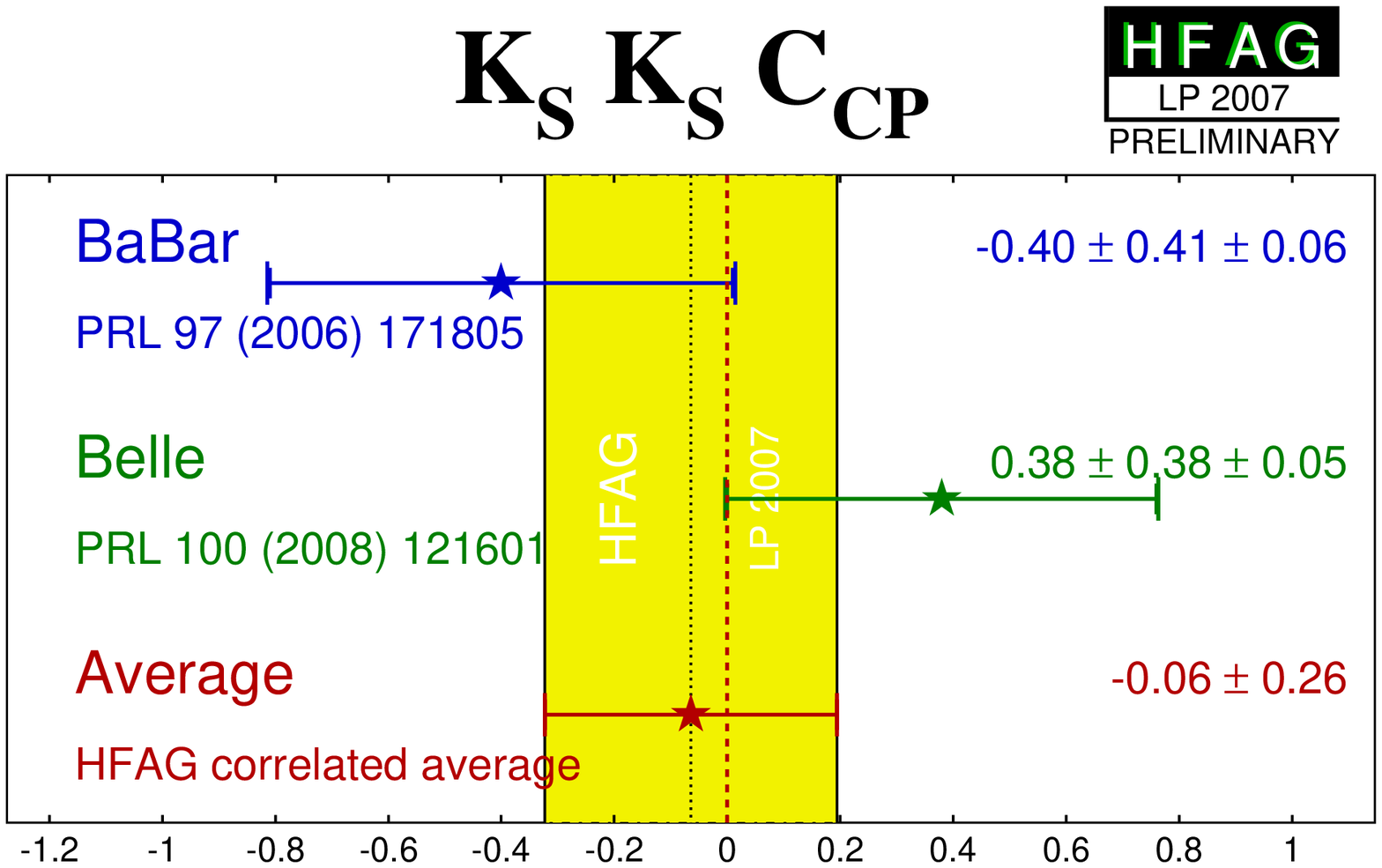}
      }
    \end{tabular}
  \end{center}
  \vspace{-0.8cm}
  \caption{
    Averages of 
    (left) $S_{b \to q\bar q d}$ and (right) $C_{b \to q\bar q d}$ 
    for the mode $\Bz \to \KS\KS$.
  }
  \label{fig:cp_uta:qqd:ksks}
\end{figure}

Both \babar~\cite{Aubert:2006gm} and \belle~\cite{Nakahama:2007dg}
have performed time-dependent analyses of $\Bz\to\KS\KS$.
The results are shown in Table~\ref{tab:cp_uta:qqd}
and Fig.~\ref{fig:cp_uta:qqd:ksks}.

\begin{table}[htb]
	\begin{center}
		\caption{
			Results for $\Bz \to \KS\KS$.
		}
		\vspace{0.2cm}
		\setlength{\tabcolsep}{0.0pc}
		\begin{tabular*}{\textwidth}{@{\extracolsep{\fill}}lrcccc} \hline
	\mc{2}{l}{Experiment} & $N(B\bar{B})$ & $S_{CP}$ & $C_{CP}$ & Correlation \\
	\hline
	\babar & \cite{Aubert:2006gm} & 350M & $-1.28 \,^{+0.80}_{-0.73} \,^{+0.11}_{-0.16}$ & $-0.40 \pm 0.41 \pm 0.06$ & $-0.32$ \\
	\belle & \cite{Nakahama:2007dg} & 657M & $-0.38 \,^{+0.69}_{-0.77} \pm 0.09$ & $0.38 \pm 0.38 \pm 0.05$ & $0.48$ \\
	\hline
	\mc{3}{l}{\bf Average} & $-1.08 \pm 0.49$ & $-0.06 \pm 0.26$ & $0.14$ \\
	\mc{3}{l}{\small Confidence level} & \mc{2}{c}{\small $0.29~(1.1\sigma)$} & \\
		\hline
		\end{tabular*}
		\label{tab:cp_uta:qqd}
	\end{center}
\end{table}

\clearpage
\mysubsection{Time-dependent asymmetries in $b \to s\gamma$ transitions
}
\label{sec:cp_uta:bsg}

The radiative decays $b \to s\gamma$ produce photons 
which are highly polarized in the Standard Model.
The decays $\Bz \to F \gamma$ and $\Bzb \to F \gamma$ 
produce photons with opposite helicities, 
and since the polarization is, in principle, observable,
these final states cannot interfere.
The finite mass of the $s$ quark introduces small corrections
to the limit of maximum polarization,
but any large mixing induced $\CP$ violation would be a signal for new physics.
Since a single weak phase dominates the $b \to s \gamma$ transition in the 
Standard Model, the cosine term is also expected to be small.

Atwood {\it et al.}~\cite{Atwood:2004jj} have shown that 
an inclusive analysis with respect to $\KS\pi^0\gamma$ can be performed,
since the properties of the decay amplitudes 
are independent of the angular momentum of the $\KS\pi^0$ system. 
However, if non-dipole operators contribute significantly to the amplitudes, 
then the Standard Model mixing-induced $\CP$ violation could be larger 
than the na\"\i ve expectation 
$S \simeq -2 (m_s/m_b) \sin \left(2\beta\right)$~\cite{Grinstein:2004uu,Grinstein:2005nu}.
In this case, 
the $\CP$ parameters may vary over the $\KS\pi^0\gamma$ Dalitz plot, 
for example as a function of the $\KS\pi^0$ invariant mass.
Explicit calculations indicate such corrections are small
for exclusive final states~\cite{Matsumori:2005ax,Ball:2006cva}.

With the above in mind, 
we quote two averages: one for $K^*(892)$ candidates only, 
and the other one for the inclusive $\KS\pi^0\gamma$ decay (including the $K^*(892)$).
If the Standard Model dipole operator is dominant, 
both should give the same quantities 
(the latter naturally with smaller statistical error). 
If not, care needs to be taken in interpretation of the inclusive parameters, 
while the results on the $K^*(892)$ resonance remain relatively clean.
Results from \babar\ and \belle\ are used for both averages;
both experiments use the invariant mass range 
$0.60 \ {\rm GeV}/c^2 < M_{\KS\pi^0} < 1.80 \ {\rm GeV}/c^2$
in the inclusive analysis.
Note that these averages include an update from \babar~\cite{Aubert:2007qj} on the 
$K^*\gamma$ parameters but not on the $\KS\pi^0\gamma$ parameters,
so that the former are measured more precisely than the latter.

\begin{table}[htb]
	\begin{center}
		\caption{
      Averages for $b \to s \gamma$ modes.
		}
		\vspace{0.2cm}
		\setlength{\tabcolsep}{0.0pc}
\begin{tabular*}{\textwidth}{@{\extracolsep{\fill}}lrcccc} \hline
	\mc{2}{l}{Experiment} & $N(B\bar{B})$ & $S_{CP} (b \to s \gamma)$ & $C_{CP} (b \to s \gamma)$ & Correlation \\
        \hline
        \mc{6}{c}{$\Kstar(892)\gamma$} \\
	\babar & \cite{Aubert:2007qj} & 431M & $-0.08 \pm 0.31 \pm 0.05$ & $-0.15 \pm 0.17 \pm 0.03$ & $0.05$ \\
	\belle & \cite{Ushiroda:2006fi} & 535M & $-0.32 \,^{+0.36}_{-0.33} \pm 0.05$ & $0.20 \pm 0.24 \pm 0.05$ & $0.08$ \\
	\mc{3}{l}{\bf Average} & $-0.19 \pm 0.23$ & $-0.03 \pm 0.14$ & $0.06$ \\
	\mc{3}{l}{\small Confidence level} & \mc{2}{c}{\small $0.43~(0.8\sigma)$} & \\
		\hline

        \mc{6}{c}{$\KS \pi^0 \gamma$ (including $\Kstar(892)\gamma$)} \\
	\babar & \cite{Aubert:2005bu} & 232M & $-0.06 \pm 0.37$ & $-0.48 \pm 0.22$ & $0.05$ \\
	\belle & \cite{Ushiroda:2006fi} & 535M & $-0.10 \pm 0.31 \pm 0.07$ & $0.20 \pm 0.20 \pm 0.06$ & $0.08$ \\
	\mc{3}{l}{\bf Average} & $-0.09 \pm 0.24$ & $-0.12 \pm 0.15$ & $0.06$ \\
        \mc{3}{l}{\small Confidence level} & \mc{2}{c}{\small $0.08~(1.8\sigma)$} & \\

		\hline
		\end{tabular*}
		\label{tab:cp_uta:bsg}
	\end{center}
\end{table}

\begin{figure}[htb]
  \begin{center}
    \begin{tabular}{cc}
      \resizebox{0.46\textwidth}{!}{
        \includegraphics{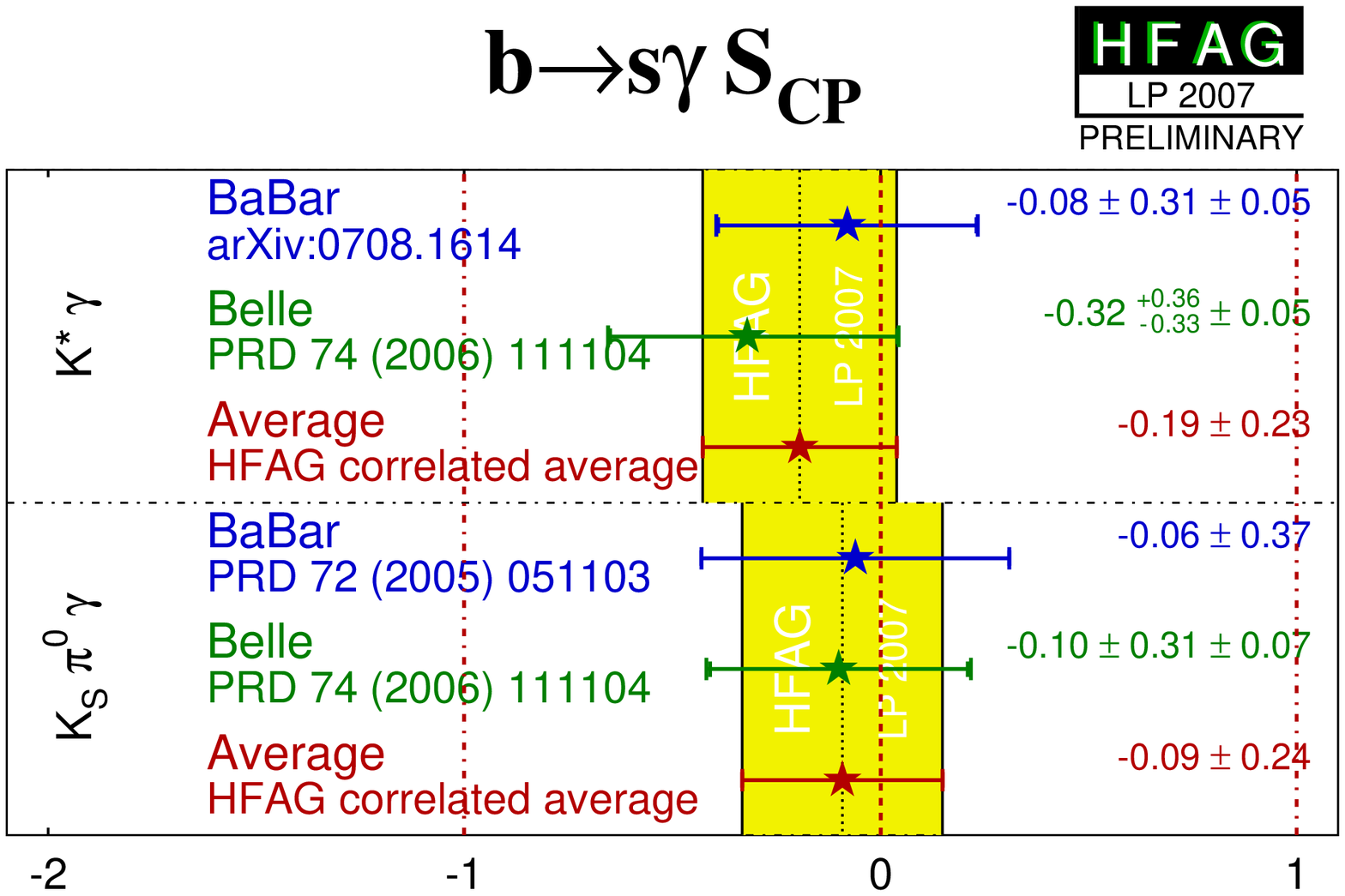}
      }
      &
      \resizebox{0.46\textwidth}{!}{
        \includegraphics{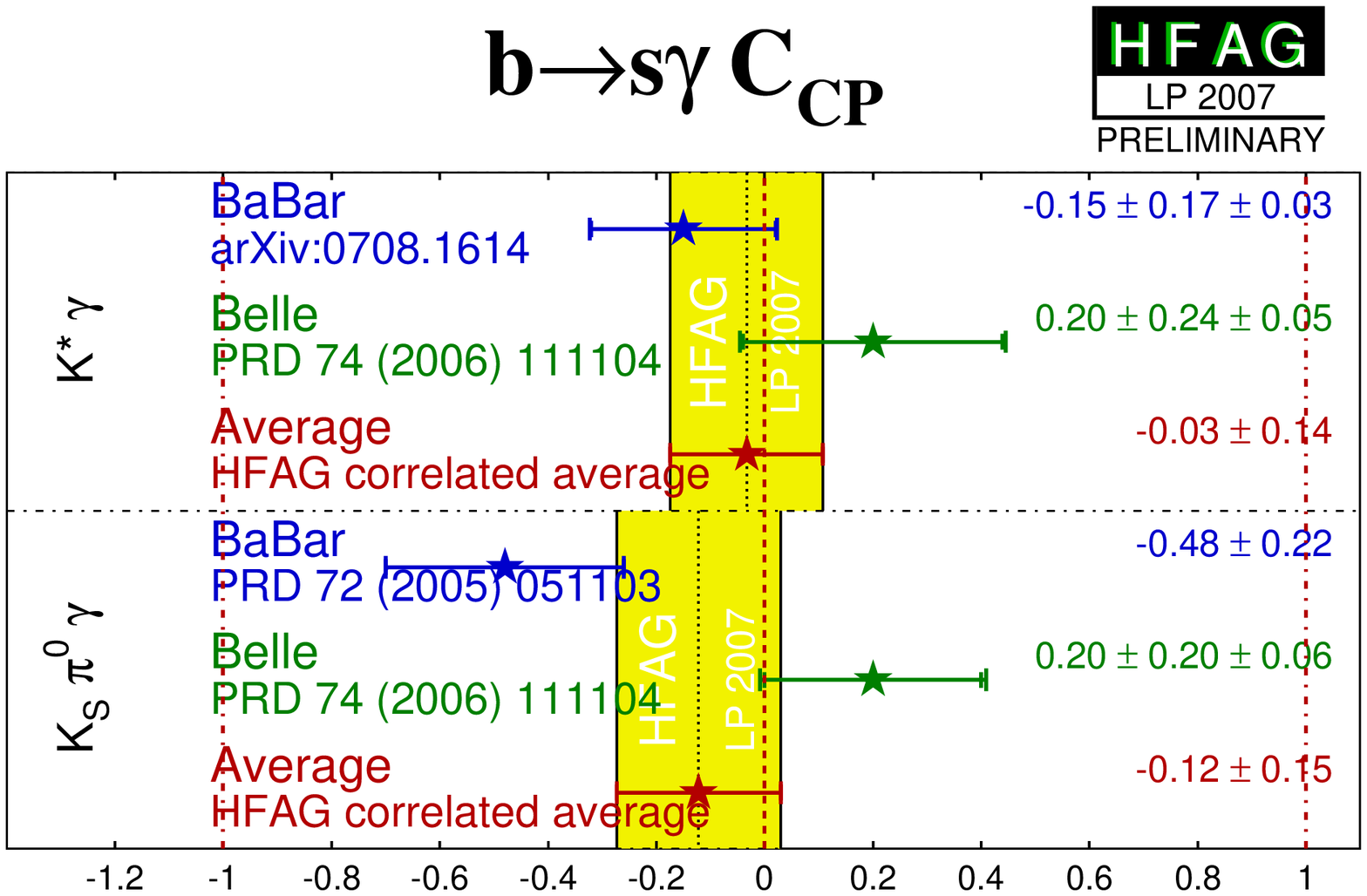}
      }
    \end{tabular}
  \end{center}
  \vspace{-0.8cm}
  \caption{
    Averages of (left) $S_{b \to s \gamma}$ and (right) $C_{b \to s \gamma}$.
    Recall that the data for $K^*\gamma$ is a subset of that for $\KS\pi^0\gamma$.
  }
  \label{fig:cp_uta:bsg}
\end{figure}

\begin{figure}[htb]
  \begin{center}
    \begin{tabular}{cc}
      \resizebox{0.46\textwidth}{!}{
        \includegraphics{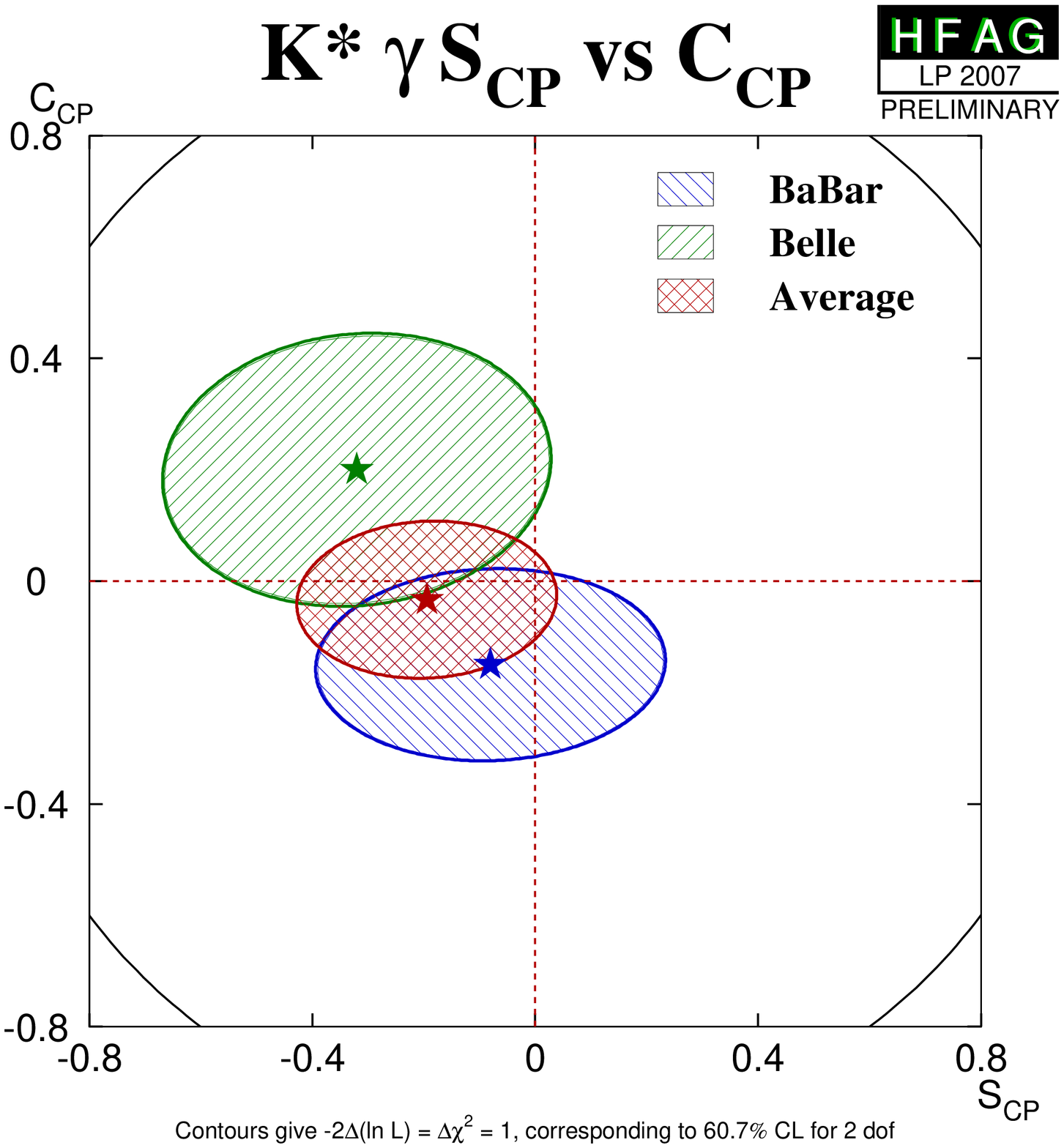}
      }
      &
      \resizebox{0.46\textwidth}{!}{
        \includegraphics{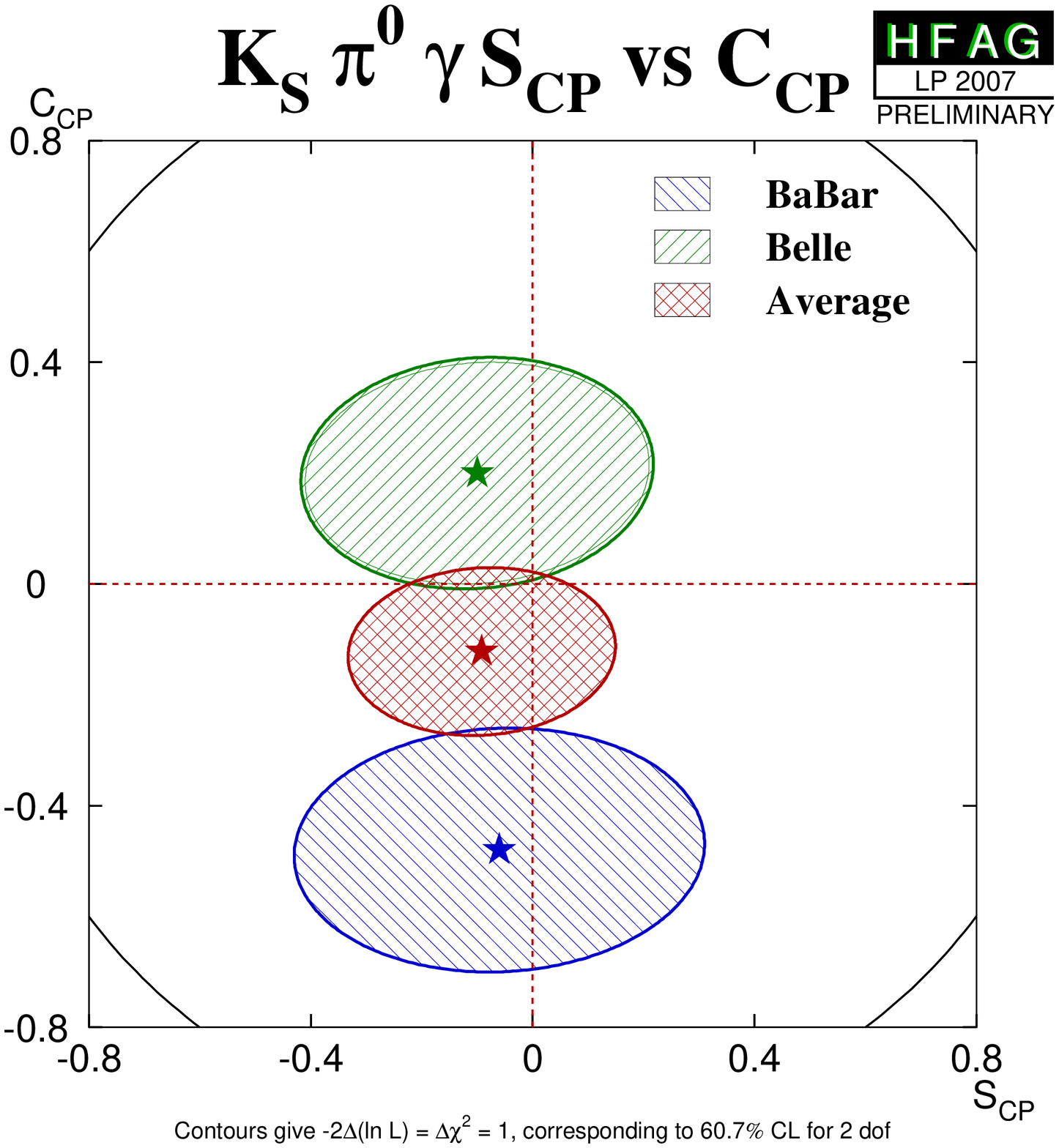}
      }
    \end{tabular}
  \end{center}
  \vspace{-0.8cm}
  \caption{
    Averages of $b \to s\gamma$ dominated channels,
    for which correlated averages are performed,
    in the $S_{\CP}$ \vs\ $C_{\CP}$ plane.
    (Left) $\Bz \to K^*\gamma$ and 
    (right) $\Bz \to \KS\pi^0\gamma$ (including $K^*\gamma$).
  }
  \label{fig:cp_uta:bsg_SvsC}
\end{figure}

The results are shown in Table~\ref{tab:cp_uta:bsg},
and in Figs.~\ref{fig:cp_uta:bsg} and~~\ref{fig:cp_uta:bsg_SvsC}.
No significant $\CP$ violation results are seen;
the results are consistent with the Standard Model
and with other measurements in the $b \to s\gamma$ system (see Sec.~\ref{sec:rare}).

\mysubsection{Time-dependent asymmetries in $b \to d\gamma$ transitions
}
\label{sec:cp_uta:bdg}

The formalism for the radiative decays $b \to d\gamma$ is much the same
as that for $b \to s\gamma$ discussed above.
Assuming dominance of the top quark in the loop,
the weak phase in decay should cancel with that from mixing,
so that the mixing-induced \CP\ violation parameter $S_{\CP}$ 
should be very small.
Corrections due to the finite light quark mass are smaller
compared to $b \to s\gamma$, since $m_d < m_s$,
and although QCD corrections may still play a role,
they cannot significantly affect the prediction $S_{b \to d \gamma} \simeq 0$.
Large direction \CP\ violation effects could, however, be seen through
a non-zero value of $C_{b \to d \gamma}$, 
since the top loop is not the only contribution.

Results using the mode $\Bz \to \rho^0\gamma$ are available from 
\belle\ and are shown in Table~\ref{tab:cp_uta:bdg}.

\begin{table}[htb]
	\begin{center}
		\caption{
			Averages for $\Bz \to \rho^{0} \gamma$.
		}
		\vspace{0.2cm}
		\setlength{\tabcolsep}{0.0pc}
		\begin{tabular*}{\textwidth}{@{\extracolsep{\fill}}lrcccc} \hline
	\mc{2}{l}{Experiment} & $N(B\bar{B})$ & $S_{CP}$ & $C_{CP}$ & Correlation \\
	\hline
	\belle & \cite{:2007jf} & 657M & $-0.83 \pm 0.65 \pm 0.18$ & $0.44 \pm 0.49 \pm 0.14$ & $-0.08$ \\
		\hline
		\end{tabular*}
		\label{tab:cp_uta:bdg}
	\end{center}
\end{table}

\clearpage
\mysubsection{Time-dependent $\CP$ asymmetries in $b \to u\bar{u}d$ transitions
}
\label{sec:cp_uta:uud}

\begin{figure}[htb]
  \begin{center}
    \begin{tabular}{cc}
      \resizebox{0.46\textwidth}{!}{
        \includegraphics{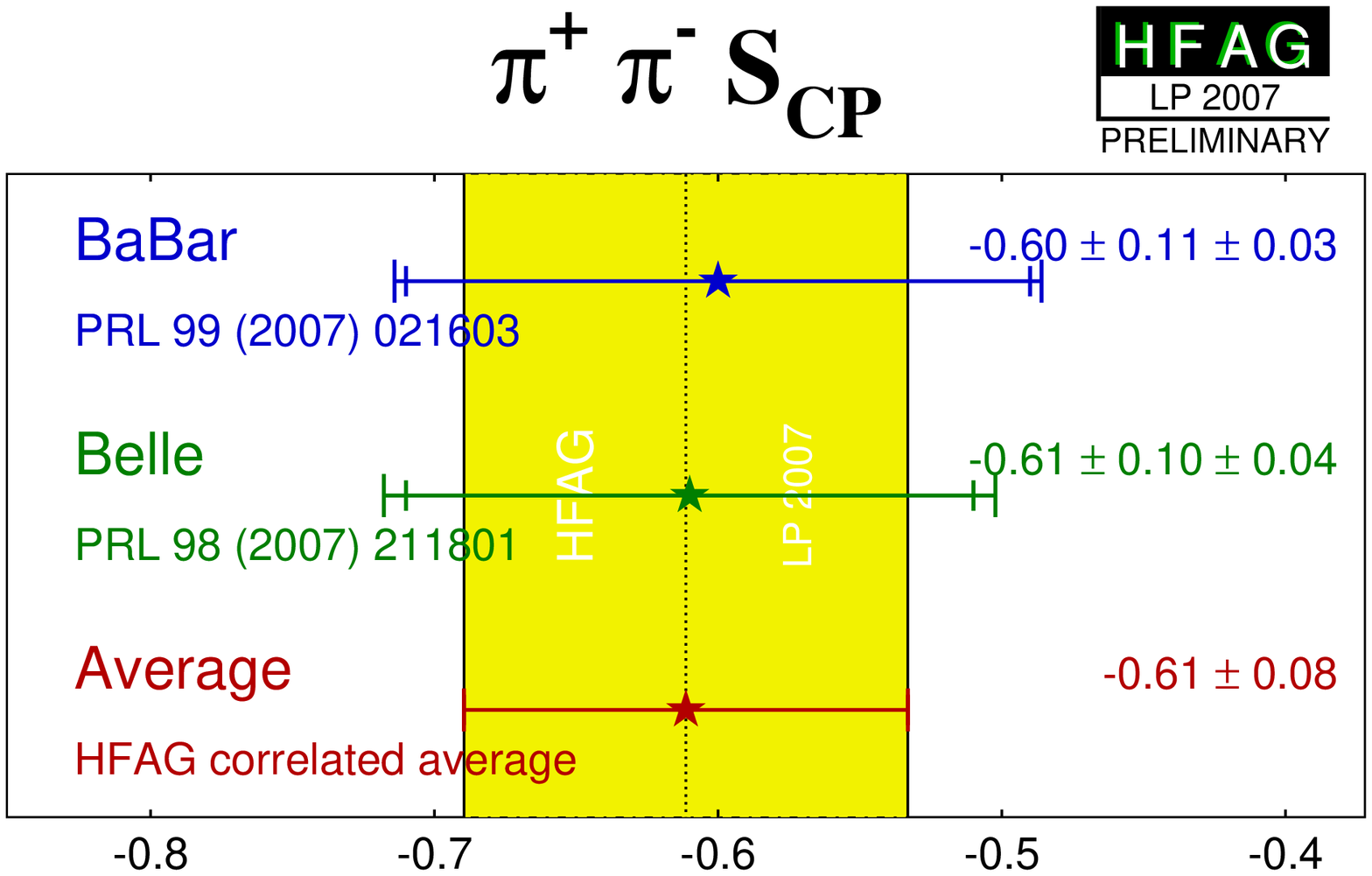}
      }
      &
      \resizebox{0.46\textwidth}{!}{
        \includegraphics{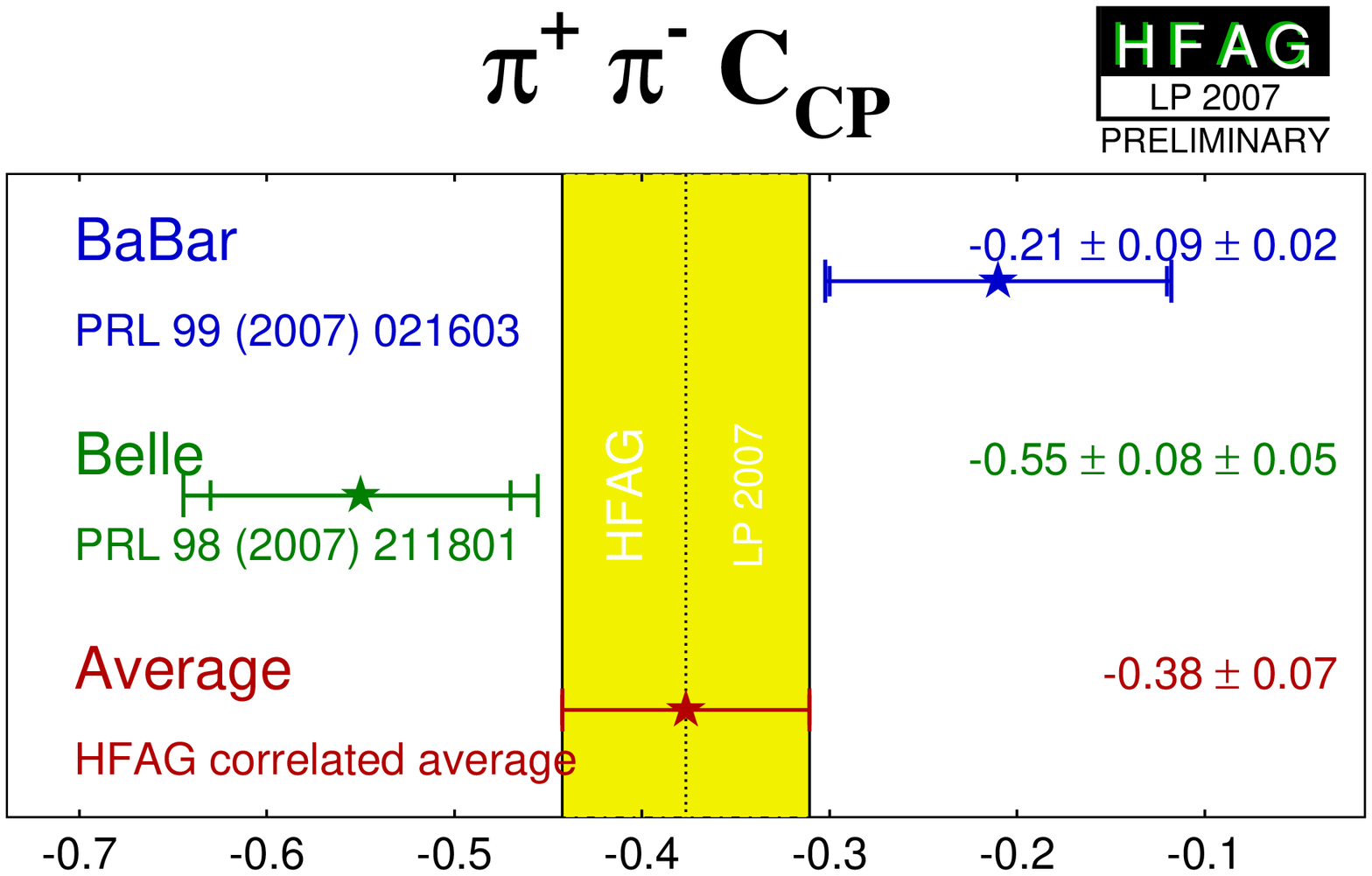}
      }
    \end{tabular}
  \end{center}
  \vspace{-0.8cm}
  \caption{
    Averages of (left) $S_{b \to u\bar u d}$ and (right) $C_{b \to u\bar u d}$
    for the mode $\Bz \to \pi^+\pi^-$.
  }
  \label{fig:cp_uta:uud:pipi}
\end{figure}

\begin{figure}[htb]
  \begin{center}
    \begin{tabular}{cc}
      \resizebox{0.46\textwidth}{!}{
        \includegraphics{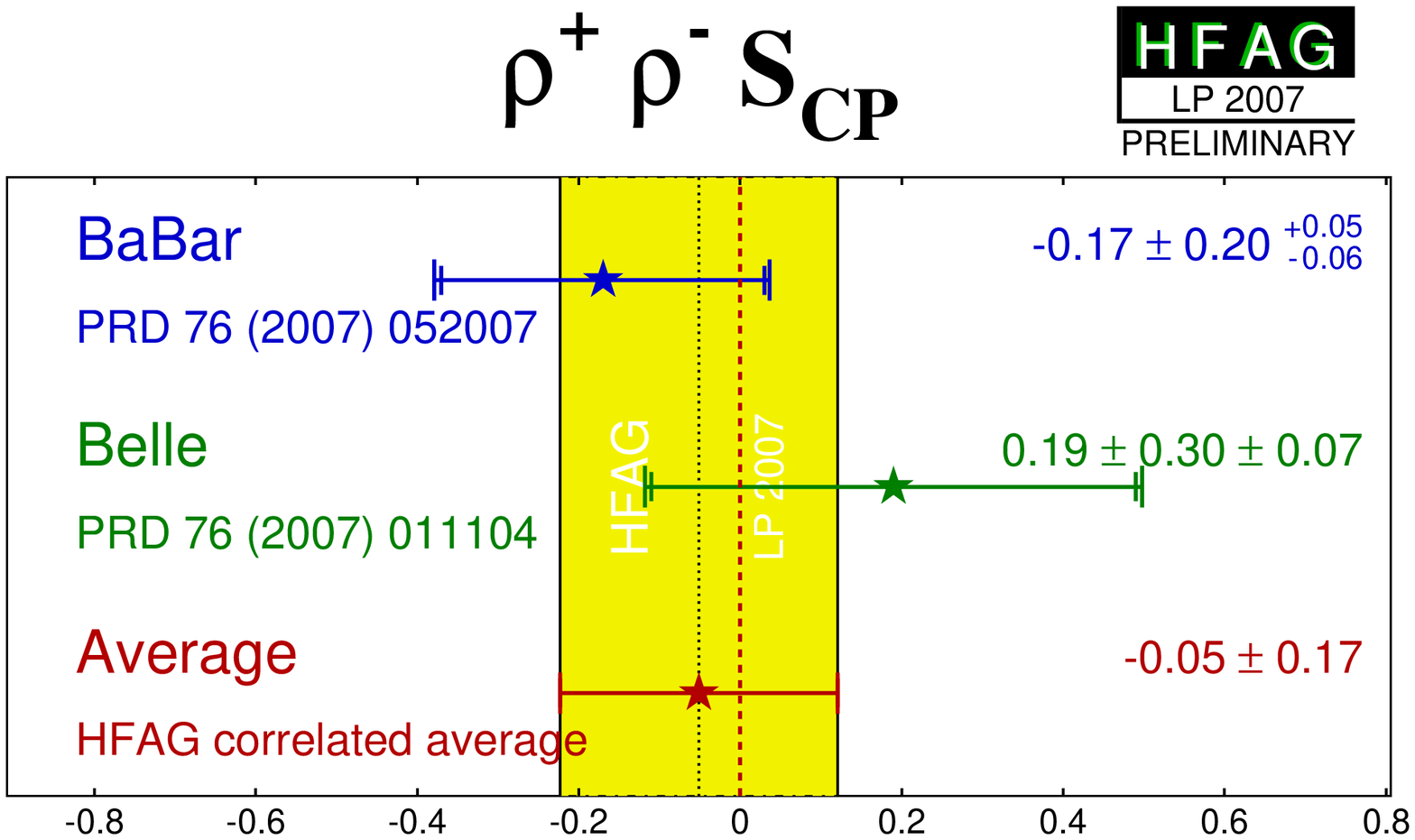}
      }
      &
      \resizebox{0.46\textwidth}{!}{
        \includegraphics{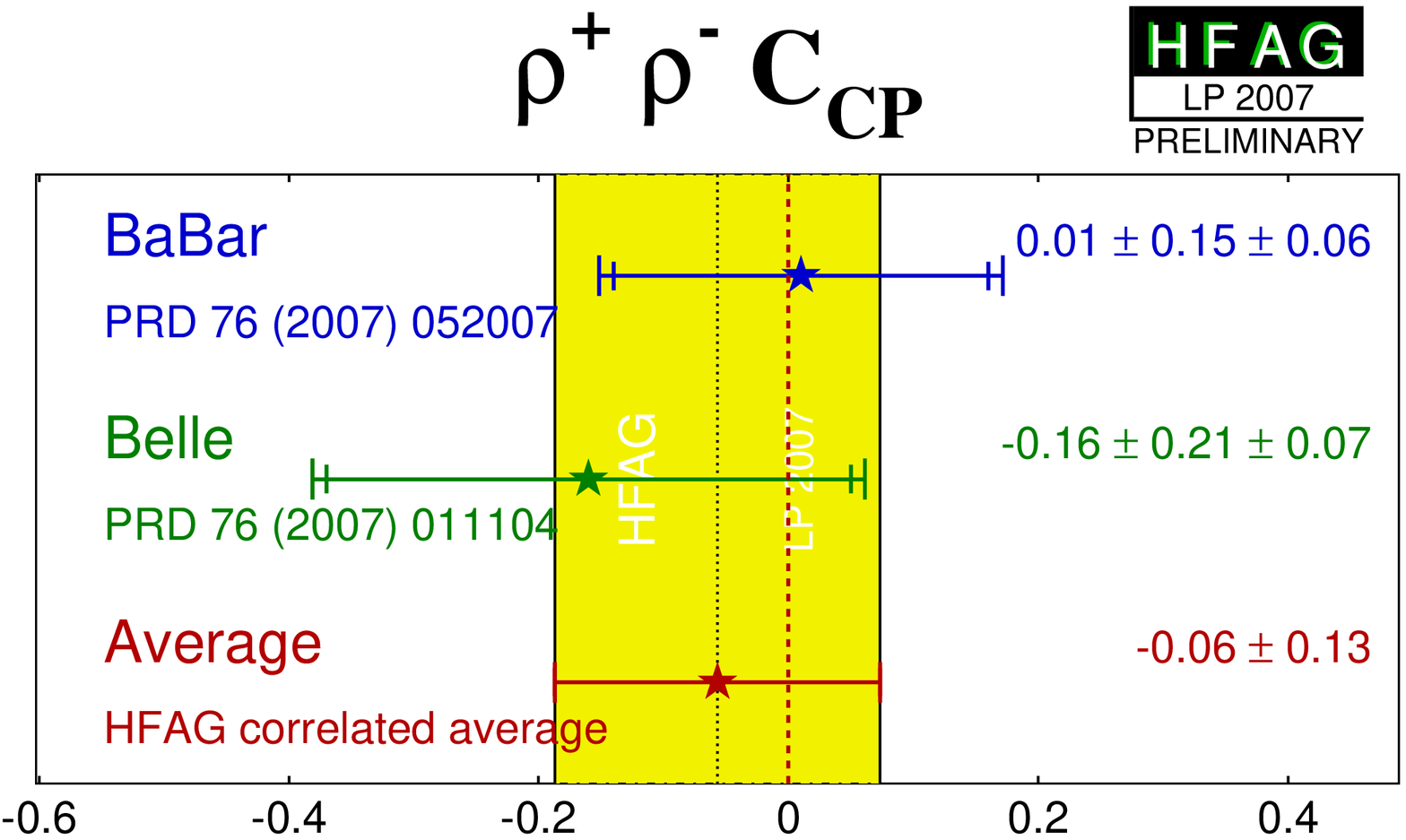}
      }
    \end{tabular}
  \end{center}
  \vspace{-0.8cm}
  \caption{
    Averages of (left) $S_{b \to u\bar u d}$ and (right) $C_{b \to u\bar u d}$
    for the mode $\Bz \to \rho^+\rho^-$.
  }
  \label{fig:cp_uta:uud:rhorho}
\end{figure}

\begin{figure}[htb]
  \begin{center}
    \begin{tabular}{cc}
      \resizebox{0.46\textwidth}{!}{
        \includegraphics{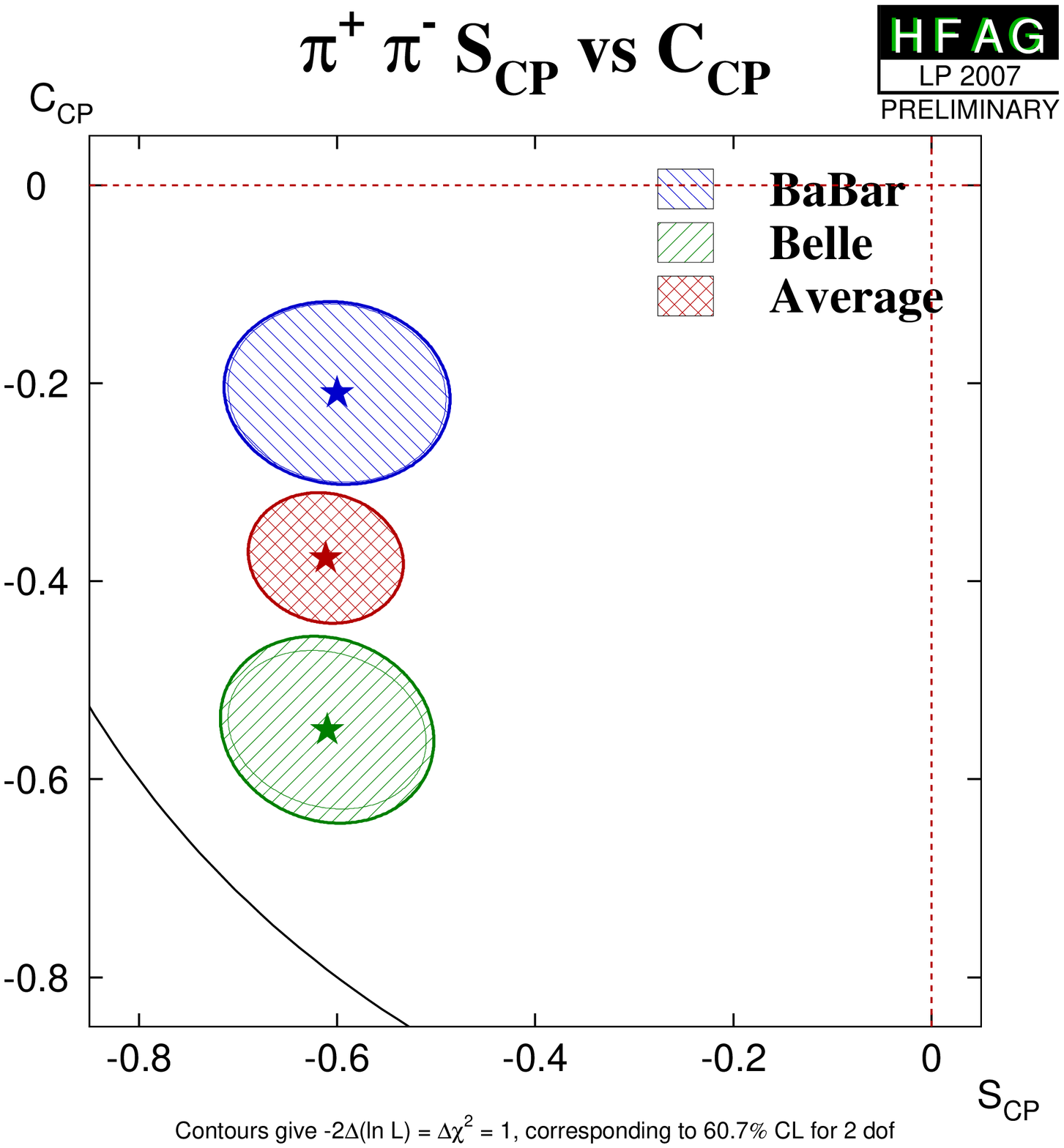}
      }      
      &
      \resizebox{0.46\textwidth}{!}{
        \includegraphics{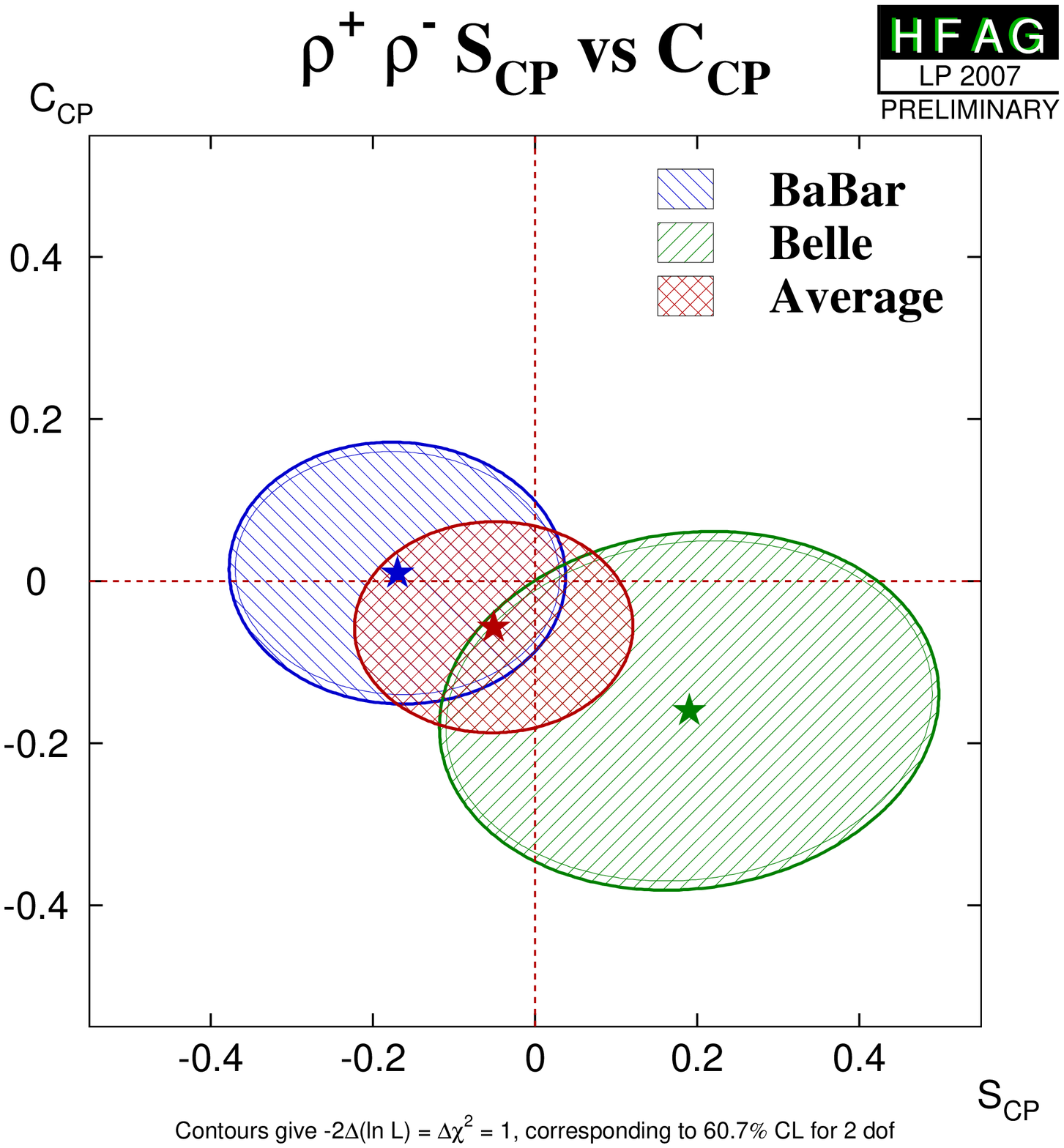}
      }
    \end{tabular}
  \end{center}
  \vspace{-0.8cm}
  \caption{
    Averages of $b \to u\bar u d$ dominated channels,
    for which correlated averages are performed,
    in the $S_{\CP}$ \vs\ $C_{\CP}$ plane.
    (Left) $\Bz \to \pi^+\pi^-$ and (right) $\Bz \to \rho^+\rho^-$.
  }
  \label{fig:cp_uta:uud_SvsC}
\end{figure}

The $b \to u \bar u d$ transition can be mediated by either 
a $b \to u$ tree amplitude or a $b \to d$ penguin amplitude.
These transitions can be investigated using 
the time dependence of $\Bz$ decays to final states containing light mesons.
Results are available from both \babar\ and \belle\ for the 
$\CP$ eigenstate ($\etacp = +1$) $\pi^+\pi^-$ final state
and for the vector-vector final state $\rho^+\rho^-$,
which is found to be dominated by the $\CP$-even
longitudinally polarized component
(\babar\ measure $f_{\rm long} = 
0.992 \pm 0.024 \, ^{+0.026}_{-0.013}$~\cite{Aubert:2007nua}
while \belle\ measure $f_{\rm long} = 
0.941 \, ^{+0.034}_{-0.040} \pm 0.030$~\cite{Somov:2006sg}).
\babar\ have also performed a time-dependent analysis of the 
vector-vector final state $\rho^0\rho^0$~\cite{Aubert:2007qsa},
in which they measure  $f_{\rm long} = 0.70 \pm 0.14 \pm 0.05$,
and furthermore have also performed a time-dependent analysis of the 
$\Bz \to a_1^\pm \pi^\mp$ decay~\cite{Aubert:2006gb}.
These results, and averages, are listed in Table~\ref{tab:cp_uta:uud}.
The averages for $\pi^+\pi^-$ are shown in Fig.~\ref{fig:cp_uta:uud:pipi},
and those for $\rho^+\rho^-$ are shown in Fig.~\ref{fig:cp_uta:uud:rhorho},
with the averages in the $S_{\CP}$ \vs\ $C_{\CP}$ plane 
shown in Fig.~\ref{fig:cp_uta:uud_SvsC}.

\begin{table}[htb]
	\begin{center}
		\caption{
      Averages for $b \to u \bar u d$ modes.
		}
		\vspace{0.2cm}
		\setlength{\tabcolsep}{0.0pc}
\begin{tabular*}{\textwidth}{@{\extracolsep{\fill}}lrcccc} \hline
	\mc{2}{l}{Experiment} & $N(B\bar{B})$ & $S_{CP}$ & $C_{CP}$ & Correlation \\
	\hline
      \mc{6}{c}{$\pi^{+} \pi^{-}$} \\
	\babar & \cite{Aubert:2007mj} & 383M & $-0.60 \pm 0.11 \pm 0.03$ & $-0.21 \pm 0.09 \pm 0.02$ & $-0.07$ \\
	\belle & \cite{Ishino:2006if} & 535M & $-0.61 \pm 0.10 \pm 0.04$ & $-0.55 \pm 0.08 \pm 0.05$ & $-0.15$ \\
	\mc{3}{l}{\bf Average} & $-0.61 \pm 0.08$ & $-0.38 \pm 0.07$ & $-0.09$ \\
	\mc{3}{l}{\small Confidence level} & \mc{2}{c}{\small $0.034~(2.1\sigma)$} & \\
		\hline

      \mc{6}{c}{$\rho^{+} \rho^{-}$} \\
	\babar & \cite{Aubert:2007nua} & 387M & $-0.17 \pm 0.20 \,^{+0.05}_{-0.06}$ & $0.01 \pm 0.15 \pm 0.06$ & $-0.04$ \\
	\belle & \cite{Abe:2007ez} & 535M & $0.19 \pm 0.30 \pm 0.07$ & $-0.16 \pm 0.21 \pm 0.07$ & $0.10$ \\
	\mc{3}{l}{\bf Average} & $-0.05 \pm 0.17$ & $-0.06 \pm 0.13$ & $0.01$ \\
	\mc{3}{l}{\small Confidence level} & \mc{2}{c}{\small $0.50~(0.7\sigma)$} & \\
		\hline

      \mc{6}{c}{$\rho^{0} \rho^{0}$} \\
	\babar & \cite{Aubert:2007qsa} & 427M & $0.5 \pm 0.9 \pm 0.2$ & $0.4 \pm 0.9 \pm 0.2$ & \textendash{} \\
 		\hline
 		\end{tabular*}

                \vspace{2ex}

    \resizebox{\textwidth}{!}{
 		\begin{tabular}{@{\extracolsep{2mm}}lrcccccc} \hline
 		\mc{2}{l}{Experiment} & $N(B\bar{B})$ & $A_{CP}$ & $C$ & $S$ & $\Delta C$ & $\Delta S$ \\
 		\hline
      \mc{8}{c}{$a_1^{\pm} \pi^{\mp}$} \\
	\babar & \cite{Aubert:2006gb} & 384M & $-0.07 \pm 0.07 \pm 0.02$ & $-0.10 \pm 0.15 \pm 0.09$ & $0.37 \pm 0.21 \pm 0.07$ & $0.26 \pm 0.15 \pm 0.07$ & $-0.14 \pm 0.21 \pm 0.06$ \\
 	\hline
		\end{tabular}
              }
		\label{tab:cp_uta:uud}
	\end{center}
\end{table}

If the penguin contribution is negligible, 
the time-dependent parameters for $\Bz \to \pi^+\pi^-$ 
and $\Bz \to \rho^+\rho^-$ are given by
$S_{b \to u\bar u d} = \etacp \sin(2\alpha)$ and
$C_{b \to u\bar u d} = 0$.
In the presence of the penguin contribution, 
direct $\CP$ violation may arise, 
and there is no straightforward interpretation 
of $S_{b \to u\bar u d}$ and $C_{b \to u\bar u d}$.
An isospin analysis~\cite{Gronau:1990ka} 
can be used to disentangle the contributions and extract $\alpha$.

\begin{figure}[htb]
  \begin{center}
    \begin{tabular}{cc}
      \resizebox{0.46\textwidth}{!}{
        \includegraphics{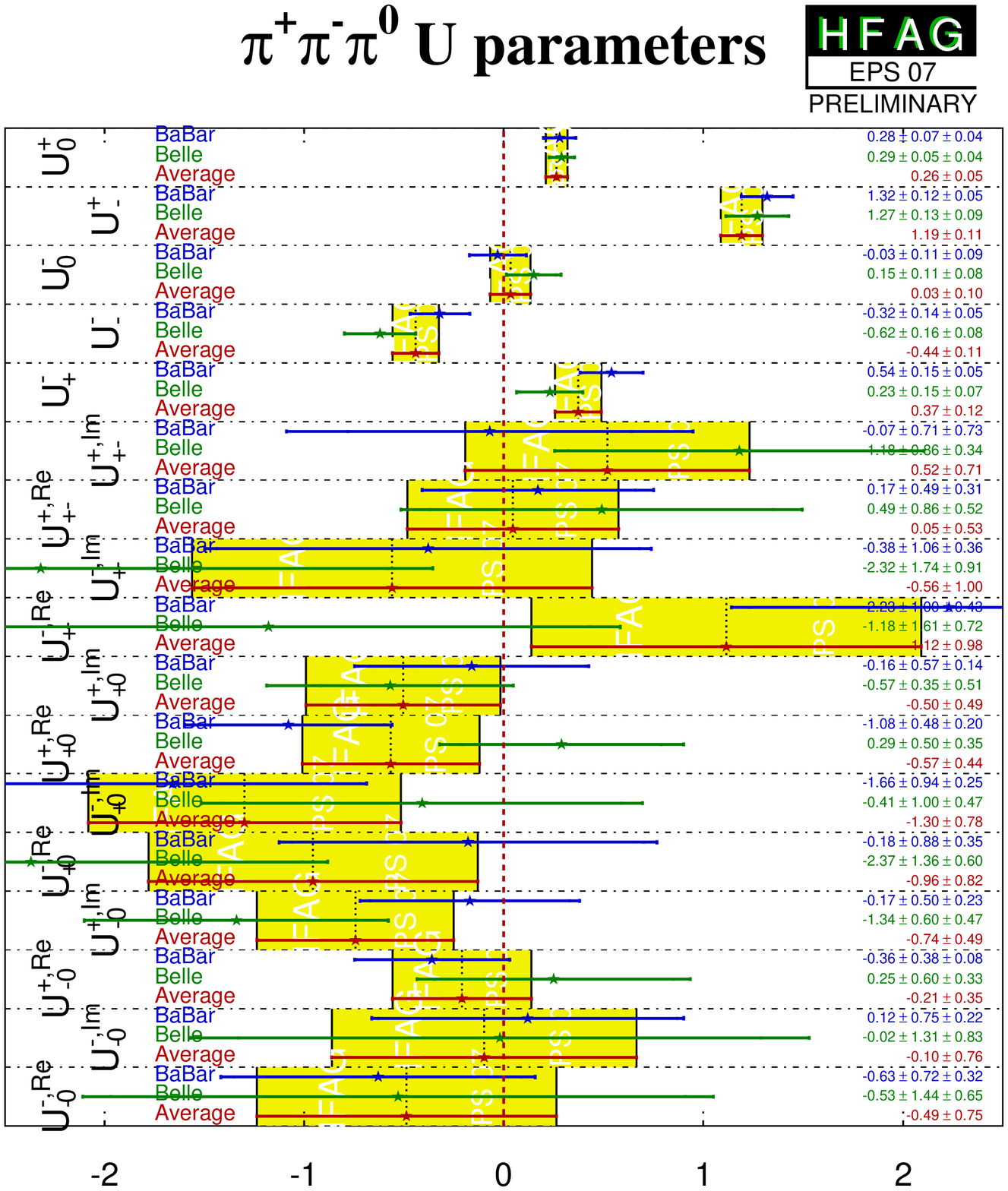}
      }
      &
      \resizebox{0.46\textwidth}{!}{
        \includegraphics{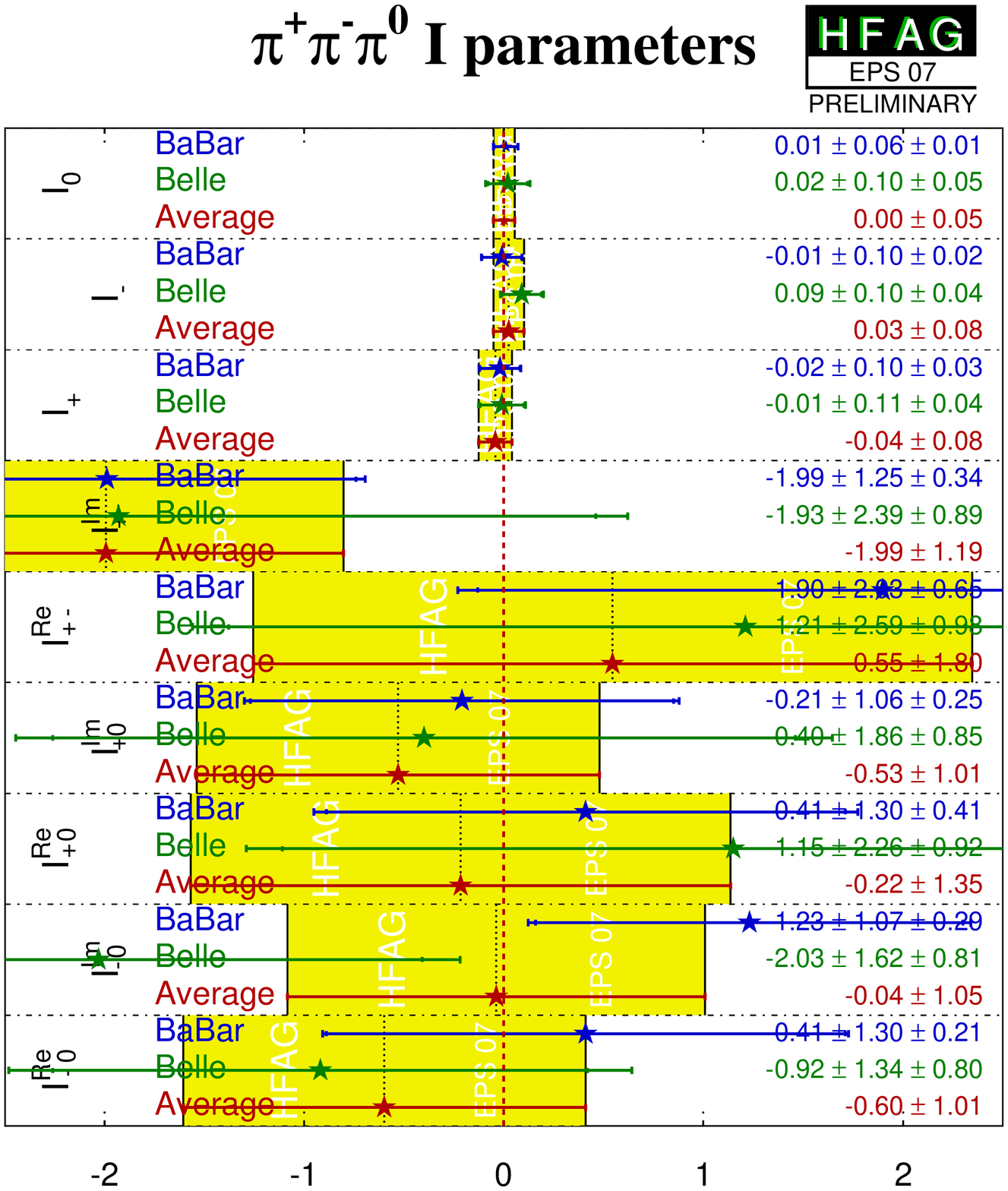}
      }
    \end{tabular}
  \end{center}
  \vspace{-0.8cm}
  \caption{
    Summary of the $U$ and $I$ parameters measured in the 
    time-dependent $\Bz \to \pi^+\pi^-\pi^0$ Dalitz plot analysis.
  }
  \label{fig:cp_uta:uud:uandi}
\end{figure}

For the non-$\CP$ eigenstate $\rho^{\pm}\pi^{\mp}$, 
both \babar~\cite{Aubert:2007jn} 
and \belle~\cite{Kusaka:2007dv,:2007mj} have performed 
time-dependent Dalitz plot (DP) analyses
of the $\pi^+\pi^-\pi^0$ final state~\cite{Snyder:1993mx};
such analyses allow direct measurements of the phases.
Both experiments have measured the $U$ and $I$ parameters discussed in 
Sec.~\ref{sec:cp_uta:notations:dalitz:pipipi0} and defined in 
Table~\ref{tab:cp_uta:pipipi0:uandi}.
We have performed a full correlated average of these parameters,
the results of which are summarized in Fig.~\ref{fig:cp_uta:uud:uandi}.

\begin{figure}[htb]
  \begin{center}
    \begin{tabular}{cc}
      \resizebox{0.46\textwidth}{!}{
        \includegraphics{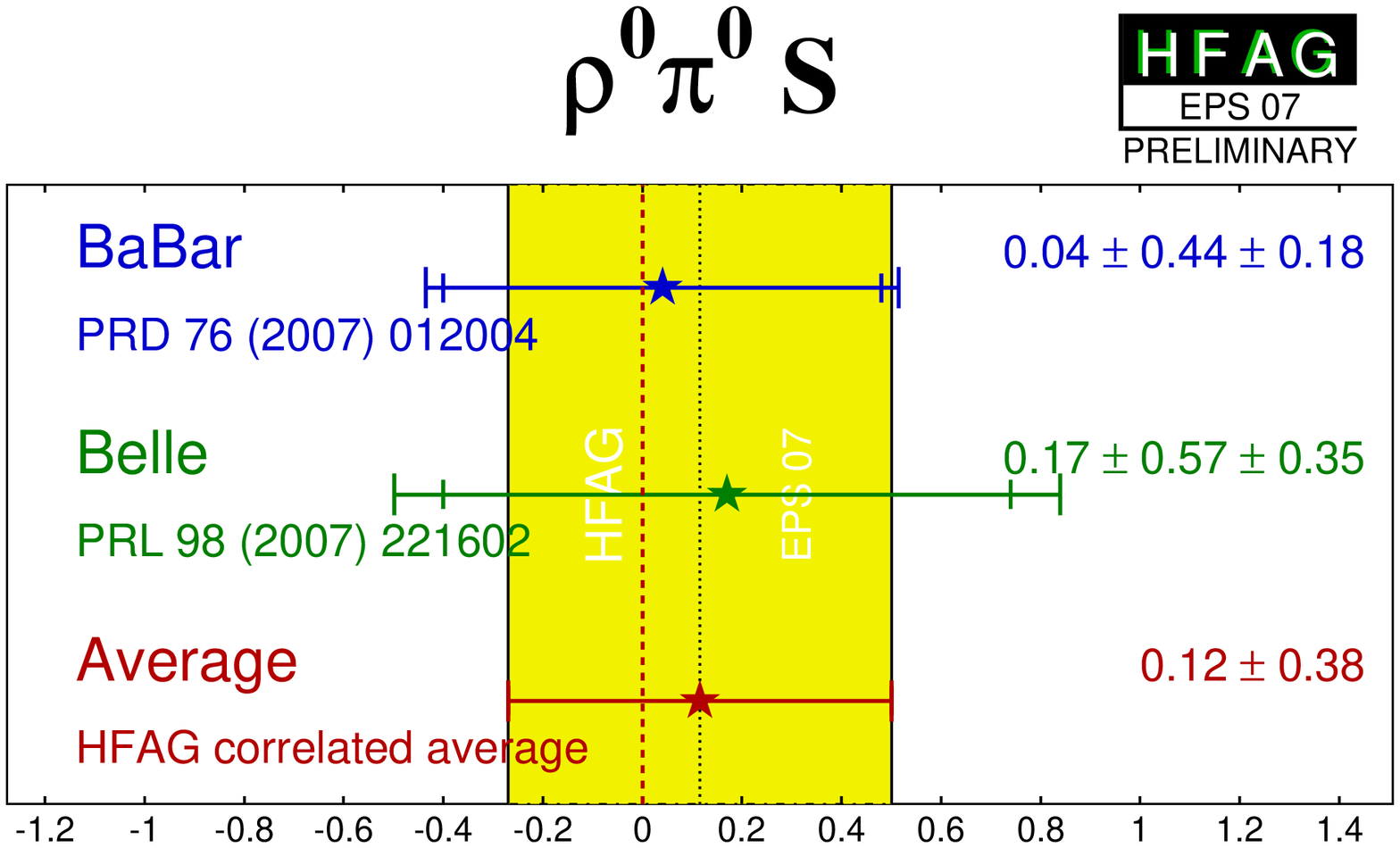}
      }
      &
      \resizebox{0.46\textwidth}{!}{
        \includegraphics{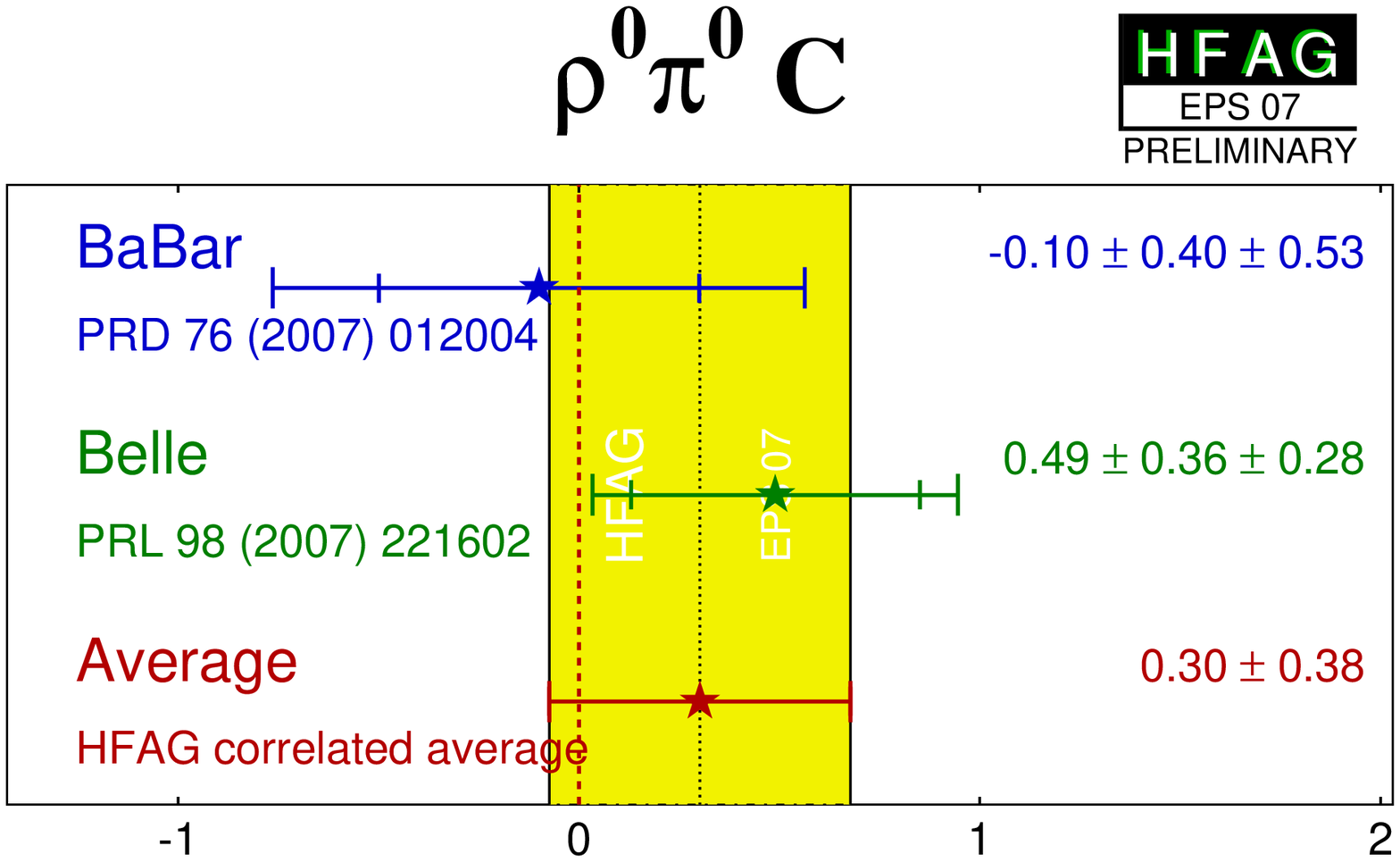}
      }
    \end{tabular}
  \end{center}
  \vspace{-0.8cm}
  \caption{
    Averages of (left) $S_{b \to u\bar u d}$ and (right) $C_{b \to u\bar u d}$
    for the mode $\Bz \to \rho^0\pi^0$.
  }
  \label{fig:cp_uta:uud:rho0pi0}
\end{figure}

\begin{figure}[htb]
  \begin{center}
    \resizebox{0.46\textwidth}{!}{
      \includegraphics{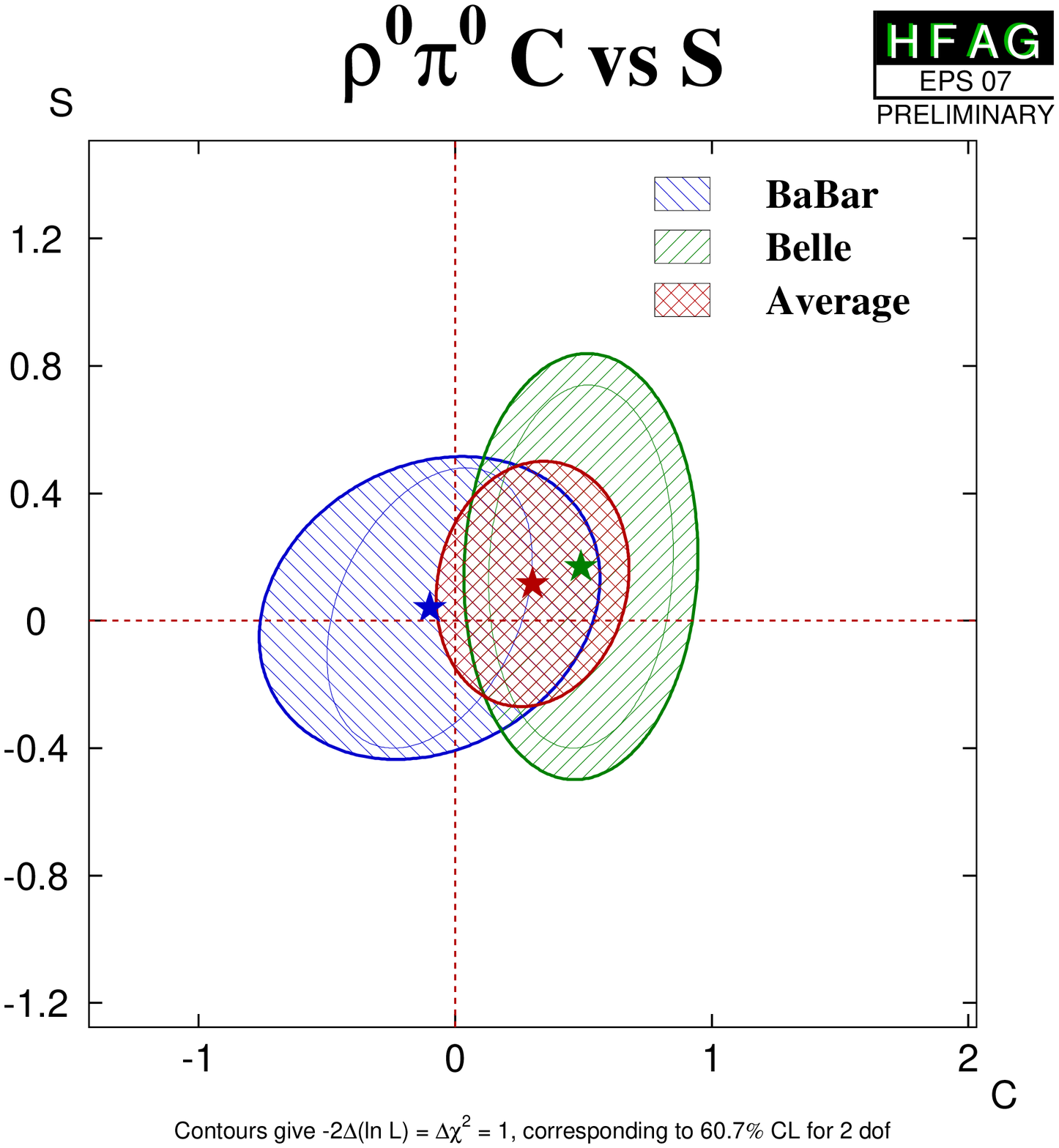}
    }      
  \end{center}
  \vspace{-0.8cm}
  \caption{
    Averages of $b \to u\bar u d$ dominated channels,
    for the mode $\Bz \to \rho^0\pi^0$
    in the $S_{\CP}$ \vs\ $C_{\CP}$ plane.
  }
  \label{fig:cp_uta:uud:rho0pi0_SvsC}
\end{figure}

Both experiments have also extracted the Q2B parameters.
We have performed a full correlated average of these parameters,
which is equivalent to determining the values from the 
averaged $U$ and $I$ parameters.
The results are shown in Table.~\ref{tab:cp_uta:uud:rhopi_q2b}.
Averages of the $\Bz \to \rho^0\pi^0$ Q2B parameters are shown in 
Figs.~\ref{fig:cp_uta:uud:rho0pi0} and~\ref{fig:cp_uta:uud:rho0pi0_SvsC}.

\begin{table}[htb]
	\begin{center}
		\caption{
                  Averages of quasi-two-body parameters extracted
                  from time-dependent Dalitz plot analysis of 
                  $\Bz \to \pi^+\pi^-\pi^0$.
		}
		\vspace{0.2cm}
		\setlength{\tabcolsep}{0.0pc}
    \resizebox{\textwidth}{!}{
\begin{tabular}{@{\extracolsep{2mm}}lrcccccc} \hline
		\mc{2}{l}{Experiment} & $N(B\bar{B})$ & ${\cal A}_{CP}^{\rho\pi}$ & $C_{\rho\pi}$ & $S_{\rho\pi}$ & $\Delta C_{\rho\pi}$ & $\Delta S_{\rho\pi}$ \\
	\hline
	\babar & \cite{Aubert:2007jn} & 375M & $-0.14 \pm 0.05 \pm 0.02$ & $0.15 \pm 0.09 \pm 0.05$ & $-0.03 \pm 0.11 \pm 0.04$ & $0.39 \pm 0.09 \pm 0.09$ & $-0.01 \pm 0.14 \pm 0.06$ \\
	\belle & \cite{Kusaka:2007dv,:2007mj} & 449M & $-0.12 \pm 0.05 \pm 0.04$ & $-0.13 \pm 0.09 \pm 0.05$ & $0.06 \pm 0.13 \pm 0.05$ & $0.36 \pm 0.10 \pm 0.05$ & $-0.08 \pm 0.13 \pm 0.05$ \\
	\mc{3}{l}{\bf Average} & $-0.13 \pm 0.04$ & $0.01 \pm 0.07$ & $0.01 \pm 0.09$ & $0.37 \pm 0.08$ & $-0.04 \pm 0.10$ \\
	\mc{3}{l}{\small Confidence level} & \mc{5}{c}{\small $0.52~(0.6\sigma)$} \\
        \hline
		\end{tabular}
              }

                \vspace{2ex}

		\begin{tabular*}{\textwidth}{@{\extracolsep{\fill}}lrcccc} \hline
		\mc{2}{l}{Experiment} & $N(B\bar{B})$ & ${\cal A}^{-+}_{\rho\pi}$ & ${\cal A}^{+-}_{\rho\pi}$ & Correlation \\
		\hline
	\babar & \cite{Aubert:2007jn} & 375M & $-0.37 \,^{+0.16}_{-0.10} \pm 0.09$ & $0.03 \pm 0.07 \pm 0.04$ & $0.62$ \\
	\belle & \cite{Kusaka:2007dv,:2007mj} & 449M & $0.08 \pm 0.16 \pm 0.11$ & $0.21 \pm 0.08 \pm 0.04$ & $0.47$ \\
	\mc{3}{l}{\bf Average} & $-0.18 \pm 0.12$ & $0.11 \pm 0.06$ & $0.40$ \\
        \mc{3}{l}{\small Confidence level} & \mc{2}{c}{\small $0.14~(1.5\sigma)$} \\
		\hline
		\end{tabular*}

                \vspace{2ex}

		\begin{tabular*}{\textwidth}{@{\extracolsep{\fill}}lrcccc} \hline
		\mc{2}{l}{Experiment} & $N(B\bar{B})$ & $C_{\rho^0\pi^0}$ & $S_{\rho^0\pi^0}$ & Correlation \\
		\hline
	\babar & \cite{Aubert:2007jn} & 375M & $-0.10 \pm 0.40 \pm 0.53$ & $0.04 \pm 0.44 \pm 0.18$ & $0.35$ \\
	\belle & \cite{Kusaka:2007dv,:2007mj} & 449M & $0.49 \pm 0.36 \pm 0.28$ & $0.17 \pm 0.57 \pm 0.35$ & $0.08$ \\
	\mc{3}{l}{\bf Average} & $0.30 \pm 0.38$ & $0.12 \pm 0.38$ & $0.12$ \\
	\mc{3}{l}{\small Confidence level} & \mc{2}{c}{\small $0.76~(0.3\sigma)$} \\
		\hline
		\end{tabular*}
		\label{tab:cp_uta:uud:rhopi_q2b}
	\end{center}
\end{table}

\begin{figure}[htb]
  \begin{center}
    \resizebox{0.46\textwidth}{!}{
      \includegraphics{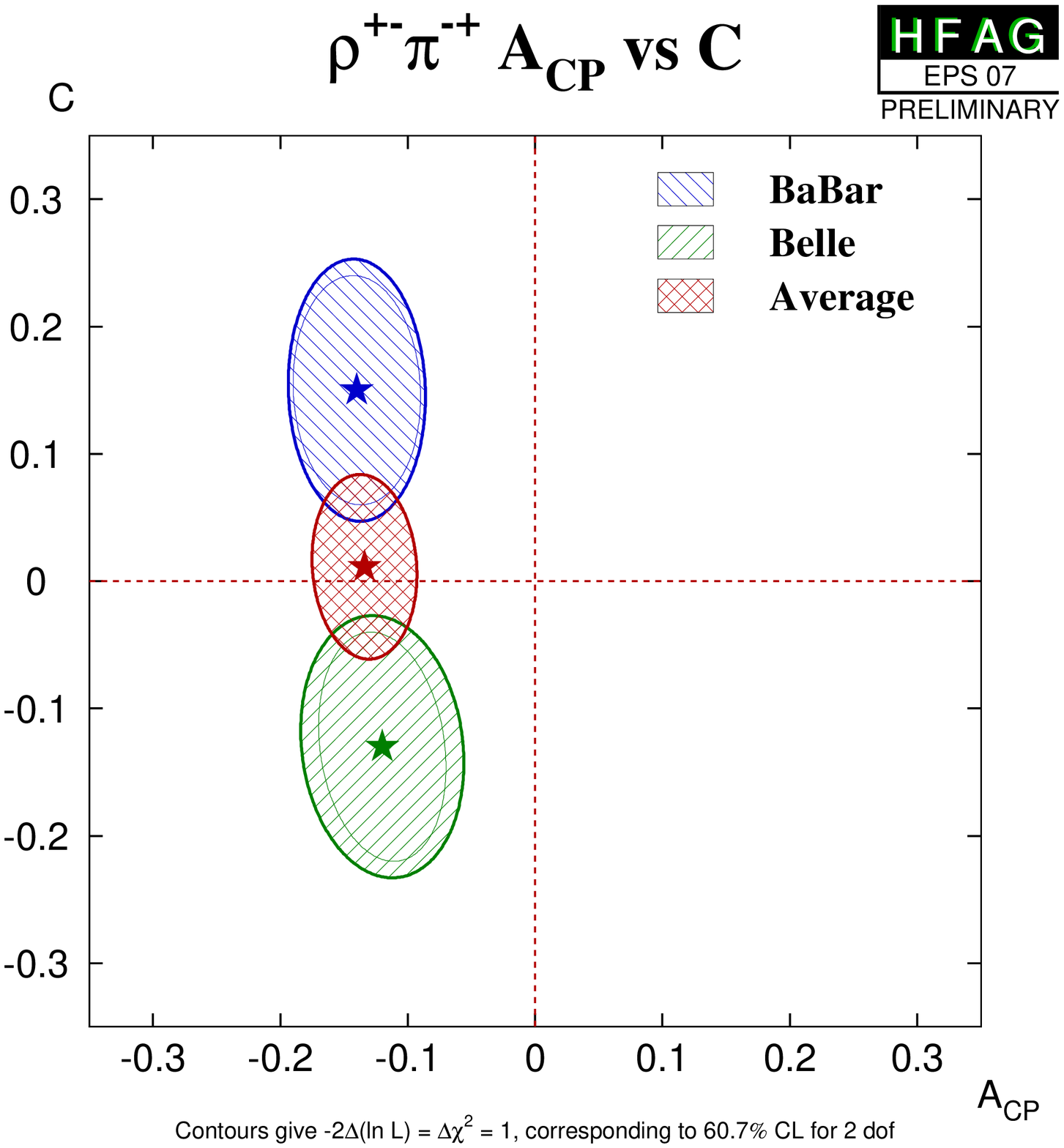}
    }
    \hfill
    \resizebox{0.46\textwidth}{!}{
      \includegraphics{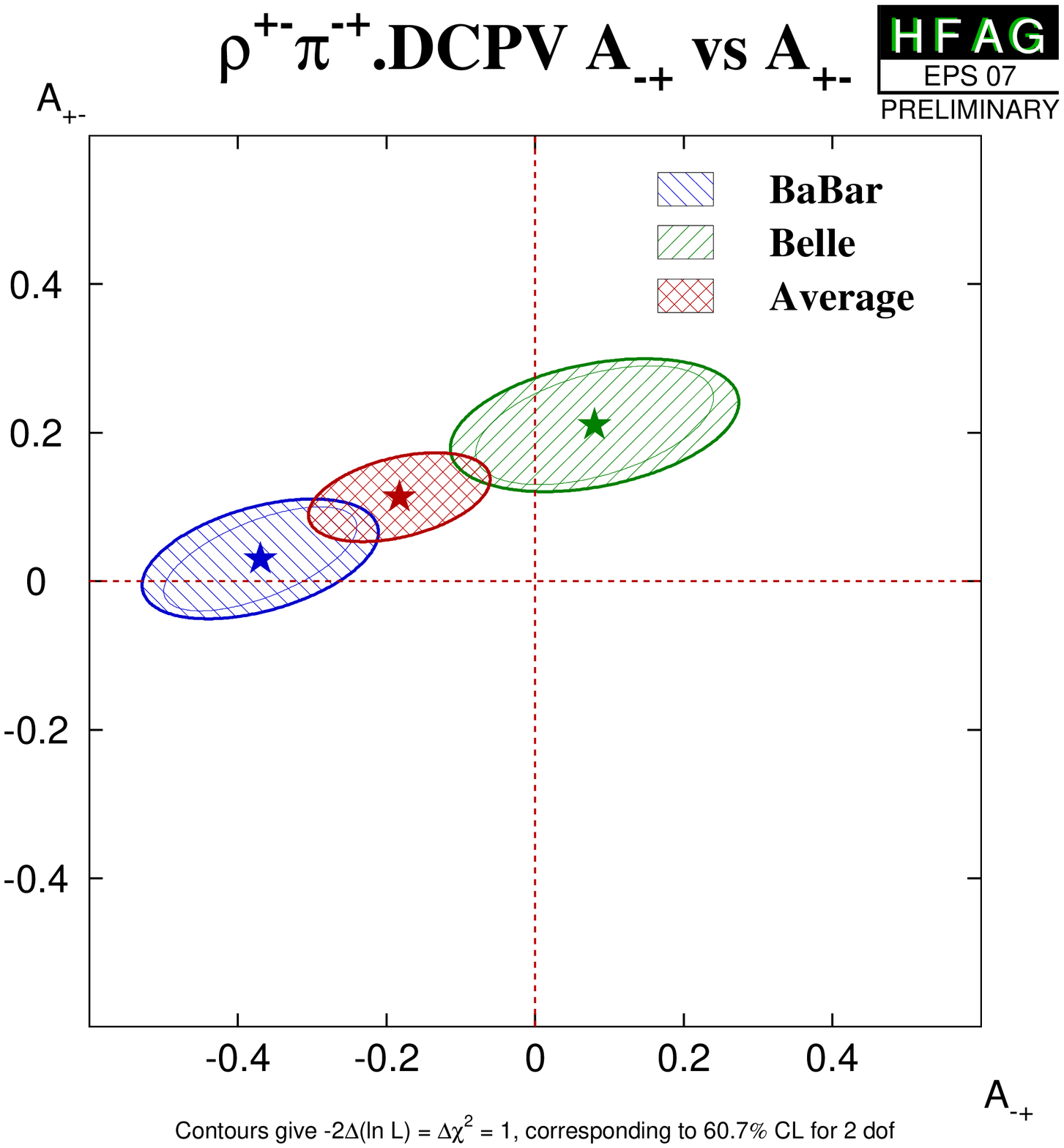}
    }
  \end{center}
  \vspace{-0.8cm}
  \caption{
    Direct $\CP$ violation in $\Bz\to\rho^\pm\pi^\mp$.
    (Left) ${\cal A}^{\rho\pi}_{\CP}$ \vs\ $C_{\rho\pi}$ space,
    (right) ${\cal A}^{-+}_{\rho\pi}$ \vs\ ${\cal A}^{+-}_{\rho\pi}$ space.
  }
  \label{fig:cp_uta:uud:rhopi-dircp}
\end{figure}

With the notation described in Sec.~\ref{sec:cp_uta:notations}
(Eq.~(\ref{eq:cp_uta:non-cp-s_and_deltas})), 
the time-dependent parameters for the Q2B $\Bz \to \rho^\pm\pi^\mp$ analysis are,
neglecting penguin contributions, given by
\begin{equation}
  S_{\rho\pi} = 
  \sqrt{1 - \left(\frac{\Delta C}{2}\right)^2}\sin(2\alpha)\cos(\delta)
  \ , \ \ \ 
  \Delta S_{\rho\pi} = 
  \sqrt{1 - \left(\frac{\Delta C}{2}\right)^2}\cos(2\alpha)\sin(\delta)
\end{equation} 
and $C_{\rho\pi} = {\cal A}_{\CP}^{\rho\pi} = 0$,
where $\delta=\arg(A_{-+}A^*_{+-})$ is the strong phase difference 
between the $\rho^-\pi^+$ and $\rho^+\pi^-$ decay amplitudes.
In the presence of the penguin contribution, there is no straightforward 
interpretation of the Q2B observables in the $\Bz \to \rho^\pm\pi^\mp$ system
in terms of CKM parameters.
However direct $\CP$ violation may arise,
resulting in either or both of $C_{\rho\pi} \neq 0$ and ${\cal A}_{\CP}^{\rho\pi} \neq 0$.
Equivalently,
direct $\CP$ violation may be seen by either of
the decay-type-specific observables ${\cal A}^{+-}_{\rho\pi}$ 
and ${\cal A}^{-+}_{\rho\pi}$, defined in Eq.~(\ref{eq:cp_uta:non-cp-directcp}), 
deviating from zero.
Results and averages for these parameters
are also given in Table~\ref{tab:cp_uta:uud:rhopi_q2b}.
Averages of the direct $\CP$ violation effect in $\Bz \to \rho^\pm\pi^\mp$
are shown in Fig.~\ref{fig:cp_uta:uud:rhopi-dircp},
both in 
${\cal A}^{\rho\pi}_{\CP}$ \vs\ $C_{\rho\pi}$ space and in 
${\cal A}^{-+}_{\rho\pi}$ \vs\ ${\cal A}^{+-}_{\rho\pi}$ space.

Some difference is seen between the 
\babar\ and \belle\ measurements in the $\pi^+\pi^-$ system.
The confidence level of the average is $0.034$,
which corresponds to a $2.1\sigma$ discrepancy.  Since there is no
evidence of systematic problems in either analysis,
we do not rescale the errors of the averages.
The averages for $S_{b \to u\bar u d}$ and $C_{b \to u\bar u d}$ 
in $\Bz \to \pi^+\pi^-$ are both more than $5\sigma$ away from zero,
suggesting that both mixing-induced and direct $\CP$ violation 
are well-established in this channel.
Nonetheless, due to the possible discrepancy mentioned above,
a slightly cautious interpretation should be made 
with regard to the significance of direct $\CP$ violation.

In $\Bz \to \rho^\pm\pi^\mp$, however,
both experiments see an indication of direct $\CP$ violation in the 
${\cal A}^{\rho\pi}_{\CP}$ parameter 
(as seen in Fig.~\ref{fig:cp_uta:uud:rhopi-dircp}).
The average is more than $3\sigma$ from zero,
providing evidence of direct $\CP$ violation in this channel.

\vspace{3ex}

\noindent
\underline{\large Constraints on $\alpha$}

The precision of the measured $\CP$ violation parameters in
$b \to u\bar{u}d$ transitions allows 
constraints to be set on the UT angle $\alpha$. 
Constraints have been obtained with various methods:
\begin{itemize}\setlength{\itemsep}{0.5ex}
\item 
  Both \babar~\cite{Aubert:2007hh}
  and  \belle~\cite{Ishino:2006if} have performed 
  isospin analyses in the $\pi\pi$ system.
  \belle\ exclude $9^\circ < \phi_2 < 81^\circ$ at the $95.4\%$  C.L. while
  \babar\ give a confidence level interpretation for $\alpha$,
  exclude the range $25^\circ < \alpha < 66^\circ$ at the $90\%$  C.L.,
  and find the solution consistent with the Standard Model to be 
  $\alpha = (96 \,^{+10}_{-6})^\circ$.
  In both cases, only solutions in $0^\circ$--$180^\circ$ are considered.

\item
  Both experiments have also performed isospin analyses in the $\rho\rho$ system.
  \babar~\cite{Aubert:2007nua} find
  $\alpha \in \left[73.1, 117.0\right]^\circ$ at $68\%$ C.L.
  while \belle~\cite{Abe:2007ez} obtain $54^\circ < \phi_2 < 113^\circ$ 
  at $90\%$ confidence level. 
  The largest contribution to the uncertainty is due to the 
  possible penguin contribution, limited by the knowledge of the 
  $\Bz \to \rho^0\rho^0$ branching fraction
  and is correlated between the measurements.  
  In the \babar~\cite{Aubert:2007qsa} study of $\Bz \to \rho^0\rho^0$,
  a constraint of $|\alpha - \alpha^{\rm eff}| < 14.5^\circ \ (16.5^\circ)$
  is obtained at 68\% (90\%) CL.
  The solution at $\alpha - \alpha^{\rm eff} = +11.3^\circ$ is preferred.

\item
  The time-dependent Dalitz plot analysis of the $\Bz \to \pi^+\pi^-\pi^0$
  decay allows a determination of $\alpha$ without input from any other 
  channels.
  \babar~\cite{Aubert:2007jn} obtain the constraint 
  $75^\circ < \alpha < 152^\circ$ at $68\%$ C.L.
  \belle~\cite{Kusaka:2007dv,:2007mj} have performed a similar analysis,
  and in addition have included information from the SU(2) partners of 
  $B \to \rho\pi$, which can be used to constrain $\alpha$
  via an isospin pentagon relation~\cite{Lipkin:1991st}. 
  With this analysis,
  \belle\ obtain the tighter constraint $\phi_2 = (83 \, ^{+12}_{-23})^\circ$
  (where the errors correspond to $1\sigma$, \ie\ $68.3\%$ confidence level).

\item 
  Each experiment has obtained a value of $\alpha$ from combining its 
  results in the different $b \to u \bar{u} d$ modes 
  (with some input also from HFAG).
  These values have appeared in talks, but not in publications,
  and are not listed here.

\item 
  The CKMfitter~\cite{Charles:2004jd} and 
  UTFit~\cite{Bona:2005vz} groups use the measurements 
  from \belle\ and \babar\ given above
  with other branching fractions and \CP asymmetries in 
  $\B\to\pi\pi$, $\rho\pi$ and $\rho\rho$ modes, 
  to perform isospin analyses for each system, 
  and to make combined constraints on $\alpha$.
\end{itemize}

Note that methods based on isospin symmetry make extensive use of 
measurements of branching fractions and direct $\CP$ asymmetries,
as averaged by the HFAG Rare Decays subgroup (Sec.~\ref{sec:rare}).
Note also that each method suffers from discrete ambiguities in the solutions.
The model assumption in the $\Bz \to \pi^+\pi^-\pi^0$ analysis 
allows to resolve some of the multiple solutions, 
and results in a single preferred value for $\alpha$ in $\left[ 0, \pi \right]$.
All the above measurements correspond to the choice
that is in agreement with the global CKM fit.

At present we make no attempt to provide an HFAG average for $\alpha$.
More details on procedures to calculate a best fit value for $\alpha$ 
can be found in Refs.~\cite{Charles:2004jd,Bona:2005vz}.

\clearpage
\mysubsection{Time-dependent $\CP$ asymmetries in $b \to c\bar{u}d / u\bar{c}d$ transitions
}
\label{sec:cp_uta:cud}

Non-$\CP$ eigenstates such as $D^\pm\pi^\mp$, $D^{*\pm}\pi^\mp$ and $D^\pm\rho^\mp$ can be produced 
in decays of $\Bz$ mesons either via Cabibbo favoured ($b \to c$) or
doubly Cabibbo suppressed ($b \to u$) tree amplitudes. 
Since no penguin contribution is possible,
these modes are theoretically clean.
The ratio of the magnitudes of the suppressed and favoured amplitudes, $R$,
is sufficiently small (predicted to be about $0.02$),
that terms of ${\cal O}(R^2)$ can be neglected, 
and the sine terms give sensitivity to the combination of UT angles $2\beta+\gamma$.

\begin{figure}[htb]
  \begin{center}
    \begin{tabular}{cc}
      \resizebox{0.46\textwidth}{!}{
        \includegraphics{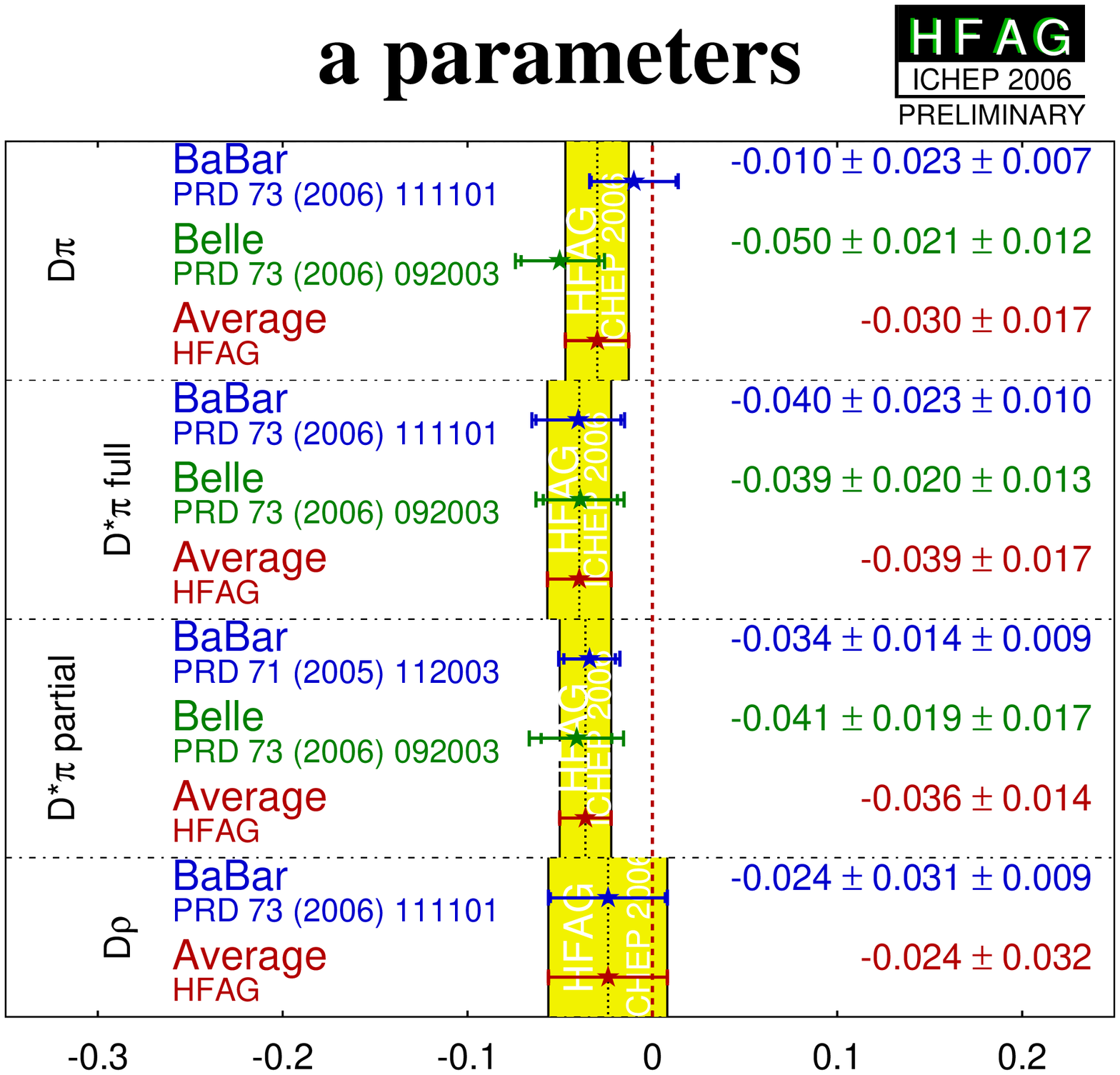}
      }
      &
      \resizebox{0.46\textwidth}{!}{
        \includegraphics{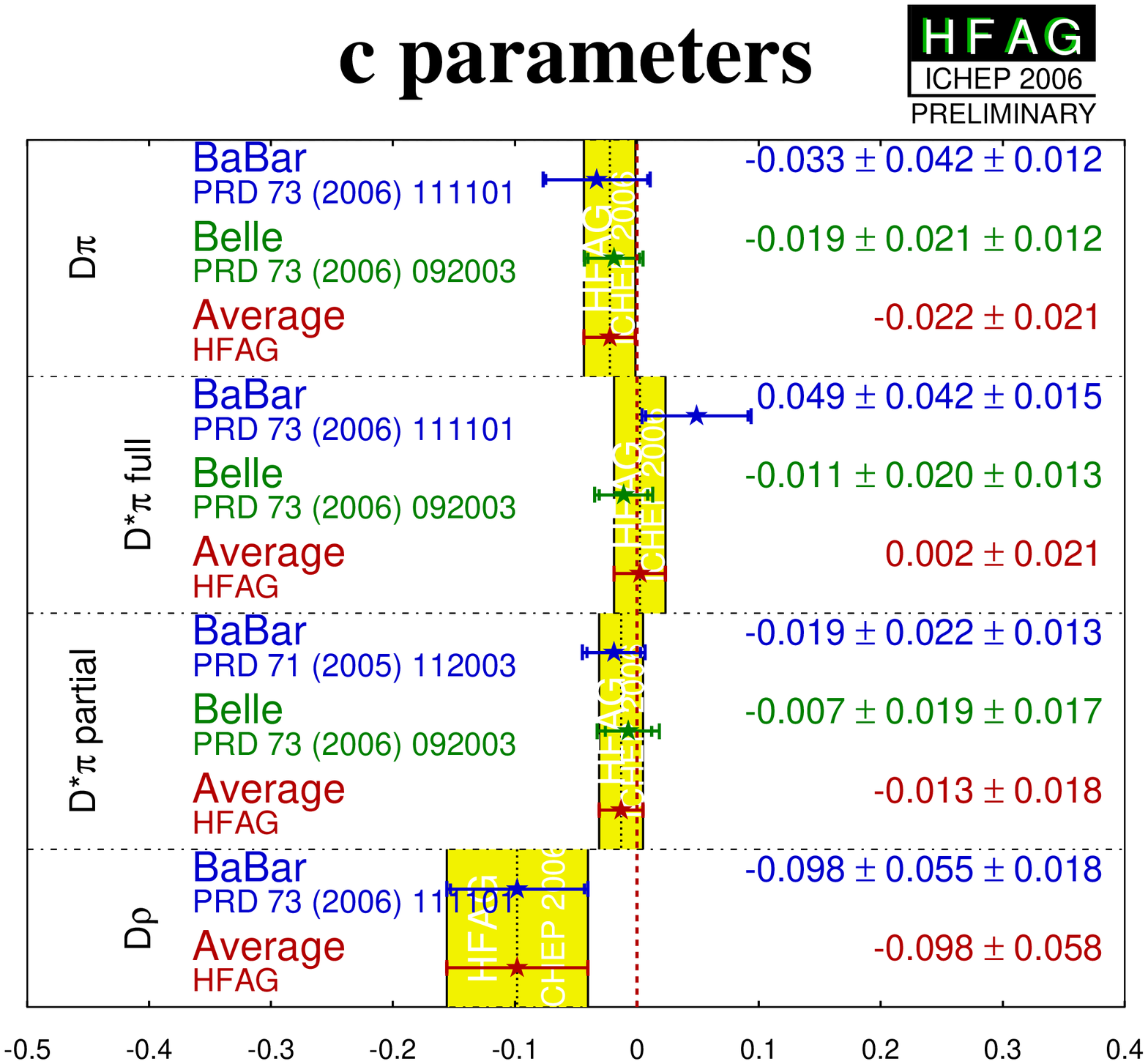}
      }
    \end{tabular}
  \end{center}
  \vspace{-0.8cm}
  \caption{
    Averages for $b \to c\bar{u}d / u\bar{c}d$ modes.
  }
  \label{fig:cp_uta:cud}
\end{figure}

\begin{figure}[htb]
  \begin{center}
    \begin{tabular}{cc}
      \resizebox{0.46\textwidth}{!}{
        \includegraphics{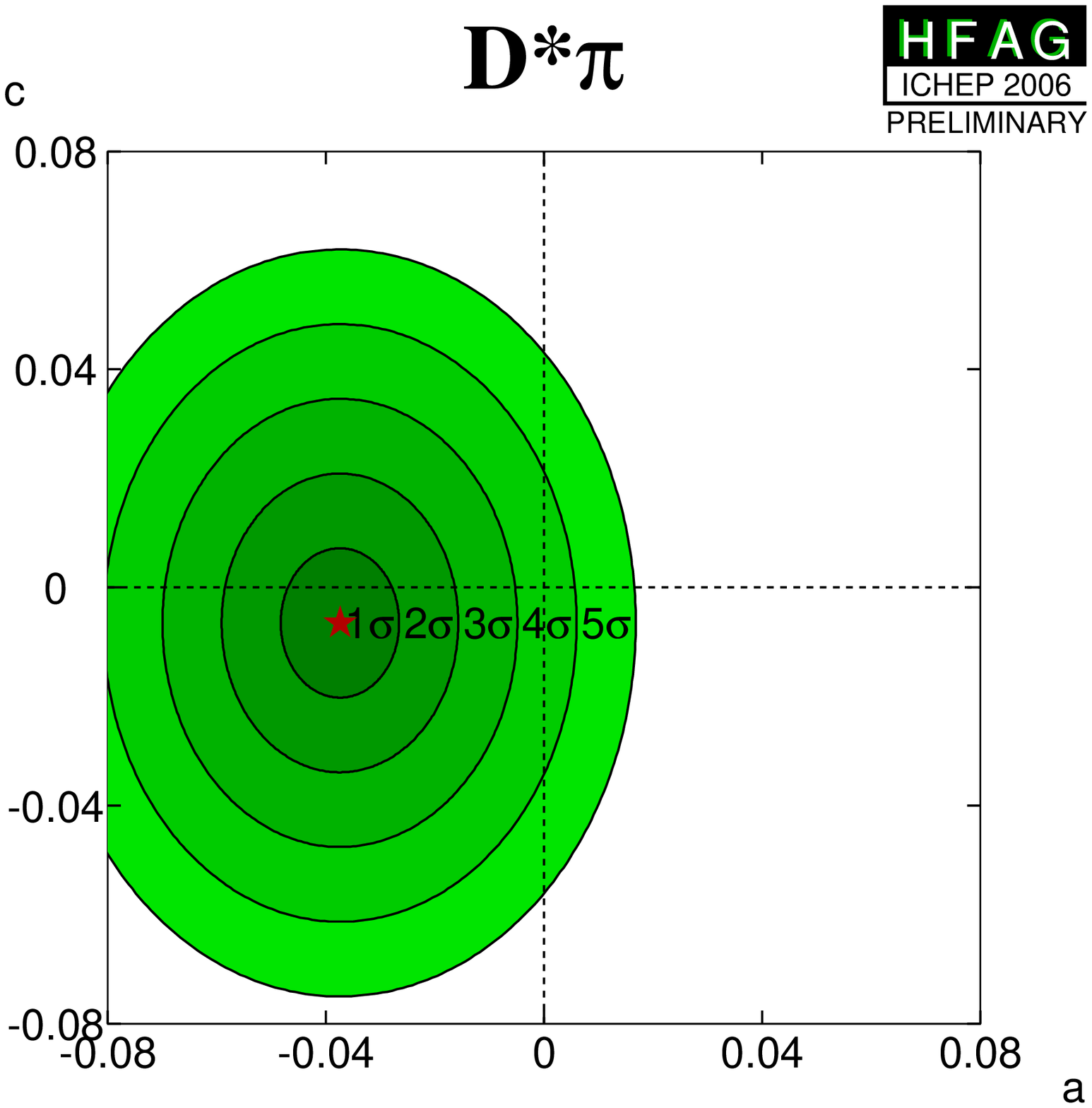}
      }
      &
      \resizebox{0.46\textwidth}{!}{
        \includegraphics{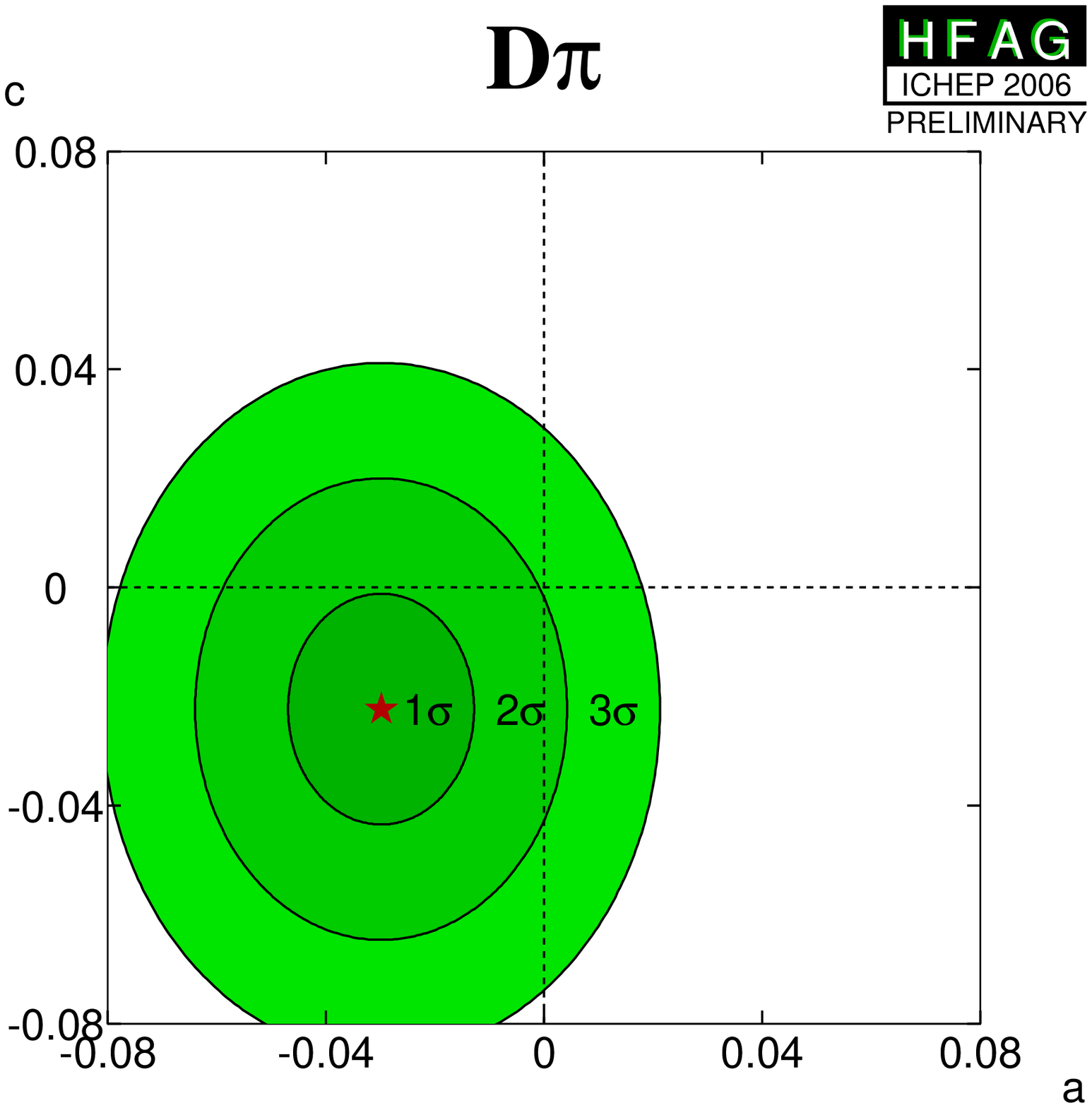}
      } \\
      \resizebox{0.46\textwidth}{!}{
        \includegraphics{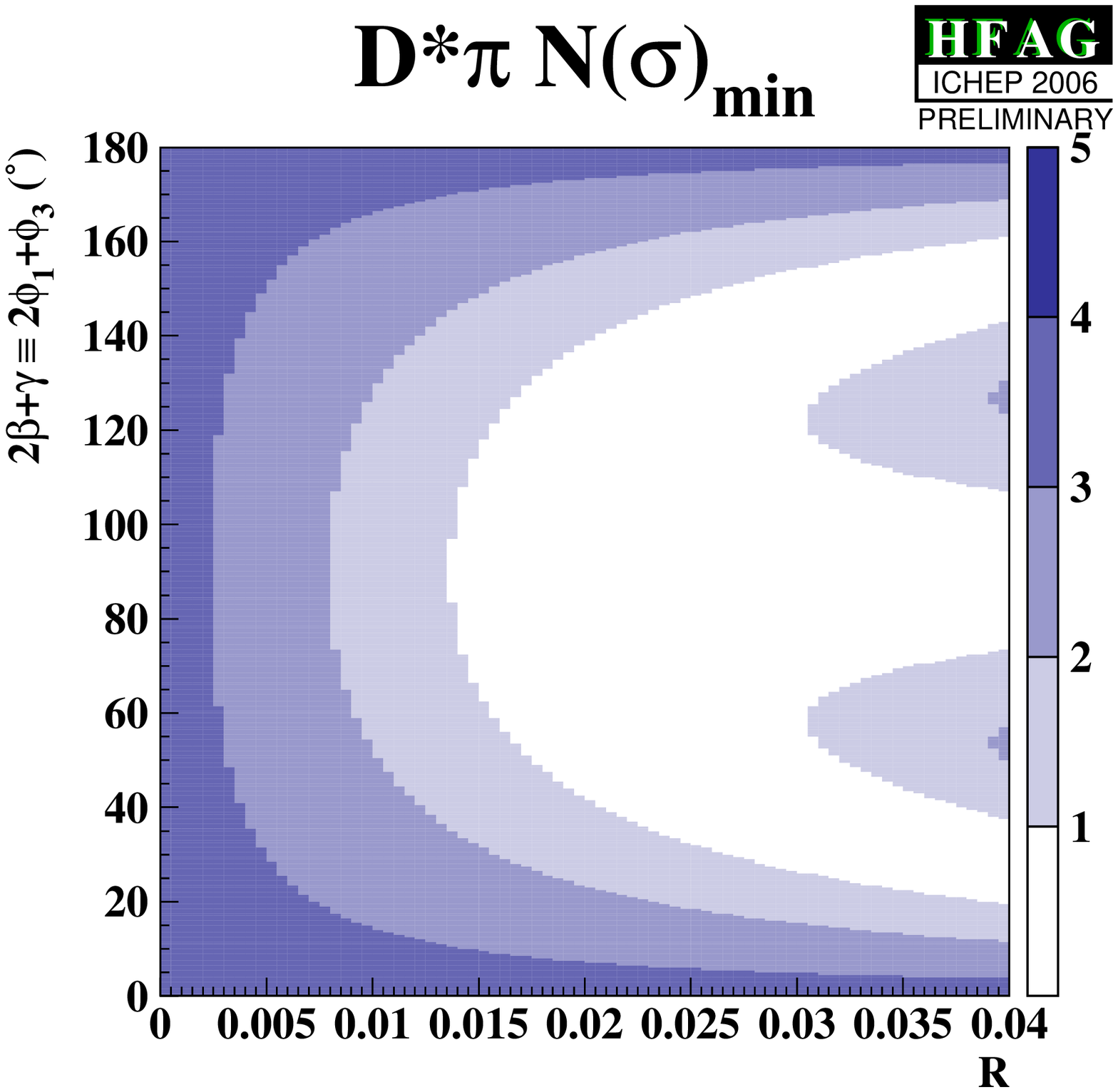}
      }
      &
      \resizebox{0.46\textwidth}{!}{
        \includegraphics{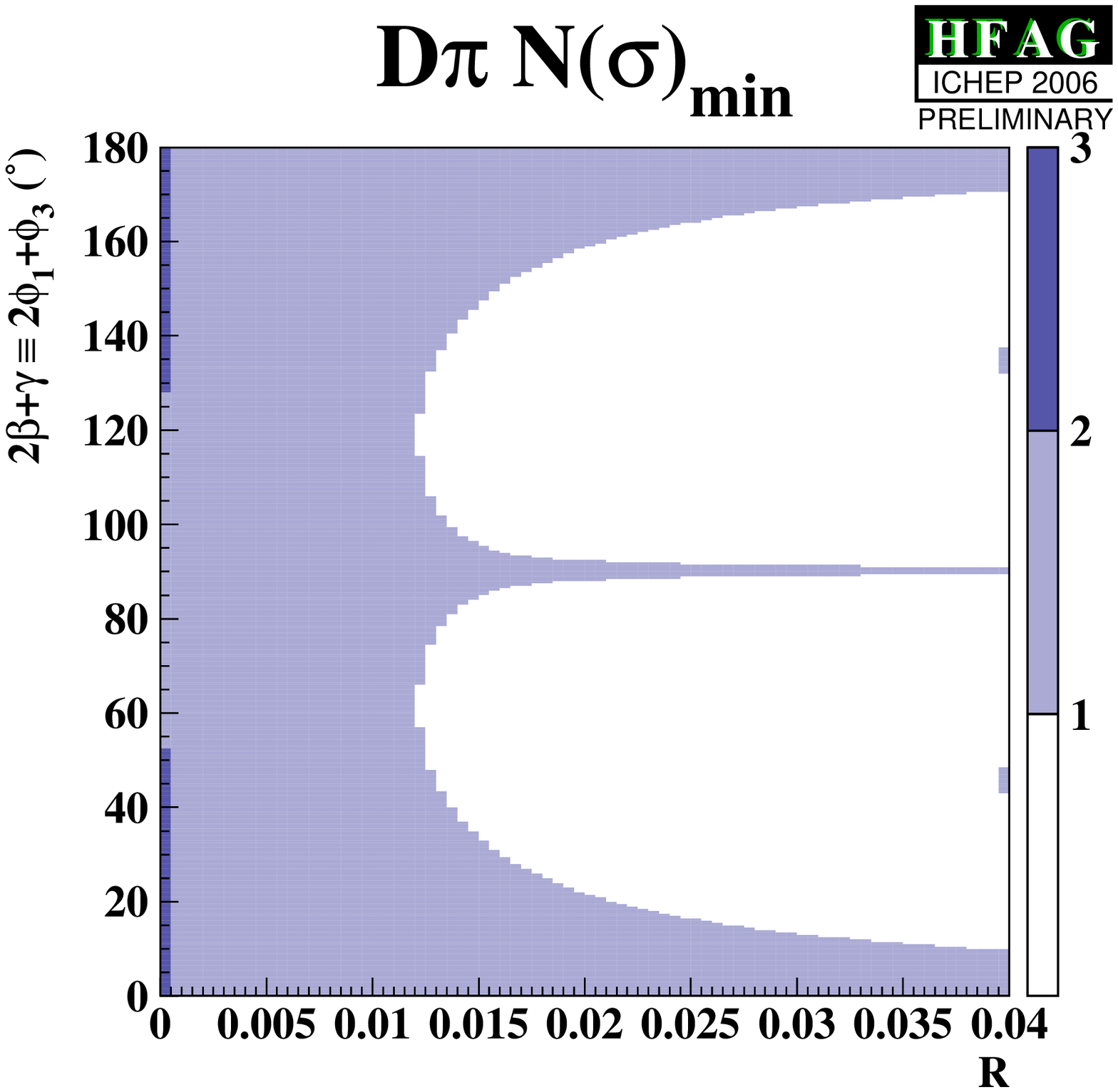}
      }          
    \end{tabular}
  \end{center}
  \vspace{-0.8cm}
  \caption{
    Results from $b \to c\bar{u}d / u\bar{c}d$ modes.
    (Top) Constraints in $c$ {\it vs.} $a$ space.
    (Bottom) Constraints in $2\beta+\gamma$ {\it vs.} $R$ space.
    (Left) $D^*\pi$ and (right) $D\pi$ modes.
  }
  \label{fig:cp_uta:cud_constraints}
\end{figure}

As described in Sec.~\ref{sec:cp_uta:notations:non_cp:dstarpi},
the averages are given in terms of parameters $a$ and $c$.
$\CP$ violation would appear as $a \neq 0$.
Results are available from both \babar\ and \belle\ in the modes
$D^\pm\pi^\mp$ and $D^{*\pm}\pi^\mp$; for the latter mode both experiments 
have used both full and partial reconstruction techniques.
(\babar\ have provided separate results with each technique,
while \belle\ have in addition provided a combined result.)
Results are also available from \babar\ using $D^\pm\rho^\mp$.
These results, and their averages, are listed in Table~\ref{tab:cp_uta:cud},
and are shown in Fig.~\ref{fig:cp_uta:cud}.
The constraints in $c$ \vs\ $a$ space for the $D\pi$ and $D^*\pi$ modes
are shown in Fig.~\ref{fig:cp_uta:cud_constraints}.
It is notable that the average value of $a$ from $D^*\pi$ is more than
$3\sigma$ from zero, providing evidence of $\CP$ violation in this channel.

\begin{table}[htb]
	\begin{center}
		\caption{
      Averages for $b \to c\bar{u}d / u\bar{c}d$ modes.
      Note that the ``\belle (combined)'' result for $D^{*\pm}\pi^{\mp}$
      is a combination of the 
      ``\belle (full rec.)'' and ``\belle (partial rec.)'' results.
                }
                \vspace{0.2cm}
                \setlength{\tabcolsep}{0.0pc}
\begin{tabular*}{\textwidth}{@{\extracolsep{\fill}}lrccc} \hline 
	\mc{2}{l}{Experiment} & $N(B\bar{B})$ & $a$ & $c$ \\
	\hline
      \mc{5}{c}{$D^{\pm}\pi^{\mp}$} \\
	\babar (full rec.) & \cite{Aubert:2006tw} & 232M & $-0.010 \pm 0.023 \pm 0.007$ & $-0.033 \pm 0.042 \pm 0.012$ \\
	\belle (full rec.) & \cite{Ronga:2006hv} & 386M & $-0.050 \pm 0.021 \pm 0.012$ & $-0.019 \pm 0.021 \pm 0.012$ \\
        \mc{3}{l}{\bf Average} & $ -0.030 \pm 0.017$ & $ -0.022 \pm 0.021 $ \\
        \mc{3}{l}{\small Confidence level} & {\small $0.24~(1.2\sigma)$} & {\small $0.78~(0.3\sigma)$} \\
        \hline
      \mc{5}{c}{$D^{*\pm}\pi^{\mp}$} \\
      \babar (full rec.) & \cite{Aubert:2006tw} & 232M & $-0.040 \pm 0.023 \pm 0.010$ & $0.049 \pm 0.042 \pm 0.015$ \\
      \babar (partial rec.)  & \cite{Aubert:2005yf} & 232M & $-0.034 \pm 0.014 \pm 0.009$ & $-0.019 \pm 0.022 \pm 0.013$ \\
      \belle (full rec.) & \cite{Ronga:2006hv} & 386M & $-0.039 \pm 0.020 \pm 0.013$ & $-0.011 \pm 0.020 \pm 0.013$ \\
      \belle (partial rec.) & \cite{Ronga:2006hv} & 386M & $-0.041 \pm 0.019 \pm 0.017$ & $-0.007 \pm 0.019 \pm 0.017$ \\
      {\small \belle (combined)} & {\small \cite{Ronga:2006hv}} & {\small 386M} & {\small $-0.040 \pm 0.014 \pm 0.011$} & {\small $-0.009 \pm 0.014 \pm 0.011$} \\
      \mc{3}{l}{\bf Average} & $-0.037 \pm 0.011$ & $-0.006 \pm 0.014$ \\
      \mc{3}{l}{\small Confidence level} & {\small $0.96~(0.0\sigma)$} & {\small $0.41~(0.8\sigma)$} \\
      \hline
      \mc{5}{c}{$D^{\pm}\rho^{\mp}$} \\
      \babar (full rec.) & \cite{Aubert:2006tw} & 232M & $-0.024 \pm 0.031 \pm 0.009$ & $-0.098 \pm 0.055 \pm 0.018$ \\
      \hline 
    \end{tabular*}
    \label{tab:cp_uta:cud}
  \end{center}
\end{table}

For each of $D\pi$, $D^*\pi$ and $D\rho$, 
there are two measurements ($a$ and $c$, or $S^+$ and $S^-$) 
which depend on three unknowns ($R$, $\delta$ and $2\beta+\gamma$), 
of which two are different for each decay mode. 
Therefore, there is not enough information to solve directly for $2\beta+\gamma$. 
However, for each choice of $R$ and $2\beta+\gamma$, 
one can find the value of $\delta$ that allows $a$ and $c$ to be closest 
to their measured values, 
and calculate the distance in terms of numbers of standard deviations.
(We currently neglect experimental correlations in this analysis.) 
These values of $N(\sigma)_{\rm min}$ can then be plotted 
as a function of $R$ and $2\beta+\gamma$
(and can trivially be converted to confidence levels). 
These plots are given for the $D\pi$ and $D^*\pi$ modes 
in Figure~\ref{fig:cp_uta:cud_constraints}; 
the uncertainties in the $D\rho$ mode are currently too large 
to give any meaningful constraint.

The constraints can be tightened if one is willing 
to use theoretical input on the values of $R$ and/or $\delta$. 
One popular choice is the use of SU(3) symmetry to obtain 
$R$ by relating the suppressed decay mode to $\B$ decays 
involving $D_s$ mesons. 
More details can be found 
in Refs.~\cite{Charles:2004jd,Bona:2005vz}.

\clearpage
\mysubsection{Rates and asymmetries in $\Bmp \to \DorDstar K^{(*)\mp}$ decays
}
\label{sec:cp_uta:cus}

As explained in Sec.~\ref{sec:cp_uta:notations:cus},
rates and asymmetries in $\Bmp \to \DorDstar K^{(*)\mp}$ decays
are sensitive to $\gamma$.
Various methods using different $\DorDstar$ final states exist.

\mysubsubsection{$D$ decays to $\CP$ eigenstates
}
\label{sec:cp_uta:cus:glw}

\begin{figure}[htb]
  \begin{center}
    \begin{tabular}{cc}
      \resizebox{0.46\textwidth}{!}{
        \includegraphics{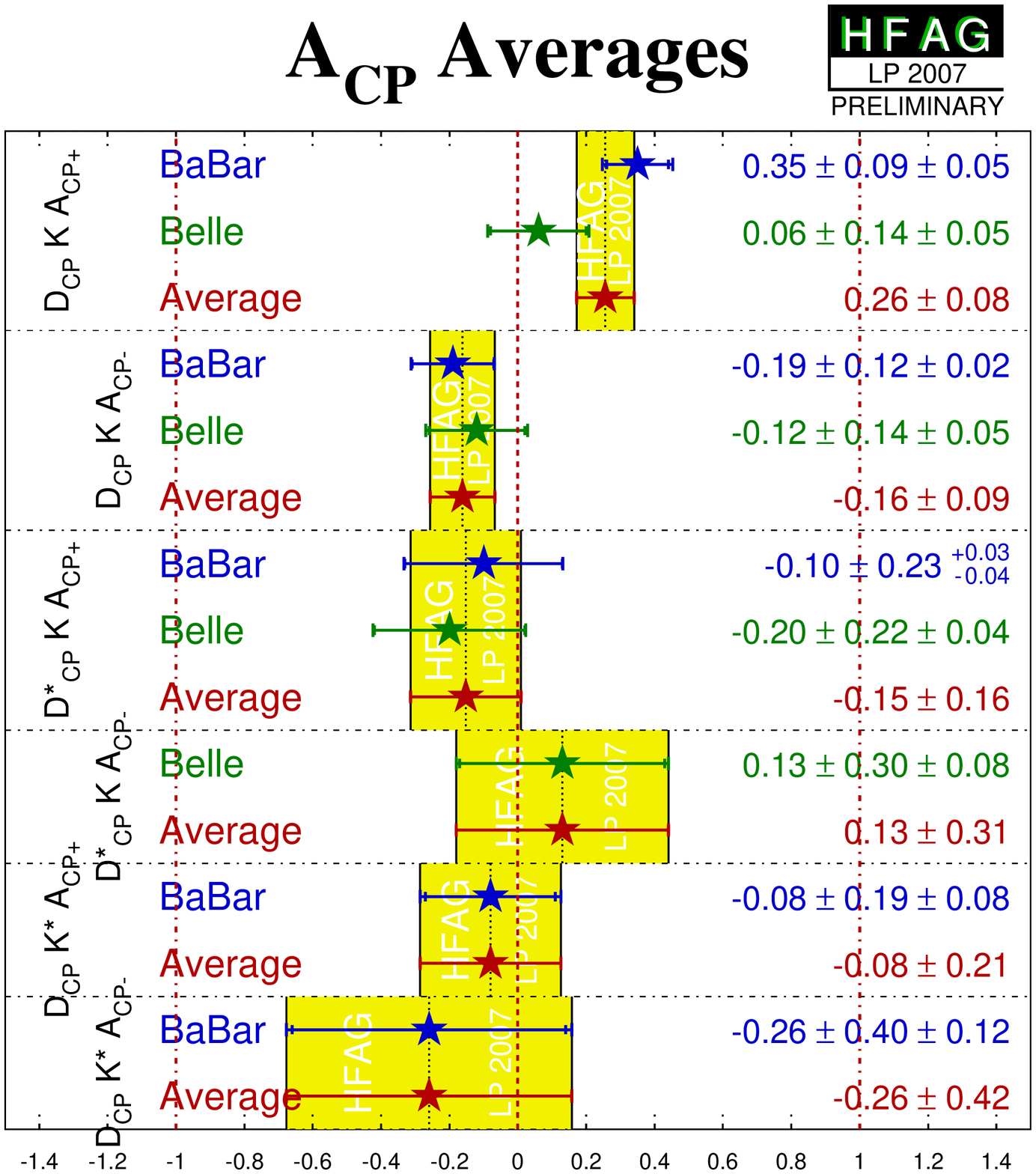}
      }
      &
      \resizebox{0.46\textwidth}{!}{
        \includegraphics{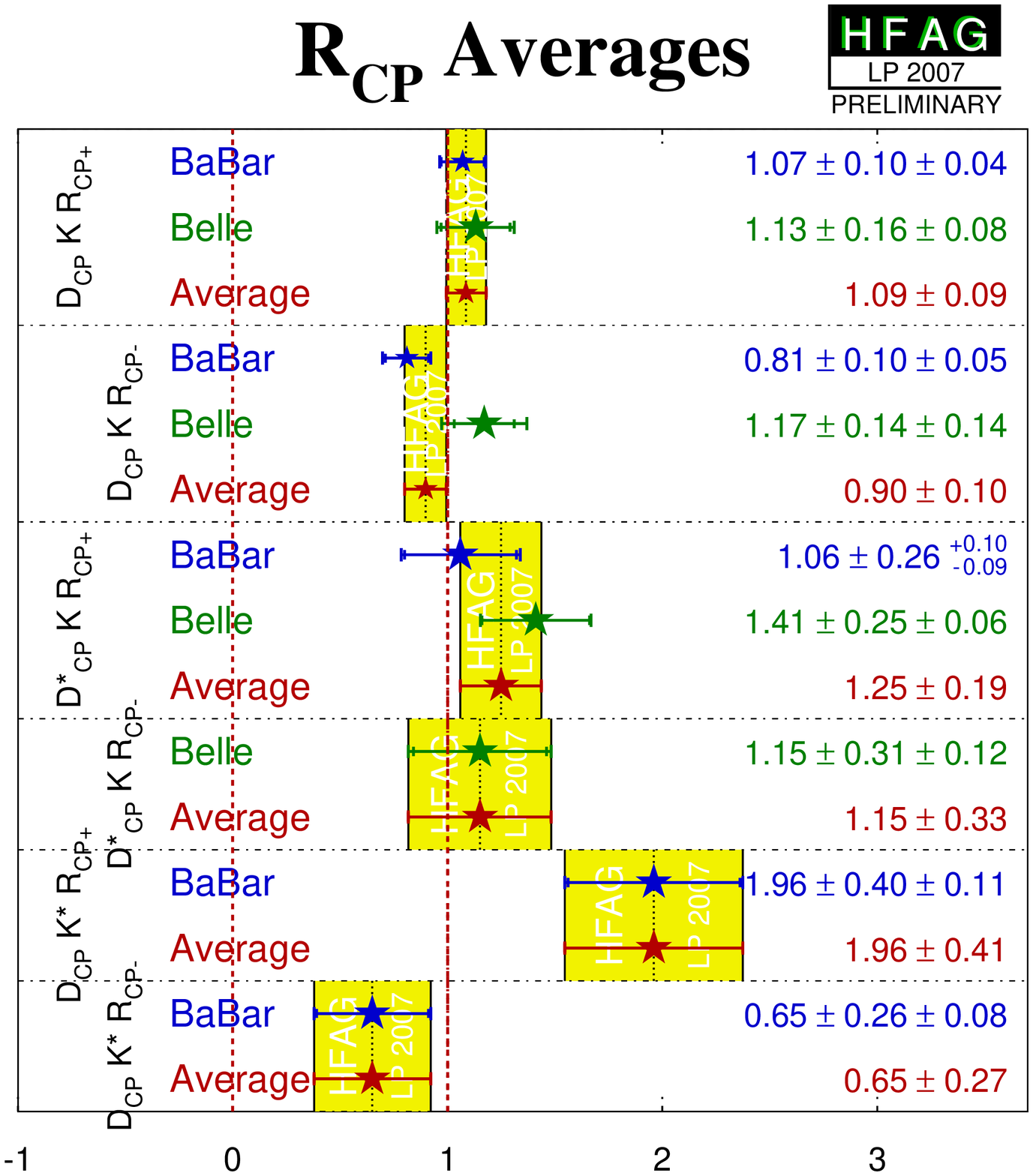}
      }
    \end{tabular}
  \end{center}
  \vspace{-0.8cm}
  \caption{
    Averages of $A_{\CP}$ and $R_{\CP}$ from GLW analyses.
  }
  \label{fig:cp_uta:cus:glw}
\end{figure}

Results are available from both \babar\ and \belle\ on GLW analyses
in the decay modes $\Bmp \to D\Kmp$, $\Bmp \to \Dstar\Kmp$ and $\Bmp \to D\Kstarmp$.
Both experiments use the 
$\CP$-even $D$ decay final states $K^+K^-$ and $\pi^+\pi^-$ in all three modes; 
both experiments also use only the $\Dstar \to D\pi^0$ decay, 
which gives $\CP(\Dstar) = \CP(D)$. 
For $\CP$-odd $D$ decay final states, 
\belle\ uses $\KS\pi^0$, $\KS\eta$ and $\KS\phi$ in all three analyses, 
and also use $\KS\omega$ in $D\Kmp$ and $\Dstar\Kmp$ analyses. 
\babar\ uses $\KS\pi^0$ only for $D\Kmp$ analysis; 
for $D\Kstarmp$ analysis they also use $\KS\phi$ and $\KS\omega$
(and assign an asymmetric systematic error due to $\CP$-even pollution 
in these $\CP$-odd channels~\cite{Aubert:2004hu}).
The results and averages are given in Table~\ref{tab:cp_uta:cus:glw}
and shown in Fig.~\ref{fig:cp_uta:cus:glw}.

\begin{table}[htb]
	\begin{center}
		\caption{
                        Averages from GLW analyses of $b \to c\bar{u}s / u\bar{c}s$ modes.
                }
                \vspace{0.2cm}
    \resizebox{\textwidth}{!}{
      \setlength{\tabcolsep}{0.0pc}
\begin{tabular}{@{\extracolsep{2mm}}lrccccc} \hline 
        \mc{2}{l}{Experiment} & $N(B\bar{B})$ & $A_{\CP+}$ & $A_{\CP-}$ & $R_{\CP+}$ & $R_{\CP-}$ \\
        \hline
        \mc{7}{c}{$D_{\CP} K^-$} \\
	\babar & \cite{Aubert:2007rj} & 382M & $0.35 \pm 0.09 \pm 0.05$ & $-0.19 \pm 0.12 \pm 0.02$ & $1.07 \pm 0.10 \pm 0.04$ & $0.81 \pm 0.10 \pm 0.05$ \\
	\belle & \cite{Abe:2006hc} & 275M & $0.06 \pm 0.14 \pm 0.05$ & $-0.12 \pm 0.14 \pm 0.05$ & $1.13 \pm 0.16 \pm 0.08$ & $1.17 \pm 0.14 \pm 0.14$ \\
	\mc{3}{l}{\bf Average} & $0.26 \pm 0.08$ & $-0.16 \pm 0.09$ & $1.09 \pm 0.09$ & $0.90 \pm 0.10$ \\
	\mc{3}{l}{\small Confidence level} & {\small $0.11~(1.6\sigma)$} & {\small $0.72~(0.4\sigma)$} & {\small $0.77~(0.3\sigma)$} & {\small $0.11~(1.6\sigma)$} \\
		\hline

        \mc{7}{c}{$\Dstar_{\CP} K^-$} \\
	\babar & \cite{Aubert:2004hu} & 123M & $-0.10 \pm 0.23 \,^{+0.03}_{-0.04}$ & \textendash{} & $1.06 \pm 0.26 \,^{+0.10}_{-0.09}$ &   \textendash{} \\
	\belle & \cite{Abe:2006hc} & 275M & $-0.20 \pm 0.22 \pm 0.04$ & $0.13 \pm 0.30 \pm 0.08$ & $1.41 \pm 0.25 \pm 0.06$ & $1.15 \pm 0.31 \pm 0.12$ \\
	\mc{3}{l}{\bf Average} & $-0.15 \pm 0.16$ & $0.13 \pm 0.31$ & $1.25 \pm 0.19$ & $1.15 \pm 0.33$ \\
	\mc{3}{l}{\small Confidence level} & {\small $0.76~(0.3\sigma)$} & \textendash{} & {\small $0.36~(0.9\sigma)$} & \textendash{} \\
		\hline

        \mc{7}{c}{$\D_{\CP} K^{*-}$} \\
	\babar & \cite{Aubert:2005cc} & 232M & $-0.08 \pm 0.19 \pm 0.08$ & $-0.26 \pm 0.40 \pm 0.12$ & $1.96 \pm 0.40 \pm 0.11$ & $0.65 \pm 0.26 \pm 0.08$ \\
		\hline
      \end{tabular}
    }
    \label{tab:cp_uta:cus:glw}
	\end{center}
\end{table}

\mysubsubsection{$D$ decays to suppressed final states}
\label{sec:cp_uta:cus:ads}

\begin{figure}[htb]
  \begin{center}
    \resizebox{0.46\textwidth}{!}{
      \includegraphics{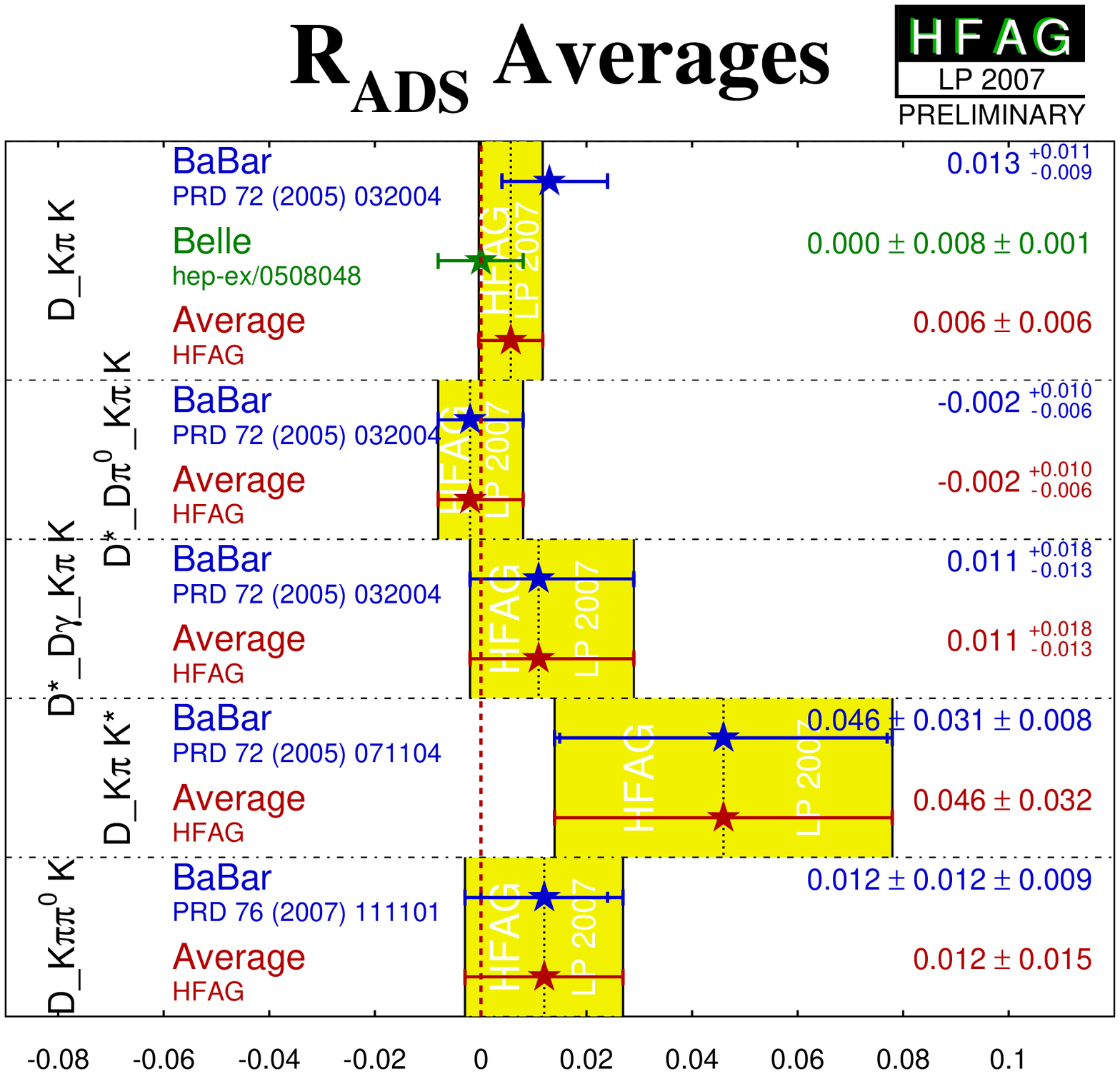}
    }
  \end{center}
  \vspace{-0.8cm}
  \caption{
    Averages of $R_{\rm ADS}$.
  }
  \label{fig:cp_uta:cus:ads}
\end{figure}

For ADS analysis, both \babar\ and \belle\ have studied
the mode $\Bmp \to D\Kmp$;
\belle\ has also studied $\Bmp \to D\pi^\mp$
and \babar\ has also analyzed the $\Bmp \to \Dstar\Kmp$ 
and $\Bmp \to D\Kstarmp$ modes
($\Dstar \to D\pi^0$ and $\Dstar \to D\gamma$ are studied separately;
$\Kstarmp$ is reconstructed as $\KS\pi^\mp$).
In all cases the suppressed decay $D \to K^+\pi^-$ has been used.
\babar\ also has results using $\Bmp \to D\Kmp$ with $D \to K^+\pi^-\pi^0$.
The results and averages are given in Table~\ref{tab:cp_uta:cus:ads}
and shown in Fig.~\ref{fig:cp_uta:cus:ads}.
Note that although no clear signals for these modes have yet been seen,
the central values are given.
In $\Bm \to \Dstar\Km$ decays there is an effective shift of $\pi$
in the strong phase difference between the cases that the $\Dstar$ is 
reconstructed as $D\pi^0$ and $D\gamma$~\cite{Bondar:2004bi}.
As a consequence, the different $D^*$ decay modes are treated separately.

\begin{table}[htb]
	\begin{center}
		\caption{
      Averages from ADS analyses of $b \to c\bar{u}s / u\bar{c}s$ and 
      $b \to c\bar{u}d / u\bar{c}d$ modes.
                }
                \vspace{0.2cm}
                \setlength{\tabcolsep}{0.0pc}
\begin{tabular*}{\textwidth}{@{\extracolsep{\fill}}lrccc} \hline 
        \mc{2}{l}{Experiment} & $N(B\bar{B})$ & $A_{\rm ADS}$ & $R_{\rm ADS}$ \\
	\hline
        \mc{5}{c}{$D K^-$, $D \to K^+\pi^-$} \\
        \babar & \cite{Aubert:2005pj} & 232M & \textendash{} & $0.013 \,^{+0.011}_{-0.009}$ \\
        \belle & \cite{Abe:2005gi} & 386M & \textendash{} & $0.000 \pm 0.008 \pm 0.001$ \\
        \mc{3}{l}{\bf Average} & \textendash{} & $ 0.006 \pm 0.006$ \\
	\mc{3}{l}{\small Confidence level} & \textendash{} & {\small $0.29~(1.1\sigma)$} \\
		\hline

        \mc{5}{c}{$\Dstar K^-$, $\Dstar \to D\pi^0$, $D \to K^+\pi^-$} \\
        \babar & \cite{Aubert:2005pj} & 232M & \textendash{} & $-0.002 \,^{+0.010}_{-0.006}$ \\
        \hline

        \mc{5}{c}{$\Dstar K^-$, $\Dstar \to D\gamma$, $D \to K^+\pi^-$} \\
        \babar & \cite{Aubert:2005pj} & 232M & \textendash{} & $0.011 \,^{+0.018}_{-0.013}$ \\
        \hline

        \mc{5}{c}{$D K^{*-}$, $D \to K^+\pi^-$, $K^{*-} \to \KS \pi^-$} \\
        \babar & \cite{Aubert:2005cr} & 232M & $-0.22 \pm 0.61 \pm 0.17$ & $0.046 \pm 0.031 \pm 0.008$ \\
        \hline

        \mc{5}{c}{$D K^{-}$, $D \to K^+\pi^-\pi^0$} \\
        \babar & \cite{Aubert:2007nu} & 226M & \textendash{} & $0.012 \pm 0.012 \pm 0.009$ \\
        \hline 

       \mc{5}{c}{$D \pi^-$, $D \to K^+\pi^-$} \\
       \belle & \cite{Abe:2005gi} & 386M &
       $ 0.10 \pm 0.22 \pm 0.06$ & $ 0.0035 \, ^{+0.0008}_{-0.0007} \pm 0.0003$ \\
       \hline 

 		\end{tabular*}
                \label{tab:cp_uta:cus:ads}
 	\end{center}
 \end{table}

\mysubsubsection{$D$ decays to multiparticle self-conjugate final states}
\label{sec:cp_uta:cus:dalitz}

For the Dalitz plot analysis, both 
\babar~\cite{Aubert:2006am} and \belle~\cite{Poluektov:2006ia} have studied
the modes $\Bmp \to D\Kmp$, $\Bmp \to \Dstar\Kmp$ and $\Bmp \to D\Kstarmp$.
For $\Bmp \to \Dstar\Kmp$,
\belle\ has used only $\Dstar \to D\pi^0$,
while \babar\ has used both $\Dstar$ decay modes and 
taken the effective shift in the strong phase difference into account.
In all cases the decay $D \to \KS\pi^+\pi^-$ has been used.
\babar\ has also performed an analysis of $\Bmp \to D\Kmp$ with 
$D \to \pi^+\pi^-\pi^0$~\cite{Aubert:2007ii}.
Results and averages are given in Table~\ref{tab:cp_uta:cus:dalitz}.
The third error on each measurement is due to $D$ decay model uncertainty.

The parameters measured in the analyses are explained in
Sec.~\ref{sec:cp_uta:notations:cus}.
Both \babar\ and \belle\ have measured the ``Cartesian''
$(x_\pm,y_\pm)$ variables,
and perform frequentist statistical procedures,
to convert these into measurements of $\gamma$, $r_B$ and $\delta_B$.
In the $\Bmp \to D\Kmp$ with $D \to \pi^+\pi^-\pi^0$ analysis,
the parameters $(\rho^{\pm}, \theta^\pm)$ are used instead.

Both experiments reconstruct $\Kstarmp$ as $\KS\pi^\mp$,
but the treatment of possible nonresonant $\KS\pi^\mp$ differs:
\belle\ assign an additional model uncertainty,
while \babar\ use a reparametrization 
suggested by Gronau~\cite{Gronau:2002mu}.
The parameters $r_B$ and $\delta_B$ are replaced with 
effective parameters $\kappa r_s$ and $\delta_s$;
no attempt is made to extract the true hadronic parameters 
of the $\Bmp \to D\Kstarmp$ decay.

We perform averages using the following procedure, 
which is based on a set of reasonable, though imperfect, assumptions.

\begin{itemize}\setlength{\itemsep}{0.5ex}
\item 
  It is assumed that effects due to the different $D$ decay models 
  used by the two experiments are negligible. 
  Therefore, we do not rescale the results to a common model.
\item 
  It is further assumed that the model uncertainty is $100\%$ 
  correlated between experiments, 
  and therefore this source of error is not used in the averaging procedure.
\item 
  We include in the average the effect of correlations 
  within each experiments set of measurements.
\item 
  At present it is unclear how to assign an average model uncertainty. 
  We have not attempted to do so. 
  Our average includes only statistical and systematic error. 
  An unknown amount of model uncertainty should be added to the final error.
\item 
  We follow the suggestion of Gronau~\cite{Gronau:2002mu} 
  in making the $DK^*$ averages. 
  Explicitly, we assume that the selection of $K^{*\pm} \to \KS\pi^\pm$
  is the same in both experiments 
  (so that $\kappa$, $r_s$ and $\delta_s$ are the same), 
  and drop the additional source of model uncertainty 
  assigned by Belle due to possible nonresonant decays.
\item 
  We do not consider common systematic errors, 
  other than the $D$ decay model. 
\end{itemize}

\begin{table}[htb]
	\begin{center}
		\caption{
      Averages from Dalitz plot analyses of $b \to c\bar{u}s / u\bar{c}s$ modes.
      Note that the uncertainities assigned to the averages do not include model errors.	
		}
		\vspace{0.2cm}
		\setlength{\tabcolsep}{0.0pc}
    \resizebox{\textwidth}{!}{

              }
		\label{tab:cp_uta:cus:dalitz}
	\end{center}
\end{table}

\begin{figure}[htb]
  \begin{center}
    \resizebox{0.30\textwidth}{!}{
      \includegraphics{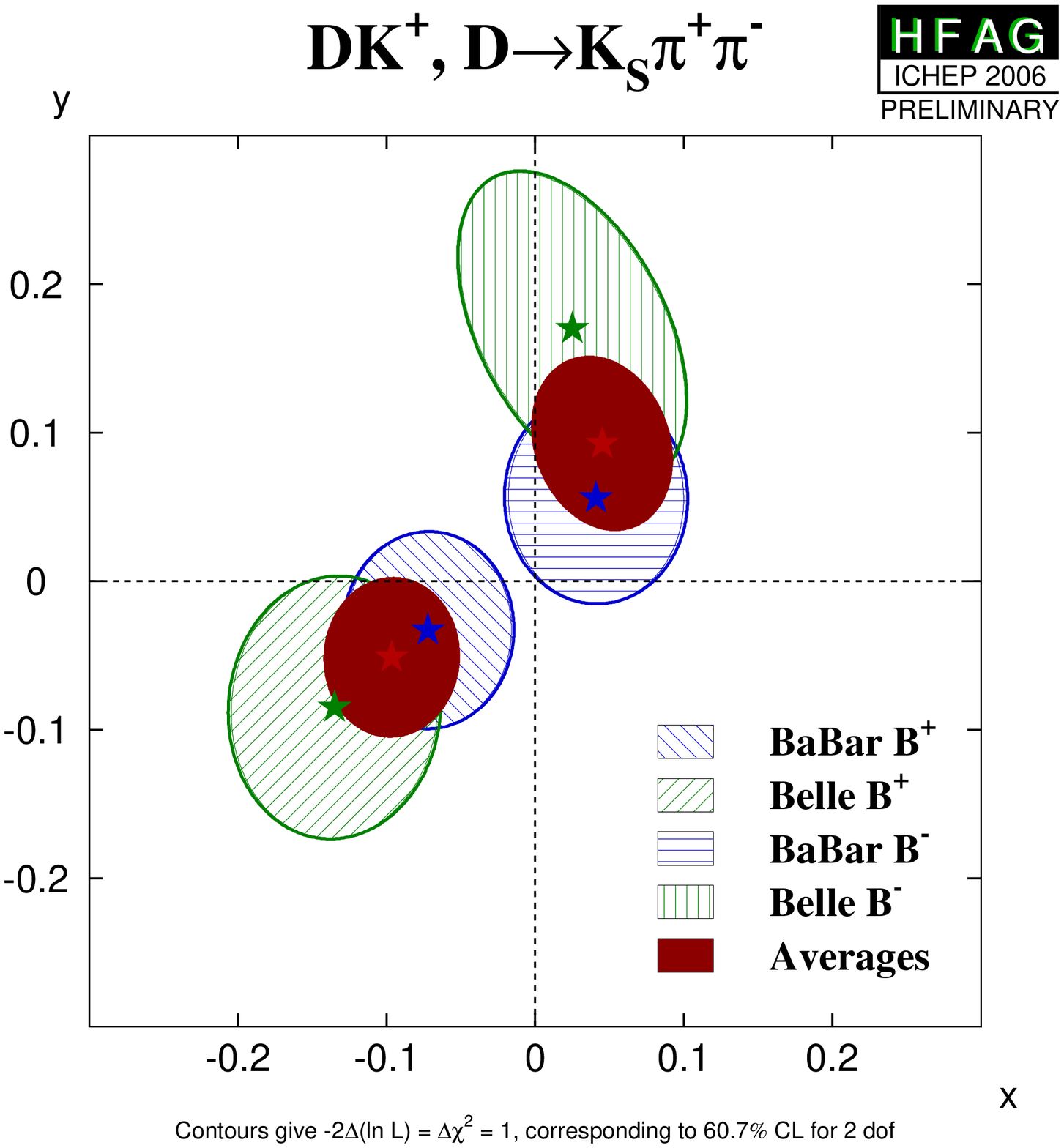}
    }
    \hfill
    \resizebox{0.30\textwidth}{!}{
      \includegraphics{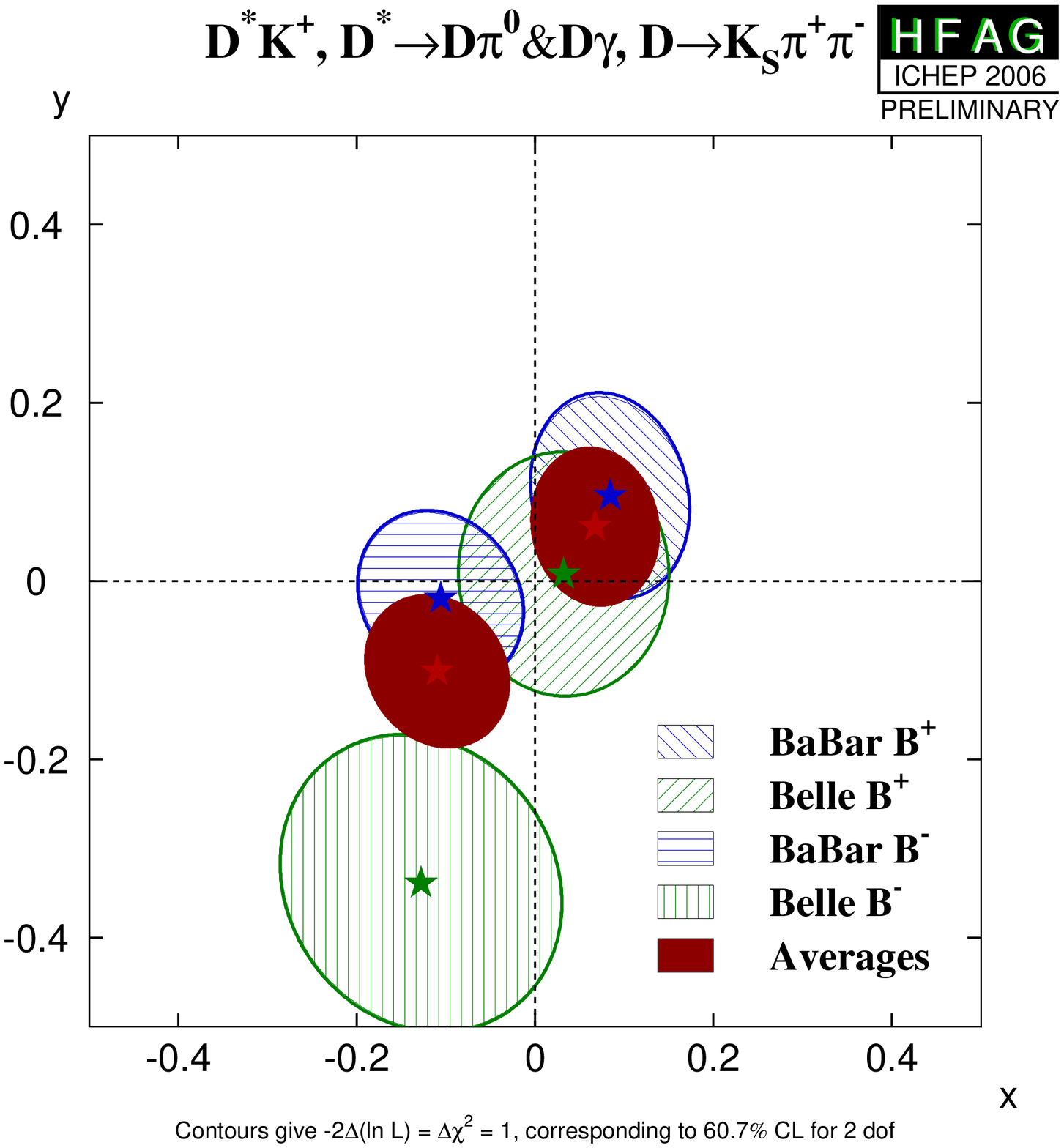}
    }
    \hfill
    \resizebox{0.30\textwidth}{!}{
      \includegraphics{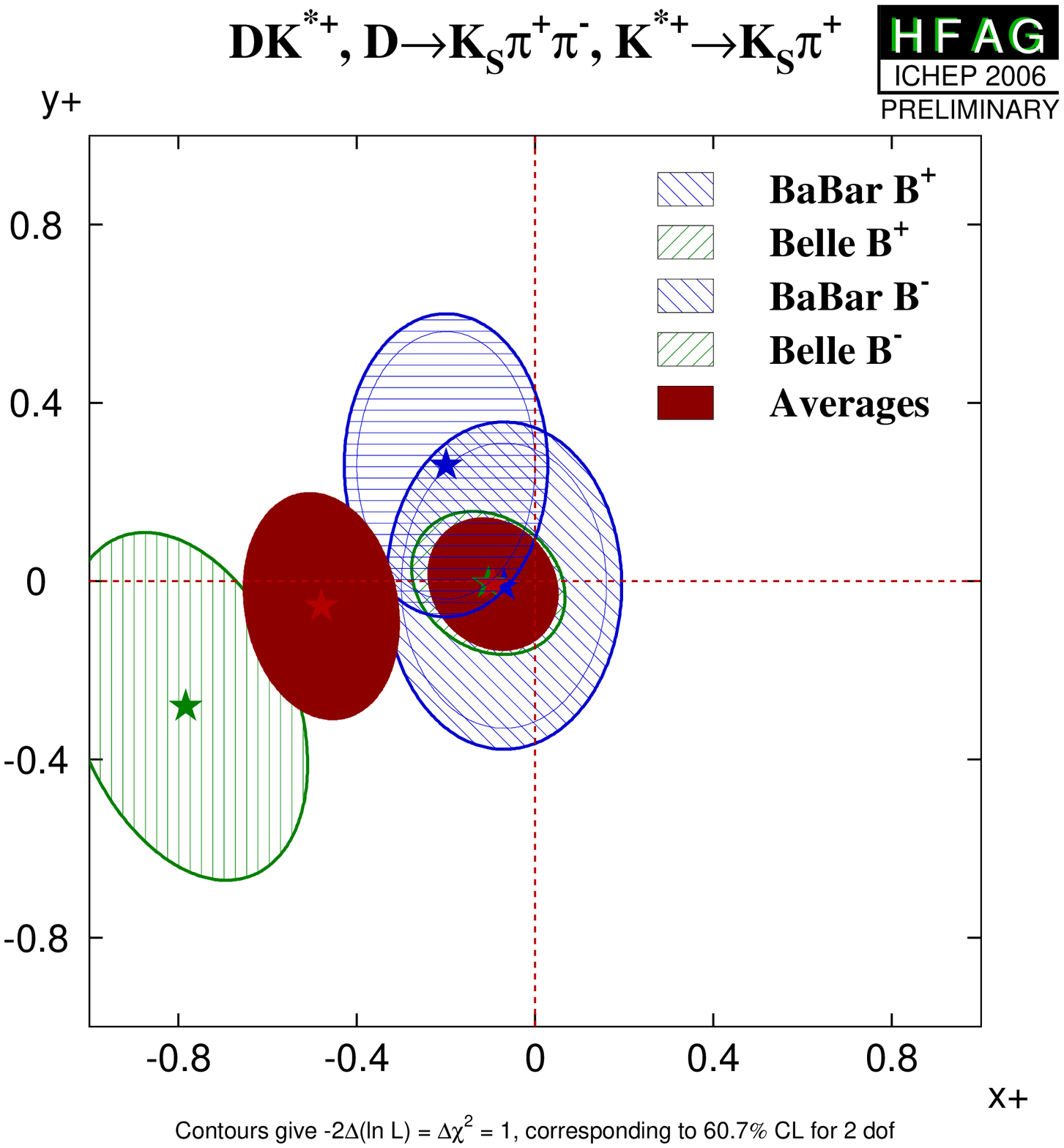}
    }
  \end{center}
  \vspace{-0.8cm}
  \caption{
    Contours in the $(x_\pm, y_\pm)$ from $\Bmp \to D^{(*)}K^{(*)\pm}$.
    (Left) $\Bmp \to D\Kmp$, 
    (middle) $\Bmp \to \Dstar\Kmp$,
    (right) $\Bmp \to D\Kstarmp$.
    Note that the uncertainities assigned to the averages given in these plots
    do not include model errors.        
  }
  \label{fig:cp_uta:cus:dalitz_2d}
\end{figure}

\begin{figure}[htb]
  \begin{center}
    \resizebox{0.40\textwidth}{!}{
      \includegraphics{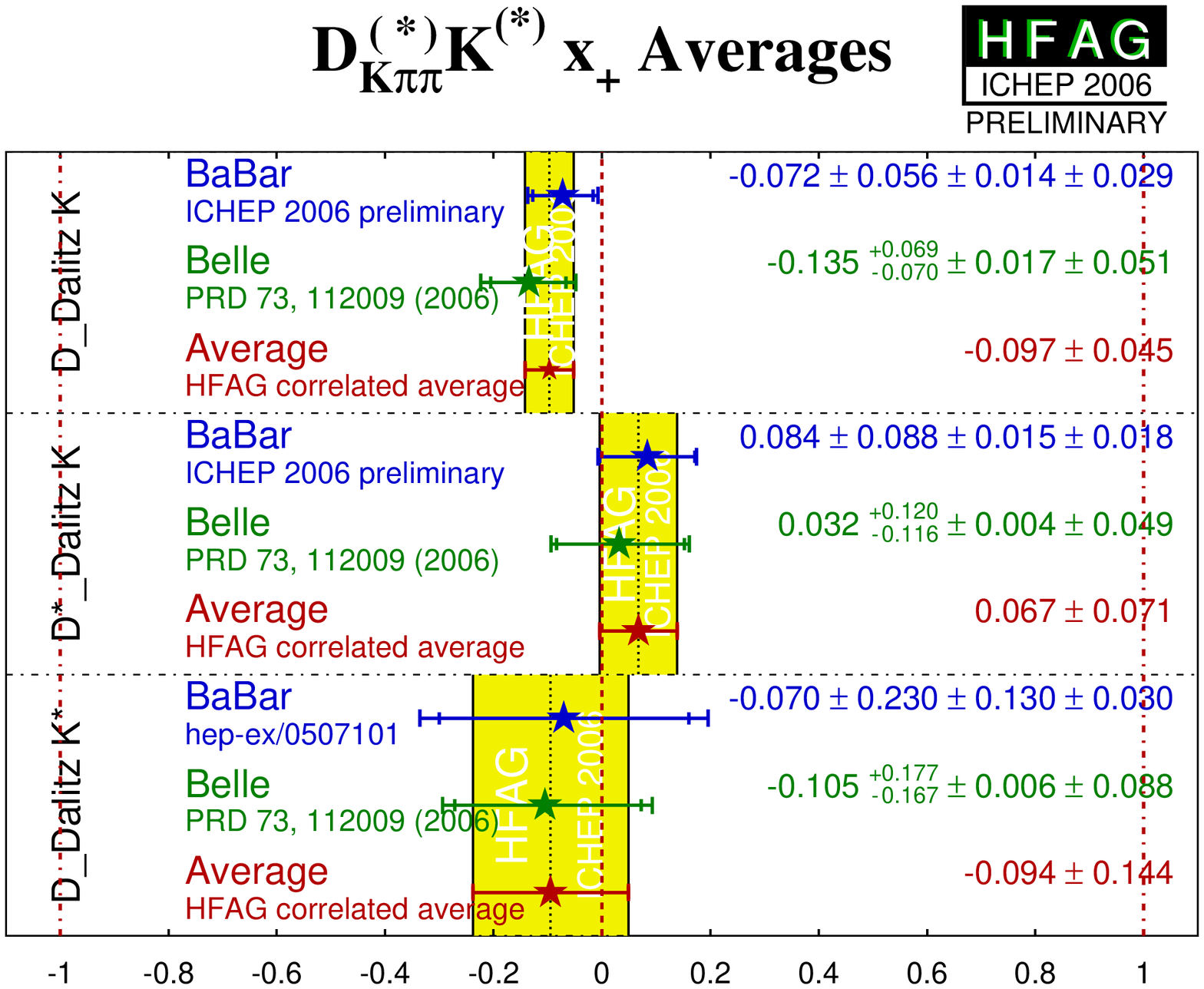}
    }
    \hspace{0.1\textwidth}
    \resizebox{0.40\textwidth}{!}{
      \includegraphics{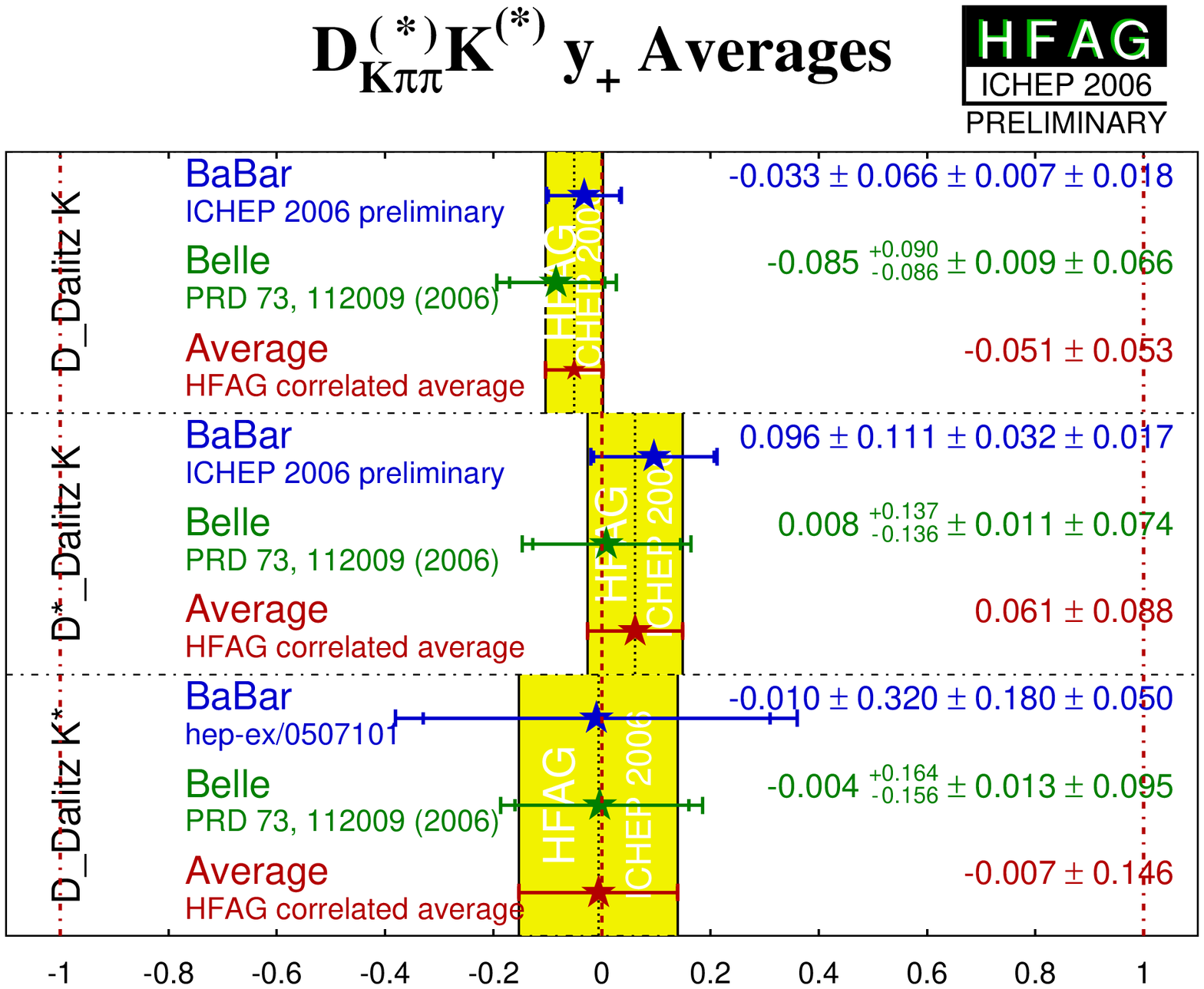}
    }
    \\
    \resizebox{0.40\textwidth}{!}{
      \includegraphics{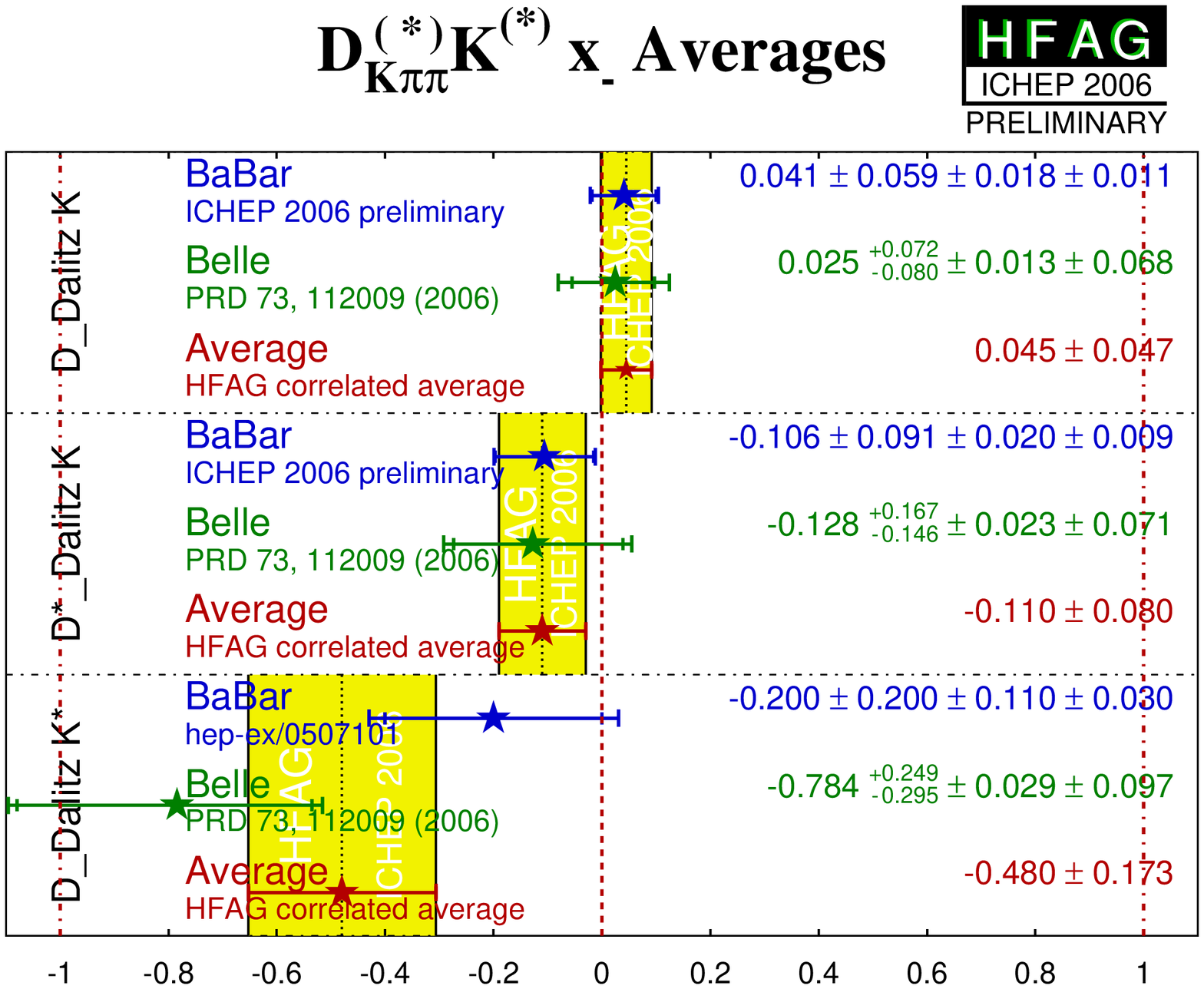}
    }
    \hspace{0.1\textwidth}
    \resizebox{0.40\textwidth}{!}{
      \includegraphics{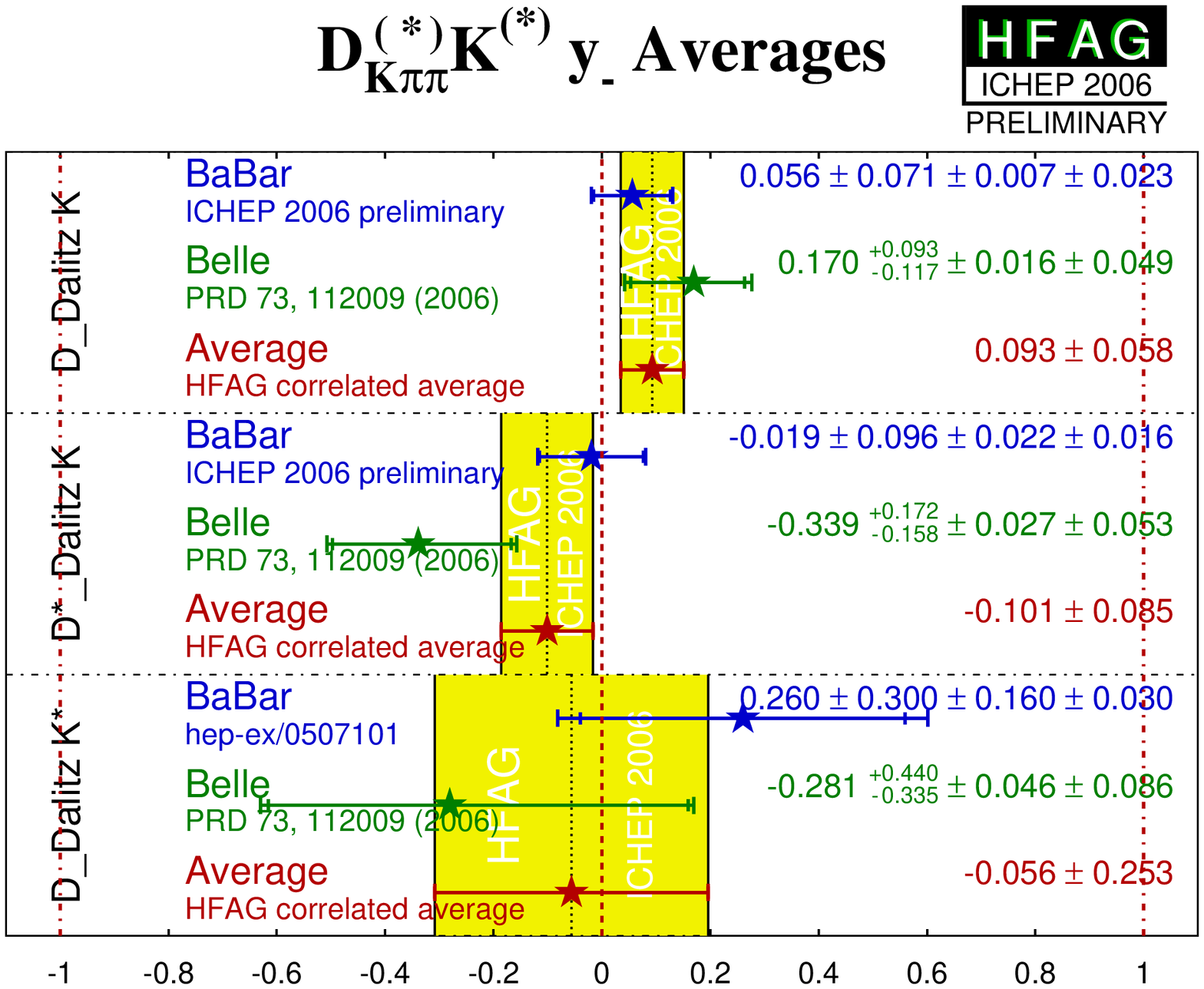}
    }
  \end{center}
  \vspace{-0.8cm}
  \caption{
    Averages of $(x_\pm, y_\pm)$ from $\Bmp \to D^{(*)}K^{(*)\pm}$.
    (Top left) $x_+$, (top right) $y_+$,
    (bottom left) $x_-$, (bottom right) $y_-$.
    Note that the uncertainities assigned to the averages given in these plots
    do not include model errors.        
  }
  \label{fig:cp_uta:cus:dalitz_1d}
\end{figure}

\vspace{3ex}

\noindent
\underline{\large Constraints on $\gamma$}

The measurements of $(x_\pm, y_\pm)$ can be used to obtain constraints on 
$\gamma$, as well as the hadronic parameters $r_B$ and $\delta_B$.
Both
\babar~\cite{Aubert:2006am} and 
\belle~\cite{Poluektov:2006ia} 
have done so using a frequentist procedure 
(there are some differences in the details of the techniques used).

\begin{itemize}\setlength{\itemsep}{0.5ex}

\item 
  \babar\ obtain $\gamma = (92 \pm 41 \pm 11 \pm 12)^\circ$
  from $D\Kpm$ and $\Dstar\Kpm$

\item
  \belle\ obtain $\phi_3 = (53 \, ^{+15}_{-18} \pm 3 \pm 9)^\circ$
  from $D\Kpm$, $\Dstar\Kpm$ and $D\Kstarpm$

\item
  The experiments also obtain values for the hadronic parameters. \\
  In $D\Kpm$ \\
  \babar\ obtain 
  $r_B (D\Kpm) < 0.140 (1\sigma)$ and
  $\delta_B (D\Kpm) = (118 \pm 63 \pm 19 \pm 36)^\circ$ \\
  \belle\ obtain 
  $r_B (D\Kpm) = 0.16 \pm 0.05 \pm 0.01 \pm 0.05$ and
  $\delta_B (D\Kpm) = (146 \, ^{+19}_{-20} \pm 3 \pm 23)^\circ$. \\
  In $\Dstar\Kpm$ \\
  \babar\ obtain   
  $0.017 < r_B (\Dstar\Kpm) < 0.203$ and
  $\delta_B (\Dstar\Kpm) = (298 \pm 59 \pm 18 \pm 10)^\circ$ \\
  \belle\ obtain 
  $r_B (\Dstar\Kpm) = 0.18 \, ^{+0.11}_{-0.10} \pm 0.01 \pm 0.05$ and
  $\delta_B (\Dstar\Kpm) = (302 \, ^{+34}_{-35} \pm 6 \pm 23)^\circ$. \\
  In $D\Kstarpm$ \\
  \belle\ obtain 
  $r_B (D\Kstarpm) = 0.56 \, ^{+0.22}_{-0.16} \pm 0.04 \pm 0.08$ and
  $\delta_B (D\Kstarpm) = (243 \, ^{+20}_{-23} \pm 3 \pm 50)^\circ$.
  \babar\ do not obtain a constraint on the hadronic parameters in 
  $D\Kstarpm$ due to the reparametrization described above.

\item 
  Improved constraints can be achieved combining the information from
  $\Bpm \to D\Kpm$ analysis with different $D$ decay modes.
  The experiments have not yet published such results,
  and none are listed here.

\item 
  The CKMfitter~\cite{Charles:2004jd} and 
  UTFit~\cite{Bona:2005vz} groups use the measurements 
  from \belle\ and \babar\ given above
  to make combined constraints on $\gamma$.

\item 
  In the \babar\ analysis of $\Bmp \to D\Kmp$ with 
  $D \to \pi^+\pi^-\pi^0$~\cite{Aubert:2007ii},
  a constraint of $-30^\circ < \gamma < 76^\circ$ is obtained 
  at the 68\% confidence level.

\end{itemize}

At present we make no attempt to provide an HFAG average for $\gamma$.
More details on procedures to calculate a best fit value for $\gamma$ 
can be found in Refs.~\cite{Charles:2004jd,Bona:2005vz}.

\clearpage


\section{Semileptonic $B$ decays}
\label{slbdecays}

Measurements of semileptonic $B$-meson decays are an important tool to
study the magnitude of the CKM matrix elements $|V_{cb}|$ and
$|V_{ub}|$, the Heavy Quark parameters (e.g. $b$ and $c$--quark masses),
QCD form factors, QCD dynamics, new physics, etc.

In the following, we provide averages of exclusive and inclusive
 branching fractions, the product of $|V_{cb}|$ and the form factor
 normalisation $F(1)$ and $G(1)$ for ${B}^{0}\to D^{*-}\ell^{+}\nu$ and
${B}^{0}\to D^{-}\ell^{+}\nu$ decays, respectively, and $|V_{ub}|$ as determined from
inclusive and exclusive measurements of $\B\to X_u \ell^+ \nul$ decays.
We will compute Heavy Quark parameters and extract QCD form factors
for ${B}^{0}\to D^{*-}\ell^{+}\nu$ decays.
Throughout this section, charge conjugate states are implicitly included, 
unless otherwise indicated.

Brief descriptions of all parameters
and analyses (published or preliminary) relevant for the
determination of the combined results are given.  The descriptions are
based on the information available on the web page at\\
 \centerline{\tt http://www.slac.stanford.edu/xorg/hfag/semi/EndOfYear07}
A description of the technique employed for calculating averages
was presented in the previous update~\cite{hfag_hepex_endof2006}. 
Asymmetric errors have been introduced in the current averages
for $\B\to X_u\ell^+\nul$ decays to take into account theoretical asymmetric errors. 


We thank U. Aglietti, I. Bigi, G. Ferrera, P. Gambino, E. Gardi, P.
Giordano, Z. Ligeti, M. Neubert, G. Ricciardi, and N. Uraltsev for 
useful discussions and for providing the theory codes.

\subsection{Common set of input parameters}
\label{sec:inputparams}
In the  combination of the  published results, the central  values and
errors are rescaled to a common set of input parameters, summarized in
Table~\ref{tab:common.param}   and   provided   in   the   file   {\tt
common.param} (accessible from the  web-page).  All measurements with a
dependence  on any  of these  parameters are  rescaled to  the central
values  given  in  Table~\ref{tab:common.param},  and their  errors are
recalculated based on the errors provided in the column ``Excursion''.
The detailed dependence for  each measurement  is contained  in files
(provided by the experiments) accessible from the web-page.
In the following tables, both the published and rescaled results
are presented.
Some of the (older) measurements are subject to considerables adjustments
due to the rescaling.
\begin{table}[!htb]
\caption{Common input  parameters for the  combination of semileptonic
$B$    decays.   Most    of    the   parameters    are   taken    from
Ref.~\cite{PDG_2007}.  This  table  is  encoded in  the  file  {\tt
common.param}. The units are picoseconds for lifetimes and percentage
for branching fractions.}
\begin{center}
\begin{tabular}{|l|c|c|l|}\hline
Parameter    &Assumed Value  &Excursion             &Description\\
\hline\hline 
rb           &$21.629$       &$\pm0.066$            &$R_b$\\
bdst         &$1.27$         &$\pm0.021$            &$\cbf(\B\to \Dstarb\tau\nu)$\\
bdsd         &$1.62$         &$\pm0.040$            &$\cbf(\B\to \Dstarb D)$\\
bdst2        &$0.65$         &$\pm0.013$            &$\cbf(b\to \Dstarb \tau)$ (OPAL incl) \\
bdsd2        &$4.2$          &$\pm1.5$              &$\cbf(b\to \Dstarb D)$ (OPAL incl)\\
bdsd3        &$0.87$         &${}^{+0.23}_{-0.19}$  &$\cbf(b\to \Dstarb D)$ (DELPHI incl)\\
xe           &$0.702$        &$\pm0.008$            &$B$ fragmentation:
$\langle E_B\rangle/E_{\rm beam}$\\
\hline
\end{tabular}
\end{center}
\label{tab:common.param}
\end{table}

\subsection{Exclusive CKM-favored decays}
\label{slbdecays_b2cexcl}
Averages are provided for the branching fractions
$\cbf(\B \to \bar{D} \ell^+{\nu}_{\ell})$ and $\cbf(\B \to \bar{D}^* \ell^+{\nu}_{\ell})$.  
For the $\bar{D}^{(*)}\pi$ excited states, averages are computed for the inclusive branching
fractions $\cbf({\B}\to \bar{D}^{(*)}\pi \ell^+{\nu}_{\ell})$, and for the 
product $\cbf(B^+ \to \bar{D}_1^0(D^{*-}\pi^+)\ell^+{\nu}_{\ell})
\times \cbf(\bar{D}_1^0 \to D^{*-}\pi^+)$ and $\cbf(B^+ \to \bar{D}_2^0(D^{*-}\pi^+)\ell^+{\nu}_{\ell})
\times \cbf(\bar{D}_2^0 \to D^{*-}\pi^+)$.
In addition,
averages are provided for
$F(1)\vcb$ vs $\rho^2$, where $F(1)$ and $\rho^2$
are the normalization and slope of the form factor at zero
recoil in ${B}^{0}\to D^{*-}\ell^{+}\nu$ decays, and for the corresponding quantities
$G(1)\vcb$ vs $\rho^2$ in ${B}^{0}\to D^{-}\ell^{+}\nu$ decays.

\mysubsubsection{$\B \to \bar{D} \ell^+{\nu}_{\ell}$}
\label{slbdecays_dlnu}

The average branching fraction $\cbf(\B \to \bar{D} \ell^+{\nu}_{\ell})$ 
is determined by the
combination of the results provided in Table~\ref{tab:dplnu} and
~\ref{tab:d0lnu}, for ${B}^0 \to D^- \ell^+{\nu}_{\ell}$ and 
$B^+ \to \bar{D}^0\ell^+{\nu}_{\ell}$, respectively.
The branching fractions are obtained from the integral over the measured differential 
 decay rates,
 apart for the \babar\ results, for which the semileptonic $B$ signal yields are
 extracted from a fit to the missing mass squared in a sample of fully
 reconstructed \BB\ events. 
 
Figure~\ref{fig:brdl} shows the measurements and the resulting average.

\begin{table}[!htb]
\caption{
Average of the branching fraction $\cbf({B}^{0}\to D^{-}\ell^{+}\nu)$ and individual
results. }
\begin{center}

\end{center}
\label{tab:dplnu}
\end{table}

\begin{table}[!htb]
\caption{Average of the branching fraction $\cbf(B^+ \to \bar{D}^0
\ell^+{\nu}_{\ell})$ and individual
results. }
\begin{center}
%
  \caption{Average branching fraction  of exclusive semileptonic $B$ decays
(a) ${B}^0 \to D^- \ell^+{\nu}_{\ell}$ and (b) 
$B^+ \to \bar{D}^0 \ell^+{\nu}_{\ell}$ and individual
  results.}
  \label{fig:brdl}
 \end{center}
\end{figure}

The average for $G(1)\vcb$ is determined by the two-dimensional
combination of the results provided in Table~\ref{tab:vcbg1}.
Figure~\ref{fig:vcbg1} (a)
provides a one-dimensional projection for illustrative purposes,
and figure~\ref{fig:vcbg1} (b) shows the average $G(1)\vcb$ and the
measurements included in the average. 

\begin{table}[!htb]
\caption{Average  of $G(1)\vcb$  determined  in the  decay ${B}^{0}\to D^{-}\ell^{+}\nu$ and
individual  results. The  fit  for the  average  has $\chi^2/\dof  =
0.3/4$.   The total  correlation between  the average  $G(1)\vcb$ and
$\rho^2$ is $0.93$.}
\begin{center}
\begin{tabular}{|l|c|c|}\hline
Experiment &$G(1)\vcb [10^{-3}]$ (rescaled)  &$\rho^2$ (rescaled) \\ 
           &$G(1)\vcb [10^{-3}]$ (published) &$\rho^2$ (published) \\
\hline\hline 
ALEPH~\hfill\cite{Buskulic:1996yq}  &$38.8 \pm11.8_{\rm stat} \pm6.2_{\rm syst}$   &$0.95 \pm0.98_{\rm stat} \pm0.36_{\rm syst}$ \\
                                    &$31.1 \pm9.9_{\rm stat}  \pm8.6_{\rm syst}$&$0.70 \pm0.98_{\rm stat} \pm0.50_{\rm syst}$ \\
\hline
CLEO ~\hfill\cite{Bartelt:1998dq}   &$44.8 \pm5.9_{\rm stat} \pm3.45_{\rm syst}$    &$1.27 \pm0.25_{\rm stat} \pm0.14_{\rm syst}$ \\
                                    &$44.8 \pm6.1_{\rm stat} \pm3.7_{\rm syst}$  &$1.30 \pm0.27_{\rm stat} \pm0.14_{\rm syst}$ \\
\hline
\belle~\hfill\cite{Abe:2001yf}       &$40.07 \pm4.4_{\rm stat} \pm5.14_{\rm syst}$    &$1.12 \pm0.22_{\rm stat} \pm0.14_{\rm syst}$ \\
                                    &$41.1 \pm4.4_{\rm stat} \pm5.1_{\rm syst}$    &$1.12 \pm0.22_{\rm stat} \pm0.14_{\rm syst}$ \\
\hline 
{\bf Average }                      &\mathversion{bold}$42.3 \pm 4.5$       &\mathversion{bold}$1.17 \pm0.18$      \\
\hline 
\end{tabular}
\end{center}
\label{tab:vcbg1}
\end{table}

For a determination of \vcb, the form factor at zero recoil $G(1)$
needs to be computed.  A possible choice is
$G(1) = 1.074\pm0.018_{\rm{stat}}\pm0.016_{\rm{syst}}$~\cite{FNALpilnu},
resulting, once corrected by a factor 1.007 for QED effect, in

\begin{displaymath}
\vcb = (39.1 \pm 4.2_{\rm exp} \pm 0.9_{\rm theo}) \times 10^{-3},
\end{displaymath}
where the errors are from experiment and theory, respectively.

\begin{figure}[!ht]
 \begin{center}
  \unitlength1.0cm 
  \begin{picture}(14.,8.) 
   \put(  8.0,  0.0){\includegraphics[width=7.8cm]{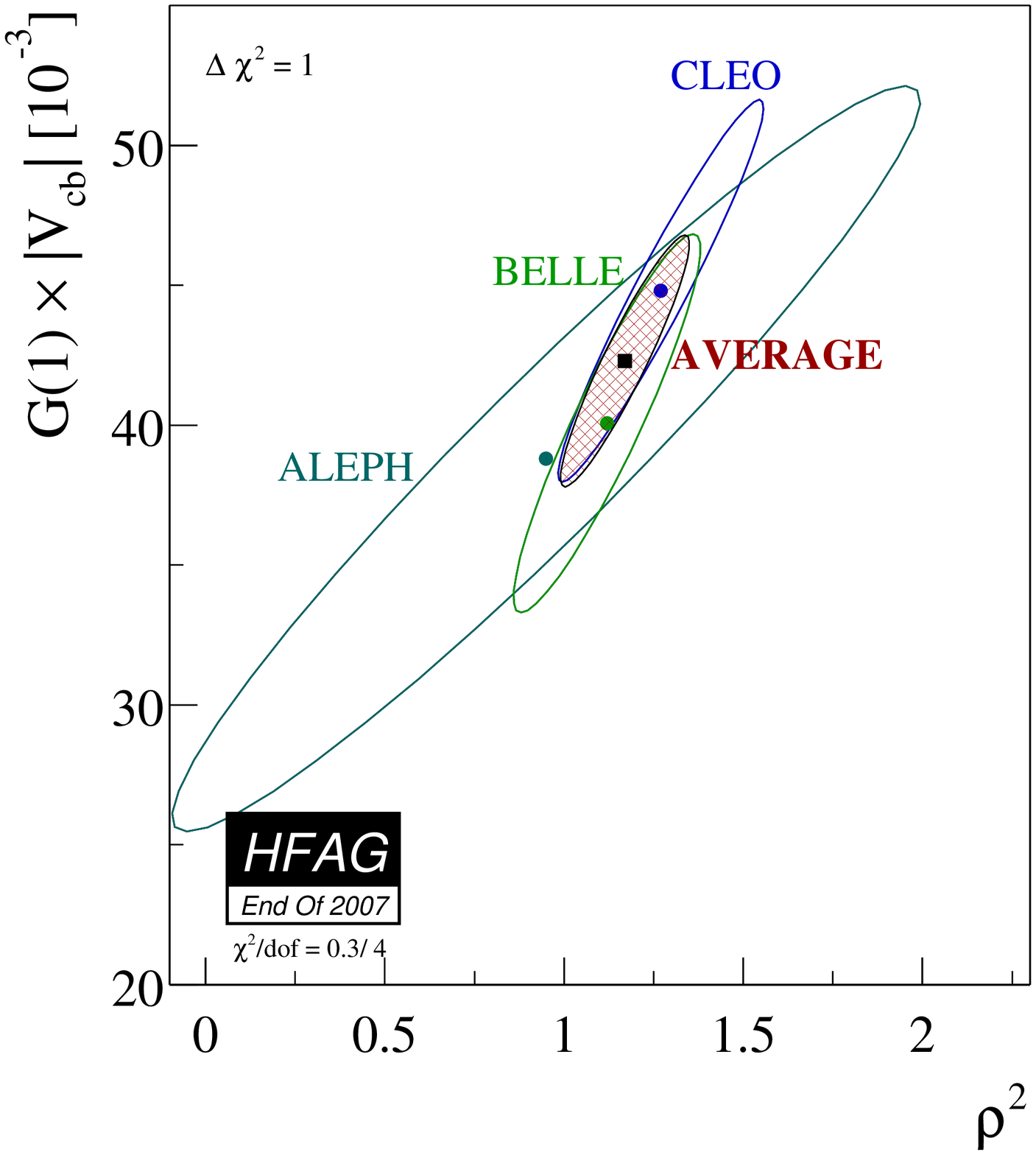}}
   \put( -0.5,  0.0){\includegraphics[width=7.8cm]{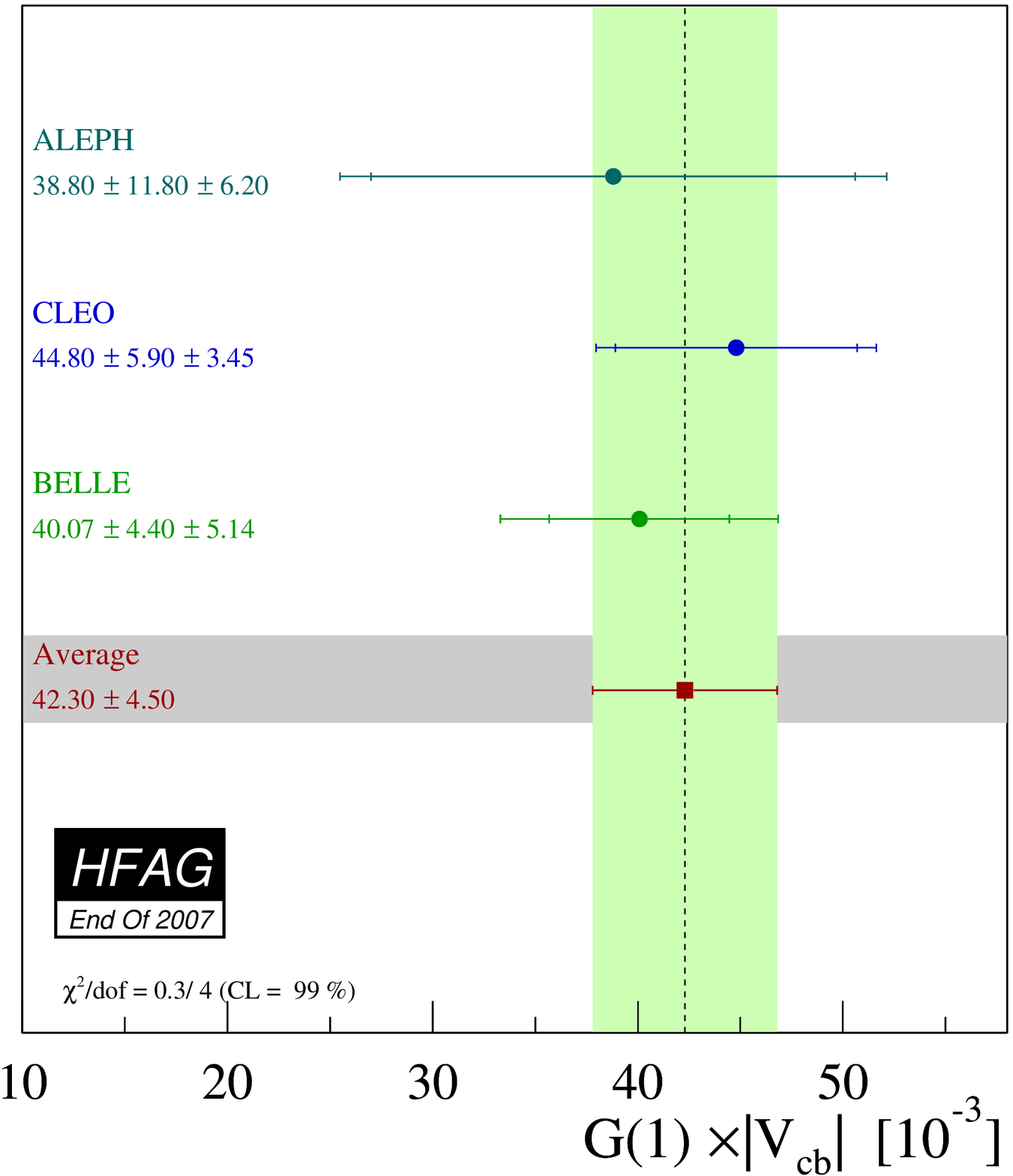}}
   \put(  5.5,  6.8){{\large\bf a)}}
   \put( 14.4,  6.8){{\large\bf b)}}
  \end{picture}
  \caption{(a)$G(1)\vcb$ average 
   and rescaled  measurements of exclusive ${B}^{0}\to D^{-}\ell^{+}\nu$ decays
   determined   in  a  two-dimensional   fit. 
   (b) $G(1)\vcb$   vs.  $\rho^2$ error ellipses corresponding to
   $\Delta\chi^2 = 1$. }
  \label{fig:vcbg1}
 \end{center}
\end{figure}

\mysubsubsection{$\B \to \bar{D}^* \ell^+{\nu}_{\ell}$}
\label{slbdecays_dstarlnu}

The average branching fraction $\cbf({\B} \to \bar{D}^{*} \ell^+{\nu}_{\ell})$ 
is determined by the
combination of the results provided in Table~\ref{tab:dstarlnu} and
~\ref{tab:dstar0lnu}, for ${B}^0 \to {D}^{*-} \ell^+{\nu}_{\ell}$ and 
$B^+ \to \bar{D}^{*0}\ell^+{\nu}_{\ell}$, respectively.
Advances have also been made in the determination of $|V_{cb}|$
from exclusive ${B} \to \bar{D}^* \ell^+ {\nu}_{\ell}$ 
decays with substantially improved
measurements of the form factor ratios $R_1$ and $R_2$.

\begin{table}[!htb]
\caption{Average branching fraction ${B}^{0}\to \bar{D}^{*+}\ell^{+}\nul$ and individual
  results, where ``excl'' and ``partial reco'' refer to full and
  partial reconstruction of the ${B}^{0}\to \bar{D}^{*+}\ell^{+}\nul$ decay, respectively.}
\begin{center} 

\end{center}
\label{tab:dstarlnu}
\end{table}

\begin{table}[!htb]
\caption{Average of the branching fraction ${B}^{0}\to \bar{D}^{*+}\ell^{+}\nul$ and individual
results. }
\begin{center}
\begin{tabular}{|l|c|c|}\hline
Experiment                                 &${B}^{0}\to \bar{D}^{*+}\ell^{+}\nul [\%]$ (rescaled) &${B}^{0}\to \bar{D}^{*+}\ell^{+}\nul [\%]$ (published) \\
\hline\hline 
ARGUS ~\hfill\cite{Albrecht:1992}        &$6.53 \pm1.60_{\rm stat} \pm0.56_{\rm syst}$ 
 &$6.6 \pm1.6_{\rm stat} \pm1.5_{\rm syst}$ \\
CLEO  ~\hfill\cite{Adam:2002uw}         &$6.64 \pm0.2_{\rm stat} \pm0.40_{\rm syst}$  &
$6.50 \pm0.20_{\rm stat} \pm0.43_{\rm syst}$ \\
\babar  ~\hfill\cite{Aubert:vcbExcl}       &$5.78 \pm0.15_{\rm stat}
\pm0.30_{\rm syst}$  &$5.83 \pm0.15_{\rm stat} \pm0.30_{\rm syst}$ \\
\hline 
{\bf Average}                              &\mathversion{bold}$6.09 \pm0.29$    
&\mathversion{bold}$\chi^2/\dof = 3.7/2$ (CL=$17\%$)  \\
\hline 
\end{tabular}
\end{center}
\label{tab:dstar0lnu}
\end{table}

For the $\cbf(B^+ \to \bar{D}^{*0} \ell^+{\nu}_{\ell})$, the average is performed
as for the ${B} \to \bar{D} \ell^+{\nu}_{\ell}$ modes, by scaling the
different measurements to a common set of input parameters. 
For the $\cbf({B}^0 \to {D}^{*-} \ell^+{\nu}_{\ell})$, the average 
is performed using a global $\chi^2$ fit built incorporating all the 
experimental determinations of $F(1)|V_{cb}|$, the slope
parameter $\rho^2$ and the other form-factor parameters $R_1$ and $R_2$.
Statistical correlations between measurements from the same experiment
are taken into account. The form-factor parametrization derived 
by Caprini, Lellouch and Neubert~\cite{CLN} is used. 

The $\chi^2$ minimization gives values for the form-factor parameters 
equal to $R_1=1.356 \pm 0.066$ and $R_2=0.851 \pm 0.039$. 
 The errors contain both the common and the experiment dependent 
 systematic uncertainties. 
With respect to the original measurements by CLEO~\cite{Duboscq:1995mv},
 $R_1=1.18\pm0.30\pm0.12$ and $R_2=0.71\pm0.22\pm0.07$, 
the accuracy on the form-factor parameters $R_1$ and $R_2$ given by the global
$\chi^2$ minimization has been
considerably improved due to a recent measurement by 
 \babar\ ~\cite{Aubert:2006mb}, with results that are consistent with the earlier ones, but
 considerably more precise.

The values extracted from the fit for $F(1)|V_{cb}|$ and the
form-factor parameters are used to obtain the 
$\cbf({B}^0 \to {D}^{*-} \ell^+{\nu}_{\ell})$  
branching fractions by computing 
the integral over the measured differential decay rates. The $\cbf({B}^0 \to
{D}^{*-} \ell^+{\nu}_{\ell})$ average is computed from these inputs, 
 apart from the \babar\ result~\cite{Aubert:vcbExcl}, 
 for which the semileptonic $B$ signal yields are
 extracted from a fit to the missing mass squared in a sample of fully
 reconstructed \BB\ events. This measurement is rescaled to the common set of
 input parameters, and then averaged with the other ones, neglecting at this
 stage remaining correlations.
Figure~\ref{fig:brdsl} shows the
measurements and the resulting average for the $\cbf({B} \to \bar{D}^{*} \ell^+
{\nu}_{\ell})$.  

The average for $F(1)\vcb$ is determined by the two-dimensional
combination of the results provided by the global $\chi^2$ minimization
described above: the corresponding values are reported 
in Table~\ref{tab:vcbf1}.  This
allows the correlation between $F(1)\vcb$ and $\rho^2$ to be
maintained. 
Figure~\ref{fig:vcbf1}(a) provides a one-dimensional
projection for illustrative purposes. 
Figure~\ref{fig:vcbf1}(b) shows the average
$F(1)\vcb$ and the measurements included in the
average. 
The largest systematic errors correlated between measurements are due to
uncertainties on: the ratio of production cross-sections
$\sigma_{b\bar{b}}/\sigma_{\rm{had}}$, the branching fractions $\cbf(D^0\to K^-\pi^+)$ and
$\cbf(D^0\to K^-\pi^+\pi^0)$, the correlated background from $D^{**}$,
and the $D^*$ form factor ratios $R_1$ and $R_2$.
Moreover, contributions from  $R_b$ and the $B^0$ fraction at
$\sqrt{s}=m_{Z^0}$ are taken into account for the measurements from 
the LEP experiments.
Together these uncertainties account for about two thirds of the
systematic error. In all the measurements the total systematic errors are
reduced with respect to the published values because the values and
uncertainties for parameters on which these measurements
depend, for example $R_1$ and $R_2$,  have since been better determined. 
 


\begin{table}[!htb]
\caption{Average of $F(1)\vcb$ determined in the decay \BzbDstarlnu\ and
individual  results, where ``excl'' and ``partial reco'' refer to full and
  partial reconstruction of the \BzbDstarlnu\ decay, respectively.
The  fit  for  the average  has  $\chi^2/\dof  =
33.9/17$ (CL=$1\%$).  The total  correlation between  the average  $F(1)\vcb$ and
$\rho^2$ is 0.24.}
\begin{center}

\end{center}
\label{tab:vcbf1}
\end{table}

For a determination of \vcb, the form factor at zero recoil $F(1)$
needs to be computed.  A possible choice is
$F(1) = 0.924{\pm 0.012}_{\rm{stat}} {\pm 0.019}_{\rm{syst}}$~\cite{Laiho:2007pn},
which takes into account the QED correction($+0.7\%$), resulting in

\begin{displaymath}
\vcb = (38.8 \pm 0.6_{\rm exp} \pm 0.9_{\rm theo}) \times 10^{-3},
\end{displaymath}
where the errors are from experiment and theory, respectively.

\noindent

\begin{figure}[!ht]
 \begin{center}
  \unitlength1.0cm 
  \begin{picture}(14.,8.0)  
   \put( -0.5,  0.0){\includegraphics[width=7.8cm]{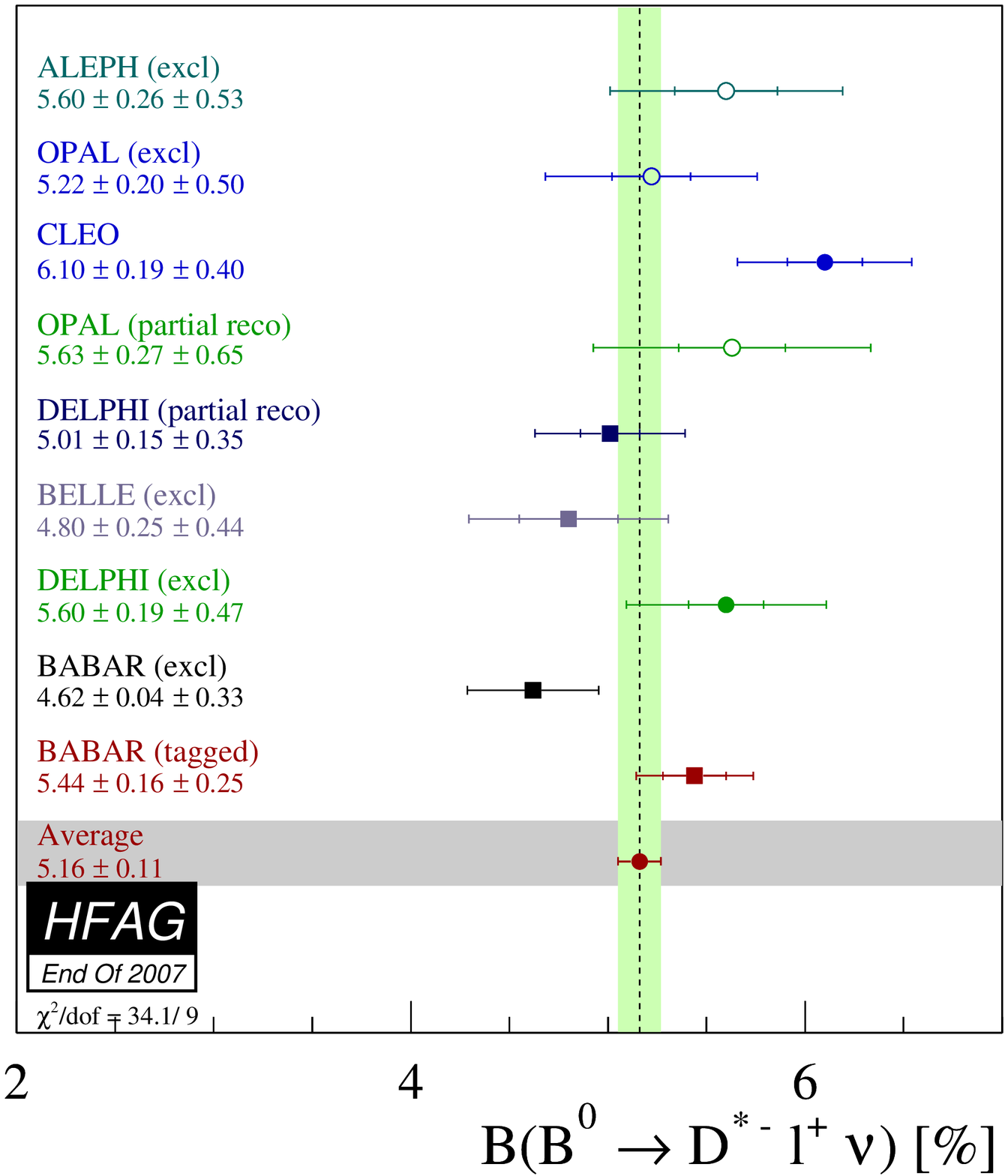}}

   \put(  8.0,  0.0){\includegraphics[width=7.5cm,height=8.4cm]{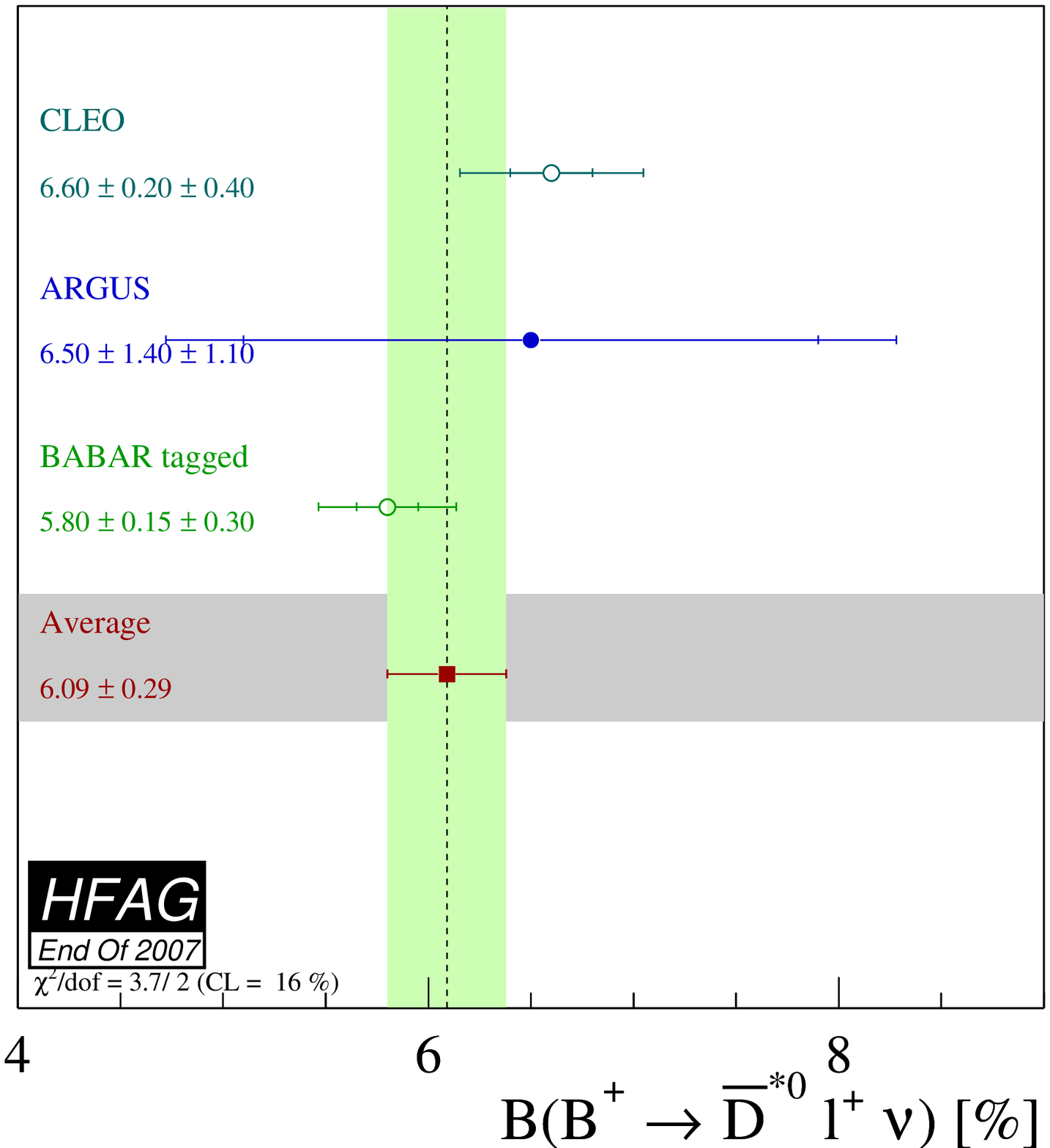}}
    
   \put(  5.5,  6.8){{\large\bf a)}}
   \put( 14.0,  6.8){{\large\bf b)}}
  \end{picture}
  \caption{Average branching fraction  of exclusive semileptonic $B$ decays
(a) ${B}^0 \to D^{*-} \ell^+ {\nu}_{\ell}$ and (b) 
$B^+ \to \bar{D}^{*0} \ell^+ {\nu}_{\ell}$ and individual
  results.  For Aleph and Delphi, the measurements of ${B}^{0}\to D^{*-}\ell^{+}\nu$ decays have been done both
with inclusive (``partial reco'') and exclusive (``excl'') analyses based 
on a partial and full reconstruction of the
${B}^{0}\to D^{*-}\ell^{+}\nu$ decay, respectively.}
  \label{fig:brdsl}
 \end{center}
\end{figure}

\begin{figure}[!ht]
 \begin{center}
  \unitlength1.0cm 
  \begin{picture}(14.,8.0)  
   \put(  8.0,  0.0){\includegraphics[width=7.8cm]{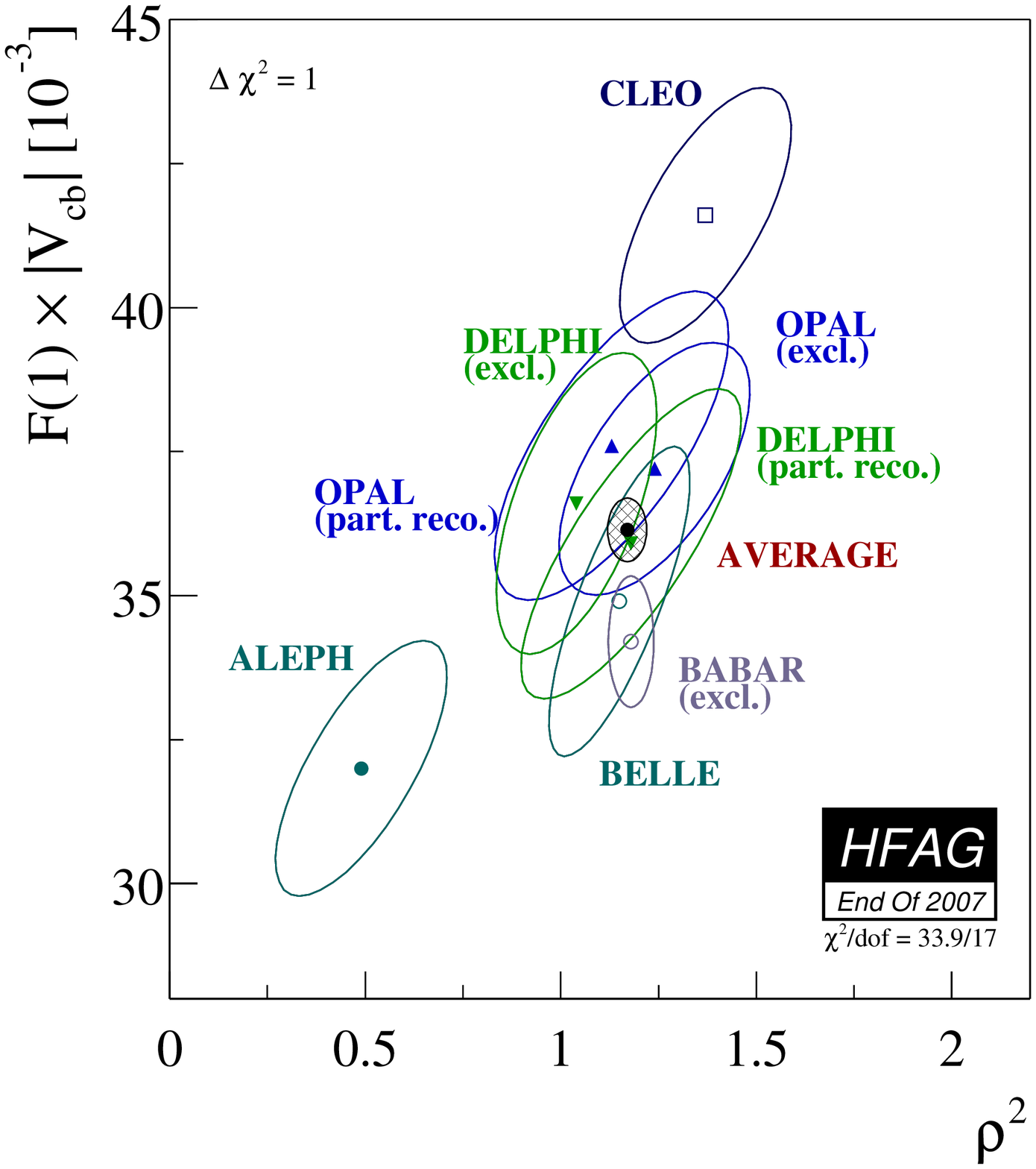}}
   \put( -0.5,  0.0){\includegraphics[width=7.8cm]{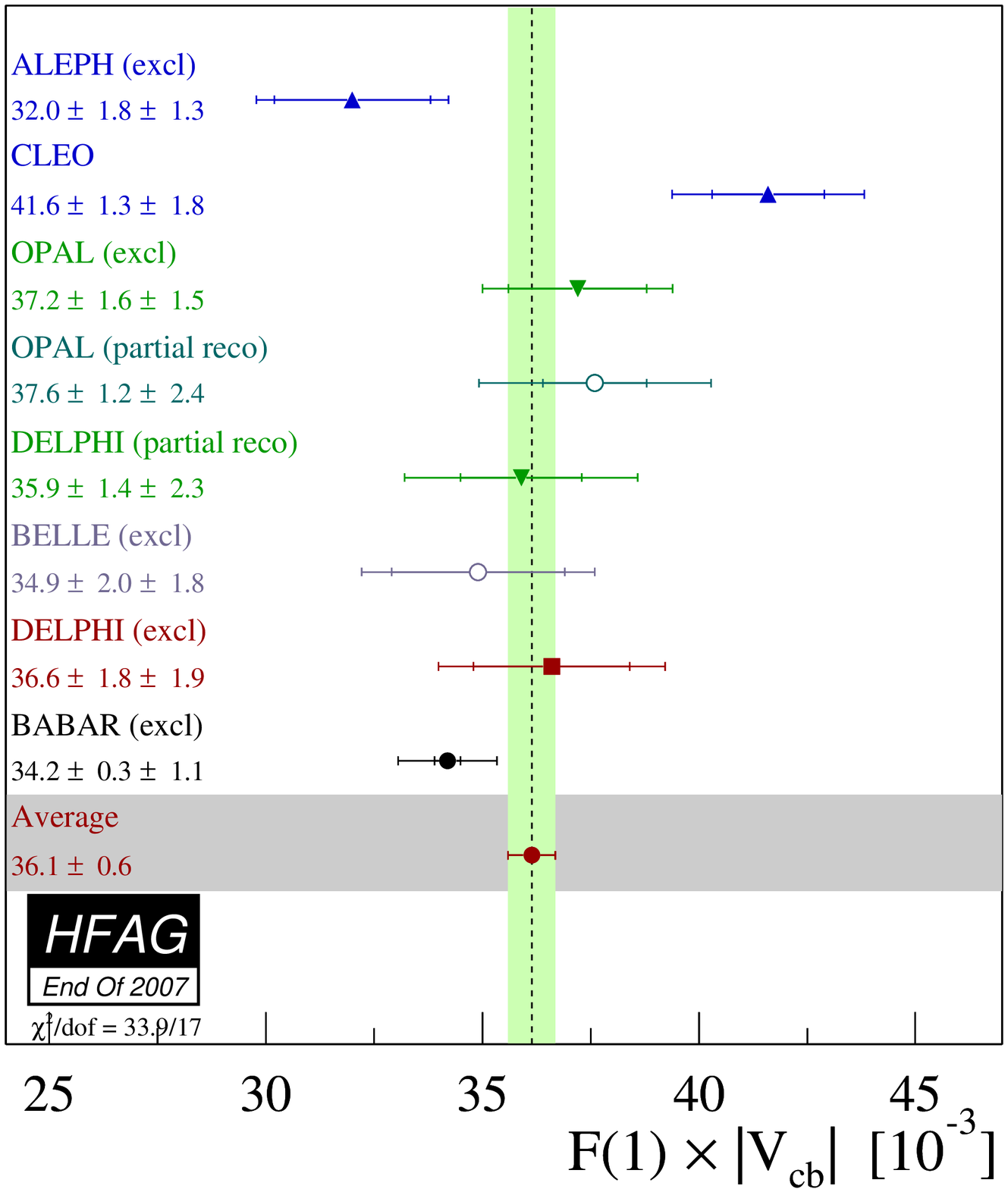}}
   \put(  5.5,  6.8){{\large\bf a)}}  
   \put( 14.4,  6.8){{\large\bf b)}}
   \end{picture} \caption{
     (a)  $F(1)\vcb$ average
   and   rescaled  measurements   of the exclusive ${B}^{0}\to D^{*-}\ell^{+}\nu$ decays
   determined   in  a  two-dimensional   fit, where ``excl'' 
   and ``partial reco'' refer to full and partial reconstruction.
    (b) $F(1)\vcb$   vs.  $\rho^2$ error ellipse for
   $\Delta\chi^2 = 1$ (CL=39\%).
  }  \label{fig:vcbf1} \end{center}
\end{figure}

\mysubsubsection{${B} \to \bar{D}^{(*)}\pi \ell^+{\nu}_{\ell}$}
\label{slbdecays_dpilnu}

The average inclusive branching fractions for ${B} \to \bar{D}^{*}\pi \ell^+{\nu}_{\ell}$
decays
, where no constrain is applied to the hadronic $D^{(*)}\pi$ system, 
are determined by the
combination of the results provided in Tables~\ref{tab:dpi1lnu} -
\ref{tab:dpi4lnu}  for 
${B}^0 \to \bar{D}^0 \pi^- \ell^+{\nu}_{\ell}$, ${B}^0 \to \bar{D}^{*0} \pi^-
\ell^+{\nu}_{\ell}$, 
$B^+ \to D^- \pi^+
\ell^+{\nu}_{\ell}$, and $B^+ \to D^{*-} \pi^+ \ell^+{\nu}_{\ell}$, respectively.
The measurements included in the average 
are scaled to a consistent set of input
parameters and their errors, see Section~\ref{sec:inputparams}.  

For both the \babar\ and Belle results, the $B$ semileptonic signal yields are
 extracted from a fit to the missing mass squared in a sample of fully
 reconstructed \BB\ events. 
 
Figure~\ref{fig:brdpil} illustrates the measurements and the
resulting average.

\begin{table}[!htb]
\caption{Average of the branching fraction ${B}^0 \to \bar{D}^0 \pi^- \ell^+{\nu}_{\ell}$ and individual
results. }
\begin{center}

\end{center}
\label{tab:dpi1lnu}
\end{table}

\begin{table}[!htb]
\caption{Average of the branching fraction ${B}^0 \to \bar{D}^{*0} \pi^-
\ell^+{\nu}_{\ell}$ and individual
results. }
\begin{center}
%
\end{center}
\label{tab:dpi2lnu}
\end{table}

\begin{table}[!htb]
\caption{Average of the branching fraction $B^+ \to D^- \pi^+
\ell^+{\nu}_{\ell}$ and individual
results. }
\begin{center}
%
\end{center}
\label{tab:dpi3lnu}
\end{table}

\begin{table}[!htb]
\caption{Average of the branching fraction $B^+ \to D^{*-} \pi^+ \ell^+{\nu}_{\ell}$ and individual
results. }
\begin{center}
%
  \caption{Average branching fraction  of exclusive semileptonic $B$ decays
(a) ${B}^0 \to \bar{D}^0 \pi^- \ell^+{\nu}_{\ell}$, (b) ${B}^0 \to \bar{D}^{*0} \pi^-
\ell^+{\nu}_{\ell}$, (c) 
$B^+ \to D^- \pi^+\ell^+{\nu}_{\ell}$, and (d) $B^+\to D^{*-} \pi^+ \ell^+{\nu}_{\ell}$.
The corresponding individual
  results are also shown.}
  \label{fig:brdpil}
 \end{center}
\end{figure}

\mysubsubsection{${B} \to \bar{D}^{**} \ell^+{\nu}_{\ell}$}
\label{slbdecays_dsslnu}

Averages are perfomed also for $B$ semileptonic decays into narrow orbitally-excited $D$
states, namely the $D_1(2420)$ and $D^*_2(2460)$~\cite{PDG_2007}. 
Due to the unknown $D_{1,2}$ branching fractions, the averages have been computed for the
product of branching fractions. The average are determined from the
combination of the results provided in Tables~\ref{tab:dss1lnu} and \ref{tab:dss2lnu} for 
$\cbf(B^+ \to \bar{D}_1^0(D^{*-}\pi^+)\ell^+{\nu}_{\ell})
\times \cbf(\bar{D}_1^0 \to D^{*-}\pi^+)$ and $\cbf(B^+ \to \bar{D}_2^0(D^{*-}\pi^+)\ell^+{\nu}_{\ell})
\times \cbf(\bar{D}_2^0 \to D^{*-}\pi^+)$, respectively. 
The measurements included in the average 
are scaled to a consistent set of input
parameters and their errors, see Section~\ref{sec:inputparams}.  

For both the B-factory and the LEP and Tevatron results, the $B$ semileptonic 
signal yields are
 extracted from a fit to the invariant mass distribution of the $D^{*-}\pi^+$ system.
Except CLEO and Belle results, the measurements 
 are for the final state ${B} \to \bar{D}_2(D^{*-}\pi^+)X \ell^+{\nu}_{\ell}$. 
Figure~\ref{fig:brdpil} shows the measurements and the
resulting average.

\begin{table}[!htb]
\caption{Average of the branching fraction $\cbf(B^+ \to \bar{D}_1^0(D^{*-}\pi^+)\ell^+{\nu}_{\ell})
\times \cbf(\bar{D}_1^0 \to D^{*-}\pi^+))$ and individual
results. }
\begin{center}

\end{center}
\label{tab:dss1lnu}
\end{table}

\begin{table}[!htb]
\caption{Average of the branching fraction $\cbf(B^+ \to \bar{D}_2^0(D^{*-}\pi^+)\ell^+{\nu}_{\ell})
\times \cbf(\bar{D}_2^0 \to D^{*-}\pi^+))$ and individual
results. }
\begin{center}
%
  \caption{Average of the product of branching fraction (a) 
  $\cbf(B^+ \to \bar{D}_1^0(D^{*-}\pi^+)\ell^+{\nu}_{\ell})
\times \cbf(\bar{D}_1^0 \to D^{*-}\pi^+)$ and (b) $\cbf(B^+ \to \bar{D}_2^0(D^{*-}\pi^+)\ell^+{\nu}_{\ell})
\times \cbf(\bar{D}_2^0 \to D^{*-}\pi^+)$
The corresponding individual
  results are also shown.}
  \label{fig:brdssl}
 \end{center}
\end{figure}

%
\subsection{Inclusive CKM-favored decays}
\label{slbdecays_b2cincl}

\subsubsection{Inclusive Semileptonic Branching Fraction for $B\to X\ell^+\nul$}
\label{ref:Xlcnu}
The branching fraction for inclusive decays $B \to X\ell^+\nul$
is presented, where $B$ corresponds to both $\Bz$ and $\Bu$.
We use measurements that require the momentum of
the prompt charged lepton $(p_{\mathrm{cms}})$ to be greater than $0.6$ GeV/$c$,
as measured in the rest frame of
either the $B$-meson or $\Upsilon(4S)$~\footnote{The difference in reference frames has a negligible impact.}. 
The measurements and average are given in Table~\ref{tab:brisltot} and
plotted in Figure~\ref{fig:b2xlnubf}. 
We determine the branching fraction over the full lepton spectrum using
an extrapolation factor derived from a global fit~\cite{Buchmuller:2005zv}
employed to extract HQET parameters in the kinetic
scheme~\cite{Gambino:2004qm,Benson:2004sg,Kolya} from measured moments of inclusive
distributions (see section~\ref{globalfitsKinetic}).
The global fit minimizes the dependence on the measurements used in this average.
The extrapolation factor is  $1.0495 \pm 0.0005 \pm 0.0010$, 
where the first error represent the experimental and theoretical
uncertainties obtained from the global fit. The second error
refers to an additional theoretical uncertainty on the ratio
$\Gamma_{\mathrm{sl}}(0.6)/\Gamma_{\mathrm{sl}}(0.0)$ of $0.1$\%,
on the recommendation of the authors of the fit.
The resultant full branching fraction is
$\cbf(B\to X\ell^+\nul) = (10.74 \pm 0.16)\%$.

\begin{table}[!htb]
\caption{Average of the partial semileptonic branching fractions
  $\cbf(B\to X\ell^+\nul)(p_{\mathrm{cms}}>0.6\,\rm{GeV}/c)$ and the
  full branching fraction extrapolated from the average. In
  parentheses we identify the type of tag used to identify $\B\Bbar$ events: 
  $e$-tag and $\ell$-tag refer to measurements with events tagged by an electron 
  and either an electron or muon, respectively,
  $B_{\rm{reco}}$-tag refers to analyses 
  with events tagged by a fully reconstructed hadronic $B$ decay.}

\begin{center}
\begin{tabular}{|l|c|c|c|}\hline
Experiment          &$\cbf(B\to X\ell^+\nul)[\%]$ (rescaled) &$\cbf(B\to X\ell^+\nul)[\%]$ (published)  \\
       & $(p_{\mathrm{cms}}>0.6\,\rm{GeV}/c)$ & $(p_{\mathrm{cms}}>0.6\,\rm{GeV}/c)$    \\
\hline\hline 
ARGUS ($e$-tag)~\hfill\cite{Albrecht:1993pu}       & $9.15  \pm0.50 \pm0.33$ & $9.10  \pm0.50 \pm0.40$ \\
\belle ($\ell$-tag)~\hfill\cite{Abe:2002du}        & $10.30 \pm0.11 \pm0.46$ & $10.24 \pm0.11 \pm0.46$ \\
CLEO  ($e$-tag)~\hfill\cite{Mahmood:2004kq}        & $10.23 \pm0.08 \pm0.22$ & $10.21 \pm0.08 \pm0.22$ \\
\babar ($e$-tag)~\hfill\cite{Aubert:2004td}        & $10.37 \pm0.06 \pm0.23$ & $10.36 \pm0.06 \pm0.23$ \\
\babar (\breco-tag)~\hfill\cite{Aubert:2006au}    & $10.03 \pm0.19 \pm0.33$  & $10.03 \pm0.19 \pm0.33$  \\
\belle (\breco-tag)~\hfill\cite{Urquijo:2006wd}    & $10.28 \pm0.18 \pm0.24$ & $10.28 \pm0.18 \pm0.24$ \\
\hline 
    {\bf Average at $(p_{\mathrm{cms}}>0.6\,\rm{GeV}/c)$}
    &\mathversion{bold}$10.23\pm0.15$ 
    & \mathversion{bold}$\chi^2/\dof = 4.2/5$ (CL=$52\%$)  \\
    \hline
    $\cbf_{tot}(\Bz/\Bu\to X\ell\nul)$ (\%)&
       $\mathbf{10.74 \pm 0.16}$ & \\ 
    \hline
\end{tabular}
\end{center}
\label{tab:brisltot}
\end{table}

\begin{figure}[!ht]
 \begin{center}
  \unitlength1.0cm 
  \begin{picture}(14.,8.0)  
   \put(  8.0,  0.0){\includegraphics[width=7.8cm]{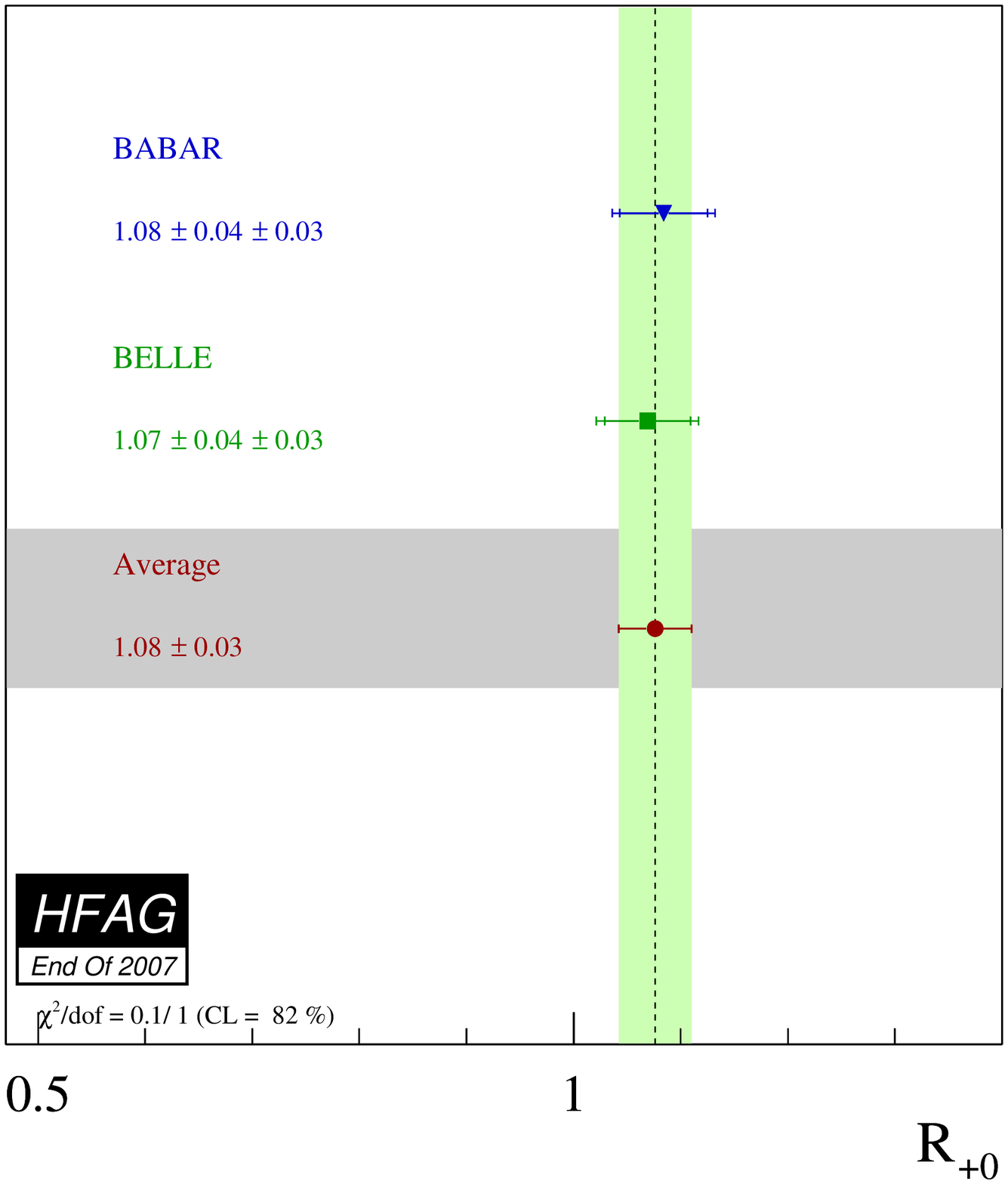}}
   \put( -0.5,  0.0){\includegraphics[width=7.8cm]{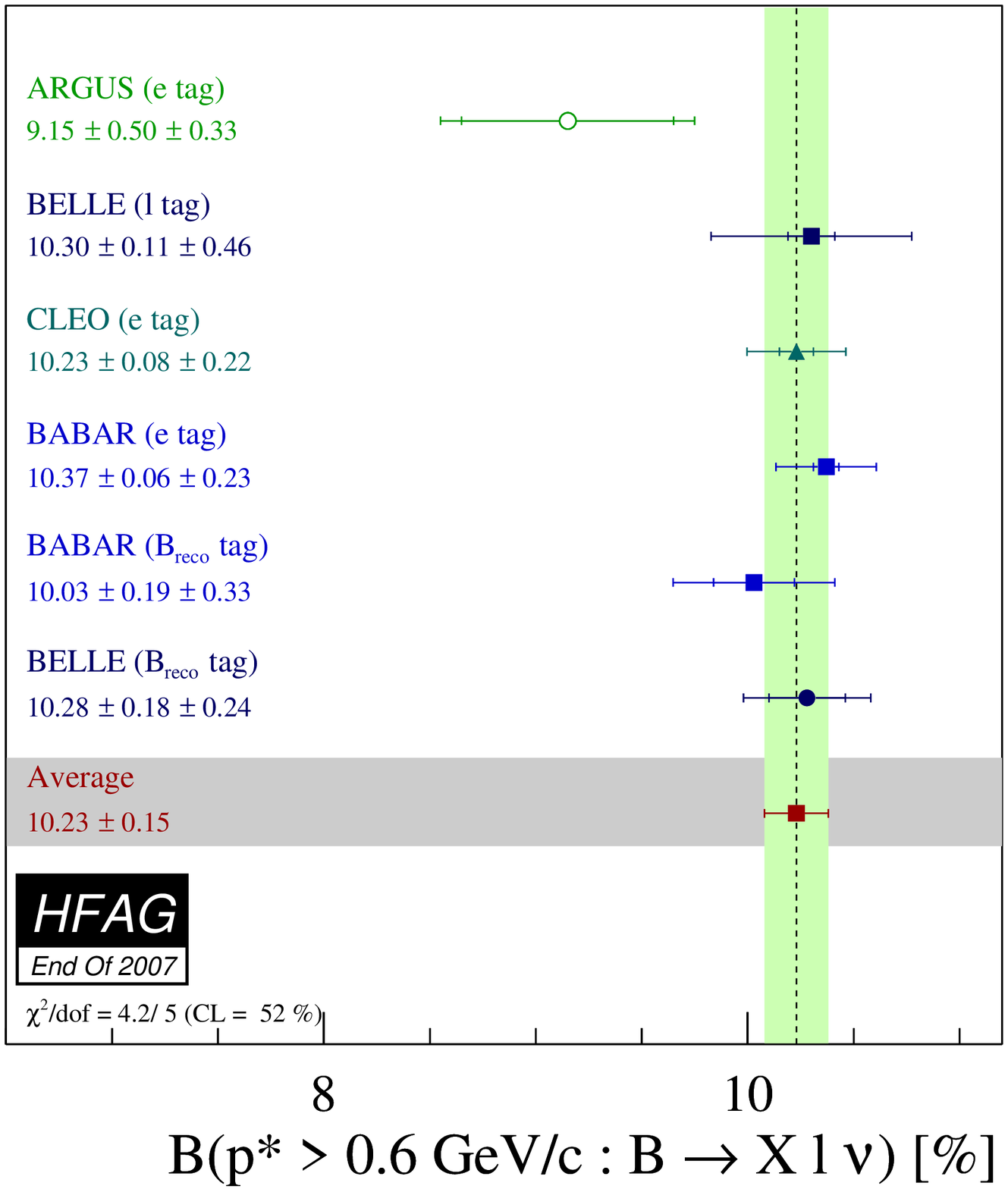}}
   \put(  5.5,  7.3){{\large\bf a)}}  
   \put( 14.4,  7.3){{\large\bf b)}}
   \end{picture} \caption{
(a) Measurements of $\cbf(B \to X\ell^+\nul)$ and their average. 
(b) Measurements of the ratio of the branching fractions $R_{+0}$ and their average.
}
\label{fig:b2xlnubf}
\end{center}
\end{figure}

\subsubsection{Ratio of $\cbf(\Bu\to X\ellp\nul)$ to $\cbf(\Bz\to X\ellp\nul)$}
The total width of semileptonic $B$-meson decays is
expected to be the same for both neutral and charged $B$ mesons.
Therefore the ratio of the branching fractions, $R_{+0}$, 
should be equivalent to the ratio of the $B$-meson lifetimes
$\tau_{B+}/\tau_{B^0}$, where
\[
R_{+0} = \frac{\cbf(\Bu\to X\ellp\nul)}{\cbf(\Bz\to X\ellp\nul)}.
\]
Recently, both \babar\ and Belle reported precise
measurements of $R_{+0}$, using a ``$B_{\rm{reco}}$''-tagged sample. The
measurements and average are listed in Table~\ref{tab:Rp0} and plotted in
Figure~\ref{fig:b2xlnubf}. The average, $1.076 \pm 0.034$, is in agreement with the
ratio of lifetimes, $1.073 \pm 0.008$ (this preprint).

\begin{table}[!htb]
\caption{Individual measurements and average of the ratio of the branching fractions $R_{+0}$.}
\begin{center}
\begin{tabular}{|l|c|}\hline
Experiment                                   &$R_{+0}$    \\
\hline\hline 
\babar~\hfill\cite{Aubert:2006au}   &$1.084\pm0.041\pm0.025$    \\
\belle~\hfill\cite{Urquijo:2006wd}   &$1.069\pm0.040\pm0.026$    \\
\hline 
    {\bf Average}
    & \mathversion{bold}$1.076 \pm 0.034$ \\
       & \mathversion{bold}$\chi^2/\dof = 0.05/1$  \\
      & {\bf CL=}\mathversion{bold}$82\%$ \\
    \hline
\end{tabular}
\end{center}
\label{tab:Rp0}
\end{table}

\subsubsection{Branching Fractions for $\Bu\to X\ellp\nul$ }
We provide averages of the branching fractions of $\Bu\to X\ellp\nul$ and
$\Bz\to X^-\ellp\nul$ separately,  
using the available measurements at the \FourS. We include the
measurements from CLEO~\cite{Artuso:1997we,Henderson:1991ks} and
ARGUS~\cite{Albrecht:1993gr},
as well as the latest measurements made by Belle and \babar\ that utilise
the full reconstruction $B$-meson tag~\cite{Aubert:2006au,Urquijo:2006wd}. 
In those cases, we extrapolate from the partial to full
branching fraction using a factor  derived from a global fit~\cite{Buchmuller:2005zv},
as in section~\ref{ref:Xlcnu}.
In contrast to the $B$
admixture average, averages are made of the full branching
fraction since CLEO and ARGUS do not present partial
branching fractions. The measurements and averages are given in
Tables~\ref{tab:brisltotbp} and~\ref{tab:brisltotb0} and plotted in
Figure~\ref{fig:brisltotbp0} for
$\Bu\to X^0\ellp\nul$ and $\Bz \to X^- \ell^+ \nul$, respectively.

\begin{table}[!htb]
  \caption{Individual measurements and average of the total
    semileptonic branching fraction
    $\cbf(\Bu\to X^0\ellp\nul)$. ``$\ell$-tag'' and ``\breco-tag'' indicate analysis experimental technique. No rescaling is applied.}
\begin{center}
\begin{tabular}{|l|c|}\hline
Experiment                                   &$\cbf_{tot}(\Bu\to X^0\ellp\nul)[\%]$    \\
\hline\hline 
CLEO  ($\ell$-tag)~\hfill\cite{Artuso:1997we}          &$10.25 \pm0.57 \pm0.66$    \\
\babar (\breco-tag)~\hfill\cite{Aubert:2006au}        &$10.90 \pm0.27 \pm0.39$    \\
\belle (\breco-tag)~\hfill\cite{Urquijo:2006wd}        &$11.17 \pm0.25 \pm0.28$     \\
\hline 
    {\bf Average}
    & \mathversion{bold}$10.99 \pm 0.28$ \\
       & \mathversion{bold}$\chi^2/\dof = 1.0/2$  \\
       &   {\bf CL=}\mathversion{bold}$61\%$ \\
    \hline
\end{tabular}
\end{center}
\label{tab:brisltotbp}
\end{table}

\begin{table}[!htb]
\caption{Individual measurements and average of the total semileptonic branching fraction $\cbf(\Bz\to X^-\ellp\nu)$. ``partial-tag'' and ``\breco-tag'' indicate analysis experimental technique. No rescaling is applied.}
\begin{center}
\begin{tabular}{|l|c|}\hline
Experiment & $\cbf_{tot}(\Bz\to X^-\ellp\nu)[\%]$   \\ 
\hline\hline 
CLEO  (partial-tag)~\hfill\cite{Henderson:1991ks}      &$9.9\pm3.0\pm0.9$ \\
ARGUS (partial-tag)~\hfill~\cite{Albrecht:1993gr}      &$9.3\pm1.1\pm 1.5$ \\
CLEO  (partial-tag)~\hfill\cite{Artuso:1997we}         &$10.78\pm0.60\pm0.69$ \\
\babar (\breco-tag)~\hfill\cite{Aubert:2006au}        &$10.14\pm0.28\pm0.33$ \\
\belle (\breco-tag)~\hfill\cite{Urquijo:2006wd}        &$10.46\pm0.30\pm0.23$ \\
\hline 
    {\bf Average} & \mathversion{bold}$10.33\pm0.28$ \\
   &  \mathversion{bold}$\chi^2/\dof =0.9/4$ \\
   &   {\bf CL=} \mathversion{bold}$92\%$ 
       \\
    \hline
\end{tabular}
\end{center}
\label{tab:brisltotb0}
\end{table}

\begin{figure}[!ht]
 \begin{center}
  \unitlength1.0cm 
  \begin{picture}(14.,8.0)  
   \put( -0.5,  0.0){\includegraphics[width=7.8cm]{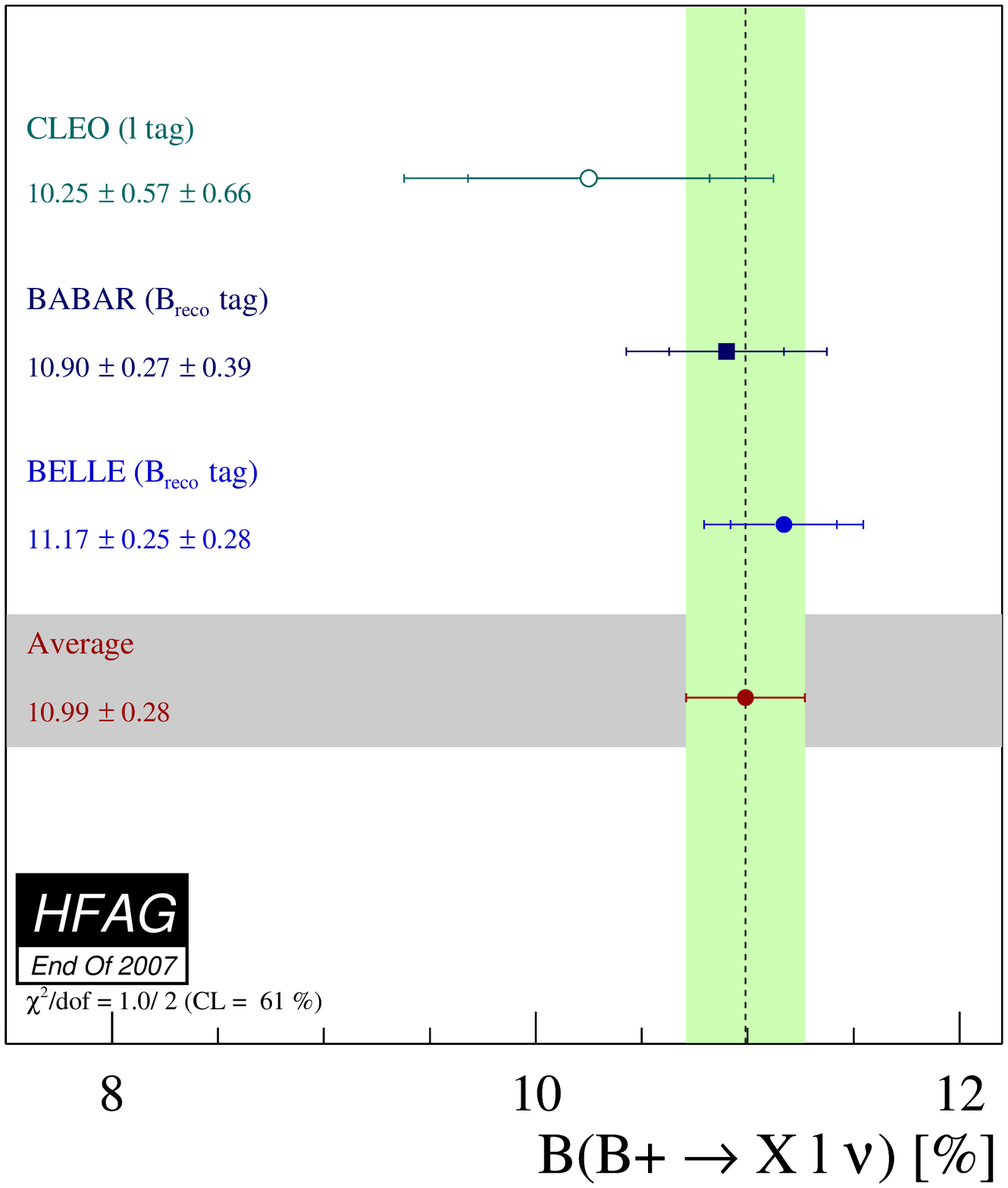}}
   \put(  8.0,  0.0){\includegraphics[width=7.8cm]{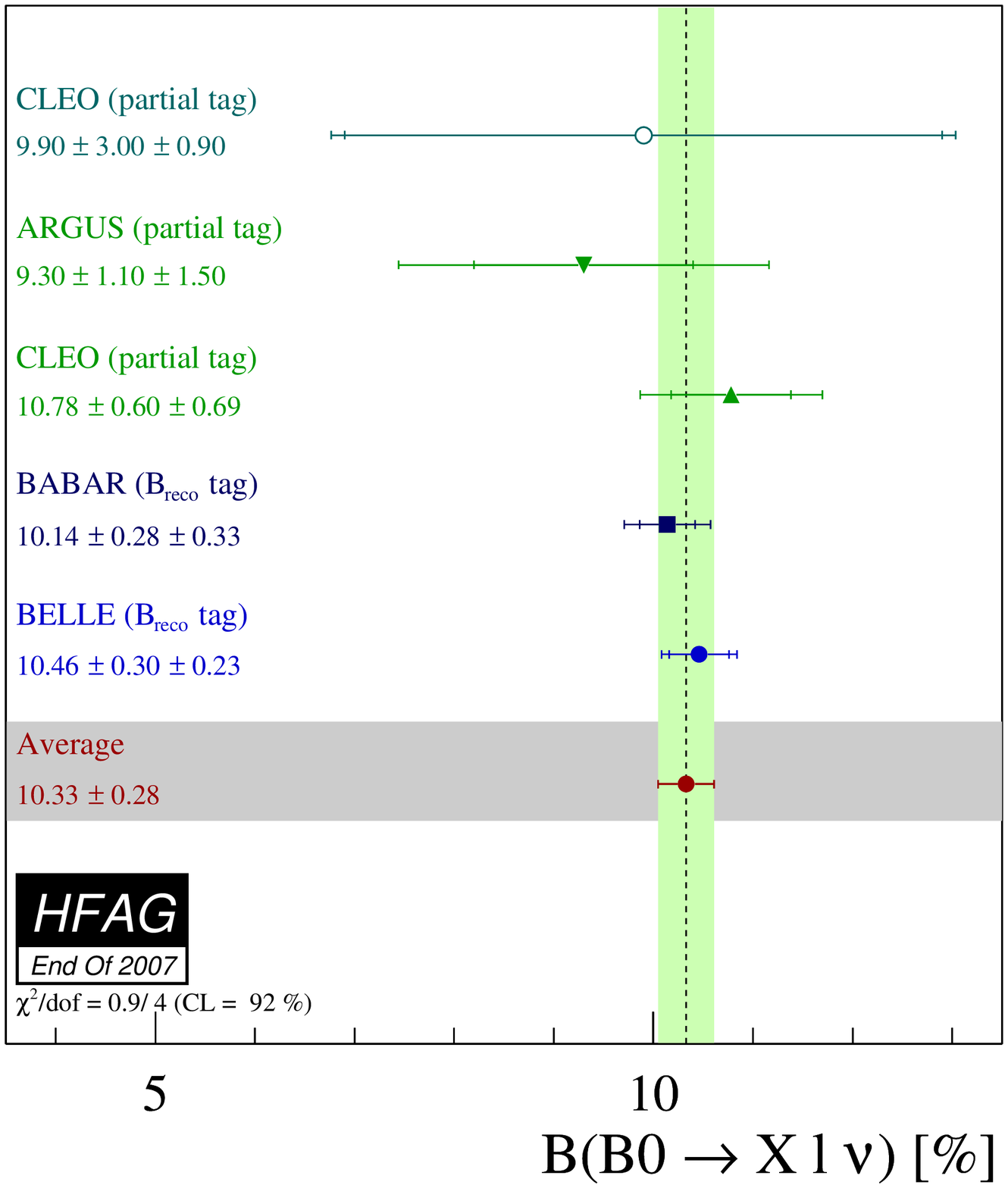}}
   \put(  5.5,  7.3){{\large\bf a)}}  
   \put( 14.4,  7.3){{\large\bf b)}}
   \end{picture} \caption{(a) Measurements of the total semileptonic 
branching fraction $\cbf(\Bu\to X\ellp\nul)$ and their average. 
(b) Individual measurements and average of the total semileptonic branching fraction 
$\cbf(\Bz\to X^-\ellp\nu)$. ``$\ell$-tag'', ``partial-tag'' and ``\breco-tag'' indicate 
the event tag used in different analyses.} 
\label{fig:brisltotbp0}
\end{center}
\end{figure}


\subsubsection{\vcb\, Determined from $B\to X \ellp \nul$}
The magnitude of the CKM matrix $|V_{cb}|$ can be determined from the branching fraction
of inclusive charmed semileptonic $B$-meson decays $B\to X \ellp \nul$ and
with parameters that describe the motion of the $b$-quark in the
$B$-meson. These parameters, within the framework of the Heavy Quark
Expansion (HQE), include the $b$-quark mass, $m_b$. Phenomenology  
of these decays is reviewed in many papers~\cite{b2xintros}. In practice $|V_{cb}|$,
$m_b$, and other parameters are determined simultaneously
from a {\it global} fit to measured moments of inclusive leptons
spectrum, the hadron mass spectrum in semileptonic decays, and the
inclusive photon spectrum in radiative $B$--meson penguin decays.
The moments are measured as a function of a minimum lepton of
photon energy. 
To date three types of global fits have been performed; they 
differ in the choice of scheme, either {\it pole}, ``1S'' or
{\it kinetic}, which are characterized by different definitions
of the $b$-quark mass $m_b$. 
However, fits in the {\it pole} scheme are disfavoured due
to the poor convergence of series expansions in that scheme.
In the following, we present the most up-to-date fits, which are performed
in the {\it kinetic} and ``1S'' schemes. They are based on a common
set of measurements. 

\subsubsection{HQE fit in the kinetic scheme}
\label{globalfitsKinetic}

This fit to the moment measurements relies on HQE calculations in the kinetic scheme 
presented in Refs.~\cite{Gambino:2004qm,Benson:2004sg,Kolya}, including lower energy dependent 
perturbative corrections to the hadron moments~\cite{Uraltsev:2004in,Trott:2004xc,Aquila:2005hq}.
The inclusive spectral moments of $B\to X_c\ell^+\nul$ decays have been calculated in the kinetic scheme up to $\mathcal{O}(1/m_b^3)$ and $\mathcal{O}(1/(\alpha_s^2))$. The theoretical expressions for the truncated moments are given in terms of HQE parameters, and coefficients determined by theory, which are functions of the lower energy cut. 
As $\mu_{G}^2$ and $\rho_{LS}^3$ can be estimated from the $B^* - B$ mass 
splitting and heavy-quark sum rules, respectively, we impose Gaussian error 
constraints of $\mu_{G}^2=0.35\pm0.07 \gev^2$ and $\rho_{LS}^3=-0.15\pm 0.10 \gev^3$ 
on these parameters as advocated in Ref.~\cite{Gambino:2004qm}.
The total of free parameters is then eight.
The result of the combined fit of the HQEs to all moment measurements 
from \babar~\cite{Aubert:2004td,BABARHAD,BABARSEMI,BABARINCL}, Belle~\cite{BELLEHAD,BELLELEP,BELLEBSGNEW}, CDF~\cite{CDF}, CLEO~\cite{CLEOHAD,CLEOBSG} and DELPHI~\cite{DELPHI}
is shown in Table~\ref{tab:def_result}.
To asses the consistency of the calculations and the measurements for the 
two different decay processes, $B \ra X_c \ellp \nul$ and 
$B \ra X_s \gamma$, we have carried out 
separate fits to $B \ra X_c \ellp \nul$ moments and to photon moments only. \\

The fit to $B \ra X_c \ellp \nul$ moments only results in:
\begin{eqnarray}
\vcb &=& 41.68 \pm 0.39_{\rm fit} \pm 0.58_{\Gamma_{SL}}  \times 10^{-3} \,, \nonumber \\
m_b  &=&  4.677 \pm 0.053_{\rm fit} \gev    \,, \nonumber \\
m_c  &=&  1.285 \pm 0.078_{\rm fit} \gev \,,\nonumber \\
\mu_{\pi}^2 &=& 0.387 \pm  0.039_{\rm fit} \gev^2 \,.\nonumber
\end{eqnarray}
From the fit to $B \ra X_s \gamma$ moments only we obtain:
\begin{eqnarray}
m_b &=  4.56 ^{+ 0.08}_{- 0.14} \gev   \,, \nonumber  \\
\mu_{\pi}^2 &= 0.44  ^{+ 0.30}_{- 0.17}  \gev^2 \,.\nonumber
\end{eqnarray}

{\small
\begin{table*}[htbp]
\caption[]{\label{tab:def_result} Results for the combined fit to all moments with experimental and theoretical uncertainties.
For \Vcb\ we add an additional theoretical error stemming from the uncertainty in the expansion for $\Gamma_{\rm SL}$ of 1.4\%. Below the fit results the correlation matrix is shown.}
\begin{center}
\begin{sideways}
\normalsize
\begin{tabular}{c|cccccccc}\hline
  & \multicolumn{8}{c}{OPE FIT RESULT: $\chi^2/N_{dof}$ =39.1/62} \\
 $B \ra X_c \ellp \nul$ & $|V_{cb}|$ & $m_b$ & $m_c$ & $\mu_{\pi}^2$ & $\rho_D^3$ & $\mu_{G}^2$ & $\rho_{LS}^3$ & $BR_{c\ell^+{\nul}}$ \\ 
 + $B \ra X_s \gamma$ & ($\times 10^{-3}$) & ($\gev$) & ($\gev$) & ($\gev^2$) & ($\gev^3$) & ($\gev^2$) & ($\gev^3$) & ($\%$)    \\ \hline 
RESULT       &41.91 &4.613 &1.187 &0.408 &0.191 &0.261 &-0.195 &10.64    \\     
$\Delta$ exp & 0.19 &0.022 &0.033 &0.017 &0.008 &0.019 & 0.052 & 0.09 \\ 
$\Delta$ HQE & 0.28 &0.027 &0.040 &0.031 &0.019 &0.035 & 0.068 & 0.07 \\ 
$\Delta$ \gsl& 0.59 &      &      &      &      &      &       &      \\ \hline 

$|V_{cb}|$    &  1.000 & -0.450 & -0.315 &  0.495 &  0.311 & -0.275 &  0.070 &  0.674 \\
$m_b$         &        &  1.000 &  0.962 & -0.525 & -0.225 & -0.226 & -0.211 &  0.121 \\
$m_c$         &        &        &  1.000 & -0.536 & -0.310 & -0.448 & -0.100 &  0.152 \\
$\mu_{\pi}^2$ &        &        &        &  1.000 &  0.750 &  0.230 &  0.071 &  0.126 \\
$\rho_D^3$    &        &        &        &        &  1.000 &  0.185 & -0.507 &  0.123 \\
$\mu_{G}^2$   &        &        &        &        &        &  1.000 & -0.034 & -0.160 \\
$\rho_{LS}^3$ &        &        &        &        &        &        &  1.000 & -0.070 \\
$BR_{c\ell^+{\nul}}$  &        &        &        &        &        &        &        &  1.000 \\ \hline
\end{tabular}
\end{sideways}
\end{center}
\end{table*}
}

A comparison of results from the combined fit with those obtained from fits
to $B \ra X_c \ellp \nul$ and $B \ra X_s \gamma$ moments separately can be 
found in Figure~\ref{fig:def_result} where the $\Delta \chi^2 =1$ contours 
for the fit results are shown in the ($m_b,\vcb$) and 
($m_b,\mu_{\pi}^2$) planes. 
However, as they are not sensitive to all the heavy quark parameters, 
all but $m_b$ and $\mu_{\pi}^2$ are fixed to the result obtained from 
the combined fit. 
It can be seen that the inclusion of the photon 
energy moments adds additional sensitivity to the $b$-quark mass $m_b$.
The values for the 
b--quark mass $m_b$  and $\mu_{\pi}^2$ from the two fits agree within
1.3$\sigma$ and 0.3$\sigma$, respectively.

It is of interest to compare the extracted $b$--quark mass in the 
kinetic scheme with other determinations in the commonly used $\overline {\rm MS}$  scheme.
The translation between the kinetic and
$\overline {\rm MS}$ masses to two loop accuracy and including 
the BLM part of the $\alpha_s^3$ corrections~\cite{Kolya} leads to
\begin{eqnarray}
\overline{m_b}(\overline{m_b})=4.22\pm 0.04\,\gev .\nonumber 
\end{eqnarray}

\begin{figure}[!htbp]
 \begin{center}
  \unitlength1.0cm 
  \begin{picture}(14.,8.0)  
   \put(  8.0,  0.0){\includegraphics[width=8.0cm]{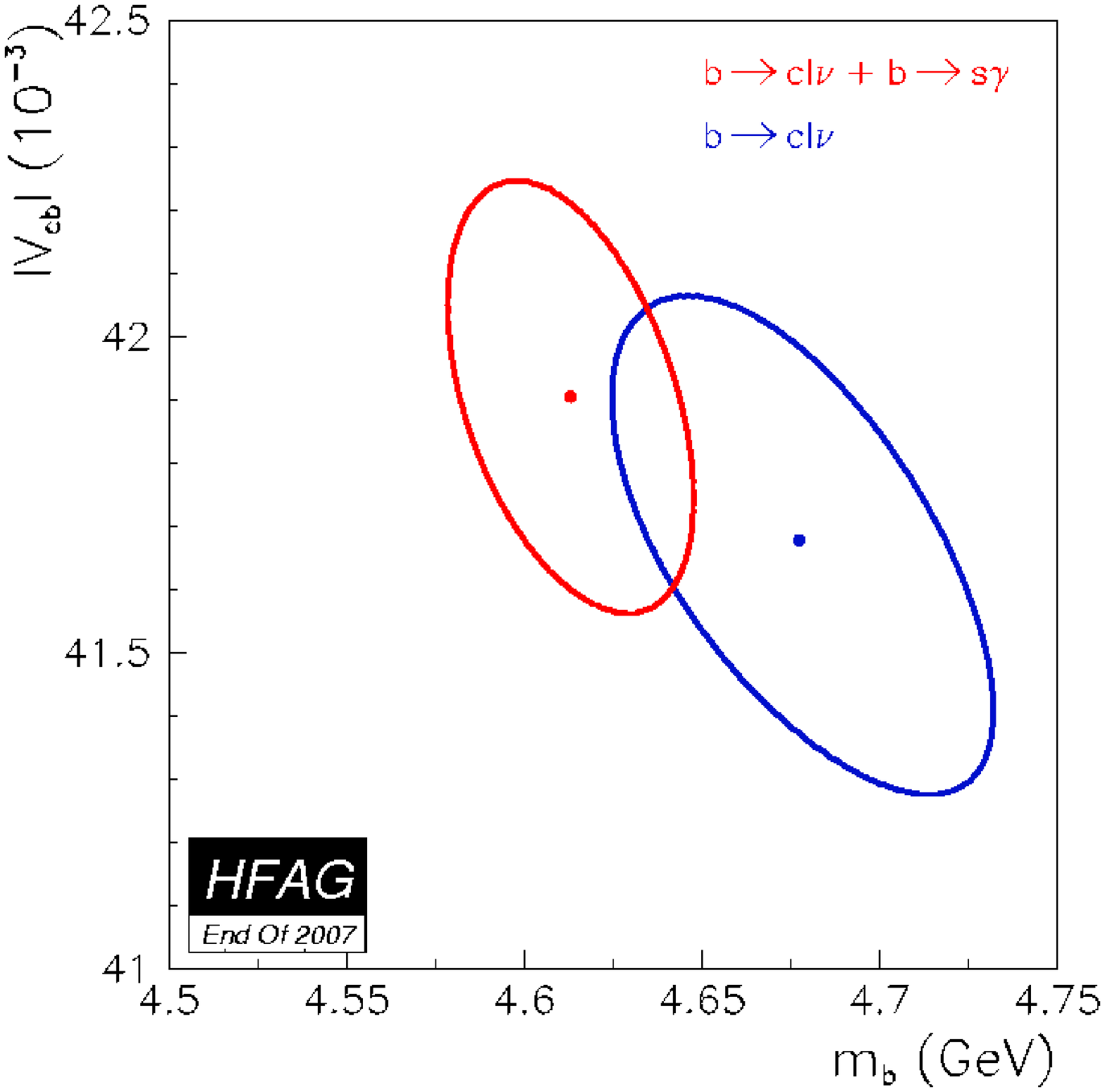}}
   \put( -0.5,  0.0){\includegraphics[width=8.0cm]{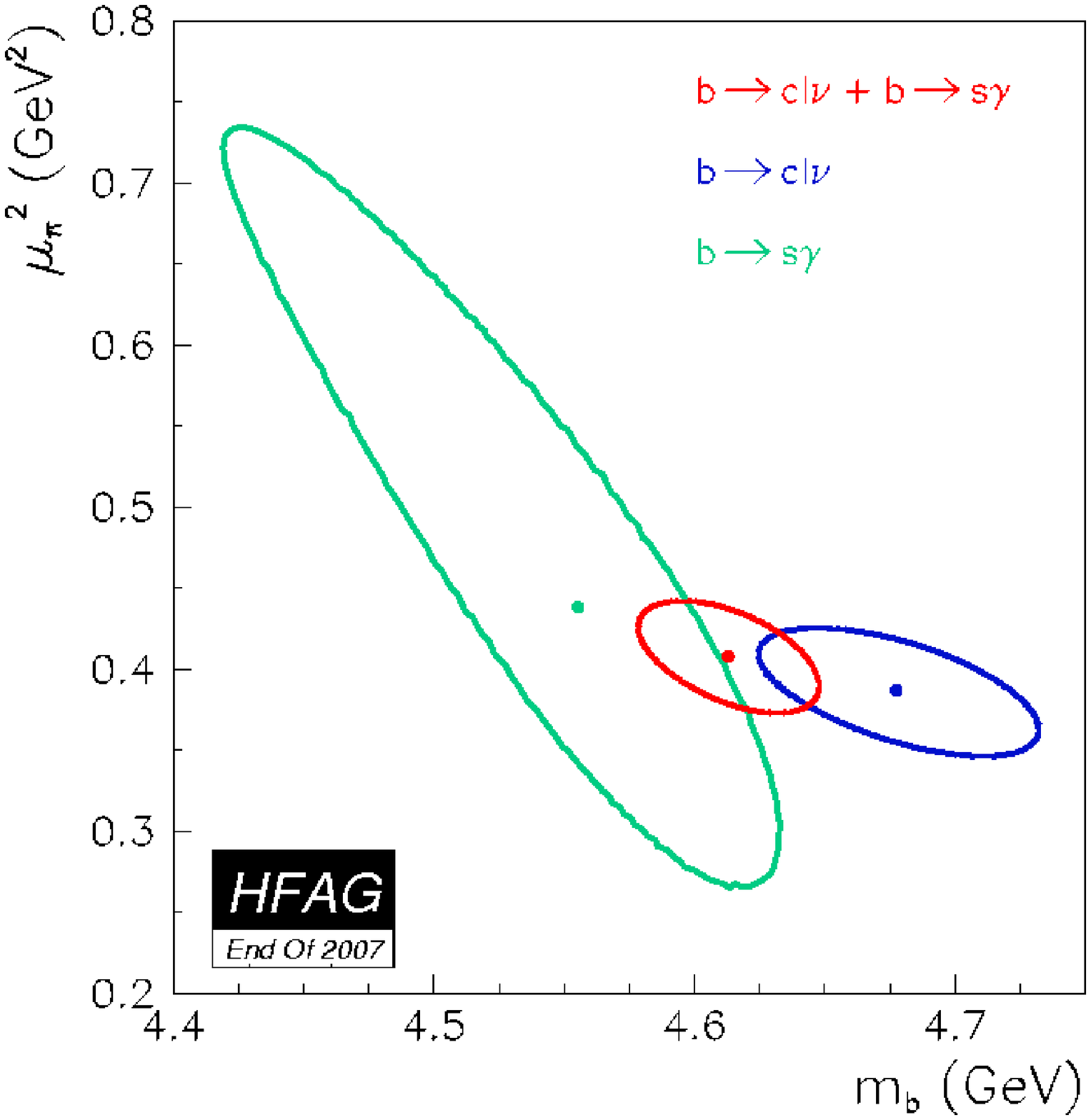}}
   \put(  5.5,  5.3){{\large\bf a)}}  
   \put( 14.4,  5.3){{\large\bf b)}}
   \end{picture} \caption{
\label{fig:def_result} 
Comparison of scenarios from fits in the kinetic mass scheme.
Figure (a) shows the $\Delta \chi^2$ = 1 contour in the ($m_b$,$\mu_{\pi}^2$) plane for the combined fit to all moments (red), the fit to hadron and lepton moments only (blue) and the fit to photon moments only (green). 
Figure (b) shows the results for the combined fit (red) and the fit to hadron and lepton moments only (blue) in the ($m_b$,\vcb) plane.
}
\end{center}
\end{figure}

\subsubsection{Global Fits in the $1S$ scheme}
\label{globalfits1S}

An independent fit in the $1S$ scheme is performed with the same set of measured lepton
energy, hadron mass and photon energy moments as in section~\ref{globalfitsKinetic}.

The inclusive spectral moments of $B\to X_c\ell\nu$ decays are derived up to $\mathcal{O}(1/m_b^3)$ in Ref.~\cite{Bauer:2004ve}. The theoretical expressions for the truncated moments are given in terms of HQE parameters, and coefficients determined by theory, which are functions of the lower lepton energy threshold $E_\mathrm{min}$. 
The non-perturbative corrections are parametrized by the parameters, ${\bar \Lambda^{1S}}$ ($\mathcal{O}(m_b)$), $\lambda_1$ and $\lambda_2$ ($\mathcal{O}(1/m_b^2)$), and $\tau_1$, $\tau_2$, $\tau_3$, $\tau_4$, $\rho_1$ and $\rho_2$ ($\mathcal{O}(1/m_b^3)$).  
One of the higher order parameters, $\tau_4$ is set to zero, and from available constraints, {\it e.g.} $B^*-B$~mass splitting, $\lambda_2=0.1227-0.0145\lambda_1$ and $\rho_2=0.1361+\tau_2$, following the prescription in Ref.~\cite{Bauer:2004ve}. 

A total of seven parameters are then determined.
We performed the 1S fit  following the method described in Ref.~\cite{Abe:2006xq}.   Measurements with higher cutoff energies ({\it i.e.} electron energy and hadron mass moments with
$E_\mathrm{min}>1.5$~GeV and  photon energy moments with $E_\mathrm{min}>2$~GeV) are not used to determine the HQE parameters, as theoretical predictions are not considered reliable in this region. Finally,  points where correlations with neighbouring points are too high have also been excluded.

The following results are obtained for the parameters in the fit to the full $B \ra X_c \ellp \nul$ and $B \ra X_s \gamma$ data set ($\chi^2/$n.d.f.$=25.3/63$):
\begin{eqnarray*}
  |V_{cb}| & = & (41. 78\pm 0.30 \pm 0.08 )\times
   10^{-3}~,\\
   m_b^\mathrm{1S} & = & (4.701\pm 0.030)~\mathrm{GeV}/c^2~,{\rm~and}\\
   \lambda_1 & = & (-0.313\pm 0.025)~\mathrm{GeV}^2~.
\end{eqnarray*}
The corresponding fit parameter correlations are given in Table \ref{tab:1s}.
The consistency between the fit results for the two different decay processes, $B \ra X_c \ellp \nul$ and $B \ra X_s \gamma$ , is assessed by performing the fit with and without the inclusion of the photon moments.  
The fit to only the $B \to X_c \ell \bar \nu$ moments results in ($\chi^2/$n.d.f.$=22.7/63$):
\begin{eqnarray*}
  |V_{cb}| & = & (41. 56\pm 0.39 \pm 0.08 )\times
   10^{-3}~,\\
   m_b^\mathrm{1S} & = & (4.751\pm 0.058)~\mathrm{GeV}/c^2~,{\rm~and}\\
   \lambda_1 & = & (-0.274\pm 0.047)~\mathrm{GeV}^2~.
\end{eqnarray*}
The first error is from the fit including experimental and theory errors, and the second error (on $|V_{cb}|$ only) is due to the uncertainty on the average $B$~lifetime.  The precision on $|V_{cb}|$ is of order 1\%.   The $\Delta \chi^2=1$ fit contour plots in the ($m_b$,$\lambda_1$) and  ($m_b$,\vcb)  planes are shown in Figure~\ref{fig:1s_result}.  All fit results are preliminary. 
 
\begin{table}
  \begin{center}
    \begin{tabular}{c@{\extracolsep{.1cm}}ccc}
      \hline 
      \rule[-1.3ex]{0pt}{4ex} & $|V_{cb}|$ & $m_b^\mathrm{1S}$ &
      $\lambda_1$\\
      \hline \hline
      \rule{0pt}{2.7ex}$|V_{cb}|$ & 1.000& $-0.379$ & $-0.232$\\
      $m_b^\mathrm{1S}$ & & $1.000$ & $0.852$\\
      \rule[-1.3ex]{0pt}{1.3ex}$\lambda_1$ & & & 1.000\\
      \hline
    \end{tabular}
  \end{center}
  \caption{Correlation coefficients of the parameters in the 1S~fit.}
  \label{tab:1s}
\end{table}

\begin{figure}[!htbp]
 \begin{center}
  \unitlength1.0cm 
  \begin{picture}(14.,8.0)  
   \put(  8.0,  0.0){\includegraphics[width=8.0cm]{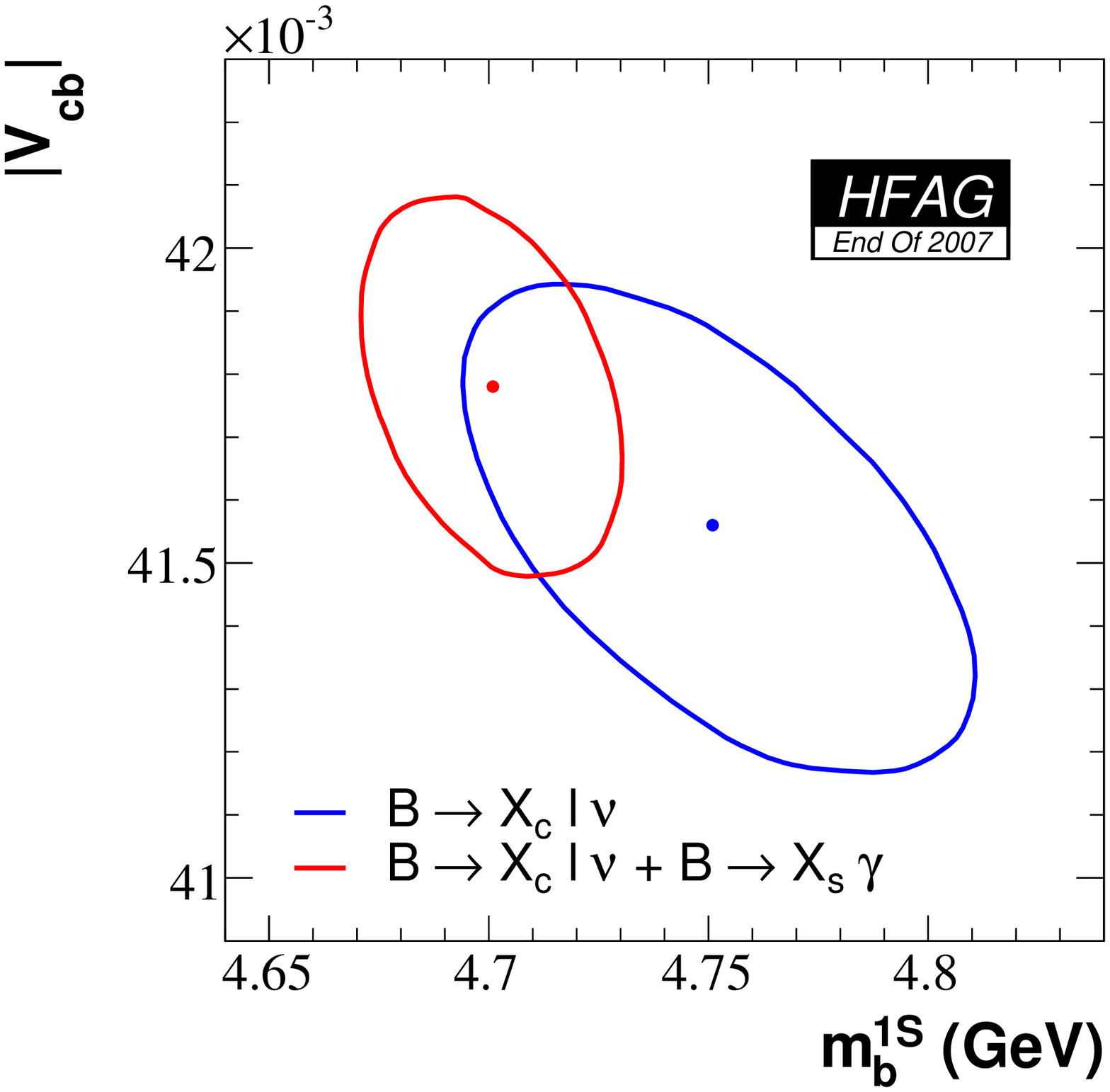}}
   \put( -0.5,  0.0){\includegraphics[width=8.0cm]{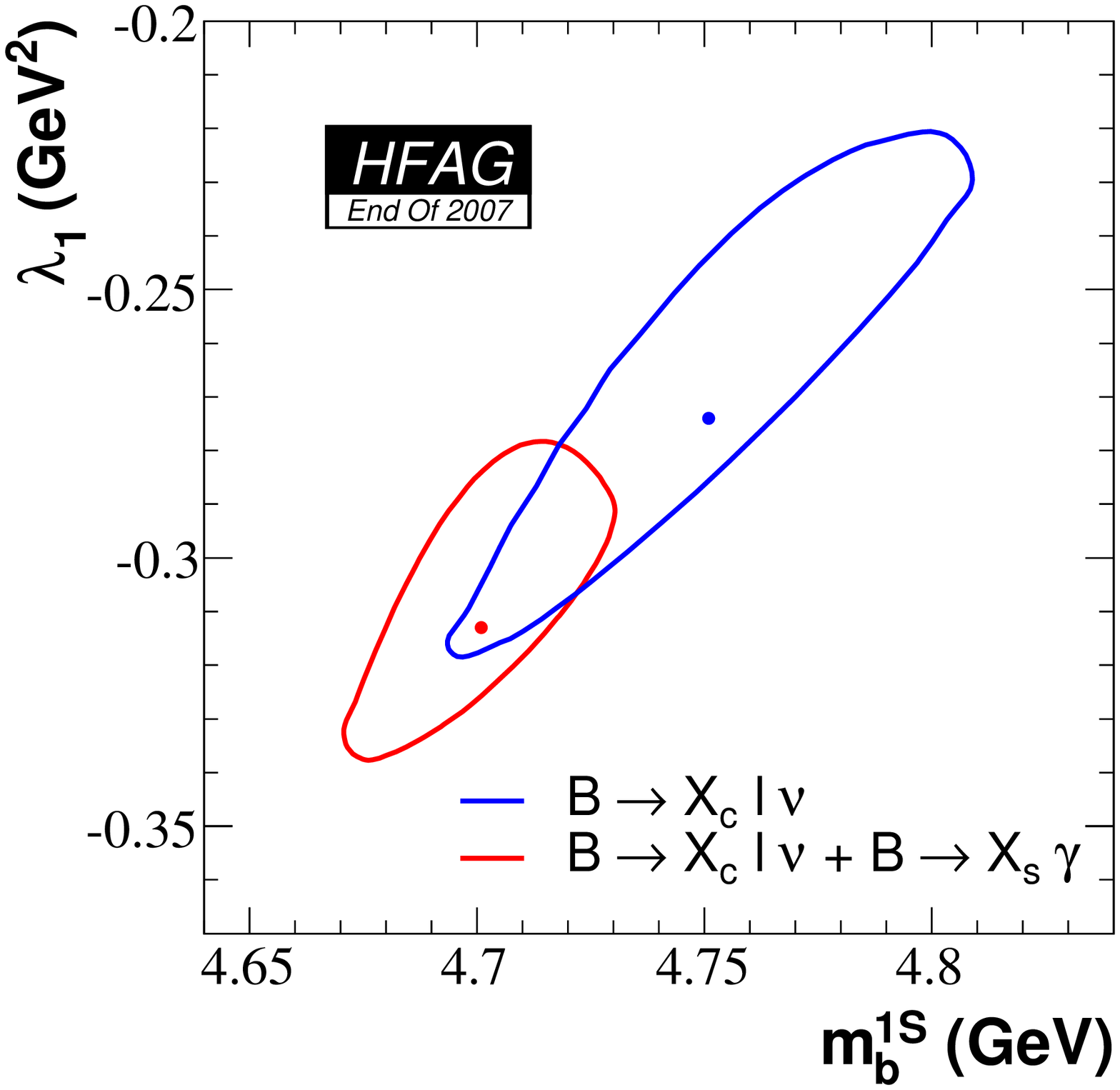}}
   \put(  5.5,  4.3){{\large\bf a)}}  
   \put( 14.4,  4.3){{\large\bf b)}}
   \end{picture} \caption{
\label{fig:1s_result} 
Comparison of scenarios from fits in the $1S$ mass scheme.
Figure (a) shows the $\Delta \chi^2$ = 1 contour in the ($m_b$,$\lambda_1$) plane for the combined fit to all moments (red), the fit to hadron and lepton moments only (blue). 
Figure (b) shows the results for the combined fit (red) and the fit to hadron and lepton moments only (blue) in the ($m_b$,\vcb) plane.
}
\end{center}
\end{figure}

\subsection{Exclusive CKM-suppressed decays}
\label{slbdecays_b2uexcl}
In this section, we list results on exclusive semileptonic branching fractions
and determinations of $\vub$ based on $\B\to\pi\ell^+\nul$ decays.
A new measurement of the exclusive decay
$\B\to\pi\ell^+\nul$ from CLEO was presented
in 2007~\cite{ref:CLEOpilnu}. 
Come previous preliminary results from the other Collaborations were finalized.
The measurements are based on two different event selections: tagged
events, in which the second $B$ meson in the event is fully
reconstructed in either a hadronic decay (``$B_{reco}$'') or in a 
CKM-favored semileptonic
decay (``SL''); and untagged events, in which case the selection infers the momentum
of the undetected neutrino from the measurement of the sum of the momenta
of all detected particles and the knowledge of the initial state.
The results for the full and partial branching fraction are given
in Table~\ref{tab:pilnubf} and shown in Figure~\ref{fig:xlnu} (a).   

When averaging these results, systematic uncertainties due to external
inputs, e.g., form factor shapes and background estimates from the
modeling of $\B\to X_c\ell^+\nul$ and $\B\to X_u\ell^+\nul$ decays, are
treated as fully correlated.
Uncertainties due to experimental reconstruction effects are treated
as fully correlated among measurements from a given experiment.  

\begin{table}[!htb]
\begin{center}
\caption{\label{tab:pilnubf}
Summary of exclusive determinations of $\cbf(\B\to\pi
\ell^+\nu_{\ell})$. The errors quoted
correspond to statistical and systematic uncertainties, respectively.
Measured branching fractions for $B^+\rightarrow \pi^0 \ell^+\nu_{\ell}$ have been
multiplied by $2\times \tau_{B^0}/\tau_{B^+}$~\cite{PDG_2007} in accordance with
isospin symmetry. The labels ``$B_{reco}$'' and ``SL'' tags refer to
the type of $B$
decay tag used in a measurement, and ``untagged'' refers to an untagged measurement.
Concerning Ref.~\cite{ref:BELLEpilnuBreco}, only the measurement in the full
$q^2$ is presented.}
\begin{small}
 \caption{
(a) Summary of exclusive determinations of $\cbf(\B\to\pi
\ell^+\nul)$ and their average.
Measured branching fractions for $B\rightarrow \pi^0 \ell^+ \nul$ have been
multiplied by $2\times \tau_{B^0}/\tau_{B^+}$~\cite{PDG_2007} in accordance with
isospin symmetry. The labels ``$B_{reco}$'' and ``SL''
refer to type of $B$ decay tag used in a measurement.
``untagged'' refers to an untagged measurement.
(b) Summary of exclusive determinations of $\cbf(\B\to\rho
\ell^+\nul)$ and their average.
}
\label{fig:xlnu}
\end{center}
\end{figure}

The determination of \vub\ from the $\B\to\pi\ell^+\nul$ decays is
shown in Table~\ref{tab:pilnuvub}, and uses the average branching
fraction given in Table~\ref{tab:pilnubf}. Two theoretical approaches are
used: QCD sum rules~\cite{BallZwicky} and Lattice QCD (unquenched~\cite{HPQCDpilnu,FNALpilnu}
and quenched~\cite{Abada:2000ty}).
Lattice calculations of the Form Factors (FF) are limited to small hadron momenta, i.e.
large $q^2$, while calculations based on light cone sum rules are restricted
to small $q^2$. More precise calculations of the FF, in particular their
normalization, are needed to reduce the overall uncertainties.


\begin{table}[hbtf]
\caption{\label{tab:pilnuvub}
Determinations of \vub\ based on the average total and partial
$\Bb\to\pi^{-}\ell^{+}\nu_{\ell}$ decay branching fraction stated in
Table~\ref{tab:pilnubf}. The first
uncertainty is experimental, and the second is from theory.  The
full or partial branching fractions are used as indicated. 
Acronysms for the calculations refer to either the method (LCSR) or
the collaboration working on it (HPQCD, FNAL, APE).
}
\begin{center}
\begin{tabular}{|lc|}
\hline
Method & $\Vub [10^{-3}]$ \\\hline\hline
LCSR, full $q^2$~\cite{BallZwicky} & $3.41\pm 0.11 {}^{+0.67}_{-0.42}$ \\ 
LCSR, $q^2<16\,\gev^2/c^2$~\cite{BallZwicky}   & $3.38\pm 0.13 {}^{+0.56}_{-0.37}$ \\ \hline
HPQCD, full $q^2$~\cite{HPQCDpilnu}& $3.11\pm 0.10 {}^{+0.73}_{-0.43}$ \\ 
HPQCD, $q^2>16\,\gev^2/c^2$~\cite{HPQCDpilnu}  & $3.47\pm 0.20 {}^{+0.60}_{-0.39}$ \\  \hline
FNAL, full $q^2$~\cite{FNALpilnu}  & $3.80\pm 0.12 {}^{+0.88}_{-0.52}$ \\ 
FNAL, $q^2>16\,\gev^2/c^2$~\cite{FNALpilnu}    & $3.69\pm 0.21 {}^{+0.64}_{-0.42}$ \\  \hline
APE, full $q^2$~\cite{Abada:2000ty}  & $3.59\pm 0.11 {}^{+1.11}_{-0.57}$ \\ 
APE, $q^2>16\,\gev^2/c^2$~\cite{Abada:2000ty}    & $3.72\pm 0.21 {}^{+1.43}_{-0.66}$ \\ 
\hline
\end{tabular}
\end{center}
\end{table}

We present for the first time also branching fractions for 
$\B\to \rho\ell^+\nul$ decays in
Table~\ref{tab:rholnu} and  Figure~\ref{fig:xlnu} (b). The determination of $\Vub$
from these other channels looks less promising than for
$\B\to\pi\ell^+\nul$, primarily because there is limited theoretical information
on the normalization and shape of the form factors, thus at the moment it is not extracted.

\begin{table}[!htb]
\begin{center}
\caption{Summary of exclusive determinations of $\cbf(\B\to\rho
\ell^+\nu_{\ell})$. The errors quoted
correspond to statistical and systematic uncertainties, respectively.}
\begin{small}
\begin{tabular}{|lc|}
\hline
& $\cbf [10^{-4}]$
\\
\hline\hline
CLEO $\rho^+$~\cite{ref:CLEOpilnu}
& $2.69\pm 0.41\pm 0.63\ $ 
\\ 
CLEO $\rho^+$~\cite{ref:CLEOpilnu}
& $2.93\pm 0.37\pm 0.37\ $ 
\\ 
\babar $\rho^+$~\cite{Aubert:2005cd}
& $2.14\pm 0.21\pm 0.55\ $
\\  \hline
{\bf Average}
& \mathversion{bold}$2.64 \pm 0.21\pm 0.33\ $
\\ 
\hline
\end{tabular}\\
\end{small}
\end{center}
\label{tab:rholnu}
\end{table}

Branching fractions for other $\B\to X_u\ell^+\nul$ decays are given in
Table~\ref{tab:xslother}. 

\begin{table}[!htb]
\caption{Summary of branching fractions to $\cbf(\B\to X\ell^+\nu_{\ell})$ decays
other than $\B\to\pi\ell^+\nu_{\ell}$ and  $\B\to\rho\ell^+\nu_{\ell}$. The errors quoted
correspond to statistical and systematic
uncertainties, respectively.  Where a third uncertainty is quoted, it
corresponds to uncertainties from form factor shapes.}
\begin{center}
\begin{small}
\begin{tabular}{|llc|}
\hline
Experiment & Mode & $\cbf [10^{-4}]$  \\\hline\hline
BELLE~\cite{ref:BELLEomegalnu} & $\Bp\to\omega\ell^+\nu_{\ell}$ & $1.3\ \,\pm 0.4\ \,\pm 0.2\ \,\pm 0.3\ \,$ \\ 
CLEO~\cite{ref:CLEOpilnu} & $\Bp\to\eta\ell^+\nu_{\ell}$ & $0.84\pm 0.31\pm 0.16 \pm 0.09$ \\ 
\babar~\cite{ref:BABARetalnuBreco} & $\Bp\to\eta\ell^+\nu_{\ell}$ & $0.84\pm 0.27\pm 0.21$ \\ 
\babar~\cite{ref:BABARetalnuBreco} & $\Bp\to\eta^\prime\ell^+\nu_{\ell}$ & $0.33\pm 0.60\pm 0.30$ \\ 
\hline
\end{tabular}\\
\end{small}
\end{center}
\label{tab:xslother}
\end{table}


%
\subsection{Inclusive CKM-suppressed decays}
\label{slbdecays_b2uincl}
The large background from $\B\to X_c\ell^+\nul$ decays is the chief
experimental limitation in determinations of $\vub$.  Cuts designed to
reject this background limit the acceptance for $\B\to X_u\ell^+\nul$
decays. The calculation of partial rates for these restricted
acceptances is more complicated and requires substantial theoretical machinery.
In this update, we had added several new theoretical calculations
to extract \vub. We do not advocate the use of one method over another.
The authors for the different calculations have provided 
codes to compute
the partial rates in limited regions of phase space covered by the measurements.

For the averages we performed, the systematic errors associated with the
modeling of $\B\to X_c\ell^+\nul$ and $\B\to X_u\ell^+\nul$ decays and the theoretical
uncertainties are taken as fully correlated among all measurements.
Reconstruction-related
uncertainties are taken as fully correlated within a given experiment.
From the three results published each by \babar\ and Belle~\cite{ref:babar-mx,ref:belle-mx}, 
only one is used in the average, because
the three measurements are all based on the same dataset and are
highly correlated. Specifically, we use the $M_X$ analysis result for
the averages.
As a consequence, the experimental results have negligible
statistical correlations. 
To make use of the theoretical calculations of Ref.~\cite{ref:BLL}, we restrict the
kinematic range in $M_X$ and $q^2$, thereby reducing the size of the data
sample significantly, but also the theoretical uncertainty, as stated by the
authors~\cite{ref:BLL}.
The dependence of the quoted error on the measured value for each source of error
is taken into account in the calculation of the averages.
Measurements of partial branching fractions for $\B\to X_u\ell^+\nul$
transitions from $\Upsilon(4S)$ decays, together with the corresponding accepted region, 
are given in Table~\ref{tab:BFbulnu}.  
The signal yields for all the measurements shown in Table~\ref{tab:BFbulnu}
are not rescaled to common input values of the $B$ meson lifetime (see table~\ref{tab:common.param})
and the semileptonic width~\cite{PDG_2007}.

It has been first suggested by Neubert~\cite{Neubert:1993um} and later detailed by Leibovich, 
Low, and Rothstein (LLR)~\cite{Leibovich:1999xf} and Lange, Neubert and Paz (LNP)~\cite{Lange:2005qn}, 
that the uncertainty of
the leading shape functions can be eliminated by comparing inclusive rates for
$\B\to X_u\ell^+\nul$ decays with the inclusive photon spectrum in $\B\to X_s\gamma$,
based on the assumption that the shape functions for transitions to light
quarks, $u$ or $s$, are the same to first order.
However, shape function uncertainties are only eliminated at the leading order
and they still enter via the signal models used for the determination of efficiency. 
For completeness, we provide a comparison of the results using 
calculations with reduced dependence on the shape function, as just
introduced, with our averages using different theoretical approaches.
Results are presented by \babar\ in Ref.\cite{Aubert:2006qi} using the LLR prescription. 
More recently, V.B.Golubev, V.G.Luth and Yu.I.Skovpen (Ref.~\cite{Golubev:2007cs})
extracted \vub\ from the 
endpoint spectrum of $\B\to X_u\ell^+\nul$ from \babar~\cite{ref:babar-endpoint}, 
using several theoretical approaches with reduced dependence on the shape function.
In both cases, the photon energy spectrum in the
rest frame of the $B$-meson by \babar~\cite{BABARSEMI} has been used.

\begin{table}[!htb]
\caption{\label{tab:BFbulnu}
Summary of inclusive determinations of partial branching
fractions for $B\rightarrow X_u \ell^+ \nu_{\ell}$ decays.
The errors quoted on $\Delta\cbf$ correspond to
statistical and systematic uncertainties.
The $P_+$ analysis is actually not used in the current 
averages since it is stringly correlated with the $M_X$ analysis. 
The $s_\mathrm{h}^{\mathrm{max}}$ variable is described in Refs.~\cite{ref:shmax,ref:babar-elq2}. }
\begin{center}
\begin{small}
\begin{tabular}{|llc|}
\hline
Measurement & Accepted region &  $\Delta\cbf [10^{-4}]$ \\
\hline\hline
CLEO~\cite{ref:cleo-endpoint}
& $E_e>2.1\,\gev$ 
& $3.3\pm 0.2\pm 0.7$  \\ 
\babar~\cite{ref:babar-elq2}
& $E_e>2.0\,\gev$, $s_\mathrm{h}^{\mathrm{max}}<3.5\,\mathrm{GeV^2}$ & $4.4\pm 0.4\pm 0.4$ \\
\babar~\cite{ref:babar-endpoint}
& $E_e>2.0\,\gev$  & $5.7\pm 0.4\pm 0.5$\\
BELLE~\cite{ref:belle-endpoint}
& $E_e>1.9\,\gev$  & $8.5\pm 0.4\pm 1.5$ \\
\babar~\cite{ref:babar-mx}
& $M_X<1.7\,\gev/c^2, q^2>8\,\gev^2/c^2$ & $8.1\pm 0.8\pm 0.7$\\
BELLE~\cite{ref:belle-mxq2Anneal}
& $M_X<1.7\,\gev/c^2, q^2>8\,\gev^2/c^2$ & $7.4\pm 0.9\pm 1.3$\\
BELLE~\cite{ref:belle-mx}
& $M_X<1.7\,\gev/c^2, q^2>8\,\gev^2/c^2$ & $8.4\pm 0.8\pm 0.4$\\
\babar~\cite{ref:babar-mx}
& $P_+<0.66\,\gev$  & $9.4\pm 1.0\pm 0.8 $ \\
BELLE~\cite{ref:belle-mx}
& $P_+<0.66\,\gev$  & $11.0\pm 1.0\pm 1.6$ \\
\babar~\cite{ref:babar-mx}
& $M_X<1.55\,\gev/c^2$ & $11.7\pm 0.9\pm 0.7 $ \\ 
BELLE~\cite{ref:belle-mx}
& $M_X<1.7\,\gev/c^2$ & $12.3\pm 1.1\pm 1.2$ \\ \hline
\end{tabular}\\
\end{small}
\end{center}
\end{table}

\subsubsection{BLNP}
Bosch, Lange, Neubert and Paz (BLNP)~\cite{ref:BLNP,
  ref:Neubert-new-1,ref:Neubert-new-2,ref:Neubert-new-3,ref:Neubert-new-4}
provide theoretical expressions for the triple
differential decay rate for $B\to X_u \ell^+ \nul$ events, incorporating all known
contributions, whilst smoothly interpolating between the 
``shape-function region'' of large hadronic
energy and small invariant mass, and the ``OPE region'' in which all
hadronic kinematical variables scale with the $b$-quark mass. BLNP assign
uncertainties to the $b$-quark mass which enters through the leading shape function, 
to sub-leading shape function forms, to possible weak annihilation
contribution, and to matching scales. The extracted values
of \vub\, for each measurement along with their average are given in
Table~\ref{tab:bulnu} and illustrated in Figure~\ref{fig:BLNP}.
The total uncertainty is $^{+8.8}_{-7.7}\%$ and is due to:
statistics ($^{+2.0}_{-2.0}\%$),
detector ($^{+2.3}_{-2.2}\%$),
$B\to X_c \ell^+ \nul$ model ($^{+1.3}_{-1.2}\%$),
$B\to X_u \ell^+ \nul$ model ($^{+1.4}_{-1.4}\%$),
heavy quark parameters ($^{+7.0}_{-5.8}\%$),
SF functional form ($^{+0.5}_{-0.5}\%$),
sub-leading shape functions ($^{+0.7}_{-0.7}\%$),
BLNP theory: matching scales $\mu,\mu_i,\mu_h$ ($^{+3.6}_{-3.3}\%$), and
weak annihilation ($^{+1.3}_{-1.3}\%$).
The error on the HQE parameters ($b$-quark mass and $\mu_\pi^2)$ 
is the source of the largest uncertainty, while the
uncertainty assigned for the matching scales is a close second.  

\begin{table}[!htb]
\caption{\label{tab:bulnu}
Summary of input parameters used by the different theory calculations,
corresponding inclusive determinations of $\vub$ and their average.
The errors quoted on \vub\ correspond to
experimental and theoretical uncertainties, respectively. Note that only the 
$M_X$ analysis is used for Refs.~\cite{ref:babar-mx,ref:belle-mx}, as the other analyses
are highly correlated.}
\begin{center}
\begin{small}
\\
\end{small}
\end{center}
\end{table}

\begin{figure}
\begin{center}
\includegraphics[width=0.48\textwidth]{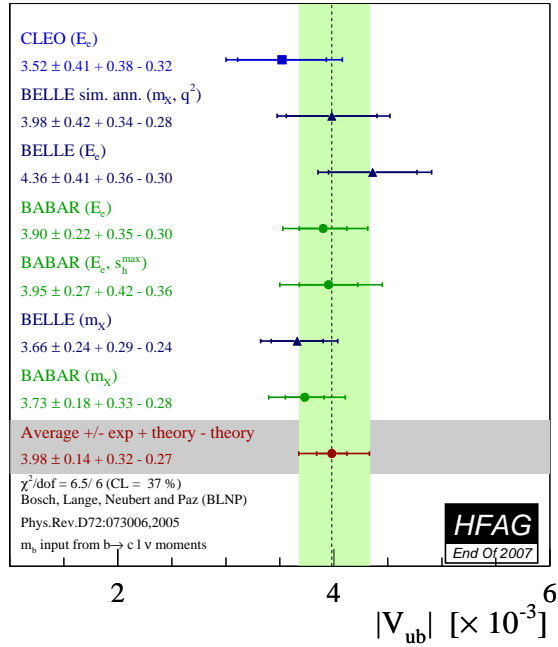}
\end{center}
\caption{Measurements of $\vub$ from inclusive semileptonic decays 
and their average based on the BLNP prescription.
``$E_e$'', ``$M_X$'', ``$(M_X,q^2)$'' and ``($E_e,s^{max}_h$)'' indicate the 
distributions and cuts used for the measurement of the partial decay rates.}
\label{fig:BLNP}
\end{figure}

\subsubsection{DGE}
J.R.~Andersen and E.~Gardi (Dressed Gluon Exponentiation, DGE)~\cite{ref:DGE} provide
a framework where the on-shell $b$-quark calculation, converted into hadronic variables, is
directly used as an approximation to the meson decay spectrum without
the use of a leading-power non-perturbative function (or, in other words,
a shape function). The on-shell mass of the $b$-quark within the $B$-meson ($m_b$) is
required as input. 
The extracted values
of \vub\, for each measurement along with their average are given in
Table~\ref{tab:bulnu} and illustrated in Figure~\ref{fig:DGE}.
The total error is $^{+6.8}_{-6.9}\%$, whose breakdown is:
statistics ($^{+1.9}_{-1.9}\%$),
detector ($^{+2.4}_{-2.3}\%$),
$B\to X_c \ell^+ \nul$ model ($^{+1.8}_{-1.7}\%$),
$B\to X_u \ell^+ \nul$ model ($^{+1.1}_{-1.1}\%$),
spectral fraction ($m_b$) ($^{+3.0}_{-3.3}\%$),
strong coupling $\alpha_s$ ($^{+0.7}_{-0.8}\%$),
total semileptonic width ($m_b$) ($^{+3.0}_{-3.0}\%$),
weak annihilation ($^{+1.9}_{-1.9}\%$),
DGE theory: matching scales ($^{+3.2}_{-3.1}\%$).
The largest contribution to the total error is due to the effect of the uncertainty 
on $m_b$  on the prediction of the event rate, closely followed by the 
specific theory error on overall DGE and the total semileptonic decay width.

\begin{figure}
\begin{center}
\includegraphics[width=0.48\textwidth]{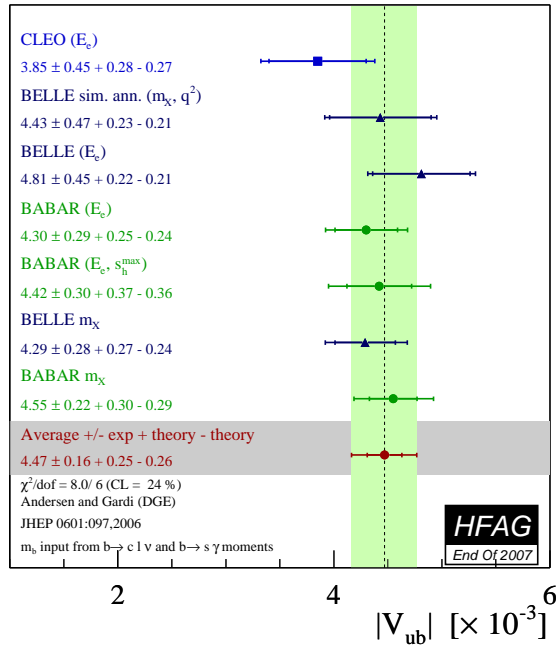}
\end{center}
\caption{Measurements of $\vub$ from inclusive semileptonic decays 
and their average based on the DGE prescription.
``$E_e$'', ``$M_X$'', ``$(M_X,q^2)$'' and ``($E_e,s^{max}_h$)'' indicate the 
analysis type.}
\label{fig:DGE}
\end{figure}

\subsubsection{GGOU}
Gambino, Giordano, Ossola and Uraltsev (GGOU)~\cite{Gambino:2007rp} 
compute the triple differential decay rates of $B \to X_u \ell^+ \nul$, 
including all perturbative and non--perturbative effects through $O(\alphas^2 \beta_0)$ 
and $O(1/m_b^3)$. 
The Fermi motion is parameterized in terms of a single light--cone function 
for each structure function and for any value of $q^2$, accounting for all subleading effects. 
The calculations are performed in the kinetic scheme, a framework characterized by a Wilsonian 
treatment with a hard cutoff $\mu \sim 1 $ GeV.
At present, GGOU have not included calculations for the ``($E_e,s^{max}_h$)'' analysis, but
this addition is planned.
The extracted values
of \vub\, for each measurement along with their average are given in
Table~\ref{tab:bulnu} and illustrated in Figure~\ref{fig:GGOU}.
The total error is $^{+6.3}_{-7.0}\%$ whose breakdown is:
statistics ($^{+2.1}_{-2.3}\%$),
detector ($^{+2.2}_{-2.2}\%$),
$B\to X_c \ell^+ \nul$ model ($^{+1.5}_{-1.2}\%$),
$B\to X_u \ell^+ \nul$ model ($^{+1.4}_{-1.6}\%$),
$\alpha_s$, $m_b$ and other non--perturbative parameters ($^{+3.9}_{-3.8}\%$), 
higher order perturbative and non--perturbative corrections ($^{+1.8}_{-1.7}\%$), 
modelling of the $q^2$ tail and choice of the scale $q^{2*}$ ($^{+2.5}_{-2.7}\%$), 
weak annihilations matrix element ($^{+0}_{-3.1}\%$), 
functional form of the distribution functions ($^{+1.3}_{-0.6}\%$), 
The leading uncertainties
on  \vub\ are both from theory, and are due to perturbative and non--perturbative
parameters and the modelling of the $q^2$ tail and choice of the scale $q^{2*}$. 

\begin{figure}
\begin{center}
\includegraphics[width=0.48\textwidth]{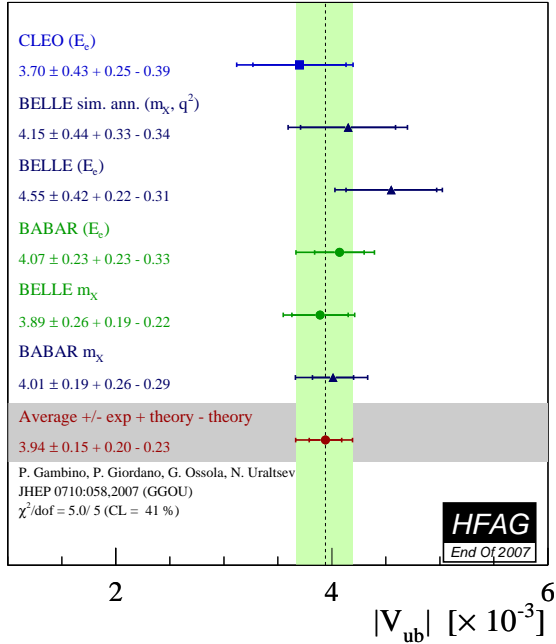}
\end{center}
\caption{Measurements of $\vub$ from inclusive semileptonic decays 
and their average based on the GGOU prescription.
``$E_e$'', ``$M_X$'', ``$(M_X,q^2)$'' and ``($E_e,s^{max}_h$)''  indicate the
analysis type.}
\label{fig:GGOU}
\end{figure}

\subsubsection{ADFR}
Aglietti, Di Lodovico, Ferrera and Ricciardi (ADFR)~\cite{Aglietti:2007ik}
use a new approach to extract \vub, which makes use of the ratio
of the  $B \to X_c \ell^+ \nul$ and $B \to X_u \ell^+ \nul$ widths. 
The normalized triple differential decay rate for 
$B \to X_u \ell^+ \nul$~\cite{Aglietti:2006yb,Aglietti:2005mb, Aglietti:2005bm, Aglietti:2005eq}
is calculated with a model based on (i) soft--gluon resummation 
to next--to--next--leading order and (ii) an effective QCD coupling without
Landau pole. This coupling is constructed by means of an extrapolation to low
energy of the high--energy behaviour of the standard coupling. More technically,
an analyticity principle is used.
The lower cut on the electron energy for the endpoint analyses is 2.3~GeV~\cite{Aglietti:2006yb}.
The extracted values
of \vub\, for each measurement along with their average are given in
Table~\ref{tab:bulnu} and illustrated in Figure~\ref{fig:AC}.
The total error is $^{+7.2}_{-7.2}\%$ whose breakdown is:
statistics ($^{+1.8}_{-1.9}\%$),
detector ($^{+2.3}_{-2.3}\%$),
$B\to X_c \ell^+ \nul$ model ($^{+1.2}_{-1.3}\%$),
$B\to X_u \ell^+ \nul$ model ($^{+1.3}_{-1.4}\%$),
$\alpha_s$ ($^{+2.0}_{-2.0}\%$), 
$|V_{cb}|$ ($^{+1.4}_{-1.4}\%$), 
$m_b$ ($^{+1.0}_{-1.0}\%$), 
$m_c$ ($^{+4.5}_{-4.3}\%$), 
semileptonic branching fraction ($^{+0.9}_{-1.0}\%$), 
theory model ($^{+3.5}_{-3.5}\%$).
The leading
uncertainties, both from theory, are due to the $m_c$ mass and the theory model.

\begin{figure}
\begin{center}
\includegraphics[width=0.48\textwidth]{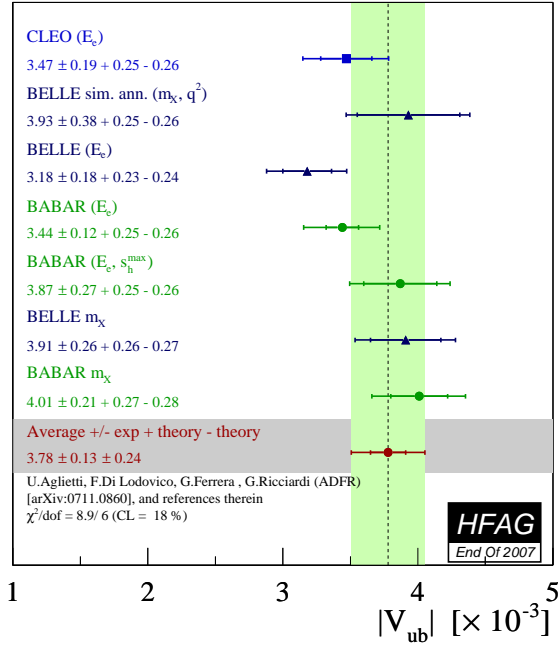}
\end{center}
\caption{Measurements of $\vub$ from inclusive semileptonic decays 
and their average based on the ADFR prescription.
``$E_e$'', ``$M_X$'', ``$(M_X,q^2)$'' and ``($E_e,s^{max}_h$)'' indicate the 
analysis type.}
\label{fig:AC}
\end{figure}

\subsubsection{BLL}
Bauer, Ligeti, and Luke (BLL)~\cite{ref:BLL} give a
HQET-based prescription that advocates combined cuts on the dilepton invariant mass, $q^2$,
and hadronic mass, $m_X$, to minimise the overall uncertainty on \vub.
In their reckoning a cut on $m_X$ only, although most efficient at
preserving phase space ($\sim$80\%), makes the calculation of the partial
rate untenable due to uncalculable corrections
to the $b$-quark distribution function or shape function. These corrections are
suppressed if events in the low $q^2$ region are removed. The cut combination used
in measurements is $M_x<1.7$ GeV/$c^2$ and $q^2 > 8$ GeV$^2$/$c^2$.  
The extracted values
of \vub\, for each measurement along with their average are given in
Table~\ref{tab:bulnu} and illustrated in Figure~\ref{fig:BLL}.
The total error is $^{+9.0}_{-9.0}\%$ whose breakdown is:
statistics ($^{+3.2}_{-3.2}\%$),
detector ($^{+3.8}_{-3.8}\%$),
$B\to X_c \ell^+ \nul$ model ($^{+1.3}_{-1.3}\%$),
$B\to X_u \ell^+ \nul$ model ($^{+2.1}_{-2.1}\%$),
spectral fraction ($m_b$) ($^{+3.0}_{-3.0}\%$),
perturbative : strong coupling $\alpha_s$ ($^{+3.0}_{-3.0}\%$),
residual shape function ($^{+4.5}_{-4.5}\%$),
third order terms in the OPE ($^{+4.0}_{-4.0}\%$),
The leading
uncertainties, both from theory, are due to residual shape function
effects and third order terms in the OPE expansion. The leading
experimental uncertainty is due to statistics. 

\begin{figure}
\begin{center}
\includegraphics[width=0.48\textwidth]{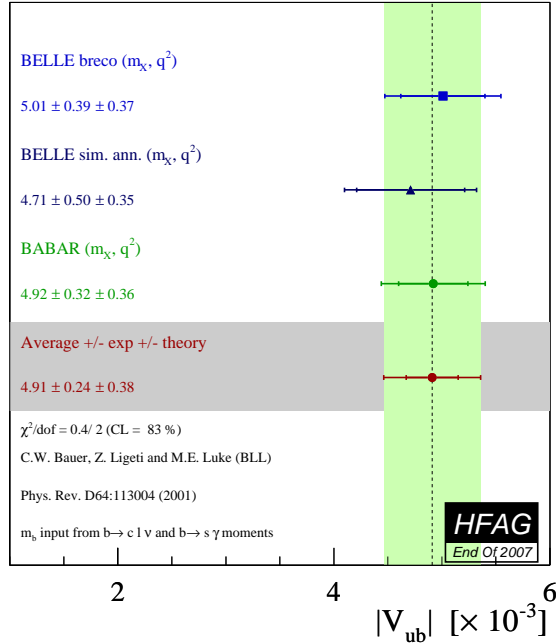}
\end{center}
\caption{Measurements of $\vub$ from inclusive semileptonic decays 
and their average in the BLL prescription.
``$(M_X, q^2)$'' indicates the analysis type.}
\label{fig:BLL}
\end{figure}

\subsubsection{Summary}
A summary of the averages presented in several different
frameworks and results by V.B.Golubev, V.G.Luth and Yu.I.Skovpen~\cite{Golubev:2007cs},
based on prescriptions by LLR~\cite{Leibovich:1999xf} and LNP~\cite{Lange:2005qn} 
to reduce the leading shape function uncertainties are presented in 
Table~\ref{tab:vubcomparison}.
A value judgement based on a direct comparison should be
avoided at the moment, experimental and theoretical uncertainties play out
differently between the schemes and the theoretical assumptions for the
theory calculations are different.

\begin{table}[!htb]
\caption{\label{tab:vubcomparison}
Summary of inclusive determinations of $\vub$.
The errors quoted on \vub\ correspond to experimental and theoretical uncertainties, except for the last two measurements where the errors are due to the \babar\ endpoint analysis, the \babar $b\to s\gamma$ analysis~\cite{Aubert:2006qi}, the theoretical errors and $V_{ts}$ for the last averages. 
}
\begin{center}
\begin{small}
\begin{tabular}{|lc|}
\hline
Framework
&  $\Vub [10^{-3}]$\\
\hline\hline
BLNP
& $3.98 \pm 0.14 ^{+0.32}_{-0.27}$ \\ 
DGE
& $4.47 \pm 0.16 ^{+0.25}_{-0.26}$ \\
GGOU
& $3.94 \pm 0.15 ^{+0.20}_{-0.23}$ \\
ADFR
& $3.78 \pm 0.13 ^{+0.24}_{-0.24}$ \\
BLL ($m_X/q^2$ only)
& $4.91 \pm 0.24 \pm 0.38$ \\ 
LLR (\babar)~\cite{Aubert:2006qi}
& $4.92 \pm 0.32 \pm 0.36$ \\
LLR (\babar)~\cite{Golubev:2007cs}
& $4.28 \pm 0.29 \pm 0.29 \pm 0.26 \pm0.28$ \\
LNP (\babar)~\cite{Golubev:2007cs}
& $4.40 \pm 0.30 \pm 0.41 \pm 0.23$ \\
\hline
\end{tabular}\\
\end{small}
\end{center}
\end{table}

%
%

\clearpage
\section{\emph{$B$}-decays to charmed hadrons}
\label{sec:BtoCharm}


\definecolor{hfviolet}{rgb}{0.5,0,0.5}
\definecolor{hflightcyan}{rgb}{0.1,1.0,1.0}

\def\hflmode{Mode}
\def\hflpdg#1{PDG #1}
\def\hflbabar{\mbox{\slshape B\kern-0.1em{\smaller A}\kern-0.1em B\kern-0.1em{\smaller A\kern-0.2em R}}}
\def\hflbelle{\mbox{Belle}}
\def\hflcdf{\mbox{CDF}}
\def\hflcleo{\mbox{CLEO}}
\def\hfld0{\mbox{D0}}
\def\hflavg{Average}

\def\hfpubhotcolor{red}
\def\hfpubcolor{magenta}
\def\hfpuboldcolor{black}

\def\hfprehotcolor{blue}
\def\hfprecolor{cyan}
\def\hfpreoldcolor{hflightcyan}

\def\hfdefcolor{black}
\def\hferrcolor{yellow}
\def\hfsuperceededcolor{hfviolet}
\def\hfwaitingcolor{green}
\def\hfinactivecolor{hfviolet}
\def\hfnoquocolor{hfviolet}
\def\hfdeftext#1{\textcolor{\hfdefcolor}{#1}}
\def\hflabel#1{\textcolor{\hfdefcolor}{$#1$}}
\def\hfavg#1{\textcolor{\hfdefcolor}{$#1$}}
\def\hfnewavg#1{\textcolor{\hfdefcolor}{\boldmath$#1$}}
\def\hfdefault#1{\textcolor{\hfdefcolor}{$#1$}}
\def\hfpdg#1{\textcolor{\hfdefcolor}{$#1$}}
\def\hfwaiting#1{\textcolor{\hfwaitingcolor}{$#1$}}
\def\hfpubhot#1{\textcolor{\hfpubhotcolor}{$#1$}}
\def\hfprehot#1{\textcolor{\hfprehotcolor}{$#1$}}
\def\hfwaitingtext#1{\textcolor{\hfwaitingcolor}{#1}}
\def\hfpubhottext#1{\textcolor{\hfpubhotcolor}{#1}}
\def\hfprehottext#1{\textcolor{\hfprehotcolor}{#1}}
\def\hfpub#1{\textcolor{\hfpubcolor}{$#1$}}
\def\hfpre#1{\textcolor{\hfprecolor}{$#1$}}
\def\hfpubold#1{\textcolor{\hfpuboldcolor}{$#1$}}
\def\hfpreold#1{\textcolor{\hfpreoldcolor}{$#1$}}
\def\hfpubtext#1{\textcolor{\hfpubcolor}{#1}}
\def\hfpretext#1{\textcolor{\hfprecolor}{#1}}
\def\hfpuboldtext#1{\textcolor{\hfpuboldcolor}{#1}}
\def\hfpreoldtext#1{\textcolor{\hfpreoldcolor}{#1}}
\def\hferror#1{\textcolor{\hferrcolor}{$#1$}}
\def\hfsuperceeded#1{\textcolor{\hfsuperceededcolor}{$#1$}}
\def\hfinactive#1{\textcolor{\hfinactivecolor}{$#1$}}
\def\hfnoquo#1{\textcolor{\hfnoquocolor}{$#1$}}
\def\hferrortext#1{\textcolor{\hferrcolor}{#1}}
\def\hfsuperceededtext#1{\textcolor{\hfsuperceededcolor}{#1}}
\def\hfinactivetext#1{\textcolor{\hfinactivecolor}{#1}}
\def\hfnoquotext#1{\textcolor{\hfnoquocolor}{#1}}
\def\hfbb#1{#1M $B\bar{B}$ pairs}

\def\hffootnotemark#1{\tiny{$^{#1}$}}
\def\hffootnotetext#1{\tiny{#1}}
\def\hffootspacing{-5pt}
\def\hffootitemsep{-0.7}

\def\hfnewp{new particles }
\def\hfdstr{strange D mesons }
\def\hfbary{baryons }
\def\hfjpsi{$J/\psi(1S)$ }
\def\hfochm{charmonium other than $J/\psi(1S)$ }
\def\hfmuld{multiple $D$, $D^{*}$ or $D^{**}$ mesons }
\def\hfsgdx{a single $D^{*}$ or $D^{**}$ meson }
\def\hfsgld{a single D meson }
\def\hfothe{charmed particles } 

\def\hfsitebase{http://hfag.phys.ntu.edu.tw/b2charm/}
\def\hfurl#1{\href{\hfsitebase#1}{\hfsitebase#1}} 
\def\hfhref#1#2{\href{\hfsitebase#1}{#2}} 
\def\hftabletype{sidewaystable}
\def\hftableposn{!htbp}
\def\hfaftercapspace{\vspace*{2mm}}

\def\hfcaption#1#2#3#4#5{\textcolor{\hfdefcolor}{#1 of #2 modes producing #3 #4, #5.}}
\def\hfnewcaption#1#2#3#4#5#6#7{\caption{ #1 of #2 modes producing #3 #4, #5. The latest version is available at: \hfurl{#6} \label{#7}  }\hfaftercapspace}
\newcommand\hftabletlcell{\rule{0pt}{2.6ex}}  
\newcommand\hftableblcell{\rule[-1.2ex]{0pt}{0pt}}

\def\hfmetadata#1{}      

\def\hfcBR{{\cal{B}}}

%
%
%
%
%
%

This section reports the updated contribution to the HFAG report from the ``$B \to$ charm" group\footnote{The HFAG/BtoCharm group was formed in the spring of 2005; it performs its work using an XML database backed web application.}. The mandate of the group is to compile measurements and perform averages of all available quantities related to $B$ decays to charmed particles, excluding CP related quantities. To date the group has analyzed a total of 492 measurements reported in 148 papers, principally branching fractions. The group aims to organize and present the copious information on $B$ decays to charmed particles obtained from a combined sample of about two billion $B$ mesons from the BABAR, Belle and CDF Collaborations. 

This huge sample of $B$ mesons allows to measure decays to states with open or hidden charm content with unprecedented precision. Branching fractions for rare $B$-meson decays or decay chains of a few $10^{-7}$ are being measured with statistical uncertainties typically below $30\%$, and new decay chains can be accessed with branching fractions down to $10^{-8}$. Results for more common decay chains, with branching fractions around $10^{-4}$, are becoming precision measurements, with uncertainties typically at the $3\%$ level. Some decays have been observed for the first time, for example $B^0 \to J/\psi\eta$ or $B^-\to \Lambda_c^- p$, with a branching fraction of $(9.6\pm1.8)\times 10^{-6}$ and $(2.1^{+1.2}_{-0.9})\times 10^{-5}$, respectively.

The large sample of $B$ mesons allows to greatly improve our understanding of recently discovered new states with either hidden or open charm content, such as the $X(3872)$, the $D_{sJ}^{*-}(2317)$ and $D_{sJ}^-(2460)$ mesons. Measurements with many different final states for these particles are reported, allowing to shed more light on their nature. The $D^0\bar{D}^{*0}(2007)$ decay of the $X(3872)$ has been observed for the first time. Using the branching fraction products $\hfcBR(B^-\to X(3872) K^-)\times\hfcBR(X(3872)\to f)$, a hierarchy can be established between the decay modes $f$: these branching fraction products are found to be $(1.67\pm0.59)\times 10^{-4}$, $(0.12\pm0.02)\times 10^{-4}$, and $(0.022\pm0.005)\times 10^{-4}$, for $D^0\bar{D}^{*0}(2007)$, $J/\psi\pi^+\pi^-$ and $J/\psi\gamma$, respectively. This is an important piece of information to discriminate between various interpretations for the $X(3872)$ state.

The measurements are classified according to the decaying particle: Charged B, Neutral B or Miscellaneous; the decay products 
and the type of quantity: branching fraction, product of branching fractions, ratio of branching fractions or other quantities. 
For the decay product classification the below precedence order is used to ensure that each measurement appears in only one category. 
\begin{itemize}\addtolength{\itemsep}{-0.4\baselineskip}
\item new particles
\item strange $D$ mesons
\item baryons
\item $J/\psi$
\item charmonium other than $J/\psi$
\item multiple $D$, $D^{*}$ or $D^{**}$ mesons
\item a single $D^{*}$ or $D^{**}$ meson
\item a single $D$ meson
\item other particles
\end{itemize}
  
Within each table the measurements are color coded according to the
publication status and age. Table~\ref{tab:hfc99999} provides a key to the
color scheme and categories used. When viewing the tables with most pdf
viewers every number, label and average provides hyperlinks to the corresponding 
reference and individual quantity web pages on the HFAG/BtoCharm group website
\hfhref{}{http://hfag.phys.ntu.edu.tw}.
The links provided in the captions of the table lead to the corresponding compilation
pages.  Both the individual and compilation webpages provide a graphical view
of the results, in a variety of formats.

Tables \ref{tab:hfc01101} to \ref{tab:hfc03300} provide either limits at 90\%
confidence level or measurements with statistical and systematic uncertainties 
and in some cases a third error corresponding to correlated systematics. 
For details on the meanings of the uncertainties and access to the references 
click on the numbers to visit the corresponding web pages.  Where there are
multiple determinations of the same quantity by one experiment the table
footnotes act to distinguish the methods or datasets used; such cases are
visually highlighted in the table by presenting the measurements on the lines
beneath the quantity label.
Where both limits and measured values of a quantity are available the limits 
are presented in the tables but are not used in the determination of the
average. Where only limits are available the most stringent is presented in
the Average column of the tables.
Where available the PDG 2006 result is also presented.

\clearpage 

\begin{\hftabletype}[\hftableposn]

\begin{center}
\caption{Key to the colors used to classify the results presented in tables \ref{tab:hfc01101} to \ref{tab:hfc03300}. When viewing these tables in a pdf viewer each number, label and average provides a hyperlink to the corresponding online version provided by the charm subgroup website \hfurl{}. Where an experiment has multiple determinations of a single quantity they are distinguished by the table footnotes.}
 \label{tab:hfc99999} 
 
\hfmetadata{  }



\end{center}

\end{\hftabletype}
\clearpage  
\begin{\hftabletype}[\hftableposn]

\begin{center}
\hfnewcaption{Branching fractions}{charged B}{\hfnewp}{in units of $10^{-3}$}{upper limits are at 90\% CL}{00101.html}{tab:hfc01101}
\hfmetadata{ [ .tml created 2008-07-05T00:46:31.96+08:00] }

%

\end{center}


\begin{center}
\hfnewcaption{Product branching fractions}{charged B}{\hfnewp}{in units of $10^{-4}$}{upper limits are at 90\% CL}{00101.html}{tab:hfc02101}
\hfmetadata{ [ .tml created 2008-07-05T00:47:25.178+08:00] }

%

\end{center}

\end{\hftabletype}
\clearpage  
\begin{\hftabletype}[\hftableposn]

\begin{center}
\hfnewcaption{Branching fractions}{charged B}{\hfdstr}{in units of $10^{-4}$}{upper limits are at 90\% CL}{00102.html}{tab:hfc01102}
\hfmetadata{ [ .tml created 2008-07-05T01:03:04.452+08:00] }

%

\begin{description}\addtolength{\itemsep}{\hffootitemsep\baselineskip}

	  \item[ \hffootnotemark{1} ] \hffootnotetext{ Evidence for the Rare Decay $B^+ \rightarrow D_s^+ \pi^0$ (\hfbb{232}) } 
     ;  \hffootnotetext{ $\bar{B}^- \rightarrow D_s^- \pi^0$ } 
	  \item[ \hffootnotemark{2} ] \hffootnotetext{ Search for $B^- \rightarrow D_s^- \pi^0$  (internal document) (\hfbb{124}) } 
     ;  \hffootnotetext{ $B^- \rightarrow D_s^- \pi^0$ } 
	  \item[ \hffootnotemark{3} ] \hffootnotetext{ Observation of the decays $B^- \rightarrow D_s^{(*)+}K^-\pi^-$ (\hfbb{324}) } 
     ;  \hffootnotemark{3a}  \hffootnotetext{ $B^- \rightarrow D_s^+K^-\pi^-$ }  ;  \hffootnotemark{3b}  \hffootnotetext{ $B^- \rightarrow D_s^{*+}K^-\pi^-$ } 
	  \item[ \hffootnotemark{4} ] \hffootnotetext{ Observation of the decays $B^- \rightarrow D_s^{(*)+}K^-\pi^-$ and $\bar{B}^0 \rightarrow D_s^{+}K^0_s\pi^-$ and Search for $\bar{B}^0 \rightarrow D_s^{*+}K^0_s\pi^-$ and $B^- \rightarrow D_s^{(*)+}K^-K^-$ (\hfbb{383}) } 
     ;  \hffootnotemark{4a}  \hffootnotetext{ $B^- \rightarrow D_s^+K^-\pi^-$ }  ;  \hffootnotemark{4b}  \hffootnotetext{ $B^- \rightarrow D_s^{*+}K^-\pi^-$ } 
\end{description}

\end{center}


\begin{center}
\hfnewcaption{Product branching fractions}{charged B}{\hfdstr}{in units of $10^{-4}$}{upper limits are at 90\% CL}{00102.html}{tab:hfc02102}
\hfmetadata{ [ .tml created 2008-07-05T01:03:37.839+08:00] }



\end{center}

\end{\hftabletype}
\clearpage  
\begin{\hftabletype}[\hftableposn]

\begin{center}
\hfnewcaption{Branching fractions}{charged B}{\hfbary}{in units of $10^{-5}$}{upper limits are at 90\% CL}{00103.html}{tab:hfc01103}
\hfmetadata{ [ .tml created 2008-07-05T01:15:07.04+08:00] }

%

\end{center}


\begin{center}
\hfnewcaption{Product branching fractions}{charged B}{\hfbary}{in units of $10^{-5}$}{upper limits are at 90\% CL}{00103.html}{tab:hfc02103}
\hfmetadata{ [ .tml created 2008-07-05T01:15:37.525+08:00] }

%

\end{center}


\begin{center}
\hfnewcaption{Ratios of branching fractions}{charged B}{\hfbary}{in units of $10^{0}$}{upper limits are at 90\% CL}{00103.html}{tab:hfc03103}
\hfmetadata{ [ .tml created 2008-07-05T01:15:46.023+08:00] }

%

\end{center}

\end{\hftabletype}
\clearpage  
\begin{\hftabletype}[\hftableposn]

\begin{center}
\hfnewcaption{Branching fractions}{charged B}{\hfjpsi}{in units of $10^{-4}$}{upper limits are at 90\% CL}{00104.html}{tab:hfc01104}
\hfmetadata{ [ .tml created 2008-07-05T01:24:00.613+08:00] }

%

\begin{description}\addtolength{\itemsep}{\hffootitemsep\baselineskip}

	  \item[ \hffootnotemark{1} ] \hffootnotetext{ MEASUREMENT OF BRANCHING FRACTIONS AND CHARGE ASYMMETRIES FOR EXCLUSIVE B DECAYS TO CHARMONIUM (\hfbb{124}) } 
     ;  \hffootnotetext{ $B^- \rightarrow J/\psi K^-$ with $J/\psi$ to leptons } 
	  \item[ \hffootnotemark{2} ] \hffootnotetext{ MEASUREMENT OF THE $B^+ \rightarrow p \overline{p} K^+$ BRANCHING FRACTION AND STUDY OF THE DECAY DYNAMICS (\hfbb{232}) } 
     ;  \hffootnotetext{ $B^- \rightarrow J/\psi K^-$ with $J/\psi \rightarrow p\overline{p}$ } 
	  \item[ \hffootnotemark{3} ] \hffootnotetext{ Measurements of the absolute branching fractions of $B^\pm \rightarrow K^\pm X_{c\overline{c}}$ (\hfbb{231.8}) } 
     ;  \hffootnotetext{ $B^- \rightarrow J/\psi K^-$ (inclusive) } 
\end{description}

\end{center}


\begin{center}
\hfnewcaption{Product branching fractions}{charged B}{\hfjpsi}{in units of $10^{-4}$}{upper limits are at 90\% CL}{00104.html}{tab:hfc02104}
\hfmetadata{ [ .tml created 2008-07-05T01:24:10.852+08:00] }



\end{center}

\end{\hftabletype}
\clearpage  
\begin{\hftabletype}[\hftableposn]

\begin{center}
\hfnewcaption{Ratios of branching fractions}{charged B}{\hfjpsi}{in units of $10^{0}$}{upper limits are at 90\% CL}{00104.html}{tab:hfc03104}
\hfmetadata{ [ .tml created 2008-07-05T01:24:32.378+08:00] }

%

\begin{description}\addtolength{\itemsep}{\hffootitemsep\baselineskip}

	  \item[ \hffootnotemark{1} ] \hffootnotetext{ Measurement of the Branching Fraction $B(B^+ \rightarrow J/\psi \pi^+)$ and Search for $B^{c+} \rightarrow J/\psi \pi^+$ } 
	  \item[ \hffootnotemark{1} ] \hffootnotetext{ Measurement of the Branching Fraction $B(B^+ \rightarrow J/\psi \pi^+)$ and Search for $B^{c+} \rightarrow J/\psi \pi^+$ } 
    
	  \item[ \hffootnotemark{2} ] \hffootnotetext{ Measurement of the Ratio of Branching Fractions B(B -- J/psi Pi)/B(B -- J/psi K) } 
     ;  \hffootnotetext{ Br(B--J/psiPi)/Br(B--J/psi K) } 
	  \item[ \hffootnotemark{3} ] \hffootnotetext{ Branching Fraction Measurements of $B \rightarrow \eta_c K$ Decays (\hfbb{86.1}) } 
     ;  \hffootnotetext{ Ratio $B^- \rightarrow \eta_{c} K^-$ to $B^- \rightarrow J/\psi K^-$ with $\eta_c \rightarrow K\overline{K}\pi$ } 
	  \item[ \hffootnotemark{4} ] \hffootnotetext{ Measurements of the absolute branching fractions of $B^\pm \rightarrow K^\pm X_{c\overline{c}}$ (\hfbb{231.8}) } 
     ;  \hffootnotetext{ Ratio $B^- \rightarrow \eta_c K^-$ to $B^- \rightarrow J/\psi K^-$  (inclusive analysis) } 
\end{description}

\end{center}

\end{\hftabletype}
\clearpage  
\begin{\hftabletype}[\hftableposn]

\begin{center}
\hfnewcaption{Branching fractions}{charged B}{\hfochm}{in units of $10^{-4}$}{upper limits are at 90\% CL}{00105.html}{tab:hfc01105}
\hfmetadata{ [ .tml created 2008-07-05T01:52:13.497+08:00] }



\begin{description}\addtolength{\itemsep}{\hffootitemsep\baselineskip}

	  \item[ \hffootnotemark{1} ] \hffootnotetext{ Dalitz-plot analysis of the decays $B^\pm \rightarrow K^\pm \pi^\mp \pi^\pm$ (\hfbb{226}) } 
     ;  \hffootnotetext{ $B^- \rightarrow \chi_{c0} K^-$ with $\chi_{c0} \rightarrow \pi^+ \pi^-$ (Dalitz analysis) } 
	  \item[ \hffootnotemark{2} ] \hffootnotetext{ MEASUREMENT OF THE BRANCHING FRACTION FOR $B^\pm \rightarrow \chi_{c0} K^\pm$. (\hfbb{88.9}) } 
     ;  \hffootnotetext{ $B^- \rightarrow \chi_{c0} K^-$ with $\chi_{c0}\rightarrow K^+ K^-, \pi^+ \pi^-$ } 
	  \item[ \hffootnotemark{3} ] \hffootnotetext{ Dalitz plot analysis of the decay $B^\pm\rightarrow K^\pm K^\pm K^\mp$ (\hfbb{226}) } 
     ;  \hffootnotetext{ $B^\pm\rightarrow K^\pm \chi_{c0}$, with $chi_c0\rightarrow K^+K^-$ (Dalitz analysis) } 
	  \item[ \hffootnotemark{4} ] \hffootnotetext{ Search for $B\rightarrow X(3872) K, X(3872)\rightarrow J/\psi\gamma$ (\hfbb{287}) } 
     ;  \hffootnotetext{ $B^- \rightarrow \chi_{c1} K^-$ with $\chi_{c1}$ to $J/\psi \gamma$ } 
	  \item[ \hffootnotemark{5} ] \hffootnotetext{ MEASUREMENT OF BRANCHING FRACTIONS AND CHARGE ASYMMETRIES FOR EXCLUSIVE B DECAYS TO CHARMONIUM (\hfbb{124}) } 
     ;  \hffootnotetext{ $B^- \rightarrow \psi(2S) K^-$ with $\psi(2S)$ to leptons } 
	  \item[ \hffootnotemark{6} ] \hffootnotetext{ Measurements of the absolute branching fractions of $B^\pm \rightarrow K^\pm X_{c\overline{c}}$ (\hfbb{231.8}) } 
     ;  \hffootnotemark{6a}  \hffootnotetext{ $B^- \rightarrow \psi(2S) K^-$ (inclusive) }  ;  \hffootnotemark{6b}  \hffootnotetext{ $B^- \rightarrow \chi_{c1} K^-$ (inclusive) }  ;  \hffootnotemark{6c}  \hffootnotetext{ $B^- \rightarrow \eta_c K^-$ (inclusive) }  ;  \hffootnotemark{6d}  \hffootnotetext{ $B^- \rightarrow \chi_{c0} K^-$ (inclusive) } 
	  \item[ \hffootnotemark{7} ] \hffootnotetext{ MEASUREMENT OF THE $B^+ \rightarrow p \overline{p} K^+$ BRANCHING FRACTION AND STUDY OF THE DECAY DYNAMICS (\hfbb{232}) } 
     ;  \hffootnotetext{ $B^- \rightarrow \eta_c K^-$ with $\eta_c \rightarrow p\overline{p}$ } 
	  \item[ \hffootnotemark{8} ] \hffootnotetext{ Branching Fraction Measurements of $B \rightarrow \eta_c K$ Decays (\hfbb{86.1}) } 
     ;  \hffootnotetext{ $B^- \rightarrow \eta_{c} K^-$ with $\eta_c \rightarrow K\overline{K}\pi$ } 
\end{description}

\end{center}

\end{\hftabletype}
\clearpage  
\begin{\hftabletype}[\hftableposn]

\begin{center}
\hfnewcaption{Product branching fractions}{charged B}{\hfochm}{in units of $10^{-4}$}{upper limits are at 90\% CL}{00105.html}{tab:hfc02105}
\hfmetadata{ [ .tml created 2008-07-05T01:52:31.648+08:00] }



\end{center}


\begin{center}
\hfnewcaption{Ratios of branching fractions}{charged B}{\hfochm}{in units of $10^{-1}$}{upper limits are at 90\% CL}{00105.html}{tab:hfc03105}
\hfmetadata{ [ .tml created 2008-07-05T01:53:00.29+08:00] }

%

\end{center}

\end{\hftabletype}
\clearpage  
\begin{\hftabletype}[\hftableposn]

\begin{center}
\hfnewcaption{Branching fractions}{charged B}{\hfmuld}{in units of $10^{-3}$}{upper limits are at 90\% CL}{00106.html}{tab:hfc01106}
\hfmetadata{ [ .tml created 2008-07-05T02:20:54.259+08:00] }

%

\begin{description}\addtolength{\itemsep}{\hffootitemsep\baselineskip}

	  \item[ \hffootnotemark{1} ] \hffootnotetext{ Observation of $B^0 \to D^+ D^-$, $B^- \to D^0 D^-$ and
$B^- \to D^0 D^{*-}$ decays (\hfbb{152}) } 
    
	  \item[ \hffootnotemark{2} ] \hffootnotetext{ Measurement of B+ - D+ D0bar branching fraction and charge asymmetry and search for B0 - D0 D0bar (\hfbb{656.7}) } 
    
\end{description}

\end{center}

\end{\hftabletype}
\clearpage  
\begin{\hftabletype}[\hftableposn]

\begin{center}
\hfnewcaption{Product branching fractions}{charged B}{\hfmuld}{in units of $10^{-4}$}{upper limits are at 90\% CL}{00106.html}{tab:hfc02106}
\hfmetadata{ [ .tml created 2008-07-05T02:21:32.152+08:00] }



\end{center}


\begin{center}
\hfnewcaption{Ratios of branching fractions}{charged B}{\hfmuld}{in units of $10^{0}$}{upper limits are at 90\% CL}{00106.html}{tab:hfc03106}
\hfmetadata{ [ .tml created 2008-07-05T02:21:54.029+08:00] }

%

\end{center}

\end{\hftabletype}
\clearpage  
\begin{\hftabletype}[\hftableposn]

\begin{center}
\hfnewcaption{Branching fractions}{charged B}{\hfsgdx}{in units of $10^{-4}$}{upper limits are at 90\% CL}{00107.html}{tab:hfc01107}
\hfmetadata{ [ .tml created 2008-07-05T02:50:01.481+08:00] }

%

\begin{description}\addtolength{\itemsep}{\hffootitemsep\baselineskip}

	  \item[ \hffootnotemark{1} ] \hffootnotetext{ Branching fraction measurements and isospin analyses for $\bar{B} \rightarrow D^{(*)}\pi^-$
decays (\hfbb{65}) } 
     ;  \hffootnotetext{ $B^- \rightarrow D^{*0}\pi^-$ } 
	  \item[ \hffootnotemark{2} ] \hffootnotetext{ Measurement of the Absolute Branching Fractions $B \rightarrow D^{(*,**)}\pi$ with a Missing Mass method (\hfbb{231}) } 
     ;  \hffootnotetext{ $B^- \rightarrow D^{*0} \pi^-$ } 
\end{description}

\end{center}

\end{\hftabletype}
\clearpage  
\begin{\hftabletype}[\hftableposn]

\begin{center}
\hfnewcaption{Branching fractions}{charged B}{\hfsgld}{in units of $10^{-4}$}{upper limits are at 90\% CL}{00108.html}{tab:hfc01108}
\hfmetadata{ [ .tml created 2008-07-05T02:57:34.662+08:00] }



\begin{description}\addtolength{\itemsep}{\hffootitemsep\baselineskip}

	  \item[ \hffootnotemark{1} ] \hffootnotetext{ Measurement of the $B^- \rightarrow D^0 K^{*-}$ branching fraction (\hfbb{232}) } 
     ;  \hffootnotetext{ Measurement of of the $B^- \rightarrow D^0 K^{*-}$ branching fraction } 
	  \item[ \hffootnotemark{2} ] \hffootnotetext{ Measurement of the Branching Fraction for $B^- \rightarrow D^0K^{*-}$ (\hfbb{86}) } 
     ;  \hffootnotetext{ $B^- \rightarrow D^0K^{*-}$ } 
	  \item[ \hffootnotemark{3} ] \hffootnotetext{ Branching fraction measurements and isospin analyses for $\bar{B} \rightarrow D^{(*)}\pi^-$
decays (\hfbb{65}) } 
     ;  \hffootnotetext{ $B^- \rightarrow D^0\pi^-$ } 
	  \item[ \hffootnotemark{4} ] \hffootnotetext{ Measurement of the Absolute Branching Fractions $B \rightarrow D^{(*,**)}\pi$ with a Missing Mass method (\hfbb{231}) } 
     ;  \hffootnotetext{ $B^- \rightarrow D^0\pi^-$ } 
\end{description}

\end{center}

\end{\hftabletype}
\clearpage  
\begin{\hftabletype}[\hftableposn]

\begin{center}
\hfnewcaption{Branching fractions}{neutral B}{\hfnewp}{in units of $10^{-3}$}{upper limits are at 90\% CL}{00201.html}{tab:hfc01201}
\hfmetadata{ [ .tml created 2008-07-05T03:02:22.236+08:00] }



\end{center}


\begin{center}
\hfnewcaption{Product branching fractions}{neutral B}{\hfnewp}{in units of $10^{-4}$}{upper limits are at 90\% CL}{00201.html}{tab:hfc02201}
\hfmetadata{ [ .tml created 2008-07-05T03:02:41.234+08:00] }

%

\end{center}


\begin{center}
\hfnewcaption{Ratios of branching fractions}{neutral B}{\hfnewp}{in units of $10^{0}$}{upper limits are at 90\% CL}{00201.html}{tab:hfc03201}
\hfmetadata{ [ .tml created 2008-07-05T03:03:10.415+08:00] }

%

\end{center}

\end{\hftabletype}
\clearpage  
\begin{\hftabletype}[\hftableposn] 
\tiny  


\begin{center}
\hfnewcaption{Branching fractions}{neutral B}{\hfdstr}{in units of $10^{-3}$}{upper limits are at 90\% CL}{00202.html}{tab:hfc01202}
\hfmetadata{ [ .tml created 2008-07-05T03:16:25.531+08:00] }

%

\begin{description}\addtolength{\itemsep}{\hffootitemsep\baselineskip}

	  \item[ \hffootnotemark{1} ] \hffootnotetext{ Observation of Decays $\bar{B}^0 \rightarrow D_s^{-(*)}\pi^+$ and $\bar{B}^0 \rightarrow D_s^{+(*)}K^-$ (\hfbb{230}) } 
     ;  \hffootnotemark{1a}  \hffootnotetext{ $\bar{B}^0 \rightarrow D_s^{*-} \pi^+$ }  ;  \hffootnotemark{1b}  \hffootnotetext{ $\bar{B}^0 \rightarrow D_s^+ K^-$ }  ;  \hffootnotemark{1c}  \hffootnotetext{ $\bar{B}^0 \rightarrow D_s^{*+} K^-$ }  ;  \hffootnotemark{1d}  \hffootnotetext{ $\bar{B}^0 \rightarrow D_s^- \pi^+$ } 
	  \item[ \hffootnotemark{2} ] \hffootnotetext{ A study of the rare decays $\bar{B}^0 \rightarrow D_s^{-(*)}\pi^+$ and 
$\bar{B}^0 \rightarrow D_s^{+(*)} K^-$ (\hfbb{84.3}) } 
     ;  \hffootnotemark{2a}  \hffootnotetext{ $\bar{B}^0 \rightarrow D_s^{*-} \pi^+$ }  ;  \hffootnotemark{2b}  \hffootnotetext{ $\bar{B}^0 \rightarrow D_s^+ K^-$ }  ;  \hffootnotemark{2c}  \hffootnotetext{ $\bar{B}^0 \rightarrow D_s^{*+} K^-$ }  ;  \hffootnotemark{2d}  \hffootnotetext{ $\bar{B}^0 \rightarrow D_s^- \pi^+$ } 
	  \item[ \hffootnotemark{3} ] \hffootnotetext{ Observation of B0bar - Ds+ Lambda pbar decay (\hfbb{449}) } 
    
	  \item[ \hffootnotemark{4} ] \hffootnotetext{ Observation of B0bar to Ds+ Lambda pbar (\hfbb{447}) } 
    
	  \item[ \hffootnotemark{5} ] \hffootnotetext{ Improved measurement of $\bar{B}^0\to D_s^-D^+$ and search for $\bar{B}0\to D_s^+D_s^-$ at Belle } 
    
	  \item[ \hffootnotemark{6} ] \hffootnotetext{ Improved measurement of B0bar - Ds-D+ and search for B0bar - Ds+Ds- (\hfbb{449}) } 
    
	  \item[ \hffootnotemark{7} ] \hffootnotetext{ Measurement of $\bar{B}^0 \rightarrow D_s^{(*)}D^*$ Branching Fractions and $D_s^*D^*$
 Polarization with a Partial Reconstruction technique (\hfbb{22.7}) } 
     ;  \hffootnotemark{7a}  \hffootnotetext{ $\bar{B}^0 \rightarrow D_s^- D^{*+}$ }  ;  \hffootnotemark{7b}  \hffootnotetext{ $\bar{B}^0 \rightarrow D_s^{*-} D^{*+}$ } 
	  \item[ \hffootnotemark{8} ] \hffootnotetext{ Study of $\bar{B} \rightarrow D^{(*)+,-}X^{-}$ and $\bar{B} \rightarrow D_s^{(*)-}X^{+,0}$ decays and measurement of $D_s^-$ and $D_{sJ}^-(2460)$ absolute branching fractions (\hfbb{230}) } 
     ;  \hffootnotemark{8a}  \hffootnotetext{ $\bar{B}^0 \rightarrow D_s^-D^{*+}$ }  ;  \hffootnotemark{8b}  \hffootnotetext{ $\bar{B}^0 \rightarrow D_s^{*-}D^{*+})$ } 
	  \item[ \hffootnotemark{9} ] \hffootnotetext{ Measurement of the $\bar{B}^0 \rightarrow D_s^{*-} D^+$ and $D_s^+ \rightarrow \phi \pi^+$ branching
fractions (\hfbb{123}) } 
     ;  \hffootnotetext{ $\bar{B}^0 \rightarrow D_s^{*-} D^{*+}$ } 
\end{description}

\end{center}

\normalsize 

\end{\hftabletype}
\clearpage  
\begin{\hftabletype}[\hftableposn]

\begin{center}
\hfnewcaption{Product branching fractions}{neutral B}{\hfdstr}{in units of $10^{-4}$}{upper limits are at 90\% CL}{00202.html}{tab:hfc02202}
\hfmetadata{ [ .tml created 2008-07-05T03:17:20.873+08:00] }



\end{center}


\begin{center}
\hfnewcaption{Ratios of branching fractions}{neutral B}{\hfdstr}{in units of $10^{0}$}{upper limits are at 90\% CL}{00202.html}{tab:hfc03202}
\hfmetadata{ [ .tml created 2008-07-05T03:18:11.81+08:00] }

%

\end{center}

\end{\hftabletype}
\clearpage  
\begin{\hftabletype}[\hftableposn]

\begin{center}
\hfnewcaption{Branching fractions}{neutral B}{\hfbary}{in units of $10^{-5}$}{upper limits are at 90\% CL}{00203.html}{tab:hfc01203}
\hfmetadata{ [ .tml created 2008-07-05T03:52:24.989+08:00] }

%

\begin{description}\addtolength{\itemsep}{\hffootitemsep\baselineskip}

	  \item[ \hffootnotemark{1} ] \hffootnotetext{ STUDY OF EXCLUSIVE B DECAYS TO CHARMED BARYONS AT BELLE. (\hfbb{31.7}) } 
    
	  \item[ \hffootnotemark{2} ] \hffootnotetext{ Study of the charmed baryonic decays $\bar{B}^0\to\Sigma_c^{++}\bar{p}\pi^-$ and $\bar{B}^0\to\Sigma_c^{0}\bar{p}\pi^+$ (\hfbb{386}) } 
     ;  \hffootnotemark{2a}  \hffootnotetext{ B0bar to Sigmac(2455)++ pbar pi } 
	  \item[ \hffootnotemark{3} ] \hffootnotetext{ Measurement of the Branching Fraction for the decays
$\bar{B}^0 \rightarrow D^{*+}p\bar{p}\pi^-$, $\bar{B}^0 \rightarrow D^+ p \bar{p} \pi^-$,
$\bar{B}^0 \rightarrow \bar{D}^{*0} p \bar{p}$, $\bar{B}^0 \rightarrow \bar{D}^0 p \bar{p}$ (\hfbb{124}) } 
     ;  \hffootnotemark{3a}  \hffootnotetext{ $\bar{B}^0 \rightarrow D^{+}p\bar{p}\pi^-$ }  ;  \hffootnotemark{3b}  \hffootnotetext{ $\bar{B}^0 \rightarrow D^{*+}p\bar{p}\pi^-$ }  ;  \hffootnotemark{3c}  \hffootnotetext{ $\bar{B}^0 \rightarrow \bar{D}^{0}p\bar{p}$ }  ;  \hffootnotemark{3d}  \hffootnotetext{ $\bar{B}^0 \rightarrow \bar{D}^{*0}p\bar{p}$ } 
	  \item[ \hffootnotemark{4} ] \hffootnotetext{ Measurements of the Decays $B^0 \rightarrow \bar{D}^0 p \bar{p}$,  $B^0 \rightarrow \bar{D}^{*0} p \bar{p}$, 
$B^0 \rightarrow D^-p\bar{p}\pi+-$, and $B^0 \rightarrow D^- p \bar{p} \pi^+$ (\hfbb{232}) } 
     ;  \hffootnotemark{4a}  \hffootnotetext{ $\bar{B}^0 \rightarrow D^{+}p\bar{p}\pi^-$ }  ;  \hffootnotemark{4b}  \hffootnotetext{ $\bar{B}^0 \rightarrow D^{*+}p\bar{p}\pi^-$ }  ;  \hffootnotemark{4c}  \hffootnotetext{ $\bar{B}^0 \rightarrow \bar{D}^{0}p\bar{p}$ }  ;  \hffootnotemark{4d}  \hffootnotetext{ $\bar{B}^0 \rightarrow \bar{D}^{*0}p\bar{p}$ } 
\end{description}

\end{center}

\end{\hftabletype}
\clearpage  
\begin{\hftabletype}[\hftableposn]

\begin{center}
\hfnewcaption{Product branching fractions}{neutral B}{\hfbary}{in units of $10^{-5}$}{upper limits are at 90\% CL}{00203.html}{tab:hfc02203}
\hfmetadata{ [ .tml created 2008-07-05T03:52:39.382+08:00] }



\end{center}

\end{\hftabletype}
\clearpage  
\begin{\hftabletype}[\hftableposn]

\begin{center}
\hfnewcaption{Branching fractions}{neutral B}{\hfjpsi}{in units of $10^{-4}$}{upper limits are at 90\% CL}{00204.html}{tab:hfc01204}
\hfmetadata{ [ .tml created 2008-07-05T04:01:46.226+08:00] }

%

\begin{description}\addtolength{\itemsep}{\hffootitemsep\baselineskip}

	  \item[ \hffootnotemark{1} ] \hffootnotetext{ Study of $B^0 \to J/\psi \pi^+ \pi^-$ decays with 449 million $B\bar{B}$ pairs at Belle (\hfbb{449}) } 
    
	  \item[ \hffootnotemark{2} ] \hffootnotetext{ MEASUREMENT OF BRANCHING FRACTIONS IN B0 --- J/PSI PI+ PI- DECAY. (\hfbb{152}) } 
    
\end{description}

\end{center}

\end{\hftabletype}
\clearpage  
\begin{\hftabletype}[\hftableposn]

\begin{center}
\hfnewcaption{Ratios of branching fractions}{neutral B}{\hfjpsi}{in units of $10^{0}$}{upper limits are at 90\% CL}{00204.html}{tab:hfc03204}
\hfmetadata{ [ .tml created 2008-07-05T04:01:59.212+08:00] }



\end{center}


\begin{center}
\hfnewcaption{Miscellaneous quantities}{neutral B}{\hfjpsi}{in units of $10^{0}$}{upper limits are at 90\% CL}{00204.html}{tab:hfc04204}
\hfmetadata{ [ .tml created 2008-07-05T04:02:10.89+08:00] }

%

\end{center}

\end{\hftabletype}
\clearpage  
\begin{\hftabletype}[\hftableposn]

\begin{center}
\hfnewcaption{Branching fractions}{neutral B}{\hfochm}{in units of $10^{-3}$}{upper limits are at 90\% CL}{00205.html}{tab:hfc01205}
\hfmetadata{ [ .tml created 2008-07-05T04:15:51.633+08:00] }

%

\begin{description}\addtolength{\itemsep}{\hffootitemsep\baselineskip}

	  \item[ \hffootnotemark{1} ] \hffootnotetext{ A study of $B$-meson decays to $\eta_c K^*$ and $\eta_c \gamma K^{(*)}$ (\hfbb{384}) } 
     ;  \hffootnotetext{ betackstar0 } 
	  \item[ \hffootnotemark{2} ] \hffootnotetext{ Branching Fraction Measurements of $B \rightarrow \eta_c K$ Decays (\hfbb{86.1}) } 
    
	  \item[ \hffootnotemark{3} ] \hffootnotetext{ Evidence for the $B^0\to p \bar{p} K^{*0}$ and $B^+\to \eta_cK^{*+}$ decays and Study of the Decay Dynamics of $B$ Meson Decays into $p\bar{p}h$ Final States. (\hfbb{232}) } 
     ;  \hffootnotemark{3a}  \hffootnotetext{ betackzero }  ;  \hffootnotemark{3b}  \hffootnotetext{ betackstarzppbar } 
\end{description}

\end{center}


\begin{center}
\hfnewcaption{Product branching fractions}{neutral B}{\hfochm}{in units of $10^{-4}$}{upper limits are at 90\% CL}{00205.html}{tab:hfc02205}
\hfmetadata{ [ .tml created 2008-07-05T04:16:05.797+08:00] }



\end{center}

\end{\hftabletype}
\clearpage  
\begin{\hftabletype}[\hftableposn]

\begin{center}
\hfnewcaption{Ratios of branching fractions}{neutral B}{\hfochm}{in units of $10^{0}$}{upper limits are at 90\% CL}{00205.html}{tab:hfc03205}
\hfmetadata{ [ .tml created 2008-07-05T04:16:25.804+08:00] }

%

\end{center}

\end{\hftabletype}
\clearpage  
\begin{\hftabletype}[\hftableposn]

\begin{center}
\hfnewcaption{Branching fractions}{neutral B}{\hfmuld}{in units of $10^{-3}$}{upper limits are at 90\% CL}{00206.html}{tab:hfc01206}
\hfmetadata{ [ .tml created 2008-07-05T04:28:48.724+08:00] }

%

\begin{description}\addtolength{\itemsep}{\hffootitemsep\baselineskip}

	  \item[ \hffootnotemark{1} ] \hffootnotetext{ Evidence for CP Violation in B0 - D+D- Decays (\hfbb{535}) } 
    
	  \item[ \hffootnotemark{2} ] \hffootnotetext{ Observation of $B^0 \to D^+ D^-$, $B^- \to D^0 D^-$ and
$B^- \to D^0 D^{*-}$ decays (\hfbb{152}) } 
    
	  \item[ \hffootnotemark{3} ] \hffootnotetext{ Measurement of Branching Fraction and CP-violating charge
asymmetried for B meson decays to $D^{(*)}D^{(*)}$ and 
implications for the CKM angle $\gamma$ (\hfbb{232}) } 
     ;  \hffootnotetext{ $\bar{B}^0 \rightarrow D^{*+}D^{*-}$ } 
	  \item[ \hffootnotemark{4} ] \hffootnotetext{ Measurement of the branching fraction and CP content for the 
decay $B^0$ to $D^*D^*$ (\hfbb{23}) } 
     ;  \hffootnotetext{ $\bar{B}^0 \rightarrow D^{*-}D^{*+}$ } 
\end{description}

\end{center}

\end{\hftabletype}
\clearpage  
\begin{\hftabletype}[\hftableposn]

\begin{center}
\hfnewcaption{Product branching fractions}{neutral B}{\hfmuld}{in units of $10^{-4}$}{upper limits are at 90\% CL}{00206.html}{tab:hfc02206}
\hfmetadata{ [ .tml created 2008-07-05T04:29:07.283+08:00] }



\end{center}


\begin{center}
\hfnewcaption{Ratios of branching fractions}{neutral B}{\hfmuld}{in units of $10^{0}$}{upper limits are at 90\% CL}{00206.html}{tab:hfc03206}
\hfmetadata{ [ .tml created 2008-07-05T04:29:21.414+08:00] }

%

\end{center}

\end{\hftabletype}
\clearpage  
\begin{\hftabletype}[\hftableposn]

\begin{center}
\hfnewcaption{Branching fractions}{neutral B}{\hfsgdx}{in units of $10^{-4}$}{upper limits are at 90\% CL}{00207.html}{tab:hfc01207}
\hfmetadata{ [ .tml created 2008-07-05T04:54:05.732+08:00] }

%

\begin{description}\addtolength{\itemsep}{\hffootitemsep\baselineskip}

	  \item[ \hffootnotemark{1} ] \hffootnotetext{ A study of the  $\bar{B}^0 \rightarrow D^{(*)0} K^{(*)0}$ decays (\hfbb{226}) } 
     ;  \hffootnotetext{ $\bar{B}^0 \rightarrow D^{*0} \bar{K}^{0}$ } 
	  \item[ \hffootnotemark{2} ] \hffootnotetext{ A study of the  $\bar{B}^0 \rightarrow D^{(*)0} K^{(*)0}$ decays (\hfbb{124}) } 
     ;  \hffootnotetext{ $B \rightarrow D^{*0} \bar{K}^{0}$ } 
	  \item[ \hffootnotemark{3} ] \hffootnotetext{ Study of $\bar{B^{0}} \to D^{(*) 0} \pi^+ \pi^-$ Decays (\hfbb{31.3}) } 
    
	  \item[ \hffootnotemark{4} ] \hffootnotetext{ Study of $\bar{B}^{0} \to  D^{(*)0} \pi^{+} \pi^{-}$ decays ; Dalitz fit analysis (\hfbb{152}) } 
    
	  \item[ \hffootnotemark{5} ] \hffootnotetext{ Measurement of the Absolute Branching Fractions $B \rightarrow D^{(*,**)}\pi$ with a Missing Mass method (\hfbb{231}) } 
     ;  \hffootnotetext{ $\bar{B}^0 \rightarrow D^{*+}\pi^-$ } 
	  \item[ \hffootnotemark{6} ] \hffootnotetext{ Branching fraction measurements and isospin analyses for $\bar{B} \rightarrow D^{(*)}\pi^-$
decays (\hfbb{65}) } 
     ;  \hffootnotetext{ $\bar{B}^0 \rightarrow D^{*+}\pi^-$ } 
\end{description}

\end{center}

\end{\hftabletype}
\clearpage  

\clearpage 
\begin{\hftabletype}[\hftableposn]

\begin{center}
\hfnewcaption{Branching fractions}{neutral B}{\hfsgld}{in units of $10^{-4}$}{upper limits are at 90\% CL}{00208.html}{tab:hfc01208}
\hfmetadata{ [ .tml created 2008-07-05T05:06:42.672+08:00] }



\begin{description}\addtolength{\itemsep}{\hffootitemsep\baselineskip}

	  \item[ \hffootnotemark{1} ] \hffootnotetext{ A study of the  $\bar{B}^0 \rightarrow D^{(*)0} K^{(*)0}$ decays (\hfbb{226}) } 
     ;  \hffootnotemark{1a}  \hffootnotetext{ $\bar{B}^0 \rightarrow D^{0} \bar{K}^{0}$ }  ;  \hffootnotemark{1b}  \hffootnotetext{ $\bar{B}^0 \rightarrow D^{0} \bar{K}^{*0}$ }  ;  \hffootnotemark{1c}  \hffootnotetext{ $\bar{B}^0 \rightarrow \bar{D}^{0} \bar{K}^{*0}$ } 
	  \item[ \hffootnotemark{2} ] \hffootnotetext{ A study of the  $\bar{B}^0 \rightarrow D^{(*)0} K^{(*)0}$ decays (\hfbb{124}) } 
     ;  \hffootnotemark{2a}  \hffootnotetext{ $B \rightarrow D^{0} \bar{K}^{0}$ }  ;  \hffootnotemark{2b}  \hffootnotetext{ $\bar{B}^0 \rightarrow D^{0} K^{*0}$ }  ;  \hffootnotemark{2c}  \hffootnotetext{ $\bar{B}^0 \rightarrow \bar{D}^{0} \bar{K}^{*0}$ } 
	  \item[ \hffootnotemark{3} ] \hffootnotetext{ Study of $\bar{B^{0}} \to D^{(*) 0} \pi^+ \pi^-$ Decays (\hfbb{31.3}) } 
    
	  \item[ \hffootnotemark{4} ] \hffootnotetext{ Study of $\bar{B}^{0} \to  D^{(*)0} \pi^{+} \pi^{-}$ decays ; Dalitz fit analysis (\hfbb{152}) } 
    
	  \item[ \hffootnotemark{5} ] \hffootnotetext{ Measurement of the Absolute Branching Fractions $B \rightarrow D^{(*,**)}\pi$ with a Missing Mass method (\hfbb{231}) } 
     ;  \hffootnotetext{ $\bar{B}^0 \rightarrow D^+\pi^-$ } 
	  \item[ \hffootnotemark{6} ] \hffootnotetext{ Branching fraction measurements and isospin analyses for $\bar{B} \rightarrow D^{(*)}\pi^-$
decays (\hfbb{65}) } 
     ;  \hffootnotetext{ $\bar{B}^0 \rightarrow D^+\pi^-$ } 
\end{description}

\end{center}

\end{\hftabletype}
\clearpage  
\begin{\hftabletype}[\hftableposn]

\begin{center}
\hfnewcaption{Product branching fractions}{neutral B}{\hfsgld}{in units of $10^{-5}$}{upper limits are at 90\% CL}{00208.html}{tab:hfc02208}
\hfmetadata{ [ .tml created 2008-07-05T05:07:13.088+08:00] }

\begin{tabular}{lccccc}  
\hflmode
&  \hflpdg2006
&  \hflbelle
&  \hflbabar
&  \hflcdf
&  \hflavg
\\\hline 

 \hfhref{BR_-511_-313+421xBR_-313_-321+211.html}{ \hflabel{D^0 \bar{K}^{*0}(892) [ K^- \pi^+ ]} } \hftabletlcell\hftableblcell 
 &  { \,}  
 &  { \,}  
 &  \hfhref{0506018.html}{ \hfpub{3.80\pm0.60\pm0.40} }  
 &  { \,}  
 &  \hfhref{BR_-511_-313+421xBR_-313_-321+211.html}{ \hfavg{3.80\pm0.72} } 
\\\hline 

\end{tabular}

\end{center}

\end{\hftabletype}
\clearpage  
\begin{\hftabletype}[\hftableposn]

\begin{center}
\hfnewcaption{Branching fractions}{miscellaneous}{charmed particles }{in units of $10^{-3}$}{upper limits are at 90\% CL}{00300.html}{tab:hfc01300}
\hfmetadata{ [ .tml created 2008-07-05T05:19:11.567+08:00] }



\end{center}


\begin{center}
\hfnewcaption{Product branching fractions}{miscellaneous}{charmed particles }{in units of $10^{-5}$}{upper limits are at 90\% CL}{00300.html}{tab:hfc02300}
\hfmetadata{ [ .tml created 2008-07-05T05:19:24.144+08:00] }

%

\end{center}


\begin{center}
\hfnewcaption{Ratios of branching fractions}{miscellaneous}{charmed particles }{in units of $10^{0}$}{upper limits are at 90\% CL}{00300.html}{tab:hfc03300}
\hfmetadata{ [ .tml created 2008-07-05T05:19:36.933+08:00] }

%

\end{center}

\end{\hftabletype}
\clearpage  
\begin{\hftabletype}[\hftableposn]

\begin{center}
\hfnewcaption{Miscellaneous quantities}{miscellaneous}{charmed particles }{in units of $10^{0}$}{upper limits are at 90\% CL}{00300.html}{tab:hfc04300}
\hfmetadata{ [ .tml created 2008-07-05T05:19:57.698+08:00] }

%

\end{center}

\end{\hftabletype}
\clearpage  


\clearpage 
\mysection{$B$ decays to charmless final states}


\label{sec:rare}

The aim of this section is to provide the branching fractions and
the partial rate asymmetries ($A_{CP}$) of charmless 
$B$ decays. The asymmetry is defined as 
$A_{CP} = \frac{N_{\Bbar} -N_B}{N_{\Bbar} +N_B}$, where $N_{\Bbar}$ 
and $N_B$ are respectively number of $\Bzb/\Bm$ and $\Bz/\Bp$ decaying
into a specific final state. 
Four different $B$ decay categories are considered: 
charmless mesonic, baryonic, radiative and leptonic. Measurements supported 
with  written documents are accepted in  
the averages; written documents could be journal papers, 
conference contributed papers, preprints or conference proceedings.  
Results from  $A_{CP}$ measurements  obtained from time dependent analyses 
are listed and described in Sec.~\ref{sec:cp_uta}. Measurements of  charmful 
baryonic $B$ decays, which were included in our previous averages
\cite{hfag_hepex_endof2006}, are now shown    
in Section 7, which deals with $B$ decays to charm.  

So far all branching fractions assume equal production of charged and
neutral $B$ pairs.  The best measurements to date show that this is
still a good approximation (see Sec.~\ref{sec:life_mix}).
For branching fractions, we provide either averages or the most stringent
90\% confidence level upper limits.  If one or more experiments have
measurements with $>$4$\sigma$ for a decay channel, all available central values
for that channel are used in the averaging.  We also give central values
and errors for cases where the significance of the average value is at
least $3 \sigma$, even if no single measurement is above $4 \sigma$. 
Since a few decay modes are sensitive to the contribution of
 new physics and the current experimental upper limits are not far from the 
Standard Model expectation, we provide the combined upper limits or
averages in these cases.
Their upper limits can be estimated assuming that the errors are 
Gaussian.  For $A_{CP}$ we provide averages in all cases.  

Our averaging is performed by maximizing the likelihood,
   $\displaystyle {\mathcal L} = \prod_i {\mathcal P}_i(x),$  
where ${\mathcal P_i}$ is the probability density function (PDF) of the
$i$th  measurement, and $x$ is the branching fraction or $A_{CP}$.
The PDF is modeled by an asymmetric Gaussian function with the measured
central value as its mean and the quadratic sum of the statistical
and systematic errors as the standard deviations. The experimental
uncertainties are considered to be uncorrelated with each other when the 
averaging is performed. No error scaling is applied when the fit $\chi^2$ is 
greater than 1 since we believe that tends to overestimate the errors
except in cases of extreme disagreement (we have no such cases).
One exception to consider the correlated systematic errors is the inclusive
$B\to X_s\gamma$ mode, which is sensitive to physics beyond the Standard Model.
In this update, we have included new measurements from both Belle and BaBar
to perform the average. The detail is  
described  in Sec. ~\ref{sec:btosg}.

At present, we have measurements of more than 350  decay modes, reported in
more than 200 papers. Because the number of references is so large, we do
not include them with the tables shown here but the full set of
references is available quickly from active gifs at the 
``Winter 2008'' link on 
the rare web page: {\tt http://www.slac.stanford.edu/xorg/hfag/rare/index.html}.
Finally many new  measurements on the scalar-tensor and vector-tensor decays 
and  direct $CP$ asymmetry on $B^0_s\to K^+\pi^-$ are   
included for the first time.  

\mysubsection{Mesonic charmless decays}

\begin{table}
\begin{center}
\caption{Branching fractions (BF) of charmless mesonic  
$B^+$ decays with kaons (in units of $\times 10^6)$). Upper limits are
at 90\% CL. Values in {\red red} ({\blue blue}) are new {\red published}
({\blue preliminary}) results since PDG2006 [as of April 15, 2008].   
}
\scriptsize

\end{center}

\vspace{-0.4cm} 
\hspace*{-0.8cm} 
\hspace{0.4cm}
$\dag$Product BF - daughter BF taken to be 100\%; 
~\S $M_{\phi\phi}<2.85$ GeV/$c^2$
\end{table}

\begin{table}
\begin{center}
\caption{Branching Fractions (BF) of charmless mesonic $B^+$ decays 
 without kaons (in units of $10^{-6}$). Upper limits are at 90\% CL.
Values in {\red red} ({\blue blue}) are new {\red published}
({\blue preliminary}) result since PDG2006  [as of April 15, 2008].
}
\scriptsize

\end{center}

\vspace{-0.4cm} 
\hspace*{0.8cm} 
$\dag$Product BF - daughter BF taken to be 100\%; 
\end{table}
\clearpage

\begin{table}
\begin{center}
\caption{Branching fractions of charmless mesonic $B^0$ decays with kaons 
(in units of $10^{-6}$).
Upper limits are at 90\% CL.
Values in {\red red} ({\blue blue}) are new {\red published}
({\blue preliminary}) result since PDG2006  [as of April 15, 2008].
}

\scriptsize

\end{center}
\vspace{-0.4cm}
$\dag$Product BF - daughter BF taken to be 100\%, 
 $\ddag$Relative BF converted to absolute BF
~\S $M_{\phi\phi}<2.85$ GeV/$c^2$
 $^1$Excludes $M(K_SK_S)$ regions [3.400,3.429] and [3.540,3.585] and $M(K_SK_L)<1.049$ GeV/$c^2$
 $^2$Includes $K\pi$ S-wave contribution and uncorrected for K*(1430) BF
\end{table}
\clearpage

\begin{table}
\begin{center}
\caption{
Branching fractions of charmless mesonic $B^0$ decays without kaons
(in units of $10^{-6}$).
Upper limits are at 90\% CL.
Values in {\red red} ({\blue blue}) are new {\red published}
({\blue preliminary}) result since PDG2006  [as of April 15, 2008].
}

\scriptsize

\end{center}
\vspace{-0.4cm}
$\dag$Product BF - daughter BF taken to be 100\%, 
 $\ddag$Relative BF converted to absolute BF
\end{table}

\begin{table}
\begin{center}
\caption{
 Relative branching fractions of $B^0\to K^+K^-, K^+\pi^-, \pi^+\pi^-$.
Values in {\red red} ({\blue blue}) are new {\red published}
({\blue preliminary}) result since PDG2006  [as of March 15, 2007].
}
\small

%
\end{center}
\end{table}
\clearpage

\mysubsection{Radiative and leptonic decays}

\begin{table}[!hb]
\begin{center}
\caption{
Branching fractions of semileptonic and radiative $B^+$ decays
(in units of $10^{-6}$). Upper limits are at 90\% CL.
Values in {\red red} ({\blue blue}) are new {\red published}
({\blue preliminary}) result since PDG2006  [as of April 15, 2008].
}
\scriptsize
%
\vspace{-0.4cm}
\end{center}
~\dag $M_{K\pi\pi} < 1.8$ GeV/$c^2$;~\ddag~$1.0 < M_{K\pi\pi} < 2.0$ GeV/$c^2$;
~\S~$M_{K\pi\pi} < 2.4$ GeV/$c^2$
\end{table}
\clearpage

\begin{table}
\begin{center}
\caption{
Branching fractions of semileptonic and radiative $B^0$ decays
(in units of $10^{-6}$).  Upper limits are at 90\% CL.
Values in {\red red} ({\blue blue}) are new {\red published}
({\blue preliminary}) result since PDG2006  [as of April 15, 2008].
}

\scriptsize
%
\vspace{-0.4cm}
\end{center}
~\dag $M_{K\pi\pi} < 1.8$ GeV/$c^2$;~\ddag~$1.0 < M_{K\pi\pi} < 2.0$ GeV/$c^2$;
~\S~$1.25$ GeV/$c^2 < M_{K\pi} < 1.6$ GeV/$c^2$

\end{table}

\begin{table}
\begin{center}
\caption{
Branching fractions of  semileptonic and radiative $B$ decays
(in units of $10^{-6}$).  Upper limits are at 90\% CL.
Values in {\red red} ({\blue blue}) are new {\red published}
({\blue preliminary}) result since PDG2006  [as of April 15, 2008].
}
\end{center}

\scriptsize
%
\vspace{-0.7cm}
 ~\dag $E_\gamma > 2.0$ GeV; ~\ddag $M(\ell^+\ell^-)>0.2$ GeV/$c^2$
\end{table}

\begin{table}
\begin{center}
\caption{
Branching fractions of  inclusive $B$ decays
(in units of $10^{-6}$). Values in {\red red} ({\blue blue}) 
are new {\red published}
({\blue preliminary}) result since PDG2006  [as of April 15, 2008].
}

\vspace{0.3cm}
%
\end{center}
\vspace{-0.3cm}
~~~~~~~~~~~~~~~~~~\dag~$p^* > 2.34$ GeV
\end{table}

\begin{table}[!htb]
\begin{center}

\caption{
 Branching fractions of leptonic  $B$ decays
(in units of $10^{-6}$). Upper limits are at 90\% CL.
Values in {\red red} ({\blue blue}) are new {\red published}
({\blue preliminary}) result since PDG2006  [as of April 15, 2008].
}

%
\end{center}
\end{table}

\mysubsection{$B\to X_s\gamma$}
\label{sec:btosg}

The decay $B \to X_s\gamma$ proceeds through a process of 
flavor changing neutral current. Since the charged Higgs or SUSY particles may
contribute in the penguin loop, the branching fraction is sensitive to physics
beyond the Standard Model. Experimentally, the branching fraction is measured
using either a semi-inclusive or an inclusive approach. A minimum 
photon energy requirement is applied in the analysis and the branching fraction
is corrected based on the theoretical model for the photon energy spectrum 
(shape function). In this average of the $B\to X_s\gamma$ branching 
fraction, we still use the extrapolation factors \cite{bf} 
obtained by O. Buchm\"uller 
and H. Fl\"acher and listed in Table~\ref{tab:factor}. The extrapolation
factors are defined as the ratios of the $B\to X_s\gamma$ branching fractions
 with minimum photon energies above and at 1.6 GeV.
 The appropriate
approach to average the experimental results is to first convert them 
according to the average extrapolation factors and then perform the average,
assuming that the errors of the extrapolation factors are 100\% correlated. 

\begin{table}[h]
\caption{Extrapolation factor in various scheme with various minimum
   photon energy requirement (in GeV).}
\begin{tabular}{lccccc} \hline\hline
Scheme & $E_\gamma < 1.7$ & $E_\gamma < 1.8$ & $E_\gamma < 1.9$ & 
$E_\gamma < 2.0$ & $E_\gamma < 2.242$ \\ \hline
Kinetic & $0.986\pm 0.001$ & $0.968\pm 0.002$ & $0.939\pm 0.005$ & $0.903\pm0.009$ & $0.656\pm 0.031$ \\
Neubert SF & $0.982\pm 0.002$ & $0.962\pm 0.004$ & $0.930\pm 0.008$ & $0.888\pm 0.014$ & $0.665\pm 0.035$ \\
Kagan-Neubert & $0.988\pm 0.002$ & $0.970\pm 0.005$ & $0.940\pm 0.009$ & $0.892\pm 0.014$ & $0.643\pm 0.033$\\ \hline
Average & $0.985\pm 0.004$ & $0.967\pm 0.006$ & $0.936\pm 0.010$ & $0.894\pm 0.016$ &  $0.655\pm 0.037$ \\ \hline
\end{tabular}
\label{tab:factor}
\end{table}        
  
After releasing our average for 2006\cite{hfag_hepex_endof2006}, 
two more measurements 
on the $B\to X_s\gamma$ branching fraction were available: the BaBar result 
\cite{babar3} using full hadronic tags and the Belle inclusive result 
\cite{belle2} with a factor of five more data than the previous measurement. 
The former used a data sample orthogonal to the lepton tag sample for the
early inclusive measurement. Therefore, the new BaBar result is included in
the average  while the Belle measurement supersedes their 
previous one. In the Belle new measurement, the $B\to X_s \gamma$ branching 
fraction was obtained with various minimum photon energy requirement, 
$1.7$ to $2.1$ GeV. The study shows that there are clearly signal events 
with photon energy between 1.7 and 1.8 GeV. 
Although lowering $E_\gamma$ to 1.7 GeV causes larger systematic error 
from the background, it will encourage a  deeper understanding of the theory
uncertainties, especially on those related to the extrapolation. Therefore,    
we choose the Belle measurement with the minimum photon energy at 1.7 GeV to 
compute the average.           
  
The six experimental measurements selected for the average are
shown in Table \ref{tab:measurement}.  They
have provided in their papers either the $B\to X_s\gamma$ branching fraction at 
a certain photon energy cut or the extrapolation factor used.
Therefore we are able to convert them to the values at $E_{\rm min}= 1.6$ 
GeV using the information in Table \ref{tab:factor}.  The errors are, 
in order, statistical, systematic and shape-function systematic,
except for the \babar\ inclusive where there is a second systematic
error (third quoted error) due to theoretical uncertainties.
Moreover, in the four inclusive analyses a possible $B\to X_d\gamma$ 
contamination has been considered according to the expectation
of $(4.0\pm 0.4)$\%. The central value is the same as used in our 2006 
average but the uncertainty shrinks by a factor of four, due to  better 
understanding of
$|V_{td}/V_{ts}|$ from the $B_S$-$\overline{B}_S$ mixing and $B\to \rho/\omega 
\;\gamma$ measurements. Compared to the other systematic uncertainties,
the error that arises from the $B\to X_d\gamma$ fraction is too small to be
considered. We perform the average assuming that the systematic errors of the 
shape function are correlated, and the other systematic
errors and the statistical errors are Gaussian and uncorrelated.     
The obtained average is 
${\cal B}(B\to X_s\gamma) = (352\pm 23\pm 9)\times 10^{-6}$ with
a $\chi^2$/DOF$= 1.00/5$, where the 
errors are combined statistical and systematic and systematic due to the shape 
function. The second error is estimated  to
be the difference of the average after simultaneously varying the central
value of each experimental result by $\pm 1\sigma$. Although  a small fraction
of events was used in both the semi-inclusive and inclusive analyses in the
same experiment, we neglect their statistical correlations. Some
other correlated systematic errors, such as photon detection and the background
suppression, are not considered in our new average. 

\begin{table}[h]
\caption{Reported branching fraction, minimum photon energy, branching fraction
at minimum photon energy  and converted branching fraction for the 
decay $b\to s\gamma$. All the branching fractions are in units of $10^{-6}$.
See text for an explanation of the errors. The CLEO measurment on the branching
fraction  at $E_{rm min}$ includes $B\to X_d \gamma$ events. The last error
of the Belle reported branching fraction in their
inclusive analysis is the systematic uncertainty due to the boost of the
$\gamma$ energy from the center of mass frame to the $B$ meson rest frame.  
} 

\label{tab:measurement}
\end{table}

\mysubsection{Baryonic decays}

\begin{table}
\begin{center}
\caption{
 Branching fractions of  baryonic $B^+$ decays (in units of $10^{-6}$).
Upper limits are at 90\% CL.
values in {\red red} ({\blue blue}) are new {\red published}
({\blue preliminary}) result since PDG2006  [as of April 15, 2008].
}
\end{center} 
\footnotesize
\begin{center}
\vspace{-0.3cm}
%
\end{center}
\vspace{-0.3cm}
\S Di-baryon mass is less than 2.85 GeV/$c^2$;  
$\dag$ Charmonium decays to $p\bar p$ have been statistically subtracted.\\
$\ddag$ The charmonium mass region has been vetoed. \\
$^*$ Product BF - daughter BF taken to be 100\%: 
$~~~~ \Theta(1540)^{++}\to K^+p$ (pentaquark candidate). \\
\end{table}

\begin{table}
\begin{center}
\caption{
Branching fractions of  baryonic $B^0$ decays
(in units of $10^{-6}$). Upper limits are at 90\% CL.
values in {\red red} ({\blue blue}) are new {\red published}
({\blue preliminary}) result since PDG2006  [as of April 15, 2008].
}
\end{center}
\footnotesize
\begin{center}
\vspace{-0.3cm}

\end{center}
\hspace*{-1.0cm}

\S Di-baryon mass is less than 2.85 GeV/$c^2$; 
$\dag$ Charmonium decays to $p\bar p$ have been statistically subtracted.
$\ddag$ The charmonium mass region has been vetoed.
$^*$ Product BF - daughter BF taken to be 100\%;
$\Theta(1540)^+\to p K^0$ (pentaquark candidate).
\end{table}

\clearpage
\mysubsection{$B_s$ decays}

\begin{table}[h]
\begin{center}
\caption{
 $B_s$  branching fractions (in units of $10^{-6}$). 
Upper limits are at 90\% CL.
Values in {\red red} ({\blue blue}) are new {\red published}
({\blue preliminary}) result since PDG2006  [as of March 15, 2007].
}
\vskip 0.25cm


\end{center}
\vspace{-0.3cm}
\hspace{1.cm} $\dag$Relative BF converted to absolute BF
\end{table}

\begin{table}[h]
\begin{center}
\caption{
 $B_s$ rare relative branching fractions.
Values in {\red red} ({\blue blue}) are new {\red published}
({\blue preliminary}) result since PDG2006  [as of April 15, 2008].
}
\vskip 0.25cm

\scriptsize
%
\end{center}
\end{table}

\mysubsection{Charge asymmetries}

\begin{sidewaystable}[!htbp]
\parbox{8.2in}{\caption{
 $CP$ asymmetries for charmless hadronic charged $B$ decays (part I).
Values in {\red red} ({\blue blue}) are 
new {\red published}
({\blue preliminary}) result since PDG2006 [as of April 15, 2008].
}}

\scriptsize
%
\end{sidewaystable}

\begin{sidewaystable}[!htbp]
\begin{center}
\parbox{7.5in}{\caption{$CP$ asymmetries for 
charmless hadronic charged  $B$ decays (part II).
Values in {\red red} ({\blue blue}) are new {\red published}
({\blue preliminary}) result since PDG2006  [as of April 15, 2008].
}}

\scriptsize
%
\end{center}

\begin{center}
\parbox{7.5in}{\caption{$CP$ asymmetries for 
charmless hadronic neutral $B$ decays.
Values in {\red red} ({\blue blue}) are new {\red published}
({\blue preliminary}) result since PDG2006  [as of April 15, 2008].
}}

\scriptsize
%
\end{center}

\vspace{-0.3cm}
\hspace{1.cm}
\dag~Measurements of time-dependent $CP$ asymmetries are listed in the section
of the Unitarity Triangle.

\end{sidewaystable}

\begin{sidewaystable}[!htbp]
\begin{center}

\parbox{7.5in}{\caption{
Charmless hadronic $CP$ asymmetries for $B^\pm/B^0$ admixtures.
Values in {\red red} ({\blue blue}) are new {\red published}
({\blue preliminary}) result since PDG2006  [as of April 15, 2008].
}}

\vspace*{ 0.4cm}
\scriptsize
\hspace*{-1.4cm}

\end{center}

\vspace{1cm}
\begin{center}
\parbox{7.5in}{\caption{
 $CP$ asymmetries for charmless hadronic neutral $B^0_S$ decays.
Values in {\red red} ({\blue blue}) are new {\red published}
({\blue preliminary}) result since PDG2006 [as of April, 2008].
}}

\vspace*{ 0.4cm}
%
\end{center}
\end{sidewaystable}

 \clearpage
\mysubsection{Polarization measurements}
%

%

\begin{table}[!htbp]
\caption{
 Longitudinal polarization fraction $f_L$ for $B^+$ decays.
Values in {\red red} ({\blue blue}) are new {\red published}
({\blue preliminary}) result since PDG2006 [as of April 15, 2008].
\vspace{0.3cm}
}

\begin{center}


\end{center}
\end{table}

\begin{table}[!htbp]
\caption{
Full angular analysis of $B^+\to\phi K^{*+}$.
Values in {\red red} ({\blue blue}) are new {\red published}
({\blue preliminary}) result since PDG2006 [as of April 15, 2008].
\vspace{0.3cm}
}

\begin{center}
%
\end{center}
\vspace{-0.3cm}
\hspace*{1.0cm}BR, $f_L$ and $A_{CP}$ are tabulated separately.
\end{table}
\clearpage


\begin{table}[!htbp]
\caption{
Longitudinal polarization fraction $f_L$ for $B^0$ decays.
Values in {\red red} ({\blue blue}) are new {\red published}
({\blue preliminary}) result since PDG2006  [as of April 15, 2008].
\vspace{0.3cm}
}

\begin{center}

\footnotesize
%
\end{center}
\end{table}

\begin{table}[!htbp]
\caption{
Full angular analysis of $B^0 \to \phi K^{*0}$.
Values in {\red red} ({\blue blue}) are new {\red published}
({\blue preliminary}) result since PDG2006  [as of April 15, 2008].
\vspace{0.3cm}
}

\begin{center}
\footnotesize
%
\end{center}
\vspace{-0.3cm}
\hspace{1.3cm}
BR, $f_L$ and $A_{CP}$ are tabulated separately.
\end{table}

\begin{table}[!htbp]
\caption{
Full angular analysis of $B^0 \to \phi K_2^*(1430)^0$.
Values in {\red red} ({\blue blue}) are new {\red published}
({\blue preliminary}) result since PDG2006  [as of April 15, 2008].
\vspace{0.3cm}
}

\begin{center}
\footnotesize
%
\end{center}
\vspace{-0.3cm}
\hspace{2.3cm}
BR, $f_L$ and $A_{CP}$ are tabulated separately.
\end{table}

\clearpage
\section{$D$ decays}
\label{sec:charm_physics}

\def\kbar{\overline{K}{}^{\,0}}
\def\dbar{\overline{D}{}^{\,0}}
\def\bbar{\overline{B}{}^{\,0}}
\def\cp{$CP$}
\def\cpv{$CPV$}
\def\ra{\!\rightarrow\!}
\def\ddbar{$D^0$-$\dbar$}

\def\dklnu{$D^0\ra K^+\ell^-\nu$}
\def\dkpi{$D^0\ra K^+\pi^-$}
\def\dkk{$D^0\ra K^+K^-$}
\def\dpipi{$D^0\ra\pi^+\pi^-$}
\def\dkkpp{$D^0\ra K^+K^-/\pi^+\pi^-$}
\def\dkspp{$D^0\ra K^0_S\,\pi^+\pi^-$}
\def\ycp{$y^{}_{\rm CP}$}

\def\gevm{~GeV/$c^2$}
\def\gevp{~GeV/$c$}
\def\geve{~GeV}
\def\mevm{~MeV/$c^2$}
\def\meve{~MeV}

\def\babar{Babar}

\def\simge{\mathrel{%
   \rlap{\raise 0.511ex \hbox{$>$}}{\lower 0.511ex \hbox{$\sim$}}}}
\def\simle{\mathrel{
   \rlap{\raise 0.511ex \hbox{$<$}}{\lower 0.511ex \hbox{$\sim$}}}}

\newcommand{\Dnan}{\ensuremath{D_0^\ast(2400)^0}}
\newcommand{\Dtan}{\ensuremath{D_2^\ast(2460)^0}}
\newcommand{\Don}{\ensuremath{D_1(2420)^{0}}}
\newcommand{\Dopn}{\ensuremath{D_1(2430)^{0}}}
\newcommand{\Dnap}{\ensuremath{D_0^\ast(2400)^\pm}}
\newcommand{\Dtap}{\ensuremath{D_2^\ast(2460)^\pm}}
\newcommand{\Dop}{\ensuremath{D_1(2420)^{\pm}}}
\newcommand{\Dopp}{\ensuremath{D_1(2430)^{\pm}}}

\newcommand{\Dsa}{\ensuremath{D_s^{\ast\pm}}}
\newcommand{\Dsna}{\ensuremath{D_{s0}^\ast(2317)^{\pm}}}
\newcommand{\Dsop}{\ensuremath{D_{s1}(2460)^{\pm}}}
\newcommand{\Dso}{\ensuremath{D_{s1}(2536)^{\pm}}}
\newcommand{\Dst}{\ensuremath{D_{s2}(2573)^{\pm}}}
\newcommand{\Dsts}{\ensuremath{D_{sJ}(2700)^{\pm}}}
\newcommand{\Dste}{\ensuremath{D_{sJ}(2860)^{\pm}}}
\newcommand{\Dstsi}{\ensuremath{D_{sJ}(2632)^{\pm}}}

\newcommand{\citep}{\cite}

\subsection{\emph{$D^0$-$\dbar$} Mixing and \emph{\cp}\ Violation}

\subsubsection{Introduction}

Mixing in the $D^0$-$\dbar$ system has been searched for for more than 
two decades without success --- until last year. Three experiments
--\,Belle\cite{belle_kk}, Babar\cite{babar_kpi}, and CDF\cite{cdf_kpi}\,-- 
have now observed evidence for this phenomenon. The measurements can 
be combined with others to yield World Average (WA) values for the 
mixing parameters 
$x\equiv(m^{}_1-m^{}_2)/\Gamma$ and 
$y\equiv (\Gamma^{}_1-\Gamma^{}_2)/(2\Gamma)$, where 
$m^{}_1,\,m^{}_2$ and $\Gamma^{}_1,\,\Gamma^{}_2$ are
the masses and decay widths for the mass eigenstates
$D^{}_1\equiv p|D^0\rangle-q|\dbar\rangle$ and
$D^{}_2\equiv p|D^0\rangle+q|\dbar\rangle$,
and $\Gamma=(\Gamma^{}_1+\Gamma^{}_2)/2$. 
Here we use the phase convention $CP|D^0\rangle=-|\dbar\rangle$
and $CP|\dbar\rangle=-|D^0\rangle$.
In the absence of \cp\ violation (\cpv), $p=q=1/\sqrt{2}$ and
$D^{}_1$ is \cp-even, $D^{}_2$ is \cp-odd.

Such WA values are calculated by the Heavy Flavor Averaging 
Group (HFAG)\cite{hfag_charm} by performing a global fit to 
measured observables for parameters $x$, $y$, $\delta$ (the 
strong phase difference between amplitudes 
${\cal A}(\dbar\ra K^+\pi^-)$ and ${\cal A}(D^0\ra K^+\pi^-)$), 
an additional strong phase $\delta^{}_{K\pi\pi}$ entering
$D^0\ra K^+\pi^-\pi^0$ decays, and 
$R^{}_D\equiv\left|{\cal A}(D^0\ra K^+\pi^-)/
              {\cal A}(\dbar\ra K^+\pi^-)\right|^2$. 
For this fit, correlations among observables are accounted for by using 
covariance matrices provided by the experimental collaborations. 
Systematic errors among different experiments are assumed uncorrelated
as, after some study, no significant correlations were identified.
The observables used are from measurements of \dklnu, 
\dkkpp, \dkpi, $D^0\ra K^+\pi^-\pi^0$, 
and \dkspp\ decays, and from double-tagged branching 
fractions measured at the $\psi(3770)$ resonance.
We have checked this method with a second method that adds
together three-dimensional log-likelihood functions 
for $x$, $y$, and $\delta$ obtained from various analyses;
this combination accounts for non-Gaussian errors.
When both methods are applied to the same set of observables, 
essentially identical results are obtained. 
The global fitting method is easily expanded to allow for \cpv. 
In this case three additional parameters are included in the fit:
$|q/p|$, $\phi\equiv {\rm Arg}(q/p)$, and
$A^{}_D\equiv (R^+_D-R^-_D)/(R^+_D+R^-_D)$, where the $+\,(-)$
superscript corresponds to $D^0\,(\dbar)$ decays. 

Mixing in heavy flavor systems such as those of $B^0$ and $B^0_s$ 
is governed by the short-distance box diagram. In the $D^0$ system,
however, this diagram is doubly-Cabibbo-suppressed relative to 
amplitudes dominating the decay width, and it is also GIM-suppressed.
Thus the short-distance mixing rate is tiny, and $D^0$-$\dbar$ 
mixing is expected to be dominated by long-distance 
processes. These are difficult to calculate reliably, and 
theoretical estimates for $x$ and $y$ range over two-three 
orders of magnitude\cite{BigiUraltsev,Petrov,Falk}.

With the exception of $\psi(3770)\ra DD$ measurements, all methods 
identify the flavor of the $D^0$ or $\dbar$ when produced by 
reconstructing the decay $D^{*+}\ra D^0\pi^+$ or $D^{*-}\ra\dbar\pi^-$; 
the charge of the pion identifies the $D$ flavor. For signal 
decays, $M^{}_{D^*}-M^{}_{D^0}-M^{}_{\pi^+}\equiv Q\approx 6$\meve, 
which is close to the threshold; thus analyses typically
require that the reconstructed $Q$ be small to suppress backgrounds. 
For time-dependent measurements, the $D^0$ decay time is 
calculated as $(d/p)\times M^{}_{D^0}$, where $d$ is
the distance between the $D^*$ and $D^0$ decay vertices and 
$p$ is the $D^0$ momentum. The $D^*$ vertex position is 
taken to be at the primary vertex\cite{cdf_kpi} ($\bar{p}p$)
or is calculated from the intersection of the $D^0$ momentum 
vector with the beamspot profile ($e^+e^-$).

\subsubsection{Input Observables}

The global fit determines central values and errors for
eight underlying parameters using a $\chi^2$ statistic
constructed from 28 observables. The
underlying parameters are $x,\,y,\,\delta,\,R^{}_D,
A^{}_D,\,|q/p|,\,\phi$, and $\delta^{}_{K\pi\pi}$.
The parameters $x$ and $y$ govern mixing, and the
parameters $A^{}_D$, $|q/p|$, and $\phi$ govern \cpv.
The parameter $\delta^{}_{K\pi\pi}$ is the strong phase 
difference between the amplitudes ${\cal A}(\dbar\ra K^+\pi^-\pi^0)$ 
and ${\cal A}(D^0\ra K^+\pi^-\pi^0)$ evaluated at 
the point $M^{}_{K^+\pi^-}\!=\!M^{}_{K^*(890)}$.

\begin{figure}
\begin{center}
\includegraphics[width=4.2in]{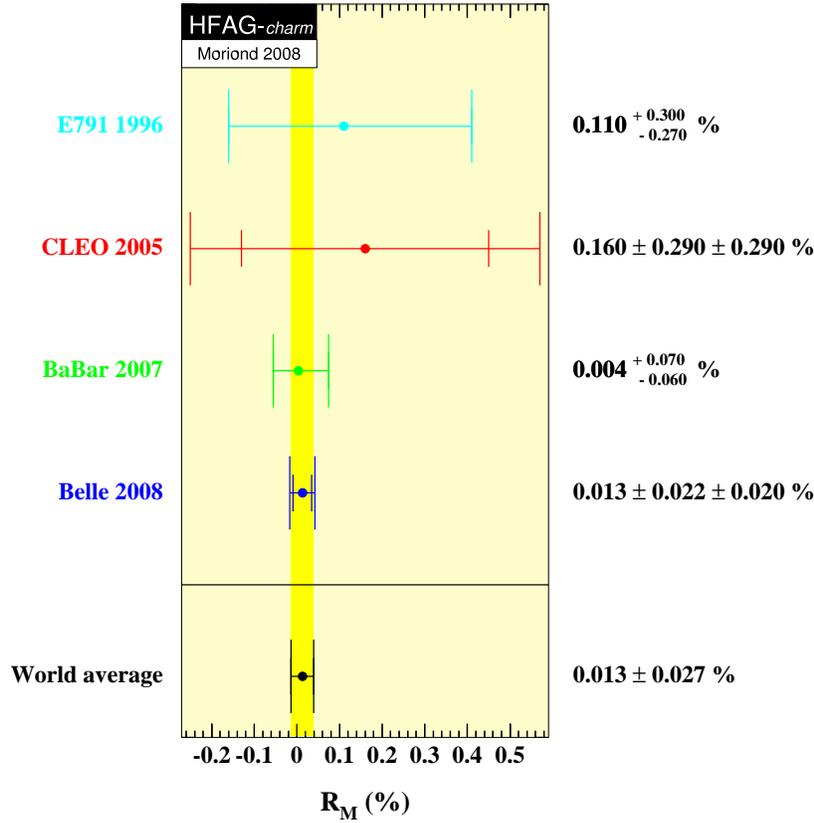}
\end{center}
\vskip-0.20in
\caption{\label{fig:rm_semi}
WA value of $R^{}_M$ from Ref.~\cite{hfag_charm},
as calculated from $D^0\ra K^+\ell^-\nu$ 
measurements\cite{semi_references}. }
\end{figure}

All input values are listed in Table~\ref{tab:observables}. 
The observable $R^{}_M=(x^2+y^2)/2$ measured in \dklnu\ 
decays\cite{semi_references} is taken to be the WA value\cite{hfag_charm} 
calculated by HFAG (see Fig.~\ref{fig:rm_semi}). The observables 
$y^{}_{CP}$ and $A^{}_\Gamma$ measured in $D^0\ra K^+K^-/\pi^+\pi^-$ 
decays\cite{belle_kk,ycp_fnal,ycp_cleo,ycp_babar} are 
also taken to be WA values\cite{hfag_charm} (see Fig.~\ref{fig:ycp}).
The observables from \dkspp\ decays\cite{kspp_references}
for no-\cpv\ are HFAG WA values\cite{hfag_charm}, but for 
the \cpv-allowed case only Belle measurements are available.
The \dkpi\ observables used are from Belle\cite{belle_kpi}, 
Babar\cite{babar_kpi}, and CDF\cite{cdf_kpi} (these measurements 
have much greater precision than earlier measurements). The 
$D^0\ra K^+\pi^-\pi^0$ results are from Babar\cite{knpi_references}, 
and the $\psi(3770)\ra DD$ results are from CLEOc\cite{cleoc}.

The relationships between the observables and the fitted
parameters are listed in Table~\ref{tab:relationships}. 
For each set of correlated observables, we construct the 
difference vector $\vec{V}$, e.g., for 
$D^0\ra K^0_S\,\pi^+\pi^-$ decays
$\vec{V}=(\Delta x,\Delta y,\Delta |q/p|,\Delta \phi)$,
where $\Delta$ represents the difference between 
the measured value and the fitted parameter value. 
The contribution of a set of measured observables 
to the overall $\chi^2$ is calculated as
$\vec{V}\cdot (M^{-1})\cdot\vec{V}^T$, where
$M^{-1}$ is the inverse of the covariance matrix 
for the measurement. All covariance matrices used 
are listed in Table~\ref{tab:observables}.

\begin{figure}
\begin{center}
\vbox{
\includegraphics[width=4.2in]{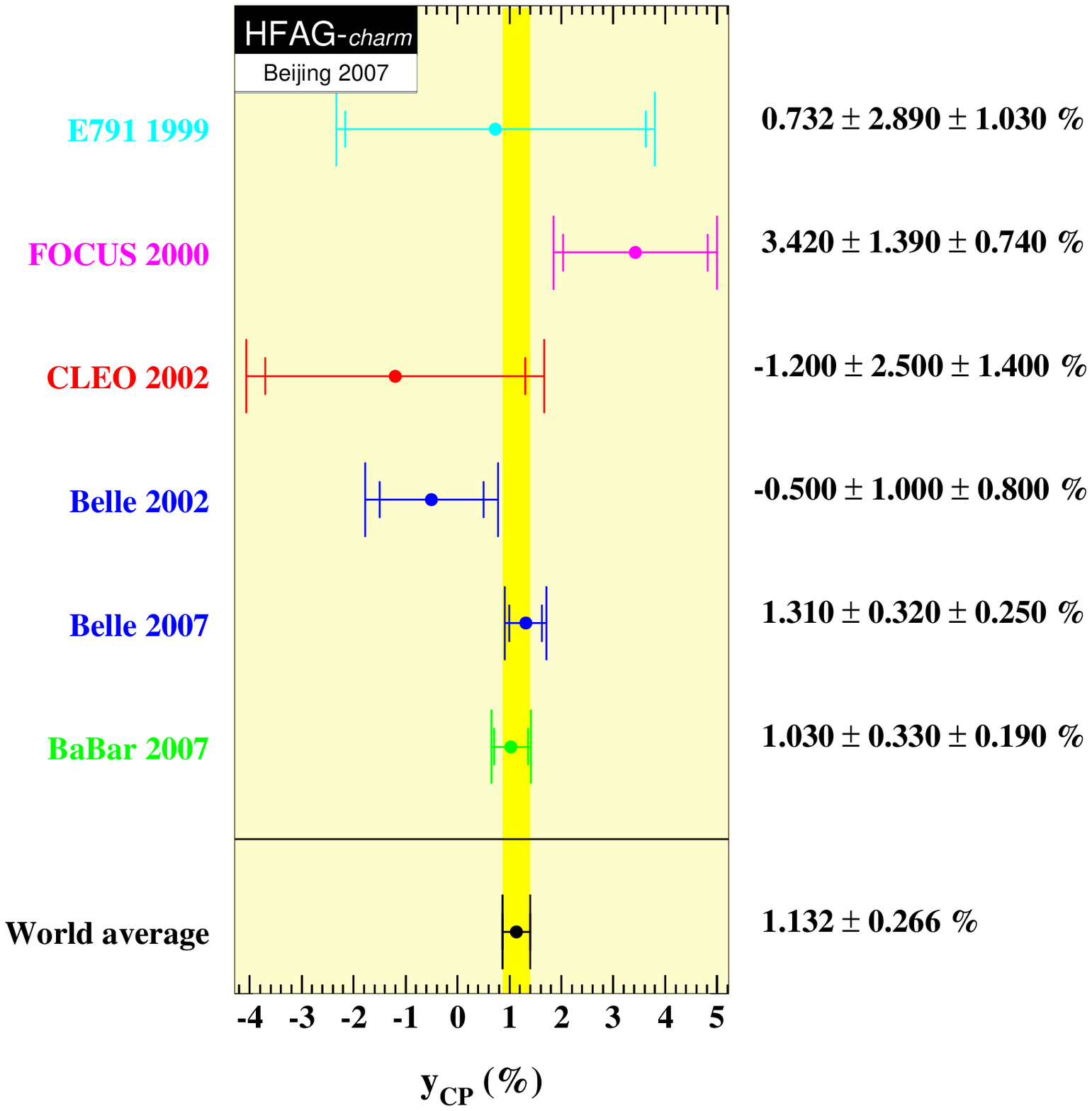}
\vskip0.10in
\includegraphics[width=4.2in]{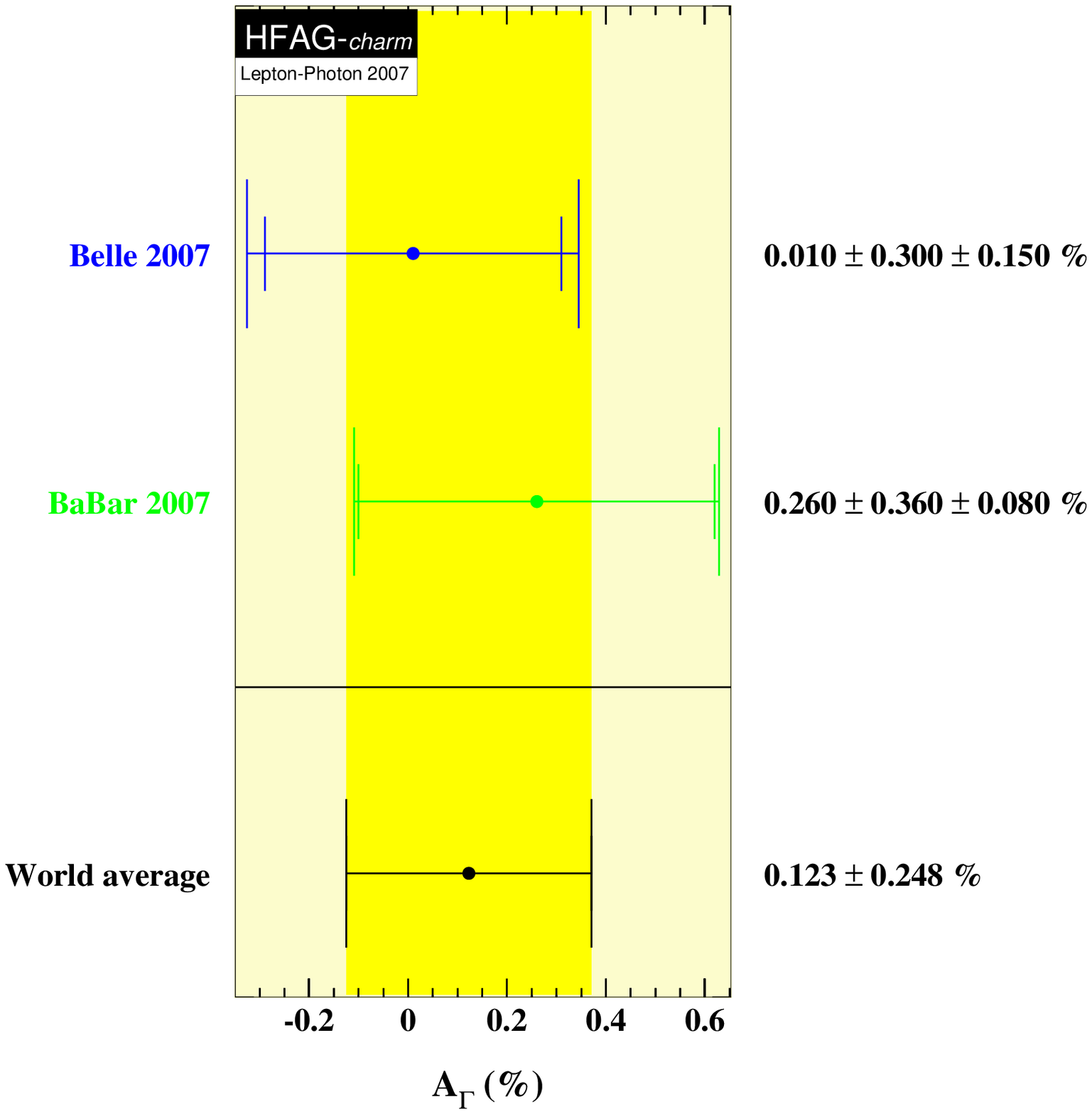}
}
\end{center}
\vskip-0.20in
\caption{\label{fig:ycp}
WA values of $y^{}_{CP}$ (top) and $A^{}_\Gamma$ (bottom)
from Ref.~\cite{hfag_charm}, as calculated from \dkkpp\ 
measurements\cite{belle_kk,ycp_fnal,ycp_cleo,ycp_babar}.  }
\end{figure}

\begin{table}
\renewcommand{\arraystretch}{1.3}
\caption{\label{tab:observables}
Observables used for the global fit, from
Refs.~\cite{semi_references,
belle_kk,ycp_fnal,ycp_cleo,ycp_babar,
kspp_references,babar_kpi,belle_kpi,cdf_kpi,knpi_references,cleoc}.}
\vspace*{6pt}
\footnotesize
\hskip-0.30in

\end{table}

\begin{table}
\renewcommand{\arraystretch}{1.3}
\begin{center}
\caption{\label{tab:relationships}
Left: decay modes used to determine fitted parameters 
$x,\,y,\,\delta,\,\delta^{}_{K\pi\pi},\,R^{}_D,\,A^{}_D,\,|q/p|$, and $\phi$.
Middle: the observables measured for each decay mode. Right: the 
relationships between the observables measured and the fitted parameters.}
\vspace*{6pt}
\footnotesize
%
$ \\
\hline
\end{tabular}
\end{center}
\end{table}

\subsubsection{Fit results}

The global fit uses MINUIT with the MIGRAD minimizer, 
and all errors are obtained from MINOS\cite{MINUIT}. 
Three separate fits are performed: 
{\it (a)}\ assuming \cp\ conservation ($A^{}_D$ 
and $\phi$ are fixed to zero, $|q/p|$ is fixed to one);
{\it (b)}\ assuming no direct \cpv\ ($A^{}_D$ is 
fixed to zero); and
{\it (c)}\ allowing full \cpv\ (all parameters
floated). The results are listed in 
Table~\ref{tab:results}. For the \cpv-allowed fit,
individual contributions to the $\chi^2$ are listed 
in Table~\ref{tab:results_chi2}. The total $\chi^2$ 
is 23.5 for $28-8=20$ degrees of freedom; this 
corresponds to a confidence level of~0.26, which 
is satisfactory.

Confidence contours in the two dimensions $(x,y)$ or 
in $(|q/p|,\phi)$ are obtained by letting, for any point in the
two-dimensional plane, all other fitted parameters take their 
preferred values. The resulting $1\sigma$-$5\sigma$ contours 
are shown in Fig.~\ref{fig:contours_ncpv} for the \cp-conserving
case, and in Fig.~\ref{fig:contours_cpv} for the \cpv-allowed 
case. The contours are determined from the increase of the
$\chi^2$ above the minimum value.
One observes that the $(x,y)$ contours for the no-\cpv\ fit are
almost identical to those for the \cpv-allowed fit. In both cases the 
$\chi^2$ at the no-mixing point $(x,y)\!=\!(0,0)$ is 91 units above 
the minimum value; for two degrees of freedom this has a confidence 
level corresponding to $9.2\sigma$. Thus, no mixing is excluded 
at this high level. In the $(|q/p|,\phi)$ plot, the point $(1,0)$ 
is within the $1\sigma$ contour; thus the data is consistent 
with \cp\ conservation.

One-dimensional confidence curves for individual parameters 
are obtained by letting, for any value of the parameter, all other 
fitted parameters take their preferred values. The resulting
functions $\Delta\chi^2=\chi^2-\chi^2_{\rm min}$ ($\chi^2_{\rm min}$
is the minimum value) are shown in Fig.~\ref{fig:1dlikelihood}.
The points where $\Delta\chi^2=3.84$ determine 95\% C.L. intervals 
for the parameters; these intervals are listed in Table~\ref{tab:results}.

\begin{figure}
\begin{center}
\includegraphics[width=4.2in]{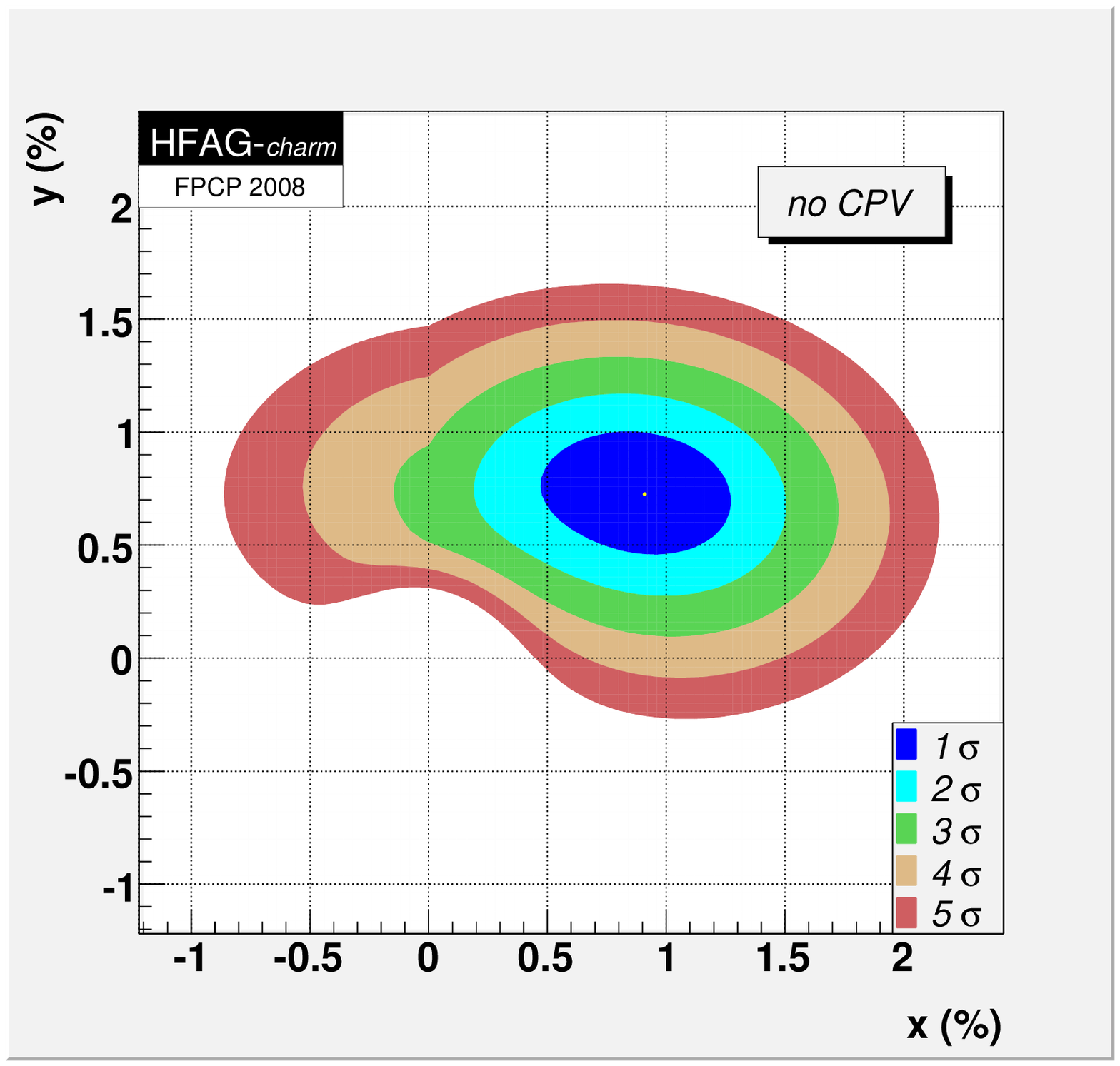}
\end{center}
\vskip-0.20in
\caption{\label{fig:contours_ncpv}
Two-dimensional contours for mixing parameters $(x,y)$, for no \cpv. }
\end{figure}

\begin{figure}
\begin{center}
\vbox{
\includegraphics[width=4.2in]{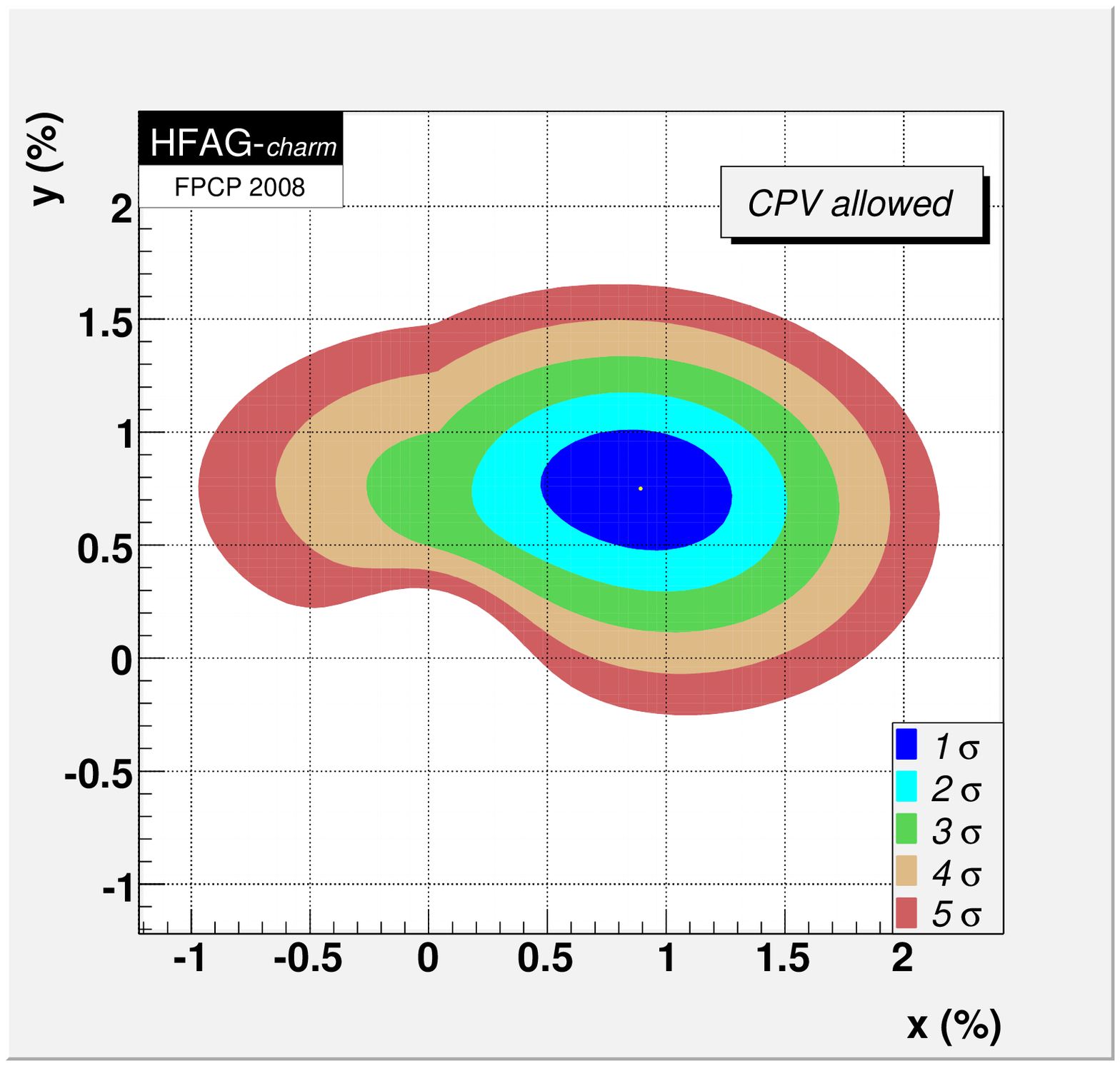}
\vskip0.10in
\includegraphics[width=4.2in]{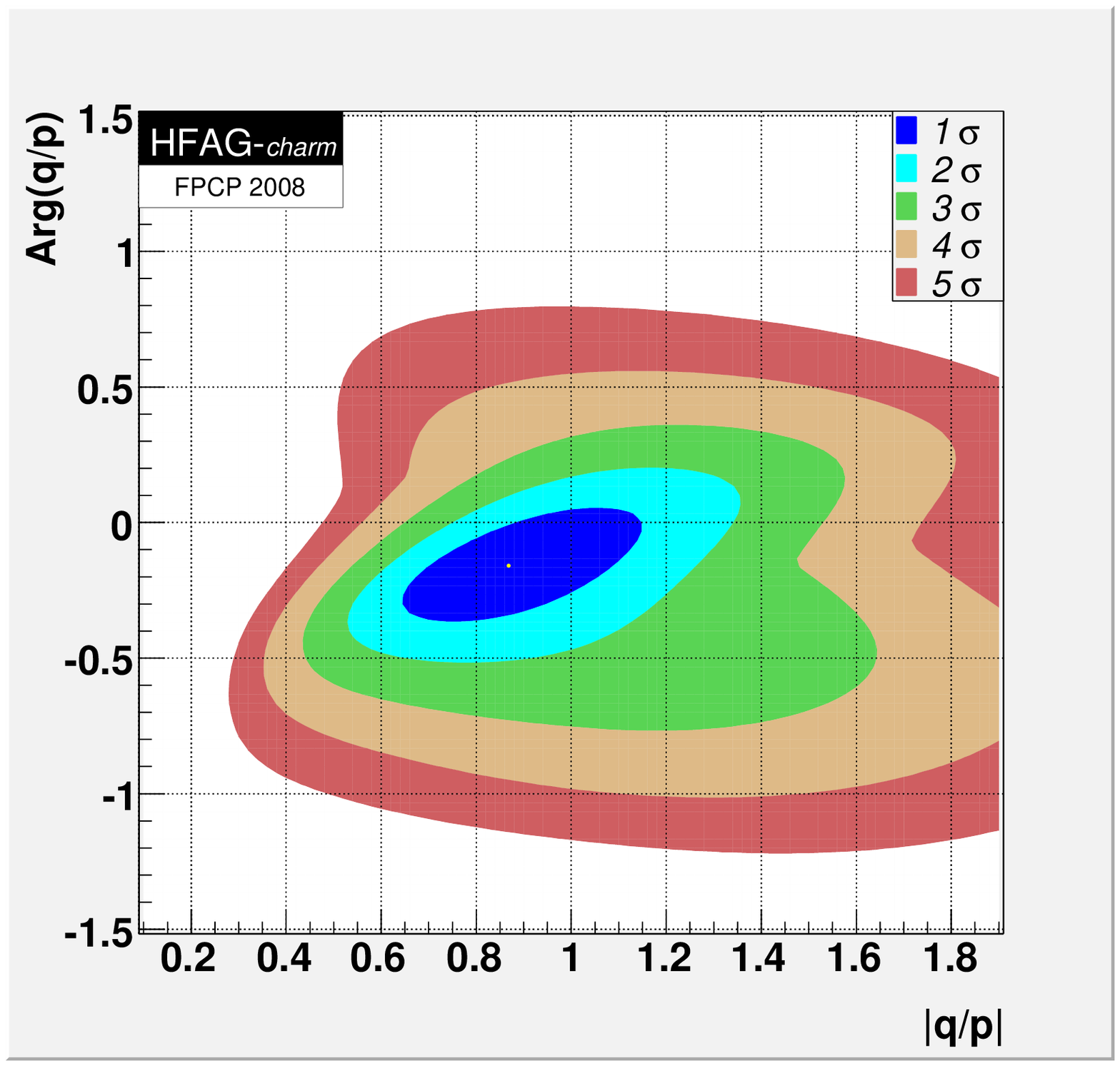}
}
\end{center}
\vskip-0.10in
\caption{\label{fig:contours_cpv}
Two-dimensional contours for parameters $(x,y)$ (top) 
and $(|q/p|,\phi)$ (bottom), allowing for \cpv.}
\end{figure}

\begin{figure}
\begin{center}
\hbox{\hskip0.50in
\includegraphics[width=72mm]{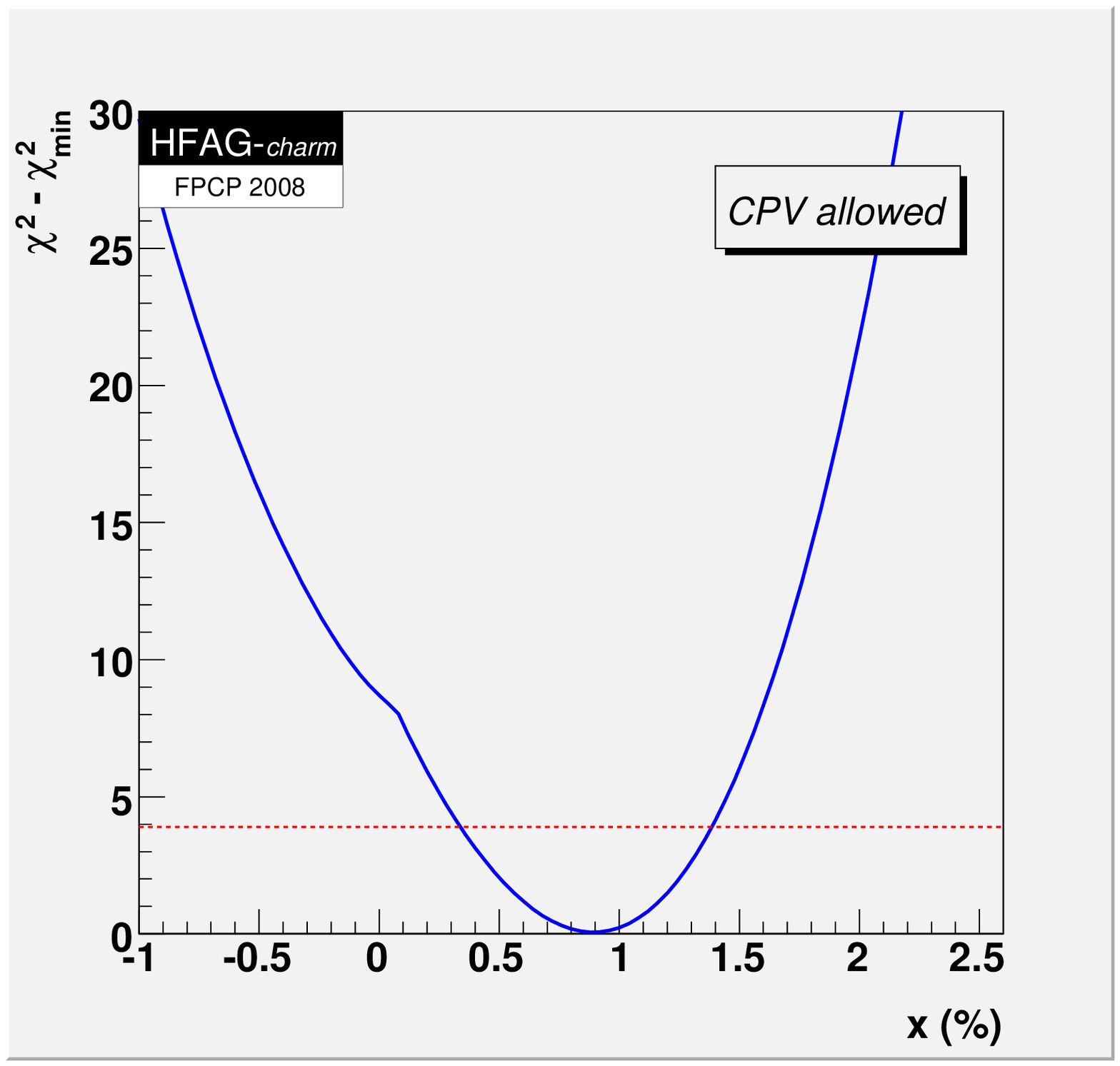}
\hskip0.20in
\includegraphics[width=72mm]{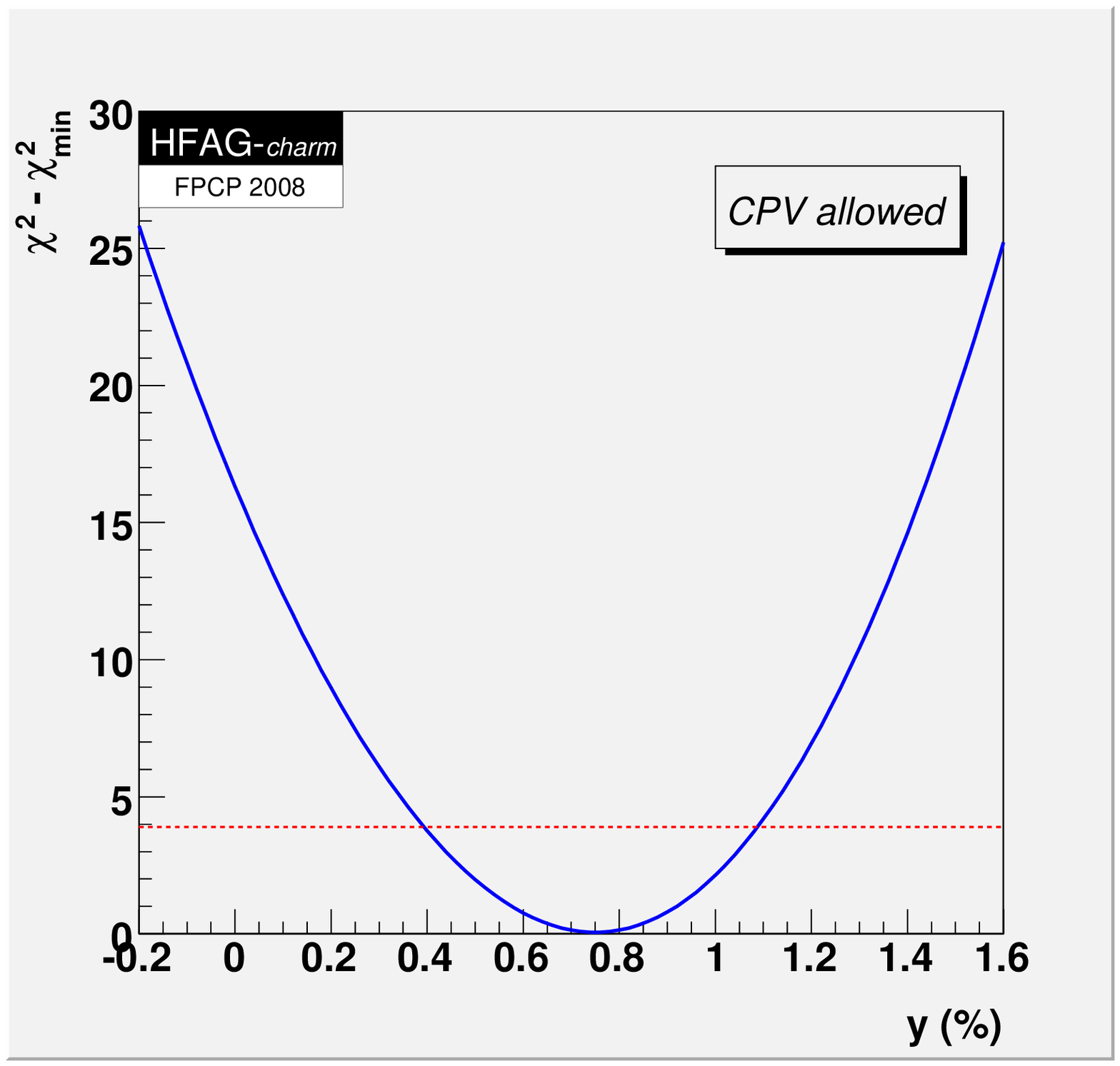}}
\hbox{\hskip0.50in
\includegraphics[width=72mm]{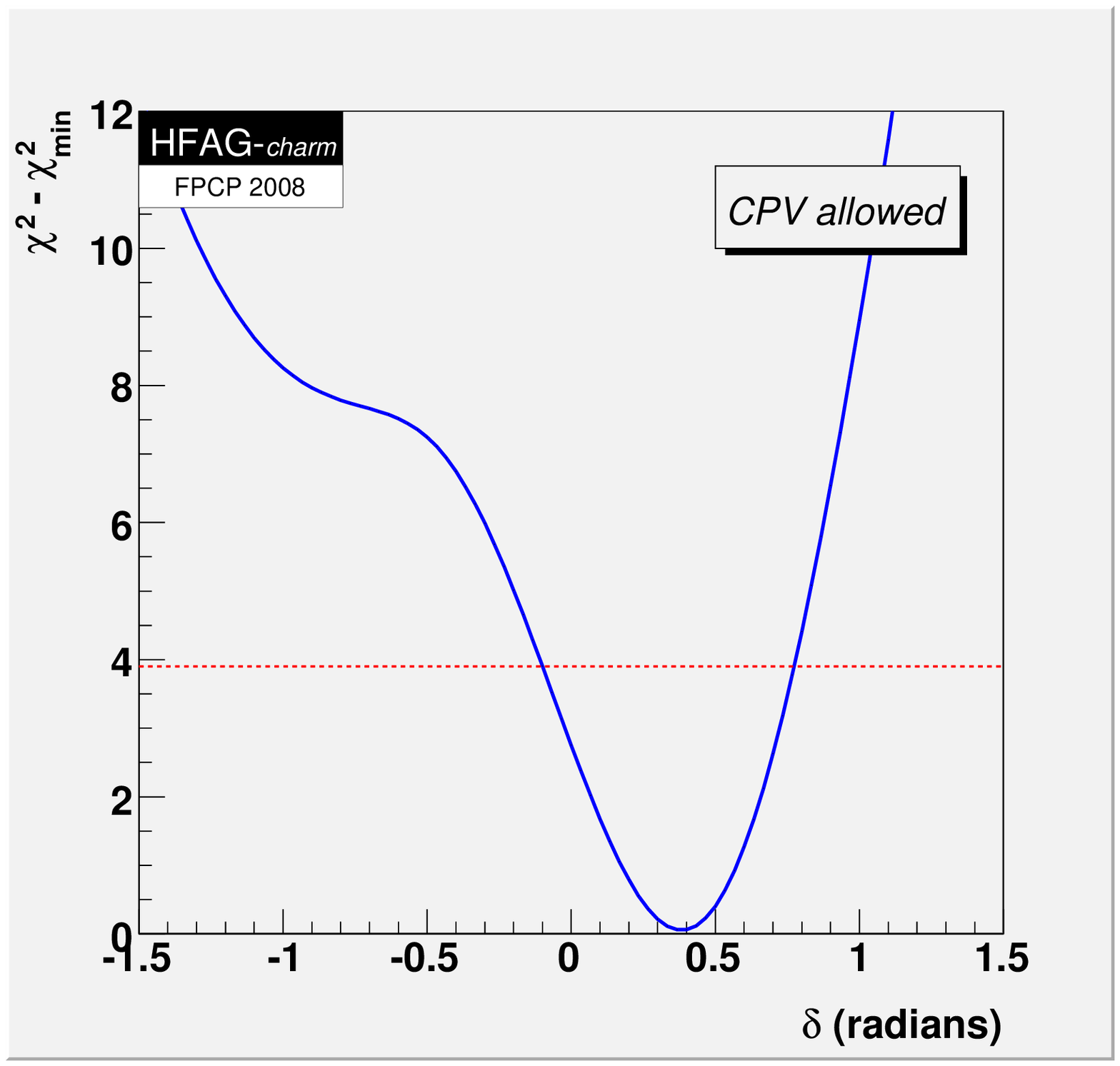}
\hskip0.20in
\includegraphics[width=72mm]{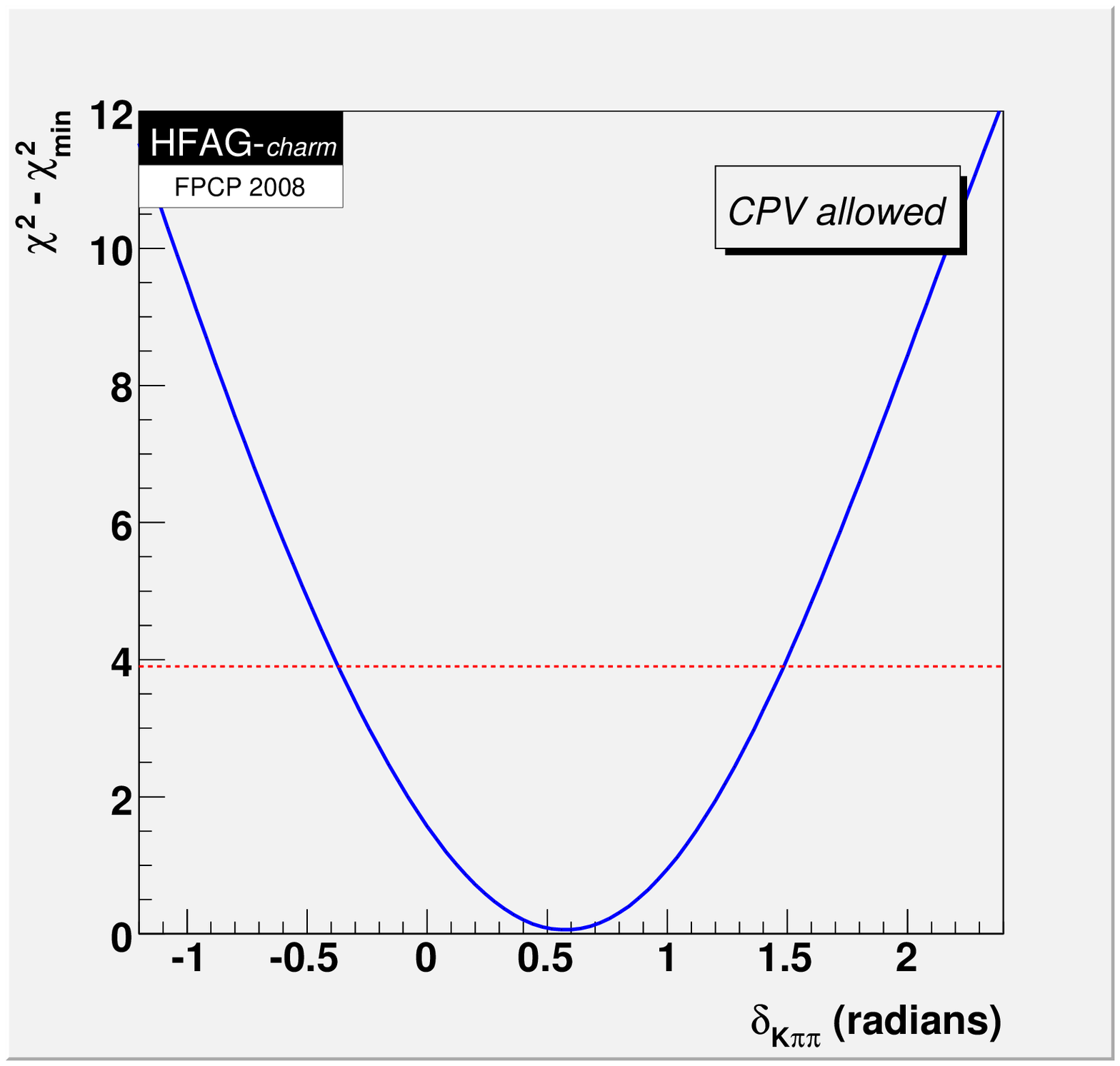}}
\hbox{\hskip0.50in
\includegraphics[width=72mm]{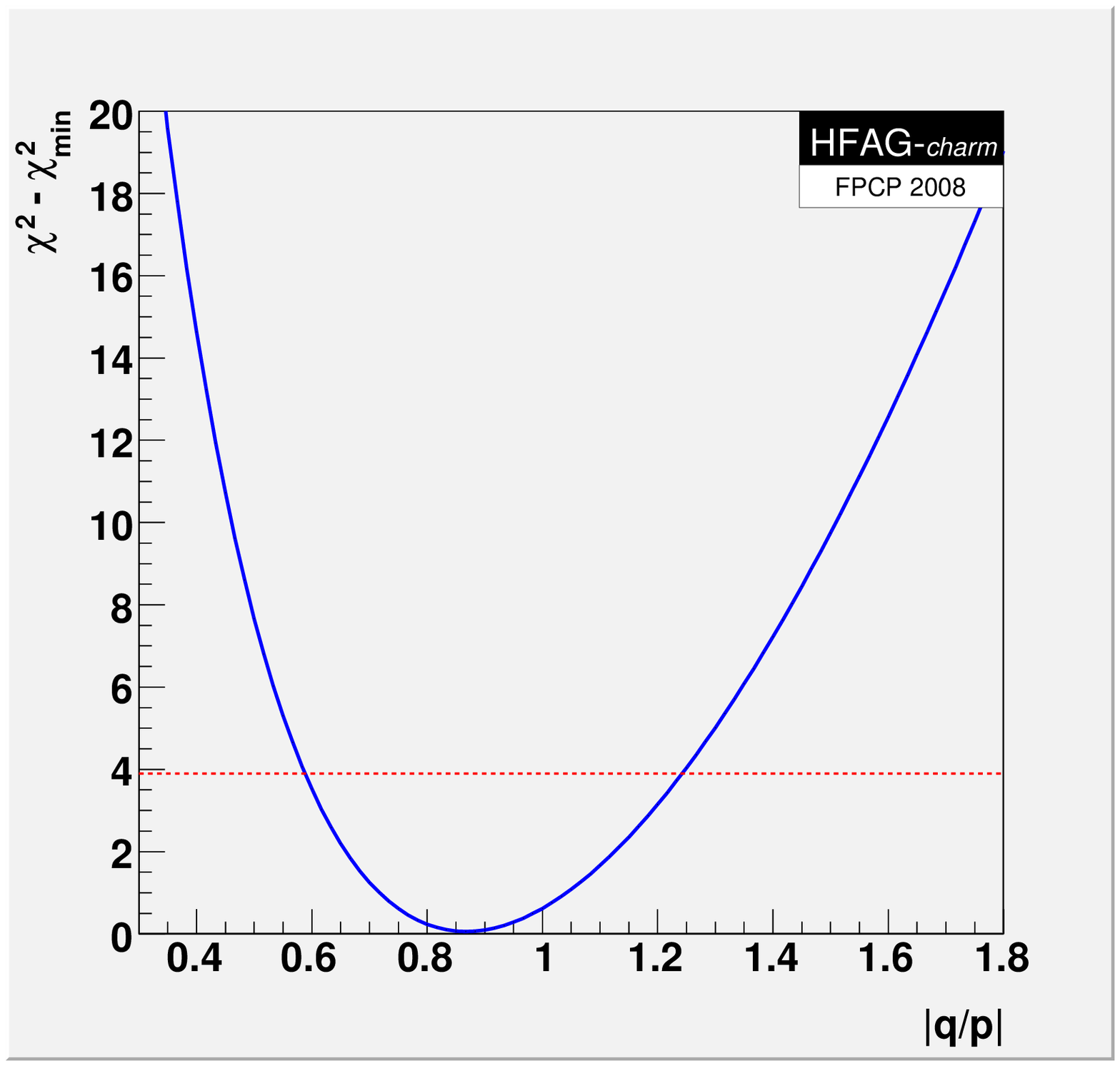}
\hskip0.20in
\includegraphics[width=72mm]{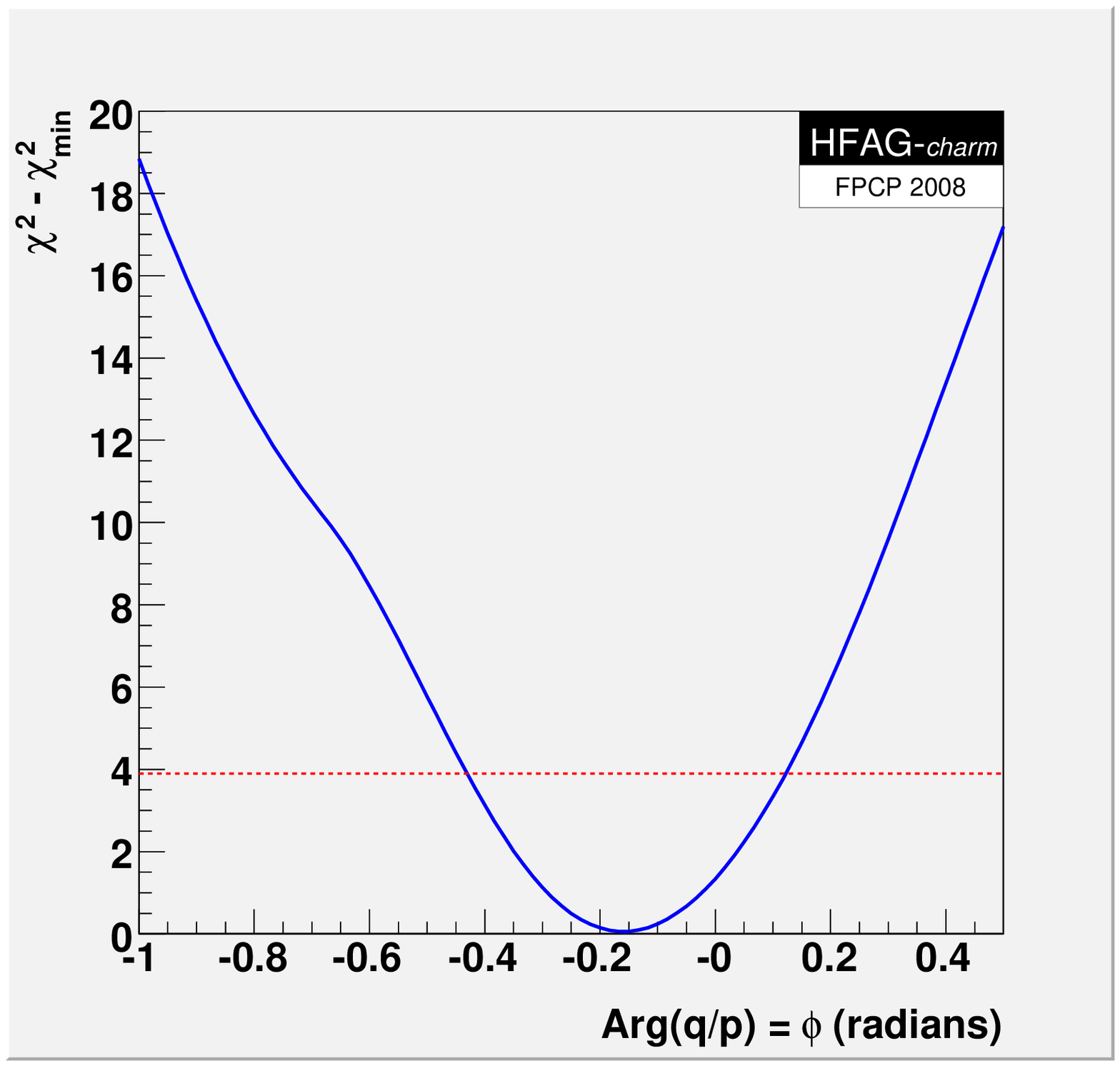}}
\end{center}
\vskip-0.30in
\caption{\label{fig:1dlikelihood}
The function $\Delta\chi^2=\chi^2-\chi^2_{\rm min}$ 
for fitted parameters
$x,\,y,\,\delta,\,\delta^{}_{K\pi\pi},\,|q/p|$, and $\phi$.
The points where $\Delta\chi^2=3.84$ (denoted by the dashed 
horizontal line) determine a 95\% C.L. interval. }
\end{figure}

\begin{table}
\renewcommand{\arraystretch}{1.4}
\begin{center}
\caption{\label{tab:results}
Results of the global fit for different assumptions concerning~\cpv.}
\vspace*{6pt}
\footnotesize

\end{center}
\end{table}

\begin{table}
\renewcommand{\arraystretch}{1.4}
\begin{center}
\caption{\label{tab:results_chi2}
Individual contributions to the $\chi^2$ for the \cpv-allowed fit.}
\vspace*{6pt}
\footnotesize
%
\end{center}
\end{table}


\subsubsection{Conclusions}

From the fit results listed in Table~\ref{tab:results}
and shown in Figs.~\ref{fig:contours_cpv} and \ref{fig:1dlikelihood},
we conclude the following:
\begin{itemize}
\item the experimental data consistently indicate that 
$D^0$ mesons undergo mixing. The no-mixing point $x=y=0$
is excluded at $9.2\sigma$. The parameter $x$ differs from
zero by $3.0\sigma$, and $y$ differs from zero by
$4.0\sigma$. This mixing is presumably dominated 
by long-distance processes, which are difficult to calculate.
Thus it may be difficult to identify new physics from mixing alone
(unless $|x|\gg |y|$ -- see Ref.~\cite{BigiUraltsev}).
\item Since \ycp\ is positive, the \cp-even state is shorter-lived,
as in the $K^0$-$\kbar$ system. However, since $x$ also appears
to be positive, the \cp-even state is heavier, 
unlike in the $K^0$-$\kbar$ system.
\item It appears difficult to accomodate a strong phase difference 
$\delta$ larger than $45^\circ$.
\item There is no evidence yet for \cpv\ in the $D^0$-$\dbar$ system.
Observing \cpv\ at the current level of sensitivity would indicate 
new physics.
\end{itemize}

\clearpage
\subsection{Excited \emph{$D_{(s)}$} Mesons}

Tables \ref{tab1}--\ref{tab3} represent a summary of recent results 
with an emphasis on information not provided in Ref.~\citep{PDG_2007}. 
For a complete list of related publications, see Ref.~\citep{PDG_2007}.
All upper limits (U.L.) correspond to 90\% confidence (C.L.) unless 
otherwise noted. The significances listed are approximate; they
are calculated as either
$\sqrt{-2\Delta\log{\cal{L}}}$ or $\sqrt{\Delta\chi^2}$, where 
$\Delta$ represents the change in the corresponding minimized 
function between two hypotheses, e.g., those for different spin states.

The broad charged $J^P\!=\!1^+$ $c\bar{d}$ state is denoted 
$D_1(2430)^+$, although it has not yet been observed. The 
masses of narrow states $\Dop$, $\Don$, $\Dtan$, $\Dtap$ and $\Dsna$, 
$\Dsop$, $\Dso$, $\Dst$ are well-measured, and thus only their averages 
are given\citep{PDG_2007}. The first observations of $\Dsop$ and $\Dsna$ 
states are described in Refs.~\citep{charm_babar1} and \citep{charm_cleo1}, 
respectively.

The masses and widths of narrow ($\Gamma\sim$ 20--40~MeV) orbitally 
excited $D$ mesons (denoted $D^{\ast\ast}$), both neutral and charged, 
are well established. Measurements of broad states ($\Gamma\sim$ 200--400~MeV)
are less abundant, as identifying the signal is more challenging. 
There is a slight discrepancy between the 
\Dnan\ masses measured by the Belle\citep{charm_belle1} and 
FOCUS\citep{charm_focus1} experiments. No data exists yet for the 
\Dopp\ state. Dalitz plot analyses of $B\to D^{(\ast)}\pi\pi$ decays
strongly favor the assignments $0^+$ and $1^+$ for the spin-parity 
quantum numbers of the \Dnan/\Dnap\ and \Dopn\ states, respectively. 
The measured masses and widths, as well as the $J^P$ values, are 
in agreement with theoretical predictions based on potential 
models\citep{theory1}. While the branching fractions for $B$
mesons decaying to a narrow $D^{\ast\ast}$ state and a pion are similar 
for charged and neutral $B$ initial states, the branching fractions 
to a broad $D^{\ast\ast}$ state and $\pi^+$ are much larger for $B^+$ 
than for $B^0$. This may be due to the fact that color-suppressed 
amplitudes contribute only to the $B^+$ decay and not to the $B^0$ 
decay (for a theoretical discussion, see Ref.~\citep{theory2}). 

The discoveries of the \Dsna\ and \Dsop\ have triggered increased 
interest in properties of, and searches for, excited $D_s$ mesons 
(here generically denoted $D_s^{\ast\ast}$). While the masses and 
widths of \Dso\ and \Dst\ states are in relatively good agreement 
with potential model predictions, 
the masses of \Dsna\ and \Dsop\ states (and consequently their 
widths, less than around 5~MeV) are significantly lower than expected 
(see Ref.~\citep{theory3} for a discussion of $c\bar{s}$ models). 
Moreover, the mass splitting between these two states greatly exceeds 
that between the \Dso\ and~\Dst. These unexpected properties have led 
to interpretations of the \Dsna\ and \Dsop\ as exotic four-quark states.

While there are few measurements with respect to the
$J^P$ values of \Dsna~and \Dsop, 
the available data favors $0^+$ and $1^+$, respectively. 
A molecule-like ($DK$) interpretation of the \Dsna\ and 
\Dsop\citep{theory4} that can account for their low masses 
and isospin-breaking decay modes is tested by searching for 
charged and neutral isospin partners of these states; thus far 
such searches have yielded negative results. 
Hence the subset of models that predict equal production rates for
different charged states is nominally excluded. The molecular picture
can also be tested by measuring the rates for the radiative processes 
$\Dsna/\Dsop\to D_s^{(\ast)}\gamma$ and comparing to theoretical 
predictions. The predicted rates, however, are below the sensitivity 
of current experiments. 
Another model successful in explaining the total widths and the
\Dsna-\Dsop\ mass splitting is based on the assumption that these 
states are chiral partners of the ground states \Ds\ and~\Dsa\citep{theory5}. 
While some measured branching fraction ratios agree with predicted
values, further experimental tests with better sensitivity are 
needed to confirm or refute this scenario.
 
In addition to the \Dsna\ and \Dsop\ states, other excited $D_s$ 
states may have been observed. SELEX has reported a \Dstsi\ candidate,
but this has not been confirmed by other experiments. Recently, 
Belle and BaBar have observed \Dsts\ and \Dste\ states, 
which may be radial excitations of the \Dsa\ and \Dsna, 
respectively. However, the \Dste\ has been searched for in
$B$ decays and not observed, which may indicate that this 
state has higher spin.

\begin{table}
\caption{\label{tab1} Recent results for properties of $D^{\ast\ast}$
  mesons.}
\vspace*{6pt}
\footnotesize
\hskip-0.25in

\end{table}

\begin{table}[h]
\caption{\label{tab2} Recent results for masses and branching
fractions of excited $D_s$ mesons.}
\vspace*{6pt}
\footnotesize
\hskip-0.35in
%
\end{table}

\begin{table}[t]
\caption{\label{tab3} Recent results for quantum numbers of 
excited $D_s$ mesons.}
\footnotesize
\begin{center}
%
\end{center}
\end{table}

\clearpage
\subsection{Semileptonic Decays}

\subsubsection{Introduction}

Semileptonic decays of $D$ mesons involve the interaction of a leptonic
current with a hadronic current. The latter is nonperturbative
and cannot be calculated from first principles; thus it is usually
parameterized in terms of form factors. The transition matrix element 
is written
\begin{eqnarray}
  {\cal M} & = & -i\,{G_F\over\sqrt{2}}\,V^{}_{cq}\,L^\mu H_\mu\,,
  \label{Melem}
\end{eqnarray}
where $G_F$ is the Fermi constant and $V^{}_{cq}$ is a CKM matrix element.
The leptonic current $L_\mu$ is evaluated directly from the lepton spinors 
and has a simple structure; this allows one to extract information about 
the form factors (in $H^{}_\mu$) from data on semileptonic decays~\cite{Becher:2005bg}.  
Conversely, because there are no final-state interactions between the
leptonic and hadronic systems, semileptonic decays for which the form 
factors can be calculated allow one to 
determine~$V^{}_{cq}$~\cite{Kobayashi:1973fv}.

\subsubsection{$D\ra P\ell\nu$ Decays}

When the final state hadron is a pseudoscalar, the hadronic 
current is given by
\begin{eqnarray}
H_\mu & = & \left< P(p) | \bar{q}\gamma^\mu c | D(p') \right> \ =\  
f_+(q^2)\left[ (p' + p)^\mu -\frac{M_D^2-m_P^2}{q^2}q^\mu\right] + 
 f_0(q^2)\frac{M_D^2-m_P^2}{q^2}q^\mu\,,
\label{eq:hadronic}
\end{eqnarray}
where $M_D$ and $p'$ are the mass and four momentum of the 
parent $D$ meson, $m_P$ and $p$ are those of the daughter meson, 
$f_+(q^2)$ and $f_0(q^2)$ are form factors, and $q = p' - p$.  
Kinematics require that $f_+(0) = f_0(0)$.
The contraction $q^\mu L_\mu$ results in terms proportional 
to $m^{}_\ell$\cite{Gilman:1989uy}, and thus for $\ell=e,\mu$
the last two terms in Eq.~(\ref{eq:hadronic}) are negligible. 
Thus, only the $f_+(q^2)$ form factor 
is relevant. The differential partial width is
\begin{eqnarray}
\frac{d\Gamma(D \to P\ell\bar\nu_\ell)}{dq^2\, d\cos\theta_\ell} & = & 
   \frac{G_F^2|V_{cq}|^2}{32\pi^3} p^{*\,3}|f_{+}(q^2)|^2\sin\theta^2_\ell\,,
\label{eq:dGamma}
\end{eqnarray}
where ${p^*}$ is the magnitude of the momentum of the final state hadron
in the $D$ rest frame.


The form factor is traditionally parametrized with an explicit pole 
and a sum of effective poles:
\begin{eqnarray}
f_+(q^2) & = & \frac{f(0)}{1-\alpha}
\left(\frac{1}{1- q^2/m^2_{\rm pole}}\right)\ +\ 
\sum_{k=1}^{N}\frac{\rho_k}{1- q^2/(\gamma_k\,m^2_{\rm pole})}\,,
\label{eqn:expansion}
\end{eqnarray}
where $\rho_k$ and $\gamma_k$ are expansion parameters. The parameter
$m_{{\rm pole}}$ is the mass of the lowest-lying $c\bar{q}$ resonance
with the appropriate quantum numbers; this is expected to provide the
largest contribution to the form factor for the $c\ra q$ transition.  
For example, for $D\to\pi$ transitions the dominant resonance is
expected to be $D^*$, and thus $m^{}_{\rm pole}=m^{}_{D^*}$.

\subsubsection{Simple Pole}

Equation~(\ref{eqn:expansion}) can be simplified by neglecting the 
sum over effective poles, leaving only the explicit vector meson pole. 
This approximation is referred to as ``nearest pole dominance'' or 
``vector-meson dominance.''  The resulting parameterization is
\begin{eqnarray}
  f_+(q^2) & = & \frac{f_+(0)}{(1-q^2/m^2_{\rm pole})}\,. 
\label{SimplePole}
\end{eqnarray}
However, values of $m_{{\rm pole}}$ that give a good fit to the data 
do not agree with the expected vector meson masses~\cite{Hill:2006ub}. 
To address this problem, the ``modified pole'' or Becirevic-Kaidalov~(BK) 
parameterization~\cite{Becirevic:1999kt} was introduced.
This parametrization assumes that gluon 
hard-scattering contributions ($\delta$) are near zero, and scaling
violations ($\beta$) are near unity~\cite{Hill:2006ub}:
\begin{eqnarray}
1 + 1\slash \beta - \delta & \equiv & 
\frac{\left(M_D^2 - m_{P}^2\right)}{f_+(0)}\ 
\left.\frac{df_+}{dq^2}\right|_{q^2=0}\ \approx\ 2\,.
\end{eqnarray}
The parameterization takes the form
\begin{eqnarray}
f_+(q^2) & = & \frac{f_+(0)}{(1-q^2/m^2_{\rm pole})}
\left(1-\alpha^{}_{\rm BK}\frac{q^2}{m^2_{\rm pole}}\right)\,.
\end{eqnarray}
To be consistent with $1 + 1\slash \beta - \delta\approx 2$, the 
parameter $\alpha^{}_{\rm BK}$ should be near the value~1.75.

This parameterization has been used by several experiments to 
determine form factor parameters.
Measured values of $m^{}_{\rm pole}$ and $\alpha^{}_{\rm BK}$ are
listed Tables~\ref{kPseudoPole} and~\ref{piPseudoPole} for
$D\to K\ell\nu$ and $D\to\pi\ell\nu$ decays, respectively.
Both tables show $\alpha^{}_{BK}$ to be substantially lower than
the expected value of~$\sim$\,1.75.


\begin{table}[htbp]
\caption{Results for $m_{\rm pole}$ and $\alpha_{\rm BK}$ from various
  experiments for $D^0\to K^-\ell^+\nu$ and $D^+\to K_S\ell^+\nu$
  decays.   The last entry is a lattice QCD prediction.
\label{kPseudoPole}}
\begin{center}

\end{center}
\end{table}

\begin{table}[htbp]
\caption{Results for $m_{\rm pole}$ and
  $\alpha_{\rm BK}$ from various experiments for 
  $D^0\to \pi^-\ell^+\nu$ and $D^+\to \pi^0\ell^+\nu$ decays.  The last 
  entry is a lattice QCD prediction.
\label{piPseudoPole}}
\begin{center}
%
\end{center}
\end{table}



\subsubsection{$z$ Expansion}

Several groups have advocated an alternative series 
expansion around some value $q^2=t_0$ to parameterize 
$f^{}_+$~\cite{Boyd:1994tt,Boyd:1997qw,Arnesen:2005ez,Becher:2005bg}.
This expansion is given in terms of a complex parameter $z$,
which is the analytic continuation of $q^2$ into the
complex plane:
\begin{eqnarray}
z(q^2,t_0) & = & \frac{\sqrt{t_+ - q^2} - \sqrt{t_+ - t_0}}{\sqrt{t_+ - q^2}
	  + \sqrt{t_+ - t_0}}\,, 
\end{eqnarray}
where $t_\pm \equiv (M_D \pm m_h)^2$ and $t_0$ is the (arbitrary) $q^2$ 
value corresponding to $z=0$. The physical region corresponds to $|z|<1$.

The form factor is expressed as
\begin{eqnarray}
f_+(q^2) & = & \frac{1}{P(q^2)\,\phi(q^2,t_0)}\sum_{k=0}^\infty
a_k(t_0)[z(q^2,t_0)]^k\,,
\label{z_expansion}
\end{eqnarray}
where the $P(q^2)$ factor accommodates sub-threshold resonances via
\begin{eqnarray}
P(q^2) & \equiv & 
\begin{cases} 
1 & (D\to \pi) \\
z(q^2,M^2_{D^*_s}) & (D\to K)\,. 
\end{cases}
\end{eqnarray}
The ``outer'' function $\phi(t,t_0)$ can be any analytic function,
but a preferred choice (see, {\it e.g.}
Refs.~\cite{Boyd:1994tt,Boyd:1997qw,Bourrely:1980gp}) obtained
from the Operator Product Expansion (OPE) is
\begin{eqnarray}
\phi(q^2,t_0) & =  & \alpha 
\left(\sqrt{t_+ - q^2}+\sqrt{t_+ - t_0}\right) \times  \nonumber \\
 & & \hskip0.20in \frac{t_+ - q^2}{(t_+ - t_0)^{1/4}}\  
\frac{(\sqrt{t_+ - q^2}\ +\ \sqrt{t_+ - t_-})^{3/2}}
     {(\sqrt{t_+ - q^2}+\sqrt{t_+})^5}\,,
\label{eqn:outer}
\end{eqnarray}
with $\alpha = \sqrt{\pi m_c^2/3}$.
The OPE analysis provides a constraint upon the 
expansion coefficients, $\sum_{k=0}^{N}a_k^2 \leq 1$.
These coefficients receive $1/M$ corrections, and thus
the constraint is only approximate. However, the
expansion is expected to converge rapidly since 
$|z|<0.051\ (0.17)$ for $D\ra K$ ($D\ra\pi$) over 
the entire physical $q^2$ range, and Eq.~(\ref{z_expansion}) 
remains a useful parameterization.

The $z$-expansion formalism has been used by 
BaBar~\cite{Aubert:2006mc} and CLEO-c~\cite{Dobbs:2007sm}.
Their fits used the first three terms of the expansion,
and the results for the ratios $r_1\equiv a_1/a_0$ and $r_2\equiv a_2/a_0$ are 
listed in Table~\ref{piPseudoZ}.  The CLEO~III\cite{Huang:2004fra} results
listed are obtained by refitting their data using the full
covariance matrix. The BaBar correlation coefficient listed is 
obtained by refitting their published branching fraction using 
their published covariance matrix.
These measurements correspond to using the standard 
outer function $\phi(q^2,t_0)$ of Eq.~(\ref{eqn:outer}) and 
$t_0=t_+\left(1-\sqrt{1-t_-/t_+}\right)$. This choice of $t^{}_0$
constrains $|z|$ to be below a maximum value within the physical region.

\begin{table}[htbp]
\caption{Results for $r_1$ and $r_2$ from various experiments, for 
$D\to \pi\/K\ell\nu$. The correlation coefficient listed is 
for the total uncertainties (statistical $\oplus$ systematic) on 
$r^{}_1$ and~$r^{}_2$.}
\label{piPseudoZ}
\begin{center}
\begin{tabular}{cccccc}
\hline
\vspace*{-10pt} & \\
Expt.     & mode &  Ref.                         & $r_1$               & $r_2$               & $\rho$        \\
\hline
 \omit    & \omit         & \omit                & \omit               & \omit               & \omit         \\
 CLEO III & $D^0\to K^+$  & \cite{Huang:2004fra} & $0.2^{+3.6}_{-3.0}$ & $-89^{+104}_{-120}$ & -0.99         \\
 BaBar    & \omit         & \cite{Aubert:2006mc} & $-2.5\pm0.2\pm0.2$  & $0.6\pm6.\pm5.$     & -0.64         \\
 CLEO-c   & \omit         & \cite{Dobbs:2007sm}  & $-2.4\pm0.4\pm0.1$  & $21\pm11\pm2$       & -0.81         \\
 Average  & \omit         &  \omit               & $-2.3\pm0.23$       & $5.9\pm6.3$         & -0.74         \\ 
\hline
 CLEO-c   & $D^+\to K_S$  & \cite{Dobbs:2007sm}  & $-2.8\pm6\pm2$      & $32\pm18\pm4$       & -0.84         \\
 CLEO-c   & $D^0\to\pi^+$ & \cite{Dobbs:2007sm}  & $-2.1\pm7\pm3$      & $-1.2\pm4.8\pm1.7$  & -0.96         \\
 CLEO-c   & $D^+\to\pi^0$ & \cite{Dobbs:2007sm}  & $-0.2\pm1.5\pm4$    & $-9.8\pm9.1\pm2.1$  & -0.97         \\
\vspace*{-10pt} & \\
\hline
\end{tabular}
\end{center}
\end{table}

Table~\ref{piPseudoZ} also lists average values for $r_1$ and $r_2$  
obtained from a simultaneous fit to CLEO~III, BaBar, and CLEO-c 
branching fraction measurements. 
To account for final-state radiation in the BaBar
measurement, we allow a bias shift between the fit parameters for
the BaBar data and those for the other measurements
(a $\chi^2$ penalty is added to the fit for any deviation 
from BaBar's central value).
Table~\ref{piPseudoZ} shows satisfactory agreement between 
the parameters measured for $D^0$ and $D^+$ decays.  


\subsubsection{$D\ra V\ell\nu$ Decays}

When the final state hadron is a vector meson, the decay can proceed through
both vector and axial vector currents, and four form factors are needed.
The hadronic current is $H^{}_\mu = V^{}_\mu + A^{}_\mu$, 
where~\cite{Gilman:1989uy} 
\begin{eqnarray}
V_\mu & = & \left< V(p,\varepsilon) | \bar{q}\gamma^\mu c | D(p') \right> \ =\  
\frac{2V(q^2)}{M_D+m_h} 
\varepsilon_{\mu\nu\rho\sigma}\varepsilon^{*\nu}p^{\prime\rho}p^\sigma \\
 & & \nonumber\\
A_\mu & = & \left< V(p,\varepsilon) | -\bar{q}\gamma^\mu\gamma^5 c | D(p') \right> 
 \ =\  -i\,(M_D+m_h)A_1(q^2)\varepsilon^*_\mu \nonumber \\
 & & \hskip2.10in 
  +\ i \frac{A_2(q^2)}{M_D+m_h}(\varepsilon^*\cdot q)(p' + p)_\mu \nonumber \\
 & & \hskip2.30in 
+\ i\,\frac{2m_h}{q^2}\left(A_3(q^2)-A_0(q^2)\right)[\varepsilon^*\cdot (p' + p)] q_\mu\,.
\end{eqnarray}
In this expression, $m_h$ is the daughter meson mass and
\begin{eqnarray}
A_3(q^2) & = & \frac{M_D + m_h}{2m_h}A_1(q^2)\ -\ \frac{M_D - m_h}{2m_h}A_2(q^2)\,.
\end{eqnarray}
Kinematics require that $A_3(0) = A_0(0)$.
The differential partial width is
\begin{eqnarray}
\frac{d\Gamma(D \to V\ell\bar\nu_\ell)}{dq^2\, d\cos\theta_\ell} & = & 
  \frac{G_F^2\,|V_{cq}|^2}{128\pi^3M_D^2}\,p^*\,q^2 \times \nonumber \\
 & &  
\left[\frac{(1-\cos\theta_\ell)^2}{2}|H_-|^2\ +\  
\frac{(1+\cos\theta_\ell)^2}{2}|H_+|^2\ +\ \sin^2\theta_\ell|H_0|^2\right]\,,
\end{eqnarray}
where $H^{}_\pm$ and $H^{}_0$ are helicity amplitudes given by
\begin{eqnarray}
H_\pm & = & \frac{1}{M_D + m_h}\left[(M_B+m_h)^2A_1(q^2)\ \mp\ 
      2M^{}_D\,p^* V(q^2)\right] \\
 & & \nonumber \\
H_0 & = & \frac{1}{|q|}\frac{M_B^2}{2m_h(M_D + m_h)}\ \times\ \nonumber \\
 & & \hskip0.01in \left[
    \left(1- \frac{m_h^2 - q^2}{M_D^2}\right)(M_D^2 + m_h^2)A_1(q^2) 
    \ -\ 4{p^*}^2 A_2(q^2) \right]\,.
\label{HelDef}
\end{eqnarray}
The left-handed nature of the quark current manifests itself as
$|H_-|>|H_+|$. The differential decay rate for $D\ra V\ell\nu$ 
followed by the vector meson decaying into two pseudoscalars is
\begin{eqnarray}
\frac{d\Gamma(D\ra V\ell\nu, V\ra P_1P_2)}{dq^2 d\cos\theta_V d\cos\theta_\ell d\chi} 
 &  = & \frac{3G_F^2}{2048\pi^4}
       |V_{cq}|^2 \frac{p^*(q^2)q^2}{M_D^2} {\cal B}(V\to P_1P_2)\ \times \nonumber \\ 
 & & \hskip0.10in \big\{ (1 + \cos\theta_\ell)^2 \sin^2\theta_V |H_+(q^2)|^2 \nonumber \\
 & & \hskip0.20in +\ (1 - \cos\theta_\ell)^2 \sin^2\theta_V |H_-(q^2)|^2 \nonumber \\
 & & \hskip0.30in +\ 4\sin^2\theta_\ell\cos^2\theta_V|H_0(q^2)|^2 \nonumber \\
 & & \hskip0.40in +\ 4\sin\theta_\ell (1 + \cos\theta_\ell) 
             \sin\theta_V \cos\theta_V \cos\chi H_+(q^2) H_0(q^2) \nonumber \\
 & & \hskip0.50in -\ 4\sin\theta_\ell (1 - \cos\theta_\ell) 
          \sin\theta_V \cos\theta_V \cos\chi H_-(q^2) H_0(q^2) \nonumber \\
 & & \hskip0.60in -\ 2\sin^2\theta_\ell \sin^2\theta_V 
                \cos 2\chi H_+(q^2) H_-(q^2) \big\}\,,
\label{eq:dGammaVector}
\end{eqnarray}
where the angles $\theta^{}_\ell$, $\theta^{}_V$, and $\chi$ are defined
in Fig.~\ref{DecayAngles}. 

\begin{figure}[htbp]
  \begin{center}
\includegraphics[width=3.50in]{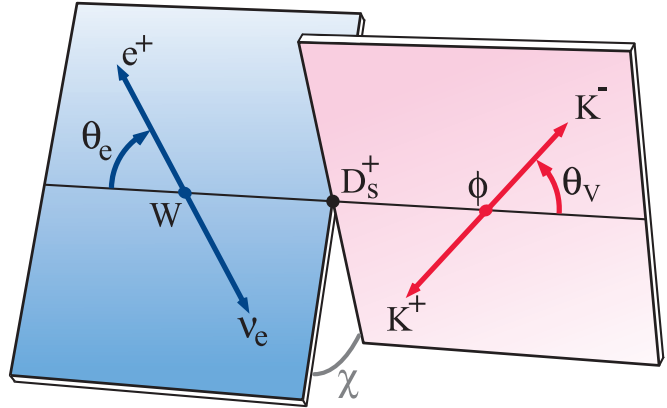}
  \end{center}
  \caption{
    Decay angles $\theta_V$, $\theta_\ell$ 
    and $\chi$. Note that the angle $\chi$ between the decay
    planes is defined in the $D$-meson reference frame, whereas
    the angles $\theta^{}_V$ and $\theta^{}_\ell$ are defined
    in the $V$ meson and $W$ reference frames, respectively.}
  \label{DecayAngles}
\end{figure}

Assuming that the simple pole form of Eq.~(\ref{SimplePole}) describes 
the $q^2$-dependence of the form factors, the 
distribution of Eq.~(\ref{eq:dGammaVector}) will depend only on the parameters
\begin{eqnarray}
r_V \equiv V(0) / A_1(0), & &  r_2 \equiv A_2(0) / A_1(0) \label{rVr2_eq}\,.
\end{eqnarray}
Table \ref{Table1} lists measurements of $r_V$ and $r_2$ from
several experiments. The average results from $D^+\ra\overline{K}^{*0}\ell^+\nu$
decays are also given. The  measurements are plotted in Figs.~\ref{fig:rv} 
and~\ref{fig:r2}, which show that the measurements are consistent
with one another.

\begin{table}[htbp]
\caption{Results for $r_V$ and $r_2$ from various experiments.
\label{Table1}}
\begin{center}
\begin{tabular}{cccc}
\hline
\vspace*{-10pt} & \\
Experiment & Ref. & $r_V$ & $r_2$ \\
\vspace*{-10pt} & \\
\hline
\vspace*{-10pt} & \\
$D^+\to \overline{K}^{*0}l^+\nu$ & \omit & \omit & \omit         \\
E691         & \cite{E691:90}     & 2.0$\pm$  0.6$\pm$  0.3  & 0.0$\pm$  0.5$\pm$  0.2    \\
E653         & \cite{E653:92}     & 2.00$\pm$ 0.33$\pm$ 0.16 & 0.82$\pm$ 0.22$\pm$ 0.11   \\
E687         & \cite{E687:93}     & 1.74$\pm$ 0.27$\pm$ 0.28 & 0.78$\pm$ 0.18$\pm$ 0.11   \\
E791 (e)     & \cite{E791:98a}    & 1.90$\pm$ 0.11$\pm$ 0.09 & 0.71$\pm$ 0.08$\pm$ 0.09   \\
E791 ($\mu$) & \cite{E791:98b}    & 1.84$\pm$0.11$\pm$0.09   & 0.75$\pm$0.08$\pm$0.09     \\
Beatrice     & \cite{BEATRICE:99} & 1.45$\pm$ 0.23$\pm$ 0.07 & 1.00$\pm$ 0.15$\pm$ 0.03   \\
FOCUS        & \cite{FOCUS:02b}   & 1.504$\pm$0.057$\pm$0.039& 0.875$\pm$0.049$\pm$0.064  \\
Average      & \omit              & 1.62$\pm$0.055           & 0.83$\pm$0.054             \\
\hline
$D^0\to \overline{K}^0\pi^-\mu^+\nu$ & \omit & \omit & \omit         \\
FOCUS        & \cite{FOCUS:05}    & 1.706$\pm$0.677$\pm$0.342& 0.912$\pm$0.370$\pm$0.104 \\
\hline
$D_s^+ \to \phi\,e^+ \nu$ &\omit  &\omit     & \omit                  \\
BaBar        & \cite{BaBar:06}    & 1.636$\pm$0.067$\pm$0.038& 0.705$\pm$0.056$\pm$0.029 \\
\hline
$D^0, D^+\to \rho\,e \nu$ & \omit  & \omit    & \omit                 \\
CLEO         & \cite{Mahlke07}    & 1.40$\pm$0.25$\pm$0.03   & 0.57$\pm$0.18$\pm$0.06    \\
\hline
\end{tabular}
\end{center}
\end{table}

\begin{figure}[htbp]
  \begin{center}
\includegraphics[width=3.1in,angle=-90]{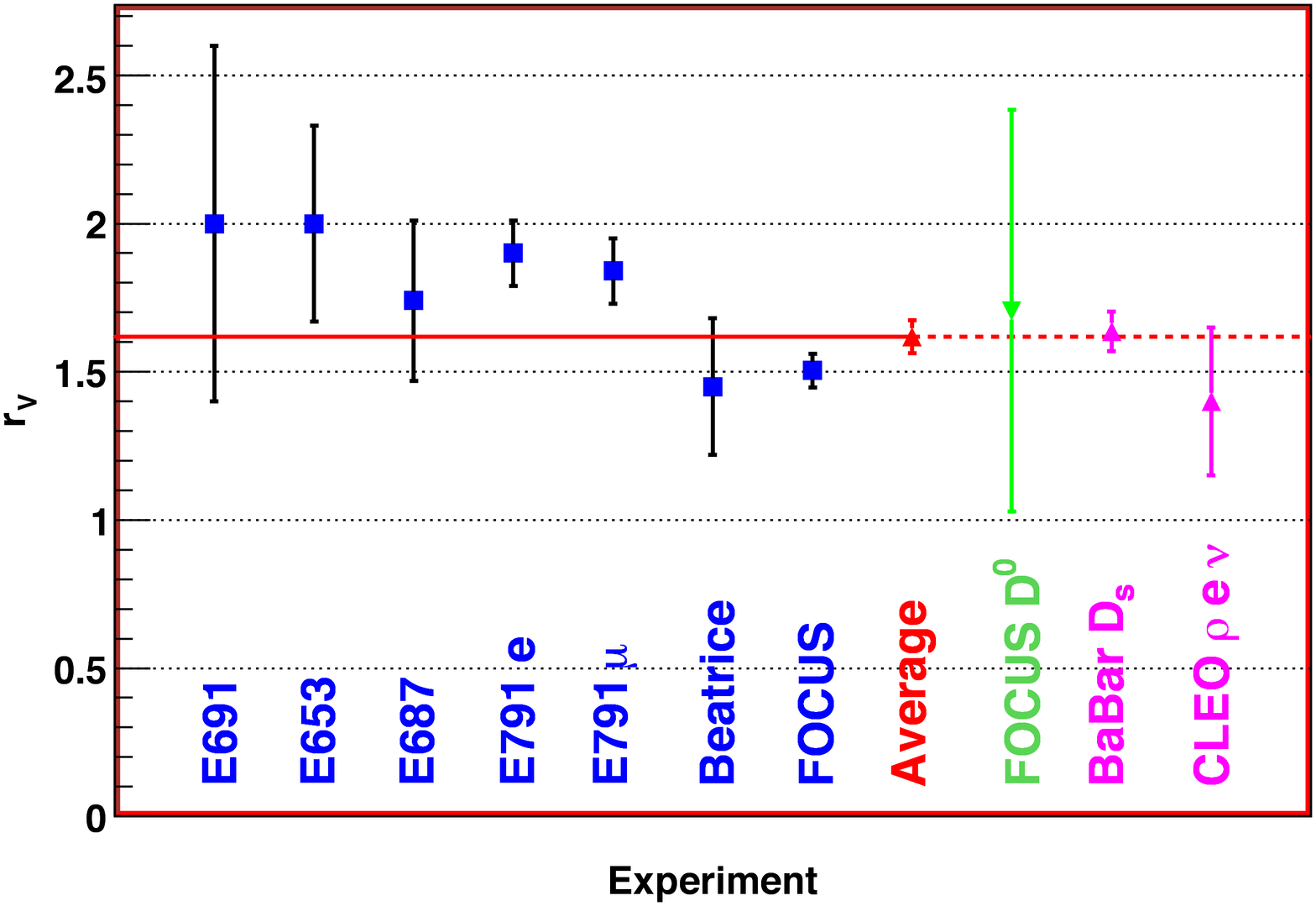}
  \end{center}
\vskip0.10in
  \caption{A comparison of $r_V$ values 
    from various experiments. The first set of measurements are for $D^+
    \to K^-\pi^+ l^+\nu_l$ decays. Plotted next is the average of these
    measurements, followed by measurements in $D^0$ decays, $D_s^+$ decays
    and Cabibbo-suppressed $D$ decays.
  \label{fig:rv}}
\end{figure}

\begin{figure}[htbp]
  \begin{center}
\includegraphics[width=3.1in,angle=-90]{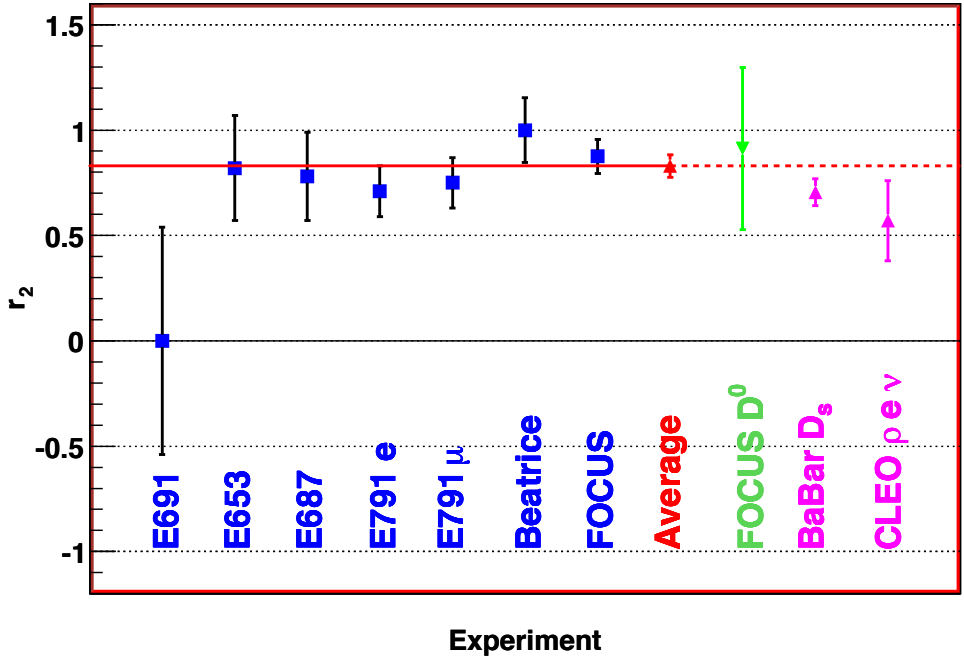}
  \end{center}
\vskip0.10in
  \caption{A comparison of $r_2$ values 
    from various experiments. The first set of measurements are for $D^+
    \to K^-\pi^+ l^+\nu_l$ decays. Plotted next is the average of these
    measurements, followed by measurements in $D^0$ decays, $D_s^+$ decays
    and Cabibbo-suppressed $D$ decays.
  \label{fig:r2}}
\end{figure}

\subsubsection{$S$-Wave Component}

In 2002 FOCUS reported~\cite{FOCUS:02a} an asymmetry in
the observed $\cos(\theta_V)$ distribution. This is interpreted as
evidence for an $S$-wave component in the decay amplitude as follows. 
Since $H_0$ typically dominates over $H_{\pm}$, the distribution given 
by Eq.~(\ref{eq:dGammaVector}) is, after integration over $\chi$,
roughly proportional to $\cos^2\theta_V$. 
Inclusion of a constant $S$-wave amplitude of the form $A\,e^{i\delta}$ 
leads to an interference term proportional to 
$|A H_0 \sin\theta_\ell \cos\theta_V|$; this term causes an asymmetry 
in $\cos(\theta_V)$.
When FOCUS fits their data including this $S$-wave amplitude, 
they obtain $A = 0.330 \pm 0.022 \pm 0.015$ GeV$^{-1}$ and 
$\delta = 0.68 \pm 0.07 \pm 0.05$ \cite{FOCUS:02b}.

\subsubsection{Model-independent Form Factor Measurement}

The CLEO-c collaboration has recently extracted model-independent 
form factors, i.e., $H^{}_+$, $H^{}_-$, and $H^{}_0$ directly as 
functions of $q^2$\cite{Wiss07}. The results are plotted in 
Fig.~\ref{fig:cleoc_mi}. The figure shows that $H^{}_0(q^2)$ 
dominates, especially at low $q^2$. CLEO-c also determined the 
$S$-wave form factor $h^{}_0(q^2)$ via the interference term, 
despite the fact that the $K\pi$ mass distribution 
appears dominated by the vector $K^*(890)$ state. The product
$H^{}_0\times h^{}_0$ is also plotted in Fig.~\ref{fig:cleoc_mi}.

\begin{figure}[htb]
  \begin{center}
\vskip0.20in
  \includegraphics[width=4.75in,angle=-90.]{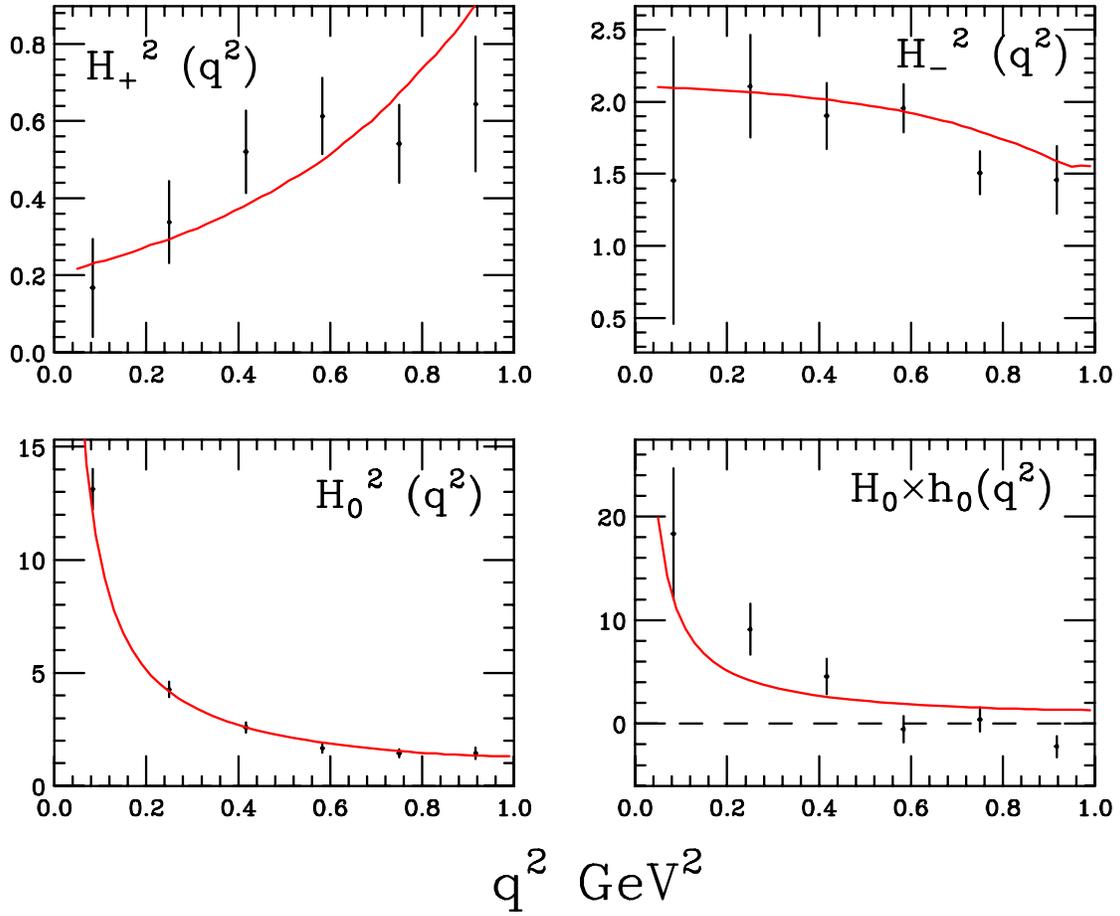}
  \end{center}
\vskip-0.20in
  \caption{Model-independent form factors measured by CLEO-c\cite{Wiss07}.
  \label{fig:cleoc_mi}}
\end{figure}

\clearpage
\subsection{\emph{CP} Asymmetries}

\emph{CP} violation occurs if the decay rate for a particle differs 
from that of its \emph{CP}-conjugate\cite{BigiSanda}. 
In general there are two classes of \emph{CP} violation, termed
{\it indirect\/} and {\it direct\/}\cite{Nir}. Indirect \emph{CP} 
violation refers to $\Delta C\!=\!2$ processes and 
arises in $D^0$ decays due to $D^0$-$\dbar$ mixing. 
It can occur as an asymmetry in the mixing itself, or it can 
result from interference between a decay 
amplitude arising via mixing and a non-mixed amplitude. 
Direct \emph{CP} violation refers to
$\Delta C\!=\!1$ processes and occurs in both charged and neutral 
$D$ decays. It results from interference between two different decay
amplitudes (e.g., a penguin and tree amplitude) that have
different weak (CKM) and strong phases\footnote{The weak 
phase difference will have opposite signs for $D\ra f$ and 
$\overline{D}\ra\bar{f}$ decays, while the strong phase difference 
will have the same sign. As a result, squaring the total amplitudes 
to obtain the decay rates gives interference terms having 
opposite sign, i.e., non-identical decay rates.}.
A difference in strong phases typically arises due to 
final-state interactions (FSI)\cite{Buccella}. A difference
in weak phases arises from different CKM vertex couplings, as 
is often the case for spectator and penguin diagrams.

\vspace{0.8cm}
The \emph{CP} asymmetry is defined as the difference between 
$D$ and $\overline{D}$ partial widths divided by their sum:
\begin{eqnarray}  
A_{CP} & = & \frac{\Gamma(D)-\Gamma(\overline{D})}
{\Gamma(D)+\Gamma(\overline{D})}\,.
\end{eqnarray}
However, to take into account differences in production rates between 
$D$ and $\overline{D}$ (which would affect the number of respective 
decays observed), experiments usually normalize to a Cabibbo-favored 
mode. In this case there is the additional benefit that most corrections 
due to inefficiencies cancel out, reducing systematic uncertainties. An 
implicit assumption is that there is no measurable \emph{CP} 
violation in the Cabibbo-favored normalizing mode. 
The \emph{CP} asymmetry is calculated as
\begin{eqnarray}
A_{CP} & = & \frac{\eta(D)-\eta(\overline{D})}{\eta(D)+\eta(\overline{D})}\,,
\end{eqnarray}
where (considering, for example, $D^0 \to K^-K^+$)
\begin{eqnarray}
 \eta(D) & = & \frac{N(D^0 \rightarrow K^-K^+)}{N(D^0 \rightarrow K^-\pi^+)}\,, \\
 \eta(\overline{D}) & = & \frac{N(\dbar\rightarrow K^-K^+)}
{N(\dbar\rightarrow K^+\pi^-)}\,.
\end{eqnarray}
In the case of $D^+$ and $D^+_s$ decays, $A^{}_{CP}$ measures 
direct \emph{CP} violation; in the case of $D^0$ decays, $A^{}_{CP}$ 
measures direct and indirect \emph{CP} violation combined.
Values of $A^{}_{CP}$ for $D^+$ and $D^0$ decays are listed in
Tables~\ref{tab:cp_charged} and \ref{tab:cp_neutral}, respectively.

\vspace{0.8cm}
The \emph{CP} lifetime asymmetry is the parameter $A_\Gamma$ discussed in 
the $D^0$-$\dbar$ mixing section:
\begin{eqnarray}
A_\Gamma & = & \frac{\tau(\dbar)-\tau(D^0)}{\tau(\dbar)+\tau(D^0)}\,.
\end{eqnarray}
It is analogous to $A^{}_{CP}$ except that the asymmetry
pertains to the {\it full\/} width rather than a partial width,
and it is calculated in terms of the reciprocal of the widths. 
The asymmetry $A^{}_\Gamma$ is measured by fitting
decay time distributions, and thus it is insensitive to a
production asymmetry between $D$ and $\overline{D}$.
Values of $A^{}_\Gamma$ for some $D^0$ decays are 
listed in Table~\ref{tab:A_gamma}.

\begin{table}
\renewcommand{\arraystretch}{1.4}
\caption{\cp\ asymmetries 
$A^{}_{CP}= [\Gamma(D^+)-\Gamma(D^-)]/[\Gamma(D^+)+\Gamma(D^-)]$
for $D^\pm$ decays.
\label{tab:cp_charged}}
\footnotesize
\begin{center}

\end{center} 
\end{table}

\begin{table}
\renewcommand{\arraystretch}{1.3}
\caption{\cp\ asymmetries 
$A^{}_{CP}=[\Gamma(D^0)-\Gamma(\dbar)]/[\Gamma(D^0)+\Gamma(\dbar)]$
for $D^0,\dbar$ decays.
\label{tab:cp_neutral}}
\footnotesize
\begin{center}
%
 \end{center} 
\end{table}

\begin{table}
\renewcommand{\arraystretch}{1.4}
\caption{Lifetime asymmetries  
$A^{}_\Gamma=[\tau(\dbar)-\tau(D^0)]/[\tau(\dbar)+\tau(D^0)]$
for $D^0,\dbar$ decays.
\label{tab:A_gamma}}
\footnotesize
\begin{center}
%
\end{center}
\end{table}

\subsection{\emph{$T$}-violating Asymmetries}
                                               
$T$-violating asymmetries are measured using triple-product
correlations and assuming the validity of the $CPT$ theorem.
Triple-product correlations of the form 
$\vec{a}\cdot(\vec{b}\times\vec{c})$, 
where $a$, $b$, and $c$ are spins or momenta, are odd 
under time reversal~(\emph{T}).
For example, for $D^0 \to K^+K^-\pi^+\pi^-$ decays, 
$C_T \equiv \vec{p}_{K^+}\cdot(\vec{p}_{\pi^+}\times \vec{p}_{\pi^-})$  
is odd under a \emph{T} transformation.
The corresponding quantity for $\dbar$ is
$\overline{C}_T \equiv 
      \vec{p}_{K^-}\cdot(\vec{p}_{\pi^-}\times \vec{p}_{\pi^+})$.
Defining  
\begin{equation}
 A_{T}  = \frac{\Gamma(C_T>0)-\Gamma(C_T<0)}{\Gamma(C_T>0)+\Gamma(C_T<0)}
\end{equation}
for $D^0$ decay and
\begin{equation}
\overline{A}_{T} = \frac{\Gamma(-\overline{C}_T>0)-\Gamma(-\overline{C}_T<0)}
                        {\Gamma(-\overline{C}_T>0)+\Gamma(-\overline{C}_T<0)}
\end{equation} 
for $\dbar$ decay, in the absence of strong phases
either $A^{}_T\neq 0$ or $\overline{A}^{}_T\neq 0$ indicates
$T$ violation. The asymmetry
\begin{eqnarray}
A^{}_{T\,{\rm viol}} & = & \frac{A_{T}-\overline{A}_{T}}{2}
\end{eqnarray}
tests for $T$ violation even with nonzero strong phases 
(see Refs.~\cite{Valencia,Bigi,Bensalem,Bensalem2}).
Values of $A_{T\,{\rm viol}}$ for some $D^+$, $D^+_s$, and
$D^0$ decay modes are listed in Table~\ref{tab:t_viol}.

\begin{table}[h]
\renewcommand{\arraystretch}{1.4}
\caption{$T$-violating asymmetries 
$A^{}_{T\,{\rm viol}} = (A_{T}-\overline{A}_{T})/2$.
\label{tab:t_viol}}
\footnotesize
\begin{center}
\begin{tabular}{|l|c|c|c|} 
\hline
{\bf Mode} & {\bf Year} & {\bf Collaboration} & {\boldmath $A^{}_{T\,{\rm viol}}$} \\
\hline
{\boldmath $D^0 \to K^+K^-\pi^+\pi^-$} &
  2005 & FOCUS~\cite{FOCUS2005}  &  $ +0.010 \pm 0.057 \pm 0.037 $ \\
\hline
{\boldmath $D^+ \to K^0_sK^+\pi^+\pi^-$} &
  2005 & FOCUS~\cite{FOCUS2005}  &  $ +0.023 \pm  0.062  \pm 0.022  $ \\
\hline
{\boldmath $D^+_s \to K^0_sK^+\pi^+\pi^-$} &
  2005 & FOCUS~\cite{FOCUS2005}  &  $ -0.036  \pm 0.067  \pm 0.023  $ \\
\hline                    
\end{tabular}
\end{center} 
\end{table}

\vskip0.30in
\begin{center}  ---------------  \end{center}
\vskip0.30in
In summary, Tables~\ref{tab:cp_charged}--\ref{tab:t_viol} show that
there is no evidence yet for \emph{CP} or $T$ violation in the charm 
sector. The most sensitive searches for \emph{CP} violation have 
reached a level of sensitivity well below~1\%.

\clearpage
\section{Summary}
\label{sec:summary}

This article provides updated world averages for 
$b$-hadron properties as of the end of 2007. Some results
that appeared at the beginning of 2008 are also included.
A small selection of highlights of the results described in Sections
\ref{sec:life_mix}-\ref{sec:charm_physics} is given in 
Table~\ref{tab_summary}.

\begin{table}
\caption{ Brief summary of the world averages at the end of 2007.}
\label{tab_summary}
\renewcommand{\arraystretch}{1.15}
\begin{center}

\end{center}
\end{table}

Concerning lifetime and mixing averages, the most significant changes
since the end of 2006~\cite{hfag_hepex_endof2006}
are due to new measurements from the Tevatron experiments, 
mainly in the areas of heavy \B meson (\Bs and \Bc) lifetimes 
and \Bs mixing parameters.
After the direct observation of \Bs oscillations 
by CDF in 2006, \dzero has now obtained an independent 
measurement of this phenomenon. Taking advantage of their 
increased data sample sizes, both experiments also 
performed new analyses of 
$\Bs \to J/\psi \phi$ untagged and tagged 
decays, leading to improvements in the 
measurement of the decay width difference in the 
\Bs system. Most notably, these analyses investigate 
mixing-induced \CP violation in
$\Bs \to J/\psi \phi$ decays and determine
the corresponding weak phase, 
although with large uncertainties and with
a two-fold ambiguity. The CDF and \dzero\ results
are consistent, and the combined result
differs from the Standard Model expectation by $2.2\,\sigma$.
We also present for the first time a combination of these 
results with searches for \CP violation in \Bs mixing,
under the hypothesis of a significant New Physics 
phase.

Measurements by \babar\ and \belle\ of the 
time-dependent \CP\ violation parameter
$S_{b \to c\bar c s}$ in \B decays to charmonium and a neutral kaon
have established \CP\ violation in \B decays, and they 
allow a precise extraction of
the Unitarity Triangle parameter $\stwob \equiv \sin\! 2\phi_1$.
Recent studies of $\B \to \jpsi \Kstar$ (Sec.~\ref{sec:cp_uta:ccs:vv}),
$\B \to D^{(*)}h^0$ where $h^0 = \pi^0$, \etc\ (Sec.~\ref{sec:cp_uta:cud_beta}),
and $\B \to D^{*+}D^{*-}\KS$ (Sec.~\ref{sec:cp_uta:ccs:DstarDstarKs})
allow one to resolve the ambiguity for $\beta \equiv \phi_1$ from
the measurement of $\stwob \equiv \sin\! 2\phi_1$.
Measurements of
time-dependent \CP asymmetries in hadronic $b \to s$ penguin decays
continue to provide insight into possible new physics.
In this area, results from both \babar\ and \belle\ have been updated.
A particularly notable change is that the \CP violation effect in 
$\B \to \eta^\prime K^0$ is now established 
with more than $5\sigma$ significance in both experiments.
First results from time-dependent Dalitz plot analyses of both 
$B \to K^+K^-K^0$ and $B \to \pi^+\pi^-K^0$ decays are now available.
Compared to the previous round of averages,
the consistency with the Standard Model expectation remains at about 
the same level in terms of significance.
Results from time-dependent analyses
of the decays $\Bz \to \pi^+\pi^-\!\!, \,\rho^\pm\pi^\mp$ and 
$\rho^+\rho^-$ provide constraints on the 
Unitarity Triangle angle $\alpha \equiv \phi_2$ (Sec.~\ref{sec:cp_uta:uud}).
Both \babar\ and \belle\ have now observed \CP violation in 
$\Bz \to \pi^+\pi^-$ with more than $5\sigma$ significance,
and both experiments have performed time-dependent Dalitz plot analyses
of $\Bz \to (\rho\pi)^0 \to \pi^+\pi^-\pi^0$.
Progress continues to constrain the third Unitarity Triangle angle 
$\gamma \equiv \phi_3$. Both \babar\ and \belle\ are using 
$\Bm \to \DorDstar \Km$ decays (Sec.~\ref{sec:cp_uta:cus}), with 
$\DorDstar$ decays to final states accessible from
both $D^{(*)0}$ and $\bar{D}{}^{(*)0}$.
At present, the most constraining results arise from the Dalitz 
plot analysis of the $D \to \KS\pi^+\pi^-$ channel.


Progress in the determination of properties of semileptonic
$B$ decays has been steady over the last year. To match the 
increasing number of results coming from \babar\ and Belle,
several new averages have been added for decays with a $D^{(*)}$
in the final state.
Regarding inclusive $B \to X_c\ell^+\nul$ decays, updated 
values of the $b$--quark mass from the kinetic and 1S schemes
were obtained.
Regarding charmless semileptonic decays, two new calculations
(GGOU and ADFR) presented in 2007 were used (together with 
the already existing calculations BLNP, DGE and BLL) 
to extract the value of $|V_{ub}|$. All this information
contributes to the current effort by theorists and experimentalists
to understand the current differences between values of $|V_{ub}|$   
extracted from inclusive and exclusive decays. Finally, the 
$\B\to\pi\ell^+\nul$ branching fraction was updated with published
(instead of preliminary) results, and for the first time the 
$\B\to\rho\ell^+\nul$ average was added.


For rare \B\ decays, branching fractions and charge asymmetries of new 
decay modes continue to be measured, mostly by \babar\ and Belle.
There are several hundred measurements in the tables in \Sec{rare}.
Particularly noteworthy is the measurement of $B^\pm\to\tau^\pm\nu$; 
Belle sees evidence for this decay, while \babar\ reports a somewhat 
smaller branching fraction.  


In the sector of $B$ decays to charmed particles, reductions 
in uncertainties and new measurements continue to be made. 
Branching fractions for rare $B$-meson decays or decay chains
of a few $10^{-7}$ are being measured with statistical uncertainties
typically below $30\%$. Results for more common decay chains, with
branching fractions around $10^{-4}$, are becoming precision
measurements, with uncertainties typically at the $3\%$ level. 
Branching fractions for several $B$ decays to
$D_{sJ}^{*-}(2317)$ and $D_{sJ}^-(2460)$ have been measured.


Mixing in the $D^0$-$\dbar$ system was finally observed last year.
Three experiments
--\,Belle\cite{belle_kk}, Babar\cite{babar_kpi}, and CDF\cite{cdf_kpi}\,--
now observe evidence for this effect. The measurements are combined with 
others to yield World Average (WA) values for mixing parameters $x$ and $y$, 
and for \cpv\ parameters $|q/p|$ and $\phi$. From this fit, the no-mixing 
point $x\!=\!y\!=\!0$ is excluded at $9.2\sigma$. The parameter $x$ differs 
from zero by $3.0\sigma$, and $y$ differs from zero by $4.0\sigma$. This 
mixing is presumably dominated by long-distance processes, which are 
difficult to calculate. Thus, it may be difficult to identify new 
physics from mixing alone. The WA value for the observable \ycp\ 
is positive, which indicates that the \cp-even state is 
shorter-lived as in the $K^0$-$\kbar$ system. However, $x$ also 
appears to be positive, which implies that the \cp-even state is 
heavier; this is unlike in the $K^0$-$\kbar$ system. It appears 
difficult to accomodate a strong phase difference $\delta$ between 
amplitudes $A(D^0\ra K^+\pi^-)$ and $A(\dbar\ra K^+\pi^-)$
larger than $45^\circ$, and there is no evidence yet for 
\cpv\ (either direct or indirect) in the $D^0$-$\dbar$ system.

\section{Acknowledgments}

We are grateful to the \babar, \belle, CDF, CLEO, \dzero,
LEP, and SLD collaborations, who have provided experimental results 
on \b-hadron and \c-hadron properties and have given assistance
to HFAG when needed for averaging.
These results would not be possible without the excellent 
operations of the PEP-II, KEKB, CESR, Tevatron, LEP, and SLC
accelerators, and fruitful collaborations between the accelerator 
groups and the experimental collaborations.



\begin{thebibliography}{999}
\bibitem{CDFBS2}
\coll{CDF} CDF note 9203, 15 February 2008, submitted to the Winter 2008 conferences; 
we consider that these new results supersede the ones from
the 3-year old CDF note 7386~\cite{CDFB5}, 
although the lifetime analysis of one of the modes ($B_s \to D_s\pi\pi\pi$) has not been updated.
\end{thebibliography}
\end{document}